\documentclass[10pt,a4paper]{book}
\usepackage{epsfig}
\usepackage[T1]{fontenc}    % Accents cods dans la fonte.
\usepackage{graphics}
\usepackage{graphicx}
\usepackage[usenames,dvipsnames]{color}
\usepackage{pstricks,pst-coil,pst-fill,pst-plot}
\usepackage[fleqn]{amsmath}    % Les symboles les plus frquents.
\usepackage{amssymb}    % Des symboles.
\usepackage{amsfonts}   % Des fontes, eg pour \mathbb.
\usepackage{verbatim}   % Pour les codes sources en informatique.
\usepackage{mathrsfs}   % Des lettres majuscules cursives (\mathscr).
\usepackage{dsfont}
\usepackage{euscript}
\usepackage{yfonts}
\usepackage{enumerate}     % met en place l'environement en umerate pour les listes
\usepackage{txfonts}
\usepackage{marvosym}
\usepackage{vmargin}        % Rgler la taille de la feuille.
\usepackage{wasysym}		% Selected Hebrew Letters
\usepackage{appendix}

\setmarginsrb{1.8cm}{2cm}{1.8cm}{2cm}{1cm}{1cm}{1cm}{1.6cm}
 \makeatletter
 \@addtoreset{equation}{section}
 \makeatother

\providecommand{\bysame}{\leavevmode\hbox to3em{\hrulefill}\thinspace}
\providecommand{\MR}{\relax\ifhmode\unskip\space\fi MR }
\providecommand{\MRhref}[2]{%
  \href{http://www.ams.org/mathscinet-getitem?mr=#1}{#2}
}
\providecommand{\href}[2]{#2}

\let\ua=\uparrow
\let\da=\downarrow
\let\tend=\rightarrow

\long\def\symbolfootnote[#1]#2{\begingroup%
\def\thefootnote{\fnsymbol{footnote}}\footnote[#1]{#2}\endgroup}

\newtheorem{theorem}{Theorem}[section]
\newtheorem{prop}[theorem]{Proposition}
\newtheorem{cor}[theorem]{Corollary}
\newtheorem{defin}[theorem]{Definition}

\newtheorem{lemme}[theorem]{Lemma}

\def\Proof{\medskip\noindent {\it Proof --- \ }}

\def\qed{\hfill\rule{2mm}{2mm}}

\newcommand\beq{\begin{equation}}
\newcommand\enq{\end{equation}}
\newcommand\bem{\begin{multline}}
\newcommand\enm{\end{multline}}

\def\beqa{\begin{eqnarray}}
\def\eeqa{\end{eqnarray}}
\def\ba{\begin{array}}
\def\ea{\end{array}}
\def\det{\operatorname{det}}

\newcommand{\f}[2]{{\ensuremath{%
    \mathchoice%
    {\dfrac{#1}{#2}}
    {\dfrac{#1}{#2}}
    {\frac{#1}{#2}}
    {\frac{#1}{#2}}
}}}
\newcommand{\tf}[2]{\ensuremath{#1/#2}}
\newcommand{\pa}[1]{\ensuremath{\left(#1\right)}}
\newcommand{\paa}[1]{\ensuremath{\left\{#1\right\}}}
\newcommand{\pac}[1]{\ensuremath{\left[#1\right]}}
\def\a{\alpha}

\def\be{\beta}
\def\ga{\gamma}
\def\Ga{\Gamma}

\def\de{\delta}

\def\De{\Delta}
\def\eps{\epsilon}
\def\veps{\varepsilon}
\def\la{\lambda}
\def\La{\Lambda}
\def\vrhp{\varrhoup}
\def\sg{\sigma}
\def\vsg{\varsigma}
\def\Sg{\Sigma}

\def\Ups{\Upsilon}
\def\ups{\upsilon}
\def\th{\theta}

\def\vth{\vartheta}
\def\vrp{\varpi}
\def\Om{\Omega}
\def\om{\omega}
\def\vp{\varphi}
\DeclareMathOperator{\cotanh}{cotanh}

\newcommand{\mc}[1]{\ensuremath{\mathcal{#1}}}
\newcommand{\mf}[1]{\ensuremath{\mathfrak{#1}}}
\newcommand{\msc}[1]{\ensuremath{\mathscr{#1}}}

\newcommand{\bs}[1]{\ensuremath{\boldsymbol{#1}}}
\newcommand{\mbb}[1]{\ensuremath{\mathbb{#1}}}

\DeclareFontFamily{OT1}{pzc}{}
\DeclareFontShape{OT1}{pzc}{m}{it}{<-> s * [1.10] pzcmi7t}{}
\DeclareMathAlphabet{\mathpzc}{OT1}{pzc}{m}{it}
\newcommand{\mpzc}[1]{\ensuremath{\mathpzc{#1}}}

\newcommand{\op}[1]{ \boldsymbol{ \texttt{#1} } }

\def \i{ \mathrm i}

\newcommand{\ov}[1]{\ensuremath{\overline{#1}}}
\newcommand{\wt}[1]{\ensuremath{\widetilde{#1}}}
\newcommand{\wh}[1]{\ensuremath{\widehat{#1}}}

\newcommand{\Int}[2]{\ensuremath{\int\limits_{#1}^{#2}}}
\newcommand{\Oint}[2]{\ensuremath{\oint\limits_{#1}^{#2}}}
\newcommand{\Fint}[2]{\ensuremath{\fint\limits_{#1}^{#2}}}

\newcommand{\sul}[2]{\ensuremath{\sum\limits_{#1}^{#2}}}
\newcommand{\pl}[2]{\ensuremath{\prod\limits_{#1}^{#2}}}

\newcommand{\R}{\ensuremath{\mathbb{R}}}
\newcommand{\Cx}{\ensuremath{\mathbb{C}}}

\newcommand{\Dp}[1]{\ensuremath{\partial_{#1}}}
\newcommand{\limit}[2]{\ensuremath{\underset{#1 \tend #2}{\longrightarrow} }}
\newcommand{\s}[1]{\ensuremath{\sinh\pa{#1}}}

\newcommand{\ex}[1]{\ensuremath{\e{e}^{#1}}}

\newcommand{\ddet}[2]{\ensuremath{\det_{#1}\pac{#2}}}

\newcommand{\abs}[1]{\ensuremath{\left| #1 \right|}}
\newcommand{\Norm}[1]{\ensuremath{\abs{\abs{#1}} }}

\newcommand{\norm}[1]{\ensuremath{|| #1 ||}}

\newcommand{\moy}[1]{\ensuremath{\langle #1 \rangle}}

\newcommand{\dd}{\mathrm{d}}
\newcommand{\e}[1]{\ensuremath{\mathrm{#1}}}

\newcommand{\intff}[2]{\ensuremath{ [  #1 \,; #2 ] }}
\newcommand{\intfo}[2]{\ensuremath{ [  #1 \,; #2 [ }}
\newcommand{\intof}[2]{\ensuremath{ ]  #1 \,; #2 ] }}
\newcommand{\intoo}[2]{\ensuremath{ ]  #1 \,; #2 [ }}

\newcommand{\intn}[2]{\ensuremath{[\![ \, #1 \,;\, #2 \,]\!]}}

\begin{document}

%\begin{flushright}

%\end{flushright}
%\par \vskip .1in \noindent

%\vspace{14pt}

\begin{center}
\begin{LARGE}
{\bf Asymptotic expansion of a partition function related to the sinh-model}
\end{LARGE}

\vspace{1cm}

{\large Ga\"etan Borot \footnote{gborot@mpim-bonn.mpg.de}}
\\[1ex]
Max Planck Institut f\"ur Mathematik, Vivatsgasse 7, 53111 Bonn, Germany.\\
Department of Mathematics, MIT, 77 Massachusetts Avenue, Cambridge, MA 02139-4307, USA.

\vspace{4mm}

{\large Alice Guionnet \footnote{guionnet@math.mit.edu}}
\\[1ex]
Department of Mathematics, MIT, 77 Massachusetts Avenue, Cambridge, MA 02139-4307, USA.

\vspace{4mm}
{\large Karol K. Kozlowski \footnote{karol.kozlowski@ens-lyon.fr}}
\\[1ex]
Laboratoire de physique, UMR 5672 du CNRS,
ENS de Lyon, Lyon, France. \\[2.5ex]

\par

\vspace{40pt}

\centerline{\bf Abstract} \vspace{1cm}
\parbox{12cm}{\small This paper develops a method to carry out the large-$N$ asymptotic analysis of a class of
$N$-dimensional integrals arising in the context of the so-called quantum separation of variables method. We push further ideas developed 
in the context of random matrices of size $N$, but in the present problem, two scales $1/N^{\a}$ and $1/N$ naturally occur. In our case, 
the equilibrium measure is $N^{\a}$-dependent and characterised by means of the solution to a $2\times 2$ Riemann--Hilbert problem, whose large-$N$ behaviour is analysed in detail.
Combining these results with techniques of concentration of measures and an asymptotic analysis of the Schwinger-Dyson equations at the distributional level, 
we obtain the large-$N$ behaviour of the free energy explicitly up to $o(1)$. The use of distributional Schwinger-Dyson is a novelty that allows us treating sufficiently differentiable 
interactions and the mixing of scales $1/N^{\a}$ and $1/N$, thus waiving the analyticity assumptions often used in random matrix theory.}

\end{center}

\newpage

\section*{An opening discussion}

The present work develops techniques enabling one to carry out the large-$N$ asymptotic
analysis of a class of multiple integrals that arise as representations for the correlation functions 
in quantum integrable models solvable by the quantum separation of variables. 
We shall refer to the general class of such integrals as the sinh model:
\beq
\mf{z}_N[W] \; = \; \Int{\R^N}{} \pl{a<b}{N} \Big\{ \sinh[\pi\om_1(y_a-y_b)]  \sinh[\pi\om_2(y_a-y_b)] \Big\}^{\be} \cdot \pl{a=1}{N}\ex{-W(y_a)}\cdot \dd^N \bs{y} \;. 
\nonumber
\enq
When $\be=1$ and for specific choices of the constants $\om_1,\om_2>0$ and of the confining potential $W$, $\mf{z}_N$
represents norms or arises as a fundamental building block of certain classes of correlation functions in quantum
integrable models that are solvable by the quantum separation of variable method. 
This method takes its roots in the works of Gutzwiller \cite{GutzwillerResolutionTodaChainSmallNPaper1,GutzwillerResolutionTodaChainSmallNPaper2} 
on the quantum Toda chain and has been developed in the mid '80s by Sklyanin \cite{SklyaninSoVFirstIntroTodaChain,SklyaninSoVGeneralOverviewFuncBA}
as a way of circumventing certain limitations inherent to the algebraic Bethe Ansatz. 
Expressions for the norms or correlation functions for various  models solvable by the quantum separation of variables method have been established, \textit{e.g.} in 
the works \cite{BabelonQuantumInverseProblemConjClosedToda,DerkachovKorchemskyManashovXXXSoVandQopNewConstEigenfctsBOp,DerkachovKorchemskyManashovXXXreelSoVandQopABAcomparaison,
GrosjeanMailletNiccoliFFofLatticeSineG,KozUnitarityofSoVTransform,KozIPForTodaAndDualEqns,SklyaninResolutionIPFromQDet,WallachRealReductiveGroupsII}.
The expressions obtained there are either directly of the form given above
or are amenable to this form (with, possibly, a change of the integration contour from $\R^N$ to $\msc{C}^N$, with $\msc{C}$ a curve in $\Cx$) upon elementary manipulations. 
Furthermore, a degeneration of $\mf{z}_N[W]$ arises as a multiple integral representation for the partition function of the 
six-vertex model subject to domain wall boundary conditions \cite{KazamaKomatsuNishimuraSoVLikeMIRepForPartFctRat6Vertex}. 
In the context of quantum integrable systems, the number $N$ of integrals defining $\mf{z}_N$ is related to the number of sites in a model 
(as, \textit{e.g.} in the case of the compact or non-compact XXZ chains or the lattice regularisations of the Sinh or Sine-Gordon models)
or to the number of particles (as, \textit{e.g.}  in the case of the quantum Toda chain). From the point of view of applications, one is mainly
interested in the thermodynamic limit of the model, which is attained by sending $N$ to $+\infty$. For instance, in the case of an integrable 
lattice discretisation of some quantum field theory, one obtains in this way an exact and non-perturbative description 
of a quantum field theory in $1+1$ dimensions and in finite volume. This limit, at the level of $\mf{z}_N[W]$, translates itself in the need to extract the large 
$N$-asymptotic expansion of  $\ln \mf{z}_N[W]$ up to $\e{o}(1)$. It is, in fact, the constant term in the expansion of $\ln \big(\mf{z}_N[W^{\prime}] / \mf{z}_N[W] \big)$
with $W^{\prime}$ some deformation of $W$ that gives rise to the correlation functions of the underlying quantum field theory in finite volume. 
These applications to physics  constitute the first motivation for our analysis. From the purely mathematical side, the motivation of our works
stems from the desire to understand better the structure of the large-$N$ asymptotic expansion of multiple integrals 
whose analysis demands to surpass the scheme developed to deal with $\be$-ensembles.

As we shall argue in \S~\ref{baby}, it is possible to understand the large-$N$ asymptotic analysis of the multiple integral $\mf{z}_N[W]$ from the one of the re-scaled multiple integral 
\beq
Z_N[V_N] \; = \; \Int{\R^N}{} \pl{a<b}{N} \Big\{ \sinh[\pi\om_1 T_N (\la_a-\la_b)]  \sinh[\pi\om_2 T_N (\la_a-\la_b)] \Big\}^{\be} \cdot 
\pl{a=1}{N}\ex{-N T_N V_N(\la_a)} \cdot \dd^N \bs{\la} \;. 
\nonumber
\enq
There $T_N$ is a sequence going to infinity with $N$ whose form is fixed by the behaviour of $W(x)$ at large $x$, and:
$V_N( \xi) \; = \; T_N^{-1} \cdot W(T_N \xi)$. 

{\bf The main task of the book is to develop an effective method of asymptotic analysis of
the rescaled multiple integral $\mc{Z}_N[V]$ in the case when $T_N=N^{\a}$, $0<\a<\tf{1}{6}$ and $V$ is a given $N$-independent strictly convex
smooth function satisfying to a few additional technical hypothesis. }

The treatment of the class of $N$-dependent potentials $V_N$ 
which would enable one to deduce the large-$N$ asymptotic expansion of $\mf{z}_N[W]$ will be the matter of a future work.

Prior to discussing in more details the results obtained in this work, 
we would like to provide an overview of the developments that took place, over the years, in the field
of large-$N$ asymptotic analysis of $N$-fold multiple integrals, as well as some motivations underlying the study of these integrals in a more general context than the focus of this book.
This discussion serves as an introduction to various ideas that appeared fruitful in such an asymptotic analysis, that we place in a more general context than the focus of this article. More importantly, it will put these 
techniques in contrast with what happens in the case of the sinh model under study. In particular, we will point out the technical aspects which complicate the large-$N$ asymptotic analysis of $\mf{z}_N[W]$ and thus highlight 
the features and  techniques that are new in our analysis. Finally, such an organisation will permit us to emphasise the main differences occurring in the structure
of the large $N$-asymptotic expansion of integrals related to the sinh-model as compared to the $\be$-ensemble like multiple integrals.

The book is organised as follows. Chapter \ref{Section Introduction} is the introduction where we  give an overview of the various methods used and results obtained
with respect to extracting the large number of integration asymptotics of integrals occurring in random matrix theory. 
Since we heavily rely on tools from potential theory, large deviations, Schwinger-Dyson equations, and Riemann-Hilbert techniques, which are often known separately in several communities but scarcely combined together, 
we thought useful to give a detailed introduction for readers with various backgrounds. 
We shall as well provide a non-exhaustive review of various kinds of $N$-fold multiple integrals that have occurred throughout the literature. Finally, we shall briefly outline the context
in which multiple integrals such as $\mf{z}_N[W]$ arise within the framework of the quantum separation of variables method approach to the analysis of quantum integrable models. 
In Chapter \ref{Section presentation des resultats}, we state and describe the results obtained in this book. 
In Chapter \ref{Section Analyse asympt eqn boucles} appears the \textit{first part of the proof}: we carry out the 
\textit{asymptotic analysis of the system of Schwinger-Dyson equations} subordinated to the sinh-model. It relies on results concerning the inversion of the master operator related with our problem. It is a singular integral operator whose
inversion enables one to construct an $N$-dependent equilibrium measure. The \textit{second part of the proof} is precisely the \textit{construction of this inverse operator}: 
it is carried out in Chapter~\ref{Section RHP Inversion sing Int Op} by solving, for $N$ large enough, an auxiliary $2 \times 2$ Riemann-Hilbert problem. 
The inverse operator itself and its main properties are described in Chapter \ref{Section construction inverse op int sing}. The \textit{third part of the proof} consists 
in obtaining \textit{fine information on the large $N$-behaviour of the inverse operator}: Chapter \ref{Section descirption cptmt unif op WN}
is devoted to deriving uniform large-$N$ local behaviour for the inverse operator. Chapter \ref{Chapitre AA integrales simples et doubles} 
deals with the asymptotic analysis of one and two fold integrals of interest to the problem. 
In Section \ref{Section asymptotic analysis of single integrals} we build on the results established so far to
carry out the large-$N$ asymptotic analysis of single integrals involving the inverse operator. Finally, in Section \ref{Section asymptotic analysis of double integrals} we establish the large-$N$ asymptotic expansion of certain 
two-fold integrals, a result that is  needed so as to obtain the final answer for the expansion of the partition function. 
The book contains four appendices. In Appendix \ref{Appendix Reminder on several useful theorems} we remind some useful results of functional analysis.
In Appendix \ref{Appendix Section preuve asympt dom part fct unrescaled} we 
establish the asymptotics for the leading order of $\ln \mf{z}_N[W]$ by adapting known large deviation techniques.
In Appendix \ref{Appendix minimisation de la mesure equilibre} we derive some properties of the $N$-dependent equilibrium measures of interest to the analysis. 
Then, in Appendix \ref{Appendix Section etude de l'integrale Gaussien}, we derive an exact expression for the partition function $\mc{Z}_N[V_{G}]$
when $\be=1$ and $V_{G}$ is a quadratic potential. We also obtain there the large-$N$ asymptotics of $\mc{Z}_N[V_{G}]$ up to $o(1)$. This result is instrumental in deriving the asymptotic expansion 
of $\mc{Z}_N[V]$ for more general potential, since the Gaussian partition function always appears as a factor of the latter. 
Finally, Appendix \ref{Appendix Section liste des formule} recapitulates all the symbols that appear throughout the book. Some basic notations are also collected in \S~\ref{nonono}.

%%%%%%%%%%%%%%%%%%%%%%%%%%%%%%%%%%%%%%%%%%%%%%%%%%%%%%%%%%%%%%%%%%%%%%%%%%%%%%%%%%%%%%%%%%%%%%%%%%%%%%%%%%%%%%%%%%%%%%%%%%%%%%%%%%%%%%%%%%%%%%%%%%%%%%%%%%%%%%%%%%%%%%%%%%%%%%%%
%%%%%%%%%%%%%%%%%%%%%%%%%%%%%%%%%%%%%%%%%%%%%%%%%%%%%%%%%%%%%%%%%%%%%%%%%%%%%%%%%%%%%%%%%%%%%%%%%%%%%%%%%%%%%%%%%%%%%%%%%%%%%%%%%%%%%%%%%%%%%%%%%%%%%%%%%%%%%%%%%%%%%%%%%%%%%%%%
%%%%%%%%%%%%%%%%%%%%%%%%%%%%%%%%%%%%%%%%%%%%%%%%%%%%%%%%%%%%%%%%%%%%%%%%%%%%%%%%%%%%%%%%%%%%%%%%%%%%%%%%%%%%%%%%%%%%%%%%%%%%%%%%%%%%%%%%%%%%%%%%%%%%%%%%%%%%%%%%%%%%%%%%%%%%%%%%

%%%%%%%%%%%%%%%%%%%%%%%%%%%%%%%%%%%%%%%%%%%%%%%%%%%%%%%%%%%%%%%%%%%%%%%%%%%%%%%%%%%%%%%%%%%%%%%%%%%%%%%%%%%%%%%%%%%%%%%%%%%%%%%%%%%%%%%%%%%%%%%%%%%%%%%%%%%%%%%%%%%%%%%%%%%%%%%%
%%%%%%%%%%%%%%%%%%%%%%%%%%%%%%%%%%%%%%%%%%%%%%%%%%%%%%%%%%%%%%%%%%%%%%%%%%%%%%%%%%%%%%%%%%%%%%%%%%%%%%%%%%%%%%%%%%%%%%%%%%%%%%%%%%%%%%%%%%%%%%%%%%%%%%%%%%%%%%%%%%%%%%%%%%%%%%%%
%%%%%%%%%%%%%%%%%%%%%%%%%%%%%%%%%%%%%%%%%%%%%%%%%%%%%%%%%%%%%%%%%%%%%%%%%%%%%%%%%%%%%%%%%%%%%%%%%%%%%%%%%%%%%%%%%%%%%%%%%%%%%%%%%%%%%%%%%%%%%%%%%%%%%%%%%%%%%%%%%%%%%%%%%%%%%%%%

%%%%%%%%%%%%%%%%%%%%%%%%%%%%%%%%%%%%%%%%%%%%%%%%%%%%%%%%%%%%%%%%%%%%%%%%%%%%%%%%%%%%%%%%%%%%%%%%%%%%%%%%%%%%%%%%%%%%%%%%%%%%%%%%%%%%%%%%%%%%%%%%%%%%%%%%%%%%%%%%%%%%%%%%%%%%%%%%
%%%%%%%%%%%%%%%%%%%%%%%%%%%%%%%%%%%%%%%%%%%%%%%%%%%%%%%%%%%%%%%%%%%%%%%%%%%%%%%%%%%%%%%%%%%%%%%%%%%%%%%%%%%%%%%%%%%%%%%%%%%%%%%%%%%%%%%%%%%%%%%%%%%%%%%%%%%%%%%%%%%%%%%%%%%%%%%%
%%%%%%%%%%%%%%%%%%%%%%%%%%%%%%%%%%%%%%%%%%%%%%%%%%%%%%%%%%%%%%%%%%%%%%%%%%%%%%%%%%%%%%%%%%%%%%%%%%%%%%%%%%%%%%%%%%%%%%%%%%%%%%%%%%%%%%%%%%%%%%%%%%%%%%%%%%%%%%%%%%%%%%%%%%%%%%%%

\section*{Acknowledgments}

The work of GB is supported by the Max-Planck Gesellschaft and the
Simons Foundation. The work of AG is supported by the Simons foundation and the NSF award DMS-1307704.
KKK is supported by the CNRS. His work has
been partly financed by the Burgundy region PARI 2013 $\&$ 2014 FABER grants
`Structures et asymptotiques d'int\'egrales multiples'. KKK also enjoys
support from the ANR `DIADEMS' SIMI 1 2010-BLAN-0120-02.

\newpage 

\tableofcontents

\newpage

%%%%%%%%%%%%%%%%%%%%%%%%%%%%%%%%%%%%%%%%%%%%%%%%%%%%%%%%%%%%%%%%%%%%%%%%%%%%%%%%%%%%%%%%%%%%%%%%%%%%%%%%%%%%%%%%%%%%%%%%%%%%%%%%%%%%%%%%%%%%%%%%%%%%%%%%%%%%%%%%%%%%%%%%%%%%%%%%
%%%%%%%%%%%%%%%%%%%%%%%%%%%%%%%%%%%%%%%%%%%%%%%%%%%%%%%%%%%%%%%%%%%%%%%%%%%%%%%%%%%%%%%%%%%%%%%%%%%%%%%%%%%%%%%%%%%%%%%%%%%%%%%%%%%%%%%%%%%%%%%%%%%%%%%%%%%%%%%%%%%%%%%%%%%%%%%%
%%%%%%%%%%%%%%%%%%%%%%%%%%%%%%%%%%%%%%%%%%%%%%%%%%%%%%%%%%%%%%%%%%%%%%%%%%%%%%%%%%%%%%%%%%%%%%%%%%%%%%%%%%%%%%%%%%%%%%%%%%%%%%%%%%%%%%%%%%%%%%%%%%%%%%%%%%%%%%%%%%%%%%%%%%%%%%%%

%%%%%%%%%%%%%%%%%%%%%%%%%%%%%%%%%%%%%%%%%%%%%%%%%%%%%%%%%%%%%%%%%%%%%%%%%%%%%%%%%%%%%%%%%%%%%%%%%%%%%%%%%%%%%%%%%%%%%%%%%%%%%%%%%%%%%%%%%%%%%%%%%%%%%%%%%%%%%%%%%%%%%%%%%%%%%%%%
%%%%%%%%%%%%%%%%%%%%%%%%%%%%%%%%%%%%%%%%%%%%%%%%%%%%%%%%%%%%%%%%%%%%%%%%%%%%%%%%%%%%%%%%%%%%%%%%%%%%%%%%%%%%%%%%%%%%%%%%%%%%%%%%%%%%%%%%%%%%%%%%%%%%%%%%%%%%%%%%%%%%%%%%%%%%%%%%
%%%%%%%%%%%%%%%%%%%%%%%%%%%%%%%%%%%%%%%%%%%%%%%%%%%%%%%%%%%%%%%%%%%%%%%%%%%%%%%%%%%%%%%%%%%%%%%%%%%%%%%%%%%%%%%%%%%%%%%%%%%%%%%%%%%%%%%%%%%%%%%%%%%%%%%%%%%%%%%%%%%%%%%%%%%%%%%%

%%%%%%%%%%%%%%%%%%%%%%%%%%%%%%%%%%%%%%%%%%%%%%%%%%%%%%%%%%%%%%%%%%%%%%%%%%%%%%%%%%%%%%%%%%%%%%%%%%%%%%%%%%%%%%%%%%%%%%%%%%%%%%%%%%%%%%%%%%%%%%%%%%%%%%%%%%%%%%%%%%%%%%%%%%%%%%%%
%%%%%%%%%%%%%%%%%%%%%%%%%%%%%%%%%%%%%%%%%%%%%%%%%%%%%%%%%%%%%%%%%%%%%%%%%%%%%%%%%%%%%%%%%%%%%%%%%%%%%%%%%%%%%%%%%%%%%%%%%%%%%%%%%%%%%%%%%%%%%%%%%%%%%%%%%%%%%%%%%%%%%%%%%%%%%%%%
%%%%%%%%%%%%%%%%%%%%%%%%%%%%%%%%%%%%%%%%%%%%%%%%%%%%%%%%%%%%%%%%%%%%%%%%%%%%%%%%%%%%%%%%%%%%%%%%%%%%%%%%%%%%%%%%%%%%%%%%%%%%%%%%%%%%%%%%%%%%%%%%%%%%%%%%%%%%%%%%%%%%%%%%%%%%%%%%

\chapter{Introduction}
\label{Section Introduction}

\section{Beta ensembles with varying weights}

One of the simplest and yet non-trivial example of an $N$-fold multiple integral that we are interested in is provided by the partition function of a $\be$-ensemble
with varying weights:
\beq
\mc{Z}_N^{(\be)}[V]\; = \; \Int{ \R^N }{} \pl{a<b}{N} |\la_a-\la_b|^{\be} \cdot \pl{a=1}{N} \ex{-N V(\la_a)} \cdot \dd^N \bs{\la}  = \Int{\R^N}{} \exp\Big\{\sum_{a < b} \beta \ln |\la_a - \la_b| - N \sul{a=1}{N} V(\la_a) \Big\}\cdot \dd^N \bs{\la}\;. 
\label{int1}
\enq
$\be>0$ is a positive parameter and $V$ is a potential growing sufficiently fast at infinity for the integral \eqref{int1} to be convergent. $\mc{Z}_N^{(\be)}$ can be interpreted as the partition function 
of the statistical-mechanical system of $N$ particles at temperature $\be^{-1}$, that interact through a two-body repulsive logarithmic interaction and are placed on the real line in an overall confining potential $V$. This logarithmic interaction is the Green function for the Laplacian in $\mathbb{R}^2$ equipped with its canonical metric. %G I added this sentence.
By "varying weights" we mean that the potential $V$ is preceded by a factor of $N$, such that the logarithmic repulsion can typically be balanced by the effect of $V$ for $\la_a$ remaining in a bounded in $N$ interval. 
This is an important feature of the model that we shall comment further on. We shall however start the discussion by explaining the origin of  $\beta$-ensembles.

The partition function \eqref{int1} can be interpreted as the result of integrating over the spectrum of certain random matrices whose distribution is invariant under one of the classical groups. Consider the real vector spaces:
$$
\mathscr{H}_{N,\, \beta} = \left\{\begin{array}{l} \beta = 1\,\,:\,\, {\rm real}\,\,{\rm symmetric} \\ \beta = 2\,\,:\,\,{\rm complex}\,\,{\rm hermitian}\,\,\\ 
\beta = 4\,\,:\,\,{\rm quaternionic}\,\,{\rm self-dual}\end{array}\right.\,\,N \times N\,\,{\rm matrices} . 
$$
We denote $\dd M$, the product of the Lebesgue measures for the linearly independent real coefficients of such matrices. The Lie groups:
$$
\mathscr{G}_{N,\, \beta} = \left\{\begin{array}{l} \beta = 1\,\,:\,\,{\rm real}\,\,{\rm orthogonal} \\ \beta = 2\,\,:\,\,{\rm complex}\,\,{\rm unitary} \\ 
\beta = 4\,\,:\,\,{\rm quaternionic}\,\,{\rm unitary} \end{array}\right.\,\,N \times N\,\,{\rm matrices}
$$
act on $\mathscr{H}_{N,\, \beta}$ by conjugation. If $M$ is a random matrix in $\mathscr{H}_{N,\, \beta}$ drawn from the distribution\symbolfootnote[3]{Such distributions are indeed invariant under conjugation by $\mathscr{G}_{N,\, \beta}$} 
$C_{N;V} \ex{ -N \e{tr}[V(M)] } \cdot \dd M$, $C_{N;V}$ being the normalisation constant, the induced distribution $\mathbb{P}_{N}^{(\be)}$ of eigenvalues must be of the form:
$$
p_{N;V}^{(\be)}\big( \bs{\la} \big) \cdot \dd^N \bs{\la} \qquad \e{with} \qquad p_{N;V}^{(\be)}(\bs{\la}) \; = \; \f{1}{ \mc{Z}_N^{(\be)}[V] }\pl{a<b}{N} |\la_a-\la_b|^{\be}\, \pl{a=1}{N}\Big\{ \ex{ - N V(\la_a) } \Big\} \;. 
$$
Hence, in this context, the partition function $\mc{Z}_N^{(\be)}[V]$ corresponds to the normalisation constant of the induced distribution of eigenvalues. 
The three cases $\beta \in \{1,2,4\}$ are very special,  since they feature a determinantal or Pfaffian structure that is unknown for general $\beta$. This additional structure 
allows one to reduce the computation of $\mc{Z}_N^{(\be)}[V]$ to one of a family of orthogonal or skew-orthogonal polynomials \cite{MehtaRandomMatrices}.

For general $\beta > 0$ and polynomial $V$, the partition function \eqref{int1}
can also be interpreted as the integral over the spectrum of a family of random tri-diagonal matrices  \cite{DumitriuEdelmanConstructionTriDiagRMInterpretationForBetaEnsemblesPartFct,KrishnapurRiderViragProofOfBkAndEdgeUniversalityByOpMethods}, 
whose entries are independent and have a well-tailored distribution depending on $V$. As there is no symmetry group acting here, this class of random matrices is very different from the invariant ensembles. It is in nature closer to stochastic 
Schr\"odinger operators.

The $\beta$-ensembles have been extensively studied for more than 20 years, see \textit{e.g.} the books
\cite{AndersonGuionnetZeitouniIntroRandomMatrices,DeiftGoievRandomMatricesUniversalityInDifferentEnsembles,MehtaRandomMatrices,PasturScherbinaEigenvalueDirstRM}, for two main reasons, that we shall develop below. 
From the probabilistic perspective, the statistical-mechanics interpretation of $\be$-ensembles makes $\mc{Z}_N^{(\be)}[V]$  and its associated probability distribution a good playground for testing the local universality of the distribution of 
repulsive particles \cite{ForresterLogGases}. From the perspective of geometry and physics, the interest in the $\beta = 2$ case --\textit{viz}. random hermitian matrices-- has been fostered, 
since the pioneering works of Br\'ezin-Itzykson-Parisi-Zuber \cite{BrezinItzyksonParisiZuberCombinatorixFormaExpansionMatrixModels}, by the insight it provides into two-dimensional quantum gravity
and the enumerative geometry of surfaces. This interest was eased by the algebraic miracles that make the 
case $\beta = 2$ quite tractable from the computational point of view, and also raised by the desire to understand the geometry (related to the integrable structure associated with the orthogonal polynomials) behind these miracles.

\subsection{Local fluctuations and universality} 

The physical idea behind universality is that the logarithmic repulsion dictates the local behaviour of the particles\footnote{By local, 
we mean "looking at intervals shrinking with $N$ so that these contain typically only a finite number of particles in the $N \rightarrow \infty$ limit".}. 
The universality classes should only depend on $\beta$ and the local environment of the chosen position on $\mathbb{R}$. Typically one expects that, when $N\tend +\infty$, the particles will localise on 
some union of segments $\cup_k \intff{a_k}{b_k}$ and that, up to a $\e{O}\big(N^{-1}\big)$ precision, the $p^{\e{th}}$ particle will localise around a "classical" position $\ga_{p}^{\e{cl}}$.

To be precise, we introduce the $k$-point density functions $\rho^{(k)}_{N}(x_1,\ldots,x_k)$. These are symmetric functions of $k$ real variables characterized (if they exist) by the property that,
for any sequence of pairwise disjoint intervals $(A_{i})_{i = 1}^k$:
\beq
\label{rhonkdef}
\mathbb{P}_{N}\Big[\exists i_1,\ldots,i_k \in \{1,\ldots,N\}\,\,:\,\,\lambda_{i_j} \in A_{j}\Big] = \int_{A_1} \cdots \int_{A_k} \rho^{(k)}(x_1,\ldots,x_k)\prod_{i = 1}^{k} \dd x_i \;. 
\enq
$\rho^{(k)}_N$ fails to be a density probability function, because of the unusual normalization:
$$
\int_{\mathbb{R}^k} \rho^{(k)}_{N}(x_1,\ldots,x_k) \prod_{i = 1}^k \dd x_i = \frac{N!}{(N - k)!}
$$
which follows from \eqref{rhonkdef} by taking a partition of $\mathbb{R}$ into $k$ pairwise disjoint intervals and symmetrising the integration range. In particular, $N^{-1}\rho_{N}^{(1)}(x)$ is the local mean density of particles.

For instance, if $\beta = 2$ and we look at intervals of size $1/N$ around a point $x_0 \in \mathbb{R}$ where the mean density of particles is smooth and positive -- \textit{i.e.} in the bulk -- 
one expects the distribution of the eigenvalues to converge to the determinantal process of the sine kernel. This means that, if $\lim_{N \rightarrow \infty} N^{-1}\rho_{N}^{(1)}(x_0) > 0$, we expect that
$$
\lim_{N \rightarrow \infty} \frac{\rho_{N}^{(k)}\Big(\Big\{x_0 + \xi_i/\rho_N^{(1)}(x_0)\Big\}_{i = 1}^k\Big)}{\big[\rho_{N}^{(1)}(x_0)\big]^k} = 
 \rho_{{\rm sin}}^{(k)}(\xi_1,\ldots,\xi_k)
$$
where $\rho_{{\rm sin}}^{(k)}$  are given by:
\beq
\label{detaa}
\rho_{{\rm sin}}^{(k)}(\xi_1,\ldots,\xi_k) = \mathop{{\rm det}}_{1 \leq i,j \leq k}\big[K_{{\rm sin}}(\xi_i,\xi_j)],\qquad K_{{\rm sin}}(x,y) = \frac{\sin\pi(x - y)}{\pi(x - y)} \;. 
\enq
Still for $\beta = 2$ and if we look at intervals of size $1/N^{2/3}$ around a point $x_0$ where the mean density vanishes like a square root, one rather expects to observe the determinantal process of the Airy kernel:
\beq
\rho_{{\rm Ai}}^{(k)}(\xi_1,\ldots,\xi_k) = \mathop{{\rm det}}_{1 \leq i,j \leq k} \big[K_{{\rm Ai}}(x_i,x_j)\big],\qquad K_{{\rm Ai}}(x,y) = \frac{{\rm Ai}(x){\rm Ai}'(y) - {\rm Ai}'(x){\rm Ai}(y)}{x - y} \;. 
\enq
To reformulate, if the condition:
\begin{equation}
\label{genedge}\lim_{x \rightarrow x_0} \frac{\lim_{N \rightarrow \infty} N^{-1} \rho_{N}^{(1)}(x)}{\sqrt{|x_0 - x|}} = A > 0
\end{equation}
holds, we expect that,
$$
\lim_{N \rightarrow \infty} N^{-k/6} \rho_{N}^{(k)}\Big(\Big\{x_0 + (\pi A)^{2/3} N^{-2/3}\xi_i \Big\}_{i = 1}^k\Big) = (\pi A)^{-2k/3}\,\rho_{{\rm Ai}}^{(k)}(\xi_1,\ldots,\xi_k)\, . 
$$
Without being too precise, let us say that \eqref{genedge} is the generic behaviour at the edge of the spectrum of random matrices of large sizes. 

The expression for the potentially universal distribution of particles for other shapes of large-$N$ local mean density of particles, and other values of $\beta$ are known 
\cite{ValkoViragSineBetaUniversalityBetaEns,RamirezRiderViragStochasticAiryBetaEnsEdgeSpectrumUniversality} -- although their understanding is currently
much more developed for $\beta \in \{1,2,4\}$. The main theme in universality problems is therefore to prove that given models exhibit these distributions for the local behaviour of particles in the large-$N$ limit. 
 As a matter of fact, the precise mode of convergence to the universal laws that one can obtain is not always
optimal from a physical point of view, namely it may hold only once integrated against a class of test functions, or only after integration on intervals of size $N^{-1 + \eta}$ for $\eta$ arbitrarily small and independent of $N$. 
We refer to the original works cited below to see which mode of convergence they establish.

First results of local universality in the bulk where obtained by Shcherbina and Pastur \cite{PasturShcherbinaFirstProofOfBulkUniversalityHermitianRMT}
at $\beta=2$. Then, at $\beta = 2$ and for polynomial $V$, Deift, Kriechenbauer, McLaughlin, Venakides and Zhou \cite{DeiftKriechMcLaughVenakZhouOrthogonalPlyVaryingExponWeights} established 
the local universality in the bulk  within the Riemann-Hilbert approach to orthogonal polynomials with orthogonality weight 
$\ex{-NV(x)}$ on the real line. These results were then extended by Deift and Gioev to $\beta \in \{1,2,4\}$ for the bulk \cite{DeiftGioevProodUniversalityBulkBeta14RHPApproach} and then 
for the generic edge \cite{DeiftGioevProodUniversalityEdgeBeta124RHPApproach} universality. 
The bulk and generic edge universality for general $\beta > 0$ were recently established by various methods and under weaker assumptions. 
Bourgade, Erd\"{o}s and Yau built on relaxation methods so as to establish the bulk \cite{BourgadeErdosYauUniversalityBulkBetaEnsGnlPot,BourgadeErdosYauUniversalityBulkBetaEnsConvPot} and
the generic edge \cite{BourgadeErdosYauUniversalityEdgeBetaEnsGnlPot} universality in the presence of generic $\mathcal{C}^{k}$ potentials. 
 Krishnapur, Rider and Vir\'{a}g \cite{KrishnapurRiderViragProofOfBkAndEdgeUniversalityByOpMethods} proved both universalities 
by means of stochastic operator methods and in the presence of convex polynomial potentials. 
Finally, the bulk universality was also established on the basis  of measure transport techniques
by Shcherbina \cite{ShcherbinaChangeVarsForProofUniversality} in the presence of real-anaytic potentials while  universality both at the bulk and
generic edge  was derived by Bekerman, Figalli and Guionnet \cite{BekermanFigalliGuionnetUniversalityFromTransportMapGeneralPot}  
for $\mathcal{C}^{k}$ potentials with $k$ large enough.

\subsection{Enumerative geometry and $N$-fold integrals}
\label{Enumer}
The motivation to study the all-order asymptotic expansion of $\ln \mc{Z}_{N}^{(\be)}[V]$ when $N \rightarrow \infty$ initially came from physics and the study of two-dimensional quantum gravity, 
especially in the case $\beta = 2$ corresponding to hermitian matrices. In the landmark article \cite{BrezinItzyksonParisiZuberCombinatorixFormaExpansionMatrixModels}, Br\'ezin, Itzykson, Parisi and Zuber have argued that,
for potentials given by formal series:
$$
V(x) = \frac{1}{u}\bigg(\frac{x^2}{2} + \sum_{j \geq 3} \frac{t_j}{j}\,x^j\bigg)
$$
the free energy $\ln \mc{Z}_{N}^{(\beta)}[V]$ has the formal expansion:
\beq
\label{mumg}\ln \Bigg( \f{ \mc{Z}_{N}^{( 2)}[V] }{  \mc{Z}_{N}^{( 2)}[V_{\mid t_{\bullet} =0}]  } \Bigg) \quad \mathop{=}^{{\rm formal}} \quad \sum_{g \geq 0} N^{2 - 2g}\,\mc{F}^{(g)}
\enq
which is to be understood as an equality between formal power series in $\{t_j\}$, and we use the notation $V|_{t_{\bullet} = 0}(x) = \frac{x^2}{2u}$. The coefficients $\mc{F}^{(g)}$ correspond to a weighted enumeration of "maps", \textit{i.e.} equivalence classes of graphs $\mathscr{G}$ embedded in a topological, 
connected, compact, oriented surface $\mathscr{S}$ of genus $g$ such that all connected components $\mathscr{C}_{i}$ of $\mathscr{S}\setminus\mathscr{G}$ are homeomorphic to disks. Each $\mathscr{C}_{i}$ which is bordered by $j$ 
edges of $\mathscr{G}$ is counted with a local weight $-t_{j}$, each vertex in $\mathscr{G}$ is counted with a local weight $u$, and the overall weight of $(\mathscr{G},\mathscr{S})$ is computed as the product of all local weights, 
divided by the number of automorphisms of $(\mathscr{G},\mathscr{S})$. For instance, choosing $t_3 \neq 0$ and $t_{j > 3} = 0$, $\mc{F}^{(g)}$ enumerates triangulations of an oriented surface of genus $g$. 
More generally, $\ln \mc{Z}_{N}^{(\beta)}[V]$ with $ \be \neq 2$ gives rise to enumerations of graphs embedded in possibly non-orientable surfaces \cite{EynardMarchalNonOrientable,MulaseYuMatrixIntegralsAndGraphsCountingNonOrientable}. 
Then,  the expansion \eqref{mumg} also contains half-integer $g$'s.
The expansions that had been obtained in \cite{BrezinItzyksonParisiZuberCombinatorixFormaExpansionMatrixModels} and in the many subsequent works in physics that have followed. These handlings were set 
in the appropriate framework of providing an equality between formal power series in $u$ and the parameters $\{t_j \}_{j \geq 3}$ in \cite{EynardDAMatrixModelsAndFormalPowerSeriesInterpretation}. 
Indeed, the fact of subtracting the free energy of the quadratic potential $V|_{t_{\bullet}}(x) = x^2/2u$ in the right-hand side of \eqref{mumg} turns the formula  into a well-defined equality between formal series; 
a combinatorial argument based on the computation of the Euler characteristics $2 - 2g$ shows that the coefficient of a given monomial $u^{k}\prod_{\ell} t_{\ell}^{n_{\ell}}$ is given by a sum over finitely many genera $g$. 
For a restricted class of potentials\symbolfootnote[2]{Roughly speaking, when there exists $1/N$ asymptotic expansion, its coefficients are the same as the formal expansion.} for which the integral \eqref{mumg}
is convergent, the formal power series also corresponds to the  $N \rightarrow \infty$ asymptotic expansion of $ \mc{Z}_{N}^{( 2)}[V]$. 
It is because of such a combinatorial interpretation that these expansions are called "topological expansions", this independently of their formal or asymptotic nature.

The random hermitian matrix model ($\beta = 2$) for finite $N$ was also interpreted as a well-defined discretised model of two-dimensional quantum gravity. For fixed $\{ t_j\}$, 
there is a finite value $u = u_{c}$ at which the model develops a critical point: the coefficients $\mathcal{F}^{(g)}$ exhibit as a singularity of the type $(u_{c} - u)^{(2 -\gamma_{{\rm str}})(1 - g)}$ with a critical 
exponent $-1/2 \leq \gamma_{{\rm str}} < 0$ depending on the universality class. One of the consequences of the appearance of such singularities is that the average or the variance of the number of faces in a map of fixed genus 
diverges when $u \rightarrow u_{c}$. This allows one to interpret the $u\tend u_c$ limit  as a continuum limit. In taking such a limit, it becomes particularly interesting to tune the $u$-parameter  in an $N$-dependent way such that
$u = u_{c} - N^{-1 + \gamma_{{\rm str}}/2}\cdot\tilde{u}$, hence making each term $N^{2 - 2g}\,\mc{F}^{(g)}$ of order $1$ when $N \rightarrow \infty$. In such a double scaling limit, the expansion \eqref{mumg} does not make sense any more. 
However, it is expected, and it can be proved in certain cases, that the double scaling limit of the appropriately rescaled partition function $\mc{Z}_{N}^{(2)}[V]$ exists. This limit was proposed as a way of defining the partition function of two-dimensional quantum gravity 
with coupling constant $\tilde{u}$.  We refer to the review \cite{DiFrancescoGinspargZinnJustinRMTAnd2DGravity} for more details. 
The investigation of these double scaling limits is similar in spirit to the investigation of universality that we already mentioned, 
with the only difference being that a continuation to complex-valued $\tilde{u}$ does have an interest from the physics point of view, whereas it is often excluded 
from mathematical study of universality given the difficulty to address it with probabilitistic techniques.

Finally, the all-order expansion (be it formal or asymptotic) of $N$-fold integrals in the $\beta$-ensembles and generalisations thereof have numerous applications at the interface of algebraic geometry and theoretical physics. 
The key point is that the coefficients of the all-order expansions have an interesting geometric interpretation, and the study of matrix models in the large $N$ limit can give some insight into topological strings, gauge theories, \textit{etc}.
Describing the exponentially small in $N$ contributions to $\ln \mc{Z}_N^{(\be)}[V]$ has also an interest of its own. It is particularly interesting 
\cite{MarinoNonPertEffectsGaugeThAndRMT} as a path towards understanding the possible non-perturbative completion(s) of the perturbative physical theories. As an illustration closer to the scope of this book, we shall give in Section~\ref{Nfoldlist} 
a non-exhaustive list of $N$-fold integrals which have a physical or geometrical interpretation.

\section{The large-$N$ expansion of  $\mc{Z}^{(\be)}_N$}

\subsection{Leading order of $\mc{Z}^{(\be)}_N$: the equilibrium measure and large deviations}

\label{LargeNNN}

Given a sufficiently regular potential $V$ growing at $x\tend \pm  \infty$ faster than $\big(\be+\eps\big) \ln |x|$
for some $\eps>0$, the leading asymptotic behaviour of the partition function $\mc{Z}_{N}^{(\be)}[V]$ takes the form :
\beq
\ln \mc{Z}^{(\be)}_N[V] \, = \; -N^2\Big( \mc{E}^{(\be)}[\mu_{\e{eq}}] \, + \, \e{o}(1) \Big)
\quad \e{with} \quad \mc{E}^{(\be)}[\mu] \; = \; \Int{}{} V(x)\,\dd\mu(x)  \,- \,  \be\Int{x < y}{} \ln|x-y|\,\dd\mu(x) \dd\mu(y) \;. 
\label{ecriture large N asymptotics beta ensembles}
\enq
In these leading asymptotics, the functional $\mc{E}^{(\be)}$ is evaluated at the so-called equilibrium measure $\mu_{\e{eq}}$, 
a probability measure on $\R$ that minimises the functional $\mc{E}^{(\be)}$. The notion of equilibrium measure arises in numerous other branches of mathematical physics, for instance 
the study of zeroes of families of polynomials or the one of the characterisation of the thermodynamic behaviour at finite temperature of quantum integrable models \cite{DorlasLewisPuleRigorousProofYangYangThermoEqnNLSE}. 
The minimiser $\mu_{\e{eq}}$ can be characterised within the framework of potential theory \cite{LandkofFoundModPotTheory}. 
One can show that the equilibrium measure associated with the functional $\mathcal{E}^{(\beta)}$ exists and is unique. We stress that $\mu_{\e{eq}}$ is characterised by \eqref{carnun} and thus 
depends on $\beta$ only via a rescaling of the potential, hence imposing the same dependence in the leading order term of the expansion \eqref{ecriture large N asymptotics beta ensembles}.

Let us explain, on a heuristic level,
the mechanism which gives rise to \eqref{ecriture large N asymptotics beta ensembles}. For this purpose, observe that the integrand of 
$ \mc{Z}^{(\be)}_N[V] $ can be recast as 
\beq
\exp\Big\{ -N^2 \mc{E}^{(\be)}[L_N^{(\bs{\la})}]  \Big\} \qquad \e{where} \qquad L_N^{(\bs{\la})} \, =  \, \f{1}{N} \sul{a=1}{N} \de_{\la_a}
\label{ecriture integrand ZN beta ens avec def mes empirique}
\enq
is the so-called empirical measure while $\de_x$ refers to the Dirac mass at $x$. 
 For finite but large $N$, $\bs{\la} \mapsto \mc{E}^{(\be)}[L_N^{(\bs{\la})}] $ with $\la_1<\dots < \la_N$ attains its minimum at a  point $\bs{\ga}_{\e{eq}}=(\ga_{\e{eq};1} , \dots  ,  \ga_{\e{eq};N})$
 whose coordinates  $\ga_{\e{eq};1}<\dots < \ga_{\e{eq};N}$ are bounded, uniformly in $N$, from above and below. This minimum results from a balance between the repulsion of the integration variables 
 induced by the logarithmic interaction and the confining nature of the potential $V$ since the entropy is negligible\footnote{ 
The Lebesgue measure does not participate to the setting of this equilibrium: the aforementioned
terms induce a $\ex{\e{O}(N^2)}$ behaviour in the light of \eqref{ecriture large N asymptotics beta ensembles}, while 
on compact subsets of $\R^N$, the Lebesgue measure produces at most a $\e{O}(\ex{cN})$
contribution, with $c$ depending on the size of the compact set.}.
It seems reasonable that the main contribution to the integral, namely the one not including exponentially small corrections, 
will issue from a small neighbourhood of the point $\bs{\ga}_{\e{eq}}$ (or those issuing from permutations of its coordinates) and hence yield, to the leading order in $N$, 
$\ln \mc{Z}_N^{(\be)}[V] \, = \,  -N^2\big( \mc{E}^{(\be)}[L_N^{(\bs{\ga}_{\e{eq} } )}] + \e{o}(1) \big) $. 
As a matter of fact, the $\ga_{\e{eq};a}$ are distributed in such a way that they densify on some compact subset of $\R$
and in such a way that, in fact, $L_N^{(\bs{\ga}_{\e{eq} } )}$ converges, in some appropriate sense, to the probability measure $\mu_{\e{eq}}$.

This reasoning thus indicates that the leading asymptotics of $\ln \mc{Z}_{N}^{(\be)}[V]$ issue from a saddle-point like estimation of the integral \eqref{int1}. This 
statement can be made precise within the framework of large deviations. Ben Arous and Guionnet \cite{BenArousGuionnetLargeDeviationForWignerLawLEadingAsymptOneMatrixIntegral} 
showed that the distribution of $L_N^{(\bs{\la})}$ under the sequence $\mathbb{P}^{(\be)}_N$ of probability measures associated with $\mc{Z}_N^{(\be)}[V]$ satisfies a large deviation principle with good rate function $\mc{E}^{(\be)}[\mu]$ at speed $N^2$. 
This means that, for any open set $\Omega$ and any closed set $F$ of the space of probability measures endowed with the weak topology, we have:
\begin{eqnarray}
\liminf_{N \rightarrow \infty} N^{-2}\,\ln \mathbb{P}_N^{(\be)}[L_N^{(\bs{\la})} \in \Omega] & \geq & - \inf_{\mu \in \Omega}\,\mathcal{E}^{(\beta)}[\mu] \nonumber \\
\limsup_{N \rightarrow \infty} N^{-2}\,\ln \mathbb{P}_N^{(\be)}[L_N^{(\bs{\la})} \in F] & \leq & - \inf_{\mu \in F} \mathcal{E}^{(\beta)}[\mu] \nonumber \;. 
\end{eqnarray}
Saying that $\mathcal{E}^{(\beta)}$ is a good rate functional means that its level sets $(\mathcal{E}^{(\beta)})^{-1}([0;M])$ are compact for any $M \geq 0$. 
As a direct consequence of this large deviation principle, the random measure $L_N^{(\bs{\la})}$ converges almost surely and in expectation, in the weak topology, towards the (deterministic) equilibrium measure $\mu_{\e{eq}}$.

The properties of the equilibrium measure $\mu_{\e{eq}}$ have been extensively studied 
\cite{DeiftKriechMcLaughEquiMeasureForLogPot,LandkofFoundModPotTheory,SaffTotikLogarithmicPotential}. Its uniqueness follows  from the strict convexity of $\mathcal{E}^{(\beta)}$. 
Indeed, given two probability measures  $\mu_0,\mu_1$ and $t \in [0,1]$, one has 
$$
\mathcal{E}^{(\beta)}[(1 - t)\mu_0 + t\mu_1] = (1 - t)\mathcal{E}^{(\beta)}[\mu_0] + t\mathcal{E}^{(\beta)}[\mu_1] - \beta\,\mathcal{Q}[\mu_1 - \mu_0]t(1-t)
$$
where, for any signed finite measure $\nu$ of zero mass, one has:
$$
\mathcal{Q}[\nu] = -\Int{x < y}{} \dd\nu(x)\dd\nu(y)\,\ln|x - y| = \int_{0}^{\infty} \frac{\dd k}{2k}\,\big|\mathcal{F}[\nu](k)\big|^2
$$
which implies that $\mathcal{Q}[\nu] \geq 0$. Furthermore, it is clear that equality holds if and only if $\nu = 0$. The latter does ensure the strictly convexity of  $\mathcal{E}^{(\beta)}$.

As a solution of a minimisation problem, $\mu_{\e{eq}}$ must satisfy an "Euler-Lagrange equation". This condition states the existence of a constant $C_{\e{eq}}$ such that:

\beq
\label{carnun}V_{{\rm eff}}(x) = V(x) - C_{\e{eq}}  -  \beta \int \ln|x - y|\dd\mu_{\e{eq}}(y),\qquad  
\left\{\begin{array}{l} V_{{\rm eff}}(x) = 0\,\,\mu_{\e{eq}}{\rm -\rm almost}\,\,{\rm everywhere} \\ 
V_{{\rm eff}}(x) \geq 0\,\,   \mu{\rm -\rm almost}\,\,{\rm everywhere}     \end{array}\right. 
\enq
where the second condition holds for any probability measure $\mu$ on $\R$ such that $\mc{E}^{(\be)}[\mu] < +\infty$. 
The inequality comes from the fact that one minimises over positive measures, 
and the constant $C_{\e{eq}}$ is a Lagrange multiplier for the constraint that the total mass should be $1$. $V_{{\rm eff}}(x)$ is the effective potential felt by a particle; it takes into account $V$ and the repulsion it feels from all other particles distributed according to $\mu_{\e{eq}}$. 
The characterisation \eqref{carnun} expresses that the effective potential is constant and minimal on the support of $\mu_{\e{eq}}$. The constant $C_{\e{eq}}$ is chosen in such a way that this minimum is zero. 
It thus appears reasonable to expect that, in the large $N$ limit, the particles should mostly likely accumulate in $\e{supp}[\mu_{\e{eq}}]$. 
More precisely, \cite{AndersonGuionnetZeitouniIntroRandomMatrices,BenArousDemboGuionnetLDPExtremeEVGaussian,BorotGuionnetAsymptExpBetaEnsOneCutRegime} 
proves a  large deviation principle for the position of individual particles at speed $N$ with good rate function $V_{{\rm eff}}$. 
This means that, for any open subset $\Omega$ and closed subset $F$ of $\mathbb{R}$, 
we have: 
\begin{eqnarray}
\liminf_{N \rightarrow \infty} \,N^{-1}\ln \mathbb{P}_{N}^{(\be)} \big[\exists a \in \{1,\ldots,N\}\,\,\lambda_a \in \Omega\big] \geq -\inf_{x \in \Omega} V_{{\rm eff}}(x) \nonumber \\
\limsup_{N \rightarrow \infty} \,N^{-1} \ln \mathbb{P}_{N}^{(\be)} \big[\exists a \in \{1,\ldots,N\},\,\,\lambda_{a} \in F\big] \leq -\inf_{x \in F} V_{{\rm eff}}(x) \nonumber \;. 
\end{eqnarray}

One can prove that if $V$ is $\mathcal{C}^{k}$ for $k \geq 2$, then $\mu_{\e{eq}}$ is Lebesgue continuous with a $\mathcal{C}^{k - 2}$ density. Besides, if $V$ is real-analytic, 
the density is the square root of an analytic function what, in its turn, implies that its support consists of a finite number of segments, called \emph{cuts}. Critical points of the model occur when the topology of the support 
becomes unstable with respect to small perturbations of the potential. Namely when there exist arbitrarily small perturbations of the potential which result
in a support of the equilibrium measure in which one of the component has split in two,
or where a new cut has appeared. When this is not the case, one says that the potential is \emph{off-critical}. For $V$ real-analytic on $\mathbb{R}$, 
a necessary condition for off-criticality is that $\mu_{{\rm eq}}$ vanishes exactly like a square root at the edges of the support:
$$
\lim_{  x \rightarrow a \in \partial{\rm supp}\mu_{ {\rm eq} }  } \frac{1}{ \sqrt{|x - a|} } \frac{ \dd\mu_{{\rm eq}} }{ \dd x }  \, >  \, 0
$$
and this is the "generic" behaviour. When $\mu_{{\rm eq}}$ vanishes like $|x - a|^{ k + \frac{1}{2} }$ with $k > 0$, a small island of particles around $a$ 
can separate from the rest and form a new cut under certain small perturbations of the potential.

The simplest example of an equilibrium measure is provided by the one subordinate to a quadratic potential $V_G(x) = x^2$. This equilibrium measure is given by the famous Wigner semi-circle distribution \cite{WignerSemiCircleLaw}:
$$
\dd\mu_{{\rm eq}}(x) = \frac{\dd x}{\beta\pi}\cdot ( \beta - 4x^2 )^{\frac{1}{2}}\cdot \bs{1}_{ \intff{-\frac{\sqrt{\beta}}{2} }{  \frac{\sqrt{\beta}}{2} } }(x)
$$
which has only one cut. Although there is no easy characterisation of the set of potentials $V$ leading to one-cut equilibrium measures, strictly convex $V$ do belong to this set \cite{MhaskarSaffLocumOfSupNormWeightedPly}. 
Indeed, since for any $y$ the function $x \mapsto -\ln|x - y|$ is strictly convex, integrating it over $y$ against the positive measure $\mu_{\e{eq}}$ still gives a convex function. As a result, if $V$ is strictly convex, 
the effective potential \eqref{carnun} is a \textit{fortiori} strictly convex. This imposes that the minimum of $V_{\e{eff}}$ is attained on a connected set, therefore the support of $\mu_{\e{eq}}$ is a segment.

A remarkable feature of $\beta$-ensembles is that, looking at the case of equality in \eqref{carnun}, the density of $\mu_{{\rm eq}}$ can be built in terms of the solution to 
a \textit{scalar} Riemann--Hilbert problem for a piecewise holomorphic function having jumps on the support of $\mu_{\e{eq}}$. 
If one assumes the support to be known, such Riemann--Hilbert problems can be solved explicitly leading to a \textit{one-fold} integral representation for the density of $\mu_{{\rm eq}}$.  
These manipulations originate in the work of Carleman \cite{CarlemanRHPSolutionTransfoHankelAxeFini}, and some aspects have also been treated in the book of Tricomi \cite{TricomiIntEqnsBook}. 
In the case where the support consists of single segment $[a;b]$ and for $V$ at least $\mathcal{C}^2$, one gets that the density of the equilibrium measure reads:
\beq
\dd\mu_{\e{eq}}(x) = \dd x\cdot \mathbf{1}_{\intff{a}{b}}(x)\cdot \Int{a}{b} \frac{\dd\xi}{\pi \beta}\cdot \frac{V'(x) - V'(\xi)}{x - \xi}\cdot \bigg\{\frac{(b - x)(x - a)}{(b - \xi)(\xi - a)}\bigg\}^{\frac{1}{2}} \;. 
\label{ecriture mesure equilibre 1 cut beta ens}
\enq
The above representation still contains two unknown parameters of the minimisation problem: the endpoints $a,b$ of the support of $\mu_{\e{eq}}$. These are determined by imposing additional non-linear consistency relations. 
In the one-cut case discussed above, the conditions on the endpoints $a$ and $b$ are:
\beq
\label{abcutfind} \Int{a}{b} \frac{\dd\xi}{\pi\beta}\cdot\frac{V'(\xi)}{\big\{(b - \xi)(\xi - a)\big\}^{\frac{1}{2}}} = 0,\qquad 
\frac{a + b}{2} \Int{a}{b} \frac{\dd \xi}{\pi\beta}\cdot\frac{\xi\,V'(\xi)\,\dd \xi}{\big\{(b - \xi)(\xi - a)\big\}^{\frac{1}{2}}} = 1 \;. 
\enq
The situation, although more involved as regards explicit expressions, is morally the same in the multi-cut case where one has to determine all the endpoints of the support. 
We stress that the very existence of a \textit{one-fold} integral representation with a \textit{fully explicit} integrand tremendously simplifies the analysis, 
be it in what concerns the description of the properties of $\mu_{\e{eq}}$, or any handling that actually involves the equilibrium measure. The one-cut case is computationally easier to deal with than the multi-cut case:
for instance when $V$ is polynomial, the conditions \eqref{abcutfind} determining the endpoints $a$ and $b$ are merely algebraic. As we will explain later, there is another, more important difference between the one-cut case and 
the multi-cut case, that pertains to the nature of the $\e{O}(1)$ corrections to the large-$N$ behaviour of the partition function $\mc{Z}^{(\be)}_N[V]$.

\subsection{Asymptotic expansion of the free energy: from Selberg integral to general potentials}
\label{Interpi}

For very special potentials, the partition function $\mc{Z}^{(\be)}_N[V]$ can be exactly evaluated in terms of a $N$-fold product. The quadratic potential $V_G(x) = x^2$ is one of these special cases
 and the associated partition function is related to the Selberg integral \cite{SelbergTheFamousIntegral}, from where it follows:
$$
\mc{Z}^{(\be)}_{N}[V_G]\;  = \;  (2\pi)^{N/2}\cdot (2N)^{-\frac{N}{2}(1 - \frac{\beta}{2} + \frac{\beta}{2}\,N)}\cdot \prod_{m = 1}^{N} \frac{\Gamma\big(1 + \frac{m\beta}{2}\big)}{\Gamma\big(1 + \frac{\beta}{2}\big)} \;. 
$$
With such an explicit representation, standard one-dimensional analysis methods lead to the large-$N$ asymptotics of the partition function:
\begin{eqnarray}
\ln \mc{Z}^{(\be)}_{N}[V_G] & = & \bigg\{\frac{\beta}{4}\cdot\ln\bigg(\frac{\beta}{4}\bigg) - \frac{3\beta}{8}\bigg\}\cdot N^2 + \frac{\beta}{2} \cdot N\ln N 
	+ \bigg\{\bigg(\frac{1}{2} + \frac{\beta}{4}\bigg)\cdot\ln\bigg(\frac{\beta}{4 \text{e} }\bigg) + \frac{\beta}{2}\cdot \ln 2 + \ln(2\pi) - \ln \Gamma\Big(1 + \frac{\beta}{2}\Big)\bigg\}\cdot N \nonumber \\
\label{ZGAUSN}& & + \frac{1}{12}\bigg(3 + \frac{\beta}{2} + \frac{2}{\beta}\bigg)\cdot \ln N + \chi'\bigg(0\,;\frac{2}{\beta},1\bigg) + \frac{\ln(2\pi)}{2} + o(1) \;. 
\end{eqnarray}
The function $\chi(s\,;b_1,b_2)$ is the meromorphic continuation in $s$ of the function  defined for ${\rm Re}\,s > 2$ by the formula:
$$
\chi(s\,;b_1,b_2) = \sum_{\substack{m_1,m_2 \geq 0 \\ (m_1,m_2) \neq (0,0)}} \frac{1}{(m_1b_1 + m_2b_2)^{s}}\;. 
$$
Note that, when $\be=2$, the constant term in \eqref{ZGAUSN} can be recast in terms of the Riemann zeta function as:
$$
\chi'(0\,;1,1) \, =  \,\zeta'(-1) \, - \,  \frac{\ln(2\pi)}{2} \; . 
$$
The $o(1)$ remainder admits an asymptotic expansion in $1/N$ whose coefficients are expressed as linear combinations with rational coefficients of Bernoulli numbers and $2/\beta$.

For generic  potentials $V$, there is no chance to obtain a simple closed formula for $\mc{Z}_{N}^{(\beta)}[V]$.
Nevertheless, the cases that are computable in closed form do play a role in the asymptotic analysis of the more general $\mc{Z}_{N}^{(\beta)}[V]$ beyond the leading order. Indeed, most of the methods of asymptotic analysis rely, 
in their final step, on an interpolation between the potential of interest $V$, and a potential of reference $V_{0}$ for which the partition function can be exactly computed. The strategy 
for obtaining the leading corrections is to conduct, first, a study of the large-$N$ corrections to 
the macroscopic distribution of eigenvalues, and in particular to the fluctuations of the linear statistics:
\beq
\label{linearstat}\mathbb{E}^{V}_{N}\bigg[\sum_{i = 1}^N f(\lambda_i) - N\Int{}{} f(x)\cdot\dd\mu_{\e{eq}}(x)\bigg] = N\cdot \mathbb{E}^{V}_{N}\bigg[\Int{}{} f(x)\cdot\dd(L_N^{(\bs{\la})} - \mu_{\e{eq}})(x)\bigg]
\equiv N \Int{ \R^N }{} p_{N;V}^{(\be)}(\bs{\la}) \Bigg\{ \Int{}{} f(x)\cdot\dd(L_N^{(\bs{\la})} - \mu_{\e{eq}})(x)\bigg] \Bigg\} \cdot \dd^N \bs{\la}
\enq
for a sufficiently large class of test functions $f$. The subscript $V$ indicates that we are considering the sequence of probability measures for which $\mc{Z}_{N}^{(\be)}[V]$ is the partition function. 
Assume that one is able to establish the large-$N$ behaviour of \eqref{linearstat} for a one parameter $t$ family $\{V_t\}_{t \in \intff{0}{1} }$
of potentials this uniformly in $t \in [0,1]$ and up to a $O(N^{-\kappa-1})$, $\kappa>0$ and fixed, remainder. Then one can build on the basic formula:
\beq
\ln\bigg(\frac{\mc{Z}_{N}^{(\be)}[V_1]}{\mc{Z}_{N}^{(\be)}[V_0]}\bigg) = -N^2\cdot \Int{0}{1} \mathbb{E}^{V_t}_N\bigg[ \Int{}{} \partial_{t} V_t(x)\cdot\dd L_N^{(\bs{\la})}(x)\bigg] \cdot \dd t 
\label{ecriture reconstruction fct part par interpolation 1 lin stat}
\enq
so as to obtain the asymptotic behaviour of the left-hand side up to a $O(N^{- \kappa})$ remainder. If by some other means  one can access to the large-$N$ asymptotics of 
$\ln \mc{Z}_{N}^{(\beta)}[V_0]$  up to $O(N^{ - \kappa})$  remainder, then one can deduce the expansion  of $\mc{Z}_{N}^{(\beta)}[V_1]$ to the same order. 
The terms of order $N \ln N$, $\ln N$, and the transcendental constant term in the asymptotic expansion of the partition function usually do not arise from the fluctuations of linear statistics, 
but rather from an "integration constant" or from some additional singularities present in the confining potential. The comparison to some known, Selberg-like integral often seems the only way of accessing to these terms,  especially in what concerns 
the highly non-trivial constant terms in such asymptotic expansions. Note that when $\be=2$ one can build on orthogonal polynomial techniques to access to the 
$N\ln N $ and $\ln N$ terms. Also, recently, some progress in an alternative approach to computing the logarithmic terms has been achieved in \cite{LebleSerfatyLDPCoulombGasInManyDimensions}.

\subsection{Asymptotic expansion of the correlators via Schwinger-Dyson equations}
\label{SDSection}
As already mentioned, for a general potential, going beyond the leading order demands taking into account the effect of fluctuations of the integration variables around their large-$N$ equilibrium distribution. 
The most effective way of doing so consists in studying the large-$N$ expansion of the multi-point expectation values of test functions versus the probability measure induced by $\mc{Z}_N^{(\be)}[V]$, \textit{i.e.} the quantities:
$$
\mathbb{E}_{N}^{V}\bigg[\Int{}{} f(x_1,\ldots,x_n)\cdot\prod_{i = 1}^n \dd L_N^{(\bs{\lambda})}(x_i)\bigg]\,,
$$
sometimes called $n$-linear statistics. Indeed the access to the sufficiently uniform in the potential large-$N$ expansion of the $1$-linear statistic allows one to deduce the asymptotic
expansion of $\mc{Z}_N^{(\be)}[V]$ by means of  \eqref{ecriture reconstruction fct part par interpolation 1 lin stat}. 
As we shall explain in the following, these linear statistics satisfy a tower of equations which allow one to express the $n$-linear statistics in terms of $k$-linear statistics with $k\leq n+1$. 
These equations are usually called the Schwinger-Dyson equations and, sometimes, also referred to as "loop equations". 

We are first going to outline the structure and overall strategy of the large-$N$ analysis of the Schwinger-Dyson equations on the example of 
a polynomial potential. The latter simplifies some expressions but also allows one to make a connection with the problem of enumerating certain maps. Then, we shall discuss
the case of a general potential which will be closer, in spirit, with the techniques developed in the core of the book.

\subsubsection{Moments and Stein's method}

For the purpose of this subsection, we shall assume that the potential is a polynomial of even degree
$V_{\e{pol}}(x)=\frac{\beta}{2}\big\{\frac{x^2}{2} +\sum_{j=1}^{2d} \frac{t_j x^j}{j}\big\}$ with $t_{2d}>0$ so that $\mathbb P_N^{(\be)}$ is well-defined. 
 \eqref{mumg} implies that for any integer number $p$, the $p$-th moment
\begin{equation}\label{momentsbeta}
m_N(p):=\mathbb E_N^{V_{\e{pol}}}\bigg[\frac{1}{N}\sum_{a = 1}^N \lambda_a^p\bigg]= -\frac{2p}{\beta N^2}\,\partial_{t_p}\ln \mc{Z}_N^{(\beta)}\big[ V_{\e{pol}} \big] |_{t_{\bullet}=0}\,. 
\end{equation}
has a power series expansion in $(t_j)_{j = 1}^{2d}$, whose coefficients enumerate maps (see Section~\ref{Enumer}) with one marked face of degree $p$. 
To explain what convergent matrix integrals and their asymptotic expansions on the one hand, and generating series of maps and their topological decomposition in (inverse) powers of $N$ on the other hand, 
have to do with each other, we shall use Schwinger-Dyson equations.

Probably, the simplest example of a Schwinger-Dyson equation can be provided by focusing on the standard Gaussian law $\ga$ on $\R$. An integration by parts shows that $\ga$ satisfies 
\begin{equation}\label{SDGauss}
\int xf(x)\,\dd\gamma(x)=\int f'(x)\,\dd\gamma(x)\,.
\end{equation}
for a sufficiently big class of test functions $f$. In fact, $\ga$ is the unique probability measure on $\R$ which satisfies \eqref{SDGauss}. Recall that the $p^{\e{th}}$-moment of the Gaussian law 
has the combinatorial interpretation of counting the number of pairings of $p$ ordered points. One can, in fact, build on the above Schwinger-Dyson equation so as to deduce
such a combinatorial interpretation by checking that the moments satisfy the same recurrence relation than the enumeration of pairings. 

The strategy to extract the large-$N$ behaviour of moments \eqref{momentsbeta} relies first on the derivation of the system of Schwinger-Dyson equations they satisfy.
As for the Gaussian law, it is obtained by an integration by parts.  To write this equation down, we denote by  $\overline{m}_N$ the quenched moments, i.e. the real-valued random variable:
$$
\overline{m}_N(p)= \frac{1}{N}\sum_{a=1}^N \lambda_a^p\,.
$$
We shall as well adopt the convention that $\overline{m}_N(p) = 0$ when $p < 0$. Then, integration by parts readily yields 
\begin{equation}\label{firstSD}
\mathbb E_N^{V_{\e{pol}}}\bigg[\sum_{l=0}^{p-1} \overline{m}_N(l) \, \overline{m}_N(p-1-l) +\frac{1}{N}\bigg(\frac{2}{\beta} - 1\Big)p \,  \overline{m}_N(p-1) - \overline{m}_N(p+1) -\sum_{j = 1}^{2d} t_j\,  \overline{m}_N(p+j-1)\bigg]=0\,.
\end{equation}
Compared to \eqref{SDGauss}, this equation depends on the dimension parameter $N$. Note that \eqref{firstSD} not only involves the observables $m_N = \mathbb{E}_{N}^{V_{\e{pol}}}[\overline{m}_{N}]$, 
but also the covariance of $\big\{ \overline{m}_{N}(p) \big\}_{p \geq 0}$. To circumvent this fact, let us assume first that these moments self-average, so that this covariance is negligible. 
Let us assume as well that the expectations $m_N(p)$ are bounded by some $C^p$ for some $C$ independent of $N$, this for all $p \le P(N)$ where $P(N)$ is some sequence going to infinity with $N$. 
Such information imply that the sequence $\big\{ m_N(p) \big\}_{p \geq 0}$ admits a limit point $\big\{ m(p) \big\}_{p \geq 0}$. Equation \eqref{firstSD} then implies that any such limit point $\big\{ m(p) \big\}_{p \geq 0}$ must satisfy
\beq
m({p+1})=\sum_{l=0}^{p-1} m(l)\,  m(p-1-l)-\sum_{j = 1}^{2d} t_j \, m({p+j-1})\,.
\label{ecriture eqn limite moment mesure}
\enq
Moreover, $m(p)\le C^p$ for all $p$. It is then not hard to see that the limiting equation \eqref{ecriture eqn limite moment mesure} has a unique solution  such that $m(0)=1$ provided the $t_i$'s are small enough: indeed this
is clear for $t_j=0$ and the result is then obtained by a straightforward perturbation argument, see \cite{GuionnetMaurelSegalaSecondOrderAsymptoticsMatrixModels} for details.  Finally, one can check that this unique solution is also given by
$$
M(p)=\sum_{\ell_1,\ldots,\ell_{2d} \geq 0} \bigg\{\prod_{j = 1}^{2d} \frac{(-t_j)^{\ell_j}}{\ell_j!}\bigg\} {\rm Map}_{0}(p,\ell_1,\ldots, \ell_{2d})\,,
$$
where ${\rm Map}_{0}(p,\ell_1,\ldots, \ell_{2d})$ is the number of connected planar maps with one marked, rooted face of degree $p$, and $\ell_j$ faces of degree $j$, for $1\le j\le 2d$.  
Indeed, Tutte surgery -- which consists in removing the root edge on the marked face, and describing all the possible maps ensuing from this removal -- reveals that these numbers satisfy the recursive relation:
\begin{eqnarray}
{\rm Map}_{0}(p+1,\ell_1,\ldots, \ell_{2k}) & = & \sum_{q=1}^{p - 1} \sum_{\ell_j \geq l_j \geq 0} \bigg\{\prod_{j = 1}^{2d}  \left(\begin{array}{c} l_j\cr
\ell_j\end{array}\right)\bigg\}\,{\rm Map}_{0}(q,l_1,\ldots,l_{2d}) {\rm Map}_{0}(p-q-1,\ell_1-l_1,\ldots,\ell_{2d}-l_{2d}) \nonumber \\
&& +\sum_{j = 1}^{2d}  \ell_j\,{\rm Map}_{0}(p+j-1,\ell_1,\ldots,\ell_{j-1}, \ell_j-1, \ell_{j+1},\ldots,\ell_{2d})\,. \nonumber
\end{eqnarray}
which turns into the Schwinger-Dyson equation for the generating function $M(p)$.

This strategy to prove convergence to the generating function of planar maps is very similar to the so-called Stein's method in classical probability, 
which is widely used to prove convergence to a Gaussian law. This method can roughly be summarised as follows. One considers a sequence of probability measures $\big\{ \mu_N \big\}_{N\geq 0 }$ on the real line and assume 
that there exists differential operators $\big\{ \mathcal{L}_N \big\}_{N \geq 0}$ such that for all $N$,
$$
\mu_N\Big[ \mathcal{L}_N[f] \Big] \, = \, 0
$$
for a set of test functions. Assume moreover that $\big\{ \mu_N \big\}_{N\geq 0}$  is tight and that $\mathcal{L}_N $ converges towards some operator $\mathcal{L}$.
Then, if this convergence holds in a sufficiently strong sense, any limit point $\mu$ of the sequence $\big\{ \mu_N \big\}_{N\geq 0}$ 
should satisfy 
$$
\mu\Big[ \mathcal{L}[f] \Big] \, = \, 0 \; .
$$
If moreover there exists a unique probability measure $\mu$ satisfying these equations, then this entails the convergence of the sequence $\big\{ \mu_N \big\}_{N}$ towards $\mu$. 
For instance, convergence to the Gaussian law is proved when $\mathcal{L}[f](x)=f'(x)-xf(x)$.  
Higher order of the expansion can be obtained similarly in the case where one knows that $\mathcal{L}_N$ admits an asymptotic expansion:
$$
\mathcal{L}_N= \mathcal{L}+ \frac{\mathcal{L}^{(1)}}{N}+ \frac{\mathcal{L}^{(2)}}{N^2} + \cdots
$$
so that if $\mathcal{L}$ is invertible one could hope to prove an asymptotic expansion of the form:
$$
\mu_N =\mu+ \frac{\mu^{(1)}}{N} + \frac{\mu^{(2)}}{N^2} +\cdots
$$
at least when integrated against a suitable class of test functions, and with:
\begin{equation}
\label{Stein1}\mu^{(1)}[f]= -\mu\Big[\mathcal{L}^{(1)} \circ \mathcal{L}^{-1}[ f] \Big],\quad \mu^{(2)}[f]= -\mu^{(1)}\Big[\mathcal{L}^{(1)} \circ \mathcal{L}^{-1}[f]\big] -\mu\big[\mathcal{L}^{(2)} \circ \mathcal{L}^{-1} [f]\Big]\,.
\end{equation}

The main difference in   $\beta$ ensembles is that the Schwinger-Dyson equation \eqref{firstSD} is not closed on the $\big\{ m_N(p) \big\} _{p \geq 0}$, 
namely that involves auxiliary quantities (covariances) which cannot be determined by the equation itself.
In order to study the large-$N$ behaviour of the moments by means of the Schwinger-Dyson equation \eqref{firstSD} it is convenient to re-centre the quenched moment 
around their mean leading to 
\bem
\mathbb E_N^{V_{\e{pol}}}\bigg[\sum_{l=0}^{p-1}\big( \overline{m}_N(l) -m_N(l) \big) \, \big( \overline{m}_N(p-1-l) -  m_N(p-1-l) \big) \bigg]
 - m_N(p+1)  \\
+ \frac{1}{N}\bigg(\frac{2}{\beta} - 1\Big)p \,  m_N(p-1) -\sum_{j = 1}^{2d} t_j\,  m_N(p+j-1) \, + \, \sul{l=0}{p-1} m_N(l) \, m_N(p-1-l) \, = \, 0\,.
\end{multline}
If one then assumes that the covariance produces $\e{o}(1/N)$ contributions and that the moments admit the expansion :
$$
m_N(p) = m(p)+ \frac{ \Delta m_N(p) }{N} + \e{o}(1/N)\,,
$$
one would find 
$$
\Xi[\Delta m_N](p+1) \sim \Big(\frac{2}{\beta}-1\Big)\,  p \,  m_N(p-1)\,,
$$
where $\Xi$ is the endomorphism of $\mathbb{R}^{\mathbb{N}}$, which associates to a sequence $\big\{ v(p)\big\}_{p \geq 0}$, the new sequence:
\begin{equation}
\label{Xioper}\Xi[v](p)= v(p)- 2\sum_{l=0}^{p-2} m(l) v(p-l-2) +\sum_{j = 1}^{2d} t_j v(p+j-2)\,.  
\end{equation}
When all $t_j$'s are equal to zero, $\Xi$ is represented by a triangular (semi-infinite) matrix with diagonal elements equal to one, and therefore it is invertible.
A perturbation argument shows that $\Xi$ is still invertible when the $t_j$'s are small enough. Hence, we deduce that
$$
\lim_{N\to\infty} \Delta m_N = \Big(\frac{2}{\beta}- 1\Big)\,p \, \Xi^{-1}[m](p-1)\,.
$$

To get the next order of the corrections, one needs to be able to characterise the leading large-$N$ behaviour of the covariance.  
To this end, following \cite{AmbjornChekovKristiansenMakeenkoAsymptoticSaddlePointMatrixIntegrals,AmbjornChekovMakeenkoFirstEffectiveFormulationLoopEqns,CheckovEynardArgumentsAndSomeCalculationsTopoExpGnrlBetaEns},
one derives a "rank $2$" Schwinger-Dyson equation, 
which will give access to the limit of the appropriately rescaled $N$ covariance in the spirit of Stein's method. 
This equation is obtained by considering the effect, to the first order in $\veps$, of an infinitesimal perturbation of the potential
$V(x) \rightarrow V(x) + \varepsilon x^k$ in the first Schwinger-Dyson equation \eqref{firstSD}, \textit{i.e.} 
$t_j \rightarrow t_j + 2\delta_{j,k} k \tf{ \epsilon }{ \be }$. 
It results in the insertion of a factor of $\overline{m}_N(k)$ in the expectation value in \eqref{firstSD}. 
If we introduced the centred random variable $\widetilde{m}_N= N(\overline{m}_N- m_N)$, this "rank 2" equation can be put in the form:
\begin{equation}
\mathbb E_N^{V_{\e{pol}}}\bigg[ \sum_{l = 0}^{p - 1} \widetilde{m}_{N}(k) \, \overline{m}_{N}(l) \, \overline{m}_{N}(p - 1 - l)  
- \widetilde{m}_{N}(k) \, \Xi[\widetilde{m}_{N}](p + 1) + \frac{1}{N}\Big(\frac{2}{\beta} - 1\Big) p \, \widetilde{m}_{N}(k)\, \widetilde{m}_{N}(p - 1)\bigg] = k \, m_N(k - 1 + p)\,.
\label{SDsecond}
\end{equation}
In this equation, we simplified some terms exploiting the fact that $\widetilde{m}_{N}$ is centred and that the first Schwinger-Dyson equation \eqref{firstSD} is satisfied. 
Again, assuming that the first term in \eqref{SDsecond} is negligible, we would deduce that:
$$
\lim_{N\to\infty} \mathbb E_N^{V_{\e{pol}}}\big[ \widetilde{m}_N(k) \, \widetilde{m}_N(p) \big]=-k\Xi^{-1}[S^{k - 2}m](p) := w(k,p)\,.
$$
where $S^{k - 2}m$ is the sequence whose $p$-th term is $m(k - 2 + p)$.

Plugging back this limit into the first Schwinger-Dyson equation yields the second order correction:
$$
m_N(p)=m(p)+ \frac{m^{(1)}(p)}{N}+ \frac{m^{(2)}(p)}{N^2} +o\Big(\frac{1}{N^2}\Big)\,,
$$
with
$$
\Xi[m^{(2)}](p+1)=\sum_{l=0}^{p-1}  \Big\{w(l,p-1-l) +m^{(1)}(l)m^{(1)}(p-1-l)\Big\} \; + \;  \Big(\frac{2}{\beta} - 1\Big)pm^{(1)}(p-1) \,.
$$
Again, one can check that when $\beta=2$, $m^{(2)}$ is the generating function for maps with genus $1$ as its derivatives at the origin satisfy the same recursion relations, which in the case of maps are derived similarly by Tutte surgery.
The same type of arguments can be carried on to all orders in $1/N$.

As a summary, to obtain the asymptotic expansion of moments, and hence of the partial derivatives of the partition function \textit{c.f.} \eqref{momentsbeta},  
we see that one needs uniqueness of the solution to the limiting equation (to obtain convergence of the observables), 
invertibility of the linearised operator $\Xi$ (to solve recursively the linearised equations), a priori estimates on covariances, or more generally of the correlators (in order to be able to get approximately closed 
linearised equations for the observables). The expansion can then be established and computed recursively, and this recursion is the topological recursion of \cite{EynardOrantinTopologicalExpansionsGeneralForm}.

\subsubsection{The Schwinger-Dyson equations for a general potential}

For the $\beta$-ensembles in presence of a general potential $V$ the Schwinger-Dyson equation also arise from an integration by parts. 
The first Schwinger-Dyson equation takes the form:
$$
\mathbb{E}_{N}^{V}\bigg[\,\frac{\beta}{2}\Int{}{} \frac{f(x) - f(y)}{x - y}\cdot\dd L_N^{(\bs{\la})}(x)\cdot \dd L_N^{(\bs{\la})}(y) 
+ \frac{1}{N}\bigg(1 - \frac{\beta}{2}\Big) \Int{}{} f'(x)\cdot \dd L_N^{(\bs{\la})}(x) - \Int{}{} V'(x)f(x)\cdot\dd L_N^{(\bs{\la})}(x)\bigg] = 0
$$
Similar equations can be derived for test functions depending on $n$-variables, although we shall not present them explicitly here. 
We see that the first Schwinger-Dyson equation relates $1$ and $2$-linear statistics. 
More generally, the $n$-th Schwinger-Dyson equations relates $n$-linear statistics to $k$-linear statistics with $k \leq (n + 1)$. 
Therefore, as such, the Schwinger-Dyson equations do not allow one for the computation of the $n$-linear statistics. However, it turns out that these equations 
are still very useful in extracting the large-$N$ asymptotic expansion of the $n$-linear statistics. For instance, the leading order as $N \rightarrow \infty$ of the first Schwinger-Dyson provides one with 
an equation satisfied by the equilibrium measure:
\beq
\label{SDleading} \frac{\beta}{2} \Int{}{} \frac{f(x) - f(y)}{x - y}\cdot \dd\mu_{\e{eq}}(x) \cdot \dd\mu_{\e{eq}}(y)\, - \Int{}{} V'(x)f(x)\cdot \dd\mu_{\e{eq}}(x) = 0 \;. 
\enq
 This equation is actually implied by differentiating the equality case in the Euler-Lagrange equation \eqref{carnun} for $\mu_{\e{eq}}$,
and then integrating the result against $f(x)\dd\mu_{\e{eq}}(x)$. The most important point, though, is that one can build on the Schwinger-Dyson 
equations so as to go beyond the leading order asymptotics. Doing so is achieved by carrying out a bootstrap 
analysis of the system of Schwinger-Dyson equations. The latter allows one to turn a rough estimate on the $(k + 1)$-linear statistics into an improved estimate of the $k$-th statistics.
One repeats such a scheme until reaching the optimal order of magnitude estimates. On the technical level, the essential step of the bootstrap method consists in the inversion of a master operator $\mathcal{K}$, 
which appears in the "centring" of the Schwinger-Dyson equation around $\mu_{\e{eq}}$, namely by substituting $L_N^{(\bs{\la})} = \mu_{\e{eq}} + \mc{L}_{N}^{(\bs{\la})}$ 
in the first Schwinger-Dyson equation what, owing to  the identity \eqref{SDleading}, leads to:
\bem
\mathbb{E}^{V}_N\Bigg[ 
\Int{}{} \mathcal{K}[f](x)\cdot \dd\mc{L}_{N}^{(\bs{\la})}(x) + \frac{\beta}{2}\Int{}{}\frac{f(x) - f(y)}{x - y}\cdot \dd \mathcal{L}_{N}^{(\bs{\la})}(x) \cdot \dd\mc{L}_{N}^{(\bs{\la})}(y) 
+ \frac{1}{N}\bigg(1 - \frac{\beta}{2}\bigg) \Int{}{} f^{\prime}(x)\cdot \dd\mc{L}_{N}^{(\bs{\la})}(x) \Bigg]  \\ 
\, = \, -\frac{1}{N}\bigg(1 - \frac{\beta}{2}\bigg) \Int{}{} f^{\prime}(x) \cdot \dd \mu_{\e{eq}}(x)
\label{equation SD centree order 1 beta ens}
\end{multline}
where:
\begin{equation}
\label{Koper}%
\mathcal{K}[f](x) = \beta \Int{}{} \frac{f(x) - f(y)}{x - y} \cdot\dd\mu_{\e{eq}}(y) - V^{\prime}(x)\,f(x) = - V_{{\rm eff}}'(x)\,f(x) - \Fint{}{} \frac{f(y)}{x - y}\cdot \dd\mu_{\e{eq}}(y) \;. 
\end{equation}
The centred around $\mu_{\e{eq}}$ $n$-th Schwinger-Dyson equations,  $n \geq 2$, all solely involve the operator $\mathcal{K}$.

Now, assuming that the operator $\mc{K}$ is invertible on some appropriate functional space, one can recast \eqref{equation SD centree order 1 beta ens} in the form 
\bem
\mathbb{E}^{V}_N\Bigg[ \Int{}{} f(x)\cdot \dd\mc{L}_{N}^{(\bs{\la})}(x) \Bigg] \; = \; - \frac{1}{N}\bigg(1 - \frac{\beta}{2}\bigg) \Int{}{} \Dp{x} \mc{K}^{-1}[f](x)  \cdot \dd \mu_{\e{eq}}(x) \\
- \frac{1}{N}\bigg(1 - \frac{\beta}{2}\bigg) \, \mathbb{E}^{V}_N\Bigg[ \Int{}{} \Dp{x}\mc{K}^{-1}[f](x)\cdot \dd\mc{L}_{N}^{(\bs{\la})}(x) \Bigg]
- \frac{\beta}{2} \mathbb{E}^{V}_N\Bigg[  \Int{}{}\frac{ \mc{K}^{-1}[f](x) - \mc{K}^{-1}[f](y)}{x - y}\cdot \dd \mathcal{L}_{N}^{(\bs{\la})}(x) \cdot \dd\mc{L}_{N}^{(\bs{\la})}(y) \Bigg] \;. 
\nonumber
\end{multline}
The term arising in the first line is deterministic and produces an $\e{O}(1/N)$ behaviour. The first term in the second line is given by a $1$-linear centred statistic that is 
preceded by a factor of $N^{-1}$. It will thus be sub-dominant in respect to the deterministic term. In fact, its contribution to the asymptotic expansion of $1$-linear statistics is the easiest
to take into account. Indeed, assume that one knows the asymptotic expansion of $1$-linear statistics up to $\e{O}\big( N^{-k}\big)$ and wants to push it one order in $N$ further. 
Then, the term we are discussing will automatically admit an asymptotic expansion up to $\e{O}\big( N^{-k-1}\big)$ what readily allows one to identify its contribution 
to the next order in the expansion of $1$-linear statistics. Taken this into account, it follows that the non-trivial part of the large-$N$ expansion of 
$1$-linear statistics will be driven by the one of $2$-linear centred statistics. In all cases, if one assumes that the $2$-linear statistics 
produce $\e{o}(1/N)$ contributions, the first Schwinger-Dyson equation yields immediately the first term in the large-$N$ expansion of $1$-linear statistics. 
In order to push the expansion further, one should access to the first term in the large-$N$ expansion of $2$-linear centred statistics. 
The latter can be inferred from the second Schwinger-Dyson equation. We shall however, not go into more details. 
\vspace{2mm}

The main point is that one can push the large-$N$ expansion of $k$-linear statistics, this to the desired order of precision in $N$, 
by picking lower order corrections out of the higher order Schwinger-Dyson equations. 
The effectiveness of such a bootstrap analysis  is due to the particular structure of the Schwinger-Dyson equations. 
It was indeed discovered in the early 90s that the coefficients of topological expansions in $1/N$ of the $n$-point correlators are determined recursively by the Schwinger-Dyson equations. The calculation of the first sub-leading correction to \eqref{ecriture large N asymptotics beta ensembles} 
based on the use of Schwinger-Dyson equations for correlators was first carried out in the seminal papers of Ambj\o{}rn, Chekhov and Makeenko \cite{AmbjornChekovMakeenkoFirstEffectiveFormulationLoopEqns}
and of these authors with  Kristjansen \cite{AmbjornChekovKristiansenMakeenkoAsymptoticSaddlePointMatrixIntegrals}. 
The approach developed in these papers allowed, in principle, for a formal\footnote{Namely based, among other things, on the assumption of the very existence of the 
asymptotic expansion.}, order-by-order computation of the large-$N$ asymptotic behaviour of $\mc{Z}_N^{(2)}[V]$.  
However due to its combinatorial intricacy, the approach was quite complicated to set in practice.  
In \cite{EynardTopologicalExpansions1MatrixIntegralsFirstIntro}, Eynard proposed a rewriting of the solutions of Schwinger-Dyson equations 
in a geometrically intrinsic form that strongly simplified the structure and intermediate calculations. Chekhov and Eynard then described the corresponding diagrammatics \cite{CheckovEynardBeta2AllGeneraFreeEnergy}, 
and it led to the emergence of the so-called topological recursion fully developed by Eynard and Orantin in
\cite{EynardOrantinTopologicalExpansions2MatrixIntegrals,EynardOrantinTopologicalExpansionsGeneralForm}. It allows,
in its present setting, for systematic order-by-order calculation of the coefficients arising in the 
large-$N$ expansions of the $\be$-ensemble partition functions, just as numerous other instances
of multiple integrals, see \textit{e}.\textit{g}. the work of Borot, Eynard and Orantin \cite{BorotEynardOrantinAbstractLoopEqnsTopoRecAndApplications}. Eynard, Chekhov, and 
subsequently these authors with Marchal have developed a similar\footnote{For potentials with logarithmic singularities and $\beta \neq 2$, some additional terms must be included \cite{CheckovLogSingPotential}.} 
theory \cite{CheckovEynardArgumentsAndSomeCalculationsTopoExpGnrlBetaEns,CheckovEynardMarchalTopoRecAndQuantumAlgGeometry1,CheckovEynardMarchalTopoRecAndQuantumAlgGeometry2}. 
For $\beta = 2$, Kostov \cite{KostovMatrixModelsAndCFTToCalculateAmplitudes} has also developed an interpretation of 
the coefficient arising in the large-$N$ expansions as conformal field theory amplitudes for a free boson living on a Riemann surface that is associated with the equilibrium measure. It was argued in \cite{KostovOrantinequivalenceKostovCFTApproachAndTopoRec} 
that Kostov's approach is indeed equivalent to the formalism of the topological recursion.

To summarise, the above works have elucidated the \textit{a priori} structure of the large $N$ expansions. We have not yet discussed the problem of actually proving the existence
of an asymptotic expansion of $\ln \mc{Z}_N^{(\be)}[V]$ to all algebraic orders in $N$, namely the fact that
\beq
\ln \mc{Z}^{(\be)}_N[V]  \; =  \; c^{(\be)}_{1}\,N\ln N + c_{0}^{(\be)}\ln N + \sul{k = 0 }{K} N^{2-k} F_k^{(\be)}[V] + O(N^{-K})
 \label{ecriture DA fct part beta ens une coupure}
\enq
for any $K \geq 0$ and with coefficients being some $\be$-dependent functionals of the potential $V$. The existence and form of the expansion up to $\e{o}(1)$ when $\be=2$ was proven by Johansson \cite{JohanssonEigenvaluesRandomMatricesLeadingAsympt}
for polynomial $V$ under the one-cut hypothesis, this by using the machinery described above and 
\textit{a priori} bounds for the correlators that were first obtained by Boutet de Monvel, Pastur et Shcherbina
\cite{BoutetdeMonvelPasturShcherbinaAPrioriBoundsOnFluctuationsAroundEqMeas}. 
Then, the existence of the all-order asymptotic expansion at $\be=2$ was proven by Albeverio, Pastur and Shcherbina \cite{AlbeverioPasturShcherbinaProof1overNExpansionForGeneratingFunctionsAtBeta=2} 
by combining Schwinger-Dyson equations and the bounds 
derived in \cite{BoutetdeMonvelPasturShcherbinaAPrioriBoundsOnFluctuationsAroundEqMeas}. In particular, this work proved that the coefficients of the asymptotic expansion coincide with the formal generating series enumerating ribbon graphs of 
\cite{BrezinItzyksonParisiZuberCombinatorixFormaExpansionMatrixModels} -- also known under the name of "maps".
Finally, Borot and Guionnet  \cite{BorotGuionnetAsymptExpBetaEnsOneCutRegime} systematised and extended to all $\beta > 0$
the approach of \cite{AlbeverioPasturShcherbinaProof1overNExpansionForGeneratingFunctionsAtBeta=2}, hence 
establishing the existence of the all-order large-$N$ asymptotic expansion of $\mc{Z}_N^{(\be)}[V]$ at arbitrary 
$\be$ and for analytic potentials under the one-cut hypothesis. This includes, in particular, the analytic convex potentials. The starting point always consists in establishing an \textit{a priori} estimate for the fluctuations of linear statistics, 
on the basis of a statistical-mechanical analysis \cite{BoutetdeMonvelPasturShcherbinaAPrioriBoundsOnFluctuationsAroundEqMeas} or of concentration of measures \cite{BorotGuionnetAsymptExpBetaEnsOneCutRegime}, 
without assumptions on the potential beyond a sufficient regularity. This estimate takes the form:
$$
\mathbb{P}_{N}^{(\be)}\Bigg[ \bigg\{ \bs{\la}\in \R^N  \; : \;  \bigg|\Int{}{} f(x)\cdot \dd(L_N^{(\bs{\la})} - \mu_{\e{eq}})(x)\bigg| > t  \bigg\} \Bigg] \; \leq \;  \exp\Big\{-C[f]\,N^2t^2 + C'N \ln N\Big\} 
$$
for some constants $C[f] > 0$, $C'$, and for $t$ large enough independently of $N$. We remind that if $\mathcal{O}_{1},\ldots,\mathcal{O}_{n}$ are random variables, their moment is:
$$
\mathbb{E}_N^{V}\Big[\prod_{i = 1}^n \mathcal{O}_{i}\Big] = \partial_{t_1 = 0}\cdots \partial_{t_n = 0} \mathbb{P}_{N}^{(\be)}\Big[\exp\Big(\sum_{i = 1}^n t_i \mathcal{O}_i\Big)\Big]
$$
while their cumulant is defined as:
$$
 \mc{C}_n\big[f_1,\dots,f_n] = \partial_{t_1 = 0}\cdots \partial_{t_n = 0} \ln \mathbb{P}_{N}^{(\be)}\Big[\exp\Big\{\sum_{i = 1}^n t_i \mathcal{O}_i\Big\}\Big] \;. 
$$
In fact, the cumulants are enough for computing all the $n$-linear statistics. Indeed, one has the reconstruction 
$$
\mathbb{E}_{N}^{V}\bigg[ \Int{}{} \pl{a=1}{n}f_a(x_a) \cdot\prod_{i = 1}^n \dd L_N^{(\bs{\la})}(x_i)\bigg] \; = \; 
\sul{s=1}{n} \sul{ \substack{\intn{1}{n} = \\ J_1 \sqcup \dots \sqcup J_{s} }   }{} \pl{a=1}{s} \mc{C}_{|J_a|}\big[ \big\{ f_k\big\}_{k\in J_a} \big] \;.
$$
The expression for $n$-linear statistics involving genuine test functions in $n$ variables belonging to the test space $\mc{T}(\R^n)$ 
is then obtained by density of, say, $\mc{T}(\R)\otimes \cdots \otimes \mc{T}(\R)$ in $\mc{T}(\R^n)$.

It is advantageous to work with $ \mc{C}_n\big[f_1,\dots,f_n] $, 
since it is a homogeneous polynomial of degree $n$ of the re-centred measure $L_N^{(\bs{\la})} - \mu_{\e{eq}}$, and concentration occurs in this re-centred measure:
$$
\Big| \mc{C}_n\big[f_1,\dots,f_n] \Big| \leq \bigg\{\prod_{i = 1}^n \mc{N}[f_i]\bigg\}\cdot \bigg(\frac{\ln N}{N}\bigg)^{\frac{n}{2}}
$$
where $\mc{N}$ is some norm. 
Then, under the one-cut assumptions, one shows that the master operator is invertible with a continuous inverse for a suitable norm $\mc{N}$. These two pieces of information 
allow one to neglect the contribution of some of the higher order cumulants in the system Schwinger-Dyson equations and lead to a successive improvement of 
the \textit{a priori} bounds up to the optimal scale  
\beq
\mc{C}_n\big[f_1,\dots,f_n]\;=\; \f{ 1 }{ N^{n-2}  } \Big\{ \mc{W}_{n}^{(0)}\big[f_1,\dots,f_n] \, + \, \e{o}(1) \Big\} \;. 
\enq
There,  $\mc{W}_{n}^{(0)}$ is some $n$-linear functional on the space of test functions that are pertinent for the analysis. 
 The same method can be pushed further and establishes recursively an all-order asymptotic expansion for the cumulants:
\beq
\label{Wnunuexp}
\mc{C}_n\big[f_1,\dots,f_n]\;=\; \left\{\begin{array}{lll}     \sum_{g  = 0}^{ [K/2]	} N^{2 - 2g-n}\,\mc{W}_n^{(g)}\big[f_1,\dots,f_n] + o(N^{2 - K}) & \quad & {\rm if}\,\,\beta = 2 \vspace{2mm} \\ 
						    \sum_{k = 0}^{K} N^{2 - k-n}\,\mc{W}_n^{[k]}\big[f_1,\dots,f_n] + o(N^{2 - K}) & \quad & {\rm if}\,\,\beta \neq 2
\end{array}\right.
\enq
for any $K \geq 0$. The asymptotic expansion \eqref{ecriture DA fct part beta ens une coupure} for the free energy can then be obtained by the interpolation method that was outlined in Section~\ref{Interpi}.

\vspace{2mm}

Although this phenomenon will not occur in the present book, we would still like to mention for completeness that, when the support of $\mu_{\e{eq}}$ has several cuts, the form \eqref{ecriture DA fct part beta ens une coupure}
of the asymptotic expansion is not valid any more: new bounded oscillatory in $N$ contributions have to be included in $F_{k}^{(\be)}[V]$ for $k \geq 0$. Heuristically speaking, 
this effect takes its roots in the possibility the particles have to tunnel from one cut to another \cite{BonnetDavidEynardTwoCutAsymptoticsForPartFct,EynardConjectureMultiCutPartFct}. On the technical level, 
this takes its origin in the fact that the master operator $\mathcal{K}$ has a kernel whose dimension is given by the number of cuts minus one. For real-analytic off-critical potentials and general $\beta > 0$, the 
form of the all-order asymptotic expansion in the multi-cut case 
was conjectured in \cite{EynardConjectureMultiCutPartFct}, and established in \cite{BorotGuionnetAsymptExpBetaEnsMultiCutRegime}.  It was also established up to $o(1)$ by Shcherbina in \cite{ShcherbinaAsymptoticsBetaEnsemblesMultiCut}
by a different technique, namely \textit{via} a coupling to Brownian motion to replace the two-body interaction between different cuts with a linear but random one. We refer the reader to \cite{BorotGuionnetAsymptExpBetaEnsMultiCutRegime} 
for a deeper discussion relative to the history of this particular problem. We also stress that, so far, the analysis of the double scaling limit around a critical potential where 
the support of the equilibrium measure changes its topology have not been addressed mainly due to difficulties such a transition induces on the level of constructing the pseudo-inverses of the master operator.

Above, we have focused our discussion solely on the approach based on the analysis of loop equations. However, when $\be=2$, the orthogonal polynomial based determinantal structure allows one to build on the Riemann--Hilbert 
problem characterisation of orthogonal polynomials \cite{FokasItsKitaevIsomonodromyPlusRHPforOrthPly} along with the non-linear steepest descent method 
\cite{DeiftZhouSteepestDescentForOscillatoryRHP,DeiftZhouSteepestDescentForOscillatoryRHPmKdVIntroMethod}
based approach to characterising the large degree asymptotic behaviour of polynomials orthogonal with respect to varying weights \cite{DeiftKriechMcLaughVenakZhouOrthogonalPlyExponWeights,DeiftKriechMcLaughVenakZhouOrthogonalPlyVaryingExponWeights}
so as to obtain the large-$N$ asymptotic expansion of $\mc{Z}^{(2)}_N[V]$. Within the Riemann--Hilbert problem approach, Ercolani and 
McLaughlin \cite{ErcolaniMcLaughlinLargeNPartFctRMTAndGraphEnumeration} established the existence of the all order 
asymptotic expansion at $\be=2$  in the case of potentials that are a perturbation of the Gaussian interaction.
Bleher and Its \cite{BleherItsAsymptoticsFreeEnergyRMTOneCutCaseRHP} obtained, up to a $\e{o}(1)$ remainder the large-$N$ expansion of $\mc{Z}^{(2)}_N[V]$ for polynomial $V$
that give rise to a one-cut potential. Also, recently, Claeys, Grava and McLaughlin \cite{ClaeysGravaMcLaughlinAERMT2CutsBeta2} developed the Riemann--Hilbert approach so as to obtain the large-$N$ expansion of $\mc{Z}^{(2)}_N[V]$
in the case of a two-cut polynomial potential $V$.

\subsection{The asymptotic expansion of the free energy up to $o(1)$}
\label{o1nextsec}
We shall now make some general remarks about the nature of the terms arising in the asymptotic expansion of $\ln \mc{Z}^{(\be)}_N[V]$.

The terms diverging in $N$ when $N \rightarrow \infty$ are not affected by the topology of the support of the equilibrium measure. In the case of regular potentials, 
the pre-factors of the $N\ln N$ and $\ln N$ corrections in \eqref{ecriture DA fct part beta ens une coupure}  take the form:
$$
c_1^{(\be)} = \frac{\beta}{2},\qquad c_0^{(\be)} = \frac{1}{12}\bigg(3 + \frac{\beta}{2} + \frac{2}{\beta}\bigg) \;. 
$$
they can be identified from the large-$N$ asymptotics of the Gaussian partition function, \textit{cf}. \eqref{ZGAUSN}. 
Their presence in \eqref{ecriture DA fct part beta ens une coupure} is only the sign that there is a more natural normalisation of the $N$-fold integral \eqref{ecriture DA fct part beta ens une coupure} which would kill the 
logarithmic corrections in the large $N$ limit. The terms of order $N^2$ and $N$ are functionals of the equilibrium measure $\mu_{\e{eq}}$, which depend in a non-local way on the density of this measure.
As we have seen in Section~\ref{LargeNNN}, the prefactor of $N^2$ is given by a double integral involving $\mu_{\e{eq}}$. The term of order $N$ is also known to be given by a single integral involving $\mu_{\e{eq}}$.
More precisely, up to a universal --\textit{viz}. $\mu_{\e{eq}}$-independent-- function of $\beta$, it is proportional to the von Neumann entropy of the equilibrium measure
\cite{CheckovEynardArgumentsAndSomeCalculationsTopoExpGnrlBetaEns,DysonStatTheEnergyLevelsII,ShcherbinaAsymptoticsBetaEnsemblesMultiCut,WiegmannZaborodinLargeNExpbetaEnsPartFct}:
\beq
\label{entropyterm}F_{1}^{(\be)} = {\rm cte}_{\beta} + \bigg(\frac{\beta}{2} - 1\bigg) \Int{}{} \ln\bigg(\frac{\dd \mu_{\e{eq}(x)}}{\dd x}\bigg)\cdot \dd\mu_{\e{eq}}(x) \;. 
\enq
The fact that the entropy only appears as a sub-leading term is a typical feature of models having varying weights. The bounded term is affected by the topology of the support: 
it is a constant in the one-cut case and it contains an additional, oscillatory contribution in the multi-cut case. Unlike the non-decaying terms, the coefficient in front of $N^{-k}$ with $k > 0$ only depends on the local behaviour 
(at an order increasing with $k$) of the equilibrium measure's density near the endpoints of its support.

Once the asymptotic expansion of the free energy is established up to $o(1)$, and in the case it does not contain oscillatory terms, one can deduce a central limit theorem for the fluctuations of linear statistics. 
The starting point is the formula for the Fourier transform of their distribution:
\begin{equation}
\label{CLTJG}\mathbb{E}^{V}_{N}\bigg[\exp\bigg({\rm i}sN\Int{}{} f(x)\cdot \dd L_N^{(\bs{\la})} \bigg)\bigg] = \frac{\mc{Z}_{N}^{(\be)}\big[V - \frac{{\rm i}sf}{N}\big]}{\mc{Z}_{N}^{(\be)}[V]} \;. 
\end{equation}
Resorting to  arguments of complex analysis, one can fairly easily extend the validity of the asymptotic expansion \eqref{ecriture DA fct part beta ens une coupure} to potentials of the form $V_0 + \Delta V/N$, %G Tu veux peut-etre changer pour faire Laplace et ne pas avoir a passer par des potentiels complexes ?
with $V$ real-valued satisfying the previous assumptions, and $\Delta V$ complex-valued and differentiable enough. Besides, since the logarithmic correction does not depend on perturbations of the initial potential 
$V_0$ and $F_{k}^{(\be)}[V]$ are smooth functionals of $V$ away from critical points of the model, one deduces that:
$$
\mathbb{E}^{V}_{N}\bigg[\exp\bigg({\rm i}sN\Int{}{} f(x)\cdot\dd(L_N^{(\bs{\la})} - \mu_{\e{eq}})(x)\bigg)\bigg] = \exp\bigg\{{\rm i}s\,\delta_{V} F_1^{(\be)}[f] - \frac{s^2}{2}\cdot \delta^2_{V} F_0^{(\be)}[f,f] + o(1)\bigg\}
$$
where the error $o(1)$ is uniform for $s$ belonging to compact subsets of $\Cx$ and $\de_{V}G[f]$ refers to the G\^{a}teaux derivative of $G$ at the point $V$ and in the direction $f$. This implies that
\beq
\label{quickfluct}\sum_{i = 1}^N f(\lambda_i) - N \Int{}{} f(x)\cdot\dd\mu_{\e{eq}}(x) = N \Int{}{} f(x)\cdot \dd(L_N^{(\bs{\la})} - \mu_{\e{eq}})(x)
\enq
converges in law to a Gaussian random variable, with covariance given by the Hessian (defined by the second-order functional derivative) of the energy functional introduced in Section~\ref{LargeNNN}, evaluated at $V$ along the direction $f$:
$$
F_0^{(\be)}[V] = -\mathcal{E}^{(\be)}[\mu_{\e{eq}}] \;. 
$$
This central limit theorem for $V$ polynomial, $f$ differentiable enough, any $\beta =2$ in the one-cut regime, was first obtained by Johansson \cite{JohanssonEigenvaluesRandomMatricesLeadingAsympt}.
The characteristics of the Gaussian variable can be computed solely from the knowledge of the functional $F_0^{(\be)}[V]$ related to the energy functional, and of $F_1^{(\be)}[V]$ related to the entropy. 
For $\beta = 2$, $F_1^{(\be)}[V]$ vanishes owing to its prefactor in \eqref{entropyterm}, and \eqref{quickfluct} converges to a centred Gaussian variable.

We observe that the fluctuations of \eqref{quickfluct} are of order $1$. In the multi-cut regime, the tunnelling of particles between different, 
far-apart segments of the support lead as well to order $1$ fluctuations, which in general destroy the gaussianity of \eqref{quickfluct} and the central limit theorem. 
This phenomenon was predicted in \cite{PasturnoCLT}, and given a precise form in \cite{BorotGuionnetAsymptExpBetaEnsMultiCutRegime,ShcherbinaAsymptoticsBetaEnsemblesMultiCut}: the Gaussian behaviour 
is in first approximation convoluted with the law of a discrete Gaussian variable, 
\textit{i.e.} supported on a lattice of an arithmetic progression on $\mathbb{R}$ with step of order $1$ and depending on $f$, and whose initial term drifts with $N$ at a speed of order $1$. 
As a result, the fluctuations display a Gaussian behaviour with interference fringes which are displaced when $N \rightarrow N + 1$.

%G Say in quantum int. systems paragraph that the terms up to o(1) do not depend on the regularisations and are the one really pertaining to the underlying field theory.

\section{Generalisations}
\label{Nfoldlist}

It is fair to say that there exists presently a pretty good understanding of large-$N$ asymptotic expansions of $\beta$ ensembles. The main remaining open questions concern the description of the asymptotic expansion uniformly around 
critical points (\textit{viz}. when the number of cuts changes) and the possibility to 
relax the regularity of the potential, for instance by allowing the existence of singularities, \textit{e.g.} of the Fisher--Hartwig type\footnote{Although, even in these two cases, some 
partial progress has been achieved at $\be=2$ where one can build on the Riemann--Hilbert approach \cite{BleherItsDoubleScalingRMTLimit,ClaeysItsKrasovskyEmergenceSingularityToeplitzAndPV,DeiftItsKrasovskyAsymptoticsofToeplitsHankelWithFHSymbols}.}. 
What we would like to stress is that the techniques of asymptotic analysis described so far are effective in the sense that they allow, upon certain more or less obvious generalisations of technical details, treating various instances of other 
multiple integrals. 

The framework of small enough perturbation of the Gaussian potential is, in general, the easiest to deal with.  
Asymptotic expansions for hermitian multi-matrix models have been obtained in such a setting. For instance, the expansion including the first sub-leading order was derived for a two-matrix model by
Guionnet and Maurel-Segala \cite{GuionnetMaurelSegalaSecondOrderAsymptoticsMatrixModels}, the one to all orders for multi-matrix models by Maurel-Segala \cite{MaurelSegalaAllOrderExpansionMatrixModelsGaussianPert} 
and the one to all orders for unitary random matrices in external fields
was then obtained by an appropriate adaptation of the analysis of Schwinger-Dyson equations in \cite{GuionnetNovakAsymptoticsOrthogonalUnitaryIntegralsAllOrders}. 
Multi-matrix models are  interesting in operator algebras, since their ring of observables in the large $N$ limit give planar algebras. 
Without claiming exhaustiveness, they also appear in the theory of random tilings of arbitrary two-dimensional domains \cite{BorodinGorinGuionnetAsymptoticsDisceteBetaEns,EynardMatrixModelforPlanePartitions}, in gauge theories \cite{BonelliMaruyoshiTanziniYagiMMandAGTAllGenra} 
and topological strings via the topological vertex \cite{EynardKashani-PoorMarchalMatrixModelForTopoStrings,SulkowskiBetaEnsMMFromNekrasovPartFct}.
Most of the time, however, the technology for their asymptotic analysis is not sufficiently developed at present 
for the purposes of theoretical physics and algebraic geometry, even in the (rare) case where the integrand is real-valued. In fact, even the mere task of 
establishing the leading order asymptotics under fairly general assumptions is an open problem. 
%G @Alice: maybe you have more to say on motivations from C^* algebras here ?

Another natural generalisation of $\be$-ensembles consists in replacing the one-particle varying potential $N\cdot V$ by a regular and varying multi-particle 
potential
\beq
N \sul{a=1}{N}V(\la_a) \; \hookrightarrow \; \sul{p=1}{r} \frac{N^{2-p}}{p!} \sul{1 \leq i_1,\ldots,i_p \leq N}{}
V_{p}\big( \la_{i_1}, \dots, \la_{i_p} \big) \;. 
\label{ecriture changement pot 1 corps vers r corps}
\enq
For $r = 2$, such interactions were studied by G\"{o}tze, Venker \cite{GotzeVenkerLocalUniversalityGeneral2bdyBetaRepInteraction} and Venker \cite{VenkerParticleSystemsWithRepBeta} 
who showed that the bulk behaviour falls in the universality class of $\beta$-ensembles. For general $r$, Borot \cite{BorotFormalLargeNAEForMultiPtInteractionsMI} has shown  that the formal asymptotic expansion of the partition function
subordinate to multi-particle potentials is captured by a generalisation of the topological recursion. The existence of the all-order asymptotic expansion 
was established by the authors in  \cite{KozBorotGuionnetLargeNBehMulIntMeanFieldTh} under certain regularity assumptions on the multi-particle interactions. 
Note that for perturbations of the Gaussian potential of the form \eqref{ecriture changement pot 1 corps vers r corps} 
the hypothesis  of \cite{KozBorotGuionnetLargeNBehMulIntMeanFieldTh} are indeed satisfied.  We give below a non-exhaustive list of physically interesting models with $r = 2$. The ones encountered in integrable models of statistical physics will be pointed out in Section~\ref{lili}.

\subsubsection*{Biorthogonal ensembles}

For $r=2$ and when $\be=2$, the structure of such models becomes determinantal in the special cases where the two-body interaction takes the form:
\beq
V_{2}(\la_1,\la_2) = \ln\Bigg(\frac{ \big(f(\la_2) - f(\la_1) \big) \big( g(\la_2)-g(\la_1) \big) }{(\la_2 - \la_1)^2}\Bigg)\;.
\enq

It is well known that, then, the associated multiple integrals can be fully characterised in terms of 
appropriate systems of bi-orthogonal polynomials in the sense of \cite{KonhauserFirstintoBiOrthPlys}. By bi-orthogonal polynomials, we mean two families of monic polynomials $\{P_n\}_{n\in \mathbb{N}}$ and $\{Q_n\}_{n\in \mathbb{N}}$
with $\e{deg}[P_n]=\e{deg}[Q_n]=n$ and which satisfy
\beq
\Int{ \R }{} P_{n}\big(f(\la)\big) \cdot \big[g(\la) \big]^{j} \cdot \ex{-N V(\la)} \cdot \dd\la \; = \; 0 \quad \e{and} \quad 
\Int{ \R }{} Q_{n}\big(g(\la)\big) \cdot \big[f(\la) \big]^{j} \cdot \ex{-N V(\la)} \cdot \dd\la \; = \; 0  \qquad \e{for} \quad j \in \{0,\ldots, n-1\} \;.
\enq
The system of bi-orthogonal polynomials subordinate to $f$ and $g$ exists and is unique for instance when $f$ and $g$ are real-valued and monotone functions. %G Apparemment, c'est monotone ou monotonic
In that case, the multiple integral of interest can be recast as a determinant which, in turn, can be evaluated in terms of the overlaps involving the polynomials $P_n$ and $Q_n$
by carrying out linear combinations of lines and columns of the determinant:
$$
\underset{j,k\in \intn{1}{N}}{\det_N}\bigg[ \Int{\R }{} \big[f(\la) \big]^{j-1} \cdot\big[g(\la) \big]^{k-1} \cdot \ex{-N V(\la)} \cdot \dd\la\bigg] \; = \; 
\pl{n=0}{N-1} \bigg\{   \Int{\R }{} P_n\big(f(\la) \big) \cdot Q_n\big(g(\la) \big) \cdot \ex{-N V(\la)} \cdot \dd\la   \bigg\} \;. 
$$

It is due to their connections to bi-orthogonal polynomials that such multiple integrals are referred to as bi-orthogonal ensembles. The case $f(\la) = \la^{\theta}$ and $g(\la)=\la$  is of special interest, 
since the bi-orthogonal polynomials can be effectively described. In \cite{BorodinBiOrthogonalEnsIntroAnStudyOfExamples} Borodin was able to establish certain universality results for specific examples of confining potentials $V$.
Furthermore, it was observed, first on a specific example by Claeys and Wang \cite{ClaeysWangIntroRHPForBioRthSinhLikePlys} and 
then in full generality by Claeys and Romano \cite{ClaeysRomanoStudyForGeneralPeriodsOfSinh-BasedPlys}
that the bi-orthogonal polynomials can be characterised by means of a Riemann-Hilbert problem. 
However, for the moment, the Riemann--Hilbert problem-based machinery still did not lead to the asymptotic evaluation of the associated partition functions.

\subsubsection*{Statistical physics in two-dimensional random lattices}

For $\beta = 2$, with the same precise meaning that was discussed in Section~\ref{Enumer}, the $N$-fold integral:
$$
\Int{ \R^N }{} \prod_{1 \leq a < b \leq N} |\la_a - \la_b|^{2}\,\exp\bigg\{\frac{1}{u}\bigg(- \sum_{a = 1}^{N} \frac{N\,\la_a^2}{2} + \sum_{p = 1}^{r} \sum_{h =0}^{H} \frac{N^{2 - 2h - p}}{p!} 
\sum_{\substack{m_1,\ldots,m_p \geq 1 \\ 1 \leq a_1,\ldots,a_p \leq N}} t_{m_1,\ldots,m_p}^{(h)} \prod_{i = 1}^p \frac{\la_{a_i}^{m_i}}{m_i}\bigg)\bigg\} \cdot \dd^N \bs{\la}
$$
enumerates maps whose faces, instead of being restricted to be homeomorphic to disks, can have any topology. Each face homeomorphic to a surface of genus $h$, with $p$ boundaries of respective perimeters $m_1,\ldots,m_p$, 
is counted with a local weight $t_{m_1,\ldots,m_p}^{(h)}$. This model was introduced in \cite{BorotFormalLargeNAEForMultiPtInteractionsMI}, and encompasses a large class of statistical physics models on two-dimensional random lattices. 
The simplest ones occur for $r = 2$, and the only other interesting cases we are aware of have $r = \infty$. The general $r = 2$ case is equivalent to the enumeration of maps carrying a configuration of self-avoiding loops crossing certain faces, 
each loop being counted with a Boltzmann weight $n$. The special case where the faces crossed by the loops are all triangles corresponds to:
\beq
\label{OnmodelV2}V_{2}(x,y) = -n\,\ln\big|1 - z\cdot (\la_a + \la_b)\big|
\enq
where $z$ is the local weight per triangle crossed by a loop. This is the $O(n)$ loop model which was first introduced by Kostov \cite{KostovOnModel}. The $6$-vertex model on a random lattice \cite{Kostov6vertexRandomlattice} is realised by:
\beq
\label{6vmodelV2} V_{2}(x,y) = -4\ln \big|\ex{ \frac{{\rm i}\gamma}{2} } \la_a - \ex{-\frac{{\rm i}\gamma}{2}}\la_b\big| \;. 
\enq
In \eqref{OnmodelV2} for $|n| < 2$ and in \eqref{6vmodelV2} for any real $\gamma$, the existence and uniqueness of the equilibrium measure is known, and the results of \cite{KozBorotGuionnetLargeNBehMulIntMeanFieldTh}
for an all-order asymptotic analysis apply within the one-cut hypothesis\footnote{It was shown in \cite{BBG1cut}, under fairly general hypothesis, that the equilibrium measures relevant for combinatorics are supported on a single segment.}.

\subsubsection*{Chern-Simons theory in Seifert spaces}

The perturbative expansions of ${\rm SU}(N)$ gauge theories lead to a weighted enumeration of Feynman graphs, which are dual to embedded graphs in surfaces, and t'Hooft observed that the $N$-dependence of the weights 
only comes from the topology of the surface. For this reason, $N$-fold integrals related to matrix models are very common in gauge theories, although the form of the probability measure might be complicated. 
For Chern-Simons theory on a simple class of three-dimensional manifolds $\msc{M}$ called Seifert spaces (which include the three-sphere), the measure can be explicitly computed 
\cite{BarNatanLawrenceRationalSurgeryLMOInvariant,MarinoChernSimonsMatrixIntegralsEtc}, 
and leads to a partition function for $\beta = 2$ and $r = 2$ with one-point and two-point interactions
$$
V_1(\la) = c_2\cdot\frac{(\ln \la)^2}{2u} + c_1\cdot \ln \la,\qquad V_{2}(\la_1,\la_2) = \sum_{i = 1}^k \ln\bigg(\frac{\la_1^{1/a_i} - \la_2^{1/a_i}}{\la_1 - \la_2}\bigg)
$$
where $(a_1,\ldots,a_k)$ are integers, $c_1,c_2$ rational numbers related to the geometry of $\msc{M}$, and $u$ is related to the coupling constant of the Chern-Simons theory. 
In this case of interest, the integration runs through $(\R^+)^N$.
For the cases $\chi = 2 - k + \sum_{i = 1}^k a_i^{-1} \geq 0$, the suitably defined energy functional is convex, and this implies existence and uniqueness of the equilibrium measure. 
The cases $2 - k + \sum_{i = 1}^k a_i^{-1} < 0$ are rather interesting: although one does not know currently how to prove uniqueness of the equilibrium measure via potential theory by lack of convexity,
Monte-Carlo simulations of the distribution of eigenvalues by Wei\ss{e} seems to indicate that it should be unique \cite[Appendix]{BorotEynardSUNOnSeifertSpaces}. The all-order asymptotic expansion 
of the correlators in these models receive an interpretation in terms of perturbative knot invariants in $\msc{M}$, and by large $N$-dualities, they can be related to topological strings amplitudes in suitable 
target spaces \cite{BorotEynardSUNOnSeifertSpaces,BorotBriniCHTheoryofsphericalSeifert}. As an example of application of the existence of the all-order large-$N$ asymptotic expansion established in \cite{KozBorotGuionnetLargeNBehMulIntMeanFieldTh}, 
Borot and Eynard derived some arithmetic properties of these perturbative knot invariants in \cite{BorotEynardSUNOnSeifertSpaces}.

\subsubsection{Multispecies $\beta$-ensembles}

The $\beta$-ensemble with the two-point interactions can be generalised to several types of particles, and appear in the study of coupling between conformal field theories with internal degrees of freedom (describing matter)
and two-dimensional quantum gravity. From the probabilistic point of view, these models are potentially the source of new universality classes that can be more usually found in multi-matrix models.
But, as they are already written as $N$-fold integrals, their study is much simpler than the multi-matrix models in which it is necessary to integrate over spaces of (non-commuting) matrices, without the possibility of simultaneous diagonalisation.

Let $\mathcal{D}$ be a graph having $M$ vertices and possibly multiple edges, and let $\mathbf{A}$ stand for its adjacency matrix --\textit{viz}. $A_{v,w}$  corresponds to the number of edges
that link the vertices $v$ and $w$--.  \cite{KharchevMarshakovMironovMorozovPakuliakConformalAltMultiMatrixModels} 
introduced the $N$-fold integral:
\beq
\label{ITEPDd}\Int{ \R^N }{} \prod_{v \in \mathcal{D}} \prod_{a = 1}^{N_{v}} \bigg\{ \dd\la_a^{(v)}\,e^{-NV_{v}(\la_{a}^{(v)})} \bigg\} \cdot  \prod_{a < b} \big|\la_a^{(v)} - \la_b^{(v)}\big|^{2} 
\prod_{\substack{\{v,w\}\,\,{\rm edge} \\ {\rm in}\,\,\mathcal{D}}} \prod_{\substack{1 \leq a \leq N_{v} \\ 1 \leq b \leq N_{w}}} \big|\la_a^{(v)} - \la_{b}^{(w)}\big|^{-\frac{A_{v,w}}{2}} \;. 
\enq
The $\big\{ \la_a^{(v)} \big\}_{a = 1}^{N_{v}}$ are thought as the positions of $N_{v}$ particles of type $v$, and $N = \sum_{v \in \mathcal{D}} N_v$ denote the total number of particles. 
These models appear in the $\mathcal{N} = 2$ supersymmetric gauge theories associated to ADE quivers \cite{DijkgraafVafaGeometryandMM}. Their study has been revived recently 
\cite{BonelliMaruyoshiTanziniYagiMMandAGTAllGenra,KharchevMarshakovMironovMorozovPakuliakConformalAltMultiMatrixModels} 
in view of the conjectures of 
Alday-Gaiotto-Tachikawa \cite{AldayGaiottoTachikawaAGTConjecture} which propose a precise relation between four-dimensional quiver gauge theories and the conformal blocks of Liouville theory of two-dimensional quantum gravity. 

Kostov introduced the slightly different model:
\beq
\label{KostovDd}\Int{ (\mathbb{R}^{+})^N}{} \prod_{v \in \mathcal{D}}\prod_{a = 1}^{N_{v}} \bigg\{ \dd\la_a^{(v)}\,e^{-NV_{v}(\la_{a}^{(v)})} \bigg\} \prod_{1 \leq a < b \leq N_v} \big|\la_{a}^{(v)} - \la_{b}^{(v)}\big|^2\cdot 
\prod_{\substack{\{v,w\}\,\,{\rm edge} \\ \in \mathcal{D}}} \prod_{\substack{1 \leq a \leq N_v \\ 1 \leq b \leq N_{w}}} \big|\la_{a}^{(v)} + \la_{b}^{(w)}\big|^{-\frac{A_{v,w}}{2}} \;. 
\enq
Its partition function enumerates maps in which each face has a colour chosen among the set of vertices of $\mathcal{D}$, restricted in such a way that faces of colour $v$ and $v'$ can be adjacent 
if and only if there is an edge between $v$ and $v'$ in $\mathcal{D}$. Each interface between a cluster of colour $v$ and a cluster of colour $v'$ is weighted by the number of edges between $v$ and $v'$ in $\mathcal{D}$. 
Up to an affine change of variables, we retrieve the $O(n)$-model \eqref{OnmodelV2} in the case in which $\mathcal{D}$ has a single vertex from which issue $n$ loops. 
The $q$-Potts model corresponds to the case where $\mathcal{D}$ is the complete graph on $q$ vertices, and the model corresponding to $\mathcal{D} = $"the Dynkin diagram of type $A_{n}$" 
is called the (restricted) height model. One can show \cite[Lemma 5.5]{BorotEynardOrantinAbstractLoopEqnsTopoRecAndApplications} 
that the suitably defined energy functional for these models is strictly convex if and only if $(2 - \mathbf{A})$ is the Cartan matrix of a Dynkin diagram of type A, D, or E, or of the extended 
Dynkin diagrams $\widehat{A}$, $\widehat{D}$, or $\widehat{E}$, or of the cyclic graph. The $N$-fold integral \eqref{KostovDd} is simpler to analyse than \eqref{ITEPDd} due to the absence
\symbolfootnote[2]{The singularities at $\la_a^{(v)} = -\la_b^{(w)}$ are absent since the integration runs through $\R^+$}
of logarithmic singularities 
when $\la_a^{(v)} = \la_b^{(w)}$.  New universality classes remembering the ADE symmetries occur precisely when the confining potentials $V_{v}$ are tuned so that the support $S_{v}$ 
of the equilibrium measure for the particles of 
type $v$ approaches $0$. Indeed, at the vicinity of $0$ the attractive interaction with $-\la^{(w)}_{a}$ of all other particles will change the local distribution of the particles. To our knowledge, the universal distributions governing 
these universality classes have not been derived (although the original work of \cite{KharchevMarshakovMironovMorozovPakuliakConformalAltMultiMatrixModels} exhibits some integrable 
structure in these models), maybe because this model is not so well-known in the community working in random matrix theory 
from the point of view of probabilities.

\subsubsection*{Conduction in disordered wires}

Experiments showed that the properties of quantum transport of electrons in chaotic cavities feature some universality, and therefore, one can expect to capture these properties as typical in an ensemble of random cavities. 
The simplest model consists of two cavities related by two wires, in which $N$ modes can propagate. Landauer theory describes the conduction in such a system by a $2N \times 2N$ scattering matrix:
$$
\mathbf{S} = \left(\begin{array}{cc} \mathbf{r} & \mathbf{t}' \\ \mathbf{t} & \mathbf{r} \end{array}\right)
$$
such that the amplitudes of the $N$ modes in the first cavity is related to the amplitudes of the $N$ modes in the second cavity by multiplication by $\mathbf{S}$. 
Conservation of the current implies that $S$ is unitary, and in turn this implies that the matrices $\mathbf{tt^{\dagger}}$, $\mathbf{t'(t')^{\dagger}}$, $\mathbf{1 - rr^{\dagger}}$ and $\mathbf{1 - r'(r')^{\dagger}}$ 
have identical spectrum consisting of $N$ eigenvalues $\la_1,\ldots,\la_N \in \intff{0}{1}$. To understand the transport properties in this setting, it is necessary to investigate the behaviour of the linear statistics $\sum_{a = 1}^N f(\la_a)$. 

In the model, the distribution of the $\la_a$ is drawn from a $\beta$-ensemble with $V(\la)$ proportional to $\ln \la$. In a model of non-ideal leads and for $\beta = 2$, the distribution depending on the mean
free path $\ell$ and the length $L$ of the wire is proportional to \cite{BeenakkerRajaeisTransmissionEVinDisorderedWire}:
\beq
\label{acdc}\prod_{1 \leq a < b \leq N} |\la_a - \la_b| \cdot \pl{a=1}{N}\ex{-N V(\la_a)}\cdot \mathop{{\rm det}}_{1 \leq a,b \leq N}  K_{a}\big[L/\ell N\,;\,\la_b\big]  \cdot \dd^N \bs{\la}
\enq
for kernels $K_{a}$ which involve Gauss hypergeometric functions, whereas for $\beta \neq 2$ it is unknown.  In the metallic regime, we have $1 \ll L/\ell \ll N$, and the distribution \eqref{acdc} 
simplifies drastically to a $\beta$-ensemble (here, for all $\beta = 1,2,4$) with a two-body interaction:
\beq
\label{nonmetalacdc}   \pl{a=1}{N}\ex{-N V(\la_a)}\cdot \prod_{1 \leq a < b \leq N} \big|\la_a - \la_b\big|\cdot \big|{\rm argsech}^2(\la_a^{1/2}) - {\rm argsech}^2(\la_b^{1/2})\big| \cdot \dd^N \bs{\la}
\enq
where ${\rm sech}(x) = \frac{1}{{\rm cosh} x}$ and ${\rm argsech}$ is its reciprocal function, and $V$ is some explicit one-body interaction. Whereas \eqref{nonmetalacdc} falls into the class of models which can be treated with the existing methods of \cite{KozBorotGuionnetLargeNBehMulIntMeanFieldTh},
the structure of \eqref{acdc} is much more involved and goes beyond the present technology based on Schwinger-Dyson equations. 

We refer to \cite{BeenakkerRMTApproach2QuantumTransport} and references therein for a justification of these distributions, 
as well as for a deeper overview of the relations between random matrix theory (notably in the form of $N$-fold integrals) and quantum transport.

\section{$\be$-ensembles with non-varying weights}
\label{Sous-Section Beta Ens avec poids non variants}

In all the examples of the multiple integrals discussed so far, the interaction potential $V$ is preceded by a power of $N$. This scaling ensures that, for typical configurations of the $\la_a$'s, 
the logarithmic repulsion is of the same order of magnitude in $N$ than the confining potential. As a consequence, 
 with overwhelming probability when $N \rightarrow \infty$, the integration variables remain in a bounded region and exhibit a typical spacing $1/N$. 
The scheme developed in \cite{AlbeverioPasturShcherbinaProof1overNExpansionForGeneratingFunctionsAtBeta=2,BorotGuionnetAsymptExpBetaEnsOneCutRegime,KozBorotGuionnetLargeNBehMulIntMeanFieldTh,
DeiftKriechMcLaughVenakZhouOrthogonalPlyVaryingExponWeights} for the asymptotic analysis was adapted to this particular 
tuning of the interactions with $N$ and, in general, breaks down if the nature of the balance between the interactions changes.

Serious problems relative to extracting the large-$N$ asymptotic behaviour already start to arise in the case of \emph{non-varying} weights, \textit{i.e.} for multiple integrals:
\beq
\Int{ \R^N }{} \pl{a<b}{N} |y_a-y_b|^{\be}\,\pl{a=1}{N} \ex{- W(y_a)} \cdot \dd^N \bs{y}  \;. 
\label{partition functions beta non varying weight}
\enq
Indeed, consider the integral \eqref{partition functions beta non varying weight} for $N$-large and focus on the contribution of a 
bounded domain of $\R^N$. In this case, the logarithmic interactions are dominant with respect to the confinement 
(and this by one order in $N$): the dominant contribution of such a region is obtained by spacing the $y_{a}$'s as far apart as possible.
Increasing the size of such a bounded region will increase the value of the dominant contribution, at least until the confining nature
of the potential kicks in. Hence, to identify the configuration maximising the value of the integral, one should rescale the integration variables
as $y_a = T_N \la_{a}$ with  $T_N \rightarrow \infty$. The sequence $T_N$ would then be chosen in such a way that the $2$-body interaction  and the confinement ensured by 
the potential have the same order of magnitude in $N$,  the ideal situation being:
\beq
W(T_N \la) \; = \; N\cdot V_N(\la) \qquad \e{with} \qquad V_N(\la) \, =  \, V_{\infty}(\la) \cdot (1+\e{o}(1) \big)
\label{ecriture potentiel rescale et approximation}
\enq
for some potential $V_{\infty}$ and pointwise almost-everywhere in $\la$. These new variables $\la$ are typically distributed in a bounded region
and have a typical spacing $1/N$.

The simplest illustration of  such a mechanism issues from the case of a polynomial potential 
$V(\la) = \sum_{a=1}^{2\ell} c_a \la^{a}$, $c_{2\ell}>0$. In this case, the sequence $T_N$ takes the form $T_N=N^{\tf{1}{(2 \ell)}}$. 
Note that, up to a trivial prefactor, the two-body interaction $\la \mapsto |\la|^{\be}$ is invariant under dilatations. 
As a consequence, for polynomial potentials, the asymptotic analysis can still be carried out by means of the previously described methods
\cite{DeiftKriechMcLaughVenakZhouOrthogonalPlyExponWeights}, with minor technical complications due to the handling of a $N$-dependent potential.  
Although illustrative, the polynomial case is by far not representative of the complexity represented by working with non-varying weights. Indeed, the genuinely hard part of the analysis stems form the fact that, in 
principle, in the expansion \eqref{ecriture potentiel rescale et approximation}:
\begin{itemize}
 \item the remainder may not be "sufficiently" uniform ;
\item the non-varying potential $W$ may have singularities in the complex plane. 
This last scenario means that the singularities of the rescaled potential $V_N$ given in \eqref{ecriture potentiel rescale et approximation}
  will collapse, with a $N$-dependent rate, on the integration domain.
\end{itemize}
In this situation, the usual scheme for obtaining sub-leading corrections breaks down. 
So far, the large-$N$ asymptotic analysis of a "non-trivial" multiple integral of the type \eqref{partition functions beta non varying weight}
were carried out only when $\be=2$ and this for only a handful of examples. 
 Zinn-Justin  \cite{ZinnJustinSixVertexDWAsMMIntegral} proposed an $N$-fold multiple integral 
representation of the type \eqref{ecriture potentiel rescale et approximation} for the partition function of the six-vertex model 
in its massless phase and subject to domain wall boundary conditions. 
By using a proper rescaling of the variables suggested in \cite{ZinnJustinSixVertexDWAsMMIntegral},
Bleher and Fokin \cite{BleherFokinAsymptoticsSixVertexFreeEnergyDWBCDisorderedRegime}  carried out the large-$N$ asymptotic analysis of the associated multiple integral within the Riemann--Hilbert problem approach to orthogonal polynomials. 
The most delicate point of their analysis was to absorb 
the contribution of the sequence of poles $\zeta_n/N$, $n=1,2,\dots$, of the rescaled potential that were collapsing on $\R$. \textit{In fine}, they
obtained the asymptotic expansion of the logarithm of the integral up to $\e{o}(1)$ corrections. 
 
The situation may, in fact, very easily be much worse than the scheme described above, simply because \eqref{ecriture potentiel rescale et approximation} might not even hold to the leading order with an $N$-independent
$V_{\infty}$. A simple example can be provided by $W(y)=y^2(3+\cos(y))$ whose rescaled large-$N$ leading behaviour has $N$-dependent oscillatory terms.  
 
To conclude, it seems fair to state that despite the considerable developments that took place over the last 20 years
in the field of large-$N$ asymptotic expansion of $N$-dimensional integrals, the techniques of asymptotic
analysis are still far from enabling one to grasp the large-$N$ asymptotic behaviour of multiple integrals
lacking the presence of a scaling of interactions.  Such integrals arise quite naturally in concrete applications. 
For instance, it is well known that correlation functions in quantum integrable models
are described by $N$-fold multiple integrals \cite{JimboKedemKonnoMiwaXXZChainWithaBoundaryElemBlcks,JimboMikiMiwaNakayashikiElementaryBlocksXXZperiodicDelta>1,JimboMiwaElementaryBlocksXXZperiodicMassless,KMTElementaryBlocksPeriodicXXZ} 
or series thereof \cite{KozKitMailSlaTerXXZsgZsgZAsymptotics}. 
Usually, for reasons stemming from the physics of the underlying model, one is interested in the large-$N$ behaviour 
of these integrals and, in particular, in the constant term arising in their asymptotics. 
However, for most cases of interest, the given $N$-fold integrals have a much too complicated integrand 
in order to apply any of the existing methods of analysis.

\section{The integrals issued from the method of quantum separation of variables}
 
\subsection{The quantum separation of variables for the Toda chain}
 
 The quantum separation of variables method refers to a technique allowing one the determination of the spectrum, eigenvectors and correlation functions of quantum integrable models. 
The method takes its roots in the 1985 work of Sklyanin \cite{SklyaninSoVFirstIntroTodaChain} and applies to a wide range of lattice quantum integrable models 
such as spin chains \cite{FaldellaKitanineNiccoliSpectrumAndScalarProductsForOpenXXZNonDiagBC,KitanineMailletNiccoliOpenXXZGenericBCCompletenessofBA,
NiccoliCompleteSpectrumAndSomeFormFactorsInhomogeneousOpenXXZChain,NiccoliCompleteSpectrumAndSomeFormFactorsDyn6VertexAnd8Vertex},
lattice discretisations of quantum field theories in 1+1 dimensions \cite{BytskoTeschnerSinhGordonFunctionalBA,GrosjeanMailletNiccoliFFofLatticeSineG,NiccoliTeschnerSineGordonRevisited} 
or multi-particle quantum Hamiltonians \cite{KharchevLebedevIntRepEigenfctsPeriodicTodaFromWhittakerFctRep,KharchevLebedevSemenovTianShanskyRTCBEigenfunctions,SklyaninSoVFirstIntroTodaChain}. 
We will outline the main ideas of the method on the example of the open quantum Toda chain Hamiltonian with $(N + 1)$-particles \cite{BBT}:
\beq
\mathtt{H}_{{\rm Td}} \; = \; \sul{a=1}{N+1} \f{ \op{p}_{a}^2 }{2} \; + \; \ex{ \op{x}_{N+1} - \op{x}_1} 
\; + \;  \sul{a=1}{N} \ex{\op{x}_a -\op{x}_{a+1} }\;.
\enq
Above, $\op{x}_a$ is to be understood as the operator of multiplication by the $a$-th coordinate $\op{x}_a \cdot \Phi( \bs{x}) \, = \, x_a \Phi(\bs{x})$ while 
$\op{p}_a$ is the canonically conjugated operator, $\op{p}_a \cdot \Phi( \bs{x}) = -{\rm i} \hbar  \tf{ \Dp{} \Phi(\bs{x}) }{ \Dp{} x_a } $, so that $\big[ \op{x}_a, \op{p}_{b} \big] \; =  \; \de_{a,b} {\rm i} \hbar$. 
Here, $\bs{x}$ denotes a $N+1$ dimensional vector  $\bs{x} = (x_1,\ldots,x_{N + 1})$. 
Within such a realisation of the operators $\op{x}_n$ and $\op{p}_n$, the Toda chain Hamiltonian is a multi-dimensional partial differential operator acting on the Hilbert space 
$\mathcal{H}_{{\rm Toda}}\,  = \,  L^2\big( \R^{N+1}, \dd^{N+1}\bs{x} \big)$. 

The quantum Toda chain Hamiltonian is a quantum integrable model. This means, among other things, that $\mathtt{H}_{{\rm Td}}$  can be embedded into a 
family $\{\op{t}_k\}_{k=0}^{N}$ of operators in involution, conveniently collected as coefficients of the polynomial:
\beq
\op{t}(\la) \; = \; \la^{N+1}\, + \, \sul{k=0}{N+1} (-1)^{N+1-k}\lambda^k\,\op{t}_k
\enq
such that $\mathtt{H}_{{\rm Td}} = \mathtt{t}_2 - \mathtt{t}_1^2$. Thus, solving the spectral problem associated with $\mathtt{H}_{   \e{Td}}$ means, in fact, solving a multi-dimensional (due to the dimensionality of the ambient space)
and multi-parameter (due to the necessity to keep track of eigenvalues $t_k$ of the operators $\op{t}_k$) spectral problem
\beq
\op{t}_k\cdot \Phi(\bs{x}) \; = \; t_k \cdot \Phi(\bs{x}) \;. 
\enq
Since the model is translation invariant, its spectrum will contain a Lebesgue continuous part corresponding to the spectrum of the 
total momentum operator $\op{P}_{\e{tot}}=\sum_{a =1}^{N+1} \op{p}_a$. However, if one puts oneself in the centre of mass frame, or if one fixes the origin of 
coordinates at $x_{N + 1}$, \textit{viz}. sets $x_{N+1} = 0$, then the restricted operator has already a purely point-wise spectrum, see \textit{e.g.} \cite{AnCompletenessEigenfunctionsTodaPeriodic}. 

The idea of quantum separation of variables is to build -- using the various symmetries stemming from the quantum integrability of the model -- a unitary transform:
\beq
\msc{U}\; : \; \mathcal{H}_{ \e{sep} } \, = \, L^2\big( \R^{N+1}, \dd \nu \big) \,\, \longrightarrow \,\, \mathcal{H}_{ \e{Toda} } \, = \, L^2\big( \R^{N+1}, \dd^{N+1} \bs{x} \big) 
\enq
such that the transformed operator $\msc{U}^{-1}\op{t}(\la) \msc{U}$ becomes "separated". 
The $L^{2}$ space $\mathcal{H}_{\e{sep}}$ is endowed with a measure $\dd \nu$ which is part of the unknowns in the problem of constructing $\msc{U}$.  
By "separated", we mean that one would like $\msc{U}$ to intertwine the $\op{t}(\la)$ operator with a direct sum of finite difference operators in \textit{one}-variable. 
It turns out that this problem can be solved. The measure  $\dd \nu$ on $\mathcal{H}_{ \e{sep} }$ factorizes $\dd\nu = \dd \mu \otimes  \dd \veps$ into a product of a "trivial" one-dimensional Lebesgue measure $\dd \veps$ that 
takes into account the spectrum $\veps$ of the total momentum operator, and a non-trivial measure $\dd \mu$ which is absolutely continuous in 
respect to the $N$-dimensional Lebesgue measure $\dd^N\bs{y}$ and takes the form 
\beq
\dd \mu(\bs{y}) \; = \; \mu(\bs{y}) \cdot \dd^N \bs{y} 
\quad \e{with} \quad  \mu(\bs{y})  \; = \; 
\f{1}{(2\pi \hbar)^{N}}  \pl{ a <  b }{ N } \bigg\{  \f{ y_a-y_b }{\pi \hbar} \cdot \sinh\Big[ \f{\pi(y_a - y_b)}{\hbar}\Big]  \bigg\}\;. 
\label{Ecriture mesure Sklyanin Partie Intro}
\enq
When applied to sufficently well behaved functions $\wh{\Phi}\in \mathcal{H}_{\e{sep}}$, the action of the unitary operator $\msc{U}$ takes the form of an integral transform 
\beq
\msc{U}\big[ \wh{\Phi} \big](\bs{x},x_{N + 1}) \; = \; 
\Int{\R^{N+1} }{}  \vp_{\bs{y}} (\bs{x})  \cdot  \ex{\f{ {\rm i} }{\hbar}(\veps-\ov{\bs{y}})x_{N+1}}     
\cdot \wh{ \Phi }(\bs{y}; \veps) \cdot   \f{ \dd\mu(\bs{y}) }{ \sqrt{N!} } \otimes \dd \veps  \qquad \e{with} \quad \ov{ \bs{y} }_{N} \; = \; \sul{a=1}{N} y_a
\label{definition transfo integrale U super cal}
\enq
where $\bs{x}$ and $\bs{y}$ are $N$-dimensional vectors and the non-trivial ingredients $\vp_{\bs{y}}(\bs{x})$ are the ${\rm GL}(N,\R)$ Whittaker functions.  The unitarity of the transform $\msc{U}$
as given by \eqref{definition transfo integrale U super cal} has been established in \cite{WallachRealReductiveGroupsII} by means of harmonic analysis on groups, and in 
\cite{KozUnitarityofSoVTransform} by means of the quantum inverse scattering method.

One can show \cite{AnCompletenessEigenfunctionsTodaPeriodic} that for the Toda chain, the original spectral problem attached to the family $\big\{ \mathtt{t}(\la) \big\}_{\la \in \mathbb{C}}$ of operators in involution, 
is in one-to-one correspondence with the problem of finding all solutions $\big( t(\la) \, , \,   q_{t}(\la) \big)$ 
to the below Baxter T-Q equation 
\beq
t(\la) \cdot  q_{t}(\la) \; = \; ({\rm i})^{N+1} q_{t}(\la + {\rm i} \hbar) \; + \;  (-{\rm i})^{N+1} q_{t}(\la-{\rm i}\hbar) 
\label{ecriture eqn TQ scalaire Toda coeur du Chapitre SoV Toda}
\enq
that, furthermore, satisfy the conditions: \vspace{1mm}
\begin{itemize}
 \item[i)] $t(\la)$ is a polynomial of the form $t(\la) \, = \, \prod_{k=1}^{N+1}(\la-\tau_k)$ with $\big\{ \tau_k \big\}_{k = 1}^{N + 1} = \big\{ \tau_k^* \big\}_{k = 1}^{N + 1}$; \vspace{1mm}
 \item[ii)]  $\la \mapsto q_{t}(\la)$ is entire and satisfies, for some $N$-dependent $C>0$, to the bound
\beq
 \abs{ q_{t}(\la) }\, \leq  \, C \cdot \exp\bigg\{-\f{(N + 1)\pi}{2\hbar}|{\rm Re}\,\la|\bigg\}  \abs{\la}^{\frac{N+1}{2\hbar}(2\abs{{\rm Im}\,\la}-\hbar)} \qquad \e{uniformly}\;\e{in}\quad 
\la \in \paa{ z \in \mathbb{C}\; : \; |{\rm Im}\,z | \leq \tf{\hbar}{2}}  \;; 
\label{ecriture borne sur fction q Baxter Toda}
\enq
\item[iii)] the roots $\{\tau_k\}_1^{N+1}$ satisfy  to $\sum_{k=1}^{N+1}\tau_k = \veps$. \vspace{1mm}
\end{itemize}
The condition iii) relates the $\tau_k$'s to the eigenvalue $\veps$ of the total momentum operator $\mathtt{P}_{{\rm tot}}$. This constraint issues from the fact that the Toda chain 
Hamiltonian is invariant under translation, hence making it more convenient
to describe the spectrum of the chain directly in a sector corresponding to a fixed eigenvalue $\veps$ of $\mathtt{P}_{{\rm tot}}$. After such a reduction, $q_{t}$ represents
the "normalisable" part of the Toda chain eigenfunction, associated with the purely point-wise spectrum. 
More precisely, if $(t(\la),q_t(\la))$ is a solution to the $T-Q$ equation \eqref{ecriture eqn TQ scalaire Toda coeur du Chapitre SoV Toda} then 
\beq
\Phi_{\veps;t}(\bs{x},x_{N + 1}) \; = \; \Int{\R^{N+1} }{}  \vp_{\bs{y}} (\bs{x})  \cdot  \ex{\f{ {\rm i} }{\hbar}(\veps-\ov{\bs{y}})x_{N+1}}     
\cdot \pl{a=1}{N} \big\{ q_t(y_a) \big\} \cdot f(\veps) \cdot   \f{ \dd\mu(\bs{y}) }{ \sqrt{N!} } \otimes \dd \veps 
\enq
represents a wave packet having a dispersion in $\veps$ momentum space given by $f \in L^{1}(\R)$. Further the function 
\beq
\Phi_{\veps;t}^{(\e{norm})}(\bs{x}) \; = \; \Int{\R^{N} }{}  \vp_{\bs{y}} (\bs{x})  \cdot \pl{a=1}{N} \big\{ q_t(y_a) \big\}  \cdot   \f{ \dd\mu(\bs{y}_{N}) }{ \sqrt{N!} } 
\enq
represents the "normalisable" part of the generalised eigenfunction of the operators $\op{t}(\la)$ associated with the eigenvalues $t(\la)$ and a total 
momentum $\veps$. One speaks of a separation of variables since the normalisable part of the generalised eigenfunction is given by a product of functions in 
one variable $q_t(\la_a)$, $a = 1,\dots,N$.

The spectral problem associated with the Baxter equation might seem under-determined, in the sense that it contains too many unknowns. 
To convince oneself of the contrary, at least heuristically, it is helpful to  make the parallel with the Sturm-Liouville 
spectral problem 
\beq
\e{find} \; \e{all} \quad  \big( E, f \big) \in \R \times H_2(\R) \; \quad \e{such} \; \e{that} \quad - f^{\prime\prime}(x) \, + \,  V(x)\,  f(x)  \; = \; E \cdot f (x) 
\enq
with $V$ sufficiently regular and growing fast enough at infinity and $H_2(\R)$ is the second Sobolev space. Although the above ordinary differential equation admits two linearly independent solutions 
for any value of $E$, only for very specific values of $E$ does one finds solutions belonging to $H_2(\R)$. 
Regarding to \eqref{ecriture eqn TQ scalaire Toda coeur du Chapitre SoV Toda}, the regularity and growth requirements on $q_{t}$ play the same role as the $H_{2}(\R)$ space in the Sturm-Liouville problem:
the T-Q equation admits solutions $(t, q_{t} )$
belonging to the desired class only for well-tuned monic polynomials of degree $N+1$. It is precisely this effect that gives rise to so-called quantisation conditions for the Toda chain.  

In light of the above discussion, the quantum separation of variables may be thought of as a way to map a multi-parameter and multi-dimensional spectral problem onto a multi-parameter (so as to keep track of the different eigenvalues of the 
$\op{t}_k$'s) but \textit{one}-dimensional spectral problem. This results in a tremendous simplification of the problem.

Nekrasov and Shatashvilii  conjectured in \cite{NekrasovShatashviliConjectureTBADescriptionSpectrumIntModels} and Kozlowski and Teschner later proved in \cite{KozTeschnerTBAToda}
that it is possible to construct all solutions to the $T-Q$ equation for the Toda chain through solutions to non-linear integral equations. 
Namely, let $\bs{\sigma} = \{\sg_k\}_{k = 1}^{N + 1}$ be complex numbers satisfying $| {\rm Im}\,\sg_k | < \tf{\hbar}{2}$ and let $\ln Y_{\bs{\sigma}} $ denote the  continuous and bounded on $\R$ solution (if it exists) to the
non-linear integral equation:
\begin{equation}\label{TBA}
\ln Y_{ \bs{\sigma} } (\la)\,=\,\int_{\R} \dd  \mu\;K(\la-\mu)\,\ln\left( 1 +  \frac{ Y_{\bs{\sigma}}(\mu) }{ \vth(\mu-{\rm i} \hbar/2)\vth(\mu+ {\rm i} \hbar/2) }\right)\,,
\end{equation}
where
\begin{equation}\label{kerneldef}
K\pa{\la} = \f{ \hbar } { \pi ( \la^2+\hbar^2 ) }  \qquad \e{and} \qquad 
\vth(\la)=\prod_{k=1}^{N+1}(\la-\sg_k) \;.
\end{equation}
 Starting from $Y_{ \bs{\sigma} }$ one constructs the functions:
\beq
\ln v_{\uparrow}\pa{\la}  = - \Int{\R}{}  \f{\dd \mu}{2{\rm i}\pi}  
\f{1}{ \la-\mu +   {\rm i} \tf{\hbar}{2} } \cdot \ln\left(1+ \f{  Y_{\bs{\sigma}} \pa{\mu} }{ \vth\pa{\mu - {\rm i}\tf{\hbar}{2}}  \vth\pa{\mu + {\rm i}\tf{\hbar}{2}}  }\right)\,,
\enq
and
\beq
\ln v_{\downarrow}\pa{\la- {\rm i} \hbar}  =  \Int{\R}{}  \f{\dd \mu}{2 {\rm i} \pi}  \f{1}{ \la-\mu - {\rm i} \tf{\hbar}{2}} \ln\left(1+ \f{ Y_{\bs{\sigma}} \pa{\mu} }{ 
\vth\pa{\mu - {\rm i}\tf{\hbar}{2}}  \vth\pa{\mu + {\rm i}\tf{\hbar}{2}}  }\right) \;.
\label{definition v up-down section vth}
\enq
These auxiliary functions $v_{\ua/\da}$ then give rise to the functions 
\begin{equation}
\mf{q}_{\bs{\sigma}}^+(\la)\, =\,    \f{  \hbar^{ {\rm i}\f{(N+1)\la}{\hbar} } \ex{-\f{(N+1)\pi}{\hbar}\la} \cdot v_{\uparrow}\pa{\la}}
{ \pl{k=1}{N+1} \bigg\{ \Ga \bigg(1- {\rm i} \f{\la-\sg_k}{\hbar} \bigg) \bigg\}  } 
\; \; ,\qquad
\mf{q}_{\bs{\sigma}}^-(\la)\, =\,  \f{ \hbar^{-{\rm i}\f{(N+1)\la}{\hbar} }  \ex{-\f{(N+1)\pi}{\hbar}\la} \cdot v_{\downarrow}\pa{\la-i\hbar} }
{\pl{k=1}{N+1} \bigg\{ \Ga \bigg(1 +  {\rm i} \f{\la-\sg_k}{\hbar} \bigg) \bigg\}   }\,.
\label{definition Q vth plus/moins} 
\end{equation}
 One can show that the ratio  
\beq
t_{\bs{\sigma}} (\la) \; =  \;
\f{ \mf{q}_{\bs{\sigma}}^{+}(\la-{\rm i} \hbar) \mf{q}_{\bs{\sigma}}^{-}(\la +  {\rm i} \hbar)  \, - \,  \mf{q}_{\bs{\sigma}}^{+}(\la + {\rm i} \hbar) \mf{q}_{\bs{\sigma}}^{-}(\la - {\rm i} \hbar)  }
{\mf{q}_{\bs{\sigma}}^+(\la) \mf{q}_{\bs{\sigma}}^-(\la+ {\rm i}\hbar) \, - \, \mf{q}_{\bs{\sigma}}^+(\la + {\rm i} \hbar) \mf{q}_{\bs{\sigma}}^-(\la)  }  
\label{t-defn}
\enq
is, in fact, a monic polynomial in $\la$  of degree $(N+1)$ that has, furthermore, a self-conjugated set of roots. All these quantities being given one constructs the function, depending on $\bs{\sigma}$ and a parameter $\zeta$:
\beq
q_{\bs{\sigma}}(\la) \; = \; \f{ \mf{q}_{\bs{\sigma}}^+(\la) - \zeta \mf{q}_{\bs{\sg} }^-(\la) }{\prod_{k=1}^{N+1} \big\{\ex{-\f{\pi \la}{\hbar}}\sinh\frac{\pi}{\hbar}(\la-\sg_k) \big\}}
\enq
which is a meromorphic solution to the T-Q equation associated with the polynomial $t_{\bs{\sigma}}(\la)$ that, furthermore, satisfies to the growth estimates  \eqref{ecriture borne sur fction q Baxter Toda}. 
 The pair $(t_{\bs{\sigma}}, q_{\bs{\sigma}})$ provides ones with a solution to the Baxter T-Q equation if and only if the parameters $\{\sg_l\}_{l = 1}^{N + 1}$ and $\zeta$ satisfy to the quantisation conditions, for any $k \in \{1,\ldots,N + 1\}$:
\begin{align}\label{ecriture conditions de quantification}
2\pi n_k & = \f{ (N+1) \sg_k}{\hbar} \ln \hbar   + {\rm i} \ln \zeta
-{\rm i} \sul{m=1}{N+1} \ln \frac{ \Gamma\big( 1+{\rm i} \tf{ (\sg_k-\sg_m) }{\hbar} \big) }{ \Gamma \big( 1 - {\rm i} \tf{ (\sg_k-\sg_m) }{\hbar} \big) } \\
& \hspace{1.5cm} +\Int{\R}{} \f{\dd \tau}{2\pi} \paa{ \f{1}{ \sg_k-\tau + {\rm i} \tf{\hbar}{2}}  +\f{1}{ \sg_k-\tau - {\rm i} \tf{\hbar}{2}}  }
\ln \pa{  1+ \f{  Y_{\bs{\sigma}}\pa{\tau} }{ \vth \pa{\tau - {\rm i} \tf{\hbar}{2} }  \vth \pa{ \tau + {\rm i}\tf{\hbar}{2} }  }   } 
\nonumber\end{align}
and $\sul{k=1}{N+1}\sg_k=\veps$.

It was shown in \cite{KozTeschnerTBAToda} that any solution to the T-Q equation gives rise to a solution $Y_{\bs{\sigma}}$ to \eqref{TBA} with a set of parameters $\bs{\sigma} = \{\sg_k\}_{k = 1}^{N + 1}$
satisfying to the quantisation conditions \eqref{ecriture conditions de quantification} and, reciprocally, that any solution to \eqref{TBA} with parameters $\bs{\sigma}$
satisfying to  \eqref{ecriture conditions de quantification} gives rise to the solution $(t_{\bs{\sigma}} (\la), q_{\bs{\sigma}}(\la)) $ to the T-Q equations.

\subsection{Multiple integral representations}

The objects of main interest to the physics of a quantum integrable model are its correlation functions, namely expectation values of products of certain physically relevant operators 
taken between two eigenstates of the Hamiltonian of the model. The simplest such objects are the \emph{form factors}, namely expectation values of so-called quasi-local operators. %G What is a quasi local operator ?
In the language of the Toda chain, such operators only act on a fixed subset of variables $(x_1,\dots, x_r)$. The knowledge of form factors allows one, in principle, to access to all correlation functions 
involving products of quasi-local operators acting on different sets of variables: it is enough to insert the closure relation in between each of the operators.  %G What is the closure relation ? Is it the same as "resolution of identity" ?
In order to compute the form factors of local operators within the quantum separation of variables method, one has to solve the inverse problem, that is to say find how the  %G local or quasi-local here ? Maybe one can just drop at this point of the paragraph the precision "quasi-local" ?
given local operator of interest is intertwined by the $\msc{U}$-transform. In other words, one should find how the given operator on $\mathcal{H}_{{\rm Toda}}$ acts on the space $\mathcal{H}_{{\rm sep}}$ where the separation of 
variables is realised. This inverse problem has been solved for different examples of quasi-local operators and for various models 
\cite{BabelonActionPositionOpsWhittakerFctions,BabelonQuantumInverseProblemConjClosedToda,GrosjeanMailletNiccoliFFofLatticeSineG,KozIPForTodaAndDualEqns,NiccoliCompleteSpectrumAndSomeFormFactorsInhomogeneousOpenXXZChain,
NiccoliCompleteSpectrumAndSomeFormFactorsAntiPeriodicXXZ,NiccoliCompleteSpectrumAndSomeFormFactorsXXXHigherSpin,SklyaninResolutionIPFromQDet}.

\subsubsection*{$\bullet$ The Toda chain}

The resolution of the inverse problem for the Toda chain has been pioneered by Babelon \cite{BabelonActionPositionOpsWhittakerFctions,BabelonQuantumInverseProblemConjClosedToda} 
in 2002 and further developed in the works \cite{KozIPForTodaAndDualEqns,SklyaninResolutionIPFromQDet}. 
These results, along with the unitarity of the separation of variables transform $\msc{U}$ lead to multiple integral representations for the form factors. 

Let $\Phi_{ \veps ; t }$ and $\Phi_{ \veps ; t^{\prime} }$ be two eigenfunctions of the Toda chain in the  sector characterised by the total momentum 
$\veps$ and built up from solutions to the Baxter T-Q equation associated with the polynomials $t(\la)$ and $t^{\prime}(\la)$. 
The associated finite part of the form factor\footnote{Namely the one built up from the normalisable part of the wave function, in contrast with the non-normalisable part associated with the continuous part of the spectrum described by $\veps$.} 
of the operator $\prod_{a=1}^{r} \big\{  \ex{ \op{x}_a - \op{x}_{N+1} }  \big\} $
takes the form 
\bem
\Big( \Phi_{ \veps ; t^{\prime} } , \prod_{a=1}^{r} \Big\{  \ex{ \op{x}_a - \op{x}_{N+1} }  \Big\}  \cdot \Phi_{ \veps ; t }  \Big)_{ \mid x_{N+1}=0 } \; = \; 
\f{ N! }{r! (N-r)! } \Int{ \R^N }{} \pl{a<b}{N} \Big\{\Big( \f{y_a-y_b}{\pi \hbar} \Big)\sinh\Big[\f{\pi(y_a - y_b)}{\hbar}\Big]  \Big\}
\cdot \pl{a=1}{N} \Big\{ \big( q_{ t^{\prime} }(y_a) \big)^* q_{ t }(y_a) \Big\} \\
\times \pl{a=1}{r} \bigg\{ \f{ q_{ t} (y_a+{\rm i} \hbar) }{ q_{ t }(y_a) } \cdot \pl{b=r+1}{N} \Big[ \f{ {\rm i} }{ y_a - y_b } \Big]    \bigg\} \cdot  \f{ \dd^N \bs{y} }{ (2\pi \hbar)^N } \;. 
\label{ecriture rep int mult correlateurs dans chaine Toda}
\end{multline}
The index $\mid x_{N+1}=0$ refers to the fact that the coordinates are chosen so that $x_{N+1}=0$.

\subsubsection*{$\bullet$ Lukyanov's conjecture for the Sinh-Gordon model}

Lukyanov \cite{LukyanovConjectureOfFieldExpValueSinhGAndRenormalization} argued that the vacuum expectation value of the exponential of the field operator $\Phi$
in the quantum Sinh-Gordon model should be obtained from the properly normalised large-$N$ limit 
\beq
\moy{ \ex{\a \Phi} }_{R} \; = \; \lim_{N\tend +\infty} \bigg\{ \Big(\f{ N }{m R }\Big)^{ \theta} \f{ \mf{z}_{\a} }{ \mf{z}_{0} }  \bigg\} \qquad \e{with} \qquad \theta \, = \,  \f{\a^2}{2(1+b^2)(1+b^{-2}) }\;. %G Changed \vartheta to \theta to avoid confusion with the polynomial introduced before
\label{ecriture conjecture de Lukyanov}
\enq
In the above expression $m$ is the Sinh-Gordon mass parameter,  $2\pi R$ is the volume, $b$ is the coupling constant, and $\mf{z}_{\a}$ is given by the $N$-fold multiple-integral representation
\beq
 \mf{z}_{\a} = \Int{\R^N}{}  \pl{k<\ell}{N} \Big\{ \sinh\big[(1+b^{2})(y_k-y_{\ell}) \big]  \cdot \sinh\big[(1+b^{-2})(y_k-y_{\ell}) \big]   \Big\} \cdot \pl{a=1}{N }\Big\{ \ex{-W_{\a}(y_a) }  \Big\} \cdot \dd^N \bs{y} \;. 
\label{ecriture integrale de Lukyanov}
\enq
 The one-body interaction is given by the confining potential 
\beq
W_{\a}(\la) \; = \; -\a \la \, + \, \f{  m R  \cosh(\la) }{ 2 \sin \bigg( \f{\pi}{1+b^2} \bigg) }  \; - \; \Int{ \R }{}  \f{ \ln \Big[ 1+\ex{-E(\mu) } \Big]  }{ \cosh(\la-\mu) } \cdot \f{ \dd \mu }{ \pi } \;. 
\label{confining potential W in SinhGordon Lukyanov approach} %G Changed \veps to E to avoid confusion with previous use of \varepsilon as momentum
\enq
Its expression involves the solution  $E$  to the non-linear integral equation 
\beq
E(\la) \, = \, 2\pi m R \cosh(\la) \, - \, \Int{ \R }{} \Phi(\la-\mu) \ln \Big[ 1+\ex{-E(\mu) } \Big] \cdot \dd \mu  
\enq
whose integral kernel takes the form
\beq
\Phi(\mu) \; = \;\f{  \cosh(\mu)\cos(\tau) }{ \cosh(\la+{\rm i} \tau) \cosh(\la-{\rm i} \tau)} \qquad \e{where} \qquad \tau=\f{ \pi (b^2-b^{-2}) }{ 2(2+ b^2+b^{-2} ) } \;.  
\enq
It is easy to see by using Banach's fixed point theorem that, at least for $R$ large enough, the non-linear integral equations admits a unique solution $E \in L^{\infty}(\R)$.

\textit{Per se} Lukyanov's conjecture \cite{LukyanovConjectureOfFieldExpValueSinhGAndRenormalization} takes its roots in semi-classical reasonings applied to the classical Sinh-Gordon model in finite volume. 
Later, Bytsko and Teschner \cite{BytskoTeschnerSinhGordonFunctionalBA} proposed a lattice discretisation version of the quantum Sinh-Gordon model in finite volume $2\pi R$ and implemented
the quantum separation of variables for this model. Teschner \cite{TeschnerSpectrumSinhGFiniteVolume} provided a characterisation of the solutions to the T-Q equation which describes the spectrum of that model. 
The results of these two papers suggest a representation for the expectation value of the exponential of the field in the quantum Sinh-Gordon model slightly more complex than \eqref{ecriture conjecture de Lukyanov}:
the confining potential \eqref{confining potential W in SinhGordon Lukyanov approach} arising in $\mf{z}_{\a}$
should be replaced by a more involved expression which, in particular, depends on $N$. We will however not provide this explicit form here. 
It is an open question whether the limit \eqref{ecriture conjecture de Lukyanov} exists and if it exists whether it takes the same value when one 
inserts  in $\mf{z}_{\a}$ the potential conjectured by Lukyanov and the one suggested by Teschner's analysis. Note that a thorough characterisation
of $\big< \ex{\a \Phi} \big>_{R}$ has been recently conjectured by Negro and Smirnov in \cite{NegroSmirnovOnePtFctsSinhGordonDiffrenceEqns}
on completely independent grounds; it is an open question whether the limit \eqref{ecriture conjecture de Lukyanov} does indeed gives rise to the same object.

\subsubsection*{$\bullet$ General structure of form factors in the quantum separation of variables method}

It is possible to obtain multiple integral representations for the form factors arising in other models solvable by the quantum separation of variables. 
Although we shall not discuss the precise form taken by these representation, the main feature is that the form factors are either directly expressed 
-- as in the case of the Lukyanov integral \eqref{ecriture integrale de Lukyanov} -- or very closely related -- as in the case of the position operator form factor of the 
Toda chain \eqref{ecriture rep int mult correlateurs dans chaine Toda} -- to multiple integrals of the type
\beq
\Int{\msc{C}^N}{} \pl{a<b}{N} \Big\{ \sinh[\pi\om_1(y_a-y_b)]  \sinh[\pi\om_2(y_a-y_b)] \Big\}^{\be} \cdot \pl{a=1}{N}\ex{-W(y_a)} \cdot \dd^N \bs{y} 
\enq
or their degenerations when some of the $\om_k$'s are send to $0$. There $\msc{C}$ is some curve in the complex plane which, in the simplest cases discussed above, coincides with $\R$.
The coefficients $\om_1, \om_2$ are related to the given model's coupling constants. The confining potential $W$, which can be $N$-dependent, contains all the informations
on the eigenfunctions and the operator involved in the form factor.

 \subsection{The goal of the book}
 
\label{lili}
This work aims at developing the main features of a theory that would enable one to 
extract the large-$N$ asymptotic behaviour out of the class of multiple integrals that naturally arises in the context of the so-called 
quantum separation of variables method: 
\beq
\mf{z}_N[W] \; = \; \Int{\R^N}{} \pl{a<b}{N} \Big\{ \sinh[\pi\om_1(y_a-y_b)]  \sinh[\pi\om_2(y_a-y_b)] \Big\}^{\be} \cdot \pl{a=1}{N}\ex{-W(y_a)} \cdot \dd^N \bs{y} \;. 
\label{ecriture MI qSoV comme exemple}
\enq
As discussed earlier on, the examples issuing from the quantum separation of variables correspond to taking $\be=1$ and specific choices of the potential $W$. 
Independently of its numerous potential applications to physics, should one only have in mind characterising the large-$N$ behaviour of $N$-fold 
multiple integrals, it is precisely the class of integrals described by \eqref{ecriture MI qSoV comme exemple}  that constitutes naturally the next one to investigate and understand 
after the $\be$-ensembles issued ones \eqref{int1} and \eqref{ecriture changement pot 1 corps vers r corps}.
Indeed, on the one hand the integrand in \eqref{ecriture MI qSoV comme exemple}  bears certain structural similarities with the one arising in $\be$-ensembles.
On the other hand, it brings two new features into the game. Therefore, $\mf{z}_N[W]$ provides one with a good playground for pushing forward the methods of 
asymptotic analysis of $N$-fold integrals and learning how to circumvent or deal with certain of the problematic features mentioned in Sub-Section \ref{Sous-Section Beta Ens avec poids non variants}. 
To be more precise, the main features of the integrand in $\mf{z}_N[W]$ which constitute an obstruction to applying the already established methods stem from the presence of 
\begin{itemize}
\item a non-varying confining one-body potential $W$;
\item  a two-body interaction that has the same local (\textit{viz}. when $\la_a\tend \la_b$)
singularity structure as in the $\be$-ensemble case, while breaking other properties of the Vandermonde interaction such as 
the invariance under a re-scaling of all the integration variables. 
\end{itemize}
Although the tools of asymptotic analysis discussed previously break down or have to be altered in a significant way, 
a certain analogy with matrix models and $\be$-ensembles persists. Indeed, upon a proper rescaling in the spirit of 
Sub-Section \ref{Sous-Section Beta Ens avec poids non variants}, one can show for certain examples of potentials that the integral localises at a configuration of
the integration variables in such a way that these condense, in the large-$N$ limit, with a density $\rho_{\e{eq}}$. 
In fact, we show in Appendix  \ref{Appendix Section preuve asympt dom part fct unrescaled} 
that it is possible to repeat, with some modifications, the large-deviation approach to $\be$-ensemble integrals so as to obtain the 
leading asymptotic behaviour of $\ln \mf{z}_N[W]$ for certain instances of confining potentials $W$.
However, in order to go beyond the leading asymptotic behaviour of the logarithm, one has to alter the picture and work directly 
at the level of the rescaled model 
\beq
Z_N[V] \; = \; 
\Int{\R^N}{}  \pl{a<b}{N} \Big\{ \sinh\big[\pi\om_1 N^{\a} (\la_a-\la_b)\big] 
				\sinh\big[\pi\om_2 N^{\a} (\la_a-\la_b)\big]  \Big\}^{\be}
\cdot \pl{a=1}{N} \Big\{  \ex{-N^{1+\a} V(\la_a)}  \Big\} \cdot \pl{a=1}{N} \dd \la_a \;. 
\label{ecriture premiere intro dans texte de fct part rescale}
\enq
This integral is related to $\mf{z}_N[W]$ by a rescaling of the integration variables. The exponent $\a$ is fixed by the growth of the original potential $W$ 
at infinity. Finally, the potential $V$ should depend on $N$ and correspond to some rescaling of the original potential $W$. In fact, 
the main result obtained in the present paper deals with the large-$N$ asymptotic expansion of the rescaled partition function $Z_N[V]$ 
and this in the case where 
\begin{itemize}
\item the potential $V$ is smooth, strictly convex, has sub-exponential growth and is $N$-independent. ;

\item $0<\a<\tf{1}{6}$;

\end{itemize}
The first assumption is more than enough to carry the large deviation analysis, which gives the leading order of $\ln Z_N[V]$ while the second assumption appears in the course of the bootstrap analysis of the Schwinger-Dyson equations. 
\textit{Per se}, the application of our technique and results to computing the asymptotics of the original integral 
$\mf{z}_N[W]$ would demand to take a $N$ dependent potential and study $Z_N[V_N]$, which is technically much more involved. 
However, this problem is \textit{not} conceptually different from the one studied in this book. Therefore, the setting we shall discuss
is more fit for developing the method of asymptotic analysis of this class of integrals. We shall address the question of $N$-dependent 
potentials $V_N$ related to specific applications to quantum integrable models in a separate publication. 

Within our setting, in order to grasp sub-leading corrections to $ \ln Z_N[V]$, one faces several difficulties:
\begin{itemize}
\item[$(i)$] owing to the scaling $N^{\alpha}$, the nature of the repulsive interaction between the $\la_a$'s changes drastically between $N = \infty$ and $N$ finite. 
Therefore, one has to keep track of the transition of scales between the \textit{per-se} leading contribution -- which feels, effectively, 
only the brute $N=\infty$ behaviour of the properly normalised two-body interaction -- and the sub-leading corrections which experience the two-body interactions at all scales.
\item[$(ii)$] The presence of two scales $N$ and $N^{\a}$ weakens a naive approach to the concentration of measures. 
\item[$(iii)$] The derivative of the two-body interaction possesses a tower of poles that collapse down to the integration line, hence making the use
of correlators and complex variables methods to study Schwinger-Dyson equations completely ineffective. 
\item[$(iv)$] The master operator arising in the Schwinger-Dyson equations is a $N$-dependent singular integral operator of truncated Wiener--Hopf type. 
One has to invert this operator effectively and derive the fine, $N$-dependent bounds on its continuity constant as an operator between spaces of 
sufficiently differentiable functions. 
\item[$(v)$] The large-$N$ behaviour of one-point functions, as fixed by a successful large-$N$ analysis of the Schwinger-Dyson equations, is expressed 
in terms of one and two dimensional integrals involving the inverse of the master operator. One has to extract the large-$N$ asymptotic behaviour of such integrals. 
\end{itemize}

The setting of methods enabling one to overcome these problems constitutes the main contributions of this work.  

First, in order to strengthen the concentration of measures and, in fact, effectively absorb part of the asymptotic expansion into a single expression, 
 one should work with $N$-dependent equilibrium measures,  that is to say 
equilibrium measures associated with a minimisation problem of a quadratic $N$-dependent functional on the space 
of probability measures on $\R$. The density of such an $N$-dependent measure can be expressed as an integral transform 
whose kernel is given by a double integral  involving the solution to an auxiliary  matrix $2\times 2$ Riemann--Hilbert problem. This very fact constitutes a crucial difference
with the matrix model case in that, in the latter case, the density of equilibrium measure can be expressed in terms of the solution to 
a scalar Riemann--Hilbert problem, hence admitting  an explicit, one-dimensional integral representation. 
On top of improving numerous bounds, the use of such $N$-dependent equilibrium measures turns out to be crucial in order to push the asymptotic
expansion of $\ln Z_N[V]$ up to $\e{o}(1)$.

Second, the \textit{per se} machinery of topological recursion mentioned earlier breaks down for this class of 
multiple integrals. In order to circumvent dealing with the collapsing of poles, we develop a distributional approach 
to the asymptotic analysis of Schwinger-Dyson equations. The latter demands, in particular, to have a much more precise
control on its constituents.

Third, the inversion of the master operator is based on handlings of the inverse of the operator driving the singular 
integral equation for the density of equilibrium measure. Obtaining fine, $N$ dependent bounds for this operator demands to
go deep into the details of the solution of the $2\times 2$ Riemann--Hilbert problem which arises as the building block of this
inverse kernel. We develop techniques enabling one to do so.

Finally, the precise control on the objects issuing from Schwinger-Dyson equations yield, through usual interpolation by means of $t$-varying potentials,
an $N$-dependent functional of the density of equilibrium measure -- itself also depending on $N$ -- as an answer for the 
large-$N$ asymptotics of $\ln Z_N[V]$. 
%The leading asymptotic behaviour thereof leads to the constant term in the asymptotic expansion of the logarithm of the 
%integral.
Setting forth methods for the asymptotic analysis of this functional demands, again, a very fine control 
of the inverse build through the Riemann--Hilbert problem approach. 
We develop such methods, in particular, by describing the new class of special functions related to our problem.

\subsection*{Putting in perspective the bi-orthogonal ensembles.}

At this point, it appears useful to make several comments with respect to the existing literature on bi-orthogonal ensembles. Indeed, the applications to the 
quantum separation of variables correspond to setting $\be$ to $1$ in $\mf{z}_N[W]$ and hence $Z_N[V]$. 
In this case, these multiple integral corresponds to a bi-orthogonal ensemble. As such, they  can be explicitly computed, at least in principle, 
by means of the system of bi-orthogonal polynomials associated with the $\i\om_1^{-1}$ or $\i\om_2^{-1}$ periodic functions $\ex{\pi \om_1 y}$, $\ex{\pi \om_2 y}$ and with respect to the weight $\ex{-W(y)}$
supported on $\R$. As shown by Claeys and Wang \cite{ClaeysWangIntroRHPForBioRthSinhLikePlys} for a specific degeneration (which corresponds 
basically to sending one of the $\om$'s in \eqref{ecriture MI qSoV comme exemple} to zero) and then in full extent by Claeys and Romano \cite{ClaeysRomanoStudyForGeneralPeriodsOfSinh-BasedPlys}, 
such a system of bi-orthogonal polynomials solves a vector Riemann-Hilbert problem.  
Furthermore, the non-linear steepest descent approach \cite{DeiftKriechMcLaughVenakZhouOrthogonalPlyExponWeights,DeiftKriechMcLaughVenakZhouOrthogonalPlyVaryingExponWeights} 
to the uniform in the variable large degree-$N$ asymptotics of orthogonal polynomials can be generalised to such a bi-orthogonal setting, 
leading to Plancherel-Rotach like asymptotics for these bi-orthogonal polynomials \cite{ClaeysWangIntroRHPForBioRthSinhLikePlys}. In principle, by adapting the steps of \cite{ErcolaniMcLaughlinLargeNPartFctRMTAndGraphEnumeration},
one should be able to derive the large-$N$ asymptotic expansion of the integral $\mf{z}_N[W]$ in presence of \textit{varying} weights, \textit{viz}.
provided the replacement $W\hookrightarrow N V$ is made. However, such a result would by no means allow one
for any easy generalisation to non-varying weights. Indeed, as we have argued, in the non-varying case, 
one rather needs to carry out the large-$N$ analysis of the rescaled model $Z_N[V]$. However, starting from such a 
multiple integral would imply that  one should study the system of bi-orthogonal polynomials associated with the functions $\ex{\pi N^{\a}\om_1 y}$, $\ex{\pi N^{\a} \om_2 y}$. 
The presence of $N^{\a}$ introduces a new scale in $N$ to the Riemann--Hilbert analysis. Taking the latter into account would probably demand a quite non-trivial
modification of the non-linear steepest descent method.

On top of all this, one needs to construct the equilibrium measure. For similar reasons of absorbing part of the asymptotic expansion, 
this measure will have to issue from the same $N$-dependent minimisation problem and hence correspond to the $N$-dependent equilibrium measure that
we construct in the present paper. However, if one goes into the details of the work \cite{ClaeysRomanoStudyForGeneralPeriodsOfSinh-BasedPlys}, 
one observes that these authors provide a one-fold integral representation for the density of the one-cut equilibrium measure arising in bi-orthogonal ensembles. 
The kernel of this transform involves the inverse of an explicit and basic transcendental function. 
Although extremely effective in the varying case, such an integral representation appears ineffective in the analysis of $Z_N[V]$. 
Indeed, then, one would have to manipulate $N$-dependent versions of this inverse and, in particular, obtain uniform in $N$
local behaviours thereof. \textit{A priori}, since this inverse does not seem to admit an explicit series expansion or a manageable integral representation,
such a characterisation seems to be quite complicated. 
Furthermore, the transform constructed in \cite{ClaeysRomanoStudyForGeneralPeriodsOfSinh-BasedPlys} does not exhibit explicitly the factorisation of 
square root singularities at the edges - in contrast to the case of the one-fold integral representation arising in $\be$ ensembles, \textit{c.f.} \eqref{ecriture mesure equilibre 1 cut beta ens}.
This means that, just as in our setting, one would have to extract the square root behaviour by hand. 
Therefore, although one dimensional, we believe that this transform, in the present state of the art, is much less effective then ours, at least
from the point of view of our perspective of asymptotic analysis. 
In fact, when specialised to the construction of the equilibrium measure, the $2\times 2$ Riemann--Hilbert analysis we use enables us, among other things, 
to provide the leading, up to exponentially small corrections in $N$, behaviour of the inverse
of the $N$-rescaled map built in \cite{ClaeysRomanoStudyForGeneralPeriodsOfSinh-BasedPlys}. Thus, indirectly, our approach solves such a problem.

\section{Notations and basic definitions}
\label{nonono}
In this section, we introduce basic notations that we shall use throughout the paper. 

\subsection*{General symbols}

\begin{itemize}

\item $\e{o}$ and $\e{O}$ refer to standard domination relations between functions. In the case of matrix function $M(z)$ and $N(z)$,
the relation $M(z)=\e{O}(N(z))$ is to be understood entry-wise, \textit{viz}. $M_{jk}(z)=\e{O}(N_{jk}(z))$.

\item $\e{O}(N^{-\infty})$ means $\e{O}(N^{-K})$ for arbitrarily large $K$'s.

\item Given a set $A\subseteq X$,  $\bs{1}_{A}$ stands for the indicator function of $A$, and $A^{\e{c}}$ denotes its complement in $X$. 

\item A Greek letter appearing in bold, \textit{e.g.} $\bs{\la}$, will always denote an 
$N$-dimensional vector: 
\beq
\bs{\la} \; = \; \big( \la_1,\dots,\la_N  \big) \in \R^N \;. 
\enq
and $\dd^{N}\bs{\la}$ denotes the product of Lebesgue measures $\prod_{a = 1}^N \dd\la_{a}$.%
\item given $x \in \R$, $  \lfloor x \rfloor$ denotes the integer satisfying $  \lfloor x \rfloor \leq x <  \lfloor x \rfloor +1$

\item Throughout the file, the curve $\msc{C}^{+}_{\e{reg}}$ will denote the curve depicted in Figure~\ref{Figure definition des courbes C pm reg} appearing in \S~\ref{Sous-section decomposition apropriee WN pour asymptotiques uniforme Wp norms}. This curve is such that $2\vsg = \e{dist}\big( \R, \msc{C}^{+}_{\e{reg}} \big)>0$. Throughout the text, this distance will always be denoted by $2\vsg$. 

\item $I_2$ is the $2\times 2$ identity matrix while $\sg^{\pm}$ and $\sg_3$ stand for the Pauli matrices:
\beq
\sg^+ \; = \; \left( \ba{cc}  0 & 1 \\ 0 & 0 \ea \right) \quad , \quad
\sg^- \; = \; \left( \ba{cc}  0 & 0 \\ 1 & 0 \ea \right)  \qquad \e{and} \qquad
\sg_3 \; = \; \left( \ba{cc}  1 & 0 \\ 0 & -1 \ea \right) \;.
\enq
\end{itemize}

\subsection*{Functional spaces}

\begin{itemize}
 \item $\mc{M}^{1}(\R)$ denotes the space of probability measures on $\R$. The weak topology on $\mc{M}^1(\mathbb{R})$ is metrized by the Vasershtein distance, defined for any two probability measure $\mu_1$ and $\mu_2$ by:
\beq
\label{Vasera}D_{V}[\mu_1,\mu_2] = \sup_{f \in {\rm Lip_{1,1}(\R)}} \bigg| \Int{\R}{}  f(\xi)\,\dd(\mu_1 - \mu_2)(\xi) \bigg| \;,
\enq
where ${\rm Lip}_{1,1}(\R)$ is the set of Lipschitz functions bounded by $1$ and with Lipschitz constant bounded by $1$. If $f$ is a bounded, Lipschitz function, its bounded Lipschitz norm is:
\beq
\label{BLnorm}\norm{f}_{{\rm BL}} = \norm{f}_{L^{\infty}(\R)} + \sup_{\xi \neq \eta \in \R} \Bigg|\frac{f(\xi) - f(\eta)}{\xi - \eta}\Bigg| \;.
\enq

\item Given an open subset $U$ of $ \Cx^n$, $\mc{O}(U)$ refers to the ring of holomorphic functions on $U$. 
If $f$ is a matrix of vector valued function, the notation $f\in \mc{O}(U)$ is to be understood entrywise, \textit{viz}. 
$\forall \; a,b$ one has $f_{ab} \in \mc{O}(U)$.

 \item $\mc{C}^k(A)$ refers to the space of function of class $k$ on the manifold $A$. $\mc{C}^k_{\e{c}}(A)$ refers to the spaces built out of 
 functions in  $\mc{C}^{k}(A)$ that have a compact support.  

 \item $L^p(A,\dd\mu)$ refers to the space of $p^{\e{th}}$-power integrable functions on a set $A$ with respect to the measure $\mu$. $L^p(A,\dd\mu)$
is endowed with the norm 
\beq
\norm{f}_{ L^p(A,\dd \mu) } \; = \; \bigg\{ \Int{A}{} | f(x) |^{p}\,\dd \mu(x) \bigg\}^{\f{1}{p}} \;. 
\enq
\item  More generally, given an $n$-dimensional manifold $A$, $W^{p}_{k} (A,\dd \mu )$ refers  to the $p^{{\rm th}}$ Sobolev space of order $k$
defined as 
\beq
W^{p}_{k} (A,\dd \mu ) \; = \; 
\bigg\{  f\in L^{p}(A,\dd \mu) \; : \; \Dp{x_1}^{a_1}\dots \Dp{x_n}^{a_n}f \in  L^{p}(A,\dd \mu) \; ,
      \, \sul{\ell=1}{n}  a_{\ell} \leq k  \quad \e{with} \quad   a_{\ell} \in \mathbb{N} \bigg\}  \;.
\enq
This space is endowed with the norm 
\beq
\label{defWninfnorm} \norm{f}_{ W^{p}_{k} (A,\dd \mu) } \; = \; \max
\Big\{  \norm{ \Dp{x_1}^{a_1}\dots \Dp{x_n}^{a_n}f  }_{  L^{p}(A,\dd \mu) }  \; : \; 
  a_{\ell} \in \mathbb{N}, \; \ell=1,\dots, n,   \; \;  \e{and}\; \e{satisfying}  \; \sul{\ell=1}{n}  a_{\ell} \leq k   \Big\}\;.
\enq
In the following, we shall simply write $L^p(A)$, $W^p_k(A)$ unless there will arise some ambiguity on the measure
chosen on $A$. 

\item We shall also need the $N$-weighted norms of order $\ell$ for a function $f\in W^{\infty}_\ell(\R^n)$, which are defined as 
\beq
\mc{N}_N^{(\ell)}[f] \; = \; \sul{p=0}{\ell} \f{\norm{ f }_{ W^{\infty}_p(\R^n) }}{ N^{p \a} } \;.
\enq
In particular, we have the trivial bound $\mc{N}_N^{(\ell)}[f] \leq (\ell+1) \norm{ f }_{ W^{\infty}_{\ell}(\R^n) }$. Also, the number of variables of $f$ is implicit in this notation.

\item The symbol $\mc{F}$ denotes the Fourier transform on $L^2(\R)$ whose expression, versus $L^1\cap L^2(\R)$
functions, takes the form
\beq
\mc{F}[\vp](\la) \; = \; \Int{ \R }{} \vp(\xi)\,\ex{ \i\xi\la}\dd \xi \;. 
\enq
Given $\mu \in \mc{M}^1(\R)$, we shall use the same symbol for denoting its Fourier transform, \textit{viz}. $\mc{F}[\mu]$.
The Fourier transform on $L^2(\R^n)$ is defined with the same normalisation. 
\item The $s^{\e{th}}$ Sobolev space on $\R^n$ is defined as  
\beq
H_s(\R^n) \; = \; \bigg\{  u \in \mc{S}^{\prime}(\R^n) \; : \;  
\norm{u}^2_{H_s(\R^n)} \; = \; \Int{\R^n}{}  \Big(1 + \big|\sul{a=1}{n} t_a^2\big|^{\f{1}{2}} \Big)^{2s} 
\big| \mc{F}[u](t_1,\dots,t_n) \big|^2 \cdot \dd^n\bs{t} \; < \; +\infty \bigg\} \; , \; 
\enq
in which $\mc{S}^{\prime}$ refers to the space of tempered distributions. 
We remind that given a closed subset $F \subseteq \R^n$, $H_s(F)$ corresponds to the subspace of 
$H_{s}(\R^n)$ of functions whose support is contained in $F$. 
\item The subspace 
\beq
\mf{X}_{s}\big(  A \big) \; = \; \Big\{  H \in H_{s}\big( A  \big) \; :  \;  
\Int{\R+\i\eps }{} \chi_{11}(\mu) \mc{F}[H](N^{\a}\mu)\ex{-\i N^{\a} \mu b_N}   \f{ \dd \mu }{ 2 i \pi }  \; = \; 0\Big\} \; 
\enq
in which $A\subseteq \R$ is closed  will play an important role in the analysis. It is defined in terms of $\chi_{11}$, the $(1,1)$ entry of the unique solution $\chi$
to the $2\times 2$ matrix valued Riemann--Hilbert problem given in Section \ref{SousSection RHP chi initial}.

\item Given a smooth curve $\Sg$ in $\Cx$, the space $\mc{M}_{\ell}\big( L^2(\Sg) \big)$ refers to $\ell\times \ell$  matrices with coefficients belonging to 
$ L^2(\Sg)$. It is endowed with the norm
\beq
\norm{M}_{\mc{M}_{\ell}(L^2(\Sg))} \; = \; 
\bigg\{ \Int{\Sg}{} \sul{a,b}{} \big[ M_{ab}(s) \big]^{*} M_{ab}(s)\,\dd \mu(s) \bigg\}^{ \f{1}{2} }\;. 
\label{definition norme L2 fonctions matricielles}
\enq
and $^*$ denotes the complex conjugation. 

\end{itemize}

\subsection*{Certain standard operators}

\begin{itemize}

\item Given an oriented curve $\Sg \subseteq \Cx$, $-\Sg$ refers to the same curve but endowed with the opposite orientation. 

\item Given a function $f$ defined on $\Cx\setminus \Sg$, with $\Sg$
an oriented curve in $\Cx$, we denote -if these exist- by $f_{\pm}(s)$  the boundary values of $f(z)$ on $\Sg$ when the argument $z$ 
 approaches the point $s \in \Sg$ non-tangentially and from the left ($+$) or the right ($-$) side of the curve. 
 Furthermore, if one deals with vector or matrix-valued function, then this notation is to be understood entry-wise. 

\item $\mathbb{H}^{\pm} = \{z \in \mathbb{C}\,\,:\,\,\mathrm{Im}\,(\pm z) > 0\}$ is the upper/lower half-plane, and $\mathbb{R}^{\pm} = \{z \in \mathbb{R}\,\,:\,\,\pm z \geq 0\}$ is the closed positive/negative real axis.

\item The symbol $\mc{C}$ refers to the Cauchy transform on $\R$:
\beq
\mc{C}[f](\la) \; = \; \Int{ \R }{}  \f{ f(s) }{ s-\la } \cdot \f{ \dd s }{ 2 \i \pi }\;.
\enq
The $\pm$ boundary values $\mc{C}_{\pm}$ define continuous operators on $H_{s}(\R)$ and admit the expression
\beq
\label{Cplusmoins}\mc{C}_{\pm}[f](\la) \; = \; \f{ f(\la) }{ 2 } \; + \;\f{ 1 }{ 2 \i } \Fint{ \R }{}  \f{ f(s)\,\dd s }{\pi(s-\la) }   
\enq
where $\Fint{}{}$ is the principal value integral. 
%and 
%
%
%
%\beq
%
%\norm{ \mc{H}[\psi] }_{H_s(\R)}^2 \; = \; \f{1}{4} \Int{ \R }{}  (1+|t|)^{2s} \cdot \big| \e{sgn}(t) \mc{F}[\psi](t) \big|^2
%
%\cdot \dd t \; = \; \f{1}{4} \norm{ \psi }_{H_s(\R)}^2\; .
%
%\enq
%
%
%

\item Given a function $f$ supported on a compact set $A$ of $\R^n$, we denote by $f_{\mf{e}}$ an extension of $f$ onto some compact set $K$
such that $A \subseteq \e{Int}(K)$. We do stress that the compact support is part of the data of the extension. As such, it can vary from 
one extension to another. However, the extension $f_{\mf{e}}$ is always assumed to be of the same class as $f$. For instance,  
if $f$ is $L^p(A), W^p_k(A)$ or $\mc{C}^{k}(A)$, then $f_{\mf{e}}$ is $L^p(K), W^p_k(K)$ or $\mc{C}^{k}(K)$.

\end{itemize}

\chapter{Main results and strategy of proof}
\label{Section presentation des resultats}

{\bf Abstract}

\textit{
In the first part of this chapter we gather the main results which follow from the analysis developed in this book. 
To start with, in Section  \ref{baby}, we discuss  an example, in Theorem~\ref{Proposition Asympt non rescaled part fct}, 
of the leading large-$N$ asymptotic expansion of $\ln \mf{z}_N[W]$ where
$\mf{z}_N[W]$ is  the unscaled partition function defined by \eqref{ecriture MI qSoV comme exemple}.
We shall also argue 
that the large-$N$ asymptotic behaviour of \eqref{ecriture MI qSoV comme exemple} -- whose integrand does not depend explicitly on $N$ -- 
can be deduced from the one of the rescaled model \eqref{definition mesure proba rescalee avec potentiel arbitraire} -- whose integrand depends explicitly on $N$-- that we propose to study.
Then, after presenting the per se model of interest and listing the assumptions on which our analysis builds in Section \ref {Section modele d interet},
we shall discuss the form of the large-$N$ asymptotic expansion of the logarithm of the rescaled partition function $\ln Z_{N}[V]$ in Section \ref{Section AE Rescaled part fct}. 
Then, in Section \ref{Section mesure equilibre et operateur master}, we shall discuss the characterisation of the $N$-dependent equilibrium measure that is 
pertinent for our study as well as the form of the inverse $\mc{W}_N$ of a fundamental singular integral operator $\mc{S}_N$ that arises naturally in the study. 
Finally, Section \ref{Section Strategie de la preuve}, we outline the main steps of the proof.}

\section{A baby integral as a motivation}
\label{baby}

Let ${\cal E}_{({\rm ply})}$ be the functional, defined in $\mathbb{R}\cup\{+\infty\}$ for any probability measure $\mu \in \mc{M}^1(\R)$ by
\beq
\mc{E}_{(\e{ply})}[ \mu ] \; = \; \Int{}{} \Big\{ \f{c_q}{2} \Big( |\xi|^q+|\eta|^q\Big) \, \; - \;  \f{ \be \pi (\om_1+\om_2) }{ 2 } |\xi-\eta| \Big\} \,\dd \mu(\xi)\dd \mu(\eta) \;. 
\label{ecriture fct taux pour LDP de zN ac pot ply}
\enq

\begin{theorem}

\label{Proposition Asympt non rescaled part fct}

$\mc{E}_{(\e{ply})}$ is a lower semi-continuous good rate function. Furthermore, given a potential  $W$ such that 
\beq
\lim_{ \abs{\xi}\tend +\infty }  \abs{\xi}^{-q} W(\xi)  \; = \; c_q >0  \quad \e{for} \; \e{some}  \;\;  q>1\;, 
\label{conditon cptmt asympt V}
\enq
it holds 
\beq
\lim_{N \tend +\infty} \f{\ln \mf{z}_N [W]}{N^{2+\frac{1}{q-1}} } \; = \; - \inf_{\mu \in \mc{M}^{1}(\R)} \mc{E}_{(\e{ply})}\big[\mu \big] \;. 
\enq
This infimum is attained at a unique probability measure $\mu_{\e{eq}}^{(\e{ply})}$. This measure is continuous with respect to the Lebesgue measure 
and has density
\beq
\label{2344}\rho_{\e{eq}}^{(\e{ply})}(\xi) \; = \; \f{ q(q-1) |\xi|^{q-2} }{ 2\pi \be (\om_1+\om_2) }\cdot \bs{1}_{\intff{a}{b}}(\xi) \;. 
\enq
$\mu_{\e{eq}}^{(\e{ply})}$ is supported on the interval $\intff{a}{b}$, with $(a,b)$ being the unique solution to the set of equations
\beq
|b|^{q-1} \; = \;  |a|^{q-1} \; = \;  \f{\pi \be(\om_1+\om_2) }{q}  \;. 
\enq
We have, explicitly:
\beq
\lim_{N \tend +\infty}  \f{\ln \mf{z}_N [W] }{N^{2+\f{1}{q-1}}} \; = \; 
\big( c_q \big)^{\f{1}{q}} \cdot \Bigg(  \f{\pi \be}{ q } (\om_1+\om_2) \Bigg)^{ \frac{q+1}{q} } \cdot 
\f{ 2q^2-9q+6 }{2(2q-1) } \;. 
\enq
\end{theorem}

The proof of this proposition is postponed to Appendix \ref{Appendix Section preuve asympt dom part fct unrescaled}, and follows similar steps to, \textit{e.g.}, \cite{AndersonGuionnetZeitouniIntroRandomMatrices}. 
We now provide heuristic arguments to justify the occurrence of scaling in $N$ in this problem. 
Just as discussed in the introduction, the repulsive effect of the $\sinh$-2 body interactions will dominate 
over the confining effect of the potential as long as the integration variables will be located in some bounded set. 
Furthermore, in the same situation, the Lebesgue measure should contribute to the integral at most as an exponential in $N$. 
We thus look for a rescaling of the variables $y_a=T_N \la_a$ where the effects of the confining potential and the $\sinh$-2 body interactions will be of the same order of magnitude in $N$. This recasts the partition function as
\beq
\mf{z}_N [W] \; = \; \Big(T_N\Big)^N
\Int{\R^N}{}  \pl{a<b}{N} \Big\{ \sinh\big[\pi\om_1 T_N (\la_a-\la_b)\big] 
				\sinh\big[\pi\om_2  T_N (\la_a-\la_b)\big]  \Big\}^{\be}
\,\pl{a=1}{N} \Big\{  \ex{- W(T_N \la_a)} \Big\} \, \dd^{N}\bs{\la} \;, 
\enq
Taking into account the large-variable asymptotics of the potential, we have:
\beq
\sul{a=1}{N} W\big( T_N \la_a \big) \; \sim \; T_N^{q} \,  N \;,
\enq
where the symbol $\sim$ means that for a "typical" distribution of the variables $\{\la_a\}_1^N$, the leading in $N$
asymptotic behaviour of the sum in the right-hand side should be of the order of the left-hand side.
 Similarly, assuming a typical distribution of the variables $\{\la_a\}_1^N$ such that most of the pairs $\{\la_a,\la_b\}$ satisfy $T_N|\la_a-\la_b|\gg 1$, one has
\beq
\sul{a<b}{N} \be \ln \Big\{ \sinh\big[\pi\om_1 T_N (\la_a-\la_b)\big] \sinh\big[\pi\om_2 T_N (\la_a-\la_b)\big]  \Big\} \; \sim \;
C\,N^2 \, T_N  \;. 
\enq
Thus, the confining potential and the two-body interaction  will generate a comparable order of magnitude in $N$ as soon as $N^2 \cdot T_N \; = \; T_N^q \cdot N$,  \textit{i.e.}
\beq
T_N \; = \; N^{ \f{1}{q-1} }  \;. 
\enq
Theorem~\ref{Proposition Asympt non rescaled part fct} indeed justifies that the empirical distribution $L_N^{(\bs{\la})}$ of $\lambda_{a} \, =\, N^{ \frac{-1}{q-1} }  y_a $ 
concentrates around the equilibrium measure, with a large deviation principle governed by the rate function \eqref{ecriture fct taux pour LDP de zN ac pot ply}.

This observation implies that, in fact, $ Z_N[V_N]$ with $V_N(\la)=N^{-\frac{q}{q-1}} \cdot W(N^{\frac{1}{q-1}} \la)$ is the good object to study
in that it involves interactions that are already tuned to the proper scale in $N$. Due to the relation $\mf{z}_N[W]=N^{\frac{N}{q-1}} \cdot Z_N[V_N]$, 
one readily has access to the large-$N$ asymptotic expansion of $\mf{z}_N[W]$.

\section{The model of interest and the assumptions}
\label{Section modele d interet}

It follows from the arguments given in the previous section that, effectively, the analysis of the 
unrescaled model boils down to the one subordinated to the partition function 
\beq
Z_N[V] \; = \; 
\Int{\R^N}{}  \pl{a<b}{N} \Big\{ \sinh\big[\pi\om_1 N^{\a} (\la_a-\la_b)\big] 
				\sinh\big[\pi\om_2 N^{\a} (\la_a-\la_b)\big]  \Big\}^{\be}\,
				\pl{a=1}{N}\ex{-N^{1+\a} V(\la_a)}\cdot \dd^{N}\bs{\la} \;, 
\label{ecriture fonction partition a la beta ensemble}
\enq
with $\a$ some parameter -- equal to $1/(q - 1)$ in the previous paragraph -- and $V$ a potential that possibly depends on $N$. Due to such an effective reduction, in this paper, we shall develop the general formalism 
to extract the large-$N$ asymptotic behaviour. Therefore, we shall keep the complexity at minimum. In particular, 
we shall \textit{not} consider the case of $N$-dependent potentials which would put the analysis of $Z_N[V]$
in complete correspondence with the one of $\mf{z}_N[W]$. Indeed, this would lead to numerous technical complication in our arguments, 
without bringing more light on the underlying phenomena. By focusing on \eqref{ecriture fonction partition a la beta ensemble}, 
we believe that the new features and ideas of our methods are better isolated and illustrated. We shall incorporate the peculiarities of the model $\mf{z}_{N}[W]$ of \eqref{ecriture MI qSoV comme exemple} 
and investigate its asymptotic behaviour up to $o(1)$ in a future publication.

In the present paper we obtain the large-$N$ asymptotic expansion of $\ln Z_N[V]$ up to $\e{o}(1)$ 
under four hypothesis
\begin{itemize}

\item the potential $V$ is confining, \textit{viz}. there exists $\eps>0$ such that 
\beq
\liminf_{ |\xi| \tend + \infty } |\xi|^{-(1 + \eps)}\,V(\xi) \; = \; + \infty \;; 
\label{Hypothesis: confinment of the potential}
\enq
\item  the potential $V$ is smooth and strictly convex on $\R$ ;
\item the potential is sub-exponential, namely there exists $\eps>0$ and $C_V>0$  such that
\beq
\label{subexpV} \forall \xi \in \mathbb{R},\qquad \sup_{\eta\in \intff{0}{\eps}} \big| V^{\prime}(\xi + \eta)\big| \, \leq \, C_V\big( |V(\xi)| +1 \big)\;,
\enq
and  given any $\kappa > 0 $ and $p\in \mathbb{N }$, there exists $C_{\kappa,p}$  such that 
\beq
 \forall \xi \in \mathbb{R},\qquad \big| V^{(p)}(\xi) \big|\,\ex{-\kappa V(\xi) } \; \leq \;  C_{\kappa,p}\;.
\label{Hypothese sous exponetialite des derivees}
\enq
\item the exponent $\a$ in $N^{\a}$ is neither too large nor too small:
\beq
0<\a <\tf{1}{6}\;.
\enq
\end{itemize}

The first hypothesis guarantees that the integral \eqref{ecriture fonction partition a la beta ensemble} is well-defined, and that the $\la$'s will typically remain in a compact region of $\mathbb{R}$ independent of $N$. 
It could be weakened to study weakly confining potentials, for the price of introducing more technicalities, similar to those already encountered for $\beta$ ensembles -- see \textit{e.g.} \cite{HardyKuijlaarsWeakyAdmissibleVectEqPblm}.

In the second assumption, $V$ could be assumed ${\cal C}^{k}$ for $k$ large enough. The convexity assumption guarantees that the support 
of the equilibrium measure is a single segment\footnote{See e.g. the expression of the $N = \infty$ equilibrium measure \eqref{denspot}. Its proof is given in Appendix \ref{Appendix minimisation de la mesure equilibre}.}.
 In principle, the multi-cut regime that may arise when the potential is not strictly convex can be addressed by importing the ideas of \cite{BorotGuionnetAsymptExpBetaEnsMultiCutRegime} to the present framework. We expect that the analysis of  %G: I modified slightly these sentence
the Riemann-Hilbert problem in the multi-cut regime is very similar to the present case, but with a larger range of degrees for the polynomial freedom appearing in the solution \eqref{ecriture solution RHP vectoriel sur Hs general s negatif}. 
Though it would certainly represent some amount of work, the ideas we develop here should also be applicable to derive the fine large $N$ analysis of the solution of the Riemann-Hilbert problem in the bulk and in the vicinity of all the 
edges of the support of the equilibrium measure.

The third assumption is not essential, but allows some simplification of the intermediate proofs concerning the equilibrium measure and the large deviation estimates, \textit{e}.\textit{g}. 
Theorem~\ref{Theorem estimation deviation mesure centree en Proba} and Corollary~\ref{Corollary bornes a priori sur les correlateurs}. It is anyway satisfied in physically relevant problems.

In the fourth assumption, $\alpha = 0$ can already be addressed with existing methods \cite{KozBorotGuionnetLargeNBehMulIntMeanFieldTh}. The upper limit $\alpha < \alpha^* = 1/6$ has a purely technical origin. 
The value of $\alpha^*$ could be increased by entering deeper into the fine structure of the analysis of the Schwinger-Dyson equation, and by finding more precise local and global bounds for the large $N$ behaviour of the inverse of 
the master operator ${\cal U}_{N}^{-1}$, in more cunning norms. Intuitively, the genuine upper limit should be $\alpha^* = 1$, since in the $\alpha > 1$ case, we reach a regime where the particles do not feel the local repulsion any more. 
However, obtaining microscopic estimates is usually a difficult question -- for $\beta$ ensembles, it has been addressed \textit{e}.\textit{g}. in 
\cite{BourgadeErdosYauUniversalityEdgeBetaEnsGnlPot,BourgadeErdosYauUniversalityBulkBetaEnsConvPot}. So, one can expect 
important technical difficulties to extend our result to values of $\alpha$ increasing up to $1$.

This set of hypothesis offers a convenient framework for our purposes, enabling us to focus on the technical aspects $(i)-(vi)$ listed in \S~\ref{lili} without adding extra complications.

\section{Asymptotic expansion of $Z_N[V]$ at $\be=1$}
\label{Section AE Rescaled part fct}

We now state one of the main results of the paper, namely the large-$N$ asymptotic expansion of the 
partition function $Z_N[V]$ which holds for any potential $V$ satisfying the hypothesis
stated above %and under the condition $0\leq \a \leq \tf{1}{6}$. 

\begin{theorem}
\label{Theorem principal du livre}
The below asymptotic expansion holds
\bem
 \ln \Bigg( \f{ Z_N[V] }{ Z_N[W_{G;N}] } \Bigg)_{ \mid \be=1}  \; = \; -N^{2+\a} \sul{p=0}{\lfloor 2/\alpha \rfloor +1} \f{\capricornus_p[V]}{N^{\a p } } 
\; + \; N^{\a}\cdot \gimel_0\cdot\Big( \leo[V,W_{G;N}](b_N) -  \leo[V,W_{G;N}](a_N)  \Big) \\ 
\; + \;  \aleph_0\cdot  \Big( \leo[V,W_{G;N}]^{\prime}(b_N) +  \leo[V,W_{G;N}]^{\prime}(a_N)  \Big)  \, + \, \e{o}(1) \;. 
\label{Ecriture DA ZN dans partie Intro resultats}
\end{multline}
The whole $V$-dependence of this expansion is encoded in the coefficients $\capricornus_p[V]$ and in the function $ \leo[V,W_{G;N}](\xi)$. 
$\gimel_0$ and $\aleph_0$ are numerical coefficients given, respectively, in terms of a single and four-fold integral. 
Also, the answer involves the Gaussian potential 
\beq
W_{G;N}(\xi) \; = \; \f{ \pi \be (\om_1+\om_2) \cdot \Big[ \xi^2 - \big( a_N+b_N  \big) \xi \Big]   }
{ b_N-a_N \, + \, \f{1}{ N^{\a} } \sul{p=1}{2} \f{1}{\pi \om_p} \ln \Big( \f{ \om_1\om_2 }{ \om_p(\om_1+\om_2) }\Big)    }
\label{definition potentiel WGN} 
\enq
and sequences $a_N$ and $b_N$ that are given in Theorem \ref{Theorem characterisation densite mesure eq a grand N}. 
If we denote $V_{N}^{\pm} = V \pm W_{G;N}$, the coefficients $\capricornus_p[V]$ take the form 
\beq
\capricornus_0[V] \; = \; \f{ - 1 }{ 4 \pi  (\om_1+\om_2) }\Int{ a_N }{ b_N } V^-_{N}(\xi) \cdot (V_{N}^-)^{\prime\prime}(\xi)\,\dd \xi
\enq
when $p=0$ and, for any $p\geq 1$:
\begin{eqnarray}
\capricornus_p[V] & = &    u_{ p+1 } \Int{a_N}{b_N} V^{-}_{N}(\xi)\cdot V^{(p+2)}(\xi)\,\dd \xi  \\
& & +  \sul{ \substack{ s+\ell=p-1 \\ s,\ell \geq 0} }{} \f{\daleth_{s,\ell}}{ s! }\Bigg\{(-1)^{\ell}\,(V^{-}_{N})^{(\ell+1)}(a_N)\cdot (V^+_{N})^{(s + 1)}(a_N) \; + \; (-1)^{s}\,(V_{N}^{-})^{(\ell+1)}(b_N)\cdot (V^+_{N})^{(s+1)}(b_N) \Bigg\}\; .  \nonumber
\end{eqnarray}
The coefficients $\daleth_{s,\ell}$ are defined by:  
\beq
\daleth_{s,\ell} \; = \; \f{ \i^{s+\ell+1} }{2\pi } \sul{r=1}{\ell+1} \f{ s!  }{r! (s+\ell+1-r)!  } \cdot 
\f{\Dp{}^r}{\Dp{}\mu^r } \Big( \f{\mu}{ R_{\da}(-\mu) } \Big)_{\mid \mu=0} \cdot 
\f{\Dp{}^{s+1+\ell-r} }{\Dp{}\mu^{s+1+\ell-r} } \Big( \f{1}{ R_{\da}(\mu) } \Big)_{\mid \mu=0}  \;,
\label{definition constantes dialeth s ellbis}
\enq
$R_{\da}$ is the $\mathbb{H}^-$  Wiener--Hopf factor of $ \tf{1}{ \mc{F}[S](\la) }$, 
with $S$ defined in \eqref{ecriture eqn int sing de depart}, that reads 
\beq
R_{\da}(\la)  \; = \; \f{ \la }{ 2\pi \sqrt{\om_1+\om_2}  } \cdot  \Big( \f{\om_2}{\om_1+\om_2} \Big)^{- \f{ \i \la}{2\pi \om_1} } 
 \cdot \Big( \f{\om_1}{\om_1+\om_2} \Big)^{- \f{\i \la}{2\pi \om_2} }     
 \cdot \f{ \Ga\bigg( \f{ \i \la }{ 2\pi \om_1 } \bigg) \cdot \Ga\bigg( \f{ \i \la }{ 2\pi \om_2 } \bigg) }
 {\Ga\bigg( \f{ \i \la (\om_1+\om_2)}{ 2\pi \om_1 \om_2 } \bigg)  } \;. 
\enq
The  function $\leo$ describing the constant term is defined as  
\beq
\leo[V,W_{G;N} ](\xi) \; = \; \f{V^{\prime}(\xi)- W_{G;N}^{\prime}(\xi) }{ V^{\prime\prime}(\xi)- W_{G;N}^{\prime \prime}(\xi)   }\,
\ln \Bigg( \f{  V^{\prime\prime}(\xi) }{ W_{G;N}^{\prime \prime}(\xi)  } \Bigg) \;. 
\label{leoDef}
\enq
The $V$-independent coefficient $\gimel_0$ in front of the term $N^{\a}$ reads 
\beq
\gimel_0 \; = \; \Int{0}{+\infty} \frac{\dd u\,J(u)}{2\pi} \big( uS^{\prime}(u) \, + \,  S(u)\big)\,\qquad \e{with} \quad 
J(u) \; = \; \Int{ \msc{C}^{(+)}_{\e{reg}}  }{} \f{2  \sinh\big[\tf{\la}{(2\om_1)} \big]  \sinh\big[\tf{\la}{(2\om_2)} \big] }
{ \sinh\big[\tf{\la(\om_1+\om_2)}{ (2\om_1\om_2) } \big]  }
\cdot \f{\ex{\i \la u} \,\dd \la }{2\i \pi}\;.
\label{gimel0def}\enq
Finally, the numerical prefactor $\aleph_0$ is expressed in terms of the four-fold integral
\bem
\aleph_0 \; = \; - \f{\om_1+\om_2}{2}{\displaystyle \Int{ \R }{} }  \f{ \dd u\,J(u)}{2\pi}   {\displaystyle \Int{ |u| }{ + \infty }  } \dd v\,\partial_{u} 
\Big\{ S(u)\cdot\Big( \mf{r}\big[\f{v - u}{2}\big] - \mf{r}\big[\f{v + u}{2}\big]  \Big) \Big\}   \\
\; + \;  \Int{ \msc{C}^{(+)}_{\e{reg}} }{}  \f{\dd \la\,\dd \mu }{ (2 \i \pi)^2 }  \f{ \mu (\om_1+\om_2) }{ (\la+\mu)R_{\da}(\la)R_{\da}(\mu)   }  
{\displaystyle  \Int{0}{+\infty} } \dd x\,\dd y\,\ex{ \i \la x  + \i \mu y}\,  \partial_{x}\Bigg\{
 S(x-y)\,\Bigg( \mf{r}(x) - \mf{r}(y) - \f{x-y}{2\pi (\om_1+\om_2)} \Bigg)\Bigg\} \;. 
\label{aleph0def} %
\end{multline}
The integrand of $\aleph_0$ involves the function $\mf{r}$ which is given by 
\beq
\mf{r}(x) \; = \;  \frac{\mf{c}_1(x) + \mf{c}_0(x)\Bigg[\sul{p=1}{2} \f{1}{2\pi \om_p} \ln \Big( \f{\om_1\om_2}{\om_p(\om_1+\om_2) }\Big)\Bigg]}{1+2\pi \be (\om_1+\om_2)  \mf{c}_0(x)}
\enq
with
\beq
\mf{c}_p(x) \, = \, \f{\i^p }{ 2{\rm i}\pi \sqrt{\om_1+\om_2} } 
\Int{ \msc{C}^{(+)}_{\e{reg}} }{} \f{ \ex{\i \la x} }{ \la } \f{\Dp{}^p }{ \Dp{}\la^p}\Big( \f{1}{ R_{\da}(\la)} \Big) \cdot \f{ \, \dd \la }{2\i \pi  }  \;. 
\enq
\end{theorem}
The result for $\beta \neq 1$ contains two more terms, and is given in the body in the book, by Proposition~\ref{Theorem DA N dependent de la fct de partition}, in terms of $N$-dependent simple and double integrals $\mathfrak{I}_{{\rm s};\beta}^{(2)}$. The final form for the asymptotics up to $o(1)$ of these extra terms can be worked out following the steps of Section~\ref{Section asymptotic analysis of double integrals}, although we decided to leave it out of the scope of this book, since $\beta \neq 1$ does not seem to appear in quantum integrable systems. %We will nevertheless keep a general $\beta$ in the proof of Proposition~\ref{Theorem DA N dependent de la fct de partition}, because assuming $\beta = 1$ does not simplify much the problem, neither gives stronger results to this stage. It is only when one wishes to obtain an explicit, $N$-independent expressions for the asymptotics of $\ln Z_N[V]$ up to $o(1)$ that the case $\beta \neq 1$ demands extra work, namely the exact computation of $Z_N[V]$ for Gaussian potential -- which we only know for $\beta = 1$ from Appendix~\ref{Appendix Section etude de l'integrale Gaussien} -- and the asymptotics of $\mathfrak{I}_{{\rm s};\beta}^{(2)}$ and $\mathfrak{I}_{{\rm d};\beta}$.

\section{The $N$-dependent equilibrium measure and the master operator}
\label{Section mesure equilibre et operateur master}

It is not hard to generalise the proof of Theorem~\ref{Proposition Asympt non rescaled part fct}
to the present setting so as to obtain the below characterisation of the leading in $N$ asymptotic behaviour for $Z_N[V]$. 

\begin{theorem}
 \label{Proposition LDP order dominant fct part rescalee ZN}

Let $\mc{E}_{\infty}$ be the lower semi-continuous good rate function 
\beq
\mc{E}_{\infty}[ \mu ] \; =\frac{1}{2} \; \Int{}{}\left(  V(\eta)+
V(\xi) \; - \;  {\pi\be (\om_1+\om_2) } 
|\xi-\eta|\right)\,\dd \mu(\xi)\dd \mu(\eta) \;.%A: cette ofrmulation set mieux define
\label{definition rate fct pour fct part ZN rescalee}
\enq
Then, one has that 
\beq
\lim_{N \tend +\infty}  \f{\ln Z_N[V] }{N^{2+\a} } \; = \; - \inf_{\mu \in \mc{M}^{1}(\R)} \mc{E}_{\infty}\big[\mu \big] \;. 
\enq
The infimum is attained at a unique probability measure $\mu_{\e{eq}}$. This measure is continuous with respect to the Lebesgue measure, and has density
\beq
\rho_{\e{eq}}(\xi) \; = \; \f{ V^{\prime \prime}(\xi) }{ 2\pi \be (\om_1+\om_2) }\cdot\bs{1}_{\intff{a}{b}}(\xi) 
\label{denspot} 
\enq
supported on the interval $\intff{a}{b}$, with $(a,b)$ being the unique solution to the set of equations
\beq
V^{\prime}(b) \; = \; - V^{\prime}(a) \; = \;  \pi \be (\om_1+\om_2) \;. 
\label{ecriture equation definissant couple a b asympt}
\enq
One has, explicitly, 
\beq
\lim_{N \tend +\infty} \f{ \ln Z_N[V]}{N^{2+\a} } \; = \; -\f{ V(a)+V(b) }{2}
\, + \, \f{ \big( V^{\prime}(b) \big)^2(b-a) \, + \,  \int_{a}^{b} \big( V^{\prime}(\xi) \big)^2\,\dd \xi   }{ 4\pi \be (\om_1+\om_2) }     \;.
\enq

\end{theorem} 

The strict convexity of $V$ guarantees that the density \eqref{denspot} is positive and that it reaches a non-zero limit at the endpoints of the support. 
This behaviour differs from the situation usually studied in $\beta$ ensembles with analytic potentials
which leads to a generic square root (or inverse square root) vanishing (or divergence) of the equilibrium density at the edges.

Note that the function $\mc{E}_{\infty}$ defined in \eqref{definition rate fct pour fct part ZN rescalee} arises as a good rate 
function in the large deviation estimates for the empirical measure $L_N^{(\bs{\la})}$, \textit{c}.\textit{f}. \eqref{ecriture integrand ZN beta ens avec def mes empirique}.
In fact, a refinement of Theorem~\ref{Proposition LDP order dominant fct part rescalee ZN} would lead to the more precise estimates
\beq
 \ln Z_N[V]  \; = \; - N^{2+\a} \mc{E}_{\infty}\big[\mu_{\e{eq}} \big] \; + \; \e{O}(N^2) \;. 
\enq
Thus, with respect to the usual varying weight $\be$-ensemble case, there is a loss of precision by a $N^{1-\a}$ factor. 
This, in fact, takes its origin in that the purely asymptotic rate function $\mc{E}_{\infty}\big[\mu_{\e{eq}} \big]$ does not absorb
enough of the fine structure of the saddle-point. As a consequence, the remainder $\e{O}(N^2)$ mixes both types of contributions:
the deviation of the saddle-point with respect to its asymptotic position and the fluctuation of the integration variables around the saddle-point. 

The fine, $N$-dependent, structure of the saddle-point is much better captured by the $N$-dependent
deformation\footnote{The property of lower semi-continuity along with the fact that $\mc{E}_N$ has compact level sets is verified exactly as in the case of $\be$-ensembles, 
so we do not repeat the proof here.} of the rate functions $\mc{E}_{\infty}$:
\beq
\mc{E}_N [ \mu  ] \; = \f{1}{2} \; \Int{}{} \left(V(\xi)\,+\,V(\eta) \,-\, \f{ \be  }{  N^{\a} } 
\ln\bigg\{ \pl{p=1}{2} \sinh\big[ \pi N^{\a} \om_p(\xi-\eta) \big] \bigg\}\right)\,\dd \mu(\xi)\dd \mu(\eta) \;. %A: je prefer cette formulation
\label{definition fnelle a minimiser N dpdte}
\enq

This $N$-dependent rate functions appear extremely effective for the purpose of our analysis. 
Namely, it allows us re-summing a whole tower of contributions into a single term. The use of $\mc{E}_N$ 
should not be considered as a mere technical simplification of the intermediate steps; it is, in fact, of prime importance. 
The use of the more classical object $\mc{E_{\infty} }$ would render the analysis of the Schwinger-Dyson equations impossible. 
This fact will become apparent in the core of the file.

As usual, this minimiser admits a  characterisation in terms of a variational problem:
\begin{prop}
\label{Proposition caracterisation rudimentaire mesure equilibre}
 For any strictly convex potential $V$, the $N$-dependent rate function $\mc{E}_N$ admits its minimum on $\mc{M}^{1}(\R)$ at a unique probability measure 
$\mu_{\e{eq}}^{(N)}$. This equilibrium measure is supported on a segment $\intff{a_N}{b_N}$ and
corresponds to the unique solution to the integral equations
\beqa
&&V(\xi) \; - \; \f{ \be  }{  N^{\a} } \Int{}{} 
\ln\bigg\{ \pl{p=1}{2} \sinh\big[ \pi N^{\a} \om_p(\xi-\eta) \big] \bigg\}\,\dd \mu_{\e{eq}}^{(N)}(\eta) \; = \; C_{\e{eq}}^{(N)}  
\qquad  \e{on}  \; \; \intff{a_N}{b_N} 
\label{definition de la cste Ceq par eqn int eq meas} \\
&& V(\xi) \; - \; \f{ \be  }{  N^{\a} } \Int{}{} 
\ln\bigg\{ \pl{p=1}{2} \sinh\big[ \pi N^{\a} \om_p(\xi-\eta) \big] \bigg\}\,\dd \mu_{\e{eq}}^{(N)}(\eta) \; >  \; C_{\e{eq}}^{(N)} 
\qquad  \e{on} \;\; \R \setminus \intff{a_N}{b_N}  \;, 
\label{ecriture condition negativite dehors support mu eq}
\eeqa
with $C_{\e{eq}}^{(N)}$ a constant whose determination is part of the problem
\eqref{definition de la cste Ceq par eqn int eq meas}-\eqref{ecriture condition negativite dehors support mu eq}.
The equilibrium measure admits a density $\rho_{\e{eq}}^{(N)}$, which 
is $\mc{C}^{k-2}$ in the interior $]a_N;b_N[$ if $V$ is $\mc{C}^{k}$. Finally, one has the behaviour at the edges:
\beq
\rho_{\e{eq}}^{(N)}(\xi) \; \mathop{=}_{\xi \rightarrow a_N^+}\; \e{O}\big(\sqrt{\xi- a_N}\big)\;,\qquad \qquad \rho_{\e{eq}}^{(N)} (\xi)\; \mathop{=}_{\xi \rightarrow b_N^-} \; \e{O}\big(\sqrt{b_N - \xi}\big)\;.
\label{ecriture du comportement en racine pour mesure equilibre}
\enq
\end{prop}

The proof of this proposition is rather classical. It follows, for instance, from \cite[Section 2.3]{KozBorotGuionnetLargeNBehMulIntMeanFieldTh} in what concerns the regularity, 
and from a convexity argument (see \cite[Theorem 2.2]{MhaskarSaffLocumOfSupNormWeightedPly}) in what concerns connectedness of the support and the strict inequality in \eqref{ecriture condition negativite dehors support mu eq}. Elements of proof are nevertheless gathered in Appendix \ref{Appendix minimisation de la mesure equilibre}. 
In fact, regarding to the equilibrium measure, we can be much more precise when $N$ is large enough:
%However, one can be much more precise when $N$ is large enough. Then, one has the 

\begin{theorem}
\label{Theorem characterisation densite mesure eq a grand N}

In the $N \rightarrow \infty$ regime, the equilibrium measure $\mu_{\e{eq}}^{(N)}$:%There exists $N_{0}$ such that, for all $N \geq N_0$ the equilibrium measure $\mu_{\e{eq}}^{(N)}$
\begin{itemize}
\item is supported on the single interval $\intff{a_N}{b_N}$ whose endpoints admit the asymptotic expansion
\beq
a_N \; =  \; a \, +\, \sul{\ell=1}{k}  \f{ a_{N;\ell}  }{ N^{\ell \a}  }  \, + \, \e{O}\Bigg( \f{1}{N^{(k+1) \a}} \Bigg)
\quad \e{and} \quad b_N  \; = \;  b \, +\, \sul{\ell=1}{k}  \f{ b_{N;\ell}  }{ N^{\ell \a}  }  \, + \, \e{O}\Bigg( \f{ 1 }{N^{ (k+1) \a}} \Bigg)  \;,
\label{ecriture DA grand N point bord du support de mu eq N}
\enq
where $k\in \mathbb{N}^*$ is arbitrary, $(a,b)$ are as defined in \eqref{ecriture equation definissant couple a b asympt} while
\beq
\left( \ba{c} b_{N;1} \\ 
	      a_{N;1} \ea \right)  \;=\;\bigg\{ \sul{p=1}{2} \f{1}{2\pi \om_p} \ln \Big( \f{ \om_1 \om_2 }{ \om_p (\om_1+\om_2) } \Big) \bigg\}  \cdot 
	      \left( \ba{c}  V^{\prime\prime}(a) \cdot \big\{ V^{\prime\prime}(b) \big\}^{-1}    \\ 
	      -  V^{\prime\prime}(b) \cdot \big\{ V^{\prime\prime}(a) \big\}^{-1}  \ea \right) \; ; 
\label{blbalb}
\enq

\item is continuous with respect to Lebesgue. Its density is $\rho_{\e{eq}}^{(N)}$ vanishes like a squareroot at the edges:
\beq
\rho_{\e{eq}}^{(N)}(\xi) \underset{\xi \tend a_N^{+} }{\sim} 
\bigg( \frac{V^{\prime\prime}(a_N) \, + \, \e{O}(N^{-\a}) }{\pi \be \sqrt{\pi(\om_1+\om_2)} }   \bigg)\, \sqrt{\xi-a_N}\;,\qquad
  \rho_{\e{eq}}^{(N)}(\xi) \underset{\xi \tend b_N^{-} }{\sim} 
 \bigg( \frac{V^{\prime\prime}(b_N) \, + \, \e{O}(N^{-\a}) }{\pi \be \sqrt{\pi(\om_1+\om_2)} }   \bigg) \, \sqrt{b_N-\xi}\;, 
\enq
and there exists a constant $C > 0$ independent of $N$ such that:
\beq
\norm{\rho_{\e{eq}}^{(N)}}_{L^{\infty}(\intff{a_N}{b_N})}  \; \leq  \;  C\,\norm{V^{\prime\prime} }_{ L^{\infty}(\intff{a_N}{b_N}) }\;.
\label{ecriture borne uniforme sur la densite}
\enq
This density takes the form $\rho_{\e{eq}}^{(N)}=\mc{W}_N[V^{\prime}]$, with $\mc{W}_N$ as defined in \eqref{definition operateur WN}.

\end{itemize}

If the potential $V$ defining the equilibrium measures satisfies  $V \in \mc{C}^{k}(\intff{a_N}{b_N})$, 
then the density is of class $\mc{C}^{k-2}$ on $\intoo{a_N}{b_N}$.

\end{theorem}

Note that the characterisation of $\rho_{\e{eq}}^{(N)}$ in the theorem above comes from the fact that it is solution to the singular integral equation $\mc{S}_N\big[ \rho_{ \e{eq} }^{(N)} \big](\xi) = V^{\prime}(\xi)$
on $\intff{a_N}{b_N}$, where 
\beq
\mc{S}_N\big[ \phi \big](\xi) \; = \; \Fint{a_N}{b_N} S\big[N^{\a}(\xi-\eta)\big] \phi(\eta)\,\dd \eta \; \qquad
\e{and} \qquad S(\xi) \; = \; \sul{p=1}{2} \beta\pi\om_p  \cotanh\big[ \pi \om_p \xi \big] \;. 
\label{ecriture eqn int sing de depart}
\enq
The unknowns in this equation $(\rho_{ \e{eq} }^{(N)}, a_N,b_N)$ should be picked in such a way that 
$\rho_{ \e{eq} }^{(N)}$ has mass 1 on $\intff{a_N}{b_N}$ and is regular at the endpoints $a_N,b_N$. 
Thus, determining the equilibrium measure boils down to an inversion of the singular integral operator $\mc{S}_N$.
In fact, the singular integral operator $\mc{S}_N$ also intervenes in the Schwinger-Dyson equations.  The precise control on its inverse $\mc{W}_N$ -- defined between appropriate functional spaces -- plays
a crucial role in the whole asymptotic analysis. 

These pieces of information can be obtained by exploiting the fact that the operator $\mc{S}_N$ is of truncated Wiener--Hopf 
type. As such, its inversion is equivalent to solving a $2\times 2$ matrix valued Riemann--Hilbert problem. 
This Riemann--Hilbert problem admits a solution for $N$ large enough that can be constructed by means of a variant
of the non-linear steepest descent method. By doing so, we are able to describe, quite explicitly, the inverse $\mc{W}_N$
by means of the unique solution $\chi$ to the $2\times 2$ matrix valued Riemann--Hilbert problem given in Section \ref{SousSection RHP chi initial}. 
We will not discuss the structure of this solution here and, rather, refer the reader to the relevant section. 
We will, however, provide the main consequence of this analysis, \textit{viz}. an explicit representation for the operator $\mc{W}_N$. 
For this purpose, we need to announce that $\chi_{11}$, the $(1,1)$ matrix entry of $\chi$, is such that 
$\mu  \mapsto \mu^{1/2} \cdot \chi_{11}(\mu) \in L^{\infty}(\R)$. 

\begin{theorem}
Let $0 < s <1/2$. The operator $\mc{S}_N \; : \; H_{s}\big( \intff{a_N}{b_N} \big) \tend 
\mf{X}_{s}\big( \R  \big)$ is continuous and invertible where, for any closed $A \subseteq \R$,  
\beq
\mf{X}_{s}\big(  A \big) \; = \; \Big\{  H \in H_{s}\big( A  \big) \; :  \;  
\Int{\R+\i \eps }{} \chi_{11}(\mu) \mc{F}[H](N^{\a}\mu)\ex{- \i N^{\a} \mu b_N}   \f{ \dd \mu }{ 2 \i \pi }  \; = \; 0\Big\} \; 
\enq
is a closed subspace of $ H_{s}\big(  A \big) $ such that $\mc{S}_N\big( H_{s}\big( \intff{a_N}{b_N} \big) \big) \, =  \, \mf{X}_s(\R)$.  The inverse is given by the integral transform $\mc{W}_N $ which takes, for $H\in \mc{C}^1(\intff{a_N}{b_N})\cap\mf{X}_s(\R)$, the form 
\beq
\mc{W}_N[H](\xi) \; = \; \f{ N^{2\a} }{ 2\pi \be}  
\Int{ \R + 2 \i \eps }{} \f{ \dd \la }{ 2 \i \pi } \Int{ \R + \i \eps }{} \f{ \dd \mu }{ 2 \i \pi } 
\f{ \ex{- \i N^{\a}(\xi-a_N) \la}  }{ \mu- \la }
\bigg\{   \chi_{11}(\la) \chi_{12}(\mu)  - \f{ \mu }{ \la }\cdot\chi_{11}(\mu) \chi_{12}(\la) \bigg\} 
\cdot \Int{a_N}{b_N} \!\! \dd \eta  \ex{ \i N^{\a} \mu (\eta-b_N) } H(\eta)  \;. 
\label{definition operateur WN}
\enq
In the above integral representations the parameter $\eps >0$ is small enough but arbitrary. 
Furthermore, for any $H \in \mc{C}^{1}\big( \intff{a_N}{b_N} \big)$, the transform $\mc{W}_N$ 
 exhibits the local behaviour 
\beq
\mc{W}_N[H] (\xi) \underset{ \xi \tend a_N^+ }{ \sim  } C_L H^{\prime}(a_N) \sqrt{ \xi - a_N} \qquad and \qquad
\mc{W}_N[H] (\xi) \underset{ \xi \tend b_N^- }{ \sim  } C_R H^{\prime}(b_N) \sqrt{ b_N - \xi } \;. 
\enq
where $C_{L/R}$ are some $H$-independent constants. 
\end{theorem}

Note that, within such a framework, the density of the equilibrium measure $\mu_{\e{eq}}^{(N)}$ is expressed in terms of the inverse as 
$\rho_{\e{eq}}^{(N)}=\mc{W}_N\big[ V^{\prime} \big]$. In this case, the pair of endpoints $(a_N, b_N)$ of the support of $\mu_{\e{eq}}^{(N)}$ 
corresponds to the unique solution to the system of equations 
\beq
\Int{a_N}{b_N} \mc{W}_N\big[ V^{\prime} \big](\xi)\,\dd  \xi \; = \; 1 \qquad \e{and} \qquad 
\Int{\R + \i \eps }{}  \f{ \dd \mu\, \chi_{11}(\mu)}{ 2 \i \pi } 
 \Int{a_N}{b_N} \ex{\i\mu N^{\a} (\eta- b_N)} V^{\prime}(\eta) \,\dd \eta   \; = \; 0 \;. 
\enq
The first condition guarantees that $\mu_{\e{eq}}^{(N)}$ has indeed mass 1, while the second one ensures 
that its density vanishes as a square root at the edges $a_N,b_N$. Using fine properties of the inverse, these conditions can be estimated more precisely in the large-$N$ limit, hence enabling one to 
fix the large-$N$ asymptotic expansion of the endpoints $a_N, b_N$ as announced in \eqref{ecriture DA grand N point bord du support de mu eq N}-\eqref{blbalb}.

\section{The overall strategy of the proof}
\label{Section Strategie de la preuve}

In the following, we shall denote by $p_N ( \bs{\la} ) $ the probability density on $ \R^N$ associated with the partition function $Z_N[V]$
defined in \eqref{ecriture fonction partition a la beta ensemble}:
\beq
p_N ( \bs{\la} ) \;  =  \;  \f{1 }{Z_N[V]} 
\pl{a<b}{N} \Big\{ \sinh\big[\pi\om_1 N^{\a}(\la_a-\la_b)\big] \sinh\big[\pi\om_2 N^{\a}(\la_a-\la_b)\big]  \Big\}^{\be}
\,\pl{a=1}{N} \ex{- N^{1+\a} V(\la_a) }   \;. 
 \label{definition mesure proba rescalee avec potentiel arbitraire}
\enq
$p_N ( \bs{\la} )$ gives rise to a probability measure $\mathbb{P}_N$ on $\R^N$. We also agree that, throughout the file,  $L_N^{(\bs{\la})}$ refers to the empirical measure
\beq
L_N^{(\bs{\la})} \; = \; \f{1}{N} \sul{a=1}{N} \de_{\la_a} 
\label{definition mesure empirique}
\enq
associated with the stochastic vector $\bs{\la}$.

\begin{defin}
\label{Definition moyenne contre mes stochastique}
Let $\nu_1,\dots, \nu_{\ell}$ be any (possibly depending on the stochastic vector  $\bs{\la}$) 
measures and $\psi$ a function in $\ell$ variables. 
Then we agree upon 
\beq
\Big<   \psi \Big>_{\nu_1 \otimes \dots \otimes \nu_{\ell} }  \equiv   
 \Big<   \psi(\xi_1,\dots,\xi_{\ell} ) \Big>_{\nu_1 \otimes \dots \otimes \nu_{\ell} }    \equiv   
\mathbb{P}_N  \Big[     \Int{\R^{\ell} }{}    \psi(\xi_1,\dots,\xi_{\ell})\,\dd \nu_1 \otimes \dots \otimes  \dd \nu_{\ell} \Big]
\label{definition moyenne contre mesures stochastiques}
\enq
whenever it makes sense. We shall add the superscript $V$ whenever the functional dependence of the probability measure on the potential $V$ needs to be made clear.%
%\textit{viz}. 
%
%
%
%\beq
%
%\Big<   \psi \Big>_{\nu_1 \otimes \dots \otimes \nu_{\ell} }^{V}  \equiv  
%
%\mathbb{P}_N ^{V} \Big[     \Int{\R^{\ell} }{}     \psi(\xi_1,\dots,\xi_{\ell}) \cdot \dd \nu_1 \otimes \dots \otimes  \dd \nu_{\ell} \Big]
%
%\enq
%
%
% 
%in which $\mathbb{P}_N ^{V}$ stands for the probability measure on $\R^N$ subordinate to the density \eqref{definition mesure proba rescalee avec potentiel arbitraire}
%defined in terms of the potential $V$ of interest. 

\end{defin}

\noindent Note that if none of the measures $\nu_1,\dots, \nu_{\ell}$ is stochastic, then the expectation versus $\mbb{P}_N$ 
in \eqref{definition moyenne contre mesures stochastiques} can be omitted.

The Schwinger-Dyson equations constitute a tower of equations which relate expectation values of functions in many, non necessarily fixed, variables
that are integrated versus the empirical measure \eqref{definition mesure empirique}. 
More precisely, the Schwinger-Dyson equations at level $k$ ($k \geq 1$)  yield exact relations between various expectation values of a 
function in $k$ variables and its transforms, this versus the empirical measure. The knowledge of these expectation values,
yields an access to the derivatives of the partition function with respect of external parameters. For instance, if $\{V_t\}_{t}$ is a smooth one parameter family of potentials, then 
\beq
\partial_{t}\ln Z_{N}[V_t] \; = \; -N^{2+\a} \moy{ \Dp{t}V_t }_{ L_N^{(\bs{\la})} }^{V_t} \;. 
\label{equation reliant derivee log fct part et corr derivee pot}
\enq
The exponent $V_t$ appearing in the right-hand side is there so as to emphasise that the expectation value is computed with respect to the probability measure subordinate to 
the $t$-dependent potential $V_t$.

Thus the problem boils down to obtaining a sufficiently precise control on the behaviour in $N$ of the one-point expectation values. 
This can be achieved on the basis of a careful analysis of the system of Schwinger-Dyson equations associated with the present model. 
Since this machinery does not simplify much in the $\beta = 1$ case, we do this for general $\beta$. The result for some sufficiently regular function $H$ and potentials $V$
satisfying to the general hypothesis, is our Proposition \ref{Poposition DA correlateur a un point}.

In the  $\beta = 1$ case, Proposition~\ref{Poposition DA correlateur a un point} reads:
\beq
 -N^{2+\a} \moy{ H }_{L_N^{(\bs{\la})}}^{V}
\; = \; -N^{2+\a} \Int{a_N}{b_N} \hspace{-1mm} H(\xi)\cdot\mc{W}_N\big[V^{\prime}\big](\xi)\,\dd \xi \;+   \; 
\f{1}{2} \mf{I}_{\e{d}}\big[ H , V  \big] \; + \; \e{o}(1) \;. 
\label{ecriture DA de ln ZV a partir eqns de boucles}
\enq
and the proof shows that the remainder $\e{o}(1)$ is uniform in $H$ and $V$ provided that $H$ is regular enough and that $V$ satisfies to the hypothesis given in 
\eqref{Hypothesis: confinment of the potential}-\eqref{Hypothese sous exponetialite des derivees}.  Furthermore,  the expansion \eqref{ecriture DA de ln ZV a partir eqns de boucles}  involves 
\beq
\mf{I}_{\e{d}}[H,V] \; = \; \Int{a_N}{b_N} \mc{W}_N\Big[ 
\partial_{\xi}\big\{ S\big(N^{\a}(\xi-*)\big) \cdot \mc{G}_N\big[H,V\big](\xi,*)     \big\}   \Big](\xi) \, \dd \xi\;,
\label{definition integrale double du DA correlateur a un point0}
\enq
with 
\beq
 \mc{G}_N\big[H,V\big](\xi,\eta) \; = \; \f{ \mc{W}_N[H](\xi) }{ \mc{W}_N[V^{\prime}](\xi)  }
 \; - \; \f{ \mc{W}_N[H](\eta) }{ \mc{W}_N[V^{\prime}](\eta)  } \;. 
\enq
Note that, in \eqref{definition integrale double du DA correlateur a un point0}, the $*$ indicates 
the variable of the function on which the operator $\mc{W}_N$ acts. Given sufficiently regular functions $H,V$, we obtain in Section \ref{Section asymptotic analysis of double integrals} and
more precisely in Proposition~\ref{Theorem DA ordre dominant integrale double} the large-$N$ asymptotic behaviour of $\mf{I}_{\e{d}}[H,V]$.  
We then have all the elements to calculate the large-$N$ asymptotic behaviour of the partition function $Z_{N}[V]$. 
For this purpose, we observe that, when $\be=1$, the partition function associated to a quadratic potential can be explicitly evaluated as shown in 
Proposition \ref{Proposition calcul explicite fct part Gaussienne}. One can also show (\textit{cf}. Lemma~\ref{Proposition pot Gaussien avec meme support mesure eq}) 
that there exists a unique, up to a constant, quadratic potential $V_{G;N}$ such that its associated equilibrium measure
has the same support $[a_N,b_N]$ as the one associated with V. Then $V_t=(1-t)V_{G;N}+t V$ is a one parameter $t$ smooth
family of strictly convex potentials, and $\mu_{{\rm eq};V_{t}}^{(N)} = (1 - t)\mu_{{\rm eq};V_{G;N}}^{(N)} + t\mu_{{\rm eq};V}^{(N)}$.  %A:ajoute N
Furthermore, if follows from the details of the analysis that led to \eqref{ecriture DA de ln ZV a partir eqns de boucles}
that the remainder $\e{o}(1)$ will be uniform in $t \in \intff{0}{1}$. As a consequence,
by combining all of the above results and integrating 
 equation \eqref{equation reliant derivee log fct part et corr derivee pot} over $t$, we get that, in the asymptotic regime,
\begin{multline}
\ln \Bigg( \f{ Z_{N}[V] }{ Z_{N}[V_{G;N}] } \Bigg) \; 
=\; -N^{2+\a} \Int{0}{1} \dd t \Int{}{} \Dp{t}V_t(\xi)\, \dd \mu_{\e{eq};V_t}^{(N)}(\xi)
\; + \; N^ {\a}\cdot\gimel_0\cdot \Big( \leo[V,V_{G;N}](b_N)- \leo[V,V_{G;N}](a_N) \Big)   \\
\; + \; \aleph_0\cdot \Big( \leo^{\prime} [V,V_{G;N}](b_N) +  \leo^{\prime} [V,V_{G;N}](a_N) \Big) \; + \; \e{o}(1) \;. 
\label{ecriture DA up to o1 de fct partition}
\end{multline}
The constants $\gimel_0$ and $\aleph_0$ were defined respectively in \eqref{gimel0def} and \eqref{aleph0def}, while $\leo$ is as given by \eqref{leoDef}.

Note also that the first integral can be readily evaluated (integration of rational functions in t) on the asymptotic
level by means of Proposition~\ref{Theorem DA tout ordre integrale 1D contre WN}. It produces an expansion into inverse powers of $N^{\a}$ and, as such, does not contribute to the constant term unless $\alpha$ is 
of the form $2/n$ for some integer $n$. Note that it is this integral that gives rise to the functional $\capricornus_p[V]$ in \eqref{Ecriture DA ZN dans partie Intro resultats}. 
Finally, the answer for the large-$N$ asymptotic behaviour of the partition function $Z_{N}[V_{G;N}]_{\mid\be=1}$  can be found in Proposition \ref{Proposition calcul explicite fct part Gaussienne}.  
As follows from Lemma ~\ref{Proposition pot Gaussien avec meme support mesure eq} the quadratic potential $V_{G;N}$ is such that $V_{G;N}-W_{G;N}=\e{O}\big(N^{-\infty}\big)$
with $W_{G;N}$ as defined in \ref{definition potentiel WGN} and where the remainder is uniform on some $N$-independent relatively compact open neighbourhood of $\intff{a_N}{b_N}$. This allows one to replace $V_{G;N}$ by $W_{G;N}$
in the right hand side of \eqref{ecriture DA up to o1 de fct partition}. Also, it is clear form the large-$N$ expansion of the partition function associated with quadratic potentials given in 
Proposition \ref{Proposition calcul explicite fct part Gaussienne} that $\ln Z_{N}[V_{G;N}]_{\mid\be=1}-\ln Z_{N}[W_{G;N}]_{\mid\be=1}=\e{o}(1)$.

For $\beta \neq 1$, \eqref{ecriture DA de ln ZV a partir eqns de boucles} is modified by the addition of two more terms $\mathfrak{I}_{{\rm s};\beta}^{(2)}$ and $\mathfrak{I}_{{\rm d};\beta}$. 
Their large $N$ behaviour can be determined without difficulty -- but with some algebra -- along the lines of Section~\ref{simpleintegrals} and \S~\ref{Section asymptotic analysis of double integrals}. 
Then, to arrive to a final answer for $Z_{N}[V]_{\beta \neq 1}$ similar to \eqref{ecriture DA up to o1 de fct partition}, we would need to compute exactly the partition function for the Gaussian potential $Z_{N}[V_{G;N}]_{\beta \neq 1}$. 
We do not know at present how to perform such a calculation. Thus, we would be able to derive the asymptotic behaviour of the partition function at $\be\not=1$ up to a universal, \textit{i.e.} not depending on the potential $V$, function of $\be$. 
However, since the values $\beta \neq 1$ do not seem to appear in quantum integrable systems, we shall limit ourselves in this book to the result of Proposition~\ref{Theorem DA N dependent de la fct de partition} for the case $\beta \neq 1$.

\chapter{Asymptotic expansion of $\ln Z_N[V]$ - the Schwinger-Dyson equation approach}
\label{Section Analyse asympt eqn boucles}

{ \bf Abstract}

\textit{ In the present chapter we develop all the necessary tools to prove the large-$N$ asymptotic 
expansion for $\ln Z_N[V]$ up to $\e{o}(1)$ terms, in the form described in \eqref{ecriture DA de ln ZV a partir eqns de boucles}. 
This asymptotic expansion contains $N$-dependent functionals of the equilibrium measure whose large-$N$ asymptotic analysis will be carried out in Sections 
\ref{Section asymptotic analysis of single integrals}-\ref{Section asymptotic analysis of double integrals}.
We shall first obtain some \textit{a priori} bounds on the fluctuations of linear statistics
around their means computed \textit{vs}. the $N$-dependent equilibrium measure $\mu_{\e{eq}}^{(N)}$. In other words, we consider observables given by integration against products of the centred measure:
$ {\cal L}_{N}^{(\bs{\la})} = L_{N}^{(\bs{\la})} - \mu_{{\rm eq}}^{(N)}\;.$
Then we shall build on a bootstrap approach to the Schwinger-Dyson equations so as to improve these \textit{a priori}
bounds.  We shall use these improved bounds so as to identify the leading and sub-leading terms 
in the Schwinger-Dyson equations what, eventually, leads to an analogue, at $\be\not=1$, of the representation \eqref{ecriture DA de ln ZV a partir eqns de boucles}
which will be given in Proposition \ref{Poposition DA correlateur a un point}.
Finally, upon integrating the relation \eqref{equation reliant derivee log fct part et corr derivee pot} so as to 
to interpolate the partition function between a Gaussian and a general potential, we will get the $N$-dependent 
large-$N$ asymptotic expansion of $\ln Z_N[V]$ in Proposition~\ref{Theorem DA N dependent de la fct de partition}. }

\section{A priori estimates for the fluctuations around $\mu_{\e{eq}}^{(N)}$}

For simplification, we use the notation:
\beq
\label{definition fonction sN noyau integral eqn mes eq} s_N(\xi) \; = \; \f{ \be }{ 2 N^{ \a } } \ln \Big[ \sinh\big(\pi \om_1 N^{\a}\xi \big) \,\sinh\big(\pi \om_2 N^{\a}\xi \big) \Big]
\enq
for the two-body interaction kernel. This kernel provides a natural way of comparing two probability measures:
\begin{defin}
\label{dddef} If $\mu,\nu \in \mathcal{M}_{1}(\mathbb{R})$, we set:
\beq
\mf{D}^2 \big[ \mu, \nu \big] \; \equiv  \; -  \Int{}{}  s_N(\xi-\eta)\,\dd (\mu-\nu)(\xi)\,\dd (\mu-\nu)(\eta)\;,
\enq
with $s_N$ as given in \eqref{definition fonction sN noyau integral eqn mes eq}. $\mf{D}^2 \big[ \mu, \nu \big] $ is a well-defined number in $\mathbb{R}\cup\{+\infty\}$.
\end{defin}
The notation is justified by the property $\mathfrak{D}^2 \geq 0$ following from:
\begin{lemme}
We have the representation:
\beq
\mf{D}^2 \big[ \mu, \nu \big]  \; = \; \Int{}{} \bigg\{ \f{ \pi \be }{ 2  N^{\a}  \vp } 
\sul{p=1}{2}   \cotanh\Big[ \f{ \vp }{ 2 \om_p N^{\a}  } \Big]   \bigg\} \cdot  \big|  \mc{F}[\mu-\nu ] ( \vp ) \big|^2  \cdot \f{ \dd \vp }{2\pi}\;,
\enq
where $\mc{F}[\mu](\xi)$ is the Fourier transform of the measure $\mu$.
\end{lemme}

\Proof  The claim follows in virtue of the formula $\mc{F}\big[ f_{t} \big]( \vp ) \; = \;  - (\pi/\vp)\big(\cotanh[\pi \vp/2t] - 2t/\pi\vp\big)$ with 
\newline $f_{t}(x) \; = \; \ln | \sinh(tx) | - t|x| +\ln 2$. \qed

\begin{defin}
\label{clapo}The classical positions $x_i^{N}$ for the measure $\mu_{{\rm eq}}^{(N)}$ are defined by 
%s of the integration variables $\bs{\la} \in \R^N$ under the large-$N$ confinement
%induced by the probability density $p_N(\bs{\la})$ given in \eqref{definition mesure proba rescalee avec potentiel arbitraire}:
%
%
%
\beq
\f{i}{N} \; = \; \Int{-\infty}{x_i^N} \dd \mu_{\e{eq}}^{(N)}(y) \qquad for \qquad i \in \intn{1}{N}  \qquad and  \qquad x_0^N = a_N \; \; \;  ,  \; \; \; x_N^N=b_N \;. 
\enq
\end{defin}

Our first task is to derive a lower bound for the partition function \eqref{ecriture fonction partition a la beta ensemble}, by restricting to configurations of points close to their classical positions:
\begin{lemme}
\label{Lemme borne inf plus fine sur Z_N} $Z_N [V] \; \geq \;   
\exp\Big\{ - N ^{2+\a}\mc{E}_N\big[ \mu_{\e{eq}}^{(N)} \big]    \; + \;  \e{O}\big( N^{1+\a} \big)   \Big\} \;.$
\end{lemme}
We stress on this occasion that using the $N$-dependent rate function $\mc{E}_{N}$ allows the gain of a factor $1/N$ in the remainder with respect to the leading term, 
while using $\mc{E}_{\infty}$ would lead to a weaker estimate $\e{O}(N^2)$ for the remainder. This is of particular importance to simplify the analysis of Schwinger-Dyson equations that will follow.

\Proof It follows from the local expressions obtained in Section \ref{Section descirption cptmt unif op WN} that 
$\mu_{\e{eq}}^{(N)}$ is continuous with respect to Lebesgue measure with density bounded by a constant $M$ independent of $N$, as shown in \eqref{ecriture borne uniforme sur la densite}. This ensures that  
\beq
\big|   x_{i+1}^N - x_i^N \big|  \geq \f{1}{M N} \;,\qquad i \in \intn{0}{N - 1}\;. 
\label{ecriture borne inf pour espacement des vp classiques}
\enq
We obtain our lower bound by keeping only configurations in 
\beq
\Om \;  =\; \big\{   \bs{\la} \in \R^N \; : \; \sup_{a} |\la_a - x_a^{N} |  \, \leq  \, \f{1}{4MN}  \big\} \; .
\nonumber
\enq
Let $\sg_{ \eps }$ be some $N$-independent $\eps$-neighbourhood of $\intff{a_N}{b_N}$. Since $V \in \mc{C}^1(\R)$, it follows that 
\beq
\big|   V(\la_a) - V(x_a^N)   \big|  \; \leq  \;   \f{  \norm{ V^{\prime} }_{L^{\infty}(\sg_{\eps})}  }{ 4MN } \qquad \quad viz.  \qquad  \quad
 - V(x_a^N)  -  \f{  \norm{ V^{\prime} }_{L^{\infty}(\sg_{\eps})}  }{ 4MN }   \; \leq  \;  -  V(\la_a) 
\enq
for $a \in \intn{1}{N}$ and for any $\bs{\la} \in \Om$. 
Thus, upon a re-centring at $x_a^N$ of the integration with respect to $\la_a$, we get 
\bem
Z_N [V] \geq \pl{a=1}{N}\Big\{  \ex{-N^{1+\a} V(x_a^N)}  \Big\}  \cdot    
\ex{  -  \f{   N^{1+\a}  }{ 4M }  \norm{ V^{\prime} }_{L^{\infty}(\sg_{\eps})} }
\times \!\!\!\!\!\!\!\!\!\!\!\!\Int{[-1/(4MN),1/(4MN)]^N}{} \!\!\!\!\!\!\!\!\!\!\!\! \dd ^N \nu\cdot   \pl{a<b}{N} 
\bigg\{ \pl{p=1}{2} \sinh\big[ \pi \om_p N^{\a} ( \nu_a -  \nu_b  +x_a^N - x_b^N ) \big]   \bigg\}^{\be}
\\
\; \geq \;\pl{a=1}{N} \Big\{ \ex{-N^{1+\a} V(x_a^N)} \Big\}   \cdot   
 \ex{  -   \f{   N^{1+\a}  }{ 4M } \norm{ V^{\prime} }_{L^{\infty}(\sg_{\eps})}  }  \times 
\pl{a<b}{N} \ex{ 2 N^{\a} s_N( x_b^N - x_a^N ) }
%
%\bigg\{ \pl{p=1}{2} \sinh\big[ \pi \om_p N^{\a} ( x_b^N - x_a^N ) \big]   \bigg\}^{\be}
%
\times \Int{    \nu_1 < \dots < \nu_N   }{} \dd ^N \nu \pl{a=1}{N} \Big\{  \bs{1}_ {|\xi| < \f{1}{4MN}}(\nu_a)     \Big\}  \;. 
\end{multline}
We remind that $s_N$ has been defined in \eqref{definition fonction sN noyau integral eqn mes eq}. The second line is obtained by keeping only the configurations 
where $i \mapsto \nu_i$ is increasing, and then using that $\sinh$ is an increasing function. Finally:
\beq
Z_N  [V] \; \geq \; \pl{a=1}{N} \Big\{ \ex{-N^{1+\a} V(x_a^N)} \Big\}   \cdot   
 \ex{  -   \f{   N^{1+\a}  }{ 4M } \norm{ V^{\prime} }_{L^{\infty}(\sg_{\eps})}  }  \cdot 
\pl{a<b}{N} \ex{ 2 N^{\a} s_N( x_b^N - x_a^N ) }
   \cdot  \f{1}{N!}  \bigg( \f{1}{4NM} \bigg)^N \;. 
\enq
We rewrite the first product involving the potential by comparison between the Riemann sum and the integral:
\beq
\f{1}{N} \sul{a=1}{N} V( x_a^N) \; = \; \Int{\R}{} V(  \xi )\,\dd   \mu_{\e{eq}}^{(N)}(\xi)   \; + \; \de_N,\qquad |\de_N| \leq \f{  \norm{V^{\prime}}_{L^{\infty}( \sg_{\eps} )  }   }{ N }   \cdot (b_{N} - a_N)\;.
\enq
%%Gaetan: shortcut \qquad  \e{with} \quad 
%
%\de_N  \; = \; \f{1}{N} \sul{a=0}{N-1} \Int{x_a^N}{x_{a+1}^N} \big[ V(  x_a^N )-  V(  \xi ) \big]    \cdot   \dd  \mu_{\e{eq}}^{(N)}(\xi)   \;.
%
%\enq
%
%
%
%The remainder can be bounded as 
%
%
%
%\beq
%
%|  \de_N  |  \; \leq \;    \sul{a=0}{N-1}   \norm{V^{\prime}}_{L^{\infty}( \sg_{\eps} ) }  
%
%  \Int{x_a^N}{x_{a+1}^N}  |  x_a^N -   \xi  |  \cdot \dd   \mu_{\e{eq}}^{(N)}(\xi)  \; \leq \; 
%
%\f{  \norm{V^{\prime}}_{L^{\infty}( \sg_{\eps} )  }   }{ N }   \cdot (b_{N} - a_N) \;. 
%
%\enq
%
%
%
It thus remains to bound from below the $\be$-exponent part. Using that $s_N$ is increasing on $\R^+$, we get:
\bem
\Int{x<y}{} s_N(y-x) \, \dd \mu_{\e{eq}}^{(N)}(x)\,\dd \mu_{\e{eq}}^{(N)}(y) \; = \; 
\sul{a,b=0}{N-1}  \Int{x_a^N}{x_{a+1}^N } \Int{x_b^N}{x_{b+1}^N }  
			 \bs{1}_{x<y}(x,y) s_N(y-x)\,\dd \mu_{\e{eq}}(x)\,\dd \mu_{\e{eq}}(y)    \\
\leq \;  \f{ 1 }{ N^2 }\sul{a=0}{N-1}  \sul{b=a+1}{N-1} s_N(x_ {b+1}^N - x_a^N)    \; + \; 
\sul{a=0}{N-1} s_N(x_{a+1}^N - x_a^N)  \cdot \f{1}{2N^2} \;. 
\end{multline}
The first sum can be recast as 
\beq
\sul{a=0}{N-1}  \sul{b=a+2}{N} s_N(x_ {b}^N - x_a^N) \; = \; 
\sul{a=1}{N-1}  \sul{b=a+1}{N} s_N(x_ {b}^N - x_a^N)
		   \; + \; \sul{b=1}{N} s_N(x_ {b}^N - x_0^N)  
		   					\; - \; \sul{a=0}{N-1} s_N (x_ {a+1}^N - x_a^N)  \;. 
\enq
 It follows from \eqref{ecriture borne inf pour espacement des vp classiques} and from $|x_a^N-x_{b}^N| < |b_N-a_N| < C$ for some $C>0$ independent of $N$, that:
\beq
\max_{0 \leq a \leq N-1} | s_N (x_ {a+1}^N - x_a^N) |  \; = \; N^{-\a}\e{O}\big(\ln N \; + \;  N^{\a} \big)  
\qquad \e{and} \qquad 
\max_{1 \leq a \leq N} | s_N(x_ {a}^N - x_0^N) |  \; = \; \e{O}\big( 1 \big)  \;. 
\enq
Hence, it follows that 
\beq
N^2 \Int{x<y}{} s_N ( y - x )\,\dd \mu_{\e{eq}}^{(N)}(x)\,\dd \mu_{\e{eq}}^{(N)}(y) 
\; \leq  \;  \e{O}\big( N \big) + \sul{a<b}{N}    s_N ( x_b^N - x_a^N )\;, \enq
thus leading to the claim. \qed

 \vspace{2mm}
 
We now estimate the fluctuations of linear statistics by using an idea introduced in \cite{MaidaMaurelSegalaInegalitesPourConcentrationDeMesures}.  
\begin{defin}
\label{definition suite lambda's tildes}
Given a configuration of points $\la_1 \leq \dots \leq \la_N$, we build a sequence $\wt{\la}_1< \dots < \wt{\la}_{N}$ defined as 
\beq
\wt{\la}_1=\la_1 \quad and \quad  \wt{\la}_{k+1} = \wt{\la}_{k} \, + \, \max \big( \la_{k+1}- \la_k  , \ex{-(\ln N) ^2 } \big) \,.
\enq
Further, for any $\bs{\la}\in \R^N$, we associate a vector $\wt{\bs{\la}} \in \R^N$ by ordering the $\la$'s with a permutation $\sigma$, apply the previous construction to obtain a $N$-uple $\wt{\bs{\la}}$,
and put them in original order with the permutation $\sigma^{-1}$. The corresponding empirical measure is: 
$$
L_{N}^{(\wt{\bs{\la}})} = \f{1}{N} \sum_{a = 1}^N \delta_{\wt{\la}_{a}}
$$
and we denote $L_{N;u}^{ ( \wt{\bs{\la}} ) } $ the convolution of $L_{N}^{ ( \wt{\bs{\la}} ) }$ with the uniform 
probability measure on $\intff{0}{ \tf{\ex{-(\ln N)^2 }}{N} }$.
%
%\label{definition suite lambda's tildes}
\end{defin}
The new configuration has been constructed such that, for $\ell \not=k$, 
\beq
\label{lalawt}%
\big| \wt{\la}_{k} - \wt{\la}_{\ell} \big| \, \geq \, \ex{-(\ln N) ^2 } \qquad , \quad
\big| \la_{k} - \la_{\ell} \big| \, \leq \, \big| \wt{\la}_{k} - \wt{\la}_{\ell} \big| \quad \e{and} \quad
\big| \la_{k} - \wt{\la}_{k} \big| \leq (k-1) \cdot \ex{-(\ln N) ^2 } \;. 
\enq
The advantage of working with $L_{N;u}^{(\wt{\bs{\la}})}$ is that it is Lebesgue continuous; as such it can appear in the argument of ${\cal E}_N$ or $\mathfrak{D}^2$ and yield finite results. 
The scale of regularisation $e^{-(\ln N)^2}$ is somewhat arbitrary, but in any case negligible compared to $N^{-\alpha}$.

% in the following way. 
%Let $\sg \in \mf{S}_N$ be such that $\la_{ \sg(1) } \leq \dots \leq \la_{ \sg(N) }$. Then we can build the sequence 
%$\wt{\la}_{ \sg(1) } \leq \dots \leq \wt{\la}_{ \sg(N) }$ according to the above procedure. 
%This defines the coordinates $\wt{\la}_{a}$ of the vector $\wt{\bs{\la}}$. 

We introduce the effective potential associated to the $N$-dependent equilibrium measure:
\beq
V_{N;\e{eff}}(\xi) \; =\; V(\xi) \; - \;  2\Int{}{} 
s_N(\xi-\eta)\,\dd \mu_{\e{eq}}^{(N)}(\eta)
 \; \;  - \;  \;  C_{\e{eq}}^{(N)} \; . 
\label{definition fonction controle deviation vp sur bord}
\enq
By the characterisation of the equilibrium measure (Theorem~\ref{Proposition caracterisation rudimentaire mesure equilibre}), $V_{N;{\rm eff}} = 0$ in the support $\intff{a_N}{b_N}$, while $V_{N;{\rm eff}} > 0$ outside $\intff{a_N}{b_N}$.

\begin{prop}
\label{Theorem estimation deviation mesure centree en Proba}
Assume that 
\begin{itemize} 
 \item the partition function $Z_{N}[V]$  satisfies a lower-bound of the form 
\beq
Z_{N}[V] \; \geq \;   \exp\Big\{  -N^{ 2 + \a } \mc{E}_N\big[ \mu_{\e{eq}}^{(N)}  \big]     \; + \; \de_N  \Big\} \; ,\qquad \de_N = \e{o}\big( N^{ 2 + \a } \big)\; ;
\enq

\item  the potential is sub-exponential, \textit{viz}. there exists $\eps>0$ and $C_V>0$ such that 
\beq
\forall x \in \R\,,\qquad \sup_{t \in \intff{0}{\eps} } \big| V^{\prime}(x+t) \big| \;  \leq \; C_V \Big(  \big| V(x) \big| + 1 \Big)\;.
\enq
\end{itemize}
Then, given any $0<\eta <1$, we have for all $\bs{\la} \in \R^N$ that 
\beq
 p_N \big( \bs{\la} \big)  \; \leq \;  
 \exp\Bigg\{  - N^{2+\a}\mf{D}^2\big[ L_{N;u}^{(\wt{\bs{\la}})},\mu_{\e{eq}}^{(N)} \big]   - \de_N - N^{2+\a }(1-\eta) \Int{ \R }{}    V_{ N;\e{eff} }(\xi)\,\dd L_{N;u}^{ (\wt{\bs{\la}}) }(\xi) \;+ \; \e{O}\big( N (\ln N)^2 \big) \Bigg\}    \;. 
\label{ecriture majorant Gaussien pour mesure de proba}
\enq
The effective potential $V_{N;\e{eff}}$ has been defined in \eqref{definition fonction controle deviation vp sur bord} while $\mf{D}^2[\mu,\nu]$ is as given in \eqref{dddef}. 
\end{prop}

\Proof  The partition function takes the form:
$$
Z_N[V] = \Int{\R^N}{} \dd^N\lambda\,\exp\Bigg\{-N^{2 + \a}\Big(\Int{}{} V(x)\,\dd L_N^{(\bs{\lambda})}(x) - \Sigma_{{\rm diag}}[L_N^{(\bs{\lambda})}]\Big)\Bigg\},\qquad \Sg_{\e{diag}}[\mu] \; = \; \Int{ x \not= y }{}  s_N(x-y)\,\dd \mu(x)\dd \mu(y) \;.
$$
where $s_N$ defined in \eqref{definition fonction sN noyau integral eqn mes eq}. We are going to estimate the cost of replacing $L_N^{(\lambda)}$ by $L_{N;u}^{(\wt{\bs{\la}})}$ in the above integration.
%
%
%
%\beq
%
%\dd L_{N;u}^{ ( \wt{\bs{\la}} ) }(\xi) \; = \; N \ex{ (\ln N)^2 } 
%
%\sul{a=1}{N} \bs{1}_{ \intff{ \wt{\la}_a }{ \wt{\la}_a + \tf{\ex{-(\ln N )^2 }}{N} } }(\xi) \cdot \dd  \xi \;.
%
%\enq
%
%
%
We start with the term involving the potential. Since we assumed $V$ sub-exponential, we have:
\bem
\bigg| \Int{}{}V(x)\,\dd L_{N}^{(\bs{\la})}(x) \, - \, \Int{}{}V(x)\,\dd L_{N}^{( \wt{\bs{\la}} )}(x)  \bigg|
\; \leq  \; \f{1}{N} \sul{a=1}{N}  \f{ (a-1) }{ \ex{(\ln N)^2} } \cdot 
\sup \bigg\{  | V^{\prime}( \wt{\la}_a + t ) | \; : \;    t \in  \Big[ 0 ; \f{ (a-1) }{\ex{(\ln N)^2}}  \Big]     \bigg\}  \\
\; \leq \;  \f{ N C_V }{  \ex{(\ln N)^2} }\bigg( \Int{}{} |V(x)|\,\dd L_N^{(\wt{\bs{\la}} )}(x) \, + \, 1 \bigg) \;. 
\end{multline}
Further, since $V(x) \rightarrow + \infty$ when $|x| \rightarrow \infty$, there exists $C^{\prime}_{\e{eff}}>0$ such that 
\beq
\forall x \in \R,\qquad C^{\prime}_{\e{eff}}\big( 1 + V_{N;\e{eff}}(x) \big) \; \geq \; C_V\big(  |V(x) | +1  \big)\;.
\enq
As a consequence,
\beq
\exp\Bigg\{ -N^{2+\a} \Int{}{} V(x)\,\dd L_{N}^{(\bs{\la})}(x)\Bigg\} \; \leq \; 
\exp\Bigg\{\frac{N^{3+\a}\,C^{\prime}_{\e{eff}}}{e^{(\ln N)^2}}\Big[1 \, + \,   \Int{}{} V_{N;\e{eff}}(x)\,\dd L_{N}^{( \wt{\bs{\la}}) }(x) \Big] -N^{2+\a} \Int{}{} V(x)\dd L_{N}^{(\wt{\bs{\la}})}(x)\Bigg\}\;.
\enq
Now, let us consider the term involving the sinh interaction. Since $s_N$ is increasing on $\mathbb{R}^+$ and the spacings between 
$\wt{\lambda}_a$'s are larger than those between the $\lambda_a$'s, it follows that $\Sg_{\e{diag}}\big[ L_{N}^{ ( \bs{\la} ) } \big] \leq \Sg_{\e{diag}}\big[ L_{N}^{ ( \wt{\bs{\la}} ) } \big]$. Furthermore, we have:
\bem
\Sg_{\e{diag}}\big[ L_{N}^{ ( \wt{\bs{\la}} ) } \big]  \, - \, \Sg_{\e{diag}}\big[ L_{N;u}^{ ( \wt{\bs{\la}} ) } \big]  \; = \; 
\Int{ x \not= y }{} \dd  L_{N}^{ ( \wt{\bs{\la}} ) }(x)\,\dd  L_{N}^{ ( \wt{\bs{\la}} ) }(y) \Int{\intff{0}{1}^2}{} \dd^2 u
\Big\{ s_N(x-y) \, - \, s_N\big( x-y + N^{-1}e^{-(\ln N)^2}(u_1-u_2)\big) \Big\} \\
\; \; - \; \; \f{1}{N} \Int{\intff{0}{1}^2}{} \dd^2u\,s_N\big[ N^{-1}e^{-(\ln N)^2}\,(u_1-u_2)\big]\;.
\label{ecriture difference entropie reg et sing}
\end{multline}
When $N$ is large enough, we can use the Lipschitz behaviour of $s_N$ on $[e^{-(\ln N)^2}/2,+\infty[$ for the first term. Indeed:
\beq
|s_N^{\prime}(x)| = \sum_{p = 1}^2 \frac{\beta\pi\omega_{p}}{2}\,\cotanh[\pi\omega_p N^{\alpha} |x|] \leq c^{\prime}N^{-\alpha}e^{(\ln N)^2}
\enq
for some $c^{\prime} > 0$. Besides, we exploit that $s_N$ is increasing to bound the second term. This leads to:
%
%
%
%\bem
%
%s_N(x-y)-s_N\big( x-y +  (u-v) \cdot \tf{\ex{- (\ln N)^2 }}{N} \big)  \; = \; 
%
%- \f{ \be }{ 2N^{\a} } \ln \pl{a=1}{2}\bigg\{  \cosh\Big[ \pi \om_a N^{\a-1} \, (u-v)  \ex{-(\ln N)^2 } \Big]    \\
%
%
%\, + \,   \sinh\Big[ \pi \om_a N^{\a} \, (u-v)   \Big]  \cdot \cotanh\Big[ \pi \om_a N^{\a-1}  (x-y)  \ex{-(\ln N)^2 } \Big] \bigg\}
%
%\end{multline}
%
%
%
%Since $x \tend |\cotanh(x)|$ is strictly decreasing and $ |x-y| \geq  \ex{-(\ln N)^2 } $ in \eqref{ecriture difference entropie reg et sing}, we get that there 
%exists $ C>0$ such that, for $N$-large enough 
%
%
%
%\beq
%
%\Big| s_N(x-y)-s_N\big( x-y +  (u-v) \tf{ \ex{-(\ln N)^2 } }{ N } \big)  \Big| \leq \; \f{ C }{ N^{\a+1} } \;. 
%
%\enq
%
%
%
%As a consequence
%
%
%
\beq
\big|  \Sg_{\e{diag}}\big[ L_{N}^{ ( \wt{\bs{\la}} ) } \big]  \; - \; \Sg_{\e{diag}}\big[ L_{N;u}^{ ( \wt{\bs{\la}} ) } \big] \big| \; \leq \;
C \Big(   N^{-\a-1} \Big) \; + \; C^{\prime}\,N^{-(1 + \a)}\,(\ln N)^2 \;.
\enq
Since the measure $L_{N;u}^{ ( \wt{\bs{\la}} ) }$ is continuous with respect to Lebesgue, it is not any more necessary to take care of the diagonal singularity in $s_N$, and we obtain:
\begin{eqnarray}
\exp\Bigg\{-N^{2 + \a}\Bigg(\Int{\R}{} V(x)\dd L_N^{(\bs{\la})}(x) - \Sigma_{{\rm diag}}[L_N^{(\bs{\la})}]\Bigg)\Bigg\} & \leq & \exp\Bigg\{-N^{2 + \a}{\cal E}_N[L_{N;u}^{(\wt{\bs{\la}})}] \;+\; \e{O}\big(N(\ln N)^2\big)\Bigg\} \\
& & \times\,\exp\Bigg\{e^{-(\ln N)^2}\,N^{3 + \alpha}\,C^{\prime}_{{\rm eff}}\Int{}{}  V_{N;{\rm eff}}(x)\dd L_{N;u}^{(\wt{\bs{\la}})}(x) \Bigg\}\;. \nonumber  
\label{jiuhgw}
\end{eqnarray}

Since $\mu_{{\rm eq}}^{(N)}$ is also continuous with respect to Lebesgue, ${\cal E}_{N}[\mu_{{\rm eq}}^{(N)}]$ is finite and we can expand the first term around $\mu_{{\rm eq}}^{(N)}$:
$$
{\cal E}_{N}\big[L_{N;u}^{(\wt{\bs{\la}})}\big] = {\cal E}_N\big[\mu_{{\rm eq}}^{(N)}\big] + \mathfrak{D}^2\big[L_{N;u}^{(\wt{\bs{\la}})},\mu_{{\rm eq}}^{(N)}\big] 
+ \Int{\R}{} \dd(L_{N;u}^{(\wt{\bs{\la}})} - \mu_{{\rm eq}}^{(N)})(x)\Bigg\{V(x) - 2\Int{\R}{} \dd \mu^{(N)}_{{\rm eq}}(y)\,s_N(x - y)\Bigg\}\;. %A: typo on mu_{eq}
$$
We recognize in the last integral $V_{N;{\rm eff}}(x) + C_{{\rm eq}}^{(N)}$ integrated against a measure of mass $0$. So, we can omit the constant $C_{{\rm eq}}$, and since $V_{N;{\rm eff}} = 0$ on the support of $\mu_{{\rm eq}}^{(N)}$, 
we actually find:
$$
{\cal E}_{N}\big[L_{N;u}^{(\wt{\bs{\la}})}\big] = {\cal E}_N\big[\mu_{{\rm eq}}^{(N)}\big] + \mathfrak{D}^2\big[L_{N;u}^{(\wt{\bs{\la}})},\mu_{{\rm eq}}^{(N)}\big] + \Int{\R}{} V_{N;{\rm eff}}(x)\,\dd L_{N;u}^{(\wt{\bs{\la}})}(x)
$$
If we plug this relation in \eqref{jiuhgw}, we obtain a similar bound but now with $V_{N;{\rm eff}}$ having the prefactor 
\newline $N^{2 + \alpha} - e^{-(\ln N)^2}N^{3 + \alpha} C^{\prime}_{{\rm eff}} \leq (1 - \eta)N^{2 + \alpha}$,
this for any $0<\eta<1$, provided that $N$ is large enough. \qed

In order to bound the one and multi-point expectation values and in particular the various terms arising in the Schwinger-Dyson equations,
we introduce the exponential regularisation of a function. This regularisation allows one to deal with functions that are unbounded at
infinity but whose expectation values are still well defined. 
\begin{defin}
\label{Definition regularisation exponentielle}
 Given a function $f$ in $n$ variables, its exponential regularisation with growth $\kappa$ is defined by 
\beq
\label{definition kappa dumped function}
\mc{K}_{\kappa}[f](\xi_1,\dots,\xi_n) \; = \; \Big\{ \pl{a=1}{n} \ex{-\kappa V(\xi_a) } \Big\} \cdot f(\xi_1,\dots,\xi_n) \;.
\enq
\end{defin}
\begin{defin}
\label{Lcurl} We define the centred empirical measure as:
\beq
\label{calLNdef} {\cal L}_{N}^{(\bs{\la})} = L_{N}^{(\bs{\la})} - \mu_{{\rm eq}}^{(N)}\;.
\enq
\end{defin}

Prior to establishing the simplest \textit{a priori} bounds on the multi-point expectation values $\big< f \big>_{\bigotimes_1^n \mc{L}^{(\bs{\la})}_N}$, we need to establish a convenient decomposition
thereof. The latter is written in such a way that the leading in $N$ behaviour comes from the part involving a restriction of $f$ to a compactly supported function. 

\begin{lemme}
\label{Lemme reduction moyenne n pts a partie cpcte}
There exists $t>0$ and a  functional $\mc{A}^{(n)}_N$ on the space of functions $f$ such that $\mc{K}_{\kappa}[f]\in W_0^{\infty}(\R^n)$ for some $\kappa>0$ satisfying 
\beq
\sup_{N\in \mathbb{N} } \Big\{ \e{supp}\big[ \mu_{\e{eq}}^{(N)}\big] \Big\} \; \subset \; \intff{ - \tf{t}{2} }{ \tf{t}{2} } \qquad  \e{and} \qquad 
\Big| \mc{A}^{(n)}_N[f] \Big| \; \leq \; c_n \ex{-c N^{1+\a} } \cdot \norm{ \mc{K}_{\kappa}[f] }_{W^{\infty}_0(\R^n) }
\enq
and such that 
\beq
\big< f \big>_{\bigotimes_1^n \mc{L}^{(\bs{\la})}_N} \; = \; \big< f_{\mid \e{c} } \big>_{\bigotimes_1^n \mc{L}^{(\bs{\la})}_N}  \; + \; \mc{A}^{(n)}_N[f] \;. 
\label{ecriture decomposition fct n pts partie cpcte et partie exp small}
\enq
In the above decomposition, 
\beq
f_{\mid \e{c}}\big( \xi_1, \dots, \xi_n \big) \, = \, f\big( \xi_1, \dots, \xi_n \big) \cdot \pl{a=1}{n}\phi(\xi_a)
\label{definition restriction compacte de f}
\enq
where $\phi\in \mc{C}^{\infty}_{\e{c}}(\R)$ is such that 
\beq
0\leq \phi \leq 1 \;\; , \qquad  \phi_{\mid \intff{-t}{t} } =1 \; \; \e{and} \quad \e{supp} [\phi] \subset \intff{ - ( t + 1 ) }{ t + 1 } \;. 
\enq

\end{lemme}

\Proof  We first claim that the constant $C_{\e{eq}}^{(N)}$ arising in the minimisation problem for the equilibrium measure \eqref{definition de la cste Ceq par eqn int eq meas}
is bounded in $N$. Indeed, it follows from \eqref{definition de la cste Ceq par eqn int eq meas} that 
\beq
C_{\e{eq}}^{(N)} \; = \; \Int{a_N}{b_N} V(\xi)\,\dd \mu_{\e{eq}}^{(N)}(\xi) \; - \; 
\f{ \be }{ N^{\a} } \Int{\intff{a_N}{b_N}^2}{}  \ln\Big\{ \pl{p=1}{2} \sinh[\pi \om_p N^{\a} (\xi-\eta) ] \Big\}\,\dd \mu_{\e{eq}}^{(N)}(\xi)  \dd \mu_{\e{eq}}^{(N)}(\eta)  \;. 
\enq
Therefore, we have: 
\beq
\big| C_{\e{eq}}^{(N)} \big| \; \leq \; \norm{ V }_{L^{\infty}(\intff{a_N}{b_N})} \; + \; \wt{C} \,\norm{ V^{\prime\prime} }_{L^{\infty}(\intff{a_N}{b_N})}^2
\Int{\intff{a_N}{b_N}^2}{}  \f{ 1 }{ N^{\a} }\Big| \ln\Big\{ \pl{p=1}{2} \sinh[\pi \om_p N^{\a} (\xi-\eta) ] \Big\} \Big|\,\dd \xi \dd \eta \;,
\label{ecriture borne en N de la constante C eq de N}
\enq
where we have used that $ \mu_{\e{eq}}^{(N)}$ is a probability measure and that its density is bounded by \eqref{ecriture borne uniforme sur la densite}. The double integral remaining in \eqref{ecriture borne en N de la constante C eq de N}
can be bounded by an $N$-independent constant. Such bounds are obtained by using that the function 
\beq
g_N(\xi) \; = \;  \f{ 1 }{ N^{\a} }\Big| \ln\Big\{ \pl{a=1}{2} \sinh[\pi \om_a N^{\a}\xi ] \Big\} \Big| - \pi (\om_1+\om_2) |\xi| 
\enq
approaches $0$ point-wise in $\xi \in \intff{a_N-b_N}{b_N-a_N}\setminus\{0\}$ and is bounded as 
$|g_N(\xi)| \, \leq \, C \big( 1 + \big|\ln |\xi|  \big| \big) $.  Since the endpoints $a_N$ and $b_N$ are 
bounded in $N$ in virtue of \eqref{ecriture DA grand N point bord du support de mu eq N}, we can 
 apply the dominated convergence theorem to  $(\xi,\eta) \mapsto g_N(\xi-\eta)$ on $\intff{a_N}{b_N}^2$.

The finiteness in $N$ of $ C_{\e{eq}}^{(N)}$ along with the confinement hypothesis \eqref{Hypothesis: confinment of the potential} on the potential implies the existence of $t > 0$ independent of $N$ such that:
\beq
\forall \xi \in \mathbb{R}\setminus[-t,t],\qquad V_{N; \e{eff} }( \xi ) \; \geq \; \f{V(\xi)}{2} \;\geq\; \f{|\xi|}{2} \,. %\f{V(\xi) }{2}  \; \leq \; \f{ |\xi|^{1+\eps} }{ 2 }  \qquad \e{for} \; \e{any} \quad \xi \in \R \setminus \intff{-t}{t} \;. 
\label{ecriture borne inf en terme de V pour pot effectif}
\enq
where the effective potential is defined by \eqref{definition fonction controle deviation vp sur bord}. In virtue of \eqref{ecriture DA grand N point bord du support de mu eq N}, 
one can always choose $t$ such that it also holds $\e{supp}[ \mu_{\e{eq}}^{(N)} ] \subset \intff{ - \tf{t}{2} }{ \tf{t}{2} }$.

Since
\beq
\big< f \big>_{\bigotimes_1^n \mc{L}^{(\bs{\la})}_N} \; = \; \big< f_{ \e{sym} } \big>_{\bigotimes_1^n \mc{L}^{(\bs{\la})}_N}  \quad \e{with} \quad 
f_{\e{sym}}(\xi_1,\dots,\xi_n)=\f{1}{n!} \sul{\sg \in \mf{S}_N }{} f\big(\xi_{\sg(1)},\dots, \xi_{\sg(n)}\big)
\enq
we may assume that $f$ is a completely symmetric function. Then, one gets 
\beq
\big< f \big>_{\bigotimes_1^n \mc{L}^{(\bs{\la})}_N} \;=\; \big< f_{\mid \e{c} } \big>_{\bigotimes_1^n \mc{L}^{(\bs{\la})}_N}  \; + \; \mc{A}^{(n)}_N[f]  \qquad,  \quad 
\mc{A}^{(n)}_N[f]\,=\,\sul{p=1}{n} {n \choose p}\,\mathcal{A}_{N;p}^{(n)} \big[ f_{\e{sym}} \big] 
\enq %G Conventionally it is (n choose p) and not (p chose n) to say C_{n}^p. I keep \mathcal{A}, no reason to use a different script.
where 
\beq
\mathcal{A}_{N;p}^{(n)} \big[ f_{\e{sym}} \big]  \; = \; \mathbb{P}_N\bigg[ \Int{ \intff{-t}{t}^{ \e{c} }  }{} \pl{a=1}{p} \dd L^{(\bs{\la})}_N(\xi_a) \Int{-(t+1) }{ (t+1) } \pl{a=p+1}{n}  \dd \mc{L}^{(\bs{\la})}_N(\xi_a)  
\pl{a=1}{p} \Big( 1\, -\,\phi(\xi_a)  \Big) \pl{a=p+1}{n}\phi(\xi_a)   f_{\e{sym}}(\xi_1,\dots,\xi_n)  \bigg] \;. 
\enq
Note that, in the intermediate steps, we have used that $\e{supp}\big[ \mu_{\e{eq}}^{(N)} \big] \cap \intff{-t}{t}^{ \e{c} }=\emptyset$.
Hence, one gets the bound
\beq
\Big|  \mathcal{A}_{N;p}^{(n)} \big[ f_{\e{sym}} \big]  \Big| \; \leq \; \norm{  \mc{K}_{\kappa} [f] }_{ W^{\infty}_{0}(\R^n) }\cdot  2^p \cdot \norm{ \ex{\kappa V} }_{L^{\infty}(\intff{-(t+1)}{t+1}) }^{n-p} \cdot A_p
\enq
%
%G I kept the same letter A to denote the constant
%
with 
\beq
A_p \; = \; \f{ 1 }{ N^{p} } \Int{\R^N}{} p_{N}\big( \bs{\la} \big)  \sul{ \substack{a_1,\dots,a_p \\ =1} }{ N } \pl{\ell=1}{p}\bigg\{  \ex{\kappa V(\la_{a_{\ell}}) } \cdot \bs{1}_{[-t;t]^{{\rm c}} }\big(\la_{a_{\ell}}\big) \bigg\} \cdot \dd^N \bs{\la}  \;.  
\enq
Observe that given any symmetric function $g$ in $p$-variables, one has the decomposition 
\beq
 \sul{ \substack{a_1,\ldots,a_p \\ =1} }{ N }  g\big(\la_{a_1},\ldots,\la_{a_p}\big) \; = \; \sul{\ell=1}{p} \sul{\substack {r_1,\ldots,r_{\ell} \geq 1 \\ r_1+\ldots+r_{\ell} = p}}{} C^{(\ell,p)}_{r_1,\dots, r_{\ell}} %G I removed \vdash, to avoid defining another notation. 
 \cdot \sul{ \substack{1 \leq b_1,\ldots,b_{\ell} \leq  N \\ {\rm pairwise}\,\,{\rm disjoint}}}{} g\Big( \underbrace{ \la_{b_1},\ldots,\la_{b_1} }_{r_1} ,\ldots, \underbrace{ \la_{b_{\ell} },\ldots,\la_{b_{\ell} } }_{ r_{\ell} }  \Big) %G You meant pairwise disjoint
\enq
where $C^{(\ell,p)}_{r_1,\dots, r_{\ell}} > 0$ are purely combinatorial coefficients. The latter implies that, for some $p$-dependent constant $C_p$:
\beq
 \sul{\ell=1}{p} \sul{\substack{r_1,\ldots,r_{\ell} \geq 1 \\ r_1+\cdots+r_{\ell}  = p}}{} C^{(\ell,p)}_{r_1,\dots, r_{\ell}}  \; \leq \;  C_p\;. 
\enq
As a consequence, we get 
\bem
A_p \; = \; \f{ 1 }{ N^{p} } \sul{\ell=1}{p} \sul{ \substack{r_1,\cdots,r_{\ell} \geq 1 \\ r_1 + \cdots + r_{\ell} = p}}{} C^{(\ell,p)}_{r_1,\dots, r_{\ell}}  
\Int{\R^N}{} p_{N}\big( \bs{\la} \big)  \sul{ \substack{ b_1,\ldots,b_{\ell} \geq 1 \\  {\rm pairwise}\,\,{\rm disjoint}} }{} \pl{s=1}{\ell}\bigg\{  \ex{\kappa r_{s} V(\la_{b_{s}} ) } \cdot \bs{1}_{ |x|>t }\big(\la_{b_{s}}\big) \bigg\} \cdot \dd^N \bs{\la}  \\
\;\leq  \; \ \sul{\ell=1}{p}  \f{N \cdots (N-\ell+1) }{ N^{p} } \sul{ \substack{ r_1+\cdots+r_{\ell} \\   \dashv p}  }{} C^{(\ell,p)}_{r_1,\dots, r_{\ell}}  
\Int{ (\intff{-t}{t}^{ \e{c} })^{\ell}   }{} \pl{a=1}{\ell}\dd\la_{a} \Int{ \R^{N-\ell}   }{} \pl{a=\ell+1}{N}\dd\la_{a} \cdot  p_{N}(\bs{\la})
\pl{a=1}{\ell} \Big\{ \ex{\kappa r_{s} V(\la_a) } \Big\} \;. 
\end{multline}
It follows from  \eqref{ecriture majorant Gaussien pour mesure de proba} with $\eta=1/2$ given in 
Proposition \ref{Theorem estimation deviation mesure centree en Proba} that:
\beq
\big| p_N(\bs{\la}) \,\bs{1}_{\Om}(\bs{\la}) \big| \; \leq \; \pl{a=1}{N} \ex{-\frac{1}{2}N^{1+\a} V_{N;\e{eff}}(\la_a) } \cdot 
\exp\Big\{ -N^{2+\a} \inf_{\bs{\la} \in \Om} \mf{D}^2\big[ L_{N;u}^{(\wt{\bs{\la}})}, \mu_{\e{eq}}^{(N)}  \big]    \Big\}   \;. 
\enq
 This bound leads to:
\beq
\label{niunsa}%
A_p \leq C_p \cdot \max_{\ell=1,\dots, p} 
\bigg\{ \Int{{\intff{-t}{t}}^{c}}{} \hspace{-1mm} \ex{-\frac{1}{2}N^{1+\a} V_{N;\e{eff}}(\xi) + \kappa V(\xi) }  \dd \xi   \bigg\}^{\ell} \cdot  
\max_{\ell=1,\dots, p}  \bigg\{ \Int{ \R  }{} \ex{-\f{1}{2}N^{1+\a} V_{N;\e{eff}}(\xi)  }  \dd \xi   \bigg\}^{N-\ell} \;. 
\enq
Further, in virtue of \eqref{ecriture borne inf en terme de V pour pot effectif} we have, for $N$ large enough,
\beq
\bigg| \Int{\intff{-t}{t}^{c}}{} \hspace{-1mm} \ex{-\frac{1}{2}N^{1+\a} V_{N;\e{eff}}(\xi) + \kappa V(\xi) }  \dd \xi   \bigg| \; \leq \;
\bigg| \Int{ \intff{-t}{t}^{c}}{} \hspace{-1mm} \ex{-\frac{1}{8}N^{1+\a} V(\xi) }  \dd \xi   \bigg|  
\; \leq \; \bigg| \Int{ \intff{-t}{t}^{c}}{} \hspace{-1mm} \ex{-\frac{1}{8}N^{1+\a} |\xi | }  \dd \xi   \bigg| 
\; = \; \e{O}\big( \ex{-c N^{1+\a} } \big) \;. 
\enq
The integral over $\mathbb{R}$ in \eqref{niunsa} is bounded uniformly by a constant $A$, since $V_{N;{\rm eff}} \geq 0$, and $V_{N;{\rm eff}}$ grows at least linearly at infinity.
All in all, for any $p \in \intn{1}{n} $, \eqref{niunsa} is bounded by $ C^{\prime} A^{N} \ex{-cN^{1 + \alpha}} = o(\ex{-c^{\prime}N^{1 + \alpha}})$, whence the result. \qed

\begin{cor}
\label{Corollary bornes a priori sur les correlateurs}

Let $ \kappa \geq 0 $. There exist constants $C_n>0$ depending on $n$ and $\kappa$ such that the below bounds hold 
for any $f$ satisfying $\mc{K}_{\kappa}[f] \in W^{\infty}_{1}(\R^n)$
\beq
\big|   \Big< f(\xi_1,\dots,\xi_n) \Big>_{ \bigotimes_{1}^{n} \mc{L}_N^{(\bs{\la})}  }  \big|  \; \leq \; 
 C_n\,\Big\{N^{-n}\,\norm{ \mc{K}_{\kappa}[f] }_{W^{\infty}_{1}(\R^n)} \; + \;  N^{(\a - 1)n/2}\,\norm{ \mc{K}_{\kappa}[f] }_{W^{\infty}_{n}(\R^n)}^{1/2}\cdot \norm{ \mc{K}_{\kappa}[f] }_{W^{\infty}_{0}(\R^n)}^{1/2} \Big\} \;. 
\label{ecriture bornes a priori correlateur n pts}
\enq

\end{cor}

\Proof  Using the decomposition \eqref{ecriture decomposition fct n pts partie cpcte et partie exp small} it is enough to obtain the bounds \eqref{ecriture bornes a priori correlateur n pts} for the 
compact restriction $f_{\mid \e{c}}$ of $f$ defined in \eqref{definition restriction compacte de f}. 
Upon decomposing ${\cal L}_N^{(\bs{\la})} = {\cal L}_{N;u}^{(\wt{\bs{\la}})} + (L_N^{(\bs{\la})} - L_{N;u}^{(\wt{\bs{\la}})})$, we can write:
\beq
   \big< f_{\mid \e{c}} \big>_{ \bigotimes_{1}^{n} \mc{L}_N^{(\bs{\la})}  }   \; = \;
 \sul{\ell=1}{n} \sul{ \substack{ i_1<\dots < i_{\ell} \\ =1 } }{ n } \mathbb{P}_{N }
 \bigg[  \Int{\R^n}{}  f_{\mid \e{c}}\big( \xi_{1},\dots, \xi_{n} \big) \pl{a=1}{\ell} \dd \mc{L}^{(\wt{\bs{\la}})}_{N;u}(\xi_{i_a})
\pl{ \substack{ a=1  \\ \not= i_1,\dots, i_{\ell} } }{ n } \dd \Big( L^{(\wt{\bs{\la}})}_{N;u} -L^{(\bs{\la})}_N     \Big)(\xi_{a})  \bigg]  
\; + \; \big< f \big>_{ \bigotimes_{1}^{n} \mc{L}_{N;u}^{(\wt{\bs{\la}})}  }  \;. 
\label{ecriture estimation valeur moyenne de f ctr lesure L cal}
\enq
Since $\wt{\la}_{a}$'s are not far from $\la_{a}$'s according to \eqref{lalawt}, we can bound for any $ \ell \leq n- 1$,  %A:change sense borne

\bem
 \bigg| \mathbb{P}_{N}\bigg[ \Int{\R^n}{}
 f_{\mid \e{c}}\big( \xi_{1},\dots, \xi_{n} \big) \pl{a=1}{\ell} \dd \mc{L}^{(\wt{\bs{\la}})}_{N;u}(\xi_{i_a})
\pl{ \substack{ a=1  \\ \not= i_1,\dots, i_{\ell} } }{ n } \dd(L^{(\wt{\bs{\la}})}_{N;u} -L^{(\bs{\la})}_N)(\xi_{a})  \bigg]  \bigg| \\
 \; \leq  \; 2^n \cdot \norm{ \mc{K}_{\kappa}[f] }_{W^{\infty}_{1}(\R^n) }   \f{ N(N-1) }{ 2}\cdot \frac{\ex{-(\ln N)^2}}{N} \;. 
\end{multline}
To get the second factor, we used the chain of bounds
\beq
\norm{ f_{\mid\e{c}} }_{W^{\infty}_{1}(\R^n) }  \; \leq \; C_1 \norm{ \mc{K}_{\kappa}[f_{\mid\e{c}}] }_{W^{\infty}_{1}(\R^n) } \; \leq \; C_2  \norm{ \mc{K}_{\kappa}[f] }_{W^{\infty}_{1}(\R^n) } \;. 
\enq

As a consequence, the first sum in \eqref{ecriture estimation valeur moyenne de f ctr lesure L cal} will only give rise to  $\norm{ \mc{K}_{\kappa}[f] }_{W^{\infty}_{1}(\R^n) } \cdot \e{O}(N^{-\infty})$ corrections. 
This being settled, Proposition \ref{Theorem estimation deviation mesure centree en Proba} ensures the existence of $M>0$  
and a constant $C > 0$ such that, for $N$ large enough:  %. Further, 
\beq
\mathbb{P}_N \Big[ \Om_{M;N} \Big]
 \; = \; \e{O}\Big(  \ex{- CM\,N^{1+\a} }  \Big)  \quad \e{with} \qquad
 \Om_{M;N} \; = \; \Big\{ \bs{\la} \in \R^N \; : \; \mf{D}^2\big[ L_{N;u}^{(\wt{\bs{\la}})}, \mu_{\e{eq}}^{(N)}  \big] \; >  \; \tf{ M  }{N} \Big\}\;. 
\enq
This ensures that 
\beq
\Big|  \big< f_{\mid\e{c}} \big>_{ \bigotimes_{1}^{n} \mc{L}_{N;u}^{(\wt{\bs{\la}})}  }  \Big| \; \leq \; C^{\prime} 
\,\norm{ \mc{K}_{\kappa}[f] }_{L^{\infty}(\R^n)}\,\ex{- C^{\prime\prime}M\,N^{1+\a} } 
\; + \; \mf{R}_{N;u}[f_{\mid\e{c}} ] 
\enq
with
\beq
\mf{R}_{N;u}[f_{\mid\e{c}} ] \; = \; \Bigg| \mathbb{P}_N\bigg[  \bs{1}_{\Om_{M;N}^{c}} \Int{\R^n}{} f_{\mid\e{c}} (\xi_1,\dots,\xi_n) \pl{a=1}{n} \dd \mc{L}_{N;u}^{(\wt{\bs{\la}})}(\xi_a)  \bigg] \Bigg|\;. 
\enq
Finally, using Cauchy-Schwarz inequality to make the distance $\mathfrak{D}$ appear:
\bem
\mf{R}_{N;u}[f] \; = \; 
\bigg| \mathbb{P}_{N}\bigg[  \bs{1}_{\Om_{M;N}^{c}}  
 \Int{\R^n}{} \mc{F}\big[ f_{\mid\e{c}}  \big](\vp_{1},\dots,\vp_{n}) 
 \pl{a=1}{n} \mc{F}\big[ \mc{L}^{(\wt{\bs{\la}})}_{N;u}\big] (-\vp_{a} ) \cdot \f{ \dd^n \bs{\vp} }{ (2\pi)^{n} }  \bigg]   \bigg| \\
\; \leq \;  \bigg\{ \Int{\R^n}{} \f{ \big| \mc{F}[ f_{\mid\e{c}} ](\vp_{1},\dots,\vp_{n}) \big|^2  }
{ \prod_{i=1}^{n}   \bigg\{ \f{ \pi \be }{ 2  N^{\a}  \vp_i } \sul{p=1}{2}   \cotanh\Big[ \f{ \vp_i }{ 2 \om_p N^{\a}  } \Big]   \bigg\}   }
 \cdot \f{ \dd^n \bs{\vp}}{ (2\pi)^{n} } \bigg\}^{ \f{1}{2} } \cdot 
 \mathbb{P}_{N}\bigg[ \; \bs{1}_{\Om_{M;N}^{c}}  \mf{D}^n\big[  L^{ (\wt{\bs{\la}}) }_{N;u} , \mu_{ \e{eq} }^{ (N) } \big]   \bigg] \;. 
\end{multline}
The last factor, because it is evaluated on the complement on $\Omega_{M;N}$, is at most $\e{O}(N^{-n/2})$. The Fourier transform part of the 
bound can be estimated with the bound:
\beq
 \prod_{i = 1}^n \Bigg|  \f{ \pi \be }{ 2  N^{\a}  \vp_i } \sul{p=1}{2}   \cotanh\Big[ \f{ \vp_i }{ 2 \om_p N^{\a}  } \Big]   \Bigg|^{-1} 
 \; \leq \; \prod_{i = 1}^n \big( C  \,N^{\a} |\vp_i|\big) \leq (C N^{\a})^{n}\,\Bigg(1 + \Big\{\sum_{i = 1}^n \vp_i^2\Big\}^{1/2}\Bigg)^n\;.
\enq
Hence, there exists a constant $C^{\prime}_n>0$ such that:
\beq
 \big| \big< f_{\mid\e{c}}  \big>_{ \bigotimes_{1}^{n} \mc{L}_N^{(\wt{\bs{\la}})}  } \big| \; \leq  \; 
C^{\prime}_n\,N^{(\a - 1)n/2}\Big( \norm{ f_{\mid\e{c}}  }_{H_{\tf{n}{2}}(\R^n) } \, + \;  \norm{ \mc{K}_{\kappa}[f]  }_{W_0^{\infty}(\R^n) } \Big)  \;. 
\label{ecriture bornes fonction a n point via normes Hs}
\enq
where the $W_{0}^{\infty}$ norm is nothing but the $L^{\infty}$ norm. In order to bound  $\norm{ f_{\mid\e{c}}  }_{H_{\tf{n}{2}}(\R^n) } $ by the $W^{\infty}_{n}$ norms (c.f. their definition \eqref{defWninfnorm}), we observe that:
\beq
\norm{f_{\mid\e{c}} }_{H_{\tf{n}{2}}(\R^n)}^2 \; \leq \; \norm{ f_{\mid\e{c}} }_{H_{n}(\R^n)}\cdot
\norm{f_{\mid\e{c}} }_{L^2(\R^n)} \;. 
\enq
The $L^2(\R^n)$ norm is bounded directly as:
\beq
 \norm{f_{\mid\e{c}} }_{L^2(\R^n)} \; \leq \; C^{\prime} \big(2t+2)^{\f{n}{2}} \cdot  \norm{ \mc{K}_{\kappa}[f]}_{W_0^{\infty}(\R^n)}\;.
\enq
Finally, in order to bound $\norm{f_{\mid\e{c}} }_{H_{n}(\R^n)}$, we remark that $(1+ |t|)^{2n} \; \leq \; 4^n (1+t^2)^n$, so that:

\beq
\Bigg( 1+ \Big\{ \sul{a=1}{n}\vp_a^2 \Big\}^{1/2}  \Bigg)^{2n} \; \leq \; C \sul{k=0}{n} P_k(\vp_1^2,\dots,\vp_n^2)
\enq
for some symmetric homogeneous polynomial of degree $k$ which has the expansion:
\beq
P_k(\vp_1^2,\dots,\vp_n^2) \; =\; \sul{k_1+\dots+k_n=k}{} p_{\{k_a\}} \cdot \vp_1^{2k_1} \cdots \vp_n^{2k_n} \qquad \e{with} \qquad p_{\{k_a\}} \geq 0\;.
\enq
This ensures that  
\begin{eqnarray}
\norm{ f_{\mid\e{c}} }_{H_{n}(\R^n)}^2 & \leq & 
 C \sul{k=0}{n} \sul{k_1+\dots+k_n=k}{} p_{\{k_a\}}   
 \Int{}{}  \Big| \pl{a=1}{n}\Dp{\xi_a}^{k_a} \cdot f_{\mid\e{c}}  (\xi_1,\dots,\xi_n) \Big|^2 \cdot \dd^n\bs{\xi} \nonumber \\
 & \leq &  C^{\prime} \cdot (2+2t)^n \cdot  \norm{ \mc{K}_{\kappa}[f]}_{W_n^{\infty}(\R^n)}^2 \;. 
\end{eqnarray}
To get the last line, we have repeatedly used the sub-exponential hypothesis \eqref{Hypothese sous exponetialite des derivees}. 
As a consequence, for some constant $C^{\prime}$
\beq
\norm{ f_{\mid\e{c}} }_{H_{\tf{n}{2}}(\R^n)} \; \leq \; C^{\prime} \cdot \norm{ \mc{K}_{\kappa}[f]  }_{W_{n}^{\infty}(\R^n) }^{\f{1}{2}} 
\cdot  \norm{ \mc{K}_{\kappa}[f]  }_{W_{0}^{\infty}(\R^n) }^{\f{1}{2}}  \;. 
\enq
Inserting the above bound in \eqref{ecriture bornes fonction a n point via normes Hs}, we obtain
\beq
\big| \big< f_{\mid\e{c}}  \big>_{ \bigotimes_{1}^{n} \mc{L}_N^{(\wt{\bs{\la}})}  } \big| \; \leq \;
C^{\prime\prime}_n\,N^{(\a - 1)n/2}\,\norm{ \mc{K}_{\kappa}[f]  }_{W_{n}^{\infty}(\R^n) }^{\f{1}{2}} 
\cdot  \norm{ \mc{K}_{\kappa}[f]  }_{W_{0}^{\infty}(\R^n) }^{\f{1}{2}} \;,
\enq
what leads to the claimed form of the bound on the average $ \big< f \big>_{ \bigotimes_{1}^{n} \mc{L}_N^{(\bs{\la})}  } $ .  \qed

\section{The Schwinger-Dyson equations}

\label{SDw1r}

 In the present section, we derive the system of Schwinger-Dyson equations in our model. 
The operator 
\beq
\mc{U}_N[\phi](\xi) \; = \; \phi(\xi)\cdot \big\{  V^{\prime}( \xi) \, -  \, \mc{S}_N[ \rho_{\e{eq}}^{(N)} ](\xi)   \big\}
 + \mc{S}_N[\phi \cdot \rho_{\e{eq}}^{(N)} ](\xi) \;, 
\label{definition noyau integral operateur S driven by mu eq}
\enq
with ${\cal S}_{N}$ defined in \eqref{ecriture eqn int sing de depart} will arise in their expression, and play a crucial role in the large-$N$ analysis. 
It will be shown in Proposition \ref{Proposition invertibilite operateur S} that the operator $\mc{S}_N$ is invertible and in Proposition \ref{Proposition characterisation operateur UN} 
that the operator $\mc{U}_N$ is invertible as well. We will build on this information until the end of this chapter.
At a later stage, we shall  use as well fine bounds on the $\mc{W}_{\ell}^{\infty}(\R)$ norms of functions $\mc{K}_{\kappa}\big[ \mc{U}_N^{-1}[\phi] \big]$ 
which will be obtained later in Proposition~\ref{Theorem bornes sur norme inverse UN via estimation fines locales}. 
 We do stress that these results on the invertibility of $\mc{S}_N$ and $\mc{U}_N$ as well as those relative to estimates involving $\mc{U}_{N}^{-1}$
will be obtained independently of the results obtained in the present chapter. By presenting this technical result only in a later Chapter 5, and using it as a tool in the present chapter, we hope to make the principles of analysis of Schwinger-Dyson equations more transparent.

Since we will be dealing with operators initially defined on functions in one variable but acting on one of the variables
of a function in many variables, it is useful to introduce the 
\begin{defin}
\label{Definition operateurs Ok} 
 Given an operator ${\cal O} : W_p^{\infty}(\R) \tend W_p^{\infty}(\R^{\ell}) $ acting on functions of one variable and $\phi\in W_p^{\infty}(\R^n)$, ${\cal O}_k[\phi]$ refers to the function
\beq
{\cal O}_k[\phi](\xi_1,\dots, \xi_{n+\ell-1}) \; = \; {\cal O}_k\big[\phi(\xi_1,\dots, \xi_{k-1}, * ,\xi_{k+\ell},\dots, \xi_{n+\ell-1} )\big](\xi_k,\dots, \xi_{k+\ell-1})\;,
\enq
in which $*$ denotes the variable of $\phi$ on which the operator ${\cal O}_k$ acts. 
\end{defin}

For instance, according to the above definition, we have $\mc{U}_{N;1}\big[ \phi \big](\xi_1,\dots, \xi_n) \; = \; \mc{U}_{N}\big[ \phi(*,\xi_2,\dots, \xi_n) \big](\xi_1).$

\begin{defin}
\label{Xipdef} If $\phi$ is a function in $n \geq 1$ variables, we denote $\partial_{p}$ the differentiation with respect to the $p^{{\rm th}}$ variable. 
We also define an operator $\Xi^{(p)}\,:\,W_{\ell}^{\infty}(\mathbb{R}^{n}) \rightarrow W_{\ell}^{\infty}(\mathbb{R}^{n - 1})$ by:
$$
\Xi^{(p)}[\phi](\xi_1,\ldots,\xi_n) = \phi(\xi_1,\ldots,\xi_{p - 1},\xi_1,\xi_p,\ldots,\xi_{n - 1})\;.
$$
\end{defin}

\begin{prop}
\label{Proposition equations des boucles}
Let $\phi_n$ be a function in $n$ real variables such that $\mc{K}_{\kappa}[\phi_n] \in W_1^{\infty}(\R^n)$, \textit{cf}. \eqref{definition kappa dumped function}, for some $\kappa \geq 0 $ that can depend on $n$. 
Then, all expectation values appearing below are well-defined. Furthermore, the level $1$ Schwinger-Dyson equation takes the form: 
\beq
\label{SDlevel1}%
- \big<   \phi_1 \big>_{  \mc{L}_N^{(\bs{\la})} }  
 \; + \;  \f{1}{2 } 
 \Big<  \mc{D}_N\circ{\cal U}_{N}^{-1}[\phi_1] \Big>_{  \mc{L}_N^{(\bs{\la})}    \otimes   \mc{L}_N^{(\bs{\la})} }
\; +  \;  \f{  (1-\be) }{ N^{1+\a}  }  \big<  \partial_{1}\mathcal{U}_{N}^{-1}[\phi_1] \big>_{  \mu_{\e{eq}}^{(N)}  }
\; + \;  \f{  (1-\be) }{ N^{1+\a}  }    \big<  \partial_{1}\mathcal{U}_N^{-1}[\phi_1] \big>_{    \mc{L}_N^{(\bs{\la})}  }   \; = \; 0 \;. 
\enq
There, $\mc{D}_N$ corresponds to the non-commutative derivative
\beq
\label{noncommet}\mc{D}_N[\phi](\xi,\eta) \; = \;  \Bigg\{\sul{p=1}{2} \beta\pi \om_p\cotanh\big[ \pi \om_pN^{\a} (\xi-\eta) \big] \Bigg\}\cdot\big(  \phi(\xi) -\phi(\eta) \big) \;.
\enq
In their turn, the Schwinger-Dyson equation at level $n$ takes the form:
\bem
 \big< \phi_n \big>_{    \bigotimes\limits^{n}   \mc{L}_N^{(\bs{\la})}    }  \; = \; 
  \f{ 1 }{ N^{2+\a} } 
 \sul{p=2}{n} \Big<   \Xi^{(p)}\circ\mc{U}_{N;1}^{-1}[\partial_{p}\phi_n] \Big>_{ \bigotimes\limits^{n-1}   \mc{L}_N^{(\bs{\la})}   } 
 \; + \; 
 \f{1}{2} \Big< \mc{D}_{N;1}\circ\mc{U}_{N;1}^{-1}[\phi_n] \Big>_{   \bigotimes\limits^{n+1}   \mc{L}_N^{(\bs{\la})}   }  \\ 
   + \; \f{  (1-\be) }{  N^{1+\a}   }  \Big<  \Dp{ 1}  \mc{U}_{N;1}^{-1}[ \phi_n ] 
   \Big>_{  \mu_{\e{eq}}^{(N)} \bigotimes\limits^{n-1}   \mc{L}_N^{(\bs{\la})}  }  
\; + \; \f{ 1 }{ N^{2+\a} } 
\sul{p=2}{n} \Big<   \Xi^{(p)}\circ \mc{U}_{N;1}^{-1}[  \Dp{p}\phi_n ] \Big>_{ \mu_{\e{eq}}^{(N)}  \bigotimes\limits^{n-2}   \mc{L}_N^{(\bs{\la})}   }  
\; + \; \f{  (1-\be) }{  N^{1+\a}   }  \Big<  \Dp{ 1}  \mc{U}_{N;1}^{-1}[ \phi_n ] \Big>_{  \bigotimes\limits^{n}   \mc{L}_N^{(\bs{\la})}  }   \;. 
\label{ecriture equation des boucles a n points}
\end{multline}

\end{prop}

\Proof  Schwinger-Dyson equations express the invariance of an integral under change of variables, or equivalently, integration by parts. Although the principle of derivation is well-known, we include the proof to be self-contained, 
following the route of infinitesimal change of variables. Let $\phi^{(a)}$, $a=1,\dots, n+1$  be a collection of smooth and compactly supported functions. We introduce an $\eps$-deformation of the probability
density $p_N$ given in \eqref{definition mesure proba rescalee avec potentiel arbitraire} 
by setting:
\beq
p_N^{(\{\eps_a\}_1^n)}\big( \bs{\la} \big)  \;  =  \;  \f{   1   }{  Z_N( \{\eps_a\} )  }
\pl{a<b}{N} \Big\{ \sinh\big[\pi\om_1 N^{\a}(\la_a-\la_b)\big] \sinh\big[\pi\om_2 N^{\a}(\la_a-\la_b)\big]  \Big\}^{\be}
\pl{a=1}{N} \ex{-N^{1+\a}  V_{(\{\eps_a\})}(\la_a)  } \;,
\label{definition fonction partiton eps shiftee}
\enq
where:
\beq
V_{(\{\eps_a\})}(\la)  \; = \; V(\la) \; + \;  \sul{a=2}{n+1} \eps_a \Bigg(\phi^{(a)}(\xi)  - \Int{}{} \phi^{(a)}(\eta)\,\dd \mu_{\e{eq}}^{(N)}(\eta)\Bigg)\;. 
\enq
The new normalisation constant $  Z_N(  \{\eps_a\}  )$ in \eqref{definition fonction partiton eps shiftee} 
is such that $p_N^{(\{\eps_a\})}$ is a still a probability density on $\R^N$. 

We then define $G_t(\mu) = \mu + t \phi^{(1)}(\mu)$. Since $\Dp{\xi} \phi^{(1)}(\xi)$ is bounded from below, for $t$ small enough $G_t$ is a
diffeomorphism of $\mathbb{R}$. Let us carry out the change of variables $\la_a = G_t(\mu_a)$ and translate the fact that $p_N^{(\{\eps_a\})}$ is a probability measure. This yields
\beq
1 \; = \; \Int{  \R^N }{  }   p_N^{(\{\eps_a\})}(\bs{\la}) \prod_{a = 1}^N \dd\la_a \; = \;   \Int{\R^N }{}  
p_N^{(\{\eps_a\})}\big(G_{t}(\la_1), \dots, G_{t}(\la_N) \big) \prod_{a = 1}^N G_t^{\prime}(\la_a)\,\dd\la_a   \;. 
\enq
 As a consequence, the change of variables yields, to the first order in $t$:
\bem
1 \; = \; \Int{ \R^N }{} \dd^N \bs{\la}  \Bigg\{ 1+ t \sul{a=1}{N} \Dp{\la_a}\phi^{(1)}(\la_a) \Bigg\}
\Bigg\{ 1 \;-\; t\,N^{1+\a} \sul{a=1}{N} \big( V_{ ( \{  \eps_a \} ) } \big)^{\prime}(\la_a)\,\phi^{(1)}(\la_a)  \Bigg \}  \\
\Bigg\{1 \;+ \; t\, N^{\a} \sul{ a < b }{ N } \Big[ \sum_{p = 1}^2 \beta\pi\om_p\cotanh[ \pi \om_p N^{\a} (\la_a-\la_b)]\Big] \Big[  \phi^{(1)}(\la_a) - \phi^{(1)}(\la_b) \Big] \Bigg\}
% \cdot  
%
 \cdot p_N^{ ( \{  \eps_a \} ) }(\bs{\la} )   \; \; +  \; \; O(t^2) \;. 
\end{multline}
Identifying the terms  linear in $t$ leads to:
\beq
-  \Big<   \phi^{(1)} \partial_{1}\big[ V_{(\{\eps_a\})}\big] \Big>_{  L_N^{(\bs{\la})}  } ^{  (\{\eps_a\}) }
\; + \; \f{1}{2}  
\Big<  \mc{D}_N[\phi^{(1)}] \Big>_{  L_N^{(\bs{\la})} \otimes L_N^{(\bs{\la})} } ^{  (\{\eps_a\}) }
\; + \;  \f{  (1-\be) }{ N^{1+\a}  }    \big<  \partial_{1}\phi^{(1)} \big>_{   L_N^{(\bs{\la})}   } ^{  (\{\eps_a\}) }   \; = \; 0 \;. 
\enq
The superscript ${  (\{\eps_a\}) }$ is there to emphasise that the averages should be taken with respect to the 
probability measure associated with the $\eps$-deformed density \eqref{definition fonction partiton eps shiftee}.
We then centralise the empirical measures with respect to $\mu_{ \e{eq} }^{(N)}$. By using the integral equation satisfied by the density of the equilibrium measure 
$V^{\prime}(\xi) \; = \;\mc{S}_N\big[ \rho_{\e{eq}}^{(N)} \big](\xi)$ for $\xi \in \intff{a_N}{b_N}$,  we obtain:
\bem
 - \Big<   \mc{U}_N[ \phi^{(1)} ]\Big>_{  \mc{L}_N^{(\bs{\la})} }  ^{  (\{\eps_a\}) }
\; - \;  \sul{p=2}{n+1}  \eps_a \bigg(  \big<  \phi^{(1)}\,\partial_{1}\phi^{(p)} \big>_{   \mu_{\e{eq}}^{(N)}  } \; + \; 
\big<   \phi^{(1)}\,\partial_{1}\phi^{(p)}  \big>_{  \mc{L}_N^{(\bs{\la})} }  ^{  (\{\eps_a\}) }  \bigg) \\
 \; + \;  \f{1}{2} 
 \Big<  \mc{D}_N[\phi^{(1)}]   \Big>_{  \mc{L}_N^{(\bs{\la})}    \otimes   \mc{L}_N^{(\bs{\la})} } ^{  (\{\eps_a\}) }  
\; +  \;  \f{  (1-\be) }{ N^{1+\a}  }  \bigg(  \big<  \partial_{1} \phi^{(1)}  \big>_{  \mu_{\e{eq}}^{(N)}  }  \; + \; 
   \big<   \partial_{1}\phi^{(1)} \big>_{    \mc{L}_N^{(\bs{\la})}  } ^{  (\{\eps_a\}) }   \bigg)  \; = \; 0 \;. 
\label{ecriture equation boucles epsilon deformee}
\end{multline}
Sending $\eps_a$'s to zero in this equation leads to the desired form of the Schwinger-Dyson equation at  level 1. 
In order to get the Schwinger-Dyson equation at level $n$, we should compute the $\eps_a$ derivatives of \eqref{ecriture equation boucles epsilon deformee} evaluated at $\epsilon_a \equiv 0$. However, first, 
it is convenient to multiply the above equation by $\tf{   Z_N( \{\eps_a\})   }{  Z_N [V] } $ so as to avoid differentiating the $\{\epsilon_a\}$-dependent partition function
entering in the definition of the density $p_N^{ (\{\eps_a\}) }( \bs{\la} )$. Doing so, however, produces additional averages in front of the averages solely involving the non-stochastic measures $\mu_{\e{eq}}$: 
\bem
 -  \Big< \mc{U}_N[ \phi^{(1)} ](\xi_1)\,\prod_{a = 2}^{n}\phi^{(a)}(\xi_a) \Big>_{    \bigotimes\limits^{n+1}   \mc{L}_N^{(\bs{\la})}    }  
\; + \;  \f{ 1 }{ N^{2+\a} } 
\sul{p=2}{n+1} \Big<   \phi^{(1)}(\xi_1)\,\partial_{1}\phi^{(p)}(\xi_1)\,  \pl{  \substack{ a=2  \\ \not= p}  }{n+1}  \phi^{(a)}(\xi^{(p)}_a)   \Big>_{ \bigotimes\limits^{n}   \mc{L}_N^{(\bs{\la})}   } \\
 \; + \;  \f{1}{2} \Big< \mc{D}_N[ \phi^{(1)}](\xi_1,\xi_2)   \pl{a=2}{n+1}   \phi^{(a)}( \xi_{a+1} )   \Big>_{   \bigotimes\limits^{n+2}   \mc{L}_N^{(\bs{\la})}   }  
   + \; \f{  (1-\be) }{  N^{1+\a}   }  \Big<  \partial_{1} \phi^{(1)} (\xi_1)  \Big>_{  \mu_{\e{eq}}^{(N)}  }  
\cdot \Big<  \pl{   a=2   }{n+1} \phi^{(a)}(\xi_{a- 1} )   \Big>_{ \bigotimes\limits^{n}   \mc{L}_N^{(\bs{\la})}   }\\
\; +\;  \f{ 1 }{ N^{2+\a} } 
\sul{p=2}{n+1} \Big<  \phi^{(1)}(\xi_1)   \partial_{1} \phi^{(p)} (\xi_1) \Big> _{ \mu_{\e{eq}}^{(N)} }
     \Big<  \pl{  \substack{ a=2  \\ \not= p}  }{n+1} \phi^{(a)}(\xi^{(p)}_{a-1})   \Big>_{ \bigotimes\limits^{n-1}   \mc{L}_N^{(\bs{\la})}   }
\; + \; \f{  (1-\be) }{  N^{1+\a}   }    \Big<  \partial_{1} \phi^{(1)} (\xi_1)  \pl{   a=2   }{n+1} \phi^{(a)}(\xi_{a} )     \Big>_{   \bigotimes\limits^{n+1}  \mc{L}_N^{(\bs{\la})}  }   \; = \; 0 \;. 
\end{multline}
To any $\bs{\xi} \in \mathbb{R}^{n - 1}$, we associated the vector $\bs{\xi}^{(p)} \in \R^n$ by $\bs{\xi}^{(p)} = (\xi_1,\ldots,\xi_{p - 1},\xi_1,\xi_p,\ldots,\xi_{n - 1})$, whose components arise in products of the type  $\prod_{  \substack{ a=2  \\ \not= p}  }^{n+1} \phi^{(a)}(\xi^{(p)}_a)$. The representation
\beq
\mc{U}_N[\phi](\xi) \; = \; \phi(\xi) V^{\prime}(\xi) \; + \; \Int{a_N}{b_N} \Bigg\{\sul{p=1}{2} \beta\pi\om_p  \cotanh\big[ \pi \om_pN^{\alpha} (\xi - \eta) \big]\Bigg\}\big( \phi(\eta)-\phi(\xi) \big)\,\rho_{\e{eq}}^{(N)}(\eta)\,\dd \eta
\enq
readily shows that the operators $\mc{U}_N$ and $\mc{D}_N$ are both continuous as operators $W_1^{\infty}(K) \tend W_0^{\infty}(K)$ for any compact $K \subseteq \mathbb{R}$. This continuity along with the finiteness of the 
measure $\mathbb{P}_N$ is then enough to conclude,  by density of $\mc{C}^{\infty}_{c}(\R)\otimes \dots \otimes \mc{C}^{\infty}_{c}(\R)$
in $\mc{C}^{\infty}_{c}(\R^n)$, that equation \eqref{ecriture equation des boucles a n points} holds 
for all functions $\phi_n \in \mc{C}^{\infty}_{c}(\R^n)$. Eventually, the assumption of compact support can be dropped. Indeed, given any  $\phi_n \in \mc{C}^{\infty}_{c}(\R^n)$, the Schwinger-Dyson equation at level $n$ can be presented as 
\bem
 \big< \mc{U}_{N;1}[\phi_n] \big>_{    \bigotimes\limits^{n}   \mc{L}_N^{(\bs{\la})}    }  \; = \; 
  \f{ 1 }{ N^{2+\a} } 
 \sul{p=2}{n} \Big<   \Xi^{(p)}[\partial_{p}\phi_n] \Big>_{ \bigotimes\limits^{n-1}   \mc{L}_N^{(\bs{\la})}   } 
 \; + \; 
 \f{1}{2} \Big< \mc{D}_{N;1}[\phi_n] \Big>_{   \bigotimes\limits^{n+1}   \mc{L}_N^{(\bs{\la})}   }  \\ 
   + \; \f{  (1-\be) }{  N^{1+\a}   }  \Big<  \Dp{ 1}  \phi_n  
   \Big>_{  \mu_{\e{eq}}^{(N)} \bigotimes\limits^{n-1}   \mc{L}_N^{(\bs{\la})}  }  
\; + \; \f{ 1 }{ N^{2+\a} } 
\sul{p=2}{n} \Big<   \Xi^{(p)}[  \Dp{p}\phi_n ] \Big>_{ \mu_{\e{eq}}^{(N)}  \bigotimes\limits^{n-2}   \mc{L}_N^{(\bs{\la})}   }  
\; + \; \f{  (1-\be) }{  N^{1+\a}   }  \Big<  \Dp{ 1}  \phi_n  \Big>_{  \bigotimes\limits^{n}   \mc{L}_N^{(\bs{\la})}  }   \;. 
\label{ecriture equation des boucles a n pointsbis}
\end{multline}
It is readily seen due to the sub-exponentiality hypothesis \eqref{Hypothese sous exponetialite des derivees} that given 
$0<\kappa<\kappa^{\prime}$ and $\phi_n$ such that $\mc{K}_{\kappa}[\phi_n]\in W_{1}^{\infty}(\R^n)$, we have: 
\beq
\norm{ \mc{K}_{\kappa^{\prime} }\big[ \mc{U}_{N;1}[ \phi_n ]  \big] }_{  W_{0}^{\infty}(\R^n) } \; \leq \; 
C  \norm{ \mc{K}_{\kappa }[ \phi_n ] ] }_{  W_{0}^{\infty}(\R^n) }
\enq
and likewise for $\mc{D}_{N;1}$. Thus, since $\mc{K}_{\kappa}[\phi_n] \in W_1^{\infty}(\R^n)$
can be approached in $W_1^{\infty}(\R^n)$ norm by functions $\mc{K}_{\kappa}[\psi_n]$ with  $ \psi_n\in \mc{C}^{\infty}_{c}(\R^n) $, 
it remains to invoke the finiteness of the measures $\mathbb{P}_{N}$ and the decomposition of the $n^{\e{th}}$ order averages obtained in 
Lemma \ref{Lemme reduction moyenne n pts a partie cpcte} so as to get \eqref{ecriture equation des boucles a n points}
in full generality. In the announcement of the result, we actually choose to write the Schwinger-Dyson equation \eqref{ecriture equation des boucles a n points} for $\mathcal{U}_{N;1}^{-1}[\phi_n]$ instead of $\phi_n$. This rewriting is possible because we construct in Chapter~\ref{Section descirption cptmt unif op WN} this inverse $\mathcal{U}_{N;1}^{-1}$, and merely anticipate the use we will make of this equation. The Schwinger-Dyson equation we have proved in the form \eqref{ecriture equation des boucles a n points} holds independent of the invertibility of $\mathcal{U}_{N;1}$.
\qed

\vspace{3mm} 

It follows from the form taken by the Schwinger-Dyson equations that, if we want to solve these equations perturbatively 
we should, in the very first place, construct the inverse to the operator $\mc{U}_N$. This should be done is such a way that one
can control explicitly or at least in a manageable way, its dependence on $N$ and its possible singularities. 
Indeed, the building blocks of $\mc{U}_{N}^{-1}$ exhibit, for instance, square root like singularities at the endpoints of the 
support $\intff{a_N}{b_N}$ of the equilibrium measure. 
In \S~\ref{Sous Section Int Rep Un moins 1 cas regulier et vague}, we shall construct a regular representation for $\mc{U}_{N}^{-1}$. 
By regularity, we mean that the various square root singularities present in its building blocks 
eventually cancel out, hence showing that $\mc{U}_N^{-1}[H]$ is smooth as long as $H$ is. Then, in \S~\ref{Sous section Sharp weighted bounds for UN moins 1},
we shall provide explicit, $N$-dependent, bounds on the $W_{\ell}^{\infty}(\R)$ norms of $\mc{U}_N^{-1}[H]$. These will play a crucial
role in the large-$N$ analysis of the Schwinger-Dyson equations.

%%%%%%%%%%%%%%%%%%%%%%%%%%%%%%%%%%%%%%%%%%%%%%%%%%%%%%%%%%%%%%%%%%%%%%%%%%%%%%%%%%%%%%%%%%%%%%%%%%%%%%%%%%%%%%%%%%%%%%%%%%%%%%%%%%%%%%%% 
%%%%%%%%%%%%%%%%%%%%%%%%%%%%%%%%%%%%%%%%%%%%%%%%%%%%%%%%%%%%%%%%%%%%%%%%%%%%%%%%%%%%%%%%%%%%%%%%%%%%%%%%%%%%%%%%%%%%%%%%%%%%%%%%%%%%%%%%

%%%%%%%%%%%%%%%%%%%%%%%%%%%%%%%%%%%%%%%%%%%%%%%%%%%%%%%%%%%%%%%%%%%%%%%%%%%%%%%%%%%%%%%%%%%%%%%%%%%%%%%%%%%%%%%%%%%%%%%%%%%%%%%%%%%%%%%%
%%%%%%%%%%%%%%%%%%%%%%%%%%%%%%%%%%%%%%%%%%%%%%%%%%%%%%%%%%%%%%%%%%%%%%%%%%%%%%%%%%%%%%%%%%%%%%%%%%%%%%%%%%%%%%%%%%%%%%%%%%%%%%%%%%%%%%%%

%%%%%%%%%%%%%%%%%%%%%%%%%%%%%%%%%%%%%%%%%%%%%%%%%%%%%%%%%%%%%%%%%%%%%%%%%%%%%%%%%%%%%%%%%%%%%%%%%%%%%%%%%%%%%%%%%%%%%%%%%%%%%%%%%%%%%%%%
%%%%%%%%%%%%%%%%%%%%%%%%%%%%%%%%%%%%%%%%%%%%%%%%%%%%%%%%%%%%%%%%%%%%%%%%%%%%%%%%%%%%%%%%%%%%%%%%%%%%%%%%%%%%%%%%%%%%%%%%%%%%%%%%%%%%%%%%

\section{Asymptotic analysis of the Schwinger-Dyson equations}
\label{s4fsfg}

The asymptotic analysis of the Schwinger-Dyson equation builds heavily on a family of $N$-weighted norms that we introduce below. 

\begin{defin}
 \label{weiei}
 For any $\phi \in W^{\infty}_n(\R^p)$, the $N$-weighted $L^{\infty}$ norm of order $\ell$ is defined by 
\beq
\mc{N}_N^{(\ell)}[\phi] \; = \; \sul{k=0}{\ell} \f{  \norm{ \phi }_{ W^{\infty}_{k}(\R^p) } }{ N^{k\a} } \; . 
\enq
This notation does not specify the number of variables of $\phi$ since this is usually clear from the context. 
\end{defin}
The weighted norm satisfies the obvious bound:
\beq
\mc{N}_{N}^{(\ell)}[\phi] \leq (\ell+1)\cdot \norm{\phi}_{W^{\infty}_{\ell}(\mathbb{R}^p)}\;,
\enq
and, respectively, the operators of differentiation and "repetition of a variable" $\Xi^{(p)}$ are bounded as :
\beq
{\cal N}_{N}^{(\ell)}[\partial_{p} \phi] \leq N^{\alpha}\,{\cal N}_{N}^{(\ell + 1)}[\phi]\;,\qquad\qquad  {\cal N}_{N}^{(\ell)}\big[\Xi^{(p)}[\phi]\big] \leq {\cal N}_{N}^{(\ell)}[\phi]\;.
\enq
Also, it is important to introduce a specific function that allows one to control the dependence on the potential in the various bounds that 
issue from the Schwinger-Dyson equations.
\begin{defin}
\label{weie2} The order $\ell$  estimate of the potential $V$ is defined as
\beq
  \mf{n}_{\ell}[V] \; = \; \f{ \max\Big\{  \pl{a=1}{\ell} \norm{ \mc{K}_{\kappa}[V^{\prime}] }_{ W^{\infty}_{k_a}(\R^n) }  \; : \;  \sul{a=1}{\ell} k_a = 2 \ell + 1  \Big\} }
{ \bigg\{ \min \Big(1\, ,\,  \inf_{ \intff{a}{b} } |V^{\prime\prime}(\xi)| \, , \,  |V^{\prime}(b+\eps)-V^{\prime}(b)| 
\, , \, |V^{\prime}(a-\eps)-V^{\prime}(a)|  \Big)\bigg\}^{\ell+1} } \;,
\label{definition fct estimatrice du potentiel}
\enq
where $\eps>0$ is small enough and fixed once for all, while $\kappa >0$. We also remind that $\mc{K}_{\kappa}$ is the exponential regularisation of Definition~\ref{Definition regularisation exponentielle}.  
\end{defin}

Since $\kappa$ only plays a minor role due to the sub-exponentiality hypothesis \eqref{Hypothese sous exponetialite des derivees}
in the estimates provided by $\mf{n}_{\ell}[V]$, we chose to keep its dependence implicit. Note also that the constants $ \mf{n}_{\ell}[V]$ satisfy 
\beq
 \mf{n}_{\ell}[V] \cdot \mf{n}_{\ell^{\prime}}[V] \; \leq \;  \mf{n}_{\ell+\ell^{\prime}+1}[V] \;. 
\label{propriete multiplicativite des constantes n ell de V}
\enq

\begin{lemme}
 \label{Lemme bornes sur normes operateurs UN et DN}
 
%and $V$ be strictly convex, such that  $\mc{K}_{ \tf{\kappa}{\ell} }[V^{\prime}] \in W^{\infty}_{2\ell+1}(\R^n)$. 
%Let $\intff{a_N}{b_N}$ be the support of the equilibrium measured defined by $V$. 

Let $\kappa>0$ . There exist constants $C_{n;\ell}, \wt{C}_{n;\ell}>0$ such that,
for any  $\phi$ satisfying 
\begin{itemize}
\item $\mc{K}_{\tf{\kappa}{\ell}}[\phi]\in W^{\infty}_{2\ell+1}(\R^n)$ 
\item $\xi\mapsto  \phi(\xi,\xi_2,\dots,\xi_n) \in \mf{X}_s(\intff{a_N}{b_N})$, $0<s<1/2$, that is to say\footnote{It is straightforward 
to check by carrying out contour deformations that, for functions $\psi$ decaying sufficiently fast at infinity with respect to its first variable, the condition \eqref{ecriture explicite condition d'appartenance a Xs sur sup mes eq} %almost surely in the other variables 
is equivalent to belonging to $ \mf{X}_s(\R)$.}
\beq 
\Int{ \R+\i \eps }{} \hspace{-2mm} \f{ \dd \mu }{ 2\i \pi }\,\chi_{11}(\mu) 
\Int{a_N}{b_N} \phi(\xi,\xi_2,\dots,\xi_n)\, \ex{\i \mu N^{\a}(\xi-b_N)}\dd \xi \; = \; 0 
\qquad  {\rm almost}\,\,\mathrm{everywhere}\,\,\e{in} \; (\xi_2,\dots,\xi_n) \in \R^{n-1} 
\label{ecriture explicite condition d'appartenance a Xs sur sup mes eq}
\enq
\end{itemize}
we have the bounds:
\begin{eqnarray}
\mc{N}_N^{(\ell)}\Big[ \mc{K}_{\kappa}\big[ \mc{U}_{N;1}^{-1}[\phi] \big] \Big]  & \leq & C_{n;\ell} \cdot \mf{n}_{\ell}[V] \cdot N^{\a}
\cdot \big( \ln N \big)^{2 \ell + 1 } \cdot \mc{N}_{N}^{(2\ell+1)}\big[ \mc{K}_{\kappa}[\phi] \big] \;, 
\label{ecriture borne Wp infty sur UN inverse} \\
\mc{N}_N^{(\ell)}\Big[  \mc{K}_{\kappa}\big[ \mc{D}_{N;1}[\phi] \big] \Big]  & \leq & \wt{C}_{n;\ell}  \cdot  (\ln N)^2 
\cdot \mc{N}_{N}^{(\ell+1)}\big[ \mc{K}_{\kappa}[\phi] \big] \;. 
\label{ecriture borne Wp infty sur op DN}
\end{eqnarray}
\end{lemme}

Note that the above lemma implies, in particular, a bound on the weighted norm of $\mc{D}_{N;1}\circ\mc{U}_{N;1}^{-1}$:
\beq
\label{normDNUN}\mc{N}_N^{(\ell)}\Big[ \mc{K}_{\kappa}\big[ \mc{D}_{N;1}\circ\mc{U}_{N;1}^{-1}[\phi] \big] \Big]  \; \leq  \;
C_{n,\ell}^{\prime} \cdot \mf{n}_{\ell+1}[V] \cdot N^{\a} \cdot \big(\ln N \big)^{2 \ell + 5 } \cdot \mc{N}_{N}^{(2\ell+3)}\big[ \mc{K}_{\kappa}[\phi] \big] \;, 
\enq
We stress for the last time that the proof of this Lemma, for the part concerning $\mathcal{U}_{N;1}^{-1}$, relies on estimates of this inverse obtained in Chapter~\ref{Section descirption cptmt unif op WN} independently of the present chapter.

\Proof We first focus on the norm of $\mc{K}_{\kappa}\big[ \mc{D}_{N;1}[\phi] \big] $. In order to obtain \eqref{ecriture borne Wp infty sur op DN}, we bound
\beq
{\cal O}_{\bs{k}_{n+1}}(\bs{\xi}_{n+1}) \, = \,  \pl{a=1}{n+1} \Dp{\xi_a}^{k_a}\, \mc{K}_{\kappa}\big[ \mc{D}_{N;1}[\phi] \big](\xi_1,\dots,\xi_{n+1}) 
\qquad \e{with} \quad
\sul{a=1}{n+1} k_a \, \leq \,  \ell \quad k_a \in \mathbb{N} 
\enq
by different means in the two cases of interest, \textit{viz}. $N^{\a}|\xi_1-\xi_2| \geq  (\ln N)^2$ and  $N^{\a}|\xi_1-\xi_2| <  (\ln N)^2$.

We first treat the case $N^{\a}|\xi_1-\xi_2| \geq  (\ln N)^2 $. Observe that for $|N^{\a} \xi | \, \geq \, ( \ln N )^{2}$, we have:
\beq
\forall \ell \geq 0,\qquad \big|\partial_{\xi}^{\ell} \big\{ S( N^{\a} \xi )  \big\} \big| \leq \delta_{\ell,0}\,c_{0}^{\prime} \, + \, 
(1 - \delta_{\ell,0})\,c^{\prime}_{\ell}\,N^{ \ell \a} \ex{-c^{\prime \prime} \ln^2 N } \; \leq \; 
c_{\ell} \big( \ln N \big)^2 
\enq
for some constants $c_{\ell}$, where $S$ is defined in \eqref{ecriture eqn int sing de depart} and $\de_{\ell,0}$ being the Kronecker symbol. Therefore:
\begin{eqnarray}
\big| {\cal O}_{\bs{k}_{n+1}}(\bs{\xi}_{n+1}) \big| & \leq & 
\sul{ \substack{p_a+\ell_a=k_a \\ a=1,2} }{} \pl{a=1}{2} {k_a \choose p_a} \cdot \Big| \Dp{\xi_1}^{p_1} \Dp{\xi_2}^{p_2} 
 \big[\phi_{\{k_a\}}(\xi_1,\xi_3, \dots, \xi_{n+1}) - \phi_{\{k_a\}}(\xi_2,\xi_3, \dots, \xi_{n+1}) \big] \Big| \cdot c_{\ell_1+\ell_2} \cdot (\ln N)^2 \nonumber \\
& \leq & C \cdot N^{[\max(k_1,k_2)]\a} \cdot (\ln N)^2 \sul{s=0}{ \max(k_1,k_2) } N^{-s \a} 
\max_{\eta \in \{\xi_1,\xi_2\}} \big| \Dp{1}^{s} \phi_{\{k_a\}}(\eta,\xi_3, \dots, \xi_{n+1}) \big| \nonumber \\
%
%
%\; \leq \; C \cdot  N^{ \ell \a} \cdot (\ln N)^2 \sul{s=0}{ \max(k_1,k_2) } 
%
%\f{ \big|\big| \mc{K}_{\kappa}[\phi] \big| \big|_{W_{s+ \ell-k_1-k_2}^{\infty}(\R^n)} }{  N^{(s+\ell-k_1-k_2) \a} }
%
\label{tjuj1}& \leq & C \cdot N^{ \ell \a} \cdot (\ln N)^2 \cdot \mc{N}_{N}^{(\ell)}\big[ \mc{K}_{\kappa}[\phi] \big] \;, 
\end{eqnarray}
where, in the intermediate calculations, we have used:
\beq
\phi_{ \{k_a\} }(\xi_1,\xi_2, \dots, \xi_{n}) \, = \,  \pl{a=3}{n+1} \Dp{\xi_{a}}^{k_a}\Big\{\mc{K}_{\kappa}[\phi](\xi_1,\xi_2, \dots, \xi_{n})\Big\}\;.
\enq

\noindent We now turn to the case when  $N^{\a}|\xi_1-\xi_2| < (\ln N)^2 $. 
Observe that  for any $\ell \in \mathbb{N}$ and $|N^{\a} \xi| \, \leq \,  (\ln N)^2 $, the function $\wt{S}$, with $\wt{S}(x)=x S(x)$,  satisfies
\beq
\forall \ell \geq 0,\qquad \big| \partial_{\xi}^{\ell}\, \big\{ \wt{S}( N^{\a} \xi ) \big\}  \big| \, \leq \, 
\de_{\ell,0} \big| N^{\a} \xi\big[ S(N^{\a} \xi) - \f{2\be}{N^{\a} \xi} \big] + 2\be \big| 
\, + \; (1 - \de_{\ell,0})\,N^{\ell \a } || \wt{S} ||_{ W_{\ell}^{\infty}(\R) }  \; \leq \; c_{\ell} N^{\a \ell} \big( \ln N \big)^2
\enq
for some constants $c_{\ell}$. Starting from the integral representation 
\beq
 {\cal O}_{\bs{k}_{n+1}}(\bs{x}_{n+1})  \;= \;  \Int{0}{1} \f{ \dd t }{ N^{\a} } \Dp{\xi_1}^{k_1}  \Dp{\xi_2}^{k_2} 
\Big\{ \Dp{1}\phi_{ \{k_a\} }(\xi_1+t(\xi_2-\xi_1),\xi_3, \dots, \xi_{n+1}) \cdot \wt{S}\big(N^{\a}(\xi_1-\xi_2)\big) \Big\}\;,
\enq
we obtain:
\begin{eqnarray}
\big|  {\cal O}_{\bs{k}_{n+1}}(\bs{x}_{n+1}) \big|  & \leq  & \sul{ \substack{p_a+\ell_a=k_a \\ a=1,2} }{} 
\f{ {k_1 \choose p_1}{k_2 \choose p_2}\, c_{\ell_1+\ell_2} } { N^{\a(1-\ell_1-\ell_2)} }
\Int{0}{1}  (1-t)^{p_1} t^{p_2}
(\Dp{1}^{p_1+p_2+1}\phi_{ \{k_a\} } )\big(\xi_1+t(\xi_2-\xi_1),\xi_3, \dots, \xi_{n+1}\big)
  \cdot    (\ln N)^2 \cdot \dd t  \nonumber \\
& \leq & C N^{(k_1+k_2)\a} (\ln N)^2 \sul{s=1}{k_1+k_2+1 } N^{-s \a} 
\max_{\eta \in \intff{\xi_1}{\xi_2} } \big| (\Dp{1}^{s} \phi_{ \{k_a\} } )(\eta,\xi_3, \dots, \xi_{n+1}) \big|  \nonumber \\
\label{tjuj}& \leq & C N^{ \ell \a} (\ln N)^2 \cdot \mc{N}_{N}^{(\ell+1)}\big[ \mc{K}_{\kappa}[\phi] \big]\;.
\end{eqnarray}
Putting together \eqref{tjuj1} and \eqref{tjuj} and taking the supremum over $\{k_a\}$ such that $\sum_{a} k_a \leq  \ell$, we deduce the desired bound \eqref{ecriture borne Wp infty sur op DN} for the weighted norm of $\mc{D}_{N}$.

The bounds for the weighted norm of $\mc{K}_{\kappa}\big[\mc{U}_{N;1}^{-1}[\phi]\big]$ are obtained quite straightforwardly by using the $W^{\infty}_{\ell}(\R)$ bounds on $ \mc{K}_{\kappa}\big[\mc{U}_{N;1}^{-1}[\phi]\big]$ 
as derived in Proposition~\ref{Theorem bornes sur norme inverse UN via estimation fines locales}. \qed

\vspace{5mm}

With the bounds on the action of the operators $\mc{U}^{-1}_{N;1}$ and $\mc{D}_{N;1}$, we can improve the \textit{a priori} bounds on the 
centred expectation values of the correlators through a bootstrap procedure.

\begin{prop}
 \label{Lemme bornes optimales sur les correlateurs}
 
Let $\a < 1/4$ and pick $\kappa>0$. There exist an increasing sequence of integers $(m_n)_n$, positive constants $(C_n)_n$, such that, for any $n \geq 1$ and 
$\phi \in \mf{X}_s(\intff{a_N}{b_N})$ in the sense of \eqref{ecriture explicite condition d'appartenance a Xs sur sup mes eq} and satisfying $\mc{K}_{\kappa}[\phi] \in W^{\infty}_{m_n}(\R^n)$,
\textit{cf}. \eqref{definition kappa dumped function}, we have:
\beq
\Big| \big< \phi \big>_{ \bigotimes^n_1 \mc{L}^{(\bs{\la})}_{N} }  \Big| \; \leq \; 
C_n \cdot \mf{n}_{m_n}[V] \cdot \mc{N}_{N}^{(m_n)}\big[ \mc{K}_{\kappa}[\phi] \big]\,N^{(\a - 1)n}\ \;.
\enq
The whole dependence of the upper bound on the potential $V$ is contained in the constant $\mf{n}_{\ell_n}[V]$, and we can take:
\beq
\label{ordernorme}
m_n = \ell_n^{(q_n)},\qquad q_n = 1 + \Big\lfloor \f{n}{1 - 4\alpha} \Big\rfloor,\qquad \ell_{n}^{(q)} = 2^{q}(n + q) + 3(2^{q} - 1)\;.
\enq
\end{prop}

\Proof The proof utilises a bootstrap-based improvement of the \textit{a priori} bounds given in Corollary \ref{Corollary bornes a priori sur les correlateurs}.
Namely, assume the existence of sequences $\eta_N \tend 0$, $\varkappa_N \in \intff{0}{1}$, and constants $C_n>0$ independent of $N$,
and integers $\ell_n$ increasing with $n$, such that, for any $\phi$ such that $\mc{K}_{\kappa}[\phi] \in W_{\ell_n}^{\infty}(\R^n)$: 
\beq
\Big|  \big< \phi \big>_{ \bigotimes^n_1 \mc{L}^{(\bs{\la})}_{N} } \Big|  \; \leq \; C_n \cdot \mf{n}_{ \ell_n }[V] \cdot 
\mc{N}_{N}^{(\ell_n)}\big[ \mc{K}_{\kappa}[\phi] \big] \cdot \Big( \eta_N^n \cdot \varkappa_N \; + \; N^{n(\a-1)} \Big) \;. 
\label{ecriture hypothese sur les ordre magntude fluctuations}
\enq
We will establish that there exists a new constants $C_n^{\prime}>0$ and integers $\ell^{\prime}_n=2\ell_{n+1}+3$ such that, 
for  $\mc{K}_{\kappa}[\phi] \in W_{ \ell_n^{\prime} }^{\infty}(\R^n)$: 
\beq
\Big| \big< \phi \big>_{ \bigotimes^n_1 \mc{L}^{(\bs{\la})}_{N} } \Big| \; \leq \; C_n  \cdot \mf{n}_{ \ell_n^{\prime} }[V]  \cdot 
\mc{N}_{N}^{(\ell^{\prime}_n)}\big[ \mc{K}_{\kappa}[\phi] \big] \cdot 
\Big( \eta_N^n \cdot \varkappa_N^{\prime} \; + \; N^{n(\a-1)} \Big) \;,
\label{ecriture amelioration des bornes sur correlateur ordre n}
\enq
where
\beq
\varkappa^{\prime}_N \, = \, \varkappa_N \,\big( \ln N \big)^{\ell^{\prime}_n+2}\, \max\big(N^{\alpha}\eta_N\,;\,N^{\alpha - 2}\eta_N^{-2}\,;\,\,N^{\alpha - 1}\eta_N^{-1}\big)\;.
\label{ecriture amelioration des bornes sur constante varkappaN}
\enq
Before justifying \eqref{ecriture amelioration des bornes sur constante varkappaN}, let us examine its consequences. The bootstrap approach can be settled if, for some $\ga_{\varkappa}>0$,
\beq
\label{conduia}\varkappa^{\prime}_{N} = N^{-\ga_{\varkappa} } \varkappa_N \;. 
\enq
Assuming momentarily that $\eta_N = N^{-\gamma}$, when $0<\a<1$, the range of $\alpha$ and $\gamma$ ensuring \eqref{conduia} is:
\beq
\label{coneqn} \alpha < \gamma < 1 - \alpha\; \qquad \e{what} \; \e{implies } \; \; \alpha < 1/2\;.
\enq
The rate $\ga_{\varkappa}$ at which $\varkappa_N^{\prime}/\varkappa_{N}$ goes to zero increases when $\gamma$ runs from $\alpha$ to $1/2$, is maximal and equal to $1/2 - \a$ when $\gamma = 1/2$, and then decreases when $\gamma$ increases between $1/2$ and $1 - \a$.

The \textit{a priori} estimate proved in Corollary \ref{Corollary bornes a priori sur les correlateurs} gives:
\beq
\Big| \big< \phi \big>_{ \bigotimes^n_1 \mc{L}^{(\bs{\la})}_{N} } \Big| \; \leq \; C_n^{\prime} \cdot 
\norm{ \mc{K}_{\kappa}[\phi]  }_{ W^{\infty}_{n}(\R^n) }^{ \f{1}{2} } \cdot  \norm{ \mc{K}_{\kappa}[\phi]  }_{ W^{\infty}_{0}(\R^n) }^{ \f{1}{2} }
\cdot N^{(\a - 1)n/2} \; \leq \; C_n^{\prime} \cdot \mc{N}_N^{(n)}\big[ \mc{K}_{\kappa}[\phi] \big]\,N^{(\a - 1/2)n}\;.
\enq
Therefore, the assumption \eqref{ecriture amelioration des bornes sur correlateur ordre n} is satisfied with $\eta_N = N^{-\gamma}$ for $\gamma = 1/2 - \alpha$, and the order $\ell_n = n$ for the weighted norm. The bootstrap condition \eqref{coneqn} then implies $\alpha < 1/4$, and in this case, we find:
\beq
\varkappa^{\prime}_N \leq \varkappa_N\,(\ln N)^{\ell^{\prime}_n}\,N^{-\f{ (1 - 4\a) }{ 2 }  }\;.
\enq

Now, we can iterate the bootstrap until the first term in \eqref{ecriture amelioration des bornes sur correlateur ordre n} becomes less or equal than the second term $N^{(\a - 1)n}$.
%which represents he "optimal" bound -- as we will see, it is the correct order of magnitude of the $n$ point correlations when $N \rightarrow \infty$. 
This is obtained in a number of steps $q_n$ determined by the equation $N^{-(1/2 - \alpha)n} N^{-(1 - 4\a)q_n/2} \ll N^{(\a - 1) n}$, therefore:
\beq
q_n = 1 + \Big\lfloor \frac{n}{1 - 4\a} \Big\rfloor \;.
\enq
The order of the weighted norm appearing in the bound of the $n$ point correlations at step $q$ of the recursion satisfies  $\ell^{(q)}_{n} = 2\ell^{(q - 1)}_{n + 1} + 3$, with initial condition $\ell^{(0)}_{n} = n$.  The solution is
\beq
\ell_n^{(q)} = 2^{q}(n + q) + 3(2^{q} - 1)\;.
\enq
Therefore, we get at the end of the recursion:
\beq
\Big| \big< \phi \big>_{ \bigotimes^n_1 \mc{L}^{(\bs{\la})}_{N} } \Big| \leq C_n \cdot N^{(\a - 1)n}\cdot \mc{N}_N^{(m_n)}\big[ \mc{K}_{\kappa}[\phi] \big],\qquad m_n = \ell_n^{(q_n)}\;.
\enq

We shall now justify the claim \eqref{ecriture amelioration des bornes sur constante varkappaN}. Starting from \eqref{ecriture hypothese sur les ordre magntude fluctuations}, we bound $ \big< \phi \big>_{ \bigotimes^n_1 \mc{L}^{(\bs{\la})}_{N} }$
given by the Schwinger-Dyson equations of Proposition \ref{Proposition equations des boucles}, using the norms of the operators ${\cal U}_{N;1}$ and ${\cal D}_{N}$ obtained in Lemma \ref{Lemme bornes sur normes operateurs UN et DN}.
We stress that it is indeed licit to apply the bound \eqref{ecriture borne Wp infty sur UN inverse} for $\mc{U}_N^{-1}$ for, if $\phi$ satisfies the condition \eqref{ecriture explicite condition d'appartenance a Xs sur sup mes eq}, then so do the functions 
$\Dp{p} \phi$ with $p=2,\dots,n$. Respecting the order of appearance of terms in \eqref{ecriture equation des boucles a n points}, we get\footnote{The third and fifth line are absent in the case $\beta = 1$, and it gives a larger range of $\alpha > 0$ for which $\eta_N$ can be chosen so that the bootstrap works. But, eventually, this does not lead to a stronger bound because we can only initialize the bootstrap with the concentration bound \eqref{Corollary bornes a priori sur les correlateurs} i.e. $\eta_N = N^{-(1/2 - \a)}$.}:
\bem
\Big| \big< \phi \big>_{ \bigotimes^n_1 \mc{L}^{(\bs{\la})}_{N} } \Big| \; \leq \; C \mf{n}_{\ell_{n-1}}^2[V] \f{ N^{2\a} }{N^{2+\a} } (\ln N)^{2\ell_{n-1}+1}
\mc{N}_N^{(2\ell_{n-1}+2)}\big[ \mc{K}_{\kappa}[\phi] \big] \cdot \Big( \eta_N^{n-1} \cdot \varkappa_N \; + \; N^{ (n-1)(\a-1) } \Big) \\
+ \, C \mf{n}_{\ell_{n+1}}[V]  \mf{n}_{\ell_{n+1}+1}[V] N^{\a}  (\ln N)^{2\ell_{n+1}+5} \cdot \mc{N}_N^{(2\ell_{n+1}+3)}\big[ \mc{K}_{\kappa}[\phi] \big] 
\cdot \Big( \eta_N^{n+1} \cdot \varkappa_N \; + \; N^{ (n+1)(\a-1) } \Big)  \\
\; + \;  C\,\mf{n}_{\ell_{n-1}}[V]  \mf{n}_{\ell_{n-1}+1}[V] \f{N^{2\a}}{N^{1+\a}}  (\ln N)^{2\ell_{n-1}+3} \cdot \mc{N}_N^{(2\ell_{n-1}+3)}\big[ \mc{K}_{\kappa}[\phi] \big] 
\cdot \Big( \eta_N^{n-1} \cdot \varkappa_N \; + \; N^{ (n-1)(\a-1) } \Big)  \\
+ \, C ( \mf{n}_{\ell_{n-2}}[V] )^2 \f{ N^{2\a} }{ N^{2+\a} }  (\ln N)^{2\ell_{n-2}+2} \cdot \mc{N}_N^{(2\ell_{n-2}+1)}\big[ \mc{K}_{\kappa}[\phi] \big] 
\cdot \Big( \eta_N^{n-2} \cdot \varkappa_N \; + \; N^{ (n-2)(\a-1) } \Big)  \\
\; + \;  C\,\mf{n}_{\ell_{n}}[V]  \mf{n}_{\ell_{n}+1}[V] \f{N^{2\a}}{N^{1+\a}}  (\ln N)^{2\ell_{n-1}+3} \cdot \mc{N}_N^{(2\ell_{n}+3)}\big[ \mc{K}_{\kappa}[\phi] \big] 
\cdot \Big( \eta_N^{n} \cdot \varkappa_N \; + \; N^{ n(\a-1) } \Big) \;, 
\end{multline}
for some constant $C > 0$ depending on $n$ and $\kappa$ only. Note that terms integrated against the probability measure $\mu_{\e{eq}}^{(N)}$ have been bounded by means of sup norms. 
The maximal powers of $N$ are exactly as in \eqref{ecriture amelioration des bornes sur constante varkappaN} -- since we assume $\eta_N \rightarrow 0$, the powers arising in the first line are negligible compared to those in the fourth line. 
We can then use \eqref{propriete multiplicativite des constantes n ell de V} to bound the products of $\mf{n}_{\ell}[V]$'s in terms of $\mf{n}_{\ell^{\prime}_n}[V]$ for a choice:
\beq
\ell_n^{\prime} \geq \max\big( 2\ell_{n-1}+2,  2\ell_{n+1}+3,  2\ell_{n-1}+3,  2\ell_{n-2}+2,  2\ell_{n}+3\big)\;.
\enq
Since $(\ell_n)_{n}$ is increasing, we can take $\ell^{\prime}_n = 2\ell_{n + 1} + 3$, and we indeed find \eqref{ecriture amelioration des bornes sur correlateur ordre n} for $N$ large enough.
Note that, the new sequence $(\ell^{\prime}_n)_n$ is, again, increasing. Then, the maximal power of $\ln N$ occurs in the second line, and is $(\ln N)^{2\ell_{n + 1} + 5} = (\ln N)^{\ell^{\prime}_{n} + 2}$. 
So, we have fully justified \eqref{ecriture amelioration des bornes sur correlateur ordre n}. \qed

The improved estimates on the multi-point correlators are almost all that is needed for obtaining the large $N$ asymptotic expansion of 
general one-point functions up to $\e{o}(N^{-(2+\a)})$ corrections. Prior to deriving such results, we still need to introduce 
an operator $\wt{\mc{X}}_N$ mapping any function $ W^{\infty}_{p}(O)$, $O$ a bounded open subset in $\R^n$, onto a function belonging to 
$\mf{X}_s(\intff{a_N}{b_N})$ in the sense of \eqref{ecriture explicite condition d'appartenance a Xs sur sup mes eq}.

\begin{defin}
 \label{defXN}
 Let $\mc{X}_N$ be the linear form on $W_1^{\infty}(\intff{a_N}{b_N})$:
\beq
\mc{X}_{N}[\phi]\; = \; \f{ \i N^{\a} }{ \chi_{11;+}(0) }\Int{\R+\i\eps }{} \f{\dd \mu}{2\i \pi }\,\chi_{11}(\mu) \Int{a_N}{b_N} \ex{\i \mu N^{\a} (\xi-b_N)}\,\phi(\xi)\,\dd\xi\  \;. 
\label{definition form lineaire contrainte X}
\enq
Then, we denote by $\wt{\mc{X}}_N$ the operator
\beq
\wt{\mc{X}}_N[\phi](\xi) \; = \; \phi(\xi) \; - \; \mc{X}_{N}[\phi] 
\label{definition operateur XN}
\enq
and also define:
\beq
\wt{\cal U}_{N}^{-1} = {\cal U}_{N}^{-1}\circ \wt{\mc{X}}_{N},\qquad \wt{\cal W}_{N} = {\cal W}_{N}\circ\wt{\mc{X}}_{N}\;.
\label{tildeopXN}
\enq
\end{defin}

It follows readily from the identity 
\beq
\Int{\R + \i \eps }{} \chi_{11}(\mu)\cdot  \f{1 - \ex{-\i\mu \ov{x}_N} }{ \mu }\cdot\f{ \dd \mu }{ 2\i \pi } \; = \;  \chi_{11;+}(0) 
\qquad \e{with} \qquad  \ov{x}_N \,= \, N^{\a}(b_N-a_N)\;,
\label{ecriture identite integrale pour chi 11}
\enq
that $\wt{\mc{X}}_N[\phi] \in \mf{X}_s(\intff{a_N}{b_N})$ in the sense of  \eqref{ecriture explicite condition d'appartenance a Xs sur sup mes eq}.
The proof of \eqref{ecriture identite integrale pour chi 11} follows from the use of the boundary conditions $\ex{-\i\la N^{\a}(b_N-a_N)} \chi_{11;+}(\la)  =  \chi_{11;-}(\la)$, $\la \in \R$
the fact that $\chi_{11} \in \mc{O}(\Cx \setminus \R)$ and that $\chi_{11}(\la)=\e{O}\big( |\la|^{-\tf{1}{2}} \big)$ at infinity. Likewise, by using the bounds  \eqref{borne sur operateur X} obtained in Corollary \ref{Corolaire bornange N independent forme lineaire X}
it is readily seen that 
\beq
\mc{N}^{(p)}_N\Big[ \mc{K}_{\kappa}\big[ \wt{\mc{X}}_N[\phi] \big]  \Big] \; \leq  \; C \cdot \mc{N}^{(p)}_N\big[ \mc{K}_{\kappa}[\phi]   \big] \;. 
\enq

\begin{prop}
 \label{Poposition DA correlateur a un point}
 
Given any $\kappa>0$, and any $\phi $ satisfying $\mc{K}_{\kappa}[\phi] \in W_{\ell}^{\infty}(\R)$, we have:
\begin{eqnarray}
\big<  \phi			\big>_{  \mc{L}_N^{(\bs{\la})} } & = &   \f{(1-\be) }{ N^{1+\a} }  \cdot 
\Big<  \Dp{1} \wt{\mc{U}}^{-1}_N\big[\phi\big]			\Big>_{\mu_{\e{eq}}^{(N)}}
\; + \; \f{1}{2 N^{2+\a} } 
\Big<   \Xi^{(2)}\Big[\Dp{2} \wt{\mc{U}}^{-1}_{N;1}\Big[  \mc{D}_N\big[ \wt{\mc{U}}^{-1}_N\big[\phi\big] \big] \Big]\Big]  \; 
			\Big>_{ \mu_{\e{eq}}^{(N)}} \nonumber \\
&& +\f{(1-\be)^2 }{2  N^{2(1+\a)}  }
\Big< \Dp{1} \Dp{2} \wt{\mc{U}}^{-1}_{N;1} \wt{\mc{U}}^{-1}_{N;2}
\Big[ \mc{D}_N\big[ \wt{\mc{U}}^{-1}_N\big[\phi\big] \big] \Big] \; 
			\Big>_{ \bigotimes^2 \mu_{\e{eq}}^{(N)}} \nonumber  \\ 
&& + \; \f{(1-\be)^2 }{  N^{2(1+\a)}  }\Big<   \Dp{1} \wt{\mc{U}}^{-1}_N\big[   \Dp{1} \wt{\mc{U}}^{-1}_N[   \phi ]  \big]	\Big>_{\mu_{\e{eq}}^{(N)}}
\; + \; \f{ \de_{N}[\phi,V] }{  N^{2+\a}  } \;.
\end{eqnarray}
The remainder $\de_{N}[\phi,V]$ is bounded as:
\beq
\big| \de_{N}[\phi,V]  \big| \; \leq  \; C \cdot \mf{n}_{\ell}[V] \cdot \mc{N}_N^{(\ell^{\prime})}\big[ \mc{K}_{\kappa}[\phi] \big] \cdot N^{6\a - 1}\,(\ln N)^{\ell^{\prime\prime}} 
\enq
for a constant $C>0$ that does not dependent on $\phi$ nor on the potential $V$, and the integers:
$$
\ell = \max(3m_3 + 5,8m_2 + 18),\qquad \ell^{\prime} = \max(4m_3 + 9,14 m_2 + 37),\qquad \ell^{\prime\prime} = \max(14m_2 + 17,6m_3 + 16)
$$
given in terms of the sequence $(m_n)_n$ introduced in \eqref{ordernorme}. %Finally, we have set $\wt{\mc{U}}_N^{-1}=\mc{U}_N^{-1}\circ $
%and analogously for their extensions as operators on functions in many variables. 

\end{prop}

\Proof  The strategy is to exploit the Schwinger-Dyson equation and get rid of expectation values of functions integrated against the random measure $ L_N^{(\bs{\la})}$. 
This can be done by replacing them approximately by integration against a deterministic measure of a transformed function, up to corrections that we can estimate.

Let $\phi$ be a sufficiently regular function of one variable. Since the signed measure $\mc{L}_N^{(\bs{\la})}$ has zero mass, it follows that $\big<  \phi \big>_{  \mc{L}_N^{(\bs{\la})} } \; = \; \big<  \wt{\mc{X}}_N[\phi] \big>_{  \mc{L}_N^{(\bs{\la})} }$. We can use the Schwinger Dyson equation at level $1$ \eqref{SDlevel1} for the function $\wt{\mc{X}}_N[\phi]$, and apply the sharp bounds of Proposition~\ref{Lemme bornes optimales sur les correlateurs} to derive:
\beq
\Big| \big<  \phi \big>_{  \mc{L}_N^{(\bs{\la})} }  \, - \,  \f{ 1-\be }{  N^{1+\a}  }\,\Big<    \partial_{1} \mc{U}^{-1}_N\big[  \wt{\mc{X}}_N[\phi] \big] \Big>_{\mu_{\e{eq}}^{(N)}} \Big|
\; \leq \; C \cdot \mf{n}_{2m_2+2}[V] \cdot \mc{N}_N^{(2m_2+3)}\big[ \mc{K}_{\kappa}[\phi] \big] \cdot  N^{3\a-2}  (\ln N)^{2m_2+5} \;. 
\label{ecriture borne crue sur correlateur 1 point}
\enq
Above, we have stressed explicitly the composition of the operator $\mc{U}^{-1}_N$ with $\wt{\mc{X}}_N$. 
This bound ensures that 
\beq
\label{susb1}\Big| \f{ 1-\be }{ N^{1+\a} } \big< \partial_{1} \wt{\mc{U}}_{N}^{-1}[\phi] \big>_{ \mc{L}_{N}^{(\bs{\la})} } \; - \; 
\f{ (1-\be)^2 }{ N^{ 2 (1+\a) } } \Big< \partial_{1}\wt{\mc{U}}_{N}^{-1}\big[ \partial_{1}\wt{\mc{U}}_{N}^{-1}[\phi] \big] \Big>_{ \mu_{\e{eq}}^{(N)} }  \Big|
\; \leq \; C^{\prime} \cdot \mf{n}_{4m_2+7}[V] \cdot \mc{N}_N^{(4m_2+9)}\big[ \mc{K}_{\kappa}[\phi] \big] \cdot  N^{4\a-3}  (\ln N)^{6m_2+14}  \;. 
\enq
where we remind that $\wt{\mc{U}}_N^{-1}=\mc{U}_N^{-1}\circ \wt{\mc{X}}_N$. Equation \eqref{susb1} can be used for a substitution of the term proportional to $(1 - \beta)$ in the Schwinger-Dyson equation at level 1 \eqref{SDlevel1}, and we get:
\bem
\Big| \big<  \phi \big>_{  \mc{L}_N^{(\bs{\la})} } \, - \, \f{ 1-\be }{ N^{1+\a} } \big< \partial_{1} \wt{\mc{U}}_{N}^{-1}[\phi] \big>_{ \mu_{\e{eq}}^{(N)} } \; - \; 
\f{ (1-\be)^2 }{ N^{ 2 (1+\a) } } \Big<\partial_{1} \wt{\mc{U}}_{N}^{-1}\big[ \partial_{1} \wt{\mc{U}}_{N}^{-1}[\phi] \big] \Big>_{ \mu_{\e{eq}}^{(N)} }   \\
\; - \; \f{1}{2}  \big< \mc{D}_N\circ \wt{\mc{U}}_{N}^{-1}[\phi] \big>_{  \bigotimes^2  \mc{L}_N^{(\bs{\la})} } \Big|
\; \leq \; C^{\prime} \cdot \mf{n}_{4m_2+7}[V] \cdot \mc{N}_N^{(4m_2+9)}\big[ \mc{K}_{\kappa}[\phi] \big] \cdot  N^{4\a-3}  (\ln N)^{6m_2+14}  \;. 
\label{ecriture bornes plus optimales sur correlateur 1 point}
\end{multline}
In order to gain a better control on the term involving ${\cal D}_{N}$ -- which is a two-point correlator -- we need to study the Schwinger-Dyson equation at level $n = 2$ \eqref{ecriture equation des boucles a n points}. 
Given a sufficiently regular function $ \psi_2$ in two variables, using the sharp bounds of Proposition~\ref{Lemme bornes optimales sur les correlateurs}, we find:
\bem
\bigg|  \big< \psi_2 \big>_{\bigotimes^2 \! \mc{L}^{(\bs{\la})}_N }\; - \;
\f{1}{ N^{2+\a} } \Big< \Xi^{(2)}\Big[\Dp{2}\wt{\mc{U}}_{N;1}^{-1}\big[ \psi_2 \big] \Big] \Big>_{ \mu_{\e{eq}}^{(N)} }  \; - \; 
\f{1-\be }{ N^{1+\a} } \Big< \Big( \partial_{1}\wt{\mc{U}}_{N;1}^{-1}[\psi_2] \Big) \Big>_{ \mu_{\e{eq}}^{(N)} \bigotimes \mc{L}^{(\bs{\la})}_N }  \bigg| \\
\; \leq \; C \cdot  \mf{n}_{2m_3+2}[V] \cdot \mc{N}_N^{(2m_3+3)}\big[ \mc{K}_{\kappa}[\psi_2] \big] \cdot  N^{4\a-3}  (\ln N)^{2m_3+5} \;.  
\label{equation bornes premieres sur valeur moyenne f}
\end{multline}
We apply this estimate to the particular choice:
\beq
\psi_2(\xi_1,\xi_2) \; = \; \mc{D}_{N}\big[ \wt{\mc{U}}_N^{-1}[\phi] \big](\xi_1, \xi_2)\,.
\enq
Thanks to the bound \eqref{normDNUN} on the norm of ${\cal D}_{N}\circ{\cal U}_{N}^{-1}$ and the sub-multiplicativity \eqref{propriete multiplicativite des constantes n ell de V} of the $\mathfrak{n}_{\ell}[V]$'s, we deduce:
\bem
\label{subs2}\bigg|  \big< \psi_2 \big>_{\bigotimes^2 \! \mc{L}^{(\bs{\la})}_N }\; - \;
\f{1}{ N^{2+\a} } \Big< \Xi^{(2)}\Big[\Dp{2}\wt{\mc{U}}_{N;1}^{-1}\big[ \psi_2 \big] \Big] \Big>_{ \mu_{\e{eq}}^{(N)} }  \; - \; 
\f{1-\be }{ N^{1+\a} } \Big< \Big( \partial_{1}\wt{\mc{U}}_{N;1}^{-1}[\psi_2] \Big) \Big>_{ \mu_{\e{eq}}^{(N)} \bigotimes \mc{L}^{(\bs{\la})}_N } \bigg| \\
\leq \, C\cdot\mathfrak{n}_{4m_3 + 7}[V]\cdot {\cal N}_{N}^{(4m_3 + 9)}\big[{\cal K}_{\kappa}[\phi]\big]\cdot N^{5\a - 3}\,(\ln N)^{6m_3 + 16}\,.
\end{multline}
This can be used for a substitution of $\big<\psi_2\big> = \big< {\cal D}_{N}\circ {\cal U}_{N}^{-1} [\phi] \big>$ in the left-hand side of \eqref{ecriture bornes plus optimales sur correlateur 1 point}. 
Before performing this substitution, we still need to control the term in \eqref{subs2} which is proportional to $(1 - \beta)$. This is a one-point correlator for the function:
\beq
\psi_1(\xi) \; = \; \f{1-\be }{ N^{1+\a} }\Int{}{}  \partial_{\eta} \wt{\mc{U}}_{N;1}^{-1}\big[ \psi_2( *  ,\xi) \big] (\eta) \,\dd  \mu_{\e{eq}}^{(N)}(\eta)  \;.
\enq
Applying the one-point estimate \eqref{ecriture borne crue sur correlateur 1 point} to the function $\psi_1$, along with the bounds \eqref{ecriture borne Wp infty sur UN inverse}-\eqref{ecriture borne Wp infty sur op DN} 
for the norms of ${\cal U}_{N}^{-1}$ and ${\cal D}_{N}$, we find:
\beq
\Big| \big<  \psi_1 \big>_{  \mc{L}_N^{(\bs{\la})} }  \, - \,  \f{ 1-\be }{  N^{1+\a}  }\,\Big<    \Dp{1}\wt{\mc{U}}^{-1}_N[ \psi_1 ] \Big>_{\mu_{\e{eq}}^{(N)}} \Big|
\; \leq \; C \cdot \mf{n}_{8m_2+18}[V] \cdot \mc{N}_N^{(8m_2+21)}\big[ \mc{K}_{\kappa}[\phi] \big] \cdot  N^{5\a-3}  (\ln N)^{14m_2+37} \;. 
\enq
This leads to:
\bem
\bigg|  \big< \psi_2 \big>_{\bigotimes^2 \! \mc{L}^{(\bs{\la})}_N }\; - \;
\f{1}{ N^{2+\a} } \Big< \Xi^{(2)}\circ \Dp{2}\wt{\mc{U}}_{N;1}^{-1}[ \psi_2] \Big>_{ \mu_{\e{eq}}^{(N)} }  \; - \; 
\f{(1-\be)^2 }{ N^{2(1+\a) } } 
\Big< \Dp{1}\wt{\mc{U}}_{N;1}^{-1}\Dp{2}\big[\wt{\mc{U}}_{N;2}^{-1}[\psi_2] \big]\Big>_{ \mu_{\e{eq}}^{(N)} \bigotimes \mu_{\e{eq}}^{(N)} }  \bigg|  \\ 
\; \leq \;  C \cdot \mf{n}_{8m_2+18}[V] \cdot \mc{N}_N^{(8m_3+21)}\big[ \mc{K}_{\kappa}[\phi] \big] \cdot  N^{5\a-3}  (\ln N)^{14m_2+37} \;. 
\label{equation bornes premieres sur valeur moyenne f2}
\end{multline}
The result follows by substituting this inequality in \eqref{ecriture bornes plus optimales sur correlateur 1 point}. \qed

%The above estimate along with \eqref{ecriture borne crue sur correlateur 1 point} shows that the one-point correlator involving the function
%
%
%
%\beq
%

%
%
%is bounded as:
%
%
%

%
%
%
%There, we made use of 
%
%
%
%\beq
%
%\mc{N}_N^{(p)}\big[ \mc{K}_{\kappa}[g] \big]\; \leq \; C \cdot \mf{n}_{3p+6}[V] \cdot \mc{N}_N^{(4p+9)}\big[ \mc{K}_{\kappa}[\phi] \big] \cdot  N^{2\a-1}  (\ln N)^{6p+14} \;.
%
%\enq
%
%
%Collecting all the terms in the expression of $\big<\phi\big>_{{\cal L}_{N}^{(\bs{\lambda})}}
%
%
%
 
%%%%%%%%%%%%%%%%%%%%%%%%%%%%%%%%%%%%%%%%%%%%%%%%%%%%%%%%%%%%%%%%%%%%%%%%%%%%%%%%%%%%%%%%%%%%%%%%%%%%%%%%%%%%%%%%%%%%%%%%%%%%%%%%%%%%%%%% 
%%%%%%%%%%%%%%%%%%%%%%%%%%%%%%%%%%%%%%%%%%%%%%%%%%%%%%%%%%%%%%%%%%%%%%%%%%%%%%%%%%%%%%%%%%%%%%%%%%%%%%%%%%%%%%%%%%%%%%%%%%%%%%%%%%%%%%%%

%%%%%%%%%%%%%%%%%%%%%%%%%%%%%%%%%%%%%%%%%%%%%%%%%%%%%%%%%%%%%%%%%%%%%%%%%%%%%%%%%%%%%%%%%%%%%%%%%%%%%%%%%%%%%%%%%%%%%%%%%%%%%%%%%%%%%%%%
%%%%%%%%%%%%%%%%%%%%%%%%%%%%%%%%%%%%%%%%%%%%%%%%%%%%%%%%%%%%%%%%%%%%%%%%%%%%%%%%%%%%%%%%%%%%%%%%%%%%%%%%%%%%%%%%%%%%%%%%%%%%%%%%%%%%%%%%

%%%%%%%%%%%%%%%%%%%%%%%%%%%%%%%%%%%%%%%%%%%%%%%%%%%%%%%%%%%%%%%%%%%%%%%%%%%%%%%%%%%%%%%%%%%%%%%%%%%%%%%%%%%%%%%%%%%%%%%%%%%%%%%%%%%%%%%%
%%%%%%%%%%%%%%%%%%%%%%%%%%%%%%%%%%%%%%%%%%%%%%%%%%%%%%%%%%%%%%%%%%%%%%%%%%%%%%%%%%%%%%%%%%%%%%%%%%%%%%%%%%%%%%%%%%%%%%%%%%%%%%%%%%%%%%%%

%%%%%%%%%%%%%%%%%%%%%%%%%%%%%%%%%%%%%%%%%%%%%%%%%%%%%%%%%%%%%%%%%%%%%%%%%%%%%%%%%%%%%%%%%%%%%%%%%%%%%%%%%%%%%%%%%%%%%%%%%%%%%%%%%%%%%%%%
%%%%%%%%%%%%%%%%%%%%%%%%%%%%%%%%%%%%%%%%%%%%%%%%%%%%%%%%%%%%%%%%%%%%%%%%%%%%%%%%%%%%%%%%%%%%%%%%%%%%%%%%%%%%%%%%%%%%%%%%%%%%%%%%%%%%%%%%

\section{The large-$N$ asymptotic expansion of $\ln Z_N[V]$ up to $\e{o}(1)$}
\label{adfsdgfsg}
We can use the large-$N$ analysis of the Schwinger-Dyson equations to establish the existence of an asymptotic expansion 
up to $\e{o}(1)$ of $\ln Z_{N}[V]$. The coefficients in this asymptotic expansion are single and double integrals whose 
integrand depends on $N$. We will work out the large-$N$ asymptotic expansion of these coefficients in Sections 
\ref{Section asymptotic analysis of single integrals}-\ref{Section asymptotic analysis of double integrals}. 
Prior to writing down this large-$N$ asymptotic expansion, we need to introduce several single and double integrals that will enter in the description of 
the result. We also remind the notation $\wt{\mc{W}}_N = \mc{W}_N\circ\wt{\mc{X}}_N$ where ${\cal W}_{N}$ is the inverse of ${\cal S}_{N}$ (\textit{cf}. \eqref{ecriture eqn int sing de depart}), 
studied in Section~\ref{Sous section construction inverse a l'operateur SN}.
Given $H,G$ sufficiently regular on $\intff{a_N}{b_N}$, we define the one-dimensional integrals: 
\beq
\label{ididi}\mf{I}_{\e{s}}\big[H, G \big]  \; = \; \Int{a_N}{b_N} H(\xi) \cdot  \mc{W}_N[G](\xi) \cdot \dd \xi\;, \qquad
\mf{I}_{\e{s};\be}^{(1)}\big[H, G \big]  \; = \; \Int{a_N}{b_N}  \mc{W}_N[G^{\prime}](\xi)\,\Dp{\xi} 
\Bigg\{ \f{  \wt{\mc{W}}_N[ H ](\xi)  }{  \mc{W}_N[G^{\prime}](\xi)  } \Bigg\}\,\dd \xi 
\enq
and
\beq
\label{ididi2}\mf{I}_{\e{s};\be}^{(2)}\big[H, G \big]  \; = \; \Int{a_N}{b_N}  \mc{W}_N[G^{\prime}](\xi)\,\partial_{\xi}\Bigg\{\f{
\wt{\mc{W}}_N\bigg[  \partial_{1}\Big(\f{  \wt{\mc{W}}_N[ H ]}{ \mc{W}_N[G^{\prime}]}\Big)\bigg]( \xi)}{\mc{W}_N[G^{\prime}](\xi) }\Bigg\}\,\dd \xi  \;. 
\enq
We also define the two-dimensional integrals:
\beq
\mf{I}_{\e{d}}\big[H, G \big]  \; = \; \Int{a_N}{b_N}  \wt{\mc{W}}_N\Bigg[ \Dp{\xi} \Big\{ S\big(N^{\a}(\xi-*)\big)\,
\Big( \f{  \wt{\mc{W}}_N[ H ](\xi)  }{  \mc{W}_N[G^{\prime}](\xi) } \, - \, \f{  \wt{\mc{W}}_N[ H ](*)  }{  \mc{W}_N[G^{\prime}](*)  } \Big) \Big\} \Bigg](\xi)
\,\dd \xi 
\label{ecriture integrale double DA a beta 1}
\enq
and 
\bem
\mf{I}_{\e{d};\be}\big[H, G \big]  \; = \; \f{1}{2} \Int{a_N}{b_N} \dd \xi \dd \eta \,  \mc{W}_N[G^{\prime}](\xi) \cdot   \mc{W}_N[G^{\prime}](\eta)    \\ 
\times  \partial_{\xi}\partial_{\eta}\left(
\f{1}{   \mc{W}_N[G^{\prime}](\xi) \cdot   \mc{W}_N[G^{\prime}](\eta) }
 \wt{\mc{W}}_{N;1}\circ \wt{\mc{W}}_{N;2}\Bigg[ S\big(N^{\a}(*_1-*_2)\big) \cdot 
\bigg\{ \f{  \wt{\mc{W}}_N[ H ](*_1)  }{  \mc{W}_N[G^{\prime}](*_1) } \, - \, \f{  \wt{\mc{W}}_N[ H ](*_2)  }{  \mc{W}_N[G^{\prime}](*_2)  } \bigg\} \Bigg](\xi,\eta)\right) \;. 
\label{definition integrale double N dpdt beta non egal 1}
\end{multline}
Above, $*$ refers to the variables on which the operators act, $*_1$, \textit{viz.} $*_2$, to the first, respectively second, 
running variable on which the product of operators $\mc{W}_{N;1}\cdot \mc{W}_{N;2}$ acts.  The subscript $\beta$ reminds that the terms concerned are absent in the case $\beta = 1$.

\begin{prop}
\label{Theorem DA N dependent de la fct de partition}
Let $V_{G;N}(\la) = g_N\la^2 + t_N \la$ be the unique Gaussian potential  associated with an equilibrium measure 
supported on $\intff{a_N}{b_N}$ as given in Lemma~\ref{Proposition pot Gaussien avec meme support mesure eq} and assume that $0< \a < 1/6$. Then
there exists $\ell \in \mathbb{N}$ such that one has the 
large-$N$ asymptotic expansion
\bem
\ln \bigg( \f{ Z_N[V] }{ Z_N\big[ V_{G;N} \big]  } \bigg) \; = \; -N^{2+\a} \Int{0}{1} \mf{I}_{ \e{s} }\big[\Dp{t}V_t, V^{\prime}_{t} \big] \cdot \dd t
\; - \,  N (1-\be) \Int{0}{1}  \mf{I}^{(1)}_{ \e{s} ;\be }\big[\Dp{t}V_t, V_{t} \big] \cdot \dd t 
\; - \,  \f{1}{2} \Int{0}{1}  \mf{I}_{\e{d}}\big[\Dp{t}V_t, V_{t} \big] \cdot \dd t   \\
\; - \, \f{ (1-\be)^2 }{ N^{\a} } \Int{0}{1} \Big\{  \mf{I}^{(2)}_{ \e{s} ; \be }\big[\Dp{t}V_t, V_{t} \big] \, +  \, 
 \mf{I}_{\e{d};\be}\big[\Dp{t}V_t, V_{t} \big]  \Big\} \cdot \dd t  \; + \; \e{O}\big(N^{6\a - 1}\,(\ln N)^{2\ell}\big)  \; .
\end{multline}
Here, $V_t=(1-t)V_{G;N}+tV$. 
\end{prop}  

\Proof The result  follows from \eqref{equation reliant derivee log fct part et corr derivee pot}.  Indeed, the remarks above  \eqref{ecriture DA up to o1 de fct partition}
allow to identify the equilibrium measures $\mu_{{\rm eq};V_{t}}^{(N)} = (1 - t)\mu_{{\rm eq};V_{G;N}}^{(N)} + t\mu_{{\rm eq};V}^{(N)}$ 
for all $t\in [0,1]$. One can then use
 Proposition \ref{Poposition DA correlateur a un point}  to expand $ \moy{ \Dp{t}V_t }_{ L_N^{(\bs{\la})} }^{V_t} $, 
along with the representation for $\mc{U}_N^{-1}$ on the support of the equilibrium measure which reads
\beq
\wt{\mc{U}}_N^{-1}[H](\xi) \; = \;  \f{ \wt{\mc{W}}_N[ H ]( \xi)  }{  \mc{W}_N[V^{\prime}](\xi) } \;. 
\enq
Taking these data into account, it solely remains to write down explicitly the one and two-dimensional integrals arising in 
Proposition \ref{Poposition DA correlateur a un point}.  
\qed

Note that the factors $\mf{I}^{(2)}_{ \e{s};\beta }\big[\Dp{t}V_t, V_{t} \big] $ and $\mf{I}_{\e{d};\beta}\big[\Dp{t}V_t, V_{t} \big]$
are preceded by the negative power of  $N^{-\a}$. Still, it does not mean that these do not contribute to the leading contribution, \textit{i.e.}
up to $\e{o}(1)$, to the asymptotics of the partition function. Indeed, the presence of derivatives in their associated integrands generates additional powers
of $N^{\a}$.

\chapter{The Riemann--Hilbert approach to the inversion of $\mc{S}_N$}
\label{Section RHP Inversion sing Int Op}

{\bf Abstract}

\textit{
In the present chapter we focus on a class of singular integral equation driven by a one parameter $\ga$-regularisation of the operator $\mc{S}_N$.
More precisely, we introduce the singular integral operator $\mc{S}_{N;\ga}$ }
\beq
\mc{S}_{N;\ga}[\phi](\xi) \; = \; \Fint{a_N}{b_N} S_{\ga}\big(N^{\a}(\xi-\eta)  \big)  \phi(\eta) \cdot  \dd \eta
\qquad where \quad \left\{ \ba{cc} 
S_{\ga}(\xi) \,=\, S(\xi) \cdot \bs{1}_{ \intff{-\ga \ov{x}_N }{ \ga \ov{x}_N } } \vspace{2mm}\\ 
\ov{x}_N \, = \, N^{\a} \cdot (b_N-a_N)  \ea \right. \;\;.
\label{definition operateur regularise S N gamma}
\enq
\textit{This operator is a regularisation of the operator $\mc{S}_N$
in the sense that, formally, $\mc{S}_{N;\infty}=\mc{S}_{N}$. This regularisation enables to set a well 
defined associated Riemann--Hilbert problem, and is such that, once all calculations have been done and the inverse of  $\mc{S}_{N;\ga}$ constructed, 
we can take send $\ga\tend+\infty$ at the level of the obtained answer. It is then not a problem to check that this limiting operator does
indeed provides one with the inverse of $\mc{S}_N$.}

\textit{We start this analysis by, first, recasting the singular integral equation into a form where the variables have been re-scaled. Then, we put the problem of 
inverting the re-scaled operator associated with $\mc{S}_{N;\ga}$ with a vector valued Riemann-Hilbert problem. The resolution of this vector problem demands
the resolution of a $2\times 2$ matrix Riemann--Hilbert problem for an auxiliary matrix $\chi$. We construct the solution
to this problem, for $N$-large enough, in \S~\ref{SousSection RHP Auxiliaire pour chi} and then exhibit some of the overall properties of the solution $\chi$
in \S~\ref{SousSection Proprietes generales de la solution chi}. 
We shall build on these results so as to invert $\mc{S}_{N;\ga}$ and then $\mc{S}_N$ in subsequent sections.  }

\section{A re-writing of the problem}

\subsection{A vector Riemann--Hilbert problem}

In the handlings that will follow, it will appear more convenient to consider a properly rescaled problem. Namely define
\beq
\vp(\xi) \; = \; \phi\big( ( \xi +  N^{\a} a_N )  N^{-\a} \big) \qquad \e{and} \qquad 
h(\xi) \; = \; \f{N^{\a}}{2 \i\pi \be } H\big( (\xi +  N^{\a} a_N )  N^{-\a} \big) \;. 
\label{ecriture changement de variable pour arriver au RHP}
\enq
It is then clear that  solutions to $\mc{S}_{N;\ga}\big[ \phi \big](\xi)=H(\xi)$ are in a one-to-one correspondence with those of 
\beq
\msc{S}_{N;\ga}[\vp](\xi) \; = \; \Fint{0}{\ov{x}_N} S_{\ga}(\xi-\eta) \vp(\eta) \cdot \f{\dd \eta}{2 \i\pi \be } \; \; = \; h(\xi) \; \;. 
\label{ecriture equation sing en variables reduites}
\enq
For any $N$ and $\ga \geq 0$, the operator $\msc{S}_{N;\ga}$ is continuous as an operator 
\beq
\msc{S}_{N;\ga}\; : \; H_s\big( \intff{0}{\ov{x}_N} \big) \quad  \longrightarrow  \quad 
H_s\big( \intff{- \ga \ov{ x }_N }{ \ga  \ov{ x }_N } \big) \; \subseteq \; H_s\big( \R \big) \;. 
\enq
Indeed, this continuity follows readily from the boundedness of the Fourier transform $\mc{F}[S_{\ga}]$ of the 
operator's integral  kernel, \textit{c}.\textit{f}.   Lemma \ref{Lemme calcul TF gN} to come. 

First, we shall start by focusing on spaces with a negative index $s<0$ and going to construct a class of its inverses 
\beq
\msc{S}_{N;\ga}^{-1} \; : \; H_s\big( \intff{- \ga \ov{ x }_N }{ \ga  \ov{ x }_N } \big)
 \quad  \longrightarrow  \quad H_s\big( \intff{0}{\ov{x}_N}\big) \;. 
\enq
What we mean here is that, \textit{per se}, the operator is non-invertible in that, 
as will be inferred from our analysis, for $-k < s < -(k-1) $
\beq
\e{dim}\,\e{ker}\, \msc{S}_{N;\ga} \; = \;  k \;. 
\enq
In fact, the analysis that will follow, provides one with a thorough characterisation of its kernel. 
Furthermore, when restricting the operator $\msc{S}_{N;\ga}$ to more regular spaces like 
$H_s\big( \intff{0}{\ov{x}_N}\big)$ with $s>0$, we get that the image $\msc{S}_{N;\ga}\big[ H_s\big( \intff{0}{\ov{x}_N}\big)  \big]$ 
is a closed, explicitly characterisable subspace of $H_s\big( \intff{- \ga \ov{ x }_N }{ (\ga+1)  \ov{ x }_N } \big)$, 
and that the operator becomes continuously invertible on it.

In the following, we shall invert the operator $\msc{S}_{N;\ga}$ by means of the resolution of an auxiliary $2\times 2$ Riemann--Hilbert problem
and then by implementing a Wiener--Hopf factorisation. The analysis is inspired by the paper of
Novokshenov \cite{NovokshenovSingIntEqnsIntervalGeneral}
where a correspondence has been built between singular integral equations on a finite segment subordinate to integral kernels depending on the difference on the one hand and 
Riemann--Hilbert problems on the other one. The large parameter analysis is, however, new. 

In fact the very setting of the Riemann--Hilbert problem-based analysis enables one to naturally construct the pseudo-inverse of 
$\msc{S}_{N;\ga}$ - \textit{i.e.} modulo elements of $\e{ker}\big[ \msc{S}_{N;\ga} \big]$ -- when the operator is understood 
to act on $H_{s}$ spaces with \textit{negative} index $s<0$. The inversion of $\msc{S}_{N;\ga} $ understood as an operator on 
$H_{s}$ spaces with \textit{positive} index $s\geq 0 $ goes, however, beyond, the "crude" construction issuing from the 
Riemann--Hilbert problem-based analysis. It is, in particular, based on an explicit characterisation, through linear constraints, of 
the image space $\msc{S}_{N;\ga}\big[H_s(\intff{0}{\ov{x}_N}])\big] $, $s\geq 0$. For $0<s<\tf{1}{2}$, which is the case of interest for us,
we show that $\msc{S}_{N;\ga}\big[H_s(\intff{0}{\ov{x}_N})\big] $ coincides with $\mf{X}_s(\intff{-\ga \ov{x}_N }{ (\ga+1)\ov{x}_N })$.

\begin{lemme}
\label{Proposition corresp operateur et WH factorisation}
Let $h \in H_s\big(\intff{0}{\ov{x}_N} \big)$, $s<0$. For any solution $\varphi \in H_s\big(\intff{0}{\ov{x}_N} \big)$ of \eqref{ecriture equation sing en variables reduites}, 
there exists a two-dimensional vector function $\Phi \in \mc{O}(\Cx\setminus\R)$ such that  $\vp \; = \; \mc{F}^{-1}[(\Phi_1)_+]$ and $\Phi$ is a solution to the boundary value problem:
\begin{itemize}
\item $\big( \Phi_a \big)_{\pm} \in \mc{F}\big[ H_s\big( \R^{\pm } \big) \big]$ for $a \in \{1,2\}$, and there exists $C>0$ such that:
\beq
\forall \mu > 0,\,\quad\forall a \in \{1,2\},\qquad \Int{ \R }{} \big| \Phi_a (\la \pm \i \mu) \big|^2 \cdot  \big(  1 + |\la | + |\mu | \big)^{2s} \cdot \dd \la \; < \; C\;.
\label{ecriture condition croissance composantes Phi}
\enq
\item We have the jump equation for $\Phi_+(\la) \; = \; G_{\chi}(\la)  \cdot \Phi_-(\la) \; + \; \bs{H}(\la)$ for $\la \in \R$, with: 
\beq
G_{\chi}(\la) \; = \; \left( \ba{cc}   \ex{ \i \la \ov{x}_N }  & 0  \\  
							\frac{ 1 }{2\i\pi \be}	\cdot \mc{F}\big[ S_{\ga} \big](\la)    &  -\ex{- \i \la \ov{x}_N }  \ea \right) \qquad and
\qquad \bs{H}(\la) \; = \; \left( \ba{c}   0 \\  - \ex{- \i \la \ov{x}_N} \mc{F}\big[ h_{\mf{e}}  \big](\la)  \ea \right)		 \;. 							 
\label{definition matrice de saut G chi}
\enq
\end{itemize}
Conversely, for any solution $\Phi \in \mc{O}(\Cx\setminus\R)$ of the above boundary value problem, $\vp \; = \; \mc{F}^{-1}\big[ \; \big( \Phi_1\big)_+ \big]$ is a solution of \eqref{ecriture equation sing en variables reduites}.

%Reciprocally, any solution $\Phi$ to the above boundary value problem gives rise to a solution 
%$\vp \in H_s(\intff{0}{\ov{x}_N})$ to $\msc{S}_{N;\ga}[\vp](\xi) = h(\xi)$  by means of the formula
%
%
%
%\beq
%
%\vp \; = \; \mc{F}^{-1}\big[ \; \big( \Phi_1\big)_+ \big] \;. 
%
%\enq
%
%
%
\end{lemme}

We do remind that $\pm$ denotes the upper/lower boundary values on $\R$ with the latter being oriented from $-\infty$ to $+\infty$ ; $ h_{\mf{e}} $ denotes any extension of $h$ to $H_s(\R)$ ; $\mc{F}\big[ S_{\ga} \big](\la)$ refers to the Fourier transform of the principal value distribution induced by $S_{\ga}$:
\beq
\mc{F}\big[ S_{\ga} \big](\la) \; = \; \Fint{-\ga \ov{x}_N}{ \ga \ov{x}_N} S(\xi)\,\ex{ \i \la \xi}\dd \xi  \;. 
\label{ecriture TF distributionnelle}
\enq

\Proof  Assume that one is given a solution $\vp$ in $H_s\big(\intff{0}{\ov{x}_N} \big)$ to 
\eqref{ecriture equation sing en variables reduites}. Then, let $\psi_L, \psi_R$
be two functions such that 
\beq
\e{supp}(\psi_R) \; = \; \intfo{\ov{x}_N}{+\infty} \qquad , \qquad
\e{supp}(\psi_L) \; = \; \intof{-\infty}{0}
\enq
and
\beq
\Fint{0}{ \ov{x}_N } S_{\ga}(\xi-\eta) \vp(\eta) \cdot  \f{ \dd \eta }{ 2\i \pi \be } \; \; - \;  h_{\mf{e}}(\xi) \; = \; 
\psi_L(\xi) \; + \; \psi_R(\xi) \; . 
\label{ecriture extension eqn integrale sing a analyser}
\enq
Then, by going to the Fourier space, we get:
\beq
\frac{ 1 }{2\i\pi \be}	\cdot \mc{F}\big[ S_{\ga} \big](\la) \cdot  \mc{F}[\vp](\la) \; - \; \mc{F}[h_{\mf{e}}](\la) \; = \; \mc{F}[\psi_L](\la) \; + \; \mc{F}[\psi_R](\la) \;. 
\enq
By Lemma \ref{Lemme calcul TF gN} that will be proved below, $\mc{F}\big[ S_{\ga} \big]  \in L^{\infty}( \R)$. Hence $\psi_R \in H_s(\R^{+})$ whereas $\psi_L \in H_s(\R^{-})$. 
Then, we introduce the vectors
\beq
\bs{F}_{\ua}(\la) \; = \; \left( \ba{c}  \mc{F}[ \vp](\la)  \\ 
									\ex{- \i \la \ov{x}_N } \mc{F}[ \psi_R ](\la )  \ea \right)
\qquad \e{and} \qquad 
\bs{F}_{\da} (\la) \; = \; \left( \ba{c}   \mc{F}[ \vp_{\ov{x}_N} ](\la)   \\ 
									\mc{F}[ \psi_L ](\la ) \ea \right)							
\enq
where we agree upon $\vp_{\ov{x}_N}(\xi)=\vp(\xi+\ov{x}_N)$. Since $\big[\bs{F}_{\ua}\big]_{a} \in \mc{F}\big[ H_s(\R^{+})\big]$, respectively 
$\big[\bs{F}_{\da}\big]_{a} \in \mc{F}\big[ H_s(\R^{-})\big]$, 
it is readily seen that 
\beq
\wt{ \bs{F} }_{\ua;a}(\la) \; = \; \big( 1-  \i \la  \big)^{s}\cdot \big[ \bs{F}_{\ua}\big]_{a}(\la) \qquad \e{resp}. \qquad
 \wt{ \bs{F} }_{\da;a}(\la) \; = \; \big( 1+ \i \la  \big)^{s}\cdot \big[ \bs{F}_{\da}\big]_{a}(\la)
\enq
defines a holomorphic function on $\mathbb{H}^+$, respectively $\mathbb{H}^-$, with $L^2(\R)$ $+$, respectively $-$, boundary values on $\R$. 
The Paley-Wiener Theorem~\ref{PaleyWie} then shows the existence of $C>0$ such that:
\beq
\forall \mu > 0,\quad \forall a \in \{1,2\},\quad \Int{ \R }{} \big| \big[ \bs{F}_{\ua/\da} \big]_{a} (\la \pm \i\mu) \big|^2 \cdot \big(  1 + |\la | + |\mu | \big)^{2s} \cdot \dd \la \; < \; C \;.
\enq
In other words the function:
\beq
\Phi \; = \;  \bs{F}_{\ua} \cdot \bs{1}_{\mathbb{H}^+}  \; + \; \bs{F}_{\da} \cdot \bs{1}_{\mathbb{H}^-}
\label{ecriture representation vecteur Phi}
\enq
solves the vector valued Riemann--Hilbert problem.

Reciprocally, suppose that one is given a solution $\Phi$ to the vector-valued Riemann--Hilbert problem in 
question. Then, set $\vp \; = \; \mc{F}^{-1}\big[ \big(\Phi_{1}\big)_+ \big]$.  We clearly have $\vp \in H_s(\R^+)$, but we now show that the support of $\vp$ is actually included in $[0,\ov{x}_N]$.
Let $(\cdot, \cdot)$ be the canonical scalar product on $L^2(\R,\Cx)$. If $\rho_{R}$ is a $\msc{C}^{\infty}$ function with compact support included in $]\ov{x}_N,+\infty[$, we have:
\beq
2\pi (\rho_{R},\vp) = \big({\cal F}[\rho_{R}]\,,\,{\cal F}[\vp]\big) = (e^{-{\rm i}\ov{x}_N *}\,{\cal F}[\rho_{R}](1 - {\rm i}*)^{-s}\,,\,(1 + {\rm i}*)^{s}(\Phi_{1})_{-}\big)\;,
\enq
where $*$ denotes the running variable. But this is zero since $(1 + {\rm i}*)^{s}(\Phi_{1})_{-} \in {\cal F}[L^2(\R^-)]$, whereas, by the Paley-Wiener Theorem~\ref{PaleyWie}, 
$\ex{ -{\rm i}\ov{x}_{N}* }{\cal F}[\rho_{R}](1 - {\rm i}*)^{-s} \in \mc{F}[L^2(\R^-)]$.
Finally, the fact that $\vp \in H_s(\intff{0}{\ov{x}_N})$ satisfies \eqref{ecriture equation sing en variables reduites} follows from taking the Fourier transform of the second line of 
the jump equation \eqref{definition matrice de saut G chi} for $\Phi$. \qed

\vspace{2mm}

For further handlings, it is useful to characterise the distributional Fourier transform $\mc{F}[S_{\ga}]$ slightly better.

\begin{lemme}
\label{Lemme calcul TF gN}
The distributional Fourier transform $\mc{F}[S_{\ga}](\la)$ defined by \eqref{ecriture TF distributionnelle} admits the representation
\beq
\frac{\mc{F}[S_{\ga}](\la)}{2{\rm i}\pi\beta} \; = \; R(\la) \;  +  \; \Big( \ex{ \i\la \ga \ov{x}_N} \; + \; \ex{-\i\la \ga \ov{x}_N} \Big)\,\f{ \kappa_N }{ \la }
\; + \; r_N(\la) \qquad where \qquad 
\kappa_N = - \sul{p=1}{2} \f{\om_p}{2} \cotanh[\pi \om_p \ga \ov{x}_N]
\label{ecriture forme precise TF gN}
\enq
\beq
\label{defrtr}R(\la) \; = \; \f{ \sinh \bigg[ \f{ \la(\omega_1 + \omega_2) }{ 2\omega_1\omega_2} \bigg] }{  2 \sinh \bigg[ \f{\la}{2\omega_1} \bigg]  \sinh \bigg[ \f{\la}{2\omega_2} \bigg] }\;,
\enq
and
\beq
r_N(\la) \; = \; \sul{p=1}{2} \f{(\pi \om_p)^2 }{\i\la( 1 - \ex{-\la/\om_p})} \Int{ 0 }{\i/\om_p}
 \bigg\{\f{ \ex{ -\i \la \ga \ov{x}_N } }{   \sinh^2[\pi \om_p (\xi-\ga \ov{x}_N) ] } - \f{ \ex{ \i \la \ga \ov{x}_N} }{   \sinh^2[\pi \om_p (\xi+\ga \ov{x}_N) ] }\bigg\} \cdot \f{\ex{ \i \la\xi}\,\dd \xi }{ 2\i\pi } \;. 
\label{definition fonction rN}
\enq
Besides, for ${\rm Im}\,\la= \eps >0$ small enough, there exists $C_{\epsilon} > 0$ independent of $N$ such that, uniformly in ${\rm Re}\,\la \in \R$:
\beq
|r_N(\la)| \; \leq \; C_{\eps}\,|\la|^{-2}\cdot 
\exp\big\{- \ga \ov{x}_N(\, 2 \pi  \min[\omega_1,\omega_2] \, - \,  \eps ) \big\} \;. 
\enq
\end{lemme}

\Proof One has that 
\begin{eqnarray}
\frac{\mc{F}[S_{\ga}](\la)}{2{\rm i}\pi\beta}\; & = & \; \f{1}{2} \sul{p=1}{2}\; \lim_{t \tend 0^+} \;
\sum_{\epsilon \in \{\pm 1\}}  \; \Int{-\gamma\ov{x}_N}{\ov{x}_N}  \pi\omega_{p}\,\cotanh[\pi \om_p (\xi + \i \eps t)]\cdot\frac{e^{\i\la\xi}\,\dd \xi}{2{\rm i}\pi} \\
& = & \frac{1}{2} \sum_{\substack{p \in \{1,2\} \\ \epsilon \in \{\pm 1\}}}\; \frac{\pi \omega_p}{1 - e^{-\la/\om_p}} \;\lim_{t \rightarrow 0^+} \;
\Int{\Gamma_{p}}{} \cotanh\big[\pi\omega_{p}(\xi + \i\epsilon t)\big]\cdot\frac{e^{{\rm i}\la \xi}\,\dd\xi}{2{\rm i}\pi}\;, \nonumber
\end{eqnarray}
where $\Gamma_{p}=\intff{-\gamma\ov{x}_N}{\gamma\ov{x}_N} \cup \intff{\gamma \ov{x}_{N} + {\rm i}/\om_p}{-\gamma\ov{x}_N + {\rm i}/\om_p}$, where the second interval is endowed with an opposite orientation. 
It then remains to add the counter-term:
\bem
r_N(\la) =  \sul{p=1}{2} \f{\pi\om_p}{ 1 - \ex{-\la/\om_p}}  \Int{0}{\f{\i}{\om_p}}
 \bigg\{ \ex{-\i\la \ga \ov{x}_N} \Big( \cotanh[\pi \om_p \ga \ov{x}_N ] +  \cotanh[\pi \om_p (\xi-\ga \ov{x}_N) ] \Big)  \\
\; + \;  \ex{ \i \la \ga \ov{x}_N} \Big( \cotanh[\pi \om_p \ga \ov{x}_N ] -  \cotanh[\pi \om_p (\xi+\ga \ov{x}_N) ] \Big)  
\bigg\}\cdot\f{ \ex{\i \la\xi}\,\dd \xi }{ 2 \i \pi}\; . 
\end{multline}
to form a closed contour $\wt{\Gamma}_{p}$. Upon integrating by parts, we find the expression \eqref{definition fonction rN} for $r_N(\la)$. Then, we pick up the residues surrounded by $\wt{\Gamma}_{p}$, and we also write aside the term behaving as $O(1/\la)$ when $\la \rightarrow \infty$. This leads to the appearence of $\kappa_N$ in \eqref{ecriture forme precise TF gN}. The bounds on the line $|{\rm Im}\,\la|=\eps>0$, with $\eps$ small enough are then obtained through straightforward
majorations.  \qed

The resolution of the vector Riemann--Hilbert problem for $\Phi$ can be done with the help of
a matrix Wiener-Hopf factorization. In order to apply this method, we first need to 
obtain a $\pm$-factorization of the matrix $G_{\chi}$ given by \eqref{definition matrice de saut G chi}. 
This leads to an $2\times 2$ matrix Riemann--Hilbert problem that we formulate and solve, for $N$
sufficiently large, in the next subsections.

\subsection{A scalar Riemann--Hilbert problem}
\label{sniung}
In order to state the main result regarding to the auxiliary $2 \times 2$ matrix Riemann--Hilbert problem, 
we first need to introduce some objects. To start with, we introduce a factorization of $R(\la)$ that separates contributions from zeroes and poles between the lower and upper half-planes $\la \in \mathbb{H}^{\pm}$. In other words, we consider the solution $\ups$
to the following scalar Riemann--Hilbert problem, depending on $\epsilon > 0$ small enough and given once for all:
\begin{itemize}
\item $\ups \in \mc{O}\big( \Cx \setminus \{\R + \i \eps \} \big) $ and has continuous $\pm$-boundary values on 
$\R+  \i \eps$ ;
\item $\ups(\la) = \left\{ 
\ba{cc}  \big(- \i \la \big)^{\f{1}{2}} \cdot   \big(1 \; + \; \e{O}\big(\la^{-1}\big) \big)   & 
			 \,\,\mathrm{if}\,\,{\rm Im}\,\la > \eps  \\
 - \i \big( \i \la \big)^{\f{1}{2}} \cdot  \big(1 \; + \; \e{O}\big(\la^{-1}\big) \big) &
				\,\,\mathrm{if}\,\,{\rm Im}\,\la < \eps  \ea\right. $ 
  when $\la \tend \infty$ non-tangentially to $\R+\i\eps$ ;
\item $ \ups_+(\la) \cdot R(\la) \; = \;   \ups_-(\la) $ \quad for \quad  $\la \in \R + \i \eps$ \;. 
\end{itemize}
This problem admits a unique solution given by 
\beq
\ups(\la) \; = \; \left\{  \ba{cc}   R_{\ua}^{-1}(\la)   &  \,\,\mathrm{if}\,\, {\rm Im}\,\la>\eps  \\ 
							  R_{\da}(\la)   &   \,\,\mathrm{if}\,\,{\rm Im}\,\la < \eps   \ea \right.   
\label{definition fonction alpha}
\enq
where
\beq
R_{\ua}(\la) =   \f{ \i }{\la} \cdot \sqrt{\om_1+\om_2} \cdot 
\bigg( \f{ \om_2 }{  \om_1+\om_2 } \bigg)^{ \f{\i \la }{2 \pi \om_1}  }  
\hspace{-3mm} \cdot \;  \bigg( \f{ \om_1 }{  \om_1+\om_2 } \bigg)^{ \f{\i \la }{2 \pi \om_2} }  
\hspace{-3mm}  \cdot\;  
\f{ \pl{p=1}{2}\Ga\bigg(1-  \f{ \i\la}{2\pi\om_p} \bigg) }{ \Ga\bigg( 1- \f{ \i\la(\om_1 + \om_2) }{2\pi\om_1\om_2} \bigg) }
  \label{ecriture explicite R plus} \\
\enq
and
\beq
R_{\da}(\la) =   \f{ \la }{2\pi \sqrt{\om_1+\om_2} }  \cdot  
 \bigg( \f{ \om_2 }{ \om_1+\om_2 } \bigg)^{ -  \f{\i \la }{2 \pi \om_1} } 
\hspace{-3mm} \cdot \; \bigg( \f{ \om_1 }{  \om_1+\om_2 } \bigg)^{ -\f{\i \la }{2 \pi \om_2} }  
\cdot\f{ \pl{p=1}{2}\Ga\bigg( \f{\i\la}{ 2\pi\om_p } \bigg) }
{ \Ga\bigg( \f{ \i\la(\om_1 + \om_2) }{ 2\pi\om_1\om_2 } \bigg) } \;.
\label{ecriture explicite R moins}
\enq
Note that 
\beq
\label{427}R_{\da}(0) \; = \; - \i  \sqrt{\om_1+\om_2}  \qquad \e{and} \qquad
\Big( \la R_{\ua}(\la) \Big)_{\mid \la=0} \; = \; \i  \sqrt{\om_1+\om_2} \;. 
\enq
Also, $R_{\ua}$ and $R_{\da}$ satisfy to the relations
\beq
R_{\ua}(-\la) \; = \; \la^{-1} \cdot R_{\da}(\la) \qquad \e{and}   \qquad  \Big( R_{\ua}(\la^*) \Big)^{*} \; = \; \la^{-1} \cdot R_{\da}(\la) \;. 
\label{ecriture identite conjugaison R plus et R moins}
\enq
Furthermore, $R_{\ua/\da}$ exhibit the asymptotic behaviour 
\beqa
R_{\ua}(\la) &  =  & \big( - \i  \la \big)^{-\f{1}{2}} \cdot  \Big( 1\; + \; \e{O}\big( \la^{-1} \big) \Big) \qquad \e{for} \quad 
\la \underset{  \la \in  \mathbb{H}^+   }{ \longrightarrow }  \infty   \\
\label{427bis} R_{\da}(\la) & = &   - \i   \big( \i \la \big)^{\f{1}{2}}  \cdot \Big( 1\; + \; \e{O}\big( \la^{-1} \big) \Big) \qquad \e{for} \quad 
\la \underset{  \la \in  \mathbb{H}^-   }{ \longrightarrow }  \infty 
\eeqa
as it should be. The notation $\ua$ and $\da$ indicates the direction in the complex plane where  $R_{\ua/\da}$ have no pole nor zeroes. 

\subsection*{Preliminary definitions} 
We need a few other definition before describing the solution to the factorisation problem for $G_{\chi}$. Let:
\beq
\label{h}\mc{R}_{\ua}(\la) \; = \; \left( \ba{cc}  0   &   -1  \\ 
									1    &    -R(\la) \ex{  \i \la \ov{x}_N }     \ea \right) \qquad \e{and} \qquad 
\mc{R}_{\da}(\la) \; = \; \left( \ba{cc}  -1   &  R(\la) \ex{- \i \la \ov{x}_N }  \\ 
									0    &   1    \ea \right)\;,
\enq
as well as their "asymptotic" versions:
\beq
\label{h2}\mc{R}_{\ua}^{(\infty)} \; = \; \left( \ba{cc}  0   &   -1  \\ 
									1    &   0    \ea \right) \qquad \e{and} \qquad 
\mc{R}_{\da}^{(\infty)} \; = \; \left( \ba{cc}  -1   &  0  \\ 
									0    &   1    \ea \right)  \;. 
\enq
We also need to introduce 
\beq
\label{h3}M_{\ua}(\la)  \; = \; \left( \ba{cc}  1   &   0  \\ 
									 -\f{ 1- R^2(\la) }{ \ups^2(\la) \cdot R(\la) } \ex{ \i \la \ov{x}_N }  & 1    \ea \right) \qquad \e{and} \qquad 
M_{\da}(\la) \; = \; \left( \ba{cc}  1   &  \ups^2(\la) \cdot \f{ 1- R^2(\la) }{  R(\la) }  \ex{- \i \la \ov{x}_N }  \\ 
									0    &   1    \ea \right)\;,
\enq
where $\ups$ is given by \eqref{definition fonction alpha}, and:
\beq
\label{PRla} 
P_R(\la) \; = \; I_2 \; + \; \f{ \th_R }{ \la} \Pi^{-1}(0)  \sg^- \Pi(0) 
\qquad \e{and} \qquad 
\left\{ \ba{c}  P_{L;\ua} (\la) \; = \; I_2 \; + \; \kappa_N \,\la^{-1}\,\ex{ \i (\ga-1)\la \ov{x}_N }\cdot  \sg^-  \vspace{2mm} \\
			P_{L;\da} (\la) \; = \; I_2 \; + \; \kappa_N\,\la^{-1}\,\ex{ - \i (\ga-1)\la \ov{x}_N }\cdot  \sg^-  \ea \right. \;\;,
\enq
in which $\Pi(0)$ is a constant matrix that will coincide later with  the value at $0$ of the matrix function $\Pi$, \textit{cf}.  \eqref{ecriture rep int matrice Pi}.

\beq
\th_R \; = \; \f{1}{\ups^2(0)}\,\frac{\kappa_N}{ 1 + \kappa_N/(\om_1+\om_2) } \;. 
\label{definition theta R}
\enq

\section{The auxiliary $2 \times 2$ matrix Riemann--Hilbert problem for $\chi$}

\subsection{Formulation and main result}

\label{SousSection RHP chi initial}

The factorisation problem for the jump matrix $G_{\chi}$ corresponds to solving the $2 \times 2$ matrix Riemann--Hilbert problem given below. 
This problem is solvable for $N$ large enough. 

\begin{prop}
\label{Theorem ecriture forme asymptotique matrice chi}

There exists $N_0$ such that, for any $N \geq N_0$, the given below $2 \times 2$ Riemann--Hilbert problem has a unique solution. 
This solution is as given in Fig. \ref{Figure definition sectionnelle de la matrice chi}

\begin{itemize}

\item the $2\times 2$ matrix function $\chi\in {\cal O}(\Cx\setminus\R)$ has continuous $\pm$-boundary values on  $\R$;
\item $\chi(\la) = \left\{ 
\ba{cc} P_{L;\ua}(\la) \cdot  \left( \ba{cc} - \e{sgn}\big( {\rm Re}\,\la \big) \cdot \ex{ \i\la \ov{x}_N }  & 1 \\
										-1 &  0     \ea \right)  
		      \cdot \big(-\i \la \big)^{-\f{\sg_3}{2}} \cdot 
  \Big(I_{2}  \; + \; \f{\chi_1}{\la} \; + \; \e{O}\big(\la^{-2}\big) \Big)   & 
			 \la \in \mathbb{H}^+ \vspace{3mm} \\
P_{L;\da}(\la) \cdot  \left( \ba{cc} -1  & \e{sgn}\big( {\rm Re}\,\la \big) \cdot \ex{- \i\la \ov{x}_N }   \\
										0  & 1    \ea \right)
			  \cdot \big( \i \la \big)^{-\f{\sg_3}{2}} \ex{ \i\f{\pi}{2}\sg_3} \cdot 
		    \Big(I_{2} \; + \; \f{\chi_1}{\la} \; + \; \e{O}\big(\la^{-2}\big) \Big) &
				\la \in \mathbb{H}^-  \ea\right. $ 

				for some constant matrix $\chi_{1}$, when $\la \tend \infty$ non-tangentially to $\R$ ;
\item $\chi_+(\la) \; = \;   G_{\chi}(\la) \cdot \chi_-(\la) $ \quad for \quad  $\la \in \R$ \;. 
\end{itemize}
Furthermore, the unique solution to the above Riemann--Hilbert problem satisfies $\det \chi(\la) = \e{sgn}\big( \e{Im}(\la) \big)$ for any $\la \in \Cx\setminus\R$.
\end{prop}
The existence of a solution $\chi$ will be established in \S~\ref{SousSectionRHP for Psi}, by a set of transformations:
\beq
\chi \rightsquigarrow \Psi \rightsquigarrow \Pi
\enq
which maps the initial RHP for $\chi$, to a RHP for $\Pi$ whose jump matrices are uniformly close to the identity when $N$ is large, and thus solvable by perturbative arguments 
\cite{BealsCoifmanScatteringInFirstOrderSystemsEquivalenceRHPSingIntEqnMention}. 
The structure of $\chi$ in terms of the solution $\Pi$ is summarized in Figure~\ref{Figure definition sectionnelle de la matrice chi}. The uniqueness of $\chi$ follows from standard arguments, 
see \textit{e}.\textit{g}. \cite{DeiftOrthPlyAndRandomMatrixRHP}, that we now reproduce.

\Proof (of uniqueness) 

Define, for $\la \in \Cx\setminus \R$, 
\beq
d(\la) \; = \; 
\det[\chi(\la)] \bs{1}_{\mathbb{H}^+}(\la) \; - \; \det[\chi(\la)] \bs{1}_{\mathbb{H}^-}(\la) \; .
\enq
Since $\chi$ has continuous $\pm$-boundary on $\R$, it follows that $d \in \mc{O}(\Cx\setminus \R)$ has continuous $\pm$ boundary values 
on $\R$ as well. Furthermore these satisfy $d_+(\la) =  d_-(\la)$. Finally, $d$ admits the asymptotic behaviour 
$d(\la) \, = \,  1 + \e{O}\big( \la^{-1}\big)$.
 $d$ can thus be continued to an entire function that is bounded at infinity. Hence, by Liouville theorem, $d \equiv 1$. 
Let $\chi_1, \chi_2$ be two solutions to the Riemann--Hilbert problem for  $\chi$. 
Since $\chi_2$ can be analytically inverted, it follows that $\wt{\chi} \; = \; \chi_2^{-1} \cdot \chi_1  $
solves the Riemann--Hilbert problem:  
\begin{itemize}
\item $\wt{\chi} \in \mc{O}(\Cx\setminus \R )$ and has continuous $\pm$-boundary values on 
$\R$;
\item $\wt{\chi}(\la) = I_{2} \; + \; \e{O}\big(\la^{-1}\big) $ \quad when $\la \tend \infty$ non-tangentially to $\R$;
\item $ \wt{\chi}_+(\la) \; = \;   \wt{\chi}_-(\la) $ \quad for \quad  $\la \in \R$ \;. 
\end{itemize}
 Thus, by analytic continuation through $\R$ and Liouville theorem $\wt{\chi}=I_2$, hence ensuring the uniqueness
of solutions.  \qed

%%%%%%%%%%%%%%%%%%%%%%%%%%%%%%%%%%%%%%%%%%%%%%%%%%%%%%%%%%%%%%%%%%%%%%%%%%%%%%%%%%%%%%%%%%%%%%%%%%%%%%%%%%%%%%%%%%%%%%%%%%%%%%%%%%%%%%%%%%%%%%%%%%%%%%%%%%%%%%%%%%%%
%%%%%%%%%%%%%%%%%%%%%%%%%%%%%%%%%%%%%%%%%%%%%%%%%%%%%%%%%%%%%%%%%%%%%%%%%%%%%%%%%%%%%%%%%%%%%%%%%%%%%%%%%%%%%%%%%%%%%%%%%%%%%%%%%%%%%%%%%%%%%%%%%%%%%%%%%%%%%%%%%%%%

%%%%%%%%%%%%%%%%%%%%%%%%%%%%%%%%%%%%%%%%%%%%%%%%%%%%%%%%%%%%%%%%%%%%%%%%%%%%%%%%%%%%%%%%%%%%%%%%%%%%%%%%%%%%%%%%%%%%%%%%%%%%%%%%%%%%%%%%%%%%%%%%%%%%%%%%%%%%%%%%%%%%
%%%%%%%%%%%%%%%%%%%%%%%%%%%%%%%%%%%%%%%%%%%%%%%%%%%%%%%%%%%%%%%%%%%%%%%%%%%%%%%%%%%%%%%%%%%%%%%%%%%%%%%%%%%%%%%%%%%%%%%%%%%%%%%%%%%%%%%%%%%%%%%%%%%%%%%%%%%%%%%%%%%%

%%%%%%%%%%%%%%%%%%%%%%%%%%%%%%%%%%%%%%%%%%%%%%%%%%%%%%%%%%%%%%%%%%%%%%%%%%%%%%%%%%%%%%%%%%%%%%%%%%%%%%%%%%%%%%%%%%%%%%%%%%%%%%%%%%%%%%%%%%%%%%%%%%%%%%%%%%%%%%%%%%%%
%%%%%%%%%%%%%%%%%%%%%%%%%%%%%%%%%%%%%%%%%%%%%%%%%%%%%%%%%%%%%%%%%%%%%%%%%%%%%%%%%%%%%%%%%%%%%%%%%%%%%%%%%%%%%%%%%%%%%%%%%%%%%%%%%%%%%%%%%%%%%%%%%%%%%%%%%%%%%%%%%%%%

%%%%%%%%%%%%%%%%%%%%%%%%%%%%%%%%%%%%%%%%%%%%%%%%%%%%%%%%%%%%%%%%%%%%%%%%%%%%%%%%%%%%%%%%%%%%%%%%%%%%%%%%%%%%%%%%%%%%%%%%%%%%%%%%%%%%%%%%%%%%%%%%%%%%%%%%%%%%%%%%%%%%
%%%%%%%%%%%%%%%%%%%%%%%%%%%%%%%%%%%%%%%%%%%%%%%%%%%%%%%%%%%%%%%%%%%%%%%%%%%%%%%%%%%%%%%%%%%%%%%%%%%%%%%%%%%%%%%%%%%%%%%%%%%%%%%%%%%%%%%%%%%%%%%%%%%%%%%%%%%%%%%%%%%%

\subsection{Transformation $\chi \rightsquigarrow \Psi \rightsquigarrow \Pi$ and solvability of the Riemann--Hilbert problem}
\label{SousSectionRHP for Psi}
\label{SousSection RHP Auxiliaire pour chi}

We construct a piecewise analytic matrix $\Psi$ out of the matrix $\chi$
according to Figure \ref{contour pour le RHP de Phi}. It is readily checked that the Riemann--Hilbert problem for $\chi$ is equivalent to the following Riemann--Hilbert problem for $ \Psi $:

\begin{itemize}

\item $\Psi \in \mc{O}(\Cx^{*}\setminus \Sg_{ \Psi } )$ and has continuous boundary values on 
$\Sg_{\Psi}$ ;
\item The matrix $\,\left( \ba{cc} -1   &  0 \\ 
				-  \kappa_N\,\la^{-1}  & 1 + \kappa_N/(\om_1+\om_2) \ea \right) 
				\cdot \big[\ups(0) \big]^{-\sg_3} \cdot \Psi(\la)$ has a limit when $\la \rightarrow 0$ ;
\item $\Psi(\la) = I_{2} \; + \; \e{O}\big(\la^{-1}\big) $ when $\la \tend \infty$ non-tangentially to $\Sg_{\Psi}$ ;
\item $\Psi_+(\la) \; = \;   G_{\Psi}(\la) \cdot \Psi_-(\la) $ \quad for \quad  $\la \in \Sg_{\Psi}$ ;
\end{itemize}
where the jump matrix $G_{\Psi}$ takes the form:
\begin{eqnarray}
\e{for}\,\,\la \in \Ga_{\ua} \qquad G_{\Psi}(\la) & = & I_2 \; + \;   \f{ \ex{ \i\la \ov{x}_N } }{  \ups^2(\la) R(\la) } \cdot \sg^-\;,\\
\label{ecriture saut Psi hors de R}  \e{for}\,\,\la \in \Ga_{\da} \qquad G_{\Psi}(\la) & = & I_2  \;  + \;    \f{  \ups^2(\la)\,\ex{-\i\la \ov{x}_N }   }{  R(\la) } \cdot \sg^+ \;, 
\end{eqnarray}
and for $\la \in \R + \i \eps$ 
\beq
\label{ecriture saut Psi sur de R}
\qquad G_{\Psi}(\la)   \; =  \;  I_2 \; +  \; \f{r_N(\la) }{ R(\la) } \cdot 
	  \left( \ba{cc}  1   &  - \ups_+(\la) \ups_-(\la) \ex{- \i \la \ov{x}_N} \\
		 \f{ \ex{ \i \la \ov{x}_N} }{ \ups_+(\la) \ups_-(\la) }	& -1     \ea \right) \;. 
\enq
%By repeating the previous reasoning, it is clear that the Riemann--Hilbert problems for $\chi$ and $\Psi$
%are in a one-to-one correspondence. 

%
%
%
\begin{figure}[h]
\begin{center}

\begin{pspicture}(12,9)

% axe reel 

\psline[linestyle=dashed, dash=3pt 2pt]{->}(0,4.5)(12,4.5)

\rput(11.5,4.2){$\R$}

% Courbe de saut R + i eps

\psline(0,5.5)(12,5.5)

\psline[linewidth=3pt]{->}(11,5.5)(11.2,5.5)

\rput(11.5,5.2){$\R + \i \eps $}

%courbe de saut Gamma up

\pscurve(0,8)(6,6.5)(12,8)

\rput(11.5,7.5){$\Ga_{\ua}  $}

\psline[linewidth=3pt]{->}(9,7)(8.8,6.95)

%courbe de saut Gamma down

\pscurve(0,0.5)(6,3)(12,0.5)

\rput(11.5,1.1){$\Ga_{\da}  $}

\psline[linewidth=3pt]{->}(10,1.65)(9.8,1.75)

\rput(6,8.8){$\circledast$}
\rput(6,8.0){$\circledast$}
\rput(6,7.2){$\circledast$}
\rput(6,0.8){$\circledast$}
\rput(6,1.6){$\circledast$}
\rput(6,2.4){$\circledast$}

%definition de la matrice Phi

\rput(6,7.5){$ \big[\ups(\la) \big]^{\sg_3} \cdot \mc{R}_{\ua }^{(\infty) } \cdot P_{L; \ua}^{-1}(\la)  \cdot \chi(\la)$}

\rput(6,6){$M_{\ua}^{-1}(\la)\cdot  \big[\ups(\la) \big]^{\sg_3} \cdot \mc{R}_{\ua}(\la) \cdot P_{L; \ua}^{-1}(\la)  \cdot \chi(\la)$}
\rput(6,5){$M_{\da}(\la)\cdot  \big[\ups(\la) \big]^{\sg_3} \cdot \mc{R}_{ \da }^{-1}(\la) \cdot P_{L; \da}^{-1}(\la) 
\cdot G^{-1}_{\chi}(\la)\cdot \chi(\la)$}
\rput(6,3.7){$M_{\da}(\la)\cdot  \big[\ups(\la) \big]^{\sg_3} \cdot \mc{R}_{ \da }^{-1}(\la) \cdot P_{L; \da}^{-1}(\la) \cdot \chi(\la)$}
\rput(5.8,1.8){$ \big[\ups(\la) \big]^{\sg_3} \cdot \Big(\mc{R}_{ \da }^{(\infty)} \Big)^{-1} \cdot P_{L; \da}^{-1}(\la) \cdot \chi(\la)$}
\end{pspicture}

\caption{$\Sg_{\Psi}=\Ga_{\ua}\cup \Ga_{\da}\cup\{ \R + \i \eps \}$ is the contour appearing in the Riemann--Hilbert problem for $\Psi$.
$\Ga_{\ua/\da}$ separates all the poles of $ R^{-1}(\la)$ from $\R$ (they are indicated by $\circledast$), and is such that $\e{dist}(\Ga_{\ua/\da}, \R) > \de$ for some $\de >0$ but sufficiently small. 
\label{contour pour le RHP de Phi}}
\end{center}
\end{figure}

\noindent The motivation underlying the construction of $\Psi$ is that its jump matrix $G_{\Psi}$ not only satisfies $G_{\Psi}-I_2 \in \mc{M}_2\Big( \big(L^{2}\cap L^{\infty} \big)\big( \Sg_{\Psi} \big) \Big)$, 
but is, in fact, exponentially in $N$ close to the identity
\beq
\label{eoim}\norm{ G_{\Psi}-I_2 }_{\mc{M}_2(L^2(\, \Sg_{\Psi} )) } \, + \,  \norm{ G_{\Psi}-I_2 }_{ \mc{M}_2(L^{\infty}(\, \Sg_{\Psi} )) }   \; = \;
 \e{O}\big( \ex{-\varkappa_{\eps} N^{\a}} \big) \;,
\enq
with
\beq
\varkappa_{\eps} \; = \; (b_N-a_N) \cdot \min \Big\{   \inf_{\la \in \Ga_{\ua}\cup \Ga_{\da} } | {\rm Im}\,\la| \,;\, 
2 \ga  \big(\, \pi \min[\omega_1,\omega_2] \, - \,  \eps \big)  \Big\} \;. 
\label{definition constante c epsilon de Pi}
\enq
Note that we have a freedom of choice of the curves $\Ga_{\ua/\da}$, provided that these
avoid (respectively from below/above) all the poles of $R^{-1}(\la)$ in $\mathbb{H}^{+/-}$. As a consequence, we have the natural bound:
\beq
\inf_{\la \in \Ga_{\ua}\cup \Ga_{\da} } | {\rm Im}\,\la|  \; \leq \;\f{ 2\pi\om_1 \om_2 }{ \om_1 + \om_2}   \;. 
\enq

These bounds are enough so as to solve the Riemann--Hilbert problem for $\Psi$. 
Indeed, introduce the singular integral operator on the space $\mc{M}_2\big( L^2( \Sg_{\Psi} ) \big)$ 
of $2\times 2$ matrix-valued $L^2\big( \Sg_{\Psi} \big)$ functions by 
\beq
\mc{C}^{(-)}_{ \Sg_{\Psi} }\big[ \Pi \big] (\la) \; = \; \lim_{ \substack{ z \tend \la \\ z \in - \e{side}  \, \e{of} \,\Sg_{\Psi} } } 
\int_{ \Sg_{\Psi} }{} \f{ (G_{\Psi}-I_2)(t) \cdot \Pi(t) }{t-z} \cdot \f{ \dd t}{ 2 \i \pi } \;. 
\enq
Since $ G_{\Psi}-I_2 \in \mc{M}_2\Big( \big(L^{\infty} \cap L^2 \big) \big( \Sg_{\Psi} \big) \Big)$ and $\Sg_{\Psi}$ is a Lipschitz curve, 
it follows from Theorem \ref{Theorem conte transfo Cauchy sur courbes Lipschitz} that ${\cal C}^{(-)}_{\Sigma_{\Psi}}$ is a continuous endomorphism on $\mc{M}_2(L^2\big( \Sg_{\Psi}))$ 
that furthermore satisfies:
\beq
\big| \big| \big|  \mc{C}^{(-)}_{\Sg_{\Psi} }   \big| \big| \big|_{ \mc{M}_2(L^2( \Sg_{\Psi} )) }  \; \leq  \; 
C \ex{-\varkappa_{\epsilon}N^{\a}} \;. 
\enq
Hence, since 
\beq
G_{\Psi}-I_2 \in \mc{M}_2\Big( L^2\big( \Sg_{\Psi} \big) \Big)  \quad \e{and} \quad 
\mc{C}^{(-)}_{\Sg_{\Psi} }[I_2] \in  \mc{M}_2\Big( L^2\big( \Sg_{\Psi} \big) \Big) 
\enq
 provided that $N$ is large enough, it follows that the singular integral equation
\beq
\Big(I_2 - \mc{C}^{(-)}_{\Sg_{\Psi} }  \Big)\big[ \Pi_- \big] \; = \; I_2
\label{ecriture eqn int sing pour matrice Pi moins}
\enq
admits a unique solution $\Pi_{-}$ such that $\Pi_{-} - I_2 \in \mc{M}_2\big( L^2( \Sg_{\Psi} ) \big)$. The bound \eqref{eoim} also implies that:
\beq
\label{nounfrw}\norm{\Pi_{-} - I_2}_{{\cal M}_{2}(L^2(\Sigma_{\Psi}))} \leq 1
\enq
for $N$ large enough. It is then a standard fact \cite{BealsCoifmanScatteringInFirstOrderSystemsEquivalenceRHPSingIntEqnMention} in the theory of Riemann--Hilbert problems 
that the matrix  
\beq
\Pi(\la) \; = \; I_2 \; + \; \Int{ \Sg_{\Psi} }{} \f{ (G_{\Psi}-I_2)(t) \cdot \Pi_-(t) }{ t-\la } \cdot \f{ \dd t }{ 2\i\pi }
\label{ecriture rep int matrice Pi}
\enq
 is the unique solution to the Riemann--Hilbert problem:
\begin{itemize}
\item $\Pi \in \mc{O}(\Cx\setminus \Sg_{ \Psi } )$ and has continuous $\pm$ boundary values on 
$\Sg_{\Psi}$ ;
\item $\Pi(\la) = I_{2} \; + \; \e{O}\big(\la^{-1}\big) $ when $\la \tend \infty$ non-tangentially to $\Sg_{\Psi}$ ;
\item $\Pi_+(\la) \; = \;   G_{\Psi}(\la) \cdot \Pi_-(\la)$ for $\la \in \Sg_{\Psi}$ \;. 
\end{itemize}
We claim that for any open neighbourhood $U$ of $\Sigma_{\Psi}$ such that $\e{dist}(\Sg_{\Psi}, \Dp{}U)> \de >0$, there exists a constant $C > 0$ such that:
\beq
\label{ecriture bornes en N pour Pi moins Id}
\forall \la \in U,\qquad \max_{a,b \in \{1,2\}}\big[\Pi(\la) - I_{2}\big]_{ab}  \leq \frac{C\,e^{-\varkappa_{\epsilon}N^{\a}}}{1 + |\la|}\;.
\enq
Indeed, we can write:
\bem
\max_{a,b \in \{1,2\}}\big[\Pi(\la) - I_2\big]_{ab} \leq \max_{a,b \in \{1,2\}} \Bigg|\Int{\Sigma_{\Psi}}{} \frac{(G_{\Psi} - I_2)_{ab}(t)}{t - \la}\cdot\frac{\dd t}{2\i\pi}\Bigg|  \\
+ \sum_{a,b \in \{1,2\}} \norm{\Pi_{-} - I_2}_{{\cal M}_2(L^2(\Sigma_{\Psi}))}\cdot\Bigg(\Int{\Sigma_{\Psi}}{} \frac{\big|(G_{\Psi} - I_2)_{ab}(t)\big|^2}{|t - \la|^2}\cdot\frac{|\dd t|}{(2\pi)^2}\Bigg)^{1/2}\;.
\label{ecriture bornage explicite comportement infini matrice P cal ab} 
\end{multline}
The second term is readily bounded with \eqref{nounfrw} and the fact \eqref{eoim} that $G_{\Psi}$ is exponentially close to the identity matrix. For the first term, we study the asymptotic behaviour of $G_{\Psi} - I_{2}$ with help of \S~\ref{sniung}:
\begin{eqnarray}
\label{moim}\mathrm{if}\,\,t \in \Gamma_{\da}\cup\Gamma_{\ua},\qquad |(G_{\Psi} - I_2)_{ab}(t)| & \leq & C\,e^{-|\mathrm{Re}\,t|}\cdot e^{-\varkappa_{\epsilon}\,N^{\a} }\;, \\
\label{mpim}\mathrm{if}\,\,t \in \R + \i\eps,\qquad |(G_{\Psi} - I_2)_{ab}(t)| & \leq & C\,|t|^{-1}\cdot e^{-\varkappa_{\epsilon}N^{\a}}\;.
\end{eqnarray}
For the contribution on $\R + \i\eps$, we split $[G_{\Psi} - I_2](t) = C_{\Psi}\cdot t^{-1} + O(t^{-2})$. We compute directly the contour integral of the term in $t^{-1}$, and find the bound bound $\max_{a,b} \big|[C_{\Psi}]_{ab}\cdot\la^{-1}\big|$ if $\mathrm{Im}\,\la > \epsilon$, and $0$ otherwise. Hence, it is bounded by $c_1/(1 + |\la|)$ for some constant $c_1 > 0$. The contribution of the remainder $O(|t|^{-2})$ to the contour integral can be bounded thanks to the lower bound ${\rm dist}(\Sigma_{\Psi},\la) \geq c_2/(1 + |\la|)$ for some constant $c_2 > 0$. Collecting all these bounds justifies \eqref{ecriture bornes en N pour Pi moins Id}.

\begin{figure}[h]
\begin{center}

\begin{pspicture}(12,9)

% axe reel 

\psline[linestyle=dashed, dash=3pt 2pt]{->}(0,4.5)(12,4.5)

\rput(11.5,4.2){$\R$}

% Courbe de saut R + i eps

\psline(0,5.5)(12,5.5)

\psline[linewidth=3pt]{->}(11,5.5)(11.2,5.5)

\rput(11.5,5.2){$\R + \i \eps $}

%courbe de saut Gamma up

\pscurve(0,8)(6,6.5)(12,8)

\rput(11.5,7.5){$\Ga_{\ua}  $}

\psline[linewidth=3pt]{->}(9,7)(8.8,6.95)

%courbe de saut Gamma down

\pscurve(0,0.5)(6,3)(12,0.5)

\rput(11.5,1.5){$\Ga_{\da}  $}

\psline[linewidth=3pt]{->}(10,1.65)(9.8,1.75)

%definition de la matrice Phi

\rput(6,7.5){$ P_{L;\ua}(\la)  \cdot \Big(\mc{R}_{ \ua }^{(\infty) } \Big)^{-1} \cdot \big[\ups(\la) \big]^{-\sg_3}  \cdot \Pi(\la) \cdot P_R(\la) $}

\rput(4.5,6.1){$P_{L;\ua}(\la)\cdot \mc{R}_{ \ua }^{-1}(\la)  \cdot \big[\ups(\la) \big]^{-\sg_3} \cdot M_{\ua}(\la)  \cdot \Pi(\la) \cdot P_R(\la)$}
\rput(5.5,5){$G_{\chi}(\la)\cdot P_{L;\da}(\la) \cdot \mc{R}_{\da }(\la) \cdot \big[\ups(\la) \big]^{-\sg_3} \cdot M_{\da}^{-1}(\la)
\cdot \Pi(\la) \cdot P_R(\la)$}
\rput(4.5,3.5){$P_{L;\da}(\la) \cdot \mc{R}_{ \da }(\la) \cdot \big[\ups(\la) \big]^{-\sg_3} \cdot M_{\da}^{-1}(\la)
\cdot \Pi(\la) \cdot P_R(\la)$}
\rput(6,1.5){$P_{L;\da}(\la)  \cdot \mc{R}_{ \da }^{(\infty)}  \cdot  \big[\ups(\la) \big]^{-\sg_3} \cdot \Pi(\la) \cdot P_R(\la)$}
\end{pspicture}

\caption{Piecewise definition of the matrix $\chi$.
The curves $\Ga_{\ua/\da}$ separate all poles of $\la \mapsto R^{-1}(\la)$ from $\R$ and are such that $\e{dist}(\Ga_{\ua/\da}, \R) > \de > \epsilon > 0$
for a sufficiently small $\de$. The matrix $\Pi$ appearing here is defined through \eqref{ecriture rep int matrice Pi}. 
\label{Figure definition sectionnelle de la matrice chi}}
\end{center}
\end{figure}
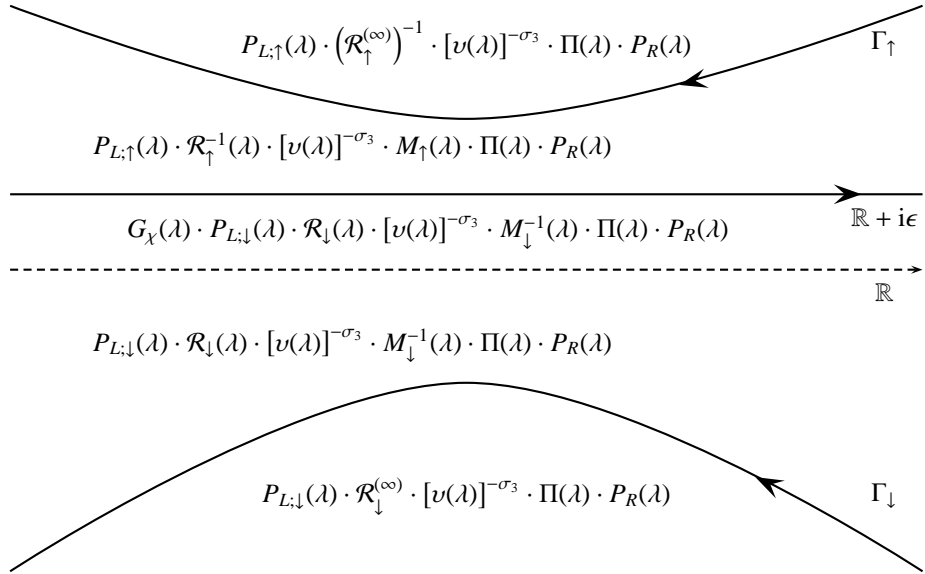

The Riemann--Hilbert problem for $\Psi$ and $\Pi$ have the same jump matrix $G_{\Psi}$, but $\Psi$ must have a zero with prescribed leading coefficient at $\la = 0$, while $\Pi$ has a finite value $\Pi(0)$.
We then see that the formula:

\beq
\label{PsiPi}\Psi(\la) \; = \; \Pi(\la) \cdot P_R(\la)
\enq
with:
\beq
P_R(\la) \; = \; I_2  \; + \; \f{ \th_R }{ \la } \cdot \Pi^{-1}(0) \sg^- \Pi(0)\;,
\qquad \e{and} \quad \th_R \; = \; \frac{1}{\ups^2(0)}\,\frac{\kappa_N}{ 1 +\kappa_N/(\om_1+\om_2)}
\label{ecriture matrice Psi finale et cste cR}
\enq
yields the unique solution to the Riemann--Hilbert problem for $\Psi$. Tracking back the transformations $\Pi \rightsquigarrow \Psi \rightsquigarrow \chi$, gives the construction of the solution $\chi$ of 
the Riemann--Hilbert problem of Proposition~\ref{Theorem ecriture forme asymptotique matrice chi}, summarized in Figure~\ref{Figure definition sectionnelle de la matrice chi}. 
This concludes the proof of Proposition~\ref{Theorem ecriture forme asymptotique matrice chi}. \qed

%%%%%%%%%%%%%%%%%%%%%%%%%%%%%%%%%%%%%%%%%%%%%%%%%%%%%%%%%%%%%%%%%%%%%%%%%%%%%%%%%%%%%%%%%%%%%%%%%%%%%%%%%%%%%%%%%%%%%%%%%%%%%%%%%%%%%%%%%%%%%%%%%%%%%%%%%%%%%%%%%%%%%
%%%%%%%%%%%%%%%%%%%%%%%%%%%%%%%%%%%%%%%%%%%%%%%%%%%%%%%%%%%%%%%%%%%%%%%%%%%%%%%%%%%%%%%%%%%%%%%%%%%%%%%%%%%%%%%%%%%%%%%%%%%%%%%%%%%%%%%%%%%%%%%%%%%%%%%%%%%%%%%%%%%%%

%%%%%%%%%%%%%%%%%%%%%%%%%%%%%%%%%%%%%%%%%%%%%%%%%%%%%%%%%%%%%%%%%%%%%%%%%%%%%%%%%%%%%%%%%%%%%%%%%%%%%%%%%%%%%%%%%%%%%%%%%%%%%%%%%%%%%%%%%%%%%%%%%%%%%%%%%%%%%%%%%%%%%

%%%%%%%%%%%%%%%%%%%%%%%%%%%%%%%%%%%%%%%%%%%%%%%%%%%%%%%%%%%%%%%%%%%%%%%%%%%%%%%%%%%%%%%%%%%%%%%%%%%%%%%%%%%%%%%%%%%%%%%%%%%%%%%%%%%%%%%%%%%%%%%%%%%%%%%%%%%%%%%%%%%%%

%%%%%%%%%%%%%%%%%%%%%%%%%%%%%%%%%%%%%%%%%%%%%%%%%%%%%%%%%%%%%%%%%%%%%%%%%%%%%%%%%%%%%%%%%%%%%%%%%%%%%%%%%%%%%%%%%%%%%%%%%%%%%%%%%%%%%%%%%%%%%%%%%%%%%%%%%%%%%%%%%%%%%
%%%%%%%%%%%%%%%%%%%%%%%%%%%%%%%%%%%%%%%%%%%%%%%%%%%%%%%%%%%%%%%%%%%%%%%%%%%%%%%%%%%%%%%%%%%%%%%%%%%%%%%%%%%%%%%%%%%%%%%%%%%%%%%%%%%%%%%%%%%%%%%%%%%%%%%%%%%%%%%%%%%%%

%%%%%%%%%%%%%%%%%%%%%%%%%%%%%%%%%%%%%%%%%%%%%%%%%%%%%%%%%%%%%%%%%%%%%%%%%%%%%%%%%%%%%%%%%%%%%%%%%%%%%%%%%%%%%%%%%%%%%%%%%%%%%%%%%%%%%%%%%%%%%%%%%%%%%%%%%%%%%%%%%%%%%
%%%%%%%%%%%%%%%%%%%%%%%%%%%%%%%%%%%%%%%%%%%%%%%%%%%%%%%%%%%%%%%%%%%%%%%%%%%%%%%%%%%%%%%%%%%%%%%%%%%%%%%%%%%%%%%%%%%%%%%%%%%%%%%%%%%%%%%%%%%%%%%%%%%%%%%%%%%%%%%%%%%%%

\subsection{Properties of the solution $\chi$}
\label{SousSection Proprietes generales de la solution chi}

\begin{lemme}
\label{Lemme Ecriture diverses proprietes solution RHP chi}
The solution $\chi$ to the Riemann--Hilbert problem given in Proposition~\ref{Theorem ecriture forme asymptotique matrice chi}
admits the following symmetries
\beq
\chi(-\la) \; = \; 
\left(  \ba{cc} 1 & 0 \\ 
							0 & -1 \ea \right) \cdot \chi(\la) \cdot 
										\left( \ba{cc} 1 & -\la \\
												0  & 1 \ea \right)  \qquad \e{and} \qquad
   \Big( \chi(\la^*) \Big)^*  \; = \; \left(  \ba{cc} 1 & 0 \\ 
							0 & -1 \ea \right) \cdot  \chi(-\la) \cdot \left( \ba{cc} -1 & 0  \\
												0  & 1 \ea \right) 		\;. 									
\label{ecriture propriete conjugaison et reflection solution RHP chi}
\enq
where $^*$ refers to the component-wise complex conjugation. 

\end{lemme}

\Proof Since $G_{\chi}(-\la) \; = \; \ex{ \f{\i\pi\sg_3}{2} } G^{-1}_{\chi} (\la) \ex{- \f{\i\pi\sg_3}{2} }$, the matrix:
\beq
\Xi(\la) \; = \; \chi^{-1}(\la) \cdot \ex{- \f{\i\pi\sg_3}{2} } \cdot \chi(-\la)
\enq
is continuous across  $\R$ and thus is an entire function. The asymptotic behaviour of $\Xi(\la)$ when $\la \rightarrow \infty$ is deduced from the growth conditions prescribed by the Riemann--Hilbert problem (\textit{cf.} 
Proposition~\ref{Theorem ecriture forme asymptotique matrice chi}):
\beq
\Xi(\la) \; = \;   \i \la \cdot \sg^+ \; - \i\big(\chi_1 \cdot \sg^+ \; + \; \sg^+ \cdot \chi_1\big)
			\; + \; \e{O}(\la^{-1}) \;.
\enq
Since $\Xi(\la)$ is entire, by Liouville theorem this asymptotic expression is exact, namely
\beq
\label{fmwu} \Xi(\la) \; = \;  \i \la \cdot \sg^+ \; - \i\big(\chi_1 \cdot \sg^+ \; + \; \sg^+ \cdot \chi_1\big)\;.
\enq
Observe that 
\beq
\chi_1 \cdot \sg^+ \; + \; \sg^+ \cdot \chi_1  \; = \; 
\left( \ba{cc}   \big[ \chi_1 \big]_{ 21}  &  \e{tr} \big[ \chi_1 \big]  \\ 
						0  &  \big[ \chi_1 \big]_{ 21} \ea \right) \;. 
\enq
By expanding the relation $\ddet{}{\chi(\la)}=1$ for $\la \in \mathbb{H}^+$ at large $\la$, we find that the matrix $\chi_1$ is actually traceless. Finally, the jump condition at $\la=0$ takes the form 
\beq
 \chi_-(0) \; = \;  \sg_3 \cdot \chi_+(0) \;. 
\enq
Using this relation and the expression for $\Xi$ given in \eqref{fmwu}, we get:
\beq
-\i \chi_+(0) \; = \; -\i \chi_+(0) \cdot \left( \ba{cc}   \big[ \chi_1 \big]_{ 21}  & 0  \\ 
						0  &  \big[ \chi_1 \big]_{ 21} \ea \right) \qquad \textit{i.e.} \qquad 
	\big[ \chi_1 \big]_{ 21} = 1 				
\enq
since $\chi_+(0)$ is invertible. This proves the first relation in \eqref{ecriture propriete conjugaison et reflection solution RHP chi}.
In order to establish the second one, we consider: 
\beq
\wt{ \Xi }(\la) \; = \; \chi^{-1}(-\la)\cdot \ex{ \f{\i\pi \sg_3}{2} } \cdot \big( \chi(\la^*) \big)^{*} \;. 
\enq
With the relation $\big(  G_{\chi}(\la^*) \big)^* \; = \; G^{-1}_{\chi}(\la)$ and the complex conjugate of the asymptotic behaviour for $\chi$, one shows that $\wt{\Xi}$ is holomorphic on 
$\Cx\setminus \R$, continuous across $\R$ and hence entire. Furthermore, since it admits the asymptotic behaviour 
%
%
%\beq
%
%\big( \chi(\la^*) \big)^{*} = \left\{ 
%
%ba{cc} \Big( I_2 + \kappa_N\la^{-1}\ex{ \i(\ga-1)\ov{x}_N\la} \cdot \sg^- \Big) \cdot  
%
%\left(\ba{cc} -1 & \e{sgn}({\rm Re}\,\la)\,\ex{ \i\la \ov{x}_N} \\ 
%
%
%						0   & 1   \ea  \right) \cdot \big(- \i\la \big)^{-\sg_3/2}  \ex{-\i\pi\sg_3/2} 
%
 % \Big(I_{2}  \; + \; \la^{-1}\cdot\chi_1^* \; + \; \e{O}\big(\la^{-2}\big) \Big)   & 
%
%			 \la \in \mathbb{H}^+ \vspace{3mm} \\
%
%
%\Big( I_2 + \kappa_N\la^{-1}\ex{-\i(\ga-1)\ov{x}_N\la} \cdot \sg^- \Big) \cdot  
%
%		\left(\ba{cc} - \e{sgn}({\rm Re}\,\la )\,\ex{\i\la \ov{x}_N}  & 1\\ 
%
%
%						-1   & 0   \ea  \right)   \cdot \big( \i\la \big)^{-\sg_3/2} 
%
%  \Big(I_{2} \; + \; \la^{-1}\cdot\chi_1^* \; + \; \e{O}\big(\la^{-2}\big) \Big) &
%
%				\la \in \mathbb{H}^-  \ea\right. 
%
%
%\enq
%
%
%	
%when $\la \tend \infty$ non-tangentially to $\R$ and the relation 
%
%
%
%\beq
%
%\big(  G_{\chi}(\la^*) \big)^* \; = \; G^{-1}_{\chi}(\la) \;, 
%
%\enq
%
%
%
%one readily gets that $\wt{\Xi}$ is analytic on $\Cx\setminus \R$, continuous across $\R$
%with a large-$\la$ asymptotic behaviour 
%
%
%
\beq
\wt{ \Xi }(\la) \; = \; \ex{- \f{ \i\pi\sg_3}{2} } \cdot \Big( I_2 \; + \; \e{O}\big( \la^{-1} \big) \Big) \; , 
\enq
by Liouville's theorem, $\wt{\Xi}(\la) = e^{- \f{ \i\pi\sigma_3}{2} }$. \qed

\begin{lemme}
\label{Lemme DA de chi ecriture explicite}

The matrix $\chi$ admits the large-$\la$, $\la \in \mathbb{H}^+$ asymptotic expansion
\beq
\label{cninfr}\chi(\la) \; \simeq \; \big(-\i\la \big)^{1/2} \cdot \sg^+ \; + \; 
\sul{k \geq 0}{}  \f{ K(\la)\cdot \chi_k -\i \sg^+ \cdot \chi_{k+1}   }{\big(-\i\la \big)^{1/2} \la^k }  \;,
\enq
where $(\chi_k)_k$ is a sequence of constant, $2\times 2$ matrices, with $\chi_{-1} = 0$ and $\chi_0 = I_2$, and:
\beq
K(\la) \; = \; \left( \ba{cc} - \e{sgn}({\rm Re}\,\la)\,\ex{ \i \la \ov{x}_N}   & 0 \\ 
					- \f{\kappa_N}{ \la } \cdot\e{sgn}({\rm Re}\,\la)\,\ex{ \i \la \ga \ov{x}_N} \, - \, 1  & 
						\; \; 	- \i \kappa_N \cdot  \e{sgn}({\rm Re}\,\la)\,\ex{ \i \la (\ga-1) \ov{x}_N} 	\ea \right)	\;. 			 
\enq
In particular, we have: 
\beqa
[\chi]_{11}(\la)  & \simeq  & \f{ 1 }{ \big(-\i\la \big)^{1/2}  } \sul{k \geq 0}{} \f{ 1 }{ \la^k }
\Big[ - \e{sgn}({\rm Re}\,\la)\,\ex{\i\la \ov{x}_N}  [\chi_k]_{11} \, - \, \i\,[\chi_{k+1}]_{21} \Big] \;, \\
\la^{-1} \cdot [\chi]_{12}(\la)  & \simeq & \f{ 1 }{ \big(-\i\la \big)^{ 1/2 }  } \sul{k \geq 0}{} \f{1}{\la^k}
\Big[ - \e{sgn}({\rm Re}\,\la)\,\ex{\i\la \ov{x}_N} [\chi_{k-1}]_{12} \, - \, \i\,[\chi_{k}]_{22} \Big] \;. 
\label{ecriture DA lambda de chi12}
\eeqa
Note that one should understand the matrix $\chi_{-1}$ occurring in \eqref{ecriture DA lambda de chi12} as 
$\chi_{-1 }:=0$. We also remind that $[\chi_{1}]_{21}=1$.

\end{lemme}

\Proof It is enough to establish that $ \Pi$ admits, for any $\ell$, the large-$\la$ asymptotic expansion of the form:
\beq
\Pi(\la) \; = \; \sul{\ell=0}{k} \la^{-\ell}\,\Pi_{\ell} \; + \; \De_{[k]}\Pi(\la) \qquad \e{with} 
\quad \De_{[k]}\Pi(\la) \; = \; \e{O}\Big( \f{1}{ \la^{k+1-\de} } \Big)  \; \; \e{for} \; \e{any} \; \de>0 \; \quad \e{and} 
\qquad \Pi_{0}=I_2 \;. 
\label{Ecriture DA grd lambda de Pi ordre k}
\enq
Indeed, once this asymptotic expansion is established for $\Pi$, the results for $\chi$ follow from matrix multiplications prescribed on the top of Figure~\ref{Figure definition sectionnelle de la matrice chi}.

Equation \eqref{ecriture bornes en N pour Pi moins Id} shows that the expansion \eqref{Ecriture DA grd lambda de Pi ordre k} holds for $k=0$ uniformly away from $\Sg_{\Psi}$. 
This is actually valid everywhere, for the jump matrix $G_{\Psi}(\la)$ is analytic in a neighbourhood of $\Sg_{\Psi}$ and asymptotically close to $I_2$ at large $\la$ in an open neighbourhood of $\Sg_{\Psi}$, 
\textit{c}.\textit{f}. \eqref{moim}-\eqref{mpim}.

Now assume that the expansion holds up to some order $k$. Consider the integral representation \eqref{ecriture rep int matrice Pi} for $\Pi$. 
We recall that $(\Pi_{-}-I_2) \in L^{2}(\Sg_{\Psi})$ and $G_{\Psi}-I_2$
decays exponentially fast along $ \Ga_{\ua} \cup \Ga_{\da}$. Thus, standard manipulations give an asymptotic expansion of the form:
\beq
\Int{ \Ga_{\ua} \cup \Ga_{\da} }{} \f{ (G_{\Psi}-I_2)(t) \cdot \Pi_-(t) }{t-\la } \cdot \f{ \dd t }{2\i \pi} \; \simeq \; \sul{\ell \geq 1}{} T_{\ell}\,\la^{-\ell} \;.
\enq

It thus remains to focus on the integral on $\R+\i\eps$. We can first move the contour to $\R + \i\epsilon^{\prime}$ for some $0<\epsilon^{\prime} < \epsilon$, and insert the assumed asymptotic expansion at order $k$:
\bem
\Int{ \R + \i \eps }{} \f{ (G_{\Psi}-I_2)(t) \cdot \Pi_-(t) }{t-\la } \cdot \f{ \dd t }{2\i \pi}   \\
\; = \; \sul{\ell=0}{k} 
\;\Int{ \R + \i \eps^{\prime} }{} \f{ (G_{\Psi}-I_2)(t) \cdot \Pi_{\ell} }{t^{\ell}(t-\la) } \cdot \f{ \dd t }{2\i \pi} \; 
\; + \; \Int{ \R + \i \eps^{\prime} }{} \f{ (G_{\Psi}-I_2)(t) \cdot \De_{[k]}\Pi(t) }{t-\la } \cdot \f{ \dd t }{2\i \pi}\;. 
\label{ecriture integrale contre Pi le long de R shifte}
\end{multline}
It follows from \eqref{definition fonction rN} that we can decompose
$r_N(\la)=r_N^{(+)}(\la) \ex{\i\la \ga \ov{x}_N} \, + \, r_N^{(-)}(\la) \ex{-\i\la \ga \ov{x}_N} $, with $r_N^{(\pm)}(\la)$
bounded in $\la$ away from its poles. This induces a decomposition $G_{\Psi}-I_2=(G_{\Psi}-I_2)^{(+)}\,+\, (G_{\Psi}-I_2)^{(-)}$ on $\R + \i\epsilon^{\prime}$. 
Inspecting the expression \eqref{ecriture saut Psi sur de R}, we can convince oneself that there exist curves $\msc{C}^{\pm}_{G_{\Psi}} \subseteq \mathbb{H}^{\pm}$ 
going to $\infty$ when ${\rm Re}\,t \tend \pm \infty$, $t \in \msc{C}^{\pm}_{G_{\Psi}}$ and such that:
\begin{itemize}
 
\item $t \mapsto  \f{ (G_{\Psi}-I_2)^{(\pm)}(t) \cdot \Pi_{\ell} }{t^{\ell} \cdot  (t-\la)}$ has no pole between $\R + \i\epsilon^{\prime}$ and $\msc{C}^{\pm}_{G_{\Psi}}$, 
\item  $(G_{\Psi} - I_2)^{(\pm)}(t)$ decays exponentially fast in $t$ when $t \rightarrow \infty$ along $\msc{C}^{\pm}_{G_{\Psi}}$. 
\end{itemize}
Therefore, we obtain:
\beq
\Int{ \R + \i \eps^{\prime} }{} \f{ (G_{\Psi}-I_2)(t) \cdot \Pi_{\ell} }{t^{\ell}(t-\la) } \cdot \f{ \dd t }{2\i \pi}  \; = \; 
 \Int{ \msc{C}^{+}_{G_{\Psi}} }{} \f{ (G_{\Psi}-I_2)^{(+)}(t) \cdot \Pi_{\ell} }{t^{\ell}(t-\la) } \cdot \f{ \dd t }{2\i \pi} \; + \;  \; 
\Int{ \msc{C}^{-}_{G_{\Psi}} }{} \f{ (G_{\Psi}-I_2)^{(-)}(t) \cdot \Pi_{\ell} }{t^{\ell}(t-\la) } \cdot \f{ \dd t }{2\i \pi}
\enq
and the properties of this decomposition ensure the existence of an all order asymptotic expansion in $\la^{-1}$ when $\la\tend \infty$. 
It thus remains to focus on the last term present in \eqref{ecriture integrale contre Pi le long de R shifte}. For $\delta > 0$ but small, we write:
\beq
\Int{ \R + \i \eps^{\prime} }{} \f{ (G_{\Psi}-I_2)(t) \cdot \De_{[k]}\Pi(t) }{t-\la } \cdot \f{ \dd t }{2\i \pi} \; = \; 
-\sul{\ell=0}{k} \f{ 1 }{ \la^{\ell+1} } \Int{ \R + \i \eps^{\prime} }{} t^{\ell} (G_{\Psi}-I_2)(t) \cdot \De_{[k]}\Pi(t) \cdot \f{ \dd t }{2\i \pi} 
\; + \; \f{  \Delta_{[k]}T(\la) }{ \la^{k+1} |\la|^{1-2\de}  }  \;. 
\enq
The decay at $\infty$ of $\De_{[k]}\Pi$ and $(G_{\Psi}-I_2)$ guarantees the existence of an asymptotic expansion of the first term in the right-hand side, this up to a $\e{O}\big( \la^{-k-2} \big)$ remainder. Finally, we have:
\beq
| \Delta_{[k]}T(\la) |\; = \; \bigg| \Int{ \R + \i \eps^{\prime} }{} t^{k+1} \f{ (G_{\Psi}-I_2)(t) \cdot \De_{[k]}\Pi(t) }
		  { |\la|^{2\de-1} \cdot ( t-\la ) } \cdot \f{ \dd t }{2\i \pi} \bigg|
\; \leq \; C  \Int{ \R + \i \eps^{\prime} }{} \f{ |\la|^{1-2\de}\,\dd t}{  |t|^{ 1-\de}\,|t-\la|  }\;, 
\enq
where we used the assumed bound \eqref{Ecriture DA grd lambda de Pi ordre k} for $\Delta_{[k]}\Pi(t)$ and the $O(1/t)$ decay \eqref{mpim} for $G_{\Psi} - I_2$. The growth of the right-hand side at large $\la$ is 
then estimated by cutting the integral into pieces:
\beq
\frac{\abs{\la}^{1 - 2\delta}}{\abs{t}^{1 - \delta}\,\abs{t - \la}} \; \leq \; 
\left\{ \begin{array}{lcl} \wt{C}\,|\la|^{-2\de}\,|t|^{-(1-\de)} &\quad & \mathrm{if}\,\,\abs{{\rm Re}\,t} \leq \abs{\la}/2   \vspace{2mm}\\ 
	  \wt{C}\,|\la|^{ -\de}\,\abs{t-\la}^{-1} & \quad & \mathrm{if}\,\,\abs{\la}/2 \leq \abs{\mathrm{Re}\,t} \leq 3\abs{\la}/2 \vspace{2mm} \\
	  \wt{C}\,|t|^{-(1+\de)} & \quad & \mathrm{if}\,\,|\mathrm{Re}\,t| \geq 3\abs{\la}/2 \end{array}\right.
\label{ecriture borne integral cauchy}
\enq
for some $\wt{C} > 0$ independent of $\la$ and $t$. The integral over $t$ of the right-hand side on each of piece is finite, and collecting all the pieces, we get $\Delta_{[k]}T(\la) =\e{o}(1)$ when $\la \rightarrow \infty$. \qed

%%%%%%%%%%%%%%%%%%%%%%%%%%%%%%%%%%%%%%%%%%%%%%%%%%%%%%%%%%%%%%%%%%%%%%%%%%%%%%%%%%%%%%%%%%%%%%%%%%%%%%%%%%%%%%%%%%%%%%%%%%%%%%%%%%%%%%%%%
%%%%%%%%%%%%%%%%%%%%%%%%%%%%%%%%%%%%%%%%%%%%%%%%%%%%%%%%%%%%%%%%%%%%%%%%%%%%%%%%%%%%%%%%%%%%%%%%%%%%%%%%%%%%%%%%%%%%%%%%%%%%%%%%%%%%%%%%%

%%%%%%%%%%%%%%%%%%%%%%%%%%%%%%%%%%%%%%%%%%%%%%%%%%%%%%%%%%%%%%%%%%%%%%%%%%%%%%%%%%%%%%%%%%%%%%%%%%%%%%%%%%%%%%%%%%%%%%%%%%%%%%%%%%%%%%%%%
%%%%%%%%%%%%%%%%%%%%%%%%%%%%%%%%%%%%%%%%%%%%%%%%%%%%%%%%%%%%%%%%%%%%%%%%%%%%%%%%%%%%%%%%%%%%%%%%%%%%%%%%%%%%%%%%%%%%%%%%%%%%%%%%%%%%%%%%%

%%%%%%%%%%%%%%%%%%%%%%%%%%%%%%%%%%%%%%%%%%%%%%%%%%%%%%%%%%%%%%%%%%%%%%%%%%%%%%%%%%%%%%%%%%%%%%%%%%%%%%%%%%%%%%%%%%%%%%%%%%%%%%%%%%%%%%%%%
%%%%%%%%%%%%%%%%%%%%%%%%%%%%%%%%%%%%%%%%%%%%%%%%%%%%%%%%%%%%%%%%%%%%%%%%%%%%%%%%%%%%%%%%%%%%%%%%%%%%%%%%%%%%%%%%%%%%%%%%%%%%%%%%%%%%%%%%%

\section{The inverse of the operator $\mc{S}_N$ }
\label{Section construction inverse op int sing}

%%%%%%%%%%%%%%%%%%%%%%%%%%%%%%%%%%%%%%%%%%%%%%%%%%%%%%%%%%%%%%%%%%%%%%%%%%%%%%%%%%%%%%%%%%%%%%%%%%%%%%%%%%%%%%%%%%%%%%%%%%%%%%%%%%%%%
%%%%%%%%%%%%%%%%%%%%%%%%%%%%%%%%%%%%%%%%%%%%%%%%%%%%%%%%%%%%%%%%%%%%%%%%%%%%%%%%%%%%%%%%%%%%%%%%%%%%%%%%%%%%%%%%%%%%%%%%%%%%%%%%%%%%%

%%%%%%%%%%%%%%%%%%%%%%%%%%%%%%%%%%%%%%%%%%%%%%%%%%%%%%%%%%%%%%%%%%%%%%%%%%%%%%%%%%%%%%%%%%%%%%%%%%%%%%%%%%%%%%%%%%%%%%%%%%%%%%%%%%%%%
%%%%%%%%%%%%%%%%%%%%%%%%%%%%%%%%%%%%%%%%%%%%%%%%%%%%%%%%%%%%%%%%%%%%%%%%%%%%%%%%%%%%%%%%%%%%%%%%%%%%%%%%%%%%%%%%%%%%%%%%%%%%%%%%%%%%%

%%%%%%%%%%%%%%%%%%%%%%%%%%%%%%%%%%%%%%%%%%%%%%%%%%%%%%%%%%%%%%%%%%%%%%%%%%%%%%%%%%%%%%%%%%%%%%%%%%%%%%%%%%%%%%%%%%%%%%%%%%%%%%%%%%%%%
%%%%%%%%%%%%%%%%%%%%%%%%%%%%%%%%%%%%%%%%%%%%%%%%%%%%%%%%%%%%%%%%%%%%%%%%%%%%%%%%%%%%%%%%%%%%%%%%%%%%%%%%%%%%%%%%%%%%%%%%%%%%%%%%%%%%%

\subsection{Solving $\msc{S}_{N;\ga}[\varphi] = h$ for $h \in H_s(\intff{ 0 }{ \ov{x}_N })$, $-1<s<0$}

With the $2 \times 2$ matrix $\chi$ in hand, we can come back to the inversion of the integral operator $\msc{S}_{N;\gamma}$ according to Lemma~\ref{Proposition corresp operateur et WH factorisation}.

\begin{prop}

Assume $-1 < s < 0$, and $h \in H_s(\intff{ 0 }{ \ov{x}_N })$. Any solution to $\msc{S}_{N;\ga}\big[ \vp \big](\xi) \; = \;  h(\xi) $ 
is of the form $\vp = \wt{\msc{W}}_{\vartheta; z_0}[h_{\mf{e}}] $ where 
\beq
\wt{\msc{W}}_{\vartheta; z_0}[h_{\mf{e}}] \;= \; \mc{F}^{-1}\Big[(*-z_0)\,\chi_{11;+} \cdot \mc{C}_+\big[  f_{ 1;z_0 } \big] \; + \; 
\chi_{12;+}\cdot \mc{C}_+\big[  f_{ 2;z_0 } \big] 
\; + \; \vartheta\cdot\chi_{11;+}  \Big] \;. 
\label{ecriture inverse sous forme abstraite Fourier}
\enq
Above, $\vartheta \in \mathbb{C}$ and $z_0 \in \mathbb{C}\setminus\R$ are arbitrary constants. 
We remind that $\chi_{+}$ is the upper boundary value of $\chi$ on $\R$, ${\cal C}$ is  the Cauchy transform \eqref{Cplusmoins}, $\mc{C}_{\pm}$ its $\pm$ boundary values and $h_{\mf{e}}$ is any extension of $h $ to $H_{s}(\R)$. %
\beq
\label{fdff}\left( \ba{c} f_{ 1;z_0 } (\la)  \\  f_{ 2;z_0 }(\la)  \ea \right)  \; = \;  
\ex{-\i\la \ov{x}_N} \mc{F}[h_{\mf{e}}](\la)  \cdot
\left( \ba{c} (\la-z_0)^{-1} \chi_{12;+}(\la)   \\  - \chi_{11;+}(\la) \ea  \right) \;. 
\enq
The  transform $\wt{\msc{W}}_{\vartheta; z_0}$ is continuous on $H_s(\R)$, $-1<s<0$:
\beq
\norm{ \wt{\msc{W}}_{0; z_0}[h_{\mf{e}}] }_{H_s(\R)} \; \leq  \; C_N\,\norm{ h_{\mf{e}} }_{H_s(\R)} \; ,  
\label{ecriture continuite Hs de transformee tildee}
\enq
the continuity constant $C_N$ being however dependent, a priori, on $N$. 
Finally, when $h\in \mc{C}^{1}(\intff{0}{\ov{x}_N})$ the transform can be recast as 
\bem
\wt{\msc{W}}_{\vartheta; z_0}[h] (\xi) \; =  
\Int{ \R + 2 \i \eps^{\prime} }{} \hspace{-2mm} \f{ \dd \la }{ 2\pi } \; \Int{ \R + \i\eps^{\prime} }{} 
\hspace{-1mm} \f{ \dd \mu }{ 2 \i \pi } \f{ \ex{- \i \xi \la - \i \ov{x}_N\mu  }  }{ \mu- \la }
\Bigg\{  \f{ \la - z_0 }{ \mu - z_0 }\chi_{11}(\la) \chi_{12}(\mu)  - \chi_{11}(\mu) \chi_{12}(\la) \Bigg\} 
%
%\cdot  \mc{F}\big[ h_{\mf{e}} \big](\mu) \\
\cdot \Int{0}{\ov{x}_N}\ex{\i \eta \mu} h(\eta) \cdot \dd \eta \\
\; + \;  \vartheta \Int{ \R + \i \eps^{\prime} }{}  \ex{- \i \la \xi} \chi_{11}(\la) \cdot \f{ \dd \la }{ 2\pi } \;. 
\label{ecriture forme generale solution tel que predite par RHP dans Hs}
\end{multline}
where $\eps^{\prime} > \eps $ is arbitrary but small enough and such that ${\rm Im}\,z_0>\eps^{\prime}$ in the case when $z_0 \in \mathbb{H}^+$. 
\end{prop}

We stress that the integrals, as written in \eqref{ecriture forme generale solution tel que predite par RHP dans Hs}, are to be understood in the Riemann
sense in that they only converge as oscillatory integrals. 
 
\Proof The proof is based on a Wiener-Hopf factorisation. For the moment, we only assume that $s<0$.
Let $ \Phi $ be any solution to the vector Riemann--Hilbert problem for $\Phi$ outlined
in Lemma~\ref{Proposition corresp operateur et WH factorisation}. Then, 
define a piecewise holomorphic function $\Ups$ by 
\beq
\label{55}\Ups(\la) \; = \; \left\{  \ba{cc}  \chi^{-1}(\la) \Phi(\la) \; - \; \wh{\bs{H}}(\la)  & \la \in \mathbb{H}^+  \vspace{2mm} \\
								  \chi^{-1}(\la) \Phi(\la) \; - \; \wh{\bs{H}}(\la)  & \la \in \mathbb{H}^-    \ea \right. 
\enq
where, for some $z_0 \in \Cx\setminus \R$
\beq
\wh{\bs{H}}(\la) \; = \; \left( \ba{c} (\la-z_0)^{\iota_s}
 \Int{\R}{} \f{  g_{1; \iota_s}(t)\,\dd t}{2{\rm i}\pi(t-\la)} \\
(\la-z_0)^{\iota_s-1} \Int{\R}{} \f{ g_{1;\iota_s - 1}(t)\,\dd t}{2{\rm i}\pi(t-\la)}   \ea  \right)  \qquad \e{with} \qquad 
\left( \ba{c}  g_{1}( \la ) \\ g_2(\la)   \ea \right)  = \chi_+^{-1}(\la) \cdot \bs{H}(\la)\;.
\enq
Above, taking into account that $s<0$, we have set 
\beq
g_{a; \iota_s}(t) \; = \; (t-z_0)^{-\iota_s}  g_{a}(t) \quad \e{with} \; \; \iota_s = k \quad \e{for}  \; -k < s < -(k-1) \;.
\enq
It follows from the asymptotic behaviour for $\chi_+(\la)$ at large $\la$ that $g_1 \in \mc{F}\big[ H_{s-1/2} \big]$
and $g_2 \in \mc{F}\big[ H_{s+ 1/2} \big]$. Recall that Theorem \ref{Theorem conte transfo Cauchy sur Fourier espaces Hs}
ensures that the $\pm$ boundary values $\mc{C}_{\pm}$ of the 
Cauchy transform on $\R$ are continuous operators on $H_{\tau}(\R)$ for any  $|\tau| < 1/2$. 
Thus, $\mc{C}_{\pm}[g_{1; \iota_s}] \in H_{s+k-1/2}(\R)$ as well as $\mc{C}_{\pm}[g_{2; \iota_s-1}]  \in H_{s+k-1/2 }(\R)$, which implies:
\beq
\label{valuse} \wh{\bs{H}}_{a;\pm} \in \mc{F}\big[ H_{s_a}(\R) \big]\qquad \e{with}\,\, s_1 = s - 1/2\,\,\e{and}\,\,s_2 = s + 1/2\;. 
\enq
Equation \eqref{ecriture condition croissance composantes Phi} ensures that, uniformly in $\mu >0$,
\beq
\forall a \in \{1,2\},\qquad \Int{\R}{} \big| \Ups_{a}(\la\pm \i \mu)   \big|^2\,\big( 1 + |\la| + |\mu| \big)^{ 2 s_a }\,\dd  \la  < C \; \;. 
\label{ecriture estimation croissance integrale Ups sur "tranches de R"}
\enq
The discontinuity equation satisfied by $\Phi$ along with $\wh{\bs{H}}_{a;+} - \wh{\bs{H}}_{a;-} \; = \; g_a$ guarantee that 
$\Ups_{a} \in \mc{O}(\Cx\setminus \R)$ admits $\mc{F}\big[ H_{s_a}(\R) \big]$ $\pm$ boundary values that are equal. 
Then, straightforward manipulations show that, in fact, $\Ups$ is entire. Furthermore, 
for any $\ell \in \mathbb{N}$ such that $s_a + \ell > -1/2$ and for any $\mu > |{\rm Im}\, z |$,  we have: 
\beq
\partial_{z}^{\ell}\Ups_a(z)  \; =  \;  \sul{\eps = \pm }{}  \eps
\Int{\R}{} \f{ \ell!\,\Ups_a(\la+\i\eps \mu)  }{ (\la + \i\eps \mu - z )^{\ell +1} }\,\f{ \dd \la }{ 2 \i \pi} \;. 
\enq
Thus
\bem
\big| \partial_{z}^{\ell}\Ups_a(z) \big|  \; \leq \; \f{1}{\pi} \max_{\eps = \pm} 
\bigg( \int_{\R }^{}  \f{ \big( 1 + |\la| + |\mu| \big)^{ -2 s_a } }{ |\la + \i\eps \mu - z |^{2(\ell+1)} }\, \dd \la  \bigg)^{1/2}
\bigg( \int_{\R }^{}  |\Ups_a(\la+ \i\eps \mu) |^2\,\big( 1 + |\la| + |\mu| \big)^{ 2 s_a }\,\dd \la  \bigg)^{1/2} 
\end{multline}
where the last integral factor is bounded.
So far, the parameter $\mu$ was arbitrary. We stress that the constant $C$ in \eqref{ecriture estimation croissance integrale Ups sur "tranches de R"} is uniform in $\mu$. Thus taking $\mu=2|z|$ and assuming that $|z|>1/2$, we find:
\beq
\big| \partial_{z}^{\ell}\Ups_a(z) \big|  \; \leq \; C^{\prime}\,|z|^{-(s_a+\ell+1/2)}\cdot  
\bigg( \int_{\R }^{}  \f{ \big(|\la| + 2 \big)^{ -2 s_a } }{ \big[ (\la-1)^2 + 1 \big]^{\ell+1} }\,\dd \la  \bigg)^{1/2} \;. 
\label{ecriture bornage plus fin de Upsa}
\enq
In particular, reminding the values of $s_a$ in \eqref{valuse}, we find that $\partial_{z}^{k - 1}\Ups_{2}(z)$ and $\partial_{z}^{k}\Ups_{1}(z)$ are entire and bounded, so they must be constant.
These constants are zero due to \eqref{ecriture bornage plus fin de Upsa}. Hence, there exist polynomials $P_1 \in \Cx_{k-1}[X]$ and $ P_2 \in  \Cx_{k-2}[X]$ such that 
\beq
\label{A55}\Ups(z) \; = \; \left(  \ba{c} P_1(z) \\ P_2(z) \ea   \right) \; . 
\enq
Reciprocally, it is readily seen that the piecewise analytic vector
\beq
\Phi(\la) \; = \;  \chi(\la) \cdot \wh{\bs{H}}(\la)  \; +  \;  \chi(\la) \cdot \left( \ba{c} P_1(z)  \\ P_2(z)   \ea \right) 
\qquad \e{with}  \quad P_a \in \Cx_{k-a}[X]\; \; \;  \e{for}\; \;  -k< s < -(k-1) 
\label{ecriture solution RHP vectoriel sur Hs general s negatif}
\enq
provides solutions to the Riemann--Hilbert problem for $\Phi$. 

From now on, we focus on the case $k = 1$, i.e. $h \in H_{s}(\intff{ 0 }{ \ov{x}_N })$ for $-1<s<0$. Then, it follows from Lemma~\ref{Proposition corresp operateur et WH factorisation} 
that any solution to $\msc{S}_{N;\ga}[\varphi] \; = \;  h$ takes the form $\varphi =  \wt{\msc{W}}_{\vartheta; z_0}[h_{\mf{e}}]$, with:
\beq
 \mc{F}\big[ \, \wt{\msc{W}}_{\vartheta; z_0}[ h_{\mf{e}} ] \big](\la) \; = \; \Phi_{1;+}(\la) \; = \; 
 \chi_{11;+}(\la) \cdot (\la-z_0) \mc{C}_+[  f_{ 1;z_0 }](\la) \; + \; 
\chi_{12;+}(\la) \cdot \mc{C}_+[f_{2;z_0 }](\la) 
\; + \; \vartheta \cdot \chi_{11;+}(\la)  
\label{ecriture TF de solution fondamentale}
\enq
with $f_{a;z_0}$'s given by \eqref{fdff}. %

\vspace{2mm} 
It is then readily inferred from the asymptotic expansion for $\chi$ at $\la \tend \infty$ given in Lemma \ref{Lemme DA de chi ecriture explicite}, and from the jump
conditions satisfied by $\chi$, that indeed  $\Phi_{1;+} \in \mc{F} \big[H_s( \intff{ 0 }{ \ov{x}_N } ) \big]$. 
Also the continuity on $\mc{F}\big[ H_{\tau}(\R) \big]$ with $|\tau|<1/2$ of the $\pm$ boundary values $\mc{C}_{\pm}$ of the Cauchy transform, \textit{cf}. 
Theorem \ref{Theorem conte transfo Cauchy sur Fourier espaces Hs}, ensures that 
\beq
\norm{\Phi_{1;+}}_{ \mc{F}[H_s(\R)] } \; \leq  \; C\,\norm{ h_{\mf{e}} }_{H_s(\R)}\;,
\enq
which in turn implies the bound \eqref{ecriture continuite Hs de transformee tildee}.

It solely remains to prove the regularised expression \eqref{ecriture forme generale solution tel que predite par RHP dans Hs}. 
Given $h\in \mc{C}^{1}(\intff{0}{\ov{x}_N})$ it is clear that $h\in H_{s}(\intff{0}{ \ov{x}_N } )$ for any $s<1/2$. 
We chose the specific extension $h_{\mf{e}}=h$. Then, it follows from the previous discussion that 
$\wt{ \msc{W} }_{\vartheta;z_0}[h]\in H_{s}( \intff{0}{ \ov{x}_N })$. The integral in the right-hand side of \eqref{ecriture forme generale solution tel que predite par RHP dans Hs}, 
considered in the Riemann sense, defines a continuous function on $\intff{0}{\ov{x}_N}$, that we denote momentarily $\wt{\msc{V}}_{\vartheta;z_0}[h]$.
Now, for any $f\in\mc{C}^{\infty}(\intff{0}{\ov{x}_N})$, starting with the expression \eqref{ecriture inverse sous forme abstraite Fourier}
for $\wt{\msc{W}}_{\vartheta; z_0}[h]$,  we have:
\begin{eqnarray}
\big(f, \wt{\msc{W}}_{\vartheta; z_0}[h] \big) & = &  \big( \mc{F}[f], \Phi_{1;+} \big) \; = \; \Int{ \R }{}  \mc{F}[f^*](-\la)\cdot\Phi_{1;+}(\la)\,\dd \la =
 \Int{ \R+2\i \eps^{\prime} }{} \hspace{-2mm} \mc{F}[f^*](-\la) \cdot  \Phi_{1}(\la)\,\dd \la \\ 
&  = & \Int{ \R}{}  \Big( \mc{F}[ \ex{2\eps^{\prime}\bullet}f](\la) \Big)^* \cdot 
\mc{F}\big[  \ex{-2\eps^{\prime} \bullet} \wt{\msc{W}}_{\vartheta; z_0}[h] \big](\la)\dd \la 
\; =\; \big( f, \wt{\msc{V}}_{\vartheta; z_0}[h] \big) \; . 
\end{eqnarray}
in $\bullet$ represents the running variable with respect to which the Fourier transform.
There, we have equality $\wt{\msc{W}}_{\vartheta; z_0}[h] = \wt{\msc{V}}_{\vartheta;z_0}[h]$ for $h \in \mc{C}^{1}\cap H_s\big(\intff{0}{\ov{x}_N} \big)$. \qed

\vspace{2mm}

A \textit{priori}, the solutions $\wt{\msc{W}}_{\vartheta; z_0}[h_{\mf{e}}]$ given in \eqref{ecriture forme generale solution tel que predite par RHP dans Hs} 
has two free parameters $\vth$ and $z_0$. This "double" freedom is, however, illusory. 
\begin{lemme}
Given $z_0, z_0^{\prime} \in \Cx\setminus\R$ and $\vartheta\in \Cx$, there exists $\vartheta^{\prime} \in \Cx$ such that 
$\wt{\msc{W}}_{\vartheta; z_0} = \wt{\msc{W}}_{ \vartheta^{\prime} ; z_0^{\prime}}$. 
\end{lemme}
\Proof  By carrying out the decomposition $\la-z_0 = \la-\mu \; + \; \mu - z_0$ in the 
first term present in the integrand of  \eqref{ecriture forme generale solution tel que predite par RHP dans Hs}, we get that 
$\wt{\msc{W}}_{\vartheta; z_0} = \wt{\msc{W}}_{ \vartheta(z_0); \infty}$
\beq
\vartheta(z_0) \; = \; \vartheta \; - \; \Int{ \R + \i \eps^{\prime} }{} \hspace{-2mm} 
\f{\chi_{12}(\mu)\cdot\mc{F}\big[ h_{\mf{e}} \big](\mu)\cdot\ex{- \i\mu\ov{x}_N }  }{ \mu- z_0 }\cdot\f{\dd \mu }{ 2\i\pi }\;,
\enq
and $\infty$ means that one should send $z_0 \tend \infty$ under the integral sign of \eqref{ecriture forme generale solution tel que predite par RHP dans Hs}. \qed

\vspace{3mm}

Hence, with the above lemma in mind, we retrieve that the kernel of $\msc{S}_{N;\ga}$ is one dimensional when considered as an 
operator on $H_{s}(\intff{0}{\ov{x}_N})$, with $-1 < s < 0$. The above lemma of course implies that we can choose $z_0$ arbitrarily in 
\eqref{ecriture forme generale solution tel que predite par RHP dans Hs}.
It is most suitable to consider the specific form of solutions obtained by taking $z_0\tend 0$ with ${\rm Im}\,z_0 <0$. 
For $h \in \mc{C}^{1}( \intff{0}{\ov{x}_N} )$,  this yields a family of solution parametrized by $\vartheta \in \mathbb{C}$:
\beq
\wt{\msc{W}}_{\vartheta}[h] (\xi) \; = 
 \Int{ \R + 2 \i \eps^{\prime} }{} \hspace{-2mm} \f{ \dd \la }{ 2\pi }  \Int{ \R + \i \eps^{\prime} }{} 
\hspace{-1mm} \f{ \dd \mu }{ 2 \i \pi } \f{ \ex{- \i  \la\xi - \i\mu\ov{x}_N}  }{ \mu- \la }
\Bigg\{  \f{ \la }{ \mu }\cdot\chi_{11}(\la) \chi_{12}(\mu)  - \chi_{11}(\mu) \chi_{12}(\la) \Bigg\} 
\,\mc{F}\big[ h \big](\mu) 
\; + \;  \vartheta \Int{ \R + \i \eps^{\prime} }{}  \chi_{11}(\la)\,\ex{-\i\la\xi}\cdot\frac{\dd\la}{2\pi} \;. 
\label{ecriture famille a un param sols}
\enq
\noindent It is possible to find real-valued solutions to $\msc{S}_{N;\gamma}[\vp] = h$ by taking $h$ purely imaginary:
%
%Given $h$ purely imaginary, one expects to be able to construct real valued solutions to 
%$\msc{S}_{N;\ga}\big[ \vp \big](\xi) = h(\xi) $. Indeed, for well suited $a$'s, $\wt{\msc{W}}_{a}[h] (\xi) $ provides one with real solutions 
%to $\msc{S}_{N;\ga}\big[ \wt{\msc{W}}_{a}[h] \big](\xi) = h(\xi) $. Namely, one has the 
%
%
%
\begin{lemme}
Let $\vartheta \in \i \R$ and let $h\in \mc{C}^{1}(\intff{0}{\ov{x}_N})$ satisfy $h^* \, = \,  - h$. Then, $\big(\wt{\msc{W}}_{\vartheta}[h_{\mf{e}}]\big)^*  \; = \;  \wt{\msc{W}}_{\vartheta}[h_{\mf{e}}]$. 
%
%\label{ecriture proprietes realite solutions en variables rescaliees}
%
%
%
\end{lemme}

\Proof  From Lemma \ref{Lemme Ecriture diverses proprietes solution RHP chi}, we have $- \chi_{11}(- \la   ) = \big(\chi_{11}( \la^{*}  )  \big)^*$ and $\chi_{12}(- \la   ) =  \big(\chi_{12}( \la^{*})\big)^*$.  Hence, under the assumptions of the present lemma
\bem
\big( \wt{\msc{W}}_{\vartheta}[h](\xi) \big)^* \; =  \; 
 \Int{ \R  }{}  \f{ \dd \la }{ 2\pi } \; \Int{ \R  }{} 
 \f{ \dd \mu }{ - 2 \i\pi } 
\f{ \ex{ \i \la\xi + \i \mu \ov{x}_N }  \ex{ 2 \eps^{\prime} \xi + \eps^{\prime}\ov{x}_N}  }{ \mu- \la + \i \eps^{\prime} }
\Bigg\{ - \f{ \la - 2 \i \eps^{\prime}}{ \mu - \i \eps^{\prime} }\cdot 
\chi_{11}(- \la + 2 \i \eps^{\prime}  ) \chi_{12}(- \mu + \i \eps^{\prime} )  \\
+  \; \chi_{11}(-\mu + \i \eps^{\prime}  ) \chi_{12}(-\la + 2 \i \eps^{\prime}  ) \Bigg\} \underbrace{\mc{F}\big[h^*]}_{-{\cal F}[h]}(- \mu + \i \eps^{\prime} )  
\;\;  - \;  \underbrace{\vartheta^*}_{-\vartheta}  \Int{ \R  }{}  \ex{ \i\la \xi} \ex{ 2 \eps^{\prime} \xi }  \chi_{11}(-\la + \i\eps^{\prime})\,\f{ \dd \la }{ 2\pi }\;. 
\end{multline}
The change of variables $(\la, \mu) \mapsto (-\la, -\mu)$ in the first integral and $\la \mapsto -\la$ 
in the second integral entails the claim. \qed

%%%%%%%%%%%%%%%%%%%%%%%%%%%%%%%%%%%%%%%%%%%%%%%%%%%%%%%%%%%%%%%%%%%%%%%%%%%%%%%%%%%%%%%%%%%%%%%%%%%%%%%%%%%%%%%%%%%%%%%%%%%%%%%%%%%
%%%%%%%%%%%%%%%%%%%%%%%%%%%%%%%%%%%%%%%%%%%%%%%%%%%%%%%%%%%%%%%%%%%%%%%%%%%%%%%%%%%%%%%%%%%%%%%%%%%%%%%%%%%%%%%%%%%%%%%%%%%%%%%%%%%

%%%%%%%%%%%%%%%%%%%%%%%%%%%%%%%%%%%%%%%%%%%%%%%%%%%%%%%%%%%%%%%%%%%%%%%%%%%%%%%%%%%%%%%%%%%%%%%%%%%%%%%%%%%%%%%%%%%%%%%%%%%%%%%%%%%
%%%%%%%%%%%%%%%%%%%%%%%%%%%%%%%%%%%%%%%%%%%%%%%%%%%%%%%%%%%%%%%%%%%%%%%%%%%%%%%%%%%%%%%%%%%%%%%%%%%%%%%%%%%%%%%%%%%%%%%%%%%%%%%%%%%

%%%%%%%%%%%%%%%%%%%%%%%%%%%%%%%%%%%%%%%%%%%%%%%%%%%%%%%%%%%%%%%%%%%%%%%%%%%%%%%%%%%%%%%%%%%%%%%%%%%%%%%%%%%%%%%%%%%%%%%%%%%%%%%%%%%
%%%%%%%%%%%%%%%%%%%%%%%%%%%%%%%%%%%%%%%%%%%%%%%%%%%%%%%%%%%%%%%%%%%%%%%%%%%%%%%%%%%%%%%%%%%%%%%%%%%%%%%%%%%%%%%%%%%%%%%%%%%%%%%%%%%

\subsection{Local behaviour of the solution $\wt{\msc{W}}_{\vartheta}[h]$ at the boundaries}

In the present subsection, we shall establish the local behaviour of $\wt{\msc{W}}_{\vartheta}[h](\xi)$ at the boundaries of the segment $\intff{0}{\ov{x}_N}$, \textit{viz}. when 
$\xi \tend 0$ or $\xi \tend \ov{x}_N$, this in the case where $h \in \mc{C}^{1}(\intff{0}{\ov{x}_N})$. We shall demonstrate that there exist constants 
$C_0, C_{\ov{x}_N}$ affine in $\vartheta$ and depending on $h$, such that $\wt{\msc{W}}_{\vartheta}[h] $ exhibits the local behaviour 
\beq
\wt{\msc{W}}_{\vartheta}[h] (\xi) \; = \;  \f{ C_{0} }{ \sqrt{\xi} } \; + \; \e{O}(1) \quad \e{for} \quad \xi \tend 0^+ 
\qquad \e{and} \qquad 
\wt{\msc{W}}_{\vartheta}[h] (\xi) \; =  \; \f{ C_{\ov{x}_N} }{ \sqrt{\,\ov{x}_N- \xi } } \; + \; \e{O}(1) 
 \quad \e{for} \quad \xi \tend (\ov{x}_N )^- \;. 
\enq

Let us recall that our motivation for studying $\wt{\msc{W}}_{\vartheta}$ takes its origin in the need to construct the density of equilibrium measure $\rho_{\e{eq}}^{(N)}$ which solves 
${\cal S}_{N}[\rho_{\e{eq}}^{(N)}] = V^{\prime}$ as well as to invert the master operator $\mc{U}_N$ arising in the Schwinger-Dyson equations described in  \S~\ref{SDw1r}. 
The density has a square root behaviour at the edges what translates itself into a square root behaviour at $\xi=0$ and $\xi=\ov{x}_N$ in the rescaled variables. 
Having this in mind, we would like to enforce $C_0=C_{\ov{x}_N}=0$. For this purpose, we can exploit the freedom of choosing $\vth$. This is however not enough and, as it will be shown
in the present section, in order to have a milder behaviour of $\wt{\msc{W}}_{\vartheta}[h]$ at the edges, one also needs to impose a linear constraint on $h$. 
In fact, we shall see later on that the latter solely translates the fact that $h \in \msc{S}_{N;\ga}[H_s(\R)]$ with $0<s<\tf{1}{2}$. 

This informal discussion only serves as a guideline and motivation for the results of this subsection, in particular:
\begin{prop}
\label{Theorem caracterisation sous class reguliere solutions}
%
%e 
%
Let 
\beq
\label{I12a}\msc{I}_{12}[h] \; = \;   \Int{\R + \i \eps }{}  \f{ \ex{- \i\mu \ov{x}_N }  }{ \mu } \chi_{12}(\mu) 
			\cdot \mc{F}[h ](\mu) \cdot  \f{ \dd \mu }{ 2 \i\pi }\;. 
\enq
Then, for any $h \in \mc{C}^1(\intff{0}{\ov{x}_N})$ such that 
\beq
\label{I12b}\msc{I}_{11}[h] \; := \; \Int{\R + \i \eps }{}  \ex{- \i\mu \ov{x}_N }  \chi_{11}(\mu) 
			\cdot\mc{F}[h ](\mu)\cdot\f{ \dd \mu }{ 2 \i\pi } \; = \; 0
\enq
we have $\wt{\msc{W}}_{\mathscr{I}_{12}[h]}[h]   \in \big(L^{1}\cap L^{\infty} \big)(\intff{0}{\ov{x}_N})$.
\end{prop}
Prior to proving the above lemma, we shall first establish a lemma characterising the local behaviour
at $0$ and $\ov{x}_N$ of functions belonging to the kernel of $\msc{S}_{N;\ga}$. 
\begin{lemme}
\label{Lemme comportement local integral vs chi11}
The function
\beq
 \psi(\xi) = \Int{\R + 2 \i \eps^{\prime} }{} \hspace{-2mm} \ex{- \i \la\xi} \chi_{11}(\la)\,\f{ \dd \la }{ 2\pi }
\qquad satisfies \qquad \msc{S}_{N;\ga}[\psi](\xi)=0 \quad \xi \in \intoo{ 0 }{ \ov{x}_N }
\enq
and admits the asymptotic behaviour 
\beq
 \psi(\xi)  \; = \; \f{1}{ \i \sqrt{\pi\xi} } \; + \; \e{O}(1) \quad when \quad 
 \xi \tend 0^+  \qquad and \qquad 
\psi(\xi)  \; = \; \f{1}{  \i\sqrt{\pi(\ov{x}_N-\xi)} } \; + \; \e{O}(1) \quad when \quad 
 \xi \tend \big(\ov{x}_N\big)^- \;.
\enq

\end{lemme}

\Proof  One has, for $\xi \in \intoo{0}{\ov{x}_N}$ and in the distributional sense, 
\begin{eqnarray}
\msc{S}_{N;\ga}[\psi](\xi) & = & \Int{ \R }{} \f{\dd \mu}{2\pi} \Int{ \R }{ } \f{ \dd \la }{2\pi }\, \f{\ex{-\i\mu \xi}\,\mc{F}[S_{\ga}](\mu) }{ 2\i \pi \be }
\cdot \chi_{11;+}(\la)  \cdot \f{ \ex{ \i(\mu-\la)\ov{x}_N}-1 }{ \i(\la-\mu) }  \nonumber \\
& = &\Int{ \R }{} \f{\dd \mu}{2\pi} \Int{ \R- \i\eps }{ } \f{ \dd \la }{2\pi }\,\f{\ex{-\i\mu \xi}\,\mc{F}[S_{\ga}](\mu) }{ 2\i \pi \be }
\cdot    \f{ \chi_{11}(\la) \ex{ \i \mu\ov{x}_N}- \ex{\i \la \ov{x}_N} \chi_{11}(\la)  }{ \i (\la-\mu) } \nonumber \\
& = & \Int{ \R }{} \f{\dd \mu}{2\pi}\,\f{\ex{-\i\mu \xi} \,\mc{F}[S_{\ga}](\mu) }{ 2\i \pi \be }\,\Bigg\{ 
 - \chi_{11;+}(\mu) \; + \;  \Int{ \R-\i\eps }{ } \f{ \dd \la }{2\pi}\,\f{\chi_{11}(\la)\,\ex{ \i \mu\ov{x}_N} }{ \i (\la-\mu) } \Bigg\}  \nonumber \\
& = & -  \Int{ \R }{} \f{\dd \mu}{2\pi} \Bigg\{ \chi_{21;-}(\mu)\ex{-\i\mu \xi}  \, + \, \ex{ \i \mu (\ov{x}_N-\xi) } \chi_{21;+}(\mu) \Bigg\}   \;.
\end{eqnarray}
Note that, in the intermediate steps, we have used that $ \chi_{11;+}(\la)=\ex{\i \la \ov{x}_N}\,\chi_{11;-}(\la)$, and deformed the integral over $\la$ to the lower half-plane. Further, we have also used that 
\beq
\chi_{21;-}(\la) \, + \, \ex{ \i \la \ov{x}_N}\,\chi_{21;+}(\la) \; = \;  \frac{\mc{F}[S_{\ga}](\la) }{2\i \pi \be }\,\chi_{11;+}(\la) \;. 
\label{ecriture condition saut Chi 21 plus moins}
\enq
Observe that, when $0 < \xi < \ov{x}_{N}$, the function $\mu \mapsto \chi_{21;-}(\mu) \ex{-\i\mu \xi}$ 
(respectively, $\mu \mapsto \chi_{21;+}(\mu) \ex{ \i \mu (\ov{x}_N-\xi)}$) admits an analytic continuation to the lower (respectively upper)
half-plane that is Riemann-integrable on $\R-\i\tau$  (respectively  $\R+\i\tau$), this for any $\tau>0$, and that decays exponentially
fast when $\tau \tend + \infty$. As a consequence, 
\beq
\forall \xi \in \intoo{0}{\ov{x}_N},\qquad  \Int{ \R }{} \f{\dd \mu}{2\pi}\,\Big( \ex{-\i\mu \xi}\,\chi_{21;-}(\mu)  \, + \, \ex{ \i \mu (\ov{x}_N-\xi) }\,\chi_{21;+}(\mu) \Big) \; = \; 0\;,
\label{ecriture nullite integrale chi 12 plus et moins}
\enq
which is equivalent to $\msc{S}_{N;\ga}[\psi](\xi) =0$.

From the large-$\la$ expansion of $\chi(\la)$ given in Lemma \ref{Lemme DA de chi ecriture explicite}, we have for $\la \in \R + 2 \i\eps^{\prime} $, 
\beq
W(\la) \; \equiv \; \chi_{11}(\la) \; + \;   \f{\e{sgn}( {\rm Re}\,\la)\,\ex{\i\la \ov{x}_N}+ \i  }{ (- \i\la)^{1/2}} 
\; = \; \e{O}\Big(  |\la|^{ -3/2}  \Big)  \;. 
\enq
Hence,
\beq
\psi(\xi) \; = \;  \Int{\R + 2\i \eps^{\prime} }{} \hspace{-2mm} W(\la)\,\ex{-\i \la\xi}\cdot\frac{\dd \la }{ 2\pi }
\; -  \Int{\R + 2\i \eps^{\prime} }{} 
\hspace{-2mm} \f{ \e{sgn}({\rm Re}\,\la)\,e^{\i\la(\ov{x}_N- \xi)}}{ (-\i\la)^{ 1/2} }\cdot\frac{\dd \la }{ 2\pi }
\;+\; \Int{\R + 2 \i \eps^{\prime} }{} 
\hspace{-2mm} \frac{\ex{- \i\la \xi}}{(- \i \la)^{ 1/2}}\cdot\frac{\dd\lambda }{ 2\i\pi}\;.
\label{decomposition fonction psi en diverse parties plus ou moins singulieres}
\enq
By dominated convergence, the first term is $\e{O}(1)$ in the limit $\xi \tend 0^+$. The second term is also a $\e{O}(1)$. This is most easily seen by deforming the contour of integration
into a loop in $\mathbb{H}_+$ around $\i \R^+ + 2 \i \eps^{\prime}$, hence making the integral strongly convergent, and then applying dominated convergence. Finally, the third term
\eqref{decomposition fonction psi en diverse parties plus ou moins singulieres} can be explicitly computed by deforming
the integration contour to $- \i\R^+$:
\beq
\label{531}\Int{\R + 2 \i \eps^{\prime} }{} \hspace{-2mm} \f{ \ex{- \i \la\xi}  }{ (- \i\la)^{1/2} }\cdot\f{ \dd \la }{ 2\i\pi }
\; = \; \f{ - 1}{ \sqrt{\xi} } \Int{0}{+\infty} 
\Bigg\{ \f{1}{ (-\ex{ \i 0^+}t)^{1/2} } \; -  \; \f{1}{ (-\ex{-\i 0^+}t)^{1/2}} \Bigg\}\,\f{\ex{-t}\,\dd t }{2\pi} 
\; = \;  \f{ \Ga(1/2) }{\i\pi\sqrt{\xi}} = \frac{1}{\i\sqrt{\pi\xi}}\;.
\enq

Similar arguments ensure that the first and last term in \eqref{decomposition fonction psi en diverse parties plus ou moins singulieres}
are a $\e{O}(1)$ in the $\xi \tend (\ov{x}_N)^-$ limit. The middle term can be estimated as
\bem
\Int{\R + 2 \i \eps^{\prime} }{} 
\hspace{-2mm} \f{ \e{sgn}\big( {\rm Re}\,\la \big) }{ (- \i \la)^{  \f{1}{2} } } \ex{ \i \la(\ov{x}_N- \xi)}\cdot\f{ \dd \la }{ 2\pi } 
\; = \;  \f{ \ex{- 2(\ov{x}_N- \xi) \eps^{\prime} } }{ 2\pi \sqrt{\, \ov{x}_N- \xi  }} 
\Int{\R}{}  \f{ \e{sgn}(\la)\,\ex{ \i \la }\,\dd\la }{ \big(- \i \la + 2\eps^{\prime}(\ov{x}_N- \xi) \big)^{1/2} } \\
\label{532} \; = \; \i \f{ \ex{- 2(\ov{x}_N- \xi) \eps^{\prime} } }{ \pi \sqrt{\, \ov{x}_N- \xi  }} 
 \Int{0}{+\infty} \f{ \ex{-t}\,\dd t }{ \big(t + 2\eps^{\prime} (\ov{x}_N- \xi) \big)^{1/2}}%
\; = \; \f{\i}{ \sqrt{\pi(\ov{x}_N- \xi)  } } \; + \; \e{O}\big( \sqrt{\, \ov{x}_N- \xi  } \big) \;. 
\end{multline}
Putting together all of the terms entails the claim. \qed 

\vspace{2mm}

Before carrying on with the proof of Proposition~\ref{Theorem caracterisation sous class reguliere solutions}
we still need to prove a technical lemma relative to the large-$\la$ behaviour of certain building blocks 
of $\wt{\msc{W}}_{\vartheta}[h]$.

\begin{lemme}
\label{Lemme comportement integrales I1a lambda infini}

Let $h \in \mc{C}^{p+1}\big( \intff{ 0 }{ \ov{x}_N } \big)$. Then, the integrals 
\beq
\msc{J}_{1a}[h](\la) \; = \; \Int{ \R \, + \,  \i \eps^{\prime} }{} 
\f{ \chi_{1a}(\mu)\cdot\mc{F}[h ](\mu)\cdot\ex{- \i \mu \ov{x}_N}}{  \mu^{\de_{2a}}  \big( \mu - \la \big) } \cdot 
\,\f{ \dd \mu }{ 2 \i \pi } \qquad with \quad
\de_{2a}= \left\{ \ba{cc} 1 & \e{if} \, a=2 \\ 
			   0 & \e{if} \, a=1 \ea \right. 
\label{definition integral J a1 caligraphe}
\enq
admit the $|\la| \rightarrow \infty$, ${\rm Im}\,\la > 2 \eps^{\prime}>0$, asymptotic behaviour:
\beq
\msc{J}_{1a}[h](\la) \; = \; - \la^{-1} \msc{I}_{1a}[h] \; + \; \sul{k=1}{p} \f{ \mpzc{w}_{k;a}^{(1/2)}(\la) }{ \big(- \i\la \big)^{1/2}  \la^{k} }
\; + \; \sul{k=1}{p} \f{ \mpzc{w}_{k;a}^{(1)} }{ \la^{k+1} } \; + \; \e{O}(\la^{-(p+3/2)})\; \,
\label{ecriture DA I1a en lambda infini}
\enq
where 
\beq
 \mpzc{w}_{k;a}^{(1/2)}(\la) \; = \; \sul{ \ell = 0 }{ k-1 } \i^{k-\ell} h^{(k-\ell-1)}\big( \ov{x}_N \big)\Bigg\{
\e{sgn}({\rm Re}\,\la)\,\ex{ \i\la \ov{x}_N}\,[\chi_{\ell-\de_{2a}}]_{1a} 
\; + \;  \i\,[\chi_{\ell + 1 -\de_{2a}}]_{2a} \Bigg\}\;,
\label{definition fonctions wak}
\enq
and $ \mpzc{w}_{k;a}^{(1)}$ are constants whose explicit expression is given in the core of the proof \;. 
\end{lemme}

\Proof  The regularity of $h$ implies the following decomposition for its Fourier transform:
\beq
\mc{F}[h](\mu) \; = \;
 - \sul{k=0}{p} \f{ h^{(k)}\big(\ov{x}_N\big)\,\ex{ \i \mu \ov{x}_N} \; - \; h^{(k)}\big( 0 \big)   }{ (- \i\mu)^{k+1} } 
 \; + \; \f{ (-1)^{p+1}  }{ \big( \i \mu \big)^{p+1} } \Int{ 0 }{ \ov{x}_N }  h^{(p+1)}(t)\,\ex{ \i \xi \mu}\,\dd \xi \;. 
\enq
It gives directly access to the large-$\mu$ expansion:
\beq
\mu^{-\de_{2a}}\chi_{1a}(\mu)\cdot\mc{F}[h](\mu) \; = \; 
\sul{k=1}{p} \f{ T_{a}^{(k)}(\mu) }{  (-\i\mu)^{1/2} \mu^k} \; + \; R_{1a}^{(p)}(\mu)  \;.
\enq
The remainder is $R_{1a}^{(p)}(\mu) = \e{O}(\mu^{-p-3/2})$ when $\mu$ is large, whereas $T_{a}^{(k)}(\mu)$
remains bounded as long as ${\rm Im}\,\mu$ is bounded. Explicitly, these functions read:
\beq
T_{a}^{(k)}(\mu) \; = \; \sul{\ell=0}{k-1} \i^{k-\ell}\Big( h^{(k-1-\ell)}(0) - \ex{\i\mu \ov{x}_N } h^{(k-1-\ell)}(\ov{x}_N)\Big)
\Bigg\{ - \e{sgn}({\rm Re}\,\mu)\,\ex{ \i \mu \ov{x}_N}\,[\chi_{\ell - \de_{2a}}]_{a1} \, - \, 
\i\,[\chi_{\ell +1 - \de_{2a}}]_{a2} \Bigg\} 
\enq
where $\chi_{m}$ are the matrices appearing in the asymptotic expansion of $\chi$, see \eqref{cninfr}.
The integral of interest can be recast as 
\begin{eqnarray}
\msc{J}_{1a}[h](\la) & = & \sul{k=1}{p}\; \Int{ \R + \i\eps^{\prime} }{} 
\f{ T_{a}^{(k)}(\mu)\,\ex{-\i\mu \ov{x}_N} }{ (-\i\mu)^{1/2} \mu^k (\mu-\la) }\cdot\f{ \dd \mu }{2 \i\pi}
-\sul{\ell=0}{p} \f{1}{\la^{\ell +1} } \Int{ \R+ \i\eps^{\prime} }{} \mu^{\ell} R_{1a}^{(p)}(\mu)\,\ex{-\i \mu \ov{x}_N}\cdot\f{ \dd \mu }{2 \i\pi} \nonumber \\
& & +  \Int{ \R+ \i\eps^{\prime} }{} \f{\mu^{p+1}\,R_{1a}^{(p)}(\mu) }{\la^{p+1} (\mu- \la) }\,\ex{-\i \mu \ov{x}_N}\cdot\f{ \dd \mu }{2 \i\pi} \;. 
\label{ecriture decomposition integrale J1a caligraphe}
\end{eqnarray}
In virtue of the bound on $R_{1a}^{(p)}$, the last term is a $\e{O}\big( \la^{-p-\frac{3}{2}} \big)$. In order to obtain the asymptotic expansion of the first term we study the model integral 
\beq
\label{Jk}J_k(\la) \; = \; \Int{ \R+ \i\eps^{\prime} }{}  
\f{ \big( c_1\e{sgn}({\rm Re}\,\mu) - c_2 \ex{-\i\mu \ov{x}_N} \big) \big(\kappa_1 \ex{\i\mu \ov{x}_N}  - \kappa_2 \big) }
{ (-\i\mu)^{1/2} \mu^k (\mu-\la) }\cdot\f{ \dd \mu }{2\i\pi} \;, 
\enq
where ${\rm Im}\,\la > \eps^{\prime}$ while $c_1,c_2$ and $\kappa_1,\kappa_2$ are free parameters. 
By deforming appropriately the contours, we get that: 
\begin{eqnarray}
J_k(\la) & = &  \kappa_1 \f{  c_1\e{sgn}({\rm Re}\,\la)\ex{\i\la \ov{x}_N} - c_2 }{(-\i\la)^{1/2} \la^k } 
\; - \;  c_1 \kappa_2 \Oint{ -\Ga( \intff{ 0 }{ \i \eps^{\prime} }) }{} 
\f{ \e{sgn}\big( {\rm Re}\,\mu \big) }{  \big( - \i \mu \big)^{1/2} \mu^{k}(\mu-\la) }\cdot\f{  \dd \mu }{ 2 \i \pi } \\
& &  + \; c_1\kappa_1 (-\i)^k \Int{\eps^{\prime} }{+ \infty } \f{ t^{- k- 1/2} \ex{- t \ov{x}_N } }{ \i t - \la }\cdot\f{ \dd t  }{ \pi }
+ \; c_2 \kappa_2 \Oint{ -\Ga( \i\R^-) }{} \f{ \ex{-\i\mu \ov{x}_N }  \mu^{-k}  }{ \big(-\i \mu \big)^{1/2} (\mu - \la) }%
\cdot\f{ \dd \mu }{ 2\i\pi }   \\
&= &  \kappa_1\,\f{  c_1\e{sgn}({\rm Re}\,\la)\,\ex{ \i\la \ov{x}_N} - c_2 }{(- \i\la)^{1/2} \la^k }  
\; - \; \sul{q = 0 }{ p } \la^{-(q+1)}\, L_{k}^{(q)}
\; + \; \la^{-(p+2)}\Delta_{[p]}\,M_k(\la) \;. 
\end{eqnarray}
The constant $L_{k}^{(q)}$ occurring above is expressed in terms of integrals
\beq
L_{k}^{(q)}  = 
- c_1 \kappa_2 \Oint{ -\Ga( \intff{ 0 }{ \i \eps^{\prime} }) }{} 
\f{ \e{sgn}\big( {\rm Re}\,\mu \big)\cdot \mu^{q-k} }{  \big( - \i \mu \big)^{1/2} } \cdot\f{  \dd \mu }{ 2 \i \pi }
\; + \; c_1\kappa_1(-\i)^{k-q} \Int{\eps^{\prime} }{+ \infty }t^{q - k- 1/2}\cdot\ex{- \ov{x}_N\,t}\cdot\f{\dd t}{\pi} + c_2 \kappa_2 \Oint{ -\Ga( \i\R^-) }{} \f{ \ex{-\i\mu \ov{x}_N }\cdot\mu^{q -k}  }{ \big(-\i \mu \big)^{1/2 } }\,
\cdot\f{ \dd \mu }{ 2\i\pi }  \;. 
\enq
and the remainder function reads:
\bem
\Delta_{[p]}M_k (\la) \;= \; 
c_1\kappa_2  \hspace{-3mm}\Oint{ -\Ga(\intff{ 0 }{ \i \eps^{\prime} }) }{}  \hspace{-4mm}
\f{ \la \cdot \e{sgn}({\rm Re}\,\mu)\cdot \mu^{p+1-k} }{ ( \mu-\la) \cdot ( - \i \mu )^{1/2}   }\cdot\f{  \dd \mu }{ 2 \i \pi }
\; - \; c_2 \kappa_2 \Oint{ \Ga( \i\R^-) }{} \f{ \la\cdot\ex{-\i\mu \ov{x}_N }\cdot\mu^{p+1 -k}}{ \big(-\i \mu \big)^{1/2} (\mu-\la)}%
\cdot\f{ \dd \mu }{ 2\i\pi }  \\
\; + \; c_1\kappa_1 (-\i)^{k-p} \Int{\eps^{\prime} }{+ \infty }  \f{ \la \cdot t^{p - k+ 1/2} \cdot \ex{- t \ov{x}_N } }{(t+\i\la) }\cdot\frac{\dd t}{\pi} \;.  
\end{multline}
If we define:
\beq
\wt{ \mpzc{w}}_{k;a}^{(1)}  \; = \; -\sul{k=1}{p}\sul{\ell=0}{k-1} \Bigg\{L_{k}^{(q)}\quad \bigg|
\begin{array}{ll} c_1 \rightarrow -[\chi_{\ell-\de_{2a}}]_{1a} & \,\, \kappa_1 \rightarrow -\i^{k-\ell} h^{(k-\ell-1)}(\ov{x}_N) \\  
	  c_2 \rightarrow  \i\,[\chi_{\ell + 1 -\de_{2a}}]_{2a}  & \,\,\kappa_2 \rightarrow  \i^{k-\ell} h^{(k-\ell-1)}(0) \end{array}\Bigg\} \;. 
\enq
we obtain:
\beq
\sul{k=1}{p} \;\Int{ \R + \i \eps^{\prime} }{} 
\f{ T_{a}^{(k)}(\mu)\,\ex{- \i \mu \ov{x}_N} }{ (- \i \mu )^{1/2} \mu^k (\mu-\la) }\cdot\f{ \dd \mu }{ 2 \i \pi }
\; = \; \sul{k=1}{p}  \f{  \mpzc{w}_{k;a}^{(1/2)} (\la) }{ ( - \i \la )^{1/2} \la^k }  \;  + \;  \sul{q=0}{p} \f{ \wt{ \mpzc{w}}_{k;a}^{(1)}   }{ \la^{q+1} } 
\; + \; \e{O}( \la^{-(p+2)})\;. 
\enq
Furthermore, the above relation and equations \eqref{definition integral J a1 caligraphe} and \eqref{ecriture decomposition integrale J1a caligraphe}, ensure that 
\beq
\Int{\R+ \i \eps^{\prime}  }{} R_{1a}^{(p)}(\mu) \ex{-\i \mu \ov{x}_N} \cdot\f{ \dd \mu  }{ 2 \i\pi }
\; = \; \mathscr{I}_{1a}[h] \; + \;\wt{ \mpzc{w}}_{0;a}^{(1)}   \;. 
\enq
Hence, putting all the terms together, we arrive to the expansion \eqref{ecriture DA I1a en lambda infini} with the 
constants $\mpzc{w}_{k;a}^{(1)} $ given by 
\beq
\mpzc{w}_{k;a}^{(1)}   \; = \; \wt{ \mpzc{w}}_{k;a}^{(1)}  \; - \; \Int{\R+i \eps^{\prime}  }{} \mu^{k} R_{1a}^{(p)}(\mu) \ex{-\i \mu \ov{x}_N}  \cdot\f{ \dd \mu  }{ 2 \i\pi } \qquad k \geq 1 \;.
\enq
\qed

\Proof ({\textit{of Proposition} \ref{Theorem caracterisation sous class reguliere solutions}}). Given $h \in \mc{C}^{1}(\intff{0}{\ov{x}_N})$ and for $\xi \in \intoo{0}{\ov{x}_N}$, we can represent $\wt{\msc{W}}_{0}$ as
an integral taken in the Riemann sense\footnote{The fact that the integral \eqref{ecriture rep int solution wide tilde cal w} is well-defined in the Riemann sense will follow from the analysis carried out in this proof.}
\beq
\wt{\msc{W}}_{0}[h](\xi) \; = \; \Int{ \R + 2 \i \eps^{\prime} }{} \ex{- \i \la \xi} 
\Big[ \la\cdot\chi_{11}(\la)\msc{J}_{12}[h](\la) \;  -  \;  \chi_{12}(\la)\msc{J}_{11}[h](\la) \Big]\,
 \f{\dd \la }{2\pi}   \;,
\label{ecriture rep int solution wide tilde cal w}
\enq
where we remind that $\msc{J}_{1a}[h](\la)$ have been defined in \eqref{definition integral J a1 caligraphe}. Using the asymptotic expansions of Lemma~\ref{Lemme DA de chi ecriture explicite} for $\chi$ 
and those of Lemma~\ref{Lemme comportement integrales I1a lambda infini} for $\msc{J}_{1a}[h]$, we can decompose:
\begin{eqnarray}
\label{fdisguiu} \la\cdot\chi_{11}(\la)\msc{J}_{12}[h](\la)  -   \chi_{12}(\la)\msc{J}_{11}[h](\la)  & = & {\msc I}_{12}[h]\cdot\frac{{\rm sgn}({\rm Re}\,\la)\,e^{\i\la\ov{x}_N} + \i}{(-{\rm i}\la)^{1/2}} - \frac{\i{\msc I}_{11}[h]}{(-{\rm i}\la)^{1/2}} \\
& & + \;\frac{\mpzc{w}_{1;2}^{(1/2)}(\la)\big\{{\rm sgn}({\rm Re}\,\la)\,e^{\i\la\ov{x}_N} + \i\big\} - \i \mpzc{w}_{1;1}^{(1/2)}(\la)}{\i\la} + \e{O}(\la^{-3/2})\;. \nonumber
\end{eqnarray}
As a matter of fact, the coefficient of $1/(-{\rm i}\la)$ in this formula vanishes, as can be seen from the expressions \eqref{definition fonctions wak} for $\mpzc{w}_{1;a}^{(1/2)}$. 
Besides, integrating the $O(\la^{-3/2})$ in \eqref{ecriture rep int solution wide tilde cal w} yields a contribution remaining finite at $\xi = 0$ and $\xi = \ov{x}_N$, that we denote 
$\wt{\msc{W}}_{0}^{\e{c}}[h] \in \mc{C}^0\big(\intff{0}{\ov{x}_N}\big)$. Eventually, the effect of the first line of \eqref{fdisguiu} once inserted in \eqref{ecriture rep int solution wide tilde cal w} 
is already described in \eqref{531}-\eqref{532}. All in all, we find:
\beq
\wt{\msc{W}}_{0}[h](\xi) = {\msc I}_{12}[h]\Bigg\{\frac{\i}{\sqrt{\pi\xi}} + \frac{\i}{\sqrt{\pi(\ov{x}_N - \xi)}} + \e{O}\big(\sqrt{\,\ov{x}_N - \xi}\big)\Bigg\} - \frac{\i{\msc I}_{11}[h]}{\sqrt{\pi\xi}} + \wt{\msc{W}}_{0}^{\e{c}}[h](\xi)\;.
\enq
Since we have $\wt{\msc{W}}_{\vartheta}[h](\xi) = \wt{\msc{W}}_{0}[h](\xi) + \vartheta\psi(\xi)$ in terms of the function $\psi$ of Lemma~\ref{Lemme comportement local integral vs chi11}, we deduce that:
\beq
\wt{\msc{W}}_{{\msc I}_{12}[h]}[h](\xi) \; =  \; - \frac{\i{\msc I}_{11}[h]}{\sqrt{\pi\xi}} + O\big(\sqrt{\,\ov{x}_N - \xi}\big) + \wt{\msc{W}}_0^{\e{c}}[h](\xi)
\enq
and this function is continuous on $\intff{0}{\ov{x}_N}$ if and only if ${\msc I}_{11}[h] = 0$. \qed

\vspace{3mm}

\subsection{A well-behaved inverse operator of ${\cal S}_{N;\gamma}$}

Since, \textit{in fine}, we are solely interested in solutions belonging to $\big(L^{1}\cap L^{\infty} \big)(\intff{0}{\ov{x}_N})$ 
we shall henceforth only focus on $\wt{ \msc{W} }_{\mathscr{I}_{12}[h]}[h]$ and denote this specific solution as $\msc{W}_{N;\ga}[h]$. Furthermore, we shall restrict our reasoning to a class of functions such that $\mathscr{I}_{11}[h]=0$.  We now establish:

\begin{prop}
\label{Proposition existence et regularite et espace pour inverse msc SN gamma}

Let $0 < s < 1/2$. The subspace
\beq
\qquad \msc{X}_s\big( \intff{-\ga \ov{x}_N}{ (\ga+1) \ov{x}_N} \big) \; = \; 
\bigg\{ h \in H_s\big( \intff{-\ga \ov{x}_N}{ (\ga+1) \ov{x}_N} \big) \; : \; \mathscr{I}_{11}[h] \; = \; 0  \bigg\} \,
\enq
is closed in $H_s\big( \intff{-\ga \ov{x}_N}{ (\ga+1) \ov{x}_N} \big)$, and the operator:
\beq
\msc{S}_{N;\ga} \; : \; H_{s}\big( \intff{0}{\ov{x}_N} \big) \longrightarrow \msc{S}_{N;\ga} \big[ H_{s}\big( \intff{0}{\ov{x}_N} \big)  \big] \; = \; \msc{X}_s\big( \intff{-\ga \ov{x}_N}{ (\ga+1) \ov{x}_N} \big)
\enq
is continuously invertible. Its inverse is the operator
\beq
\msc{W}_{N;\gamma}\,:\,\msc{X}_s\big( \intff{-\ga \ov{x}_N}{ (\ga+1) \ov{x}_N} \big) \longrightarrow H_{s}\big(\intff{0}{\ov{x}_N}\big)\;.
\enq
On functions $h\in \mc{C}^{1}(\intff{0}{\ov{x}_N})$, it is defined as:
\beq
\msc{W}_{N;\ga}[h](\xi) \; =  \Int{ \R + 2\i\eps^{\prime} }{} \hspace{-2mm} \f{ \dd \la }{ 2\pi } \; \Int{ \R + \i\eps^{\prime} }{} 
\hspace{-1mm} \f{ \dd \mu }{ 2\i\pi } \f{ \ex{- \i\la\xi -\i \mu \ov{x}_N }  }{ \mu- \la }
\bigg\{   \chi_{11}(\la) \chi_{12}(\mu)  - \f{ \mu }{ \la }\cdot\chi_{11}(\mu) \chi_{12}(\la) \bigg\}  \mc{F}[h](\mu)	\;.
\label{definition de wN espace transforme}
\enq
For $h\in \mc{C}^{1}(\intff{0}{\ov{x}_N})$, $\msc{W}_{N;\ga}[h](\xi)$ is a continuous function on $\intff{0}{\ov{x}_N}$, which vanishes at least like a square root at $0$ and $\ov{x}_N$.
The  operator $\msc{W}_{N;\ga}$ extends continuously to $H_s(\intff{0}{\ov{x}_N})$, $0<s<\tf{1}{2}$ although the constant of continuity of $\msc{W}_{N;\ga}$ depends, a priori, on $N$. 
\end{prop}

Comparing \eqref{definition de wN espace transforme} with the double integral defining $\msc{W}_{N;\theta}$ in \eqref{ecriture famille a un param sols}, one observes that $\la/\mu$ in front of $\chi_{11}(\la)$ is absent 
and that there is an additional pre-factor $\mu/\la$ in front of $\chi_{12}(\la)$.

\Proof \textit{Continuity of $\msc{W}_{N;\ga}$.} 

Take $h\in \mc{C}^{1}\big(\intff{-\ga \ov{x}_N}{ (\ga+1) \ov{x}_N} \big)$. We first establish that $\msc{W}_{N;\ga}[h]$, as defined by \eqref{definition de wN espace transforme}, 
extends as a continuous operator from $H_s\big( \intff{-\ga \ov{x}_N}{ (\ga+1) \ov{x}_N} \big)$ to $H_s(\R)$. We observe that:
\beq
\mc{F}\big[ \ex{-2\eps^{\prime}*} \msc{W}_{N;\ga}[h] \big](\la) \; = \; \chi_{11}(\la+2\i\eps^{\prime})\cdot\mc{C}\big[ \wh{\chi}_{12}\mc{F}[h_{\epsilon^{\prime}}] \big](\la+\i\eps^{\prime}) 
\, - \,\f{ \chi_{12}(\la+2\i\eps^{\prime}) }{ \la+2\i\eps^{\prime} }\cdot\mc{C}\big[ \wh{\chi}_{11}\mc{F}[h_{\epsilon^{\prime}}] \big](\la+\i\eps^{\prime})  
\enq
with $h_{\epsilon^{\prime}}(\xi) \, = \, \ex{-\eps^{\prime}\xi}\,h(\xi)$, 
\beq
			\wh{\chi}_{11}(\mu) \, = \, (\mu+\i\eps^{\prime})\,\chi_{11}(\mu+\i\eps^{\prime})\,\ex{-\i (\mu+\i\eps^{\prime}) \ov{x}_N}  
\qquad \e{and} \qquad 			\wh{\chi}_{12}(\mu) \, = \, \chi_{12}(\mu+\i\eps^{\prime})\,\ex{-\i (\mu+\i\eps^{\prime})\ov{x}_N}     \; . 
\enq
It thus follows from the growth at infinity of $\chi_{11}$ and $\chi_{12}$ and the continuity on $H_{\tau}(\R)$, $| \tau | \leq 1/2$, 
of the transforms $\mc{C}_{\eps}$, where $\mc{C}_{\eps}[f](\la)= C[f](\la+\i \eps^{\prime})$,  \textit{cf}. Proposition \ref{Proposition continuite operateurs cauchy shiftes}, 
that 
\begin{eqnarray}
\norm{ \msc{W}_{N;\ga}[h] }_{H_{s}(\R)} \; & \leq & \;
 C\Bigg\{   \Norm{ \mc{C}_{\eps^{\prime} }\big[\wh{\chi}_{12}\cdot\mc{F}[h_{\epsilon^{\prime}}] \big] }_{\mc{F}[H_{s-1/2}(\R) ] }
\; + \;    \Norm{ \mc{C}_{\eps^{\prime} }\big[ \wh{\chi}_{11}\cdot\mc{F}[h_{\epsilon^{\prime}} ] \big] }_{\mc{F}[H_{s-1/2}(\R) ] }  \Bigg\}  \\
& \leq & C^{\prime}\Bigg\{ \,\Norm{\,\wh{\chi}_{12}\cdot\mc{F}[h_{\epsilon^{\prime}}] }_{\mc{F}[H_{s-1/2}(\R) ] }
\; + \;    \Norm{\,\wh{\chi}_{11}\cdot\mc{F}[h_{\epsilon^{\prime}}]  }_{\mc{F}[H_{s-1/2}(\R) ] }  \Bigg\} \\
& \leq & C^{\prime \prime}\,\norm{ h_{\epsilon^{\prime}} }_{ H_{s}( \R )  } 
\; \leq \;  C^{\prime\prime\prime}\,\norm{ h }_{ H_{s}( \intff{-\ga \ov{x}_N}{ (\ga+1) \ov{x}_N})  }\;. 
\end{eqnarray}

\Proof \textit{ The space $\msc{X}_s\big( \intff{-\ga \ov{x}_N}{ (\ga+1) \ov{x}_N} \big)$}. 

Given $h \in H_s\big( \intff{-\ga \ov{x}_N}{ (\ga+1) \ov{x}_N} \big)$, we have:
\beq
\big| \mathscr{I}_{11}[h] \big| \;  \leq \;  \bigg( \Int{ \R }{} (1+|\mu|)^{-2s} |\chi_{11}(\mu)|^2\,\dd \mu   \bigg)^{1/2}
\cdot \norm{  h }_{  H_s( \intff{-\ga \ov{x}_N}{ (\ga+1) \ov{x}_N}) }\;. 
\label{ecriture continuite sur Hs de I11}
\enq
As a consequence, $\mathscr{I}_{11}$ is a continuous linear form on $H_s\big( \intff{-\ga \ov{x}_N}{ (\ga+1) \ov{x}_N} \big)$. 
In particular, its kernel is closed, what ensures that $\msc{X}_s\big( \intff{-\ga \ov{x}_N}{ (\ga+1) \ov{x}_N} \big)$ is a  closed
subspace of  $H_s\big( \intff{-\ga \ov{x}_N}{ (\ga+1) \ov{x}_N} \big)$. We now establish that:
\beq
\msc{S}_{N;\ga}\big[ H_s\big( \intff{0}{  \ov{x}_N} \big) \big] \subseteq \msc{X}_s\big( \intff{-\ga \ov{x}_N}{ (\ga+1) \ov{x}_N} \big)\;.
\label{Ecriture Image de S inclue dans Xs}
\enq
Let $\vp \in \mc{C}^1\big( \intff{0}{  \ov{x}_N} \big) $ and define $h=\msc{S}_{N;\ga}[\vp]$. Then, using the jump condition \eqref{ecriture condition saut Chi 21 plus moins}:
\beq
\mathscr{I}_{11}[h] = \Int{0}{\ov{x}_N} \dd \eta\,\vp(\eta) \Int{ \R }{} \ex{- \i \mu \ov{x}_N}\,\chi_{11;+}(\mu)\,
\frac{\mc{F}[S_{\ga}](\mu)}{2{\rm i}\pi\beta}\,\ex{ \i \mu \eta} =  \Int{0}{\ov{x}_N} \dd \eta\,\vp(\eta) 
 \Int{ \R }{} \Big(\chi_{21;-}(\mu) \ex{\i\mu(\eta-\ov{x}_N)} \; + \; \chi_{21;+}(\mu) \ex{\i\mu\eta} \Big)\,\dd \mu
\enq
and this quantity vanishes according to \eqref{ecriture nullite integrale chi 12 plus et moins}. 
The equality can then be extended to the whole of $H_s\big( \intff{0}{  \ov{x}_N} \big)$, $0<s<\tf{1}{2}$ since $\mathscr{I}_{11}$ and $\msc{S}_{N;\ga}$
are continuous on this space and $\mc{C}^1\big( \intff{0}{  \ov{x}_N} \big)$ is dense in $H_s\big( \intff{0}{  \ov{x}_N} \big)$.

\Proof \textit{Relation to the inverse.}  

By definition, for any $h\in \big( H_s \cap \mc{C}^1 \big)(\intff{- \ga \ov{x}_N }{ (1+\ga) \ov{x}_N })$,  we have: 
\bem
 \wt{\msc{W}}_{\mathscr{I}_{12}[h]}[h] (\xi) \; =  \Int{ \R + 2\i\eps^{\prime} }{} \hspace{-2mm} \f{ \dd \la }{ 2\pi } \; \Int{ \R + \i\eps^{\prime} }{} 
\hspace{-1mm} \f{ \dd \mu }{ 2 \i \pi } \f{ \ex{- \i \la\xi - \i \mu \ov{x}_N }  }{ \mu- \la }
\bigg\{  \f{ \la }{ \mu } \cdot \chi_{11}(\la) \chi_{12}(\mu)  - \chi_{11}(\mu) \chi_{12}(\la) \bigg\} 
\cdot \mc{F}[h](\mu) \\
+ \bigg( \Int{\R + 2 \i \eps^{\prime} }{} \hspace{-2mm} \f{ \dd \la }{ 2\pi }\ex{- \i \la\xi} \chi_{11}(\la)  \bigg)
\cdot \bigg( \Int{\R + \i \eps^{\prime} }{} \hspace{-2mm} \f{ \dd \mu }{ 2 \i \pi } \f{ \ex{- \i \mu \ov{x}_N }  }{ \mu } \chi_{12}(\mu) 
			\cdot \mc{F}[h](\mu)      \bigg)   \\ 
\; =  \Int{ \R + 2 \i \eps^{\prime} }{} \hspace{-2mm} \f{ \dd \la }{ 2\pi } \; \Int{ \R + \i\eps^{\prime} }{} 
\hspace{-1mm} \f{ \dd \mu }{ 2 \i \pi } \f{ \ex{- \i \la\xi - \i \mu \ov{x}_N }  }{ \mu- \la }
\bigg\{   \chi_{11}(\la) \chi_{12}(\mu)  - \chi_{11}(\mu) \chi_{12}(\la) \bigg\} 
\cdot \mc{F}[h](\mu)	\\ 
- \bigg( \Int{\R + 2 \i \eps^{\prime} }{} \hspace{-2mm} \f{ \dd \la }{ 2\pi }\,\ex{- \i \la\xi}\,\f{ \chi_{12}(\la) }{ \la }  \bigg)
\cdot \underbrace{ 
\bigg( \Int{\R + \i \eps^{\prime} }{} \hspace{-2mm} \f{ \dd \mu }{ 2 \i \pi }\, \ex{- \i\mu \ov{x}_N }\,\chi_{11}(\mu) 
			\cdot \mc{F}[h](\mu)    \bigg)  
			}_{=0}  \; = \; \msc{W}_{N;\ga}[h](\xi) \;.  
\label{equation avec le regularisation trick}
\end{multline}
In the last line, we used the freedom to add a term proportional to ${\msc I}_{11}[h] = 0$, so that the combination retrieves the announced expression \eqref{definition de wN espace transforme}. 
The continuity of the linear functional ${\msc I}_{12}$ on $H_s(\intff{0}{\ov{x}_N})$ is proven analogously to \eqref{ecriture continuite sur Hs de I11}, hence ensuring the continuity of the operator
$\wt{\msc{W}}_{\msc{I}_{12}[h]}$. Since  both operators $\msc{W}_{N;\ga}$ and $\wt{\msc{W}}_{\msc{I}_{12}[h]}$ are continuous on $H_s(\intff{0}{\ov{x}_N})$ and coincide on ${\cal C}^1$ functions which form a dense subspace, they coincide on the whole $H_s(\intff{0}{\ov{x}_N})$. From there we deduce two facts:
\begin{itemize}
\item we indeed have $\msc{S}_{N;\ga}\big[\msc{W}_{N;\ga}[h] \big]\,  = \, h$, as a consequence of $\msc{S}_{N;\ga}\big[  \wt{\msc{W}}_{\msc{I}_{12}[h]}[h] \big]\,  = \, h$. This shows that the reverse inclusion 
to \eqref{Ecriture Image de S inclue dans Xs} holds as well. 

\item  The function $\msc{W}_{N;\ga}[h]$ is supported on  $ \intff{0}{\ov{x}_N} $ (and thus belongs to $H_s(\intff{0}{\ov{x}_N})$) since
Lemma~\ref{Proposition corresp operateur et WH factorisation} ensures that  $ \wt{\msc{W}}_{\msc{I}_{12}[h]}[h]$
is supported on $\intff{0}{\ov{x}_N}$ this for any $h \in H_s\big( \intff{-\ga \ov{x}_N}{ (\ga+1) \ov{x}_N} \big)\subseteq 
H_{\tau}\big( \intff{-\ga \ov{x}_N}{ (\ga+1) \ov{x}_N} \big)$ with $0<s< 1/2 $ and $-1<\tau<0$.  
\end{itemize}

\Proof \textit{Local behaviour for $\mc{C}^1(\intff{0}{\ov{x}_N})$ functions.}

It follows from a slight improvement of the local estimates carried out in the proof of Proposition~\ref{Theorem caracterisation sous class reguliere solutions} that, 
given $h\in \mc{C}^{1}(\intff{0}{\ov{x}_N})$, we have:
\beq
\msc{W}_{N;\ga}[h](\xi) \; = \;  C_{L}^{(0)} + C_{L}^{(1/2)}\sqrt{\xi} + \e{O}\big( \xi \big)\,\qquad \msc{W}_{N;\ga}[h](\xi) \; = \;  C_{R}^{(0)} + C_{R}^{(1/2)}\sqrt{\,\ov{x}_N-\xi} + \e{O}\big(\ov{x}_N-\xi \big) \;  ,
\nonumber
\enq
form some constants $C_{L/R}^{(a)}$ with $a \in \{0,1/2\}$. It thus remains to check that 
$C_{L}^{(0)} = C_{R}^{(0)} =0$. It follows also from the proof of Proposition~\ref{Theorem caracterisation sous class reguliere solutions} that $\msc{W}_{N;\ga}[h]$
is, in fact, continuous on $\R$. Since $\e{supp}\big[ \msc{W}_{N;\ga}[h] \big]=\intff{ 0 }{ \ov{x}_N } $, the function has to vanish at $0$ and $\ov{x}_N$
so as to ensure its continuity. Thence, $ C_{L}^{(0)}= C_{R}^{(0)}=0$. \qed

%%%%%%%%%%%%%%%%%%%%%%%%%%%%%%%%%%%%%%%%%%%%%%%%%%%%%%%%%%%%%%%%%%%%%%%%%%%%%%%%%%%%%%%%%%%%%%%%%%%%%%%%%%%%%%%%%%%%%%%%%%%%%%%%%%%%%%
%%%%%%%%%%%%%%%%%%%%%%%%%%%%%%%%%%%%%%%%%%%%%%%%%%%%%%%%%%%%%%%%%%%%%%%%%%%%%%%%%%%%%%%%%%%%%%%%%%%%%%%%%%%%%%%%%%%%%%%%%%%%%%%%%%%%%%

%%%%%%%%%%%%%%%%%%%%%%%%%%%%%%%%%%%%%%%%%%%%%%%%%%%%%%%%%%%%%%%%%%%%%%%%%%%%%%%%%%%%%%%%%%%%%%%%%%%%%%%%%%%%%%%%%%%%%%%%%%%%%%%%%%%%%%
%%%%%%%%%%%%%%%%%%%%%%%%%%%%%%%%%%%%%%%%%%%%%%%%%%%%%%%%%%%%%%%%%%%%%%%%%%%%%%%%%%%%%%%%%%%%%%%%%%%%%%%%%%%%%%%%%%%%%%%%%%%%%%%%%%%%%%

%%%%%%%%%%%%%%%%%%%%%%%%%%%%%%%%%%%%%%%%%%%%%%%%%%%%%%%%%%%%%%%%%%%%%%%%%%%%%%%%%%%%%%%%%%%%%%%%%%%%%%%%%%%%%%%%%%%%%%%%%%%%%%%%%%%%%%
%%%%%%%%%%%%%%%%%%%%%%%%%%%%%%%%%%%%%%%%%%%%%%%%%%%%%%%%%%%%%%%%%%%%%%%%%%%%%%%%%%%%%%%%%%%%%%%%%%%%%%%%%%%%%%%%%%%%%%%%%%%%%%%%%%%%%%

%%%%%%%%%%%%%%%%%%%%%%%%%%%%%%%%%%%%%%%%%%%%%%%%%%%%%%%%%%%%%%%%%%%%%%%%%%%%%%%%%%%%%%%%%%%%%%%%%%%%%%%%%%%%%%%%%%%%%%%%%%%%%%%%%%%%%%
%%%%%%%%%%%%%%%%%%%%%%%%%%%%%%%%%%%%%%%%%%%%%%%%%%%%%%%%%%%%%%%%%%%%%%%%%%%%%%%%%%%%%%%%%%%%%%%%%%%%%%%%%%%%%%%%%%%%%%%%%%%%%%%%%%%%%%

\subsection{${\cal W}_{N}$: the inverse operator of $\mc{S}_N$}
\label{Sous section construction inverse a l'operateur SN}

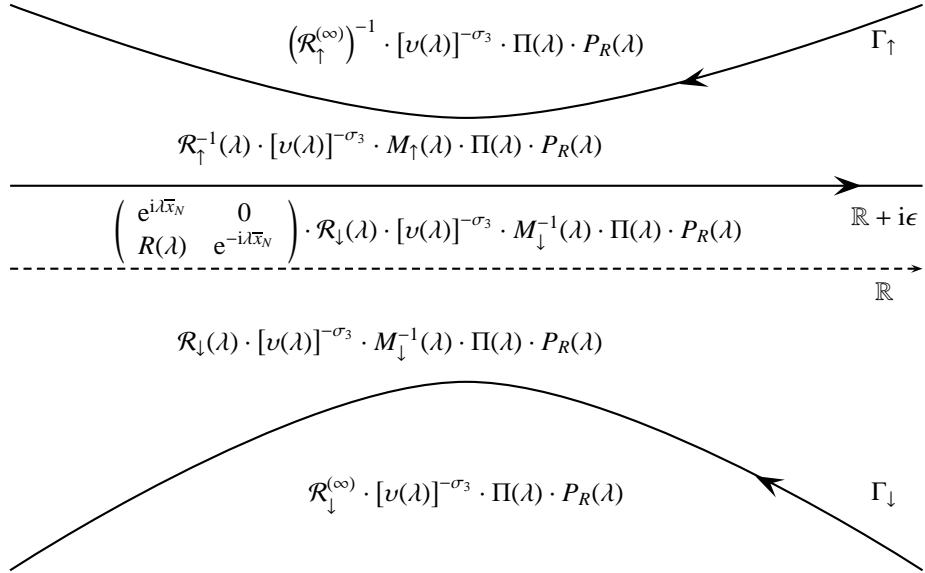
\begin{figure}[h]
\begin{center}

\begin{pspicture}(12,9)

% axe reel 

\psline[linestyle=dashed, dash=3pt 2pt]{->}(0,4.5)(12,4.5)

\rput(11.5,4.2){$\R$}

% Courbe de saut R + i eps

\psline(0,5.6)(12,5.6)

\psline[linewidth=3pt]{->}(11,5.6)(11.2,5.6)

\rput(11.5,5.2){$\R + \i \eps $}

%courbe de saut Gamma up

\pscurve(0,8)(6,6.5)(12,8)

\rput(11.5,7.5){$\Ga_{\ua}  $}

\psline[linewidth=3pt]{->}(9,7)(8.8,6.95)

%courbe de saut Gamma down

\pscurve(0,0.5)(6,3)(12,0.5)

\rput(11.5,1.5){$\Ga_{\da}  $}

\psline[linewidth=3pt]{->}(10,1.65)(9.8,1.75)

%definition de la matrice Phi

\rput(6,7.5){$    \Big(\mc{R}_{ \ua }^{(\infty) } \Big)^{-1} \cdot \big[\ups(\la) \big]^{-\sg_3}  \cdot \Pi(\la) \cdot P_R(\la) $}

\rput(5,6.1){$ \mc{R}_{ \ua }^{-1}(\la)  \cdot \big[\ups(\la) \big]^{-\sg_3} \cdot M_{\ua}(\la)  \cdot \Pi(\la) \cdot P_R(\la)$}
\rput(5.5,5){$\left( \ba{cc}  \ex{\i\la \ov{x}_N} & 0 \\ 
				  R(\la)     & \ex{-\i \la \ov{x}_N} \ea \right)  \cdot \mc{R}_{\da }(\la) \cdot \big[\ups(\la) \big]^{-\sg_3} \cdot M_{\da}^{-1}(\la)
\cdot \Pi(\la) \cdot P_R(\la)$}
\rput(5,3.5){$ \mc{R}_{ \da }(\la) \cdot \big[\ups(\la) \big]^{-\sg_3} \cdot M_{\da}^{-1}(\la)
\cdot \Pi(\la) \cdot P_R(\la)$}
\rput(6,1.5){$ \mc{R}_{ \da }^{(\infty)}  \cdot  \big[\ups(\la) \big]^{-\sg_3} \cdot \Pi(\la) \cdot P_R(\la)$}
\end{pspicture}

\caption{Piecewise definition of the matrix $\chi$ (at $\gamma \rightarrow +\infty$).
The curves $\Ga_{\ua/\da}$ separate all poles of $\la \mapsto \la R(\la)$ from $\R$ and are such that $\e{dist}(\Ga_{\ua/\da}, \R) > \de$
for some $\de >0$ but sufficiently small. 
\label{Figure definition sectionnelle de la matrice chi a gamma infty} }
\end{center}
\end{figure}

In order to construct the inverse to $\msc{S}_N$, we should take the limit $\ga \tend +\infty$ in the previous formulae. It so happens
that this limit is already well-defined at the level of the  
solution to the Riemann--Hilbert problem for $\chi$ as defined through Figure~\ref{Figure definition sectionnelle de la matrice chi}. 
More precisely, from now on, let $\chi$ be as defined in Figure~\ref{Figure definition sectionnelle de la matrice chi a gamma infty}
where the matrix $\Pi$ is as defined through \eqref{ecriture eqn int sing pour matrice Pi moins}-\eqref{ecriture rep int matrice Pi}
with the exception that one should send $\ga \tend +\infty$ in the jump matrices for $\Psi$ 
\eqref{ecriture saut Psi hors de R}-\eqref{ecriture saut Psi sur de R}. Note that, in this limit, $G_{\Psi}=I_2$ on $\R+\i \eps$, 
\textit{viz}. $\Psi$ is continuous across $\R+\i \eps$.  Then, we can come back to the inversion of the initial operator ${\cal S}_{N}$ in unrescaled variables -- compare 
\eqref{definition operateur regularise S N gamma}, \eqref{ecriture changement de variable pour arriver au RHP} and \eqref{ecriture equation sing en variables reduites}.

\begin{prop}
\label{Proposition invertibilite operateur S}

Let $0 < s < 1/2$. The operator 
$\mc{S}_N \; : \; H_{s}\big( \intff{a_N}{b_N} \big) \longrightarrow H_{s}\big( \R \big)$
is continuous and invertible on its image:
\beq
\label{x}
\mf{X}_{s}\big( \R \big) = \Bigg\{H \in H_{s}(\R)\;\; :\;\;\Int{\R+\i \eps }{} \chi_{11}(\mu) \mc{F}[H](N^{\a}\mu)\ex{- \i N^{\a} \mu b_N}\cdot\f{\dd\mu}{ 2 \i \pi }  \; = \; 0\Bigg\}\;.
\enq 
The inverse is then given by the operator $\mc{W}_N\,:\,\mf{X}_{s}(\R) \longrightarrow H_s(\intff{a_N}{b_N})$ defined in \eqref{definition operateur WN}:
\beq
\label{forumule explicite pour WN de 1bis}\mc{W}_N[H](\xi) \; = \; \f{ N^{2\a} }{ 2\pi \be}  
\Int{ \R + 2 \i \eps }{} \f{ \dd \la }{ 2 \i \pi }\,\Int{ \R + \i \eps }{} \f{ \dd \mu }{ 2 \i \pi }\, 
\f{ \ex{- \i N^{\a}\la(\xi-a_N) }}{ \mu- \la }
\bigg\{   \chi_{11}(\la) \chi_{12}(\mu)  - \f{ \mu }{ \la }\cdot\chi_{11}(\mu) \chi_{12}(\la) \bigg\} 
\,\ex{- \i N^{\a} \mu b_N} \mc{F}\big[H\big](N^{\a}\mu) 
\enq
with $\chi$ being understood as defined in Figure~\ref{Figure definition sectionnelle de la matrice chi a gamma infty}.
\end{prop}

\Proof Starting from the expression for the inverse operator to $\msc{S}_{N;\ga}$ and
 carrying out the change of variables, one obtains an operator $\mc{W}_{N;\ga}$ which corresponds to the 
inverse of the operator $\mc{S}_{N;\ga}$. Then, in this expression we replace $\chi$ at finite $\ga$
by the solution $\chi$ at $\ga \tend +\infty$, as it is given in Figure~\ref{Figure definition sectionnelle de la matrice chi a gamma infty}. 
This corresponds to the operator  $\mc{W}_N$, as defined in \eqref{definition operateur WN}. 
One can then verify explicitly on the integral representation for $\mc{W}_N$ by using certain elements of the 
Riemann--Hilbert problem satisfied by $\chi$ that the equation 
$\mc{S}_N\big[\mc{W}_N[H] \big]=H$ does hold on $\intff{a_N}{b_N}$. All the other conclusions of the theorem can be 
proved similarly to Proposition \ref{Proposition existence et regularite et espace pour inverse msc SN gamma}. \qed

\vspace{0.2cm}

\noindent We describe a symmetry of the integral transform $\mc{W}_N$ that
will appear handy in the remaining of the text.

\begin{lemme}
\label{Lemme propriete symetrie R et L op WN}
The operator $\mc{W}_N$ has the reflection symmetry: 
\beq
\mc{W}_N[H](a_N+b_N-\xi) \; =\; -\mc{W}_N[H^{\wedge}](\xi) 
\enq
where we agree upon $H^{\wedge}(\xi) = H(a_N+b_N-\xi)$. 

\end{lemme}

\Proof It follows from the jump conditions satisfies by $\chi$ and from Lemma \ref{Lemme Ecriture diverses proprietes solution RHP chi}
that, for $\la \in \R$,
\beq
\chi_{11;+}(-\la) \; = \; \ex{-\i\la\ov{x}_N}\cdot \chi_{11;+}(\la)  \quad \e{and} \quad 
\chi_{12;+}(-\la) \; = \; \ex{-\i\la\ov{x}_N}\cdot \Big( \chi_{12;+}(\la) -\la \, \chi_{11;+}(\la)  \Big) \;. 
\enq
Upon squeezing the contours of integration in the integral representation for $\mc{W}_N$
to $\R$ we get, in particular, the $+$ boundary values of $\chi_{1a}$. It is then enough to implement the change of variables
$(\la,\mu,\eta) \mapsto (-\la,-\mu,b_N+a_N-\eta)$ and observe that, all in all, the unwanted terms
cancel out. \qed

\vspace{0.2cm} 
\noindent In the case of a constant argument (which clearly does \textit{not} belong to $\mf{X}_s(\R)$) the expression for $\mc{W}_N$ simplifies:

\begin{lemme}
 
The function $\mc{W}_N[1](\xi)$ admits the one-fold integral representation 
\beq
\mc{W}_N[1](\xi) \; = \; -\frac{N^{\a}\,\chi_{12;+}(0)}{2\i\pi\be} \Int{ \R +\i \eps^{\prime} }{} \f{ \chi_{11}(\la) }{ \la }\ex{-\i N^{\a} \la (\xi-a_N) }
\cdot \f{ \dd \la }{ 2\i \pi } \;. 
\label{forumule explicite pour WN de 1ter}
\enq

\end{lemme}

\Proof  Starting from the representation \eqref{definition operateur WN} we get, for any $\xi \in \intoo{a_N}{b_N}$,  
\beq
\mc{W}_N[1](\xi) \; = \; \f{ N^{\a} }{ 2\pi \i \be}  
\Int{ \R + 2 \i \eps }{} \f{ \dd \la }{ 2 \i \pi } \Int{ \R + \i \eps }{} \f{ \dd \mu }{ 2 \i \pi } 
\f{ \ex{- \i N^{\a}(\xi-a_N) \la}  }{ \mu- \la }
\bigg\{   \f{1}{\mu}\cdot\chi_{11}(\la) \chi_{12}(\mu)  - \f{ 1}{ \la }\cdot \chi_{11}(\mu) \chi_{12}(\la) \bigg\} 
\cdot \Big(1\, - \, \ex{-\i \mu \ov{x}_N}  \Big) \;. 
\label{forumule explicite pour WN de 1fdag}
\enq
One should then treat the terms involving the function $1$ and $\ex{-\i \mu \ov{x}_N} $ arising in the right-hand side differently. The part 
involving $1$ is zero as can be seen by deforming the $\mu$-integral up to $+\i \infty$. In what concerns the 
part involving  $\ex{-\i \mu \ov{x}_N} $, we deform the $\mu$-integral up to $-\i \infty$ by using the jump conditions 
$\ex{-\i \la \ov{x}_N} \chi_{1a;+}(\la)\, =\, \chi_{1a;-}(\la) $. Solely the pole at $\mu=0$ contributes, hence leading to \eqref{forumule explicite pour WN de 1ter}. \qed

%%%%%%%%%%%%%%%%%%%%%%%%%%%%%%%%%%%%%%%%%%%%%%%%%%%%%%%%%%%%%%%%%%%%%%%%%%%%%%%%%%%%%%%%%%%%%%%%%%%%%%%%%%%%%%%%%%%%%%%%%%%%%%%%%%%%
%%%%%%%%%%%%%%%%%%%%%%%%%%%%%%%%%%%%%%%%%%%%%%%%%%%%%%%%%%%%%%%%%%%%%%%%%%%%%%%%%%%%%%%%%%%%%%%%%%%%%%%%%%%%%%%%%%%%%%%%%%%%%%%%%%%%

%%%%%%%%%%%%%%%%%%%%%%%%%%%%%%%%%%%%%%%%%%%%%%%%%%%%%%%%%%%%%%%%%%%%%%%%%%%%%%%%%%%%%%%%%%%%%%%%%%%%%%%%%%%%%%%%%%%%%%%%%%%%%%%%%%%%
%%%%%%%%%%%%%%%%%%%%%%%%%%%%%%%%%%%%%%%%%%%%%%%%%%%%%%%%%%%%%%%%%%%%%%%%%%%%%%%%%%%%%%%%%%%%%%%%%%%%%%%%%%%%%%%%%%%%%%%%%%%%%%%%%%%%

%%%%%%%%%%%%%%%%%%%%%%%%%%%%%%%%%%%%%%%%%%%%%%%%%%%%%%%%%%%%%%%%%%%%%%%%%%%%%%%%%%%%%%%%%%%%%%%%%%%%%%%%%%%%%%%%%%%%%%%%%%%%%%%%%%%%
%%%%%%%%%%%%%%%%%%%%%%%%%%%%%%%%%%%%%%%%%%%%%%%%%%%%%%%%%%%%%%%%%%%%%%%%%%%%%%%%%%%%%%%%%%%%%%%%%%%%%%%%%%%%%%%%%%%%%%%%%%%%%%%%%%%%

%%%%%%%%%%%%%%%%%%%%%%%%%%%%%%%%%%%%%%%%%%%%%%%%%%%%%%%%%%%%%%%%%%%%%%%%%%%%%%%%%%%%%%%%%%%%%%%%%%%%%%%%%%%%%%%%%%%%%%%%%%%%%%%%%%%%
%%%%%%%%%%%%%%%%%%%%%%%%%%%%%%%%%%%%%%%%%%%%%%%%%%%%%%%%%%%%%%%%%%%%%%%%%%%%%%%%%%%%%%%%%%%%%%%%%%%%%%%%%%%%%%%%%%%%%%%%%%%%%%%%%%%%

%%%%%%%%%%%%%%%%%%%%%%%%%%%%%%%%%%%%%%%%%%%%%%%%%%%%%%%%%%%%%%%%%%%%%%%%%%%%%%%%%%%%%%%%%%%%%%%%%%%%%%%%%%%%%%%%%%%%%%%%%%%%%%%%%%%%
%%%%%%%%%%%%%%%%%%%%%%%%%%%%%%%%%%%%%%%%%%%%%%%%%%%%%%%%%%%%%%%%%%%%%%%%%%%%%%%%%%%%%%%%%%%%%%%%%%%%%%%%%%%%%%%%%%%%%%%%%%%%%%%%%%%%

%%%%%%%%%%%%%%%%%%%%%%%%%%%%%%%%%%%%%%%%%%%%%%%%%%%%%%%%%%%%%%%%%%%%%%%%%%%%%%%%%%%%%%%%%%%%%%%%%%%%%%%%%%%%%%%%%%%%%%%%%%%%%%%%%%%%
%%%%%%%%%%%%%%%%%%%%%%%%%%%%%%%%%%%%%%%%%%%%%%%%%%%%%%%%%%%%%%%%%%%%%%%%%%%%%%%%%%%%%%%%%%%%%%%%%%%%%%%%%%%%%%%%%%%%%%%%%%%%%%%%%%%%

\chapter{The operators $\mc{W}_N$ and $\mc{U}_N^{-1}$}
\label{Section descirption cptmt unif op WN}

{\bf Abstract}

\textit{
In this chapter we derive a local (in $\xi$), uniform (in $N$), behaviour of the inverse $\mc{W}_N[H](\xi)$. 
This will allow an effective simplification, in the large-$N$ limit, of the various integrals involving 
$\mc{W}_N[H]$ arising from the Schwinger-Dyson equations of Proposition~\ref{Proposition equations des boucles}.  
Furthermore, these local asymptotics will provide a base for estimating the $W_p^{\infty}$ norms of the inverse of the master operator $\mc{U}_N$, \textit{cf}. \eqref{definition noyau integral operateur S driven by mu eq}. 
In fact, such estimates demand to  have a control on the leading and sub-leading contributions 
issuing from $\mc{W}_N$ with respect to $W_{p}^{\infty}$ norms. We shall demonstrate 
in \S~\ref{Sous-section decomposition apropriee WN pour asymptotiques uniforme Wp norms} that the operator $\mc{W}_N$  can be  decomposed as }
\beq
\mc{W}_N \; = \; \mc{W}_R \, + \, \mc{W}_{\e{bk}} \, + \, \mc{W}_L \; + \; \mc{W}_{\e{exp}}\;.
\label{ecriture decomposition operateur WN}
\enq
\textit{ The operator $\mc{W}_{\e{exp}}$ represents an exponentially small remainder in $W_{p}^{\infty}$ norm, while the three other operators contribute to the leading order 
asymptotics when $N \rightarrow \infty$. Their expression is constructed solely out of the leading asymptotics in $N$ of the solution $\chi$ to the Riemann--Hilbert problem given in 
Proposition~\ref{Theorem ecriture forme asymptotique matrice chi}. }

\textit{ In \S~\ref{Sous-section asymptotiques uniformes locales de WN} we shall build on this decomposition so as to show that there arise two regimes for the large-$N$ asymptotic behaviour 
of $\mc{W}_N[H]$ namely when 
\begin{itemize}
\item $\xi$ is in the "bulk" of $\intff{a_N}{b_N}$, \textit{i.e.} uniformly in $N$ away from the endpoints $a_N$ and $b_N$.
\item $\xi$ is close to the boundaries, \textit{viz}. in the vicinity of the endpoints  $a_N$ (resp  $b_N$).
\end{itemize}
In addition to providing the associated asymptotic expansions, we shall also establish certain properties of the 
remainders which will turn out to be crucial for our further purposes. }

\section{Local behaviour of $\mc{W}_N[H](\xi)$ in $\xi$, uniformly in $N$}

\subsection{An appropriate decomposition of $\mc{W}_N$}
\label{Sous-section decomposition apropriee WN pour asymptotiques uniforme Wp norms}

We remind that any function $H \in \mc{C}^{k}\big( \intff{a_N}{b_N} \big)$ admits a continuation 
into a function $\mc{C}^{k}_{\e{c}}\big(\intoo{a_N - \eta}{b_N + \eta} \big)$ for some $\eta > 0$.
We denote any such extension by $H_{\mf{e}}$, as it was already specified in the notation and basic definition section.  
In the present subsection we establish a decomposition that is adapted for deriving the local and uniform in $N$
asymptotic expansion for $\mc{W}_N$. 

In this section and the next ones, we will use extensively the following notations:
\begin{defin}
\label{xrxl}To a variable $\xi$ on the real line, we associate $x_{R} = N^{\a}(b_N - \xi)$ and $x_{L} = N^{\a}(\xi - a_N)$ the corresponding rescaled and centred around the right (respectively left) boundary variables. 
Similarly, for a variable $\eta$, we denote $y_{R}$ and $y_{L}$ its rescaled and centred variable.
\end{defin}
\begin{defin}
\label{definition tilde fonction} If $H$ is a function of a variable $\xi$, we denote $H^{\wedge}(\xi) = H(a_N + b_N  - \xi)$ its reflection around the centre 
of $\intff{a_N}{b_N}$ (as already met in Lemma~\ref{Lemme propriete symetrie R et L op WN}). This exchanges the role of the left and right boundaries. 
If $H$ is a function of many variables, by $H^{\wedge}$ we mean that all variables are simultaneously reflected. If $\mc{O}$ is an operator, we define the reflected operator by:
\beq
\mc{O}^{\wedge}[H] = \big(\mc{O}[H^{\wedge}]\big)^{\wedge}
\enq
\end{defin}

\begin{defin}
\label{defKKK}Let  $\mathscr{C}_{{\rm reg}}^{(+)}$ (respectively $\mathscr{C}_{{\rm reg}}^{(-)}$) be a contour such that:
\begin{itemize}
 \item  it passes between $\R$ and $\Ga_{\ua}$ (respectively $\Ga_{\da}$).
\item it comes from infinity in the direction of angle $\ex{\pm3\i\pi/4}$ and goes to infinity in the direction of angle $\ex{\pm\i\pi/4}$.
\end{itemize}
These contours are depicted in Figure~\ref{Figure definition des courbes C pm reg}, and we denote $\vsg/2 = {\rm dist}(\mathscr{C}_{{\rm reg}}^{(+)},\R) > 0$. We also introduce an odd function $J$ by setting, for $x > 0$:
\beq
J(x) \; = \; \Int{ \msc{C}^{(+)}_{\e{reg}} }{} \f{ \ex{\i \la x}  }{ R(\la) } \, \f{ \dd \la }{ 2 \i \pi }\;.
\label{definition  fonction J}
\enq
\end{defin}

\begin{figure}[h!]
\begin{center}

\begin{pspicture}(12,9)

% axe reel 

\psline[linestyle=dashed, dash=3pt 2pt]{->}(0,4.5)(12,4.5)

\rput(11.5,4.2){$\R$}

% Courbe de saut R + i eps

\psline(0,5.6)(12,5.6)

\psline[linewidth=3pt]{->}(11,5.6)(11.2,5.6)

\rput(11.5,5.2){$\R + \i \eps $}

%courbe de saut Gamma up

\pscurve(0,9)(6,7.5)(12,9)

\rput(11.5,8.5){$\Ga_{\ua}  $}

\psline[linewidth=3pt]{->}(9,8)(8.8,7.95)

\rput(6,8.7){ poles $\&$ zeroes of $\la \mapsto  R_{\da}(\la) $ }

%courbe de saut Gamma down

\pscurve(0,0.5)(6,3)(12,0.5)

\rput(11.5,0.5){$\Ga_{\da}  $}

\psline[linewidth=3pt]{->}(10,1.65)(9.8,1.75)

\rput(6,1.8){poles $\&$ zeroes of $\la \mapsto \la R_{\ua}(\la)$}

%courbe se saut C+reg

\pscurve(0,7.5)(6,6)(12,7.5)

\rput(10.5,6.5){$\msc{C}^{(+)}_{\e{reg}}  $}

\psline[linewidth=3pt]{<-}(9,6.5)(8.8,6.45)

%courbe de saut C-reg

\pscurve(0,1.5)(6,4)(12,1.5)

\rput(11.5,2.5){$\msc{C}^{(-)}_{\e{reg}}  $}

\psline[linewidth=3pt]{<-}(10,2.65)(9.8,2.75)

\end{pspicture}

\caption{The curves $\msc{C}^{(\pm)}_{\e{reg}}$. 
\label{Figure definition des courbes C pm reg} }
\end{center}
\end{figure}
\begin{prop}
\label{Proposition decomposition op WN en diverses sous parties}
Given any function $H \in \mc{C}^{k}\big( \intff{a_N}{b_N} \big)$ with $k\geq 1$ belonging $\mathfrak{X}_{s}(\R)$ (the image of $\mc{S}_N$, see \eqref{x}), the function $\mc{W}_N[H]$ is $\mc{C}^{k-1}\big( \intoo{a_N}{b_N} \big)$ and admits the representation 
\beq
\mc{W}_N[H](\xi) \; =  \; \mc{W}_R[H_{\mf{e}}](x_R,\xi) \; + \; \mc{W}_{\e{bk}}[H_{\mf{e}}](\xi) \; + \; \mc{W}_{L}[H_{\mf{e}}](x_L,\xi) \; + \; 
\mc{W}_{\e{exp}}[H](\xi)  
\label{ecriture decomposition WN en drte gche bk exp rudimentaire}  
\enq
with:
\begin{eqnarray}
\mc{W}_{\e{bk}}[H_{\mf{e}}](\xi) & = &
 \f{N^{\a} }{2\pi \be } \Int{ \R }{} \big[H_{\mf{e}}\big(\xi + N^{-\a}y\big) - H_{\mf{e}}(\xi) \big]\,J(y) \cdot \dd y\;, \\
\mc{W}_R[H_{\mf{e}}](x,\xi) & = & 
\label{definition fonctionnelle WR}-\f{ N^{\a} }{ 2\pi \be  } \Int{x}{+\infty} \big[H_{\mf{e}}\big(\xi + N^{-\a}y \big) - H_{\mf{e}}(\xi) \big]\,J(y) \cdot \dd y\;,  \\
&& \; - \; \f{ N^{2\a} }{ 2\pi \be  } \Int{ \msc{C}^{(+)}_{\e{reg}} }{} \f{ \dd \la }{ 2 \i \pi } 
\Int{  \msc{C}^{(-)}_{\e{reg}}  }{} \f{ \dd \mu }{ 2 \i \pi}
\f{ \ex{\i\la x}  }{ (\mu-\la) R_{\da}(\la) R_{\ua}(\mu)  }\Bigg\{  \Int{a_N}{b_N} H_{\mf{e}}(\eta) \ex{\i \mu N^{\a}(\eta - b_N)}  \dd \eta 
\; - \; \f{ H_{\mf{e}}(\xi) }{ \i \mu N^{\a} }   \Bigg\} \nonumber \\
\mc{W}_{L}[H_{\mf{e}}](x,\xi) & = & -\mc{W}_{R}[H_{\mf{e}}^{\wedge}](x,a_N+b_N-\xi)\;.
\end{eqnarray}
The remainder operator $\mc{W}_{\e{exp}}[H_{\mf{e}}]$ reads:
\beq
\mc{W}_{\e{exp}} \; = \; \mc{W}_N^{(++)} \, - \, \big(\mc{W}_N^{(++)}\big)^{\wedge}\, + \,
\mc{W}_{\e{res}} \, - \, \big(\mc{W}_{\e{res}}\big)^{\wedge} +  \Delta\mc{W}_N^{(+-)}\,-\, \big(\Delta\mc{W}_N^{(+-)}\big)^{\wedge}\;,
\enq
where the operators $\mc{W}_N^{(++)}$ and $\Delta\mc{W}_N^{(+-)}$ are given by 
\begin{eqnarray}
\label{definition op WN++} \mc{W}_N^{(+ +)}[H](\xi) & \!\!\!=\!\!\! &  \f{ N^{2\a} }{ 2\pi \be } \Int{ \msc{C}^{(+)}_{\e{reg}} }{} \hspace{-1mm}\f{\dd \la }{ 2\i \pi } 
\Int{ \msc{C}^{(+)}_{\e{reg}} }{} \hspace{-1mm}\f{\dd \mu }{ 2\i \pi } \Int{a_N}{b_N} \hspace{-1mm} \dd \eta\,
\f{ \ex{-\i N^{\a}\la(\xi-b_N) + \i N^{\a}\mu(\eta-a_N)} }{  (\mu - \la)R_{\da}(\la) R_{\da}(\mu)}
\Bigg\{  \Psi_{11}(\la) \Psi_{12}(\mu) - \f{\mu}{\la}\cdot\Psi_{11}(\mu) \Psi_{12}(\la)   \Bigg\} 
H(\eta)\;, \nonumber \\
\Delta \mc{W}_N^{(+-)}[H](\xi) & \!\!\!=\!\!\! &   \f{ N^{2\a} }{ 2\pi \be } \Int{ \msc{C}^{(+)}_{\e{reg}} }{} \hspace{-1mm}\f{\dd \la }{ 2\i \pi } 
\Int{ \msc{C}^{(-)}_{\e{reg}} }{} \hspace{-1mm}\f{\dd \mu }{ 2\i \pi } \Int{a_N}{b_N} \hspace{-1mm} \dd \eta%
\f{ \ex{-\i N^{\a}\la(\xi-b_N) + \i N^{\a}\mu(\eta-b_N)} }{ (\mu - \la)  R_{\da}(\la) R_{\ua}(\mu)} \Bigg\{1 + \f{\mu}{\la}\cdot\Psi_{21}(\mu) \Psi_{12}(\la)  - \Psi_{11}(\la) \Psi_{22}(\mu)\Bigg\} 
H(\eta) \nonumber \\
\label{definition op de W +-}
\end{eqnarray}
while ${\cal W}_{{\rm res}}$ is the one-form:
\beq
\mc{W}_{\e{res}}[H] \; = \; - \frac{N^{2\a}}{2\pi\beta}\,\f{ \Pi_{12}(0)\th_R }{R_{\da}(0) }
\Int{ \msc{C}^{(+)}_{\e{reg}} }{} \hspace{-1mm}\f{\dd \mu }{ 2\i \pi }\, \f{\Psi_{11}(\mu) }{ R_{\da}(\mu) }
\Int{a_N}{b_N} \hspace{-1mm} \dd \eta\,H(\eta)\,\ex{\i N^{\a} \mu (\eta-a_N)} \;. 
\label{definition op W res}
\enq
The piecewise holomorphic matrix $\Psi(\mu)$ corresponds to the solution to the Riemann--Hilbert problem for $\Psi$
described in Section \ref{SousSectionRHP for Psi} in which we have taken the limit $\ga \tend +\infty$.
\end{prop}
In the expressions above, we have used an extension $H_{\mf{e}}$ of $H$ whenever it was necessary to integrate $H$ over the whole real line, but we can keep $H$ when only the integrals over $\intff{a_N}{b_N}$ are involved -- e.g. in \eqref{definition op WN++}.
The decomposition  given in Proposition \ref{Proposition decomposition op WN en diverses sous parties} 
splits $\mc{W}_N$ into a sum of four operators. The operator $\mc{W}_{\e{bk}}$ takes into
account the purely bulk-type contribution of the inverse, namely those which do not feel
the presence of the boundaries $a_N, b_N$ of the support of the equilibrium measure. 
This operator does not single out a specific point but rather takes values which are of the 
same order of magnitude throughout the whole of the interval $\intff{a_N}{b_N}$.  
In their turn the operators $\mc{W}_{R/L}$ represent the contributions of the right/left boundaries of the support of the equilibrium
measure. These operators localise, with exponential precision, on their respective left or right boundary. 
Namely, they decay exponentially fast in $x_{R/L}$ when $x_{R/L} \tend +\infty$. This fact is a consequence
of the exponential decay at $\pm \infty$ of $J(x)$ in what concerns the first integral in \eqref{definition fonctionnelle WR} and an immediate bound of the second one which 
follows from $\inf\big\{{\rm Im}\,\la,\,\,{ \la \in \msc{C}^{(+)}_{\e{reg}} }\big\} >0$.

\Proof  We remind that since we are considering the $\ga \tend +\infty$ limit, the matrix $\Psi$
has no jump across $\R+\i\eps$. A straightforward calculation based on the identity:
\beq
\chi(\la)\; = \; \left(  \ba{cc}  -R^{-1}_{\da}(\la)\,\ex{\i \la \ov{x}_N} & R_{\ua}^{-1}(\la)  \\ 
				-R_{\ua}(\la) &       0 \ea \right) \cdot  \Psi(\la) \qquad \e{valid} \; \e{for} \; \la \; \e{between} \; \R \; \e{and} \; \Ga_{\ua}
\enq
shows that, for such $\la$'s and $\mu$'s,
\beq
\f{N^{2\a}}{2\pi \be } \cdot \ex{-\i N^{\a}\la(\xi-a_N)}\,
\bigg\{ \chi_{11}(\la) \chi_{12}(\mu) \; - \; \f{\mu}{\la}\cdot\chi_{11}(\mu) \chi_{12}(\la)  \bigg\}\,
\ex{\i N^{\a}\mu(\eta-b_N)} \; = \; \sul{\eps_1,\eps_2\in \{ \pm  \} }{} K_{\eps_1,\eps_2}\big( \la,\mu \mid \xi, \eta \big) \;. 
\label{ecriture decomp noyau integral en ss noyaux appropries}
\enq
The above decomposition contains four kernels 
\beqa
K_{--}\big( \la,\mu \mid \xi,\eta \big) & =   & \f{N^{2\a}}{2\pi \be }\,
	\f{ \ex{-\i N^{\a}\la(\xi-a_N) + \i N^{\a}\mu(\eta-b_N)} }{ R_{\ua}(\la) R_{\ua}(\mu) }
	  \bigg\{  \Psi_{21}(\la) \Psi_{22}(\mu) \; - \; \f{\mu}{\la} \Psi_{21}(\mu) \Psi_{22}(\la)   \bigg\}\;,   \\
K_{++}\big( \la,\mu \mid \xi,\eta \big) & =   & \f{N^{2\a}}{2\pi \be }\,\f{ \ex{-\i N^{\a}\la(\xi-b_N) + \i N^{\a}\mu(\eta-a_N)} }{ R_{\da}(\la) R_{\da}(\mu) }
\bigg\{  \Psi_{11}(\la) \Psi_{12}(\mu) \; - \; \f{\mu}{\la} \Psi_{11}(\mu) \Psi_{12}(\la)   \bigg\}\;, \\
K_{+-}\big( \la,\mu \mid \xi,\eta \big) & =   & - \f{N^{2\a}}{2\pi \be } \,
				    \f{ \ex{-\i N^{\a}\la(\xi-b_N) + \i N^{\a}\mu(\eta-b_N)} }{ R_{\da}(\la) R_{\ua}(\mu) }
\bigg\{  \Psi_{11}(\la) \Psi_{22}(\mu) \; - \; \f{\mu}{\la} \Psi_{21}(\mu) \Psi_{12}(\la)   \bigg\}\;,   \\
K_{-+}\big( \la,\mu \mid \xi,\eta \big) & =   & - \f{N^{2\a}}{2\pi \be } \,
	\f{ \ex{-\i N^{\a}\la(\xi-a_N) + \i N^{\a}\mu(\eta-a_N)} }{ R_{\ua}(\la) R_{\da}(\mu) }
 \bigg\{  \Psi_{21}(\la) \Psi_{12}(\mu) \; - \; \f{\mu}{\la} \Psi_{11}(\mu) \Psi_{22}(\la)   \bigg\} \;. 
\eeqa
The labeling of the kernels $ K_{\eps_1,\eps_2}\big( \la,\mu \mid \xi, \eta \big)$ by the subscripts $\eps_1,\eps_2$ refers to the half-planes  
$\mathbb{H}^{\eps_1}\times \mathbb{H}^{\eps_2}$ in which they are exponentially small when $N \rightarrow \infty$, provided that the variables $\xi,\eta \in \intff{a_N}{b_N}$ are uniformly away from the 
boundaries $a_N$ or $b_N$.

One should note that the above kernels $K_{\eps_1,\eps_2}$ have a simple pole at $\la=0$. In particular, 
\beqa
\mathop{{\rm Res}} \Big(K_{-+}\big( \la,\mu \mid \xi, \eta \big)\, \dd \la, \la=0 \Big)  & = &
\f{ \mu \Psi_{11}(\mu) }{ R_{\da}(\mu) } \cdot \ex{\i N^{\a}\mu(\eta-a_N)}\,
 \f{ N^{2\a} \cdot \th_R \cdot \Pi_{12}(0) }{ 2\pi \be  \cdot \big( \la R_{\ua}(\la) \big)_{\mid \la=0}  }  \; \; , \\
\mathop{{\rm Res}} \Big(K_{--}\big( \la,\mu \mid \xi, \eta \big)\, \dd \la, \la=0 \Big) & = &
- \f{ \mu \Psi_{21}(\mu) }{ R_{\ua}(\mu) } \cdot \ex{\i N^{\a}\mu(\eta-b_N)}\,
 \f{ N^{2\a} \cdot \th_R \cdot \Pi_{12}(0) }{ 2\pi \be  \cdot \big( \la R_{\ua}(\la) \big)_{\mid \la=0}  } \; \; . 
\eeqa
Furthermore, the kernels are related. Indeed, according to the definition of $\Psi$ in terms of $\chi$ in Figure~\ref{contour pour le RHP de Phi}, we have for $\la$ between $\Ga_{\da}$ and $\R$:
\beq
\chi(\la)\; = \; \left(  \ba{cc}  -R^{-1}_{\da}(\la)  & R_{\ua}^{-1}(\la)\,\ex{-\i \la \ov{x}_N}  \\ 
				0 &     R_{\da}(\la)  \ea \right) \cdot  \Psi(\la)   
\enq
and by invoking the reflection relation for $\chi$ obtained in Lemma  \ref{Lemme Ecriture diverses proprietes solution RHP chi}, we can show that:
\beq
\Psi(-\la)  \; = \; \left( \ba{cc}  0  & \la   \\  
									-\la^{-1} & 0 \ea \right) \cdot  \Psi(\la)  \cdot 
						 \left( \ba{cc}  1  & -\la   \\  
									0 & 1 \ea \right)		 \;. 	
\enq
The above equation ensures that 
\begin{eqnarray}
K_{-+}\big( -\la,-\mu\mid a_N+b_N-\xi , a_N+b_N-\eta \big) & = & K_{+-}\big( \la,\mu \mid \xi,\eta \big)\;,  \\
K_{--}\big( -\la, -\mu \mid a_N + b_N - \xi , a_N + b_N - \eta \big) & = & K_{++}\big( \la,\mu \mid \xi,\eta \big)\;.
\end{eqnarray}
The decomposition \eqref{ecriture decomp noyau integral en ss noyaux appropries} of the integral kernel allows one recasting
the operator $\mc{W}_{N}$ as:
\beq
\mc{W}_N[H](\xi) \; = \; \sul{\eps_1,\eps_2\in \{\pm 1 \} }{} \wt{\mc{W}}_N^{(\eps_1 \eps_2)}[H](\xi) 
\enq
where 
\beq
\wt{\mc{W}}_N^{(\eps_1 \eps_2)}[H](\xi) \; =  \; \Int{ \R +2\i \eps }{}\f{ \dd \la }{2\i \pi} 
\Int{\R + \i \eps }{} \f{ \dd \mu }{2\i \pi} \Int{a_N}{b_N} \dd \eta\, 
\f{  K_{\eps_1  \eps_2}\big( \la,\mu \mid \xi,\eta \big)  }{ \mu - \la } H(\eta) \;. 
\enq
The next step consists in deforming the contours arising in the definition of $\wt{\mc{W}}_N^{(\eps_1 \eps_2)}[H]$.
We shall discuss these handlings on the example of $\wt{\mc{W}}_N^{(-+)}[H]$. In this case, one should deform the 
$\la$-integration to $\R-2\i\eps$. In doing so, we pick the residues at the poles at $\la=0$ and $\la=\mu$ leading to 
\beq
\wt{\mc{W}}_N^{(-  +)}[H](\xi) \; =  \; \mc{W}_{\e{res}}[H]\; + \; 
 \Int{ \R  }{}\! \f{ \dd \la }{2 \i \pi} \Int{a_N}{b_N}\! \dd \eta\, K_{-  +}\big( \la,\la \mid \xi,\eta \big)  H(\eta)
  \; + \; \Int{ \R -2\i \eps }{} \hspace{-2mm} \f{ \dd \la }{2\i \pi} 
\Int{\R - \i \eps }{} \hspace{-2mm} \f{ \dd \mu }{2\i \pi} \Int{a_N}{b_N} \dd \eta\, 
\f{  K_{-  +}\big( \la,\mu \mid \xi,\eta \big)  }{ \mu - \la } H(\eta)\;. \nonumber
\enq
It remains to implement the change of variables $(\la,\mu) \mapsto (-\la,-\mu)$ in the last integral and observe that 
\beq
 K_{- +}\big( \la,\la \mid \xi,\eta \big) \; = \; \frac{N^{2\a}}{2\pi\beta}\,\f{ \ex{ \i N^{\a} \mu (\eta - \xi) }  }{R(\la) }
\quad \e{since} \quad \det \Psi (\la)  \, = \, 1 \;,
\enq
so as to obtain
\beq
\wt{\mc{W}}_N^{(-  +)}[H](\xi) \; =  \; \mc{W}_{\e{res}}[H]\; + \; \mc{W}^{(0)}_{\e{bk}}[H](\xi) \; - \; 
\wt{\mc{W}}_N^{(+  -)}\big[H^{\wedge}]\big( a_N + b_N - \xi \big) \;,
\enq
with $\mc{W}_{\e{res}}$ being given by \eqref{definition op W res} and
\beq
\mc{W}^{(0)}_{\e{bk}}[ H ](\xi)  \; = \; \f{ N^{2\a} }{ 2\pi \be } \Fint{a_N}{b_N} J\big(N^{\a}(\eta-\xi) \big)\,H(\eta)\,\dd \eta 
\enq
with the function $J$ given in \eqref{definition  fonction J}. A similar reasoning applied to the case of $\wt{\mc{W}}_N^{(--)}[H](\xi)$ yields 
\beq
\wt{\mc{W}}_N^{(--)}[H](\xi) \; =  \; - \wt{\mc{W}}_N^{(++)}\big[H^{\wedge}\big](a_N+b_N-\xi) \, - \, \mc{W}_{\e{res}}[H^{\wedge} ] \;.
\enq
Hence, eventually, upon deforming the contours to $\msc{C}^{(+)}_{\e{reg}}$ or $\msc{C}^{(-)}_{\e{reg}} $ in the $\mc{W}^{ (\eps,\eps^{\prime}) }_N$ operators,
\bem
\mc{W}_N[H](\xi) \; = \; \mc{W}_N^{(++)}[H](\xi) \, - \, \mc{W}_N^{(++)}\big[H^{\wedge}\big](a_N+b_N-\xi)  \; + \;  \mc{W}_N^{(+-)}[H](\xi)  \\ 
\, - \, \mc{W}_N^{(+-)}\big[H^{\wedge}\big](a_N+b_N-\xi)
\; + \; \mc{W}_{\e{res}}[H] \, - \, \mc{W}_{\e{res}}[H^{\wedge} ] \; + \; \mc{W}_{\e{bk}}^{(0)}[H](\xi) \;. 
\end{multline}
The operator $\mc{W}_N^{(++)}$ appearing above has been defined in \eqref{definition op WN++} whereas 
\beq
\mc{W}_N^{(+-)}[H](\xi)  \; =  \;  \Int{ \msc{C}^{(+)}_{\e{reg}} }{}\f{ \dd \la }{2\i \pi} 
\Int{ \msc{C}^{(-)}_{\e{reg}} }{} \f{ \dd \mu }{2\i \pi} \Int{a_N}{b_N} \dd \eta\,\f{  K_{+-}\big( \la,\mu \mid \xi,\eta \big)  }{ \mu - \la }  
\,H(\eta)  \; . 
\enq
At this stage, it remains to observe that 
\beq
\mc{W}_N^{(+-)}[H](\xi)  \; =  \; \mc{W}_{R}^{(0)}[H](x_R) \; + \; \Delta \mc{W}_N^{(+-)}[H](\xi) \;,
\enq
where $\Delta\mc{W}_N^{(+-)}$ is as defined in \eqref{definition op de W +-}, while 
\beq
\mc{W}_R^{(0)}[H](x) \; = \; - \f{N^{2\a} }{ 2 \pi \be} \Int{ \msc{C}^{(+)}_{\e{reg}} }{  } \f{\dd \la }{2 \i \pi} 
\Int{ \msc{C}^{(-)}_{\e{reg}}  }{} \f{\dd \mu }{2\i \pi} \Int{a_N}{b_N} \! \dd \eta \, 
 \f{ H(\eta)\,\ex{\i \la x + \i N^{\a}\mu(\eta- b_N)} }{ (\mu-\la)\,R_{\da}(\la)R_{\ua}(\mu) } \;. 
\enq
As a consequence, we obtain the decomposition:
\beq
\mc{W}_N[H](\xi) \; = \; \mc{W}_L^{(0)}[H](x_{L}) \; + \; \mc{W}_{\e{bk}}^{(0)}[H](\xi)
 \; + \; \mc{W}_R^{(0)}[H](x_{R}) \; + \; \mc{W}_{\e{exp}}[H](\xi) \;,
\label{ecriture decomposition W infty sur ops type 0}
\enq
where we have set $\mc{W}_L^{(0)}[H]\big( x \big) \, = \,  - \mc{W}_R^{(0)}[H^{\wedge}]\big( x \big)$. In order to obtain the representation \eqref{ecriture decomposition WN en drte gche bk exp rudimentaire} 
it is enough to incorporate certain terms present in $\mc{W}_{\e{bk}}^{(0)}[H](\xi)$
into the $R$ and $L$-type operators. Namely, we can recast $\mc{W}_{\e{bk}}^{(0)}[H](\xi)$ as 
\bem
\mc{W}_{\e{bk}}^{(0)}[H](\xi)   %\f{N^{2\a} }{ 2 \pi \be} \Int{ \R  }{} \f{\dd \mu }{2i\pi \cdot R(\mu)} 
%
%\Int{a_N}{b_N} \! \dd \eta \,  H(\eta) \ex{\iN^{\a}\mu(\eta- \xi)}  \\ 
%
%
   =  \f{N^{2\a} }{ 2 \pi \be} \Int{a_N}{b_N} J\big(N^{\a}(\eta-\xi)\big)\,\big[H(\eta)-H(\xi) \big]\,\dd \eta
   \; - \; N^{\a}  H(\xi) 
\big[ \vrhp_0(x_R) \, - \,  \vrhp_0(x_L) \big]  \\ 
=  \mc{W}_{\e{bk}}[H_{\mf{e}}](\xi)  \; - \; N^{\a}  \big[ \vrhp_0(x_R) \, - \,  \vrhp_0(x_L) \big]\,H_{\mf{e}}(\xi)\,
   \, - \, \f{ N^{\a} }{ 2\pi \be  }\,\Bigg\{ \Int{x_R}{+\infty} + \Int{-\infty}{-x_L} \Bigg\} 
   \big[ H_{\mf{e}}\big(\xi + N^{-\a}y \big) - H_{\mf{e}}(\xi) \big]\,J(y)\,\dd y    \;. 
\nonumber
\end{multline}
There, we have introduced 
\beq
 \vrhp_0(x) \; = \; \f{-1}{ 2\i \pi \be } 
\Int{ \msc{C}^{(+)}_{\e{reg}} }{} \f{ \ex{ \i \la x}  }{ \la R(\la) }\,\f{ \dd \la }{ 2 \i \pi }  
\qquad i.e. \qquad  \vrhp_0^{\prime}(x)=-\f{ J(x) }{ 2\pi \be } \;. 
\label{definition fonction varrho0}
\enq
The representation  \eqref{ecriture decomposition WN en drte gche bk exp rudimentaire} 
for $\mc{W}_N[H]$ follows by redistributing the terms.  This decomposition also ensures that $\mc{W}_N[H]\in \mc{C}^{k-1}(\intoo{a_N}{b_N})$. 
Indeed, this regularity follows from the exponential 
decay of the integrands in Fourier space when $\xi \in \intoo{a_N}{b_N}$ and derivation under the integral theorems. 

\qed 

\vspace{3mm}

\noindent Note that the integral defining $\vrhp_0$ in \eqref{definition fonction varrho0} can be evaluated explicitly leading to:
\beq
\label{explitrho}\vrhp_0(x) \; = \;  \f{ 1 }{ 2\pi^2 \be }  \cdot \ln \Bigg| \f{ 1-\ex{- \f{ 2  \pi |x| \om_1 \om_2 }{ \om_1+\om_2 }}  }
				{ 1+\ex{- \f{ 2  \pi |x| \om_1 \om_2 }{ \om_1+\om_2 }  \, + \, \i \pi \f{ \om_2-\om_1 }{ \om_1+\om_2 } }   }  \Bigg| \;. 
\enq
In particular, it exhibits a logarithmic singularity at the origin meaning that $J(x)$ has a $\tf{1}{x}$ behaviour around $0$.

%%%%%%%%%%%%%%%%%%%%%%%%%%%%%%%%%%%%%%%%%%%%%%%%%%%%%%%%%%%%%%%%%%%%%%%%%%%%%%%%%%%%%%%%%%%%%%%%%%%%%%%%%%%%%%%%%%%%%%%%%%%%%%%%%%%%
%%%%%%%%%%%%%%%%%%%%%%%%%%%%%%%%%%%%%%%%%%%%%%%%%%%%%%%%%%%%%%%%%%%%%%%%%%%%%%%%%%%%%%%%%%%%%%%%%%%%%%%%%%%%%%%%%%%%%%%%%%%%%%%%%%%%

%%%%%%%%%%%%%%%%%%%%%%%%%%%%%%%%%%%%%%%%%%%%%%%%%%%%%%%%%%%%%%%%%%%%%%%%%%%%%%%%%%%%%%%%%%%%%%%%%%%%%%%%%%%%%%%%%%%%%%%%%%%%%%%%%%%%
%%%%%%%%%%%%%%%%%%%%%%%%%%%%%%%%%%%%%%%%%%%%%%%%%%%%%%%%%%%%%%%%%%%%%%%%%%%%%%%%%%%%%%%%%%%%%%%%%%%%%%%%%%%%%%%%%%%%%%%%%%%%%%%%%%%%

%%%%%%%%%%%%%%%%%%%%%%%%%%%%%%%%%%%%%%%%%%%%%%%%%%%%%%%%%%%%%%%%%%%%%%%%%%%%%%%%%%%%%%%%%%%%%%%%%%%%%%%%%%%%%%%%%%%%%%%%%%%%%%%%%%%%
%%%%%%%%%%%%%%%%%%%%%%%%%%%%%%%%%%%%%%%%%%%%%%%%%%%%%%%%%%%%%%%%%%%%%%%%%%%%%%%%%%%%%%%%%%%%%%%%%%%%%%%%%%%%%%%%%%%%%%%%%%%%%%%%%%%%

\subsection{Local approximants for $\mc{W}_N$}
\label{Sous-section asymptotiques uniformes locales de WN}

In this subsection, we obtain uniform -- in the running variable -- asymptotic expansions for the 
operators $\mc{W}_{\e{bk}}$, $\mc{W}_{R}$ and $\mc{W}_{\e{exp}}$.
In particular, we shall establish that 
if $\xi$ is uniformly away from $b_N$ (respectively $a_N$), $\mc{W}_{R}$ (respectively $\mc{W}_L$) 
will only generate exponentially small (in $N$) corrections. Finally, this exponentially small bound
will hold uniformly after a finite number of $\xi$-differentiations. Prior to discussing these matters 
we need to introduce two families of auxiliary functions on $\R^+$ and constants that come into play during the description of these behaviours. 

\begin{defin}
\label{oinoin}For any integer $\ell \geq 0$:
\begin{eqnarray}
\vrp_{\ell}(x) & = & \f{1}{2\pi \be } \Int{x}{+\infty} y^{\ell}\,J(y)\,\dd y\;,   \\
\label{bsaepi}\vrhp_{\ell}(x) & = & \f{ \i^{\ell + 1} }{2\pi \be } \Int{ \msc{C}^{(+)}_{\e{reg}} }{} \hspace{-2mm} \f{ \dd \la }{2 \i\pi} 
 \Int{ \R -\i\eps^{\prime} }{} \hspace{-2mm} \f{ \dd \mu }{2 \i\pi}
 \f{ \ex{\i \la x }  }{ \mu^{\ell+1}R_{\ua}(\mu)(\mu-\la)R_{\da}(\la) } \;, \\
u_{\ell} & = & \f{\i^{\ell} }{2\i\pi \be\,\ell! } 
 \f{ \Dp{}^{\ell} }{ \Dp{}\la^{\ell} } \bigg( \f{1}{ R(\la) } \bigg)_{\mid \la=0} \;.
\end{eqnarray}
Note that $u_{2p}=0$ since $R$ is an odd function -- given in \eqref{defrtr}.
\end{defin}
For $\ell = 0$, this definition of $\varrho_0$ coincide with \eqref{definition fonction varrho0}, whose explicit expression is \eqref{explitrho}.
Indeed, we remember from \S~\ref{sniung} that $\ua$ means that we can move the contour of integration over $\mu$ up to $+\i\infty$ without hitting a pole of $R_{\ua}^{-1}(\mu)$. 
According to \eqref{427}, $\mu\,R_{\ua}(\mu)$ has a non-zero limit when $\mu \rightarrow 0$, so we just pick up the residue at $\mu = \lambda$, which leads to the expression \eqref{definition fonction varrho0}. 
For $\ell \geq 1$, the function $\vrp_{\ell}$ is continuous at $x=0$. Furthermore,  for any $\ell \geq 0$, $\vrhp_\ell(x)$ and $\vrp_p(x)$ decay exponentially fast in $x$ when $x \tend +\infty$. 
Indeed, it is readily seen on the basis of their explicit integral representations that there exists $C_{\ell} > 0$ such that:
\beq
|\, \vrhp_\ell(x) | \, + \, |\, \vrp_\ell(x) | \; \leq \; C_\ell\,\ex{-C^{\prime}_\ell x} \qquad \e{for} \; \ell \geq 1 \;.
\enq

\begin{prop}
\label{Proposition Ecriture reguliere uniforme des divers const de WN}

Let $k \geq 0$ be an integer, $H \in \mc{C}^{2k+1}\big(\intff{a_N}{b_N} \big)$, and define:
\begin{eqnarray}
\label{641} \mc{W}_{R;k}[H](x,\xi) & = &  \f{ H(\xi) - H(b_N) }{ \xi- b_N }\cdot  x \vrhp_0(x)
\; -\;  \sul{ \ell = 1}{k}  \f{ H^{(\ell)}(\xi)  }{ N^{(\ell-1) \a} \cdot \ell! } \cdot \vrp_{\ell}(x)
\; + \; \sul{\ell = 1 }{k}  \f{  H^{(\ell)}(b_N) }{  N^{(\ell-1) \a} }\cdot \vrhp_{\ell}(x)\;, \\
\mc{W}_{\e{bk};k}[H](\xi) & = &  \sul{\ell = 1 }{ k } \f{  H^{(\ell)}(\xi) }{ N^{\a (\ell-1) } } \cdot u_{\ell}\;.
\label{definition W bk ordre k et ctes u ell}
\end{eqnarray}
These operators provide the asymptotic expansions, uniform for $\xi \in \intff{a_N}{b_N}$:
\begin{eqnarray}
\label{iuniun1}\mc{W}_R[H_{\mf{e}}](x_R,\xi) & = & \mc{W}_{R;k}[H](x_R,\xi) \; + \; \De_{[k]}\mc{W}_R[H_{\mf{e}}](x_R,\xi)\;, \\
 \label{iuniun2}\mc{W}_{\e{bk}}[H_{\mf{e}}](\xi) & = & \mc{W}_{\e{bk};k}[H](\xi) \; + \; \De_{[k]}\mc{W}_{\e{bk}}[H_{\mf{e}}](\xi) \;. 
\end{eqnarray}
The remainder in \eqref{iuniun1} takes the form:
\beq
\De_{[k]}\mc{W}_R[H_{\mf{e}}](x,\xi) \; = \; \mc{R}^{(0)}_{R;[k]}[H_{\mf{e}}](x,\xi) \; + \; 
\sul{\ell=0}{k} x^{\ell + 1/2}\,\mc{R}^{(1/2)}_{R;[k];\ell }[H_{\mf{e}}](x)\;,
\label{ecriture decomposition reste droit}
\enq
with $\mc{R}^{(0)}_{R;[k]}[H_{\mf{e}}]\in W_{k}^{(\infty)}\big( \R^+ \times \intff{a_N}{b_N} \big)$ and $\mc{R}^{(1/2)}_{R;[k];\ell}[H_{\mf{e}}] \in W_{k}^{(\infty)}\big( \R^+ \big)$, and the more precise bound:
\beq
\forall m \in \intn{0}{k},\qquad \max_{ \substack{ p \in \intn{0}{m} \\ \ell \in \intn{0}{k} } }\bigg\{ 
\big|  \partial_{\xi}^{p}\mc{R}^{(0)}_{R;[k]}[H_{\mf{e}}](x_R,\xi) \big| \, + \, 
\big|  \partial_{\xi}^p\mc{R}^{(1/2)}_{R;[k];\ell}[H_{\mf{e}}](x_R) \big|    \bigg\} \; \leq \;
\f{ C \ex{-C^{\prime} x_R} }{ N^{ (k-m)\a }  }\,\norm{ H^{(k+1)}_{\mf{e}} }_{ W^{\infty}_{m}( \R ) } 
\label{ecriture bornes sur reste droit}
\enq
for some $C,C^{\prime} > 0$ independent of $N$ and $H$. The remainder in \eqref{iuniun2} is bounded by:
\beq
 \Norm{ \De_{[k]}\mc{W}_{\e{bk}}[H_{\mf{e}}] }_{ W^{\infty}_{m}( \intff{a_N}{b_N} ) } \; \leq \; 
C\,N^{-k\a}\,\norm{ H_{\mf{e}}^{(k+1)} }_{ W^{\infty}_{m}( \R ) } \;.
\label{ecriture bornes sur rest bulk}
\enq
\end{prop}

\begin{prop}
\label{Proposition Ecriture reguliere uniforme des divers const de WN2} Let $k \geq 0$ be an integer, and $H \in \mc{C}^{2k+1}\big(\intff{a_N}{b_N} \big)$. The operator $\mc{W}_{\e{exp}}$ takes the form:
\begin{eqnarray}
\mc{W}_{\e{exp}}[H](\xi) & = & \mc{R}^{(0)}_{\e{exp};R}[H](x_R,\xi) \; + \; 
\sul{\ell=0}{k} x_R^{\ell + 1/2}\,\mc{R}^{(1/2)}_{\e{exp};R;\ell }[H](x_R)  \nonumber \\ 
& &  + \mc{R}^{(0)}_{\e{exp};L}[H](x_L,\xi) \; + \; 
\sul{\ell=0}{k} x_L^{\ell + 1/2}\,\mc{R}^{(1/2)}_{\e{exp};L;\ell }[H](x_L)
\label{ecriture decomposition Wexp et restes locaux}
\end{eqnarray}
with $\mc{R}^{(0)}_{\e{exp};R/L}[H_{\mf{e}}] \in W_{k}^{(\infty)}\big(\R^+ \times \intff{a_N}{b_N} \big)$  
and $\mc{R}^{(1/2)}_{\e{exp};R/L;\ell}[H_{\mf{e}}] \in W_{k}^{(\infty)}\big( \R^+ \big)$, and the more precise bound:
\beq
\forall m \in \intn{0}{k},\qquad \max_{ \substack{ p \in \intn{0}{m} \\ \ell \in \intn{0}{k} } }\bigg\{ 
\big|  \partial_{\xi}^{p}\mc{R}^{(0)}_{ \e{exp};R/L }[H_{\mf{e}}](x_{R/L},\xi) \big| \, + \, 
\big|  \partial_{\xi}^p\mc{R}^{(1/2)}_{ \e{exp} ; R/L ;\ell}[H_{\mf{e}}](x_{R/L}) \big|    \bigg\} \; \leq \;
C N^{m\a} \ex{-C^{\prime} N^{\a} }\,\norm{ H^{(k+1)}_{\mf{e}} }_{ W^{\infty}_{m}( \R ) } 
\label{ecriture bornes sur reste exponentiel}
\enq
for some $C,C^{\prime} > 0$ independent of $N$ and $H$.
\end{prop}

\vspace{5mm}
The idea for obtaining the above form of the asymptotic expansions is to represent $H$ in terms of its Taylor-integral expansion of order $k$. 
We can then compute explicitly the contributions issuing from the polynomial part of the Taylor series
expansion for $H$ and obtain sharp bounds  on the remainder by exploiting the structure of the integral remainder in the Taylor integral 
series. In particular, the analysis of this integral remainder allows uniform bounds for the remainder as given in \eqref{ecriture bornes sur reste droit}, \eqref{ecriture bornes sur rest bulk}
and \eqref{ecriture bornes sur reste exponentiel}. The reason for such handlings instead of more direct bounds issues from the fact that the 
integrals we manipulate are only weakly convergent. One thus has first to build on the analytic structure of the integrand so as to obtain
the desired bounds and expressions and, in particular, carry out some contour deformations. Clearly, such handlings cannot
be done anymore upon inserting the absolute value under the integral sign, as then the integrand is no more analytic.

\Proof  We carry out the analysis, individually, for each operator. 

\subsubsection*{The operator $\mc{W}_{\e{bk}}$}

The Taylor integral expansion of $H$ up to order $k$ yields the representation
\beq
\label{Wbjdew}\mc{W}_{\e{bk}}[H_{\mf{e}} ](\xi) \; = \; \sul{p=1}{k} \frac{1}{2\pi \beta\,N^{(p-1)\a}}\,\f{H^{(p)}(\xi) }{p!}\,
 \Int{\R}{} y^p J(y)\,\dd y  \; + \; 
\De_{[k]}\mc{W}_{\e{bk}}[H_{\mf{e}} ](\xi) \;,  
\enq
where 
\beq
\De_{[k]}\mc{W}_{\e{bk}}[H_{\mf{e}}](\xi)\; = \;   \f{ 1 }{2\pi \be N^{k\a} }\Int{0}{1}\hspace{-1mm} \dd t\,\frac{(1-t)^{k}}{k!}
\Int{ \R }{} \hspace{-1mm} \dd y\,y^{k+1}J(y)\,H^{(k+1)}_{\mf{e}}(\xi + N^{-\a}ty)\;. 
\enq
In the first terms of \eqref{Wbjdew} we identify:
\beq
\Int{\R}{} y^{\ell} J(y) \dd y \; = \; \i^{\ell - 1} \f{ \Dp{}^{\ell} }{ \Dp{}\la^{\ell} } \bigg( \f{1}{ R(\la) } \bigg)_{\mid \la=0} = 2\pi\beta\,\ell!\,u_{\ell}\;,
\enq
and we remind that this is zero when $\ell$ is even. Finally, we get that the remainder is a ${\cal C}^{k}$ function of $\xi$, and:
\beq
\forall m \in \intn{0}{k},\qquad \norm{ \De_{[k]}\mc{W}_{\e{bk}}[H_{\mf{e}}] }_{ W^{\infty}_{m}(\intff{a_N}{b_N}) } \; \leq \; 
\frac{\norm{ H^{(k+1)}_{\mf{e}} }_{ W^{\infty}_{m}( \R ) }}{N^{k\a}}
\Int{ \R }{} |y|^{k+1} |J(y)|\,\frac{\dd y}{2\pi \be}\;.
\enq
Since $J$ decays exponentially at $\infty$ (see \eqref{definition fonction varrho0} and \eqref{explitrho}), the last integral gives a finite, $k$-dependent constant.

\subsubsection*{The operator $\mc{W}_R$}

The contribution arising in the first line of \eqref{definition fonctionnelle WR} can be treated analogously to 
$\mc{W}_{\e{bk}} $, what leads to 
\beq
-\f{ N^{\a} }{ 2\pi \be  } \Int{x}{+\infty} J(y)\,\big[ H_{\mf{e}}\big(\xi + N^{-\a}y \big) - H_{\mf{e}}(\xi) \big]\,\dd y \; = \;
-\;  \sul{ \ell = 1}{k}  \f{H^{(\ell)}_{\mf{e}}(\xi)}{ N^{(\ell-1) \a}\,\ell! }\, \vrp_{\ell}(x)
\, + \, \De_{[k]}\mc{W}_{R}^{(1)}[H_{\mf{e}}](x,\xi) 
\enq
with 
\beq
\De_{[k]}\mc{W}_{R}^{(1)}[H_{\mf{e}}](x,\xi)  \; = \; \f{-1}{ 2\pi \be\,N^{ k\a }  } \Int{x}{+\infty} \hspace{-1mm}  \dd y
\,y^{k+1} J(y)
\Int{0}{1} \hspace{-1mm} \dd t\,\f{ (1-t)^{k} }{ k! } \,  H^{(k+1)}_{\mf{e}}(\xi + N^{-\a}\,ty)\;.
\enq
Since $J$ decays exponentially at infinity, we clearly have: 
\beq
\max_{p \in \intn{0}{m} } \big|  \partial_{\xi}^p  \cdot \De_{[k]}\mc{W}_{R}^{(1)}[H_{\mf{e}}](x_R,\xi) \big| \; \leq \;
C\, \ex{-C^{\prime} x_R}\cdot\f{\norm{ H^{(k+1)}_{\mf{e}} }_{ W^{\infty}_{m}( \R ) }
 }{ N^{ (k-m)\a }  } %
\enq
for some constants $C, C^{\prime}$ independent of $H$ and $N$. We remind that the $\xi$-derivative can act on both variables $\xi$ \textit{and} $x_{R} = N^{\a}(b_N - \xi)$.

We now focus on the contributions issuing from the second line of  \eqref{definition fonctionnelle WR}. 
For this purpose, observe that the Taylor-integral series representation for $H$ yields the following
representation for the Fourier transform of $H$:
\beq
\Int{a_N}{b_N}H(\eta) \ex{\i \mu N^{\a}(\eta-b_N)}\,\dd \eta \; = \; 
\mc{F}_{1;k}[H](\mu) \, + \, \mc{F}_{2;k}[H_{\mf{e}}](\mu) \, + \, \mc{F}_{3}[H_{\mf{e}}](\mu)\;,
\enq
where we complete the integral over $\intff{a_N}{b_N}$ to $\intof{-\infty}{b_N}$ in the first term, while the two last terms come from subtracting the right and left contributions:
\beq
\mc{F}_{1;k}[H](\mu) \; = \; -\sul{p=0}{ k } \Bigg( \f{ \i }{ N^{\a} \mu  } \Bigg)^{p+1}\!\!\!\!\cdot H^{(p)}(b_N) 
\qquad , \quad 
\mc{F}_{3}[H_{\mf{e}}](\mu) \; = \; \Int{-\infty}{a_N}H_{\mf{e}}(\eta)\,\ex{\i \mu N^{\a}(\eta-b_N)}\,\dd \eta
\enq
and
\beq
\mc{F}_{2;k}[H_{\mf{e}}](\mu)  \; = \; \Int{-\infty}{b_N} \dd \eta  \Int{0}{1}\dd t  
\,\f{ (1-t)^{k} }{ k!  } \ex{\i \mu N^{\a}(\eta-b_N)} 
(\eta-b_N)^{k+1}\,H^{(k+1)}_{\mf{e}}\big(b_N+t(\eta-b_N) \big)  \;. 
\enq
Thus, 
\bem
\; - \; \f{ N^{2\a} }{ 2\pi \be  } \Int{ \msc{C}^{(+)}_{\e{reg}} }{} \f{ \dd \la }{ 2 \i \pi } \Int{ \msc{C}^{(-)}_{\e{reg}} }{} \f{ \dd \mu }{ 2 \i \pi}
\f{ \ex{ \i \la x}  }{ (\mu-\la) R_{\da}(\la) R_{\ua}(\mu)  } \Int{a_N}{b_N} H(\eta) \ex{ - \i \mu y_R }  \dd \eta   \\
\label{iniung}\; = \; \sul{\ell=0}{k} \f{ H^{(\ell)}(b_N) }{N^{(\ell-1)\a}}\,\vrhp_{\ell}(x) \; + \; 
\mc{L}_{\La_0}\big[ \mc{F}_{2;k}[H_{\mf{e}}] \, + \,  \mc{F}_{3}[H_{\mf{e}}] \big](x)\; .
\end{multline}
$\mc{L}_{\La_0}$ is an operator with integral kernel -- see later equation \eqref{definition op int N}:
\beq
\La_0(\la,\mu) \, = \,  \f{-1}{ R_{\ua}(\mu) R_{\da}(\la) }
\enq
which satisfies the assumptions of Lemma~\ref{Lemme decomp et cont op N} appearing below. Thence, 
Lemma~\ref{Lemme decomp et cont op N} entails the decomposition:
\bem
\mc{L}_{\La_0}\big[ \mc{F}_{2;k}[H_{\mf{e}}] \, + \,  \mc{F}_{3}[H_{\mf{e}}] \big](x) \; = \; 
\sul{ \ell = 0 }{ k } \bigg\{ x^{\ell + 1/2} \ex{-\varsigma x} 
\mc{L}_{ \La_0 ; \ell }\big[ \mc{F}_{2;k}[H_{\mf{e}}] \, + \,  \mc{F}_{3}[H_{\mf{e}}] \big](x)  \bigg\} \\
		  \, + \,  (\De_{[k]}\mc{L}_{\La_0})\big[ \mc{F}_{2;k}[H_{\mf{e}}] \, + \,  \mc{F}_{3}[H_{\mf{e}}] \big](x) 
\end{multline}
in which both $\mc{L}_{ \La_0 ; \ell }\big[ \mc{F}_{2;k}[H_{\mf{e}}] \, + \,  \mc{F}_{3}[H_{\mf{e}}] \big](x)$ and 
$(\De_{[k]}\mc{L}_{\La_0})\big[ \mc{F}_{2;k}[H_{\mf{e}}] \, + \,  \mc{F}_{3}[H_{\mf{e}}] \big](x) $ belong to $W_k^{\infty}(\R^+)$
and are as given in \eqref{definition operateur mc L Lambda k}-\eqref{definition operateur De k mc L Lambda}

By using the bounds:
\beq
\big| \mc{F}_{2;k}[H_{\mf{e}}] (\mu)  \big| \; \leq \; \frac{c_{k} \norm{H^{(k+1)}_{\mf{e}}}_{L^{\infty}(\R)}}{(N^{\a}|\mu|)^{k + 2}}\qquad \e{since}\quad 
\f{1}{| {\rm Im}\,\mu | } \, \leq \, \f{ c^{\prime} }{ |\mu | } \;\;\mathrm{for}\;\;\mu \in \msc{C}_{\e{reg}}^{(-)} \;,  
\enq
and 
\beq
\big| \mc{F}_{3}[H_{\mf{e}}] (\mu)  \big| \; \leq \; c \f{ \norm{H_{\mf{e}}}_{L^{\infty}(\R)}  }{ |\mu | N^{\a} } 
 \cdot \ex{-\ov{x}_N |{\rm Im}\,\mu| } \;.
\enq
we get that there exists $N$-independent constants $C, C^{\prime}$ such that 
\beq
 \max_{ \substack{ p \in \intn{0}{m} \\ \ell \in \intn{0}{k} }} \Big|  \partial_{\xi}^{p} \cdot \Big\{
 \Big(  \ex{-\varsigma x_R} \mc{L}_{ \La_0 ; \ell }\, + \, \De_{[k]}\mc{L}_{\La_0} \Big)\Big[ \mc{F}_{2;k}[H_{\mf{e}}] \, + \,  \mc{F}_{3}[H_{\mf{e}}] \Big](x_R)  \Big\} \Big| 
 \; \leq \; C\,\ex{-C^{\prime} x_R}\, \f{\norm{ H_{\mf{e}}^{(k+1)} }_{ L^{\infty}(\R) } }{ N^{(k-m)\a} } \;. 
\enq

\noindent We have relied on:
\begin{lemme}
\label{Lemme decomp et cont op N}

Let $\La(\la,\mu)$ be a holomorphic function of $\la$ and $\mu$ belonging to the region of the 
complex plane delimited by $\msc{C}^{(+)}_{\e{reg}}$ and $\msc{C}^{(-)}_{\e{reg}}$ and such that it admits 
an asymptotic expansion 
\beq
\La(\la,\mu) \; = \; \sul{\ell=0}{k}  \f{ \La_{\ell}(\mu)  }{ \big[ \i(\la- \i\varsigma ) \big]^{\ell + 1/2}  }
 \; + \; \De_{[ k ]}\La(\la,\mu) \quad \e{with} \quad
\left\{ \ba{lcl} 
| \La_{ \ell }(\mu)| &=& \e{O}\big( |\mu|^{1/2} \big)  \vspace{2mm} \\
| \De_{[k]}\La(\la,\mu) | &=& \e{O}\big( |\la|^{-(k+3/2)}\cdot |\mu|^{1/2} \big)  \ea \right.  \;. 
\enq
Then, the integral operator on 
$\mu \cdot L^{\infty}\big( \msc{C}_{\e{reg}}^{(-)} \big) \, \equiv \, 
\Big\{ f \; : \; \mu \mapsto \mu f(\mu) \in L^{\infty}\big( \msc{C}_{\e{reg}}^{(-)} \big) \Big\}$
\beq
\mc{L}_{\La}\big[ f \big](x) \; = \; \f{ N^{2\a} }{ 2\pi \be }  \Int{  \msc{C}_{\e{reg}}^{(+)} }{} \f{ \dd \la }{ 2 \i \pi }
\Int{  \msc{C}_{\e{reg}}^{(-)} }{} \f{ \dd \mu }{ 2 \i \pi }  \f{ \La(\la,\mu) }{ \mu - \la } \ex{\i \la x} f(\mu) 
\label{definition op int N}
\enq
can be recast as 
\beq
\mc{L}_{\La}\big[ f \big](x) \; = \; \sul{ \ell = 0 }{ k } x^{\ell + 1/2} \ex{-\varsigma x} \mc{L}_{ \La ; \ell }[f](x) 
		  \, + \,  \De_{[k]}\mc{L}_{\La}\big[ f \big](x) \;,
\enq
where the operators 
\beqa
\mc{L}_{\La; k}[f](x) & = & \f{ N^{2\a}  }{ 2\pi \be }  
\Int{ \Ga(\i \R^{+}) }{ } \hspace{-2mm} \f{ \dd \la }{ 2 \i \pi }
\Int{  \msc{C}_{\e{reg}}^{(-)} }{}  \hspace{-2mm} \f{ \dd \mu }{ 2 \i \pi }  
\f{ \La_k(\mu) \ex{\i \la } f(\mu) }{ \big[ x\big( \mu-\i\varsigma \big) - \la \big] (\i \la)^{\ell+\f{1}{2}}  } \;, \label{definition operateur mc L Lambda k}\\
\De_{[k]}\mc{L}_{\La}[ f](x) & = & \f{ N^{2\a} }{ 2\pi \be } 
 \Int{  \msc{C}_{\e{reg}}^{(+)} }{}\hspace{-2mm} \f{ \dd \la }{ 2 \i \pi }
\Int{  \msc{C}_{\e{reg}}^{(-)} }{} \hspace{-2mm} \f{ \dd \mu }{ 2 \i \pi }  
\f{ \De_{[k]}\La(\la,\mu) }{ \mu - \la } \ex{\i \la x} f(\mu)   \label{definition operateur De k mc L Lambda}  \;,
\eeqa
are continuous as operators $ \mu \cdot L^{\infty}\big( \msc{C}_{\e{reg}}^{(-)} \big) \tend W^{\infty}_{k}\big( \R^+ \big)$. 
Note that, above, $\Ga\big( \i \R^+ \big)$ corresponds to a small counterclockwise loop around $\i\R^+$. 

\end{lemme}

\Proof  It is enough to insert the large-$\mu$ expansion of $\La$ and then, in the part subordinate to the
inverse power-law expansion, deform the $\la$-integrals to $\Ga\big( \i \R +\i\varsigma\big)$, translate by 
$+\i\varsigma$ and, finally, rescale by $x$. The statements about continuity are evident. \qed

\subsubsection*{The operator $\mc{W}_{\e{exp}}$ (Proposition~\ref{Proposition Ecriture reguliere uniforme des divers const de WN2})}

The analysis relative to the structure of $\mc{W}_{\e{exp}}[H_{\mf{e}}]$ follows basically the same steps as above so we shall not detail them here
again. 
The main point, though, is the presence of an exponential prefactor $\ex{-cN^{\a}}$ which issues from the 
bound  \eqref{ecriture bornes en N pour Pi moins Id} on $\Pi-I_2$. \qed

\subsection{Large $N$ asymptotics of the approximants of $\mc{W}_N$}

The results of Propositions~\ref{Proposition Ecriture reguliere uniforme des divers const de WN} and \ref{Proposition Ecriture reguliere uniforme des divers const de WN2} induce the representation 
\beq
\mc{W}_N[H](\xi) \; = \; \mc{W}_{R;k}[H](x_R,\xi) \; + \; \mc{W}_{\e{bk};k}[H](\xi) \; - \;  \mc{W}_{R;k}[H](x_L,a_N+b_N-\xi)
\; + \; \De_{[k]}\mc{W}_N[H_{\mf{e}}](\xi) \;,
\label{definition reste asympt ordre k pour WN}
\enq
with all remainders at order $k$ are collected in the last term. In this subsection, we shall derive asymptotic expansion (in $N$) of the approximants $\mc{W}_{\e{bk};k}$ and $\mc{W}_{R;k}$ in the case when their unrescaled variable $\xi$ scales 
towards $b_N$ as $\xi = b_N - N^{-\a}\,x$ with $x$ being independent of $N$. 
We, however, first need to establish properties of certain auxiliary functions that appear in this analysis.

\begin{defin}
\label{giu}Let $\ell \geq 0$ be an integer. As a supplement to Definition~\ref{oinoin}, we introduce, for any integer $\ell \geq 0$:
\beq
\mf{b}_{\ell}(x) \; = \; \vrhp_{\ell+1}(x) \, - \, \f{ (-x)^{\ell+1} }{ (\ell+1)! }\vrhp_0(x) \; - \; 
\sul{\substack{ s+p=\ell \\ s,p\geq 0}  }{}  \f{ (-x)^p \vrp_{s+1}(x)}{ p! (s+1)! } \qquad and \qquad
\mf{u}_{\ell}(x) \; = \; \sul{\substack{ s+p=\ell \\ s,p\geq 0}  }{} \f{(-x)^p u_{s+1} }{ p! }
\label{definition fcts b et r goth}
\enq
and:
\beq
\mf{a}_0(x)\; = \; \mf{b}_{0}(x) + \mf{u}_{0}(x)\;, \qquad %
\mf{a}_{\ell}(x)\; = \; \f{ \mf{b}_{\ell}(x) + \mf{u}_{\ell}(x) }{ \mf{a}_0(x) }\quad \mathrm{for}\,\,\ell \geq 1\;.
\label{definition fcts a goth}
\enq
\end{defin}

It will be important for the estimates of \S~\ref{Sous section Sharp weighted bounds for UN moins 1} to remark that $x^{-1/2}\mathfrak{a}_{0}(x)$ is a smooth and positive function:
\begin{lemme}
\label{Lemme comportement fonction a goth}
%Given the combinations of the functions $\vrhp_{\ell}$ and $\vrp_{\ell}$ below
%
%
%
%
%
%
%with $u_s$ given by \eqref{definition W bk ordre k et ctes u ell}, define the functions $\mf{a}_{\ell}$, with $\ell \geq 0$, as
%
%
Let $\ell,n,m \geq 0$ be three integers such that $n > m$. There exist polynomials $p_{\ell;m,n}$ of degree at most $n+\ell$ and functions 
$f_{\ell;m,n} \in W^{\infty}_{n-m}\big(\R^+\big)$
such that, for any $x > 0$:
\beq
\label{propriete fct a bornes polynomiales}
\mf{a}_0(x) \; = \; \sqrt{x}\,p_{0;m,n}(x) \ex{-\vsg x } \; + \;  x^m\,f_{0;m,n}(x) \qquad and \qquad 
\mf{a}_0(x) \cdot \mf{a}_{\ell}(x)\; = \; \sqrt{x}\,p_{\ell;m,n}(x) \ex{-\vsg x } \; + \;  x^m\,f_{\ell;m,n}(x)\;.
\enq
The function $\mf{a}_0(x)$ is positive for $x > 0$ and satisfies 
\beq
\label{keyzero}\mf{a}_0(x) \mathop{=}_{x \rightarrow 0} \frac{1}{\pi\beta}\,\sqrt{\frac{x}{\pi(\omega_1 + \omega_2)}} \; + \; \e{O}(x)
\enq
Finally, one has, in  the  $x \rightarrow +\infty$ regime,
\beq
\mf{a}_0(x) \; = \; u_1 \; + \;  \e{O}(\ex{-\varsigma x}) \qquad , \qquad 
\mf{a}_0(x) \cdot \mf{a}_{\ell}(x)\; = \;  \mf{u}_{\ell}(x) \; + \;  \e{O}( \ex{-\vsg x})
\label{propriete fct a cpt polynomial a infini}
\enq
and the bound on the remainder is stable with respect to finite-order differentiations. \end{lemme}

\Proof  By using the integral representation \eqref{definition  fonction J} for the function $J$, we can readily recast $\vrp_{\ell}(x)$, for $x>0$  as:
\beq
\vrp_{\ell}(x) \; = \; \f{ \i^{\ell+1} }{ 2\pi \be  } \Int{ \msc{C}^{(+)}_{\e{reg}} }{} \f{ \ex{ \i \la x} }{ \la }
 \f{\Dp{}^{\ell} }{ \Dp{}\la^{\ell} } \bigg( \f{1}{R(\la)} \bigg) \,\f{\dd \la }{2 \i \pi }\;. 
\enq
The $\mu$-integral arising in the definition \eqref{bsaepi} of $\vrhp_{\ell}$ can be computed by moving the contour of integration over $\mu$ up to $+\i\infty$, and picking the residues at $\mu=\la$ and $\mu=0$:
\beq
\vrhp_{\ell}(x) \; = \; \f{ \i^{\ell+1} }{ 2\pi \be  }\Int{ \msc{C}^{(+)}_{\e{reg}} }{}\hspace{-2mm} 
 \f{ \ex{ \i\la x} }{ \la^{\ell+1} R(\la) }\,\f{ \dd \la }{2 \i \pi}
\; + \; \tau_{\ell}(x) \quad \e{with} \quad 
\tau_{\ell}(x) \; = \; \f{ \i^{\ell+1} }{ 2\pi \be  }\Int{ \msc{C}^{(+)}_{\e{reg}} }{}\hspace{-2mm} 
 \f{\ex{ \i\la x}}{\ell ! R_{\da}(\la) } \cdot \f{ \Dp{}^{\ell} }{ \Dp{}\mu^{\ell} } \bigg( \f{1}{(\mu-\la) R_{\ua}(\mu)} \bigg)_{\mid \mu=0} 
		\f{ \dd \la }{2 \i \pi}   \; . 
\nonumber
\enq
The first term can be related to the functions $\vrhp_0$ and $\vrp_{s}$ of Definition~\ref{oinoin} by an $\ell$-fold integration by parts based on the identities:
\beq
\f{1}{ \la^{\ell + 1 } } \; = \; \f{ \Dp{}^{\ell} }{ \Dp{}\la^{\ell} } \bigg\{ \f{ (-1)^{\ell} }{\la\,\ell!} \bigg\}
\qquad \e{and} \qquad
\f{ \Dp{}^{\ell} }{ \Dp{}\la^{\ell} } \bigg\{ \f{ \ex{ \i\la x} }{ R(\la) }  \bigg\} \; = \; \sul{ \substack{ s+p = \ell \\ s,p \geq 0 } }{}
\f{\ell ! }{s! p !} (\i x)^{p} \ex{\i \la x} \cdot\f{ \Dp{}^{s} }{ \Dp{}\la^{s} } \bigg\{ \f{ 1 }{ R(\la) } \bigg\} \;.
\enq
Namely, we obtain -- writing the identity for $\ell + 1$ instead of $\ell$ -- that: 
\beq
\vrhp_{\ell+1}(x) \, - \, \tau_{\ell+1}(x) \; = \; \f{ (-x)^{\ell+1} }{ (\ell+1)!}\,\vrhp_0(x) \; + \; 
\sul{\substack{ s+p=\ell \\ s,p\geq 0}  }{}  \f{ (-x)^p \vrp_{s+1}(x)}{ p! (s+1)! } \;. 
\enq
According to Definition~\ref{giu}, we can thus identify $\tau_{\ell + 1}(x) = \mathfrak{b}_{\ell}(x)$ -- in this proof, we will nevertheless keep the notation $\tau_{\ell}$.
Hence, it remains to focus on $\tau_{\ell}(x)$. Computing the $\ell^{\e{th}}$-order $\mu$-derivative appearing in its integrand  
and then repeating the same integration by parts trick, we obtain that: 
\beq
\label{taudefgfsg}\tau_{\ell}(x)  \; = \; - \f{\i^{\ell+1} }{ 2\pi \be  }\sul{s+r+p=\ell }{}
\f{ (\i x)^{r} }{s!p!r!} \f{ \Dp{}^{s} }{ \Dp{}\mu^{s} } \bigg( \f{1}{ R_{\ua}(\mu)} \bigg)_{\mid \mu=0} 
\Int{ \msc{C}^{(+)}_{\e{reg}} }{}  \f{ \ex{ \i\la x} }{  \la  } \cdot 
\f{ \Dp{}^{p} }{ \Dp{}\la^{p} } \bigg\{ \f{1}{ R_{\da}(\la) } \bigg\}\,\f{ \dd \la }{2 \i \pi}   \; . 
\enq
In the second integral, let us move a bit the contour $\mathscr{C}_{{\rm reg}}^{(+)}$ to a contour $\mathscr{C}_{{\rm reg},0}^{(+)}$ which passes below $0$ while keeping the same asymptotic directions as $\mathscr{C}_{{\rm reg}}^{(+)}$. Doing so, we pick up the residue at $\la = 0$:
\beq
\label{jon}\Int{ \msc{C}^{(+)}_{\e{reg}} }{}  \f{ \ex{ \i\la x} }{  \la  } \cdot 
\f{ \Dp{}^{p} }{ \Dp{}\la^{p} } \bigg\{ \f{1}{ R_{\da}(\la) } \bigg\}\,\f{ \dd \la }{2 \i \pi} = -  \f{ \Dp{}^{p} }{ \Dp{}\la^{p} } \bigg\{ \f{1}{ R_{\da}(\la) } \bigg\}_{\mid \la=0} + \Int{ \msc{C}^{(+)}_{\e{reg},0} }{}  \f{ \ex{ \i\la x} }{  \la  } \cdot 
\f{ \Dp{}^{p} }{ \Dp{}\la^{p} } \bigg\{ \f{1}{ R_{\da}(\la) } \bigg\}\,\f{ \dd \la }{2 \i \pi}
\enq
We observe that there exist constants $c_{p;q}$ such that:
\beq
\label{jiugw}\f{ 1 }{  \la  } \cdot \f{ \Dp{}^{p} }{ \Dp{}\la^{p} } \bigg\{ \f{1}{ R_{\da}(\la) } \bigg\} \; = \; 
\sul{q=p+1}{n} \f{ c_{p;q} }{ \big[ \i (\la - \i \vsg) \big]^{q+1/2} }
\; + \; \Delta_{[n]}^{(p)}\big[ R_{\da}^{-1}\big](\la)\;,
\enq
This decomposition ensures that $\Delta_{[n]}^{(p)}\big[ R_{\da}^{-1}\big](\la)$ is holomorphic in $\mathbb{H}^{-}$, has a simple pole at $\la=0$ and 
satisfies $\Delta_{[n]}^{(p)}\big[ R_{\da}^{-1}\big](\la) =\e{O}\Big( \la^{-(n+\tf{3}{2})}\Big)$.

Since $\vsg/2$ is the distance between $\mathscr{C}_{\rm reg}^{(+)}$ and $\R$, we can choose this contour -- for a fixed $\vsg$ -- such that the branch cut of the denominators in \eqref{jiugw} is located on a vertical half-line above $\mathscr{C}_{{\rm reg},0}^{(+)}$. This implies that the remainder in \eqref{jiugw} is holomorphic below $\mathscr{C}_{{\rm reg},0}^{(+)}$. So, in the second integral of \eqref{jon}, we obtain with the first sum contributions involving:
\beq
\Int{\mathscr{C}_{\rm reg}^{(+)}}{} \frac{e^{\i\la x}}{\big[\i(\lambda - \i\vsg)\big]^{q + 1/2}}\,\frac{\dd\la}{2\i\pi} = \frac{e^{-\varsigma x}\,x^{q - 1/2}}{\i\Gamma( q+ \tf{1}{2})}
\enq
in which (after the change of variable $t = -\i x(\la - \i\vsg)$) we have recognised the Hankel contour integral representation of $\big\{ \Gamma(q + 1/2) \big\}^{-1}$. 
In its turn, the contribution of the remainder in \eqref{jiugw} can be written:
\beq
\Int{ \msc{C}^{(+)}_{\e{reg},0} }{} \ex{ \i\la x} \Delta_{[n]}^{(p)}\big[R_{\da}^{-1}\big](\la)\,\f{ \dd \la }{2 \i \pi} \; = \; 
\Int{\msc{C}^{(+)}_{\e{reg}}}{} \Bigg(\ex{ \i\la x} - \sum_{r = 0}^{m - 1} \frac{(\i\la)^{r}x^{r}}{r!}\Bigg)\cdot\Delta_{[n]}^{(p)}\big[R_{\da}^{-1}\big](\la)\cdot\f{ \dd \la }{2 \i \pi} 
+ \sum_{r = 0}^{m - 1} \underbrace{\Int{\msc{C}^{(+)}_{\e{reg},0}}{} \frac{(\i\la)^{r}x^{r}}{r!}\cdot\Delta_{[n]}^{(p)}\big[R_{\da}^{-1}\big](\la)\cdot \f{ \dd \la }{2 \i \pi}}_{= 0} \;. 
\enq
Note that  the last sum vanishes since we can deform the contour of integration to $-\i\infty$ provided $m \leq n$. 
Also, we could deform $\msc{C}^{(+)}_{\e{reg},0}$ back to $\msc{C}^{(+)}_{\e{reg}}$ in the first term since the integrand has no pole at $\la=0$. All-in-all, we get 
\bem
\Int{\mathscr{C}_{{\rm reg}}^{(+)}}{}  \f{ \ex{ \i \la x } }{ \la } \,
\f{ \Dp{}^{p} }{ \Dp{}\la^{p} } \bigg\{ \f{1}{ R_{\da}(\la) } \bigg\}\,\f{ \dd \la }{2 \i \pi} \, = \, -  \f{ \Dp{}^{p} }{ \Dp{}\la^{p} } \bigg\{ \f{1}{ R_{\da}(\la) } \bigg\}_{\mid \la=0} 
+ \sum_{q = p + 1}^n \frac{c_{p;q}\,e^{-\varsigma x}\,x^{q-1/2 }}{\i\Gamma( q + 1/2) } \\
 + \; \Int{ \msc{C}^{(+)}_{\e{reg}} }{}  \Delta_{[n]}^{(p)}\big[ R_{\da}^{-1}\big](\la)\,\Bigg(\ex{ \i\la x} \; - \; \sul{r=0}{m-1} \f{ (\i x)^{r} \la^r }{ r! }\Bigg)\,\f{ \dd \la }{2 \i \pi}  \;.
\label{estimation integrale fct 1 sur R moins}
\end{multline}

\noindent With the bound 
\beq
\Big|  \ex{ \i\la x} \; - \; \sul{r=0}{m-1} \f{ (\i x)^{r} \la^r}{ r! } \Big| \leq x^m |\la|^m 
\enq
and theorems of derivation under the integral, we can conclude that the last 
term in \eqref{estimation integrale fct 1 sur R moins} is at least $n-m$ times differentiable and
that it has, at least, an $m$-fold zero at $x=0$. With the decomposition~\eqref{estimation integrale fct 1 sur R moins}, we can come back to $\tau_{\ell}$ given by \eqref{taudefgfsg}. 
The second term in \eqref{estimation integrale fct 1 sur R moins} -- which contain derivatives of $1/R_{\da}$ -- can be recombined with its prefactor -- containing derivatives of $1/R_{\ua}$ -- by using the Leibniz 
rule backwards for the representation of the derivative at $0$ of  $1/R = 1/(R_{\ua}R_{\da})$. Subsequently, we find there exist a polynomial $p_{\ell;m,n}$ of degree at most $n+\ell$ 
and a function $f_{\ell;m,n} \in W^{\infty}_{n-m}\big(\R^+)$ such that 
\beq
\tau_{\ell+1}(x) \; = \; \sqrt{x}\,p_{\ell;m,n}(x) \ex{-\vsg x} \; + \; x^m\, f_{\ell;m,n}(x) \; - \; 
\f{ \i^{\ell} }{  2 \pi \be  }
 \sul{s+p=\ell}{} \f{ (\i x)^p }{ (s+1)! p! }\,\f{ \Dp{}^{s+1} }{ \Dp{}\la^{s+1} } \bigg\{ \f{1}{R(\la) }  \bigg\}_{\mid \la=0} \;. 
\label{ecriture representation canonique pour tau ell+1}
\enq
The claim then follows upon adding up all of the terms. Finally, the estimates at $x \tend +\infty$
of $\mf{a}_{\ell}$ follow readily from the exponential decay at $x \tend + \infty$ of the functions $\vrhp$ and $\vrp$.

To compute the behaviour at $x \rightarrow 0$, we remind that:
\beq
\mathfrak{a}_0(x) \; = \;  \mathfrak{b}_0(x) + \mathfrak{u}_1  \; = \;  \tau_1(x) + \mathfrak{u}_1 \;. 
\enq
We already know from \eqref{ecriture representation canonique pour tau ell+1} that $\mathfrak{a}_0(0) = 0$, and we just have to look in \eqref{taudefgfsg}-\eqref{estimation integrale fct 1 sur R moins} 
for the coefficient of $\sqrt{x}$ in the case $\ell = 1$. For this purpose, it is enough to write \eqref{taudefgfsg} with $n = 1$. Then, the squareroot behaviour occur for $p = r = 0$ and $s = 1$ in the sum, and gives:
\beq
\mathfrak{a}_0(x) = \frac{c_{0;1}\,x^{1/2}\,\ex{-\varsigma x}}{2 \i \pi\beta\cdot\Gamma(3/2)}\,\partial_{\mu} R_{\ua}^{-1}(\mu)|_{\mu = 0} \; + \;  \e{O}(x) \;. 
\enq
The coefficient $c_{0;1}$ is given by the large $\la$ asymptotics in \eqref{jiugw}, coming from that of $R_{\da}(\la)$ given by \eqref{427bis}:
\beq
c_{0;1} = -1 \;. 
\enq
On the other hand, we know from \eqref{427} that:
\beq
\partial_{\mu} R_{\ua}^{-1}(\mu)|_{\mu = 0} = \frac{1}{\i\sqrt{\omega_1 + \omega_2}}
\enq
Therefore:
\beq
\mathfrak{a}_0(x) = \frac{1}{\pi\beta}\,\sqrt{\frac{x}{\pi(\omega_1 + \omega_2)}} \; + \; \e{O}(x) \;. 
\enq
We finally turn to proving that $\mf{a}_0>0$ on $ \R^+$. It follows from the previous calculations that 
\beq
\mf{a}_{0}(x) \; = \; \f{1}{2\pi \be } \bigg( \f{1}{\mu \cdot R_{\ua}(\mu)}  \bigg)_{ \mid \mu=0} \Int{ \msc{C}^{(+)}_{\e{reg}} }{} 
\f{ \ex{\i\la x}-1 }{ \la R_{\da}(\la) }\,\f{ \dd \la }{ 2\i \pi } \;. 
\label{expression integrale pour a goth 0}
\enq
The integral can be computed by deforming the contour up to $+\i\infty$ and, in doing so, we pick up the residues of the poles
located at 
\beq
 \la = \f{2\i\pi n\,\om_1 \om_2  }{ \om_1 +  \om_2 }\;,  \qquad n \geq 1\;.
\enq
All-in-all this yields
\beq
\mf{a}_0(x) \; = \;  \sum_{n \geq 1} \mf{a}_{0;n}\Big( 1 \, - \,  \ex{- \frac{ 2  \pi   \om_1 \om_2 }{ \om_1 +  \om_2} n x } \Big) \qquad \e{with} \qquad
\mf{a}_{0;n} \; = \;  
\f{(\om_1 +  \om_2 )  \cdot (-1)^{n-1} \cdot \kappa^{-n\kappa} \cdot (1-\kappa)^{-n(1-\kappa)} }
{2 \pi \be \om_1 \om_2 \cdot n^{2} \cdot  n! \cdot \Ga\big(  -\kappa n \big) \cdot \Ga\big( -(1-\kappa) n \big)   }
\label{formule rep serie pour mathfrak a0}
\enq
and $\kappa = \om_2/(\om_1+\om_2)<1$. 
By using the Euler reflection formula, we can recast $\mf{a}_{0;n}$ into a manifestly strictly positive form 
\beq
\mf{a}_{0;n} \; = \; \f{ (\om_1 +  \om_2 ) }{ 2\pi \be \om_1 \om_2  }  \cdot \bigg( \f{ \sin[\pi \kappa n] }{ \pi } \bigg)^{2} \cdot 
\f{ \Ga\big(  1+\kappa n \big) \cdot \Ga\big( 1+(1-\kappa) n \big)  }{ n^{2} \cdot n! \cdot  \kappa^{n\kappa} \cdot (1-\kappa)^{n(1-\kappa)}  } \;. 
\enq
The asymptotics of $\mf{a}_{0;n}$ then takes the form 
\beq
\mf{a}_{0;n} \underset{n \tend  + \infty}{\sim} \f{ (\om_1 +  \om_2 ) }{ 2 \be \om_1 \om_2  } \cdot \sqrt{ \f{2 \kappa (1-\kappa)}{ \pi n^{3}} } 
\cdot \bigg( \f{ \sin[\pi \kappa n] }{ \pi } \bigg)^{2} \;.
\enq
Thus the series \eqref{formule rep serie pour mathfrak a0} defining $\mf{a}_0(x)$ converges uniformly for $x \in \R^+$.
Since the series only contains positive summands, $\mf{a}_0(x)$ is positive for $x > 0$.\qed

\vspace{3mm}
The main reason for investigating the properties of the functions $\mf{a}_{\ell}(x)$ lies in the fact that they describe the large-$N$ asymptotics 
of the function $\mc{W}_{R;k}[H]( x,b_N-N^{-\a}\,x) \, + \, \mc{W}_{\e{bk};k}[H]( b_N-N^{-\a}\,x)$. 
In particular, $\mf{a}_{0}(x)$ arises as the first term and plays a particularly
important role in the analysis that will follow. Let us remind Definition~\ref{weiei} for the weighted norm:
\beq
{\cal N}_{N}^{(\ell)}[H] = \sum_{k = 0}^{\ell} \frac{\norm{H}_{W^{\infty}_{k}(\R)}}{N^{k\a}}\;.
\enq
\begin{lemme}
\label{Lemme structure locale au bord pour WR et Wbk}
Let $k \geq 0$ be an integer, $H \in \mc{C}^{2k+1}\big(\intff{a_N}{b_N} \big)$. Define the functions:
\beqa
\label{692}\mc{W}_{R;k}^{(\e{as})}[H](x) & = & H^{\prime}(b_N)\,\mf{b}_0(x)
\; + \; \sul{\ell = 1 }{k-1} \f{ H^{(\ell + 1)}(b_N)\,\mf{b}_{\ell}(x) }{ N^{\ell \a} }  \;,\\
\label{693}\mc{W}_{\e{bk};k}^{(\e{as})}[H](x) & = & H^{\prime}(b_N) \, u_1 + \; \sul{\ell = 1 }{  k-1 }
\f{H^{(\ell + 1)}(b_N)\,\mathfrak{u}_{\ell}(x)}{N^{\ell \a} } \;.
\eeqa
The approximants at order $k$, $\mc{W}_{R;k}[H](x,b_N - N^{-\a}\,x)$ and $\mc{W}_{\e{bk};k}[H](b_N - N^{-\a}\,x)$, admit the large-$N$ asymptotic expansions:
\begin{eqnarray}
\mc{W}_{R;k}[H](x,b_N - N^{-\a}x) & = & \mc{W}_{R;k}^{(\e{as})}[H](x)
\; + \; \Delta_{[k]}\mc{W}_{R}^{(\e{as})}[H](x) \; , 
\label{ecriture DA localise W asymp R k} \\
\label{ecriture DA localise W asymp bk k} \mc{W}_{\e{bk};k}[H](b_N - N^{-\a}x) & = & \mc{W}_{\e{bk};k}^{(\e{as})}[H](x) \; + \; 
\Delta_{[k]}\mc{W}_{\e{bk}}^{(\e{as})}[ H ](x) \;. 
\end{eqnarray}
The remainders have the following structure:
%These hold uniformly in $x \in \intff{0}{\eps N^{\a} }$ for some $\eps>0$ small enough. 
%
%
%
\begin{eqnarray}
\Delta_{[k]}\mc{W}_{R}^{(\e{as})}[ H ](x) & = & N^{-k\a}\cdot \ex{-\varsigma x}\,\bigg\{(\ln x)\,\mc{R}_{\e{as};k}^{(1)}[ H ](x) 
\; + \; \mc{R}_{\e{as};k}^{(2)}[ H ](x) \bigg\} \;,\\
\Delta_{[k]}\mc{W}_{\e{bk}}^{(\e{as})}[ H ](x) & = & N^{-k\a}\cdot \mc{R}_{\e{as};k}^{(3)}[ H ](x)\;,
\label{ecriture decomposition delta asympt en reste bord et bulk fins}
\end{eqnarray}
where  $\mc{R}_{\e{as};k}^{(a)}[ H ] \in W^{\infty}_{\ell}(\R^+)$ for $a = 1,2,3$. For $a = 1$, we have:
\beq
\label{gfmdum}\big|\mc{R}_{\e{as};k}^{(1)}[ H ](x)\big| = \e{O}(x^{k+1})
\enq
uniformly in $N$. Moreover, we have uniform bounds for $x \in \intff{0}{\eps N^{\a}}$, namely for $\ell \in \intn{0}{k}$:
\begin{eqnarray}
\label{ecriture bornes fines sur reste asympt}  \qquad \big|\partial_{\xi}^{\ell}\mc{R}_{\e{as};k}^{(1)}[H](x_R)\big| & \leq & 
									    C_{k,\ell} \cdot  x^{k - \ell + 1}_{R} \cdot N^{\ell \a} \cdot {\cal N}^{(\ell)}_{N}\big[H_{\mf{e}}^{(k + 1)}\big] \;,\\
\label{ecriture bornes fines sur reste asympt2} a = 2,3,\qquad \big|\partial_{\xi}^{\ell}\mc{R}_{\e{as};k}^{(a)}[H](x_R)\big| & \leq & C_{k,\ell}\cdot N^{\ell \a} \cdot {\cal N}^{(\ell)}_{N}\big[H_{\mf{e}}^{(k + 1)}\big]\;,
\end{eqnarray}
where we remind $x_{R} = N^{\a}(b_N - \xi)$.
%
%Finally, for  $x \in \intff{0}{\eps N^{\a} }$, one can bound explicitly the derivatives as
%
%
%
%\beq
%
% \Big|  \f{ \Dp{}^{\ell} }{ \Dp{}\xi^{\ell} } \mc{R}_{\e{as};k}^{(a)}[ H ] (x_R)  \Big| \; \leq \; 
%
%C_{k,\ell}   \Big( N^{\ell \a} \cdot \big(x_R)^{ k+1 -\ell} +\de_{a2} + \de_{a3} \Big) \cdot \sul{p=0}{\ell} \f{ 1 }{ N^{p \a} } 
%
%\norm{ H_{\mf{e}}^{(k+1)} }_{W_{p}^{\infty}(\R)}  \;. 
%
%\label{ecriture bornes fines sur reste asympt}
%\enq
%
%
%
\end{lemme}
\noindent Note that we can combine the operators into the asymptotic expansion
\beq
\mc{W}_{R;k}^{(\e{as})}[H](x) \; + \; \mc{W}_{\e{bk};k}^{(\e{as})}[H](x) \; = \; 
 H^{\prime}(b_N) \, \mf{a}_0(x)\, \Bigg\{ 1+\sul{\ell=1}{k}\f{H^{(\ell + 1)}(b_N)\,\mathfrak{a}_{\ell}(x)}{H^{\prime}(b_N)\,N^{\a \ell} }\Bigg\} \;. 
\label{ecriture DA localise W bk et asymp R k}
\enq
%
%
%
%There, we agree upon 
%
%
%
%\beq
%
% \wh{\mf{a}}_{\ell}[H](x) \; = \; \f{ H^{(\ell+1)}(b_N)}{ H^{\prime}(b_N) } \cdot  \mf{a}_{\ell}(x)  
%
%\enq
%
%
%
%in which $\mf{a}_{\ell}[H]$ has been defined in \eqref{definition fcts a goth}. 

\Proof The form of the large-$N$ asymptotic expansion follows from straightforward manipulations on the Taylor
integral representation for $H^{(\ell)}(\xi)$ around $\xi=b_N$ for $\ell \in \intn{0}{k}$. The control on the remainder arising 
in \eqref{ecriture DA localise W asymp R k}, \eqref{ecriture DA localise W asymp bk k} and \eqref{ecriture DA localise W bk et asymp R k}
follows from the explicit integral representation for the remainder in the Taylor-integral series:
\begin{eqnarray}
\Delta_{[k]}\mc{W}_{R}^{(\e{as})}[H](x) & = & N^{-k\a} \Int{0}{1}\dd t\,H^{(k+1)}(b_N - N^{-\a}\,tx)\,
\Bigg\{ - \f{  (1-t)^k\,(-x)^{k+1} \varrho_0(x) }{ k!} \,-\,\sul{\ell=1}{k} \f{  (1-t)^{k - \ell}(-x)^{1+k-\ell}\,\varpi_{\ell}(x) }{ \ell!  (k-\ell)!}   \Bigg\}\;, \nonumber \\
\Delta_{[k]}\mc{W}_{\e{bk}}^{(\e{as})}[H](x) & = & N^{-k\a}   \sul{\ell=1}{k} u_{\ell}\,\f{ (-x)^{k+1-\ell} }{ (k-\ell)! }
\Int{0}{1} \dd t\,(1-t)^{k-\ell}\,H^{(k+1)}(b_N - N^{-\a}\,tx)\;. 
\end{eqnarray}
and we remark that $\varrho_0(x)$ -- given by \eqref{explitrho} -- has a logarithmic singularity when $x \rightarrow 0$. The details to arrive to \eqref{ecriture bornes fines sur reste asympt}-\eqref{ecriture bornes fines sur reste asympt2} are left to the reader. \qed

\vspace{0.2cm}

\noindent Collecting the bounds, we have obtained in sup norms, we find in particular $\mc{W}_N[H]$ is bounded when $H$ is $\mc{C}^1$:

\begin{cor}
 
 There exists $C>0$ independent of $N$ such that, for any $H \in \mc{C}^{1}(\intff{a_N}{b_N})$,
\beq
\norm{\mc{W}_N[H]}_{W_0^{\infty}(\intff{a_N}{b_N})} \; \leq  \;  C\,\norm{ H_{\mf{e}} }_{W_1^{\infty}(\R) } \;. 
\enq
\end{cor}

\section{The operator $\mc{U}_N^{-1}$}

Let us remind the definition of the operators ${\cal U}_{N}$ and ${\cal S}_{N}$:
\begin{eqnarray}
\label{mim}\mc{U}_N[\phi](\xi) & = & \phi(\xi)\cdot \Big\{  V^{\prime}( \xi) \, -  \, \mc{S}_N[ \rho_{\e{eq}}^{(N)} ](\xi)   \Big\}
 + \mc{S}_N[\phi \cdot \rho_{\e{eq}}^{(N)} ](\xi)  \\
\label{82} \mc{S}_N\big[ \phi \big](\xi) & = & \Fint{a_N}{b_N} S\big[N^{\a}(\xi-\eta)\big] \phi(\eta)\,\dd \eta \; \qquad
\e{and} \qquad S(\xi) \; = \; \sul{p=1}{2} \beta\pi\om_p  \cotanh\big[ \pi \om_p \xi \big] \;. 
\end{eqnarray}
and the fact that ${\cal W}_{N}$ defined in \S~\ref{Sous section construction inverse a l'operateur SN} is the inverse operator to ${\cal S}_{N}$. 
We also remind that the density $\rho_{\e{eq}}^{(N)}$ of the $N$-dependent equilibrium measure satisfies the integral equation:
\beq
\forall \xi \in \intff{a_N}{b_N},\qquad {\cal S}_{N}[\rho_{\e{eq}}^{(N)}](\xi) = V^{\prime}(\xi)\;.
\enq
This makes the first term of \eqref{mim} vanish for $\xi \in \intff{a_N}{b_N}$, but it can be non-zero outside of this segment.

In this section we obtain an integral representation for the inverse of ${\cal U}_{N}$, which shows that $\mc{U}_N^{-1}[H]$ is smooth as long as $H$ is. Then, in \S~\ref{Sous section Sharp weighted bounds for UN moins 1},
we shall provide explicit, $N$-dependent, bounds on the $W_{\ell}^{\infty}(\R)$ norms of $\mc{U}_N^{-1}[H]$. This technical result is crucial in the analysis of the Schwinger-Dyson equation performed in \S~\ref{s4fsfg}.

\subsection{An integral representation for $\mc{U}_{N}^{-1}$}
\label{Sous Section Int Rep Un moins 1 cas regulier et vague} 

\begin{prop}
\label{Proposition characterisation operateur UN}
The operator $\mc{U}_N$ is invertible on $\big( \mf{X}_s\cap \mc{C}^{1}\big) (\R)$, $0<s< 1/2$, and its inverse admits the representation
\beq
\mc{U}_N^{-1}[H](\xi) \; = \; \f{  \mc{V}_N[H](\xi)     }{   \mc{V}_N[ V^{\prime} ](\xi)   }\;,
\label{ecriture representation integrale agreable pour UN inverse}
\enq
where $\mc{V}_N \, = \, \mc{V}_N^{[1]} \,  +  \, \mc{V}^{[2]}_N$ with 
\beq
 \mc{V}_N^{[1]}[H](\xi) \; = \; \Int{a_N}{b_N}  \f{[H(\xi) - H(s)]\,\dd s}{ (\xi-s) \sqrt{(s-a_N)(b_N-s)}  } %
\qquad and \qquad 
 \mc{V}_N^{[2]}[H](\xi) \; = \; \Int{a_N}{b_N}  V_N^{[2]}(\xi,\eta)\cdot\mc{W}_N[ H ](\eta)\, \dd \eta \;. 
\label{definition operateurs V1 et V2}
\enq
and the integral kernel of the operator $ \mc{V}_N^{[2]}$ reads:
\beq
V_N^{[2]}(\xi,\eta)\; = \;  \Int{a_N}{b_N} \f{ S_{\e{reg}}\big[N^{\a}(s-\eta) \big] \, - \, S_{\e{reg}}\big[N^{\a}(\xi-\eta) \big] }
{ (\xi-s) \sqrt{(s-a_N)(b_N-s)}   } \,\dd s 
\qquad with  \qquad 
S_{\e{reg}}(\xi)   \, = \,  S(\xi) \,  - \,  \f{ 2\be }{ \xi } \;.  
\label{definition noyau integral VN}
\enq
Finally, we have that, for any $ \xi \in \intff{a_N}{b_N}$, $\mc{V}_N[V^{\prime}](\xi) \not= 0$. 
\end{prop}
Note that the above representation is not completely fit for obtaining a fine bound of the $W_{\ell}^{\infty}(\R)$ norm of 
$\mc{U}_{N}^{-1}[H]$ in the large-$N$ limit. Indeed, we will show in Appendix~\ref{Appendix minimisation de la mesure equilibre} that $\mc{V}_N[V^{\prime} ](\xi) > c_N >0$ for $N$ large enough. 
Unfortunately, the constant $c_N\tend 0$
and thus does not provide an optimal bound for the $W_{\ell}^{\infty}(\R)$ norm.  
Gaining a more precise control on $c_N$ (\textit{eg}. its dependence on $N$) is much harder, but a more precise control
is one of the ingredients that are necessary for obtaining sharp $N$-dependent bounds for the $W_{\ell}^{\infty}(\R)$ norm of 
$\mc{U}_{N}^{-1}[H]$. We shall obtain such a more explicit control on $c_N$ in the course of the proof of 
Theorem \ref{Theorem bornes sur norme inverse UN via estimation fines locales}. 

\Proof  Given $H \in (\mf{X}_s\cap \mc{C}^{1}_{\e{c}})(\R)$, let $\phi$ be the unique solution to the equation $\mc{S}_{N}[\phi](\xi) = H(\xi)$ on $\intff{a_N}{b_N}$. 
Reminding the definition of ${\cal S}_N$ in \eqref{ecriture eqn int sing de depart}, it means that, for $\xi \in \intoo{a_N}{b_N}$:
\beq
\Fint{a_N}{b_N} \f{ \phi(\eta)\, \dd \eta }{ (\xi-\eta) \i \pi }\; = \; U(\xi)
\qquad \e{where} \qquad 
U(\xi)\; = \; \f{ N^{\a} }{2 \i \pi \be } 
\Bigg\{ H(\xi) \, - \, \Int{a_N}{b_N}  \! \! S_{\e{reg}}\big[ N^{\a}(\xi- \eta) \big] \phi(\eta)\,\dd \eta  \Bigg\} \;. 
\label{expressions PP de solution eqn SN}
\enq
As a consequence, the function 
\beq
\label{sqtu}F(z) \; = \; \f{1}{q(z)}  \Int{a_N}{b_N} \f{\phi(\eta) }{ z-\eta } \cdot \f{\dd \eta }{2 \i \pi} \qquad \e{with} \qquad 
q(z) \; = \; \sqrt{(z-a_N)(z - b_N)}
\enq
solves the scalar Riemann--Hilbert problem
\begin{itemize}
\item $F \in \mc{O}( \Cx \setminus \intff{a_N}{b_N})$ and admits $\pm$ $L^{p}\big(\intff{a_N}{b_N} \big)$ boundary values for $p \in \intoo{1}{2}$ ;
\item $F(z) \; = \; \e{O}\big( z^{-1} \big)$ when $z \tend \infty$ ;
\item $F_+(x)-F_-(x) \; = \; U(x)/q_{+}(x)$ for any $x \in \intoo{a_N}{b_N}$ . 
\end{itemize}
Note that the $L^p$ character of the boundary values follows from the fact that both $\phi$ and the principal value 
integral are continuous on $\intff{a_N}{b_N}$. The former follows from Propositions 
\ref{Proposition decomposition op WN en diverses sous parties}-\ref{Proposition Ecriture reguliere uniforme des divers const de WN}
whereas the latter is a consequence of \eqref{expressions PP de solution eqn SN}. By uniqueness of the solution to such a Riemann--Hilbert problem, it follows that 
\beq
F(z) \; = \; \Int{a_N}{b_N}\f{U(s)}{q_+(s)(s-z) }\,\f{ \dd s }{ 2 \i\pi } \qquad \e{for} \qquad 
z \in \Cx \setminus \intff{a_N}{b_N}\;. 
\enq
By using that, for $\xi \in \intoo{a_N}{b_N}$, 
\beq
-\phi(\xi) \; = \; q_{+}(\xi) \cdot \Big(  F_+(\xi) \, + \,  F_-(\xi) \Big) \qquad \e{and}  \qquad
\Fint{a_N}{b_N}  \f{1}{q_{+}(s) \cdot (s-\xi)} \cdot \f{ \dd s }{ \i\pi} \; = \; 0\;,
\enq
we obtain that:
\beq
\phi(\xi) \; = \; \f{ \sqrt{ N^{2\a}(\xi-a_N)(b_N-\xi) }  }{ 2\pi^2 \be }\,\mc{V}_{N}[H](\xi) 
\label{ecriture rep. sol. ds sup mes eq}
\enq
with the expression of $\mc{V}_N$ given by \eqref{definition operateurs V1 et V2}. Further, given any $\xi \in \R \setminus \intff{a_N}{b_N}$, we have: 
\beq
\mc{S}_N[\phi](\xi) \; = \; \Int{a_N}{b_N} S_{\e{reg}}\big[ N^{\a} (\xi-\eta) \big] \phi(\eta)\,\dd \eta
\; + \;  \f{ 4 \i \pi \be  }{ N^{\a}} q(\xi) F(\xi) \;. 
\enq
It then remains to use that, for such $\xi$'s
\beq
\Int{a_N}{b_N}  \f{ 1 }{ q_{+}(s)(s-\xi) } \cdot \f{ \dd s }{ \i \pi} \; = \; \f{1}{q(\xi)} 
\enq
so as to get the representation 
\beq
\mc{S}_N[\phi](\xi) \; = \; H(\xi) \; -\f{ q(\xi) }{\pi}\cdot\mc{V}_N[H](\xi) \;. 
\label{ecriture rep alt action op SN en dehors sup mes eq}
\enq

We can now go back to the original problem. Let $\psi$ be any solution to $\mc{U}_N[\psi]=H$. 
Due to the integral equation satisfied by the density of equilibrium measure
on $ \intff{a_N}{b_N}$, it follows that, for any $\xi \in \intff{a_N}{b_N}$ such that $\mc{W}_N[V^{\prime}](\xi) \not=0$, 
\beq
\label{8155}\psi(\xi) \; = \; \f{ \mc{W}_N[H](\xi) }{ \mc{W}_N[V^{\prime}](\xi) } \;. 
\enq
and we can conclude thanks to the relation \eqref{ecriture rep. sol. ds sup mes eq}. 
For $\xi \in \R \setminus \intff{a_N}{b_N}$, we rather have:
\beq
\psi(\xi) \; = \; \f{ \mc{S}_N\big[ \mc{W}_N[H] \big](\xi) \, - \,  H(\xi)    }
					{ \mc{S}_N\big[ \mc{W}_N[V^{\prime}] \big](\xi) \, - \,  V^{\prime}(\xi)   } 
\enq
at any point where the denominator does not vanish. It then solely remains to invoke the relation 
\eqref{ecriture rep alt action op SN en dehors sup mes eq}. Note that 
\beq
\mc{V}_{N}[V^{\prime}](\xi) \; = \; \f{ 2\pi^2 \be\,\rho_{\e{eq}}^{(N)}(\xi) }{ \sqrt{ N^{2\a}(\xi-a_N)(b_N-\xi) } }   \;. 
\enq
It is shown in proof of Theorem~\ref{Proposition caracterisation rudimentaire mesure equilibre} given in Appendix~\ref{Appendix minimisation de la mesure equilibre}, point $(ii)$, 
that $\rho_{{\rm eq}}^{(N)}(\xi) > 0$ for $\xi \in \intoo{a_N}{b_N}$ for $N$ large enough and that it vanishes as a square root at the edges. 
Furthermore, it is also shown in that appendix, Equation~\eqref{ecriture positivite stricte de V eff en dehors support mu eq}, that 
$   V^{\prime}(\xi) \, - \, \mc{S}_N\big[ \mc{W}_N[V^{\prime}] \big](\xi) \neq 0$ %A: change >0 en \neq 0
on $\R \setminus \intff{a_N}{b_N}$. 
Thus the denominator in \eqref{ecriture representation integrale agreable pour UN inverse} never vanishes and thus  holds for any $\xi \in \R$ and any $H \in \mf{X}_s\cap\mc{C}^{1}_{\e{c}}(\R)$. 
The result then follows by density of $\mf{X}_s\cap\mc{C}^{1}_{\e{c}}(\R)$ in $\mf{X}_s\cap\mc{C}^{1}(\R)$.  \qed

%%%%%%%%%%%%%%%%%%%%%%%%%%%%%%%%%%%%%%%%%%%%%%%%%%%%%%%%%%%%%%%%%%%%%%%%%%%%%%%%%%%%%%%%%%%%%%%%%%%%%%%%%%%%%%%%%%%%%%%%%%%%%%%%%%%%%%%%%%%%%%
%%%%%%%%%%%%%%%%%%%%%%%%%%%%%%%%%%%%%%%%%%%%%%%%%%%%%%%%%%%%%%%%%%%%%%%%%%%%%%%%%%%%%%%%%%%%%%%%%%%%%%%%%%%%%%%%%%%%%%%%%%%%%%%%%%%%%%%%%%%%%%

%%%%%%%%%%%%%%%%%%%%%%%%%%%%%%%%%%%%%%%%%%%%%%%%%%%%%%%%%%%%%%%%%%%%%%%%%%%%%%%%%%%%%%%%%%%%%%%%%%%%%%%%%%%%%%%%%%%%%%%%%%%%%%%%%%%%%%%%%%%%%%
%%%%%%%%%%%%%%%%%%%%%%%%%%%%%%%%%%%%%%%%%%%%%%%%%%%%%%%%%%%%%%%%%%%%%%%%%%%%%%%%%%%%%%%%%%%%%%%%%%%%%%%%%%%%%%%%%%%%%%%%%%%%%%%%%%%%%%%%%%%%%%

%%%%%%%%%%%%%%%%%%%%%%%%%%%%%%%%%%%%%%%%%%%%%%%%%%%%%%%%%%%%%%%%%%%%%%%%%%%%%%%%%%%%%%%%%%%%%%%%%%%%%%%%%%%%%%%%%%%%%%%%%%%%%%%%%%%%%%%%%%%%%%
%%%%%%%%%%%%%%%%%%%%%%%%%%%%%%%%%%%%%%%%%%%%%%%%%%%%%%%%%%%%%%%%%%%%%%%%%%%%%%%%%%%%%%%%%%%%%%%%%%%%%%%%%%%%%%%%%%%%%%%%%%%%%%%%%%%%%%%%%%%%%%

\subsection{Sharp weighted bounds for $\mc{U}_N^{-1}$}
\label{Sous section Sharp weighted bounds for UN moins 1}

The aim of the present subsection is to prove one of the most important technical propositions of the paper, namely
sharp $N$-dependent bounds on the $W_{\ell}^{\infty}(\R)$ norm of $\mc{U}_N^{-1}[H]$.  Part of the difficulties of the proof consists in obtaining lower bounds for $\mc{W}_N[V^{\prime}]$ in the vicinity of $a_N$ and $b_N$ as well as in gaining a sufficiently precise control on the square root behaviour of $\mc{W}_N[H]$ at the edges.

Proposition~\ref{Theorem bornes sur norme inverse UN via estimation fines locales} below is the key tool for the 
large-$N$ analysis of the Schwinger-Dyson equations. We insist that although our result is effective in what concerns our purposes,
it is \textit{not} optimal. More optimal results can be obtained with respect to 
local $W_{\ell}^{\infty}$ norms, \textit{viz}. $W_{\ell}^{\infty}(J)$ with $J$ being specific subintervals of $\R$, or 
with respect to milder ones such as the $W_{\ell}^{p}(\R)$ ones. 
However, obtaining these results demands more efforts on the one hand and, on the other hand, requires 
much more technical handlings so as to make the best of them when dealing with the Schwinger-Dyson equations. 
We therefore chose not to venture further in these technicalities. 

%This theorem will build on the Riemann--Hilbert based analysis, and on the sharp estimates for the large-$N$
%behaviour of the operator $\mc{W}_N$.

Before stating the theorem, we remind the expression for the weighted norm (Definition~\ref{weiei}):
\beq
\label{weights}\mc{N}_N^{(\ell)}[\phi] \; = \; \sul{k=0}{\ell} \f{  \norm{ \phi }_{ W^{\infty}_{\ell}(\R) } }{ N^{\ell\a} } \; . 
\enq
and the \textit{ad hoc} norms on the potential (Definition~\ref{weie2}):
\beq
  \mf{n}_{\ell}[V] \; = \; \f{ \max\Big\{  \pl{a=1}{\ell} \norm{ \mc{K}_{\kappa}[V^{\prime}] }_{ W^{\infty}_{k_a}(\R^n) }  \; : \;  \sul{a=1}{\ell} k_a = 2 \ell + 1  \Big\} }
{ \bigg\{ \min \Big(1\, ,\,  \inf_{ \intff{a}{b} } |V^{\prime\prime}(\xi)| \, , \,  |V^{\prime}(b+\eps)-V^{\prime}(b)| 
\, , \, |V^{\prime}(a-\eps)-V^{\prime}(a)|  \Big)\bigg\}^{\ell+1} } 
\label{definition fct estimatrice du potentiel2}
\enq
for some $\epsilon > 0$ small enough but independent of $N$. We also remind that ${\cal K}_{\kappa}[H]$ is an exponential regularisation of $H$, 
see Definition~\ref{Definition regularisation exponentielle}.

\begin{prop}
\label{Theorem bornes sur norme inverse UN via estimation fines locales}

Let $\ell \geq 0$ be an integer, and $C_{V}$, $\kappa$  be positive constants. There exist a constant $C_{\ell}>0$ such that for any $H$ and $V$ satisfying 
\begin{itemize}
\item $\mc{K}_{\tf{ \kappa }{ \ell} }[H] \in W_{2\ell+1}^{\infty}(\R)$ and $\mc{K}_{ \tf{\kappa}{\ell} }[V] \in W_{2\ell+2}^{\infty}(\R)$ ;
\item $\norm{ V }_{W_3^{\infty}(\intff{a-\de}{b+\de})} <C_V$ \; for some $\de>0$ where $(a,b)$ are such that  
$(a_N,b_N) \underset{N \tend +\infty}{\tend} (a,b)$ ;
\item $H \in \mf{X}_s(\intff{a_N}{b_N})$ ;
\end{itemize}
we have the following bound:
\beq
\Norm{  \mc{K}_{\kappa}\big[ \mc{U}_{N}^{-1}[H] \big] }_{ W^{\infty}_{\ell}(\R) }  \; \leq  \; C_{\ell} \cdot \mf{n}_{\ell}[V] \cdot N^{(\ell+1)\a}
\cdot (\ln N)^{2 \ell + 1 } \cdot \mc{N}_{N}^{(2\ell+1)}\big[ \mc{K}_{\kappa}[H] \big] \;.
\enq
%
%
%
%We remind that 
%
%
%
%\beq
%
%  \mf{n}_{\ell}[V] \; = \;  
%
 % \f{ \max\Big\{ \pl{a=1}{\ell} \norm{ \mc{K}_{\kappa}[ V^{\prime} ]}_{ \mc{W}^{\infty}_{k_a}(\R) }  \; : \;  \sul{a=1}{\ell} k_a \, = \,  2 \ell + 1  \Big\}  }
%  
%{ \bigg\{ \min \Big( 1, \inf_{ \intff{a}{b} } |V^{\prime\prime}(\xi)| , |V^{\prime}(b+\eps)-V^{\prime}(b)| ,  |V^{\prime}(a-\eps)-V^{\prime}(a)| \Big)\bigg\}^{\ell+1} } 
%
%\enq
%
%
%
% where $\eps>0$ is taken small enough. 
\end{prop}

\Proof  As discussed in the proof of Proposition \ref{Proposition characterisation operateur UN}, the operator $\mc{U}_N^{-1}$ can be recast as
\beq
\label{mimim}\mc{U}_N^{-1}[H](\xi) \; = \; \f{ \mc{W}_N[H](\xi) }{ \mc{W}_N[V^{\prime}](\xi) } \cdot \bs{1}_{\intff{a_N}{b_N}}(\xi) \; + \; 
\f{ \mc{S}_N\big[ \mc{W}_N[H] \big](\xi) \, - \,  H(\xi)    }
					{ \mc{S}_N\big[ \mc{W}_N[V^{\prime}] \big](\xi) \, - \,  V^{\prime}(\xi)   } 
	\cdot  \bs{1}_{\intff{a_N}{b_N}^c}(\xi) \;. 
\enq
Therefore, obtaining sharp bounds on $\mc{U}_N^{-1}[H]$ demands to control, with sufficient accuracy, both ratios appearing in the 
formula above. Observe that the same Proposition \ref{Proposition characterisation operateur UN} and, in particular, 
equations \eqref{ecriture rep. sol. ds sup mes eq}-\eqref{ecriture rep alt action op SN en dehors sup mes eq}
ensure that, given $\eps>0$ small enough and $H$ of class $\mc{C}^{k+1}$, the functions
\beq
\xi \mapsto \f{ \mc{W}_N[H](\xi) }{  q_{R}(\xi) }\qquad \e{and} \qquad 
\xi \mapsto \f{\mc{S}_N\big[\mc{W}_N[H]\big](\xi)-H(\xi) }{ q_{R}(\xi) } \; 
 \label{ecriture fct in et ext sur qb}
\enq
with:
\beq
\label{Rgith}q_{R}(\xi) \, = \, \sqrt{N^{\a} (b_N-\xi)} = x_R^{1/2}
\enq
are respectively $\mc{C}^k(\intff{b_N-\eps}{b_N})$ and $\mc{C}^k(\intff{b_N}{b_N+\eps})$. A similar statement holds at the left boundary. 
Furthermore, the same proposition readily ensures that both functions are clearly $\mc{C}^{k+1}$ uniformly away from the boundaries.

The large-$N$ behaviour of both functions in \eqref{ecriture fct in et ext sur qb} is not uniform on $\R$ and depends
on whether one is in a vicinity of the endpoints $a_N,b_N$ or not. Therefore, we will split the analysis for $\xi$ in one of the four regions, from right to left on the real axis:

\begin{eqnarray}
\label{definition intervalles locaux 1}\mathbb{J}_N^{(R;\e{out})} & = & \intfo{b_N +\eps (\ln N)^2\cdot N^{-\a}}{ + \infty} \\
\label{definition intervalles locaux 2}\mathbb{J}_N^{(R;\e{ext})} & = & \intff{ b_N}{b_N +\eps (\ln N)^2\cdot N^{-\a} } \\
\label{definition intervalles locaux 3}\mathbb{J}_N^{(R;\e{in})} & = & \intff{b_N -\eps (\ln N)^2\cdot N^{-\a} }{ b_N} \\
\label{definition intervalles locaux 4}\mathbb{J}_N^{(\e{bk})} & = & \intff{a_N +\eps (\ln N)^2\cdot N^{-\a} }{ b_N-\eps (\ln N)^2\cdot N^{-\a} } \;.
\end{eqnarray}
Indeed, the behaviour on the three other regions:
\begin{eqnarray}
\label{definition intervalles locaux 5}\mathbb{J}_N^{(L;\e{in})} & = & \intff{ a_N }{ a_N +\eps (\ln N)^2\cdot N^{-\a} } \\
\label{definition intervalles locaux 6}\mathbb{J}_N^{(L;\e{ext})} & = & \intff{ a_N -\eps (\ln N)^2\cdot N^{-\a} }{ a_N } \\
\label{definition intervalles locaux 7}\mathbb{J}_N^{(L;\e{out})} & = & \intof{-\infty}{a_N -\eps (\ln N)^2\cdot N^{-\a}} 
\end{eqnarray}
can be deduced by the reflection symmetry 
\beq
\mc{W}_N[H](\xi) \; = \; -\mc{W}_N\big[H^{\wedge} \big](a_N+b_N-\xi) \; 
\enq
from the analysis on the local intervals \eqref{definition intervalles locaux 1}-\eqref{definition intervalles locaux 3}. 

The proof consists in several steps. First of all, we bound the $W_{\ell}^\infty(\mathbb{J}_N^{(*)} ) $ norm of 
the functions in \eqref{ecriture fct in et ext sur qb}, this depending on the interval of interest.
Also, we obtain \textit{lower} bounds for the same functions with $H \leftrightarrow V^{\prime}$. 
Finally, we use the partitioning of $\R$ into the local intervals \eqref{definition intervalles locaux 1}-\eqref{definition intervalles locaux 3}
so as to raise the local bounds into global bounds on $\mc{U}_N^{-1}[H]$ issuing from those on 
$\mc{W}_N[H]\cdot q_{R}^{-1}$
and $\big\{ \mc{S}_N\big[\mc{W}_N[H]\big]-H \big\} \cdot q_{R}^{-1}$.

\subsubsection*{Lower and upper bounds on $\mathbb{J}_N^{(R;\e{out})}$ }

Let us decompose $S$ given in \eqref{82} into:
\beq
\label{fmu} S(x) = S_{\infty}(x) + (\Delta S)(x),\qquad \mathrm{with}\quad S_{\infty}(x) = \be\pi(\omega_1 + \omega_2){\rm sgn}(x)
\enq
We observe that when $\xi \in \mathbb{J}_{N}^{(R;\e{out})}$ and $\eta \in \intff{a_N}{b_N}$ one avoids the simple pole in the kernel functions $S[N^{\a}(\xi - \eta)]$ of the integral operator ${\cal S}_{N}$. 
Besides, the decomposition \eqref{fmu} has the property that, for any integer $\ell \geq 0$, there exists constants $c,C_{\ell} > 0$ independent of $N$ such that:
\beq
\label{emium}\forall \xi \in \mathbb{J}_{N}^{(R;\e{out})},\,\,\forall \eta \in \intff{a_N}{b_N},\qquad \big| \partial_{\xi}^{\ell} (\Delta S)[N^{\a}(\xi-\eta)] \big| 
%*
\; \leq \;C_{\ell}\,N^{\ell \a}\,\ex{-c(\ln N )^2}\,.
\enq
We have proved in Lemma~\ref{Lemme bornage integrale simple de WN de H} and \ref{Lemme bornage norme L1 de WN} that 
\beq
\label{estimt} \bigg| \Int{a_N}{b_N} \mc{W}_N[H](\xi)\,\dd \xi \bigg| \; \leq \; C\,\norm{H_{\mf{e}}}_{W^{\infty}_0(\R)}\,,\qquad \norm{  \mc{W}_N[H] }_{ L^{1}(\intff{a_N}{b_N}) } 
\; \leq \;  C \, \norm{ H_{\mf{e}} }_{W_1^{\infty}(\R) }
\enq
for some $C > 0$ independent of $N$. Subsequently:
\begin{eqnarray}
\Norm{ \mc{S}_N\big[\mc{W}_N[H]\big] }_{ W_{\ell}^{\infty}(\mathbb{J}_N^{(R;\e{out})})} & \leq & \de_{\ell 0}\,C\,\norm{H_{\mf{e}}}_{W^{\infty}_0(\R)} 
\; + \; C_{\ell}\,N^{\ell \a}\,\ex{-c(\ln N )^2}\,\norm{ \mc{W}_N[H] }_{L^1(\intff{a_N}{b_N})} \\ 
& \leq & \de_{\ell 0}\,C^{\prime}\,\mc{N}^{(0)}_N\big[ \mc{K}_{\kappa}[H] \big] \; + \; 
C^{\prime}_{\ell}\,N^{(\ell+1) \a}\,\ex{-c(\ln N )^2}\,(b_N-a_N)\,\mc{N}^{(1)}_N\big[ \mc{K}_{\kappa}[H] \big] \;. 
\label{ecriture bornage operateur integral SN loin en dehors support} 
\end{eqnarray}
We have used: in the first line, the estimates \eqref{estimt} ; in the second line, the definition \eqref{weights} of the weighted norm, and we have included exponential regularisations 
via ${\cal K}_{\kappa}$, whose only effect is to change the value of the constant prefactors. Since $(a_N, b_N) \tend (a,b)$ in virtue of Corollary \ref{Corollaire DA des bornes aN et bN},
we can write for $N$ large enough:
\beq
\label{fsisss}\norm{ \mc{K}_{\kappa}\big[ \mc{S}_N\big[\mc{W}_N[H]\big] -H  \big] }_{ W_{\ell}^{\infty}(\mathbb{J}_N^{(R;\e{out})}) }
\; \leq \; \wt{C}_{\ell}\cdot N^{\ell\a }\cdot \mc{N}_N^{(\ell)}\big[ \mc{K}_{\kappa}[H] \big] \;. 
\enq
Indeed, a bound from the left-hand side is obtained by adding the $W_{\ell}^{\infty}$ norm of $H$ to \eqref{ecriture bornage operateur integral SN loin en dehors support}, 
which is itself bounded by a multiple of $N^{\ell\a}{\cal N}_{N}^{(\ell)}\big[{\cal K}_{\kappa}[H]\big]$.

\noindent Thanks to the decomposition \eqref{fmu} using that ${\rm sgn}(\xi - \eta) = 1$ for $\xi \in \mathbb{J}_{N}^{R;(\e{out})}$ and $\eta \in \intff{a_N}{b_N}$, as well as 
the exponential estimate \eqref{emium} and the $L^1$ bound of $\mc{W}_N$ from Lemma~\ref{Lemme bornage norme L1 de WN}, we can also write:
\beq
\mc{S}_{N}\big[\mc{W}_{N}[V']\big](\xi) - V'(\xi) = \underbrace{\pi \beta(\omega_1 + \omega_2)}_{= V'(b)} \underbrace{\Int{a_N}{b_N} \mc{W}_{N}[V'](\xi)\,\dd\xi}_{=1}  \; - \; 
V'(\xi) + \e{O}\Big(e^{-c(\ln N)^2}\,\norm{V^{\prime}}_{W_{1}^{\infty}(\intff{a_N}{b_N})}\Big)\;.
\enq
The identification of the first term comes from \eqref{definition fonction controle deviation vp sur bord}.
%It also readily follows from \eqref{ecriture bornage operateur integral SN loin en dehors support} that we have, for $N$ large enough and for any $\xi \in \mathbb{J}_N^{(R;\e{out})}$:
%
%
%
%\beq
%
%\mc{S}_N\big[\mc{W}_N[ V^{\prime} ]\big](\xi) \, - \,  V^{\prime}(\xi) \; = \; \underbrace{\pi \be (\om_1+\om_2)}_{=V^{\prime}(b)} 
%
%\underbrace{ \int_{a_N}^{b_N} \mc{W}_N[V^{\prime}](\xi) \dd \xi }_{ = 1 }  \, - \,  V^{\prime}(\xi) 
%
%\; + \; \e{O}\Big( \ex{-c (\ln N)^2} N^{\a} \norm{ V^{\prime} }_{W^{\infty}_1(\intff{a_N}{b_N})}  \Big) \; . 
%
%
%\; = \; \underbrace{ \big( V^{\prime}(b_N)-V^{\prime}(\xi) \big) }_{\geq N^{-\a} (\ln N)^2 \eps V^{\prime\prime}(b_N)}
%
%\; + \;   \underbrace{ \big( V^{\prime}(b)-V^{\prime}(b_N) \big)  }_{ \e{O}(N^{-\a}) }
%
%\; + \; \e{O}\Big( \ex{-c (\ln N)^2} N^{\a} \norm{ V^{\prime} }_{W^{\infty}_1(\intff{a_N}{b_N})}  \Big) \\
%
%
%\; \geq \; \f{\eps}{2} \f{  (\ln N)^2 }{ N^{\a} } V^{\prime\prime}(b_N) \;. 
%
%\enq
%
%
%
Further, we have for $|\xi-b| \leq \eps$ and $\xi \in \mathbb{J}_N^{(R;\e{out})}$:
\beq
\big| V^{\prime}(b)-V^{\prime}(\xi) | \; \geq \; |\xi-b| \cdot \inf_{\xi \in \intff{b}{b+\eps} }|V^{\prime\prime}(\xi)|  \; \geq \; 
\f{ \eps }{ 2 } \f{  (\ln N)^2 }{ N^{\a} } V^{\prime\prime}(b) \geq \frac{\epsilon}{2}\,\frac{V^{\prime\prime}(b)}{N^{\a}}
\enq
To obtain the last bound we have assumed that $\eps$ was small enough -- but still independent of $N$ -- and made use of $|b-b_N|=\e{O}(N^{-\a})$ as well as of 
$\norm{V}_{ W^{\infty}_3(\intff{a-\de}{b+\de}) } < +\infty$ and $N$ large enough. Finally, it is clear from the strict convexity of $V$ that in the case $|b - \xi| > \eps$:
\beq
|V^{\prime}(b)-V^{\prime}(\xi)| \; \geq  \; |V^{\prime}(b+\eps)-V^{\prime}(b)| \geq \frac{\epsilon}{2}\,\frac{V^{\prime}(b + \eps) - V^{\prime}(b)}{N^{\a}}\;,
\enq
where the last inequality is a trivial one. Therefore, in any case, for $N$ large enough:
\beq
\label{mimi}\big|{\cal S}_{N}\big[\mc{W}_{N}[V']\big] - V'(\xi)\big| \geq \frac{\epsilon}{4N^{\a}}\,\min \Big\{ \inf_{\xi \in \intff{a}{b}} V''(\xi)\,,\,|V'(b + \epsilon) - V'(b)|\Big\}\;.
\enq
The combination of the numerator upper bound \eqref{fsisss} applied to $H = V^{\prime}$ (using that the weighted norm is dominated by the $W^{\infty}$ norm) and 
the denominator lower bound \eqref{mimi} implies that, for any $\kappa>0$ such that both sides below are well-defined:
\beq
\label{842}\f{ \norm{ \mc{K}_{\kappa}\big[ \mc{S}_N\big[\mc{W}_N[V^{\prime}]\big] -V^{\prime}\big]  }_{ W_{\ell}^{\infty}(\mathbb{J}_N^{(R;\e{out})}) } }
{\big| \mc{S}_N\big[\mc{W}_N[ V^{\prime} ]\big](\xi) \, - \,  V^{\prime}(\xi) \big|  } \; \leq \;  
\f{N^{(\ell + 1)\a}\cdot  C_{\ell} \cdot \norm{V^{\prime}}_{W_{\ell}^{\infty}(\R)}}
{ \min \Big\{  \inf_{\xi \in \intff{a}{b} } |V^{\prime\prime}(\xi)| \, , \,   | V^{\prime}(b+\eps)-V^{\prime}(b)|   \Big\} } \;. 
\enq
Implicitly, we have treated $\epsilon$ from \eqref{mimi} like a constant.

\subsubsection*{Lower and upper bounds on $\mathbb{J}_N^{(\e{bk})}$ }

Consider the decomposition of ${\cal W}_{N}$ from \eqref{definition reste asympt ordre k pour WN}:
\bem
\mc{W}_N[H](\xi) \; = \; \mc{W}_{\e{bk};k}  [H](\xi) \; + \; \De_{[k]}\mc{W}_{\e{bk};k}[H_{\mf{e}}](\xi) \; + \; 
 \mc{W}_{R} [H_{\mf{e}}](x_R)  \\ 
\, -\,  \mc{W}_{R}\big[H^{\wedge} \big](x_L,b_N+a_N-\xi)\, + \, \mc{W}_{\e{exp}}[H_{\mf{e}}](\xi)\;.
\end{multline}
From the expression of ${\cal W}_{{\rm bk};k}$ in \eqref{definition W bk ordre k et ctes u ell}, we have the bound:
\beq
 \norm{ \mc{W}_{\e{bk};k}  [H] }_{  W_{\ell}^{\infty}(\mathbb{J}_N^{ (\e{bk}) }) }  \; \leq \;  c_{k;\ell}\cdot \max_{s \in \intn{0}{\ell}} {\cal N}_{N}^{(k-1)}[H^{(s + 1)}] 
 %c_{k;\ell}  \cdot \max_{s=0,\dots, \ell} \Big\{ \mc{N}_{N}^{(k-1)}[H^{(s+1)}] \Big\}
%
\label{ecriture borne sur Wbk dans bulk support}
\enq
and recollecting the estimates of the other terms from Propositions~\ref{Proposition decomposition op WN en diverses sous parties} and ~\ref{Proposition Ecriture reguliere uniforme des divers const de WN}, we also find:
\beq
  \Big|\Big|  \De_{[k]}\mc{W}_{\e{bk};k}[H_{\mf{e}}] \, + \, \mc{W}_{R} [H_{\mf{e}}] \, - \, (\mc{W}_{R})^{\wedge}[H_{\mf{e}}] \, + \, \mc{W}_{\e{exp}}[H_{\mf{e}}] \Big|\Big|
  _{ W_{\ell}^{\infty}(\mathbb{J}_N^{(\e{bk})}) } \; \leq \; c_{\ell}\,N^{-k\a}  \norm{ H^{(k+1)}_{\mf{e}} }_{W_{\ell}^{\infty}(\R)} \;, 
 \label{ecriture borne sur reste contre Wk bulk}
\enq
with the reflected operator $\mc{W}_{R}^{\wedge}$ as introduced in Definition~\ref{definition tilde fonction}. %\big[ H \big] (x_R,\xi) = \mc{W}_{R}\big[ H \big](x_L,b_N+a_N-\xi)$. 
We do stress that, in the present context, $H_{\mf{e}}$ denotes a compactly supported extension of $H$ from $\intff{a_N}{b_N}$ to $\R$
that, furthermore, satisfies the same regularity properties as $H$. All in all, the bounds 
\eqref{ecriture borne sur Wbk dans bulk support}-\eqref{ecriture borne sur reste contre Wk bulk} yield
\beq
\label{8422}  \norm{  \mc{W}_{N}[H](\xi) }_{ W_{\ell}^{\infty}(\mathbb{J}_N^{(\e{bk})}) } \; \leq \; 
c_{k;\ell}^{\prime}   \cdot \max_{s \in \intn{0}{\ell}} \Big\{ \mc{N}_{N}^{(k)}[H^{(s+1)}] \Big\} \;. 
\enq
Besides, for $k = 1$ we have from \eqref{definition W bk ordre k et ctes u ell}:
\beq
\mc{W}_{{\rm bk};k}[ V^{\prime} ](\xi) \;  = \;  u_1\,V^{\prime\prime}(\xi)\;.
\enq
The constant $u_1$ was introduced in Definition~\ref{oinoin}, and according to the expression of $R(\la)$ in \eqref{defrtr}, it takes the value:
\beq
\label{u1eq} u_1 = \frac{1}{2\pi\beta (\omega_1 + \omega_2)} > 0\;. 
\enq
So, using the bound \eqref{ecriture borne sur reste contre Wk bulk} for $k=1$ and $\ell = 0$ to control the extra terms in $\mc{W}_{N}$ in sup norm, we get 
\beq
\big| \mc{W}_{N}[V^{\prime}](\xi)  \big| \; \geq \; u_1
\inf_{ \xi \in \intff{a}{b} } V^{\prime\prime}(\xi) \; - \; \f{C}{N^{\a}} \norm{ V_{\mf{e}} }_{W^{\infty}_3(\R) }
\; \geq \; \f{u_1}{2} \cdot \inf_{ \xi \in \intff{a}{b} } \big\{ V^{\prime\prime}(\xi) \big\} 
\enq
where the last lower bound holds for $N$ large enough. The above lower bound leads to 
\beq
\label{850}\f{ \norm{  \mc{W}_{\e{bk};k}[V^{\prime}]  }_{ W_{\ell}^{\infty}(\mathbb{J}_N^{(\e{bk})}) }  }{ \big| \mc{W}_{N}[V^{\prime}](\xi)  \big| } \; \leq \; 
 \f{ C_{\ell} \cdot \norm{ V^{\prime} }_{ W_{k+\ell+1}^{\infty}( \mathbb{J}_N^{(\e{bk})}) }  }
 { \underset{ \xi \in \intff{a}{b} }{\inf} V^{\prime\prime}(\xi) } \;. 
\enq

\subsubsection*{Lower and upper bounds on $\mathbb{J}_N^{(R;\e{in})}$ }

In virtue of Lemma \ref{Lemme structure locale au bord pour WR et Wbk} and Proposition \ref{Proposition Ecriture reguliere uniforme des divers const de WN},
given $k \in \mathbb{N}^*$, we have the decomposition
\beq
\mc{W}_N[H](\xi) \; = \; \big( \mc{W}^{(\e{as})}_{R;k} \, + \, \mc{W}^{(\e{as})}_{\e{bk};k} \big) [H](x_R) \; + \; 
\Om_{R;k}[H_{\mf{e}}](x_R,\xi)
\enq
\beq 
\Om_{R;k}[H_{\mf{e}}](x_R,\xi) \; = \;   \Delta_{[k]} \mc{W}_{R}^{(\e{as})} [H](x_R) \, + \, \Delta_{[k]}\mc{W}_{\e{bk}}^{(\e{as})}  [H](x_R)
\, - \, \mc{W}_{R;k}[H^{\wedge}] (x_L,b_N+a_N-\xi)\, + \, \De_{[k]}\mc{W}_N[H_{\mf{e}}](\xi)
\enq
where $\De_{[k]}\mc{W}_N[H_{\mf{e}}]$ has been introduced in \eqref{definition reste asympt ordre k pour WN}. 
We remind from \eqref{ecriture DA localise W bk et asymp R k} that:
\beq
\big( \mc{W}^{(\e{as})}_{R;k} \, + \, \mc{W}^{(\e{as})}_{\e{bk};k} \big) [H](x_R)  = H'(b_N)\mathfrak{a}_0(x_R) 
\; +  \;  \sum_{r = 1}^{k} \frac{H^{(r + 1)}(b_N)}{N^{r\a}}\,(\mathfrak{a}_0\cdot\mathfrak{a}_{r})(x_R)
\enq
For any integers $n,\ell$ such that $n \geq \ell + 2$, Lemma~\ref{Lemme comportement fonction a goth} applied to $(\ell,m,n) \hookrightarrow (r,\ell + 1,n)$ tells us:
\beq
\frac{\mathfrak{a}_0(x)}{\sqrt{x}}  \; = \;  p_{0;\ell + 1,n}(x) \ex{-\vsg x }\,  + \,  x^{\ell + 1/2}f_{0;\ell + 1,n}(x),\qquad 
\frac{(\mathfrak{a}_0\cdot \mathfrak{a}_{r})(x)}{\sqrt{x}} \; = \;  p_{r;\ell + 1,n}(x)  \ex{-\vsg x } + x^{\ell + 1/2}f_{r;\ell + 1,n}(x)
\enq
for some polynomials $p_{k;\ell + 1,n}(x)$ of degree at most $n+k$ and functions $f_{k;\ell + 1,n} \in W_{n - (\ell + 1)}^{\infty}(\R_+)$. We therefore get:
\beq
\big| \big|   q^{-1}_{R} \big( \mc{W}^{(\e{as})}_{R;k}  + \mc{W}^{(\e{as})}_{\e{bk};k} \big) [H]     \big| \big|_{ W_{\ell}^{\infty}( \mathbb{J}_N^{(R;\e{in})}) } 
\leq c_{k;\ell}\cdot N^{\ell\a}\cdot (\ln N)^{2\ell + 1}\cdot \mc{N}_{N}^{(k - 1)}[H^{\prime}_{\mf{e}}] \; . 
\enq
In this inequality, one power of $N^{\a}$ pops up at each action of the derivative of $x_{R} = N^{\a}(b_N - \xi)$.
%the properties of the functions $\mf{a}_{\ell}$, \eqref{propriete fct a bornes polynomiales}-\eqref{propriete fct a cpt polynomial a infini} that 
%
%
%
%\beq
%
%\big| \big|   q^{-1}_{R} \big( \mc{W}^{(\e{as})}_{R;k}  + \mc{W}^{(\e{as})}_{\e{bk};k} \big) [H]     \big| \big|_{ W_{\ell}^{\infty}( \mathbb{J}_N^{(R;\e{in})}) }
%
%\; \leq \; c_{k;\ell} \cdot (\ln N )^{ 2 k-3} \cdot  N^{\ell\a}  \cdot  \mc{N}^{(k-1)}_N[ H_{\mf{e}}^{\prime} ] \;. 
%
%\enq
%
%
%
Furthermore, by putting together the control of the remainders in Proposition~\ref{Proposition Ecriture reguliere uniforme des divers const de WN} and Lemma~\ref{Lemme structure locale au bord pour WR et Wbk}, we get that:
\beq
\Om_{R;k}[H_{\mf{e}}](x_R,\xi) \; = \; \sul{m=0}{k} \Big\{ c_{k;m}^{(0)}  x_R^{m} \; + \; c_{k;m}^{(1/2)}  x_R^{m+\frac{1}{2}}   \Big\}
\; +\; f_k(x_R) 
\label{ecriture expression locale pour Psi R k}
\enq
where, for any $0 \leq \ell \leq k$, the function $f_k$ satisfies:
\beq
\big|\partial_{\xi}^{\ell}\big(x_R^{-1/2} f_k(x_R)\big)\big| \; \leq \; C_{k;\ell} \cdot x_R^{k+\f{1}{2}-\ell} 
\cdot N^{(\ell-k)\a} \cdot  \mc{N}_N^{(\ell)}[H_{\mf{e}}^{(k+1)}] \cdot \big( \ln(x_R)\,\ex{-C x_R}\, + \, 1 \big) \; . 
\enq
%
%
% 
%as guaranteed by the aforecited regularity of the remainders. 
Since the functions $ \big( \mc{W}^{(\e{as})}_{R;k}  + \mc{W}^{(\e{as})}_{\e{bk};k} \big) [H] \cdot q^{-1}_{R}$ and 
$\mc{W}_N[H] \cdot q^{-1}_{R}$ are smooth on $\mathbb{J}^{(R;\e{in})}_{N}$, so must be $\Om_{R;k}[H_{\mf{e}}] \cdot q^{-1}_{R}$. 
As a consequence, we necessarily have $c_{k;m}^{(0)}=0$. The properties of the remainders then ensure that, for any $0 \leq \ell \leq k$,
\beq
 \big| c_{k;m}^{(1/2)} \big| \; \leq \; 
C_{k;m}  \cdot N^{-k\a} \cdot \big| \big| H_{\mf{e}}^{(k+1)} \big| \big|_{W^{\infty}_{m}(\R)} \;. 
\enq
Thus, all-in-all, by choosing properly the compactly supported regular extension $H_{\mf{e}}$ of $H$ from $\intff{a_N}{b_N}$ to $\R$ we get 
\beq
\label{8474}\norm{ q_{R}^{-1}\cdot \mc{W}_N[H] }_{W_{\ell}^{\infty}(\mathbb{J}_N^{(R;\e{in})}) }\; \leq \; C_{\ell}\cdot (\ln N )^{2\ell+1} \cdot  
N^{(\ell+1)\a} \cdot \mc{N}_N^{(2\ell+1)}[ \mc{K}_{\kappa}[ H_{\mf{e}} ] ]
\enq
upon choosing $ k=\ell$. This holds for any $\kappa>0$, the right-hand side being possibly $+\infty$.

In what concerns the lower bounds, observe that 
\beq
 x_R^{ - 1/2 } \cdot \big(\mc{W}^{(\e{as})}_{R;1}  + \mc{W}^{(\e{as})}_{\e{bk};1} \big) [H](x_R)  \; = \; 
\f{ \mf{a}_0(x_R) }{ \sqrt{x_R} } V^{\prime \prime}(b_N) \Bigg( 1 \, + \, \f{ V^{(3)}(b_N)  }{ V^{\prime \prime}(b_N) } \cdot 
 \f{ \mf{a}_1(x_R) }{ N^{\a} }  \Bigg) 
\enq
as well as 
\beq
\big| c_{1;0}^{(1/2)} \, + \,  c_{1;1}^{(1/2)}x_R \, + \,  x_R^{-1/2 } f_{1}(x_R) \big| \; \leq \; 
C \cdot \Big\{  N^{-\a}\cdot (x_R+1) \norm{V^{\prime\prime}_{\mf{e}}}_{W_1^{\infty}(\R)} 
\; + \;  N^{-\a}\,x_R^{\f{3}{2}}\big( \ln x_R\,\ex{-Cx_R}+1 \big)\norm{V^{\prime\prime}_{\mf{e}}}_{W_0^{\infty}(\R)}  \Big\} \;. 
\enq
These estimates imply, for $N$ large enough:
\beq
\Big|  \f{ \mc{W}_N[V^{\prime}](\xi) }{ q_{R}(\xi) } \Big| \; > \; \f{ \mf{a}_0(x_R)  }{ \sqrt{x_R} } V^{\prime \prime}(b_N)
\; - \; \f{ (\ln N )^3  }{ N^{\a} } \norm{ V_{\mf{e}} }_{ W_{3}^{\infty}(\R) } \;. 
\enq
The function $x \tend \tf{ \mf{a}_0(x) }{ \sqrt{x} }$ is bounded from below on $\R^+$, \textit{cf}. Lemma \ref{Lemme comportement fonction a goth}
and $(a_N, b_N) \tend (a,b)$ in virtue of Corollary \ref{Corollaire DA des bornes aN et bN}. As a consequence, for any potential $V$
such that $\norm{ V_{\mf{e}} }_{ W_{3}^{\infty}(\intff{a}{b}) }<C$, there exists $N_0$ large enough  and $c>0$ such that 
\beq
\label{863}\Big|  \f{ \mc{W}_N[V^{\prime}](\xi) }{ q_{R}(\xi)  } \Big| \; > \; c \inf_{\intff{a}{b}} \big\{ V^{\prime \prime}( \xi ) \big\}  \;. 
\enq
We can deduce from the above bounds that, for any $\xi \in \mathbb{J}_N^{(R;\e{in})}$,
\beq
\label{864}\f{  \norm{ q_{R}^{-1}\cdot \mc{W}_N[V^{\prime}] }_{W_{\ell}^{\infty}(\mathbb{J}_N^{(R;\e{in})}) }  }{ q_{R}^{-1}(\xi) \cdot \mc{W}_N[V^{\prime}](\xi)  } 
\; \leq \; C_{\ell} \cdot  (\ln N )^{ 2 \ell +1} \cdot  
N^{\ell\a} \cdot \f{ \norm{ V^{\prime} }_{W_{2\ell+1}^{\infty}(\mathbb{J}_N^{(R;\e{in})}) }  }{ \inf_{\intff{a}{b}} \big\{ V^{\prime \prime}( \xi ) \big\} } \;. 
\enq

\subsubsection*{Lower and upper bounds on $\mathbb{J}_N^{(R;\e{ext})}$ }

Let us go back to the vector Riemann--Hilbert problem discussed in Lemma~\ref{Proposition corresp operateur et WH factorisation}. The representation \eqref{ecriture extension eqn integrale sing a analyser} and 
the fact that the solution $\Phi$ to this vector Riemann--Hilbert problem allows one the reconstruction of
the functions $\psi_1$ and $\psi_2$ arising in  \eqref{ecriture extension eqn integrale sing a analyser}
through \eqref{ecriture representation vecteur Phi}. Using the reconstruction formula \eqref{ecriture solution RHP vectoriel sur Hs general s negatif} with $P_1=P_2=0$ and $z_0=\infty$ 
and applying the regularisation trick exactly as in \eqref{equation avec le regularisation trick}, we get $\xi \in \intfo{b_N}{+\infty}$:
\beq
\mc{S}_N\big[ \mc{W}_N[H] \big](\xi) \;= \; N^{\a} \hspace{-2mm} \Int{ \R + 2\i \eps}{} \hspace{-2mm} \f{\dd \la }{ 2\pi }  \hspace{-1mm}
\Int{ \R + \i \eps}{} \hspace{-1mm} \f{\dd \mu }{ 2  \i \pi }
\Int{a_N}{b_N} \hspace{-1mm} \dd \eta  H(\eta) \f{ \ex{\i \la N^{\a}(b_N-\xi)- \i \mu N^{\a}(b_N-\eta) } }{ \mu - \la } \cdot 
\Big\{ \chi_{21}(\la) \chi_{12}(\mu) \, - \, \f{ \mu }{ \la }\cdot  \chi_{11}(\mu) \chi_{22}(\la)  \Big\} \;. 
\enq
The local behaviour of the above integral representation can be studied with the set of tools
developed throughout Section \ref{Section descirption cptmt unif op WN}. We do not reproduce this reasoning again. 
All-in-all, we obtain: 
\beq
\label{866}\norm{ q_{R}^{-1}\cdot \mc{K}_{\kappa}\big[ \mc{S}_N\big[ \mc{W}_N[H] \big] -H \big] }_{W_{\ell}^{\infty}(\mathbb{J}_N^{(R;\e{ext})}) }\; \leq \; C_{\ell} (\ln N )^{2\ell+1} \cdot  
N^{(\ell+1)\a} \cdot \mc{N}_N^{(2\ell+1)}\big[ \mc{K}_{\kappa}[H] \big]
\enq
and, for any $\xi \in \mathbb{J}_N^{(R;\e{ext})}$, 
\beq
\big| q^{-1}_{R}(\xi) \cdot \big\{ \mc{S}_N\big[ \mc{W}_N[ V^{\prime} ](\xi) \big] -V^{\prime}(\xi) \big\} \big| \; > \; c \inf_{\intff{a}{b}} V^{\prime \prime}(b_N)  
\enq
provided that $N$ is large enough. Likewise, we have the bounds:
\beq
\label{873}\f{  \norm{ q_R^{-1} \mc{K}_{\kappa}\big[ \mc{S}_N\big[ \mc{W}_N[V^{\prime}] \big] -V^{\prime} \big]  }_{W_{\ell}^{\infty}(\mathbb{J}_N^{(R;\e{ext})}) }  }
{ q_{R}^{-1}(\xi) \cdot \Big\{ \mc{S}_N\big[ \mc{W}_N[V^{\prime}] \big] -V^{\prime}  \Big\} } 
\; \leq \; C_{\ell} \cdot  (\ln N )^{ 2 \ell +1} \cdot  
N^{\ell\a} \cdot \f{ \norm{ \mc{K}_{\kappa}[V^{\prime}] }_{W_{2\ell+1}^{\infty}(\mathbb{J}_N^{(R;\e{ext})}) }  }{ \inf_{\intff{a}{b}} \big\{ V^{\prime \prime}( \xi ) \big\} } \;. 
\enq

\subsubsection*{Synthesis}

Let us now write:
\bem
\mc{U}_N^{-1}[H](\xi) \; = \; \sul{A=L,R}{} \Bigg\{
\f{ \mc{S}_N\big[ \mc{W}_N[ H ]\big] (\xi) - H(\xi)    }
{\mc{S}_N\big[ \mc{W}_N[ V^{\prime} ]\big] (\xi) - V^{\prime} (\xi)    }  \cdot \bs{1}_{ \mathbb{J}_N^{(A;\e{out})} }(\xi)
 \; + \; \f{ \mc{W}_N[ H ] (\xi)  }{ \mc{W}_N[ V^{\prime} ] (\xi)   }  \cdot \bs{1}_{ \mathbb{J}_N^{(A;\e{in})} }(\xi) \\
 \; + \; \f{ q^{-1}_{R}(\xi) \cdot \big\{ \mc{S}_N\big[ \mc{W}_N[ H ]\big] (\xi) - H(\xi) \big\}   }
{ q^{-1}_{R}(\xi) \cdot \big\{ \mc{S}_N\big[ \mc{W}_N[ V^{\prime} ]\big] (\xi) - V^{\prime} (\xi) \big\}   }  \cdot \bs{1}_{ \mathbb{J}_N^{(A;\e{ext})} }(\xi)    \Bigg\} 
 \; + \; \f{ \mc{W}_N[ H ] (\xi)  }{ \mc{W}_N[ V^{\prime} ] (\xi)   }  \cdot \bs{1}_{ \mathbb{J}_N^{(\e{bk})} }(\xi)   \;, 
\end{multline}
The piecewise bounds \eqref{fsisss}-\eqref{842} on $\mathbb{J}_{N}^{(R;\e{out})}$, \eqref{8422}-\eqref{850} on $\mathbb{J}_{N}^{(\e{bk})}$, \eqref{8474}-\eqref{864} on $\mathbb{J}_{N}^{(\e{bk})}$, 
\eqref{866}-\eqref{873} on $\mathbb{J}_{N}^{(R;\e{in})}$, and those which can be deduced by reflection symmetry on the three other segments defined in \eqref{definition intervalles locaux 5}-\eqref{definition intervalles locaux 7}, 
can now be used together with the Fa\'a d{i} Bruno formula
\beq
\f{\dd^{\ell}  }{  \dd \xi^{\ell} } \Big( \f{f}{g} \Big)(\xi) \; = \; \sul{n+m=\ell}{} \sul{ \sum k n_k=n}{}
\f{ \ell! \big( \sum_{k=1}^{n} n_k \big)!  }{ m!  } \cdot \f{ f^{(m)}(\xi) }{ g(\xi) } \cdot 
\pl{j=1}{n} \bigg\{ \f{1}{n_j!} \Big( \f{- g^{(j)}(\xi) }{j!g(\xi)}   \Big)^{n_j}   \bigg\} 
\enq
to establish the global bound. Note that, in the intermediate bounds, one should use the obvious property of the exponential regularisation:%
\beq
K_{\kappa}[f_1\cdots f_p] \; = \; \pl{a=1}{p} \mc{K}_{ \tf{\kappa}{p} }[f_a] \;.  
\enq
The details are left to the reader. \qed

\chapter{Asymptotic analysis of integrals}
\label{Chapitre AA integrales simples et doubles}

{\bf Abstract}

\textit{ In this Chapter we carry out the large-$N$ asymptotic analysis of the single and double integrals that arise in the problem. 
First, in \S~\ref{Soussection AA de integrale contraintes}, we deal with the one-fold integrals that arise in the characterisation of the image space 
$\mf{X}_{s}(\R)$ of $H_{s}\big( \intff{a_N}{b_N}\big)$ under the operator $\mc{S}_N$. Then, in \S~\ref{simpleintegrals} we evaluate 
asymptotically in $N$ one-dimensional integrals of  $\mc{W}_N[H]$ versus test functions $G$. 
This provides the first set of results that were necessary in \S~\ref{adfsdgfsg} for a thorough calculation of the large-$N$ expansion of the partition function.
Then, in Section \ref{Section characterisation support mesure equilibre} we build on the obtained large-$N$ expansion of the two types of single integrals
so as to characterise of the support of the equilibrium measure.  Finally, in Section \ref{Section asymptotic analysis of double integrals} we obtain the 
large-$N$ expansion, up to  a vanishing with $N$ remainder, of the double integral \eqref{ecriture integrale double DA a beta 1} arising in the large-$N$ expansion of the partition function
at $\be=1$.}

\section{Asymptotic analysis of single integrals}
\label{Section asymptotic analysis of single integrals}

\subsection{Asymptotic analysis of the constraint functionals $\mc{X}_{N}[H]$}
\label{Soussection AA de integrale contraintes}

Recall that for any $H \in \mc{C}^{1}(\intff{a_N}{b_N})$ the linear form $\mc{X}_N[H]$ defined in \eqref{definition form lineaire contrainte X}:
\beq
\label{fmiun}\mc{X}_N[H] \; = \; \f{ \i N^{\a} }{ \chi_{11;+}(0) } \Int{ \R  + \i \eps^{\prime} }{} \hspace{-2mm} \f{ \dd \mu }{2 \i \pi } 
\; \chi_{11}(\mu)  \Int{a_N}{b_N} H(\eta) \ex{ \i N^{\a}\mu(\eta-b_N)}\,\dd \eta 
\enq
is related to the constraint $\msc{I}_{11}[h]$ defined in \eqref{I12b} where $H$ and $h$ are related by the rescaling \eqref{ecriture changement de variable pour arriver au RHP}:
\beq
\mathscr{I}_{11}[h] \; = \; -\frac{N^{\a}\,\chi_{11;+}(0)}{2\pi\beta}\,{\cal X}_{N}[H]\;,\qquad h(x) = \frac{N^{\a}}{2\i\pi\beta}\,H(a_N + N^{-\a}x) \;.
\enq 
In the following, we shall obtain the large-$N$ expansion  of the linear form ${\cal X}_{N}[h]$ introduced in \eqref{definition form lineaire contrainte X} 
and defining the hyperplane $\mathfrak{X}_{s}$ where we inverse operators. We first need to define new constants:
\begin{defin}
\label{firstdalet}If $p \geq 0$ is an integer, we define:
\beq
\daleth_p \; = \; - \frac{R_{\da}(0) }{2} \Int{ \R + \i \eps^{\prime} }{} \f{1}{\mu^{p+1} R_{\da}(\mu) } \cdot \f{ \dd \mu }{ 2 \i \pi }
 \; = \; (-1)^{p+1}\,\frac{R_{\da}(0) }{2} \Int{ \R - \i \eps^{\prime} }{} \f{1}{\mu^{p+2} R_{\ua}(\mu) } \cdot \f{ \dd \mu }{ 2 \i \pi }\;.
\label{definition constante cp pour DA integrales}
\enq
\end{defin}
The equality between the two expressions of $\daleth_{p}$ follows from the symmetry \eqref{ecriture identite conjugaison R plus et R moins}.

\begin{lemme}
\label{Proposition DA integrale contrainte}
Let $k \geq 1$ be an integer, and $H \in \mc{C}^{k}\big( \intff{a_N}{b_N} \big)$. We have an asymptotic expansion:%
\beq
\mc{X}_N[H]  =     \sul{p =  0}{k-1}
\f{ \i^{p}\,\daleth_{ p }}{ N^{\a p } }\,\Big\{ H^{(p)}(a_N) \, + \, (-1)^p  H^{(p)}(b_N) \Big\}  
\;\; + \; \; \De_{[k]} \mc{X}_N[H]  \;,
\label{ecriture DA contrainte I11} 
\enq
where:
\beq
\big| \De_{[k]}\mc{X}_N[H]  \big| \; \leq \; C\,N^{-k\a}\,\norm{ H }_{ W^{\infty}_{k}(\intff{a_N}{b_N})}\;.
\enq
\end{lemme}

%Furthermore, the expansion \eqref{ecriture DA contrainte I11}  up to order $k$ holds for $H \in \mc{C}^k(\intff{a_N}{b_N})$.

\Proof For $\la $ between $\Ga_{\ua}$ and $\R  $, we decompose $\chi$ into:
\beq
\chi(\la) \; = \; \chi^{({\rm as})}_{\ua}(\la) \; + \; \chi^{(\e{exp})}_{\ua}(\la)
\label{ecriture decompositon asymptotique chi}
\enq
In terms of the various matrices used \S~\ref{SousSection RHP Auxiliaire pour chi}, the main part is:
\beq
\label{777}\chi^{(\e{as})}_{\ua}(\la) \; = \; \mc{R}_{\ua}^{-1}(\la) \cdot \big[\ups(\la)\big]^{-\sg_3} \cdot 
M_{\ua}(\la) \cdot \Bigg(I_2 + \frac{\sigma^{-}}{\la}\Bigg) %% I set P_{R}^{(0)}(\la) = I_2 + \sigma^{-}/\la
= \; \left( \ba{cc} -\f{ \ex{\i\la \ov{x}_N} }{ R_{\da}(\la) }  \; + \; \f{ 1 }{ \la R_{\ua}(\la)  }  & \f{ 1 }{ R_{\ua}(\la) }   
\vspace{2mm}\\ 
	 - \; R_{\ua}(\la)     &     \; 0			    \ea \right) 
\enq
and is such that the remainder is exponentially small in $N$:
\beq
\label{778}\chi^{(\e{exp})}_{\ua}(\la) \; = \; \chi^{(\e{as})}_{\ua}(\la) \cdot [\delta\Pi](\la) %
 \qquad \e{with} \qquad [\delta\Pi](\la) \, = \, \Bigg( I_2 \, + \, \f{  \sg^{-} }{ \la }  \Bigg)^{-1}\!\!\! \cdot \Pi(\la) \cdot P_{R}(\la) \; - \;  I_2\;.
\enq
Indeed, the large-$N$ behaviour of $\th_R$ inferred from \eqref{ecriture forme precise TF gN} and \eqref{definition theta R} as well as 
the estimate \eqref{ecriture bornes en N pour Pi moins Id} on the matrix $\Pi - I_2$ imply that, for $\epsilon^{\prime}$ fixed but small enough, and uniformly in $\la \in \R + \i \tau$,
$0<\tau<\eps^{\prime}$:
\beq
 \big|  [\delta\Pi]_{ab}(\la)  \big| \; \leq \; \f{ C\,\ex{- \varkappa_{\epsilon^{\prime}}\,N^{\a} }}{ 1 + |\la | }\;.   
%
%\e{where} \quad \varkappa_{\epsilon}=x_N \inf_{\la\in \Ga_{\ua}\cup\Ga_{\da}}|{\rm Im}\,\la | \, > \,  2 x_N \eps^{\prime} 
%
\enq
Furthermore, a direct calculation shows that
\beq
[\chi_{\ua}^{(\e{exp})}]_{11}(\la)   \; = \;
 \Bigg( \f{ 1 }{\la R_{\ua}(\la)} - \f{ \ex{ \i \la \ov{x}_N} }{ R_{\da}(\la) }   \Bigg)\,[\delta\Pi]_{11}(\la)
\; + \; \f{ [\delta\Pi]_{21}(\la)}{ R_{\ua}(\la) } \;, 
\enq
and taking into account the large-$\la$ behaviour of $R_{\ua/\da}$ given in \eqref{ecriture explicite R plus}- \eqref{ecriture explicite R moins}, we also get a uniform bound for$\la \in \R + \i \tau$,
$0<\tau<\eps^{\prime}$:
\beq
\label{ijnin} \big| [\chi_{\ua}^{(\e{exp})}]_{11}(\la)   \big| \; \leq \;
\f{ C^{\prime}\,\ex{- \varkappa_{\epsilon^{\prime}} N^{\a} } }{ \sqrt{ 1+ |\la|} }  \;. 
\enq
In particular, this estimate \eqref{ijnin} implies:
\beq
\label{aaasf}\frac{1}{\chi_{11;+}(0)}\, = \, -\frac{R_{\da}(0) }{ 2 }\, + \, \e{O}(e^{-\varkappa_{\epsilon^{\prime}}N^{\a}})\;.
\enq
The decomposition \eqref{ecriture decompositon asymptotique chi} in formula~\eqref{fmiun} induces a decomposition:%
\beq
\mc{X}_N[H] \; = \; \mc{X}^{(\e{as})}_N[H] \; + \;  \mc{X}^{(\e{exp})}_N[H]
\enq
where
\beq
\mc{X}^{(\e{exp})}[H] \; = \; \f{ \i N^{\a} }{ \chi_{11;+}(0) }
 \Int{ \wt{\msc{C}}^{(-)} }{} \hspace{-2mm} \f{ \dd \mu }{2 \i \pi } \, 
 \big[ \chi_{\ua}^{(\e{exp})}(\mu) \big]_{11} \Int{a_N}{b_N} H(\eta) \ex{ \i N^{\a}\mu(\eta-b_N)} \cdot \dd \eta  
\enq
and $\wt{\msc{C}}^{(-)}$ is a contour surrounding $0$ from above, going to $\infty$ in $\mathbb{H}^-$ along the rays $t\ex{-\f{3\i\pi}{4}}$ and $t\ex{-\f{\i\pi}{4}}$ 
and such that $\max\big\{ \e{Im}(\la) \; : \; \la \in \wt{\msc{C}}^{(-)} \big\} = \eps^{\prime}$. Note that we could have carried out this contour deformation since $\Pi(\la)$
is holomorphic in the domain delimited by $\R+\i\eps^{\prime}$ and $\wt{\msc{C}}^{(-)}$. 

\noindent Since for $\la \in \wt{\msc{C}}^{(-)}$, we have:
\beq
\bigg|  \Int{a_N}{b_N} H(\eta) \ex{ \i N^{\a}\la(\eta-b_N)}\,\dd \eta  \bigg| \; \leq \;
 \f{ C\,\ex{   \ov{x}_N \eps^{\prime} }\,}{ |\la | }\,\norm{ H }_{ L^{\infty}(\intff{a_N}{b_N})}\; ,  
\enq
it is readily seen that 
\beq
\big| \mc{X}^{(\e{exp})}_{N}[H]  \big| \; \leq  \;  C^{\prime} \cdot  N^{\a} \ex{ - \f{ \varkappa_{\epsilon^{\prime}} }{2} N^{\a} }\,  
\norm{ H }_{ L^{\infty}( \intff{a_N}{b_N}) } \;. 
\label{ecriture borne exp petite sur I cal pert de H}
\enq
It thus remains to estimate 
\beq
\mc{X}^{(\e{as})}_N[H] = \mc{X}_{R}^{(\e{as})}[H] \; + \; \mc{X}_{R}^{(\e{as})}[H^{\wedge} ] 
\enq
where 
\beq
\label{717}\mc{X}_{R}^{(\e{as})}[H] \;  =  \;  \f{\i N^{\a} }{ \chi_{11;+}(0) }
 \Int{ \msc{C}^{(-)}_{\e{reg}} }{} \hspace{-2mm} \f{ \dd \mu }{2 \i \pi } \, 
\f{ 1 }{\mu R_{\ua}(\mu)} \Int{a_N}{b_N} H(\eta) \ex{ \i N^{\a}\mu(\eta-b_N) }\,\dd \eta  \;,
\enq
and the second term arises upon the change of variables $(\mu,\eta) \mapsto (-\mu,a_N+b_N-\eta)$ in the initial expression. The dependence in $N$ is implicit in these new notations. Note that we could deform
the contour from $\R+\i \eps^{\prime}$ up to $\R - \i\epsilon^{\prime}$ or $\msc{C}_{\e{reg}}^{(-)}$ since the integrand is holomorphic in the 
domain swapped in between. Replacing $H$ by its Taylor series with integral remainder at order $k$, we get:
\beq
\mc{X}_{R}^{(\e{as})}[H] \; = \; \mc{X}_{R;k}^{(\e{as})}[H]  \; + \; \De_{[k]} \mc{X}_{R}^{(\e{as})}[H] \;.
\enq
The first term is:
\beq
\mc{X}_{R;k}^{(\e{as})}[H] \; = \; \i N^{\a}   \sul{p=0}{k-1} \f{ H^{(p)}(b_N) }{ p! \,\chi_{11;+}(0) } \Int{ \R - \i\epsilon^{\prime} }{} \f{ \dd \mu }{ 2\i \pi}
\f{1}{\mu R_{\ua}(\mu) } \Int{- \infty }{ 0 } \eta^p \ex{\i N^{\a} \mu \eta }\, \dd \eta 
\; = \;  \f{ -2 }{ R_{\da}(0) \chi_{11;+}(0)  } \sul{p=0}{k-1} \frac{(-\i)^{p}\,\daleth_p\,H^{(p)}(b_N) }{ N^{ p\a } }
\enq
where we have recognised the constants $\daleth_p$ of Definition~\ref{firstdalet}. The remainder is:
\bem
\De_{[k]} \mc{X}_{R}^{(\e{as})}[H] \; = \; \f{1}{\i\chi_{11;+}(0) }\Bigg\{\sul{p=0}{k-1} \f{ H^{(p)}(b_N) }{ p!\,N^{p\a} } 
\Int{ \msc{C}_{\e{reg}}^{(-)} }{} \f{ \dd \mu }{ 2\i \pi} \f{1}{\mu R_{\ua}(\mu) } \Int{- \infty }{ -\ov{x}_N } \eta^p \ex{ \i \mu \eta } \,\dd \eta  \\
\; + \; \Int{ \msc{C}_{\e{reg}}^{(-)} }{} \f{ \dd \mu }{ 2\i \pi}
\f{ N^{\a} }{\mu R_{\ua}(\mu) } \Int{ a_N }{ b_N } \dd \eta\,(\eta-b_N)^k \Int{0}{1} \dd t\,\f{ (1-t)^{k-1} }{ (k-1)! } \ex{ \i N^{\a} \mu (\eta-b_N) }\,
H^{(k)}\big( b_N+t(\eta-b_N) \big)\Bigg\} \;. 
\label{ecriture rest integral IRinfty}
\end{multline}
$\mc{X}_{R;k}^{(\e{as})}[H] $ yields  the leading terms of the asymptotic expansion announced in \eqref{ecriture DA contrainte I11}. Hence, it remains to bound  
$\De_{[k]} \mc{X}_{R}^{(\e{as})}[H]$. The first line in \eqref{ecriture rest integral IRinfty} is exponentially small and bounded by 
a term proportional to $\norm{ H }_{W^{\infty}_{k-1}(\intff{a_N}{b_N})}$. The second line is bounded by 
\beq
N^{\a} \cdot |R_{\da}(0) |\cdot \norm{ H }_{W^{\infty}_{k}(\intff{a_N}{b_N})}\,\Int{ \msc{C}_{\e{reg}}^{(-)} }{} \hspace{-1mm} \f{ |\dd \mu| }{ 2\pi\,k!}
\f{1}{ \big|\mu R_{\ua}(\mu)\big| } \Int{ -\infty }{ b_N } \! \dd \eta  \, (b_N - \eta)^k\,\ex{-N^{\a} \big[{\rm Im}\,\mu    (\eta-b_N) \big] }  \leq 
C\,N^{-k\a}\,\norm{ H }_{W^{\infty}_{k}(\intff{a_N}{b_N})}\;. 
\enq
It thus solely remains to put all the pieces together. \qed

\vspace{0.2cm}

\noindent Using these estimates, we obtain the continuity of the linear form $\mc{X}_N$ in sup norms:

\begin{cor}
\label{Corolaire bornange N independent forme lineaire X} 
 There exists $C > 0$ independent of $N$, such that:
\beq
\big| \wt{\mc{X}}_N[H] \big| \; \leq \; C \Norm{H}_{ W_0^{\infty}(\intff{a_N}{b_N}) }\;.
\label{borne sur operateur X}
\enq

\end{cor}

\Proof  We have shown in the proof of Lemma~\ref{Proposition DA integrale contrainte} a decomposition:
\beq
\mc{X}^{(\e{as})}_N[H] \; =  \; \mc{X}_{R}^{(\e{as})}[H] \; + \; \mc{X}_{R}^{(\e{as})}[H^{\wedge} ] +\mc{X}^{(\e{exp})}_N[H]\;.
\enq
$\mc{X}_{R}^{(\e{as})}[H] $ is given in \eqref{717}. It has $\chi_{11;+}(0)$ as prefactor, and we have seen in \eqref{aaasf} that this quantity takes the non-zero value $-2/R_{\da}(0)$ up to exponential small (in $N$) corrections. So, we have the bound:
\beq
\big| \mc{X}_{R}^{(\e{as})}[H] \big| \; \leq \; \frac{|R_{\da}(0)|}{2}\cdot\Norm{H}_{ W_0^{\infty}(\intff{a_N}{b_N}) } \cdot 
\Int{ \msc{C}^{(-)}_{\e{reg}} }{} \frac{1}{|\mu||{\rm Im}\,\mu|\,R_{\ua}(\mu)|}\,\frac{|\dd\mu|}{2\pi}  
\enq
where the inverse power of $|\mathrm{Im}\,\mu|$ and the loss of the prefactor $N^{\a}$ resulted from integrating the decaying exponential $|e^{\i N^{\a}\mu(\eta - b_N)}|$ over $\intff{a_N}{b_N}$, given that ${\rm Im}\,\mu < 0$ for $\mu \in \mathscr{C}_{{\rm reg}}^{(-)}$. We conclude by combining this estimate with \eqref{ecriture borne exp petite sur I cal pert de H} which shows that the remainder is exponentially small. \qed

%%%%%%%%%%%%%%%%%%%%%%%%%%%%%%%%%%%%%%%%%%%%%%%%%%%%%%%%%%%%%%%%%%%%%%%%%%%%%%%%%%%%%%%%%%%%%%%%%%%%%%%%%%%%%%%%%%%%%%%%%%%%%%%%%%%%
%%%%%%%%%%%%%%%%%%%%%%%%%%%%%%%%%%%%%%%%%%%%%%%%%%%%%%%%%%%%%%%%%%%%%%%%%%%%%%%%%%%%%%%%%%%%%%%%%%%%%%%%%%%%%%%%%%%%%%%%%%%%%%%%%%%%

%%%%%%%%%%%%%%%%%%%%%%%%%%%%%%%%%%%%%%%%%%%%%%%%%%%%%%%%%%%%%%%%%%%%%%%%%%%%%%%%%%%%%%%%%%%%%%%%%%%%%%%%%%%%%%%%%%%%%%%%%%%%%%%%%%%%
%%%%%%%%%%%%%%%%%%%%%%%%%%%%%%%%%%%%%%%%%%%%%%%%%%%%%%%%%%%%%%%%%%%%%%%%%%%%%%%%%%%%%%%%%%%%%%%%%%%%%%%%%%%%%%%%%%%%%%%%%%%%%%%%%%%%

%%%%%%%%%%%%%%%%%%%%%%%%%%%%%%%%%%%%%%%%%%%%%%%%%%%%%%%%%%%%%%%%%%%%%%%%%%%%%%%%%%%%%%%%%%%%%%%%%%%%%%%%%%%%%%%%%%%%%%%%%%%%%%%%%%%%
%%%%%%%%%%%%%%%%%%%%%%%%%%%%%%%%%%%%%%%%%%%%%%%%%%%%%%%%%%%%%%%%%%%%%%%%%%%%%%%%%%%%%%%%%%%%%%%%%%%%%%%%%%%%%%%%%%%%%%%%%%%%%%%%%%%%

\subsection{Asymptotic analysis of simple integrals}
\label{simpleintegrals}

In the present subsection, we obtain the large-$N$ asymptotic expansion of one-dimensional integrals involving $\mc{W}_N[H]$. 
This provides the first set of results that were necessary in \S~\ref{adfsdgfsg} for a thorough calculation of the large-$N$ expansion of the partition function.

\begin{defin}
\label{simpleJ} If $G$ and $H$ are two functions on $\intff{a_N}{b_N}$, we define:
\beq
\mf{I}_{\e{s}}\big[G,H\big] \; = \; \Int{a_N}{b_N} G(\xi) \cdot \mc{W}_N[H](\xi)\,\dd \xi 
\enq
where the ${\cal W}_N$ is the operator defined in \eqref{definition operateur WN}.
\end{defin}
To write the large $N$-expansion of $\mf{I}_{\e{s}}$, we need to introduce some more constants:
\begin{defin}
\label{secondal}If $s,\ell \geq 0$ are integers, we set:
\beq
\daleth_{s,\ell} \; = \; \Int{ 0 }{ +\infty} \hspace{-1mm} x^{s}\,\mf{b}_{\ell}(x)\,\dd x 
\label{definition constantes dialeth s ell}
\enq
where the function $\mathfrak{b}_{\ell}$ has been introduced in Definition~\ref{giu}.
\end{defin}

\begin{prop}
\label{Theorem DA tout ordre integrale 1D contre WN}

Let $k \geq 1$ be an integer, $G \in \mathcal{C}^{k - 1}(\intff{a_N}{b_N})$ and $H \in \mathcal{C}^{k + 1}(\intff{a_N}{b_N})$. We have the asymptotic expansion:
\bem
\mf{I}_{\e{s}}\big[G,H\big] \; = \; u_1 \Int{a_N}{b_N} G(\xi)\cdot H^{\prime}(\xi)\,\dd \xi 
 \; + \; \sul{ p=1 }{ k-1 } \f{ 1 }{ N^{\a p}  } \Bigg\{  u_{ p+1 } \Int{a_N}{b_N} G(\xi)  H^{(p+1)}(\xi)\,\dd \xi  \\
 \; + \; \sul{ \substack{ s+\ell=p-1 \\ s,\ell \geq 0} }{} \f{\daleth_{s,\ell}}{ s! }\,\,\Big[ (-1)^{s}\,H^{(\ell+1)}(b_N)\cdot G^{(s)}(b_N) \, + \, (-1)^{\ell}H^{(\ell+1)}(a_N) G^{(s)}(a_N) \Big] 
\Bigg\}  \; + \; \De_{[k]}\mf{I}_{\e{s}}\big[G,H\big] \;. 
\label{ecriture integrale 1D de G contre WN de H}
\end{multline}
where we remind that $u$'s are the constants appearing in Definition~\ref{oinoin}. The remainder is bounded as
\beq
\big| \De_{[k]}\mf{I}_{\e{s}}\big[G,H\big]  \big| \; \leq \; C\,N^{-k\a}\,\norm{ G }_{W^{\infty}_{k-1}(\intff{a_N}{b_N}) }\,
\norm{ H_{\mf{e}} }_{W^{\infty}_{k+1}(\intff{a_N}{b_N}) }  
\enq
for some constant $C > 0$ independent of $N$, $G$ and $H$.
\end{prop}

Note  that the leading asymptotics of $\mf{I}_{\e{s}}\big[G,H\big]$, \textit{i.e}. up to the $\e{o}(1)$ remainder, correspond precisely to 
the contribution obtained by replacing the integral kernel $S\big( N^{\a}(\xi-\eta) \big)$ of $\mc{S}_N$ by the sign function-- which corresponds to the almost sure pointwise limit of $S\big(N^{\a}(\xi-\eta)\big)$, see \eqref{ecriture eqn int sing de depart} -- and then inverting the formal limiting operator.  The corrections, however, are already more complicated as they stem from the fine behaviour at 
the boundaries. 

\Proof  Recall from Propositions \ref{Proposition decomposition op WN en diverses sous parties} and \ref{Proposition Ecriture reguliere uniforme des divers const de WN} that $\mc{W}_N[ H ]$ decomposes as
\beq
\label{deciudgf}\mc{W}_N[H](\xi) \; = \; \mc{W}_{R;k}[H](x_R,\xi) \; +  \; \mc{W}_{\e{bk};k}[H](\xi) \; - \; \mc{W}_{R;k}[H^{\wedge}](x_L,a_N+b_N-\xi) \; + \; 
\De_{[k]}\mc{W}_N[H_{\mf{e}}](\xi) 
\enq
where 
\beq
 \label{estmium} \Norm{ \De_{[k]}\mc{W}_N[H_{\mf{e}}]  }_{L^{\infty}(\intff{a_N}{b_N})} \; \leq \; C\,N^{-k\a}\,\norm{ H_{\mf{e}}^{ (k+1) } }_{ L^{\infty}(\R) } \;. 
\enq
This leads to the decomposition 
\beq
\mf{I}_{\e{s}}[G,H] \; = \; \mf{I}_{\e{s};k}^{(\e{bk})}[G,H]  \; + \;  \mf{I}_{\e{s};k}^{(\partial)}[G,H] \; - \; 
\mf{I}_{\e{s};k}^{(\partial)}[G^{\wedge} , H^{\wedge}] \; + \; \De_{[k]}\mf{I}_{\e{s}}[G,H_{\mf{e}}]
\enq
where:
\begin{eqnarray}
\mf{I}_{\e{s};k}^{(\e{bk})}[G,H] & = &  \Int{a_N}{b_N} G(\xi) \cdot \mc{W}_{\e{bk};k}[H](\xi)\,\dd \xi\;, \nonumber \\
\mf{I}_{\e{s};k}^{(\partial)}[G,H] & = & 
\f{1}{ N^{\a} } \Int{ 0 }{ \ov{x}_N } G\big( b_N - N^{-\a} x \big) \cdot \mc{W}_{R;k}[H]\big( x, b_N - N^{-\a} x \big)\,\dd x \;,\nonumber \\
 \De_{[k]}\mf{I}_{\e{s}}[G,H_{\mf{e}}] & = &  \Int{a_N}{b_N} G(\xi) \cdot \De_{[k]}\mc{W}_N[H_{\mf{e}}](\xi)\,\dd \xi \;. 
\end{eqnarray}
Clearly from the estimate \eqref{estmium}, there exist a constant $C^{\prime} > 0$ such that:
\beq
 \big|   \De_{[k]}\mf{I}_{\e{s}}[G,H_{\mf{e}}]  \big| \; \leq  \; C^{\prime}\,N^{-k\a} \cdot \norm{ G }_{L^{\infty}(\intff{a_N}{b_N})} \cdot 
 \norm{ H_{\mf{e}}^{ (k+1) } }_{ L^{\infty}(\R) }  \;. 
\enq
The asymptotic expansion of $\mf{I}_{\e{s};k}^{(\e{bk})}$ follows readily from the expression \eqref{definition W bk ordre k et ctes u ell} 
for $\mc{W}_{\e{bk};k}[H]$. It produces the first line of \eqref{ecriture integrale 1D de G contre WN de H}. As a consequence, it remains to focus on $\mf{I}_{\e{s};k}^{(\partial)}$. 
Recall from Proposition~\ref{Lemme structure locale au bord pour WR et Wbk} the decomposition 
\beq
 \label{iiun}\mc{W}_{R;k}[H]\big( x, b_N - N^{-\a} x \big) \; = \;  \mc{W}_{R;k}^{(\e{as})}[H]( x )\; + \; \Delta_{[k]} \mc{W}_{R}^{(\e{as})}[H](x) 
\enq
and especially the bounds \eqref{gfmdum}-\eqref{ecriture bornes fines sur reste asympt2} on the remainder, which imply:
\beq
\label{ccdsd}\big| \Delta_{[k]}  \mc{W}_{R}^{(\e{as})}[H](x)  \big| \; \leq \; C\,\ex{-\varsigma x}\,x^{k + 1}\,\ln x \cdot N^{-k \a }\cdot 
\norm{ H_{\mf{e}} }_{ W_{k+1}^{\infty}(\R) } \;. 
\enq
The contribution of the first term of \eqref{iiun} involves the functions $\mf{b}_{\ell}$. it remains to replace $G$ by its Taylor series with integral remainder of appropriate order so as to get 
\beq
\mf{I}_{\e{s};k}^{(\partial)}[G,H] \; = \; \sul{p=0}{k-1}\f{ 1 }{ N^{(p + 1)\a} } \sul{\substack{s+\ell =p \\ s,\ell \geq 0}}{} 
\f{ (-1)^s }{ s! } H^{(\ell+1)}(b_N) \cdot G^{(s)}(b_N) \Int{0}{ \ov{x}_N } x^s\,\mf{b}_{\ell}(x)\,\dd x 
 \; + \; \Delta_{[k]} \mf{I}_{\e{s}}^{(\partial)}[G,H] 
\label{ecriture Js ordre k bd sous forme presque asympt}
\enq
where 
\begin{eqnarray}
\Delta_{[k]}\mf{I}_{\e{s}}^{(\partial)}[G,H] & = & \f{ 1 }{ N^{k\a} } \sul{\ell = 0}{k-1} \f{ H^{(\ell+1)}(b_N) }{(k-\ell-2)!}
\Int{0}{\ov{x}_N} \! \dd x\,\mf{b}_{\ell}(x)\,(-x)^{k-\ell-1} \Int{0}{1} \! \dd t \;  (1-t)^{k-2-\ell} G^{(k-\ell-1)}( b_N - N^{-\a} t x) \\
& & \; + \; \f{ 1 }{ N^{\a} } \Int{0}{\ov{x}_N} G( b_N -  N^{-\a}x) \cdot \Delta_{[k]}\mc{W}_{R}^{(\e{as})}[H](x)\,\dd x \;. 
\end{eqnarray}
Clearly from \eqref{ccdsd}, there exists $C^{\prime\prime} > 0$ such that: 
\beq
\big| \Delta_{k}\mf{I}_{\e{s};k}^{(\partial)}[G,H] \big| \; \leq  \;  C^{\prime \prime}\,N^{-k\a}\,\norm{ H_{\mf{e}} }_{ W_{k+1}^{\infty}(\R) }
\cdot  \norm{ G }_{ W_{k-1}^{\infty}(\intff{a_N}{b_N}) } \;. 
\enq
Moreover, one can extend the integration in \eqref{ecriture Js ordre k bd sous forme presque asympt}
from $\intff{0}{\ov{x}_N}$ up to $\R^+$, this for the price of exponentially small corrections in $N$.  Adding up all the pieces 
leads to \eqref{ecriture integrale 1D de G contre WN de H}. \qed

\vspace{2mm} 

In the case when $G = 1$,  \textit{i.e.} to estimate the magnitude of the total integral of ${\cal W}_{N}[H]$, we can obtain slightly better bounds, solely involving the sup norm.
\begin{lemme}
\label{Lemme bornage integrale simple de WN de H}
There exists $C>0$ independent of $N$ such that, for any $H\in \mc{C}^{1}(\intff{a_N}{b_N})$, 
\beq
\bigg| \Int{a_N}{b_N} \mc{W}_N[H](\xi)\,\dd \xi \bigg| \; \leq \; C\,\norm{ H_{\mf{e}} }_{ W_0^{\infty}(\R) } \;. 
\enq
\end{lemme}
\Proof   Recall from Propositions \ref{Proposition decomposition op WN en diverses sous parties}  the decomposition:
\beq
\mc{W}_N[H](\xi) \; = \; \mc{W}_{R}[H_{\mf{e}}](x_R,\xi) \; +  \; \mc{W}_{\e{bk}}[H_{\mf{e}}](\xi) \; - \; \mc{W}_{R}[H^{\wedge}_{\mf{e}} ](x_L,a_N+b_N-\xi) \; + \; 
\mc{W}_{\e{exp}}[H](\xi) \;. 
\enq
We focus on the integral of each of the terms taken individually.  We have:
\beq
\Int{ a_N }{ b_N } \mc{W}_{\e{bk}}[H_{\mf{e}}](\xi)\,\dd \xi \; = \; \f{ N^{\a} }{ 2\pi \be } \Int{ \R }{} \hspace{-1mm}  \dd y\, J(y) \Int{0}{ N^{-\a}y}
\hspace{-2mm}  \big[ H_{\mf{e}}(b_N+t)\, - \, H_{\mf{e}}(a_N+t)   \big]\,\dd t\;,
\enq
thus leading to 
\beq
\bigg| \Int{a_N}{b_N} \mc{W}_{\e{bk}}[H_{\mf{e}}](\xi) \,\dd \xi \bigg| \; \leq \; C\,\norm{ H_{\mf{e}} }_{ W_0^{\infty}(\R) } \;. 
\enq
Next, we have:
\bem
\Int{ a_N }{ b_N }  \mc{W}_{R}[H_{\mf{e}}](x_R,\xi)\,\dd \xi \; = \; 
- \f{ N^{\a} }{2\pi \be} \Int{ \ov{x}_N }{ + \infty } \hspace{-1mm} \dd y\,J(y) \Int{a_N}{b_N} \hspace{-1mm}  \dd \xi \, \big[ H_{\mf{e}}(\xi+N^{-\a}y)-H_{\mf{e}}(\xi)  \big]\\
- \f{ N^{\a} }{2\pi \be} \Int{0}{\ov{x}_N } \hspace{-1mm} \dd y\,J(y) \Int{0}{N^{-\a}y}\hspace{-1mm} \big[ H_{\mf{e}}(b_N+t)-H_{\mf{e}}( b_N-N^{-\a}y+t )  \big]\,\dd t \\
+ \f{ N^{\a} }{2 \i \pi  \be}  \Int{ \msc{C}^{(+)}_{\e{reg}} }{} \hspace{-1mm} \f{ \dd \la }{ 2\i \pi } \Int{ \msc{C}^{(-)}_{\e{reg}} }{} \hspace{-1mm} \f{ \dd \mu }{ 2\i \pi }
\f{ 1 }{ (\mu-\la) R_{\da}(\mu) R_{\ua}(\la) } \Bigg\{  \f{ \ex{\i \la \ov{x}_N}-1 }{ \la } \Int{a_N}{b_N}H_{\mf{e}}(\eta) \ex{-\i \mu y_R} \dd \eta
\; + \;  \f{1}{ \mu } \Int{a_N}{b_N}H_{\mf{e}}(\xi) \ex{\i \la x_R} \dd \xi \Bigg\} \;. 
\label{ecriture estimation integrale WR}
\end{multline}
The exponential decay of $J$ at $ +\infty $ ensures that the first two lines of \eqref{ecriture estimation integrale WR} are indeed bounded 
by $C\,\norm{ H_{\mf{e}} }_{ W_0^{\infty}(\R) }$ for some $N$-independent $C>0$.
The last line is bounded similarly by using 
\beq
\forall \la \in \msc{C}^{(\pm)}_{\e{reg}},\qquad \bigg| \Int{a_N}{b_N} H_{\mf{e}}(\xi) \ex{ \pm \i \la N^{\a}(b_N-\xi)}\,\dd \xi   \bigg| \; \leq \;
\f{ C^{\prime}\,\norm{ H_{\mf{e}} }_{ W_0^{\infty}(\R) } }{ |\la| N^{\a} }\;.
\enq

It thus solely remains to focus on the exponentially small term $\mc{W}_{\e{exp}}[H]$. 
In fact, we only discuss the operator $\mc{W}_N^{(++)}$ as all others can be treated in a similar fashion. Thanks to the bound \eqref{ecriture bornes en N pour Pi moins Id} for $\Pi(\la) - I_2$ and the expression \eqref{PsiPi} of the matrix $\Psi$ in terms of $\Pi$, we have:
\beq
\Psi(\la) \; = \; \left( \ba{cc} 1   & 0  \\  \tf{1}{\la}  &  1 \ea \right) \; + \; \e{O}\Bigg(  \f{ \ex{-\varkappa_{\eps}N^{\a}} }{ 1+|\la| } \Bigg)
\enq
which is valid for $\la$ uniformly away from the jump contour $\Sigma_{\Psi}$ (see Figure~\ref{contour pour le RHP de Phi}). Therefore, using the definition \eqref{definition op WN++} of $\mc{W}_N^{(++)}$:
\beq
\bigg| \Int{a_N}{b_N} \mc{W}_N^{(++)}[H_{\mf{e}}](\xi)\,\dd \xi \bigg| \; \leq \; C^{\prime \prime}\,\norm{ H_{\mf{e}} }_{ W_0^{\infty}(\R) }\,
\ex{-\varkappa_{\epsilon} N^{\a}}\,
\Int{ \msc{C}^{(+)}_{\e{reg}} }{} \f{|\dd \la \dd \mu |}{(2\pi)^2} \f{ 1 }{ |\la-\mu|\,|R_{\da}(\la) R_{\da}(\mu) \, \la  | } \;. 
\enq
 Adding up all the intermediate bounds readily leads to the claim. \qed

By a slight modification of the method leading to Lemma~\ref{Lemme bornage integrale simple de WN de H}, we can likewise control the $L^{1}(\intff{a_N}{b_N})$ norm of $\mc{W}_N$ in terms of the $W_1^{\infty}$ norm of (an extension of) $H$. 
\begin{lemme}
 \label{Lemme bornage norme L1 de WN}
For any $H\in \mc{C}^{1}(\intff{a_N}{b_N})$ it holds 
\beq
\norm{  \mc{W}_N[H] }_{ L^{1}(\intff{a_N}{b_N}) } \; \leq \;  C \, \norm{ H_{\mf{e}} }_{W_1^{\infty}(\R) }
\quad \e{and} \quad 
\norm{  \mc{W}_{\e{exp}}[H] }_{ L^{1}(\intff{a_N}{b_N}) } \; \leq \;  C \, \ex{-C^{\prime}N^{\a} }\,\norm{H_{\mf{e}}}_{W_1^{\infty}(\R)}\;.
\label{ecriture bornage norme L1 de WN}
\enq
\end{lemme}

\section{The support of the equilibrium measure}
\label{Section characterisation support mesure equilibre}
In the present subsection we build on the previous analysis so as to prove the existence of the endpoints $(a_N,b_N)$ of the support of the equilibrium 
measure and thus the fact that 
\beq
 \rho_{{\rm eq}}^{(N)}(\xi) \; = \; \bs{1}_{\intff{a_N}{b_N}}(\xi) \cdot \mc{W}_N[V^{\prime}](\xi)\,\dd \xi\;,
\enq
where $\mc{W}_N$ is as defined in \eqref{definition operateur WN}.

\begin{lemme}
 There exists a unique sequence $(a_N,b_N)$ -- defining the support of the Lebesgue-continuous equilibrium measure which corresponds to the unique solution 
 to the minimisation problem \eqref{definition de la cste Ceq par eqn int eq meas}-\eqref{ecriture condition negativite dehors support mu eq}. 
The sequences $a_N$ and $b_N$ are bounded in $N$. 
 
\end{lemme}

\Proof  The existence and uniqueness of the solution to the minimisation problem 
\eqref{definition de la cste Ceq par eqn int eq meas}-\eqref{ecriture condition negativite dehors support mu eq} is obtained
through a straightforward generalisation of the proof arising in the random matrix case, see \textit{e.g.} \cite{DeiftOrthPlyAndRandomMatrixRHP}. 

The endpoint of the support of the equilibrium measure should be chosen in such a way that, on the one hand, the density of equilibrium 
measure admits at most a square root behaviour at the endpoints and, on the other hand, that it indeed defines a probability measure. 
In other words, the endpoints are to be chosen so that the two constraints are satisfied
\beq
\mc{X}_{N}[ V^{\prime}] \; = \; 0 \qquad \e{and} \qquad 
\mf{I}_{s}[ 1, V^{\prime}] = \Int{a_N}{b_N} \mc{W}_{N}[V^{\prime}](\xi)\,\dd\xi \; = \; 1\;.  
\label{ecriture system equations pour endpoints}
\enq
The asymptotic expansion of $\mc{X}_N[ V^{\prime}]$ and $\mf{I}_{s}[ 1, V^{\prime}]$ is given, respectively, in 
Lemma~\ref{Proposition DA integrale contrainte} and Proposition~\ref{Theorem DA tout ordre integrale 1D contre WN}.
However, the control on the remainder obtained there does depend on $a_N$ and $b_N$. Should $a_N$ or $b_N$ be unbounded in $N$
this could brake the \textit{a priori} control on the remainder. 
Still, observe that if $(a_N,b_N)$ solve the system of equations \eqref{ecriture system equations pour endpoints} then 
$\xi \mapsto \mc{W}_N[V^{\prime}](\xi)$ with $\mc{W}_N$ associated with the support $\intff{a_N}{b_N}$ provides one with a solution to the minimisation 
problem of $\mc{E}_N$ defined in \eqref{definition fnelle a minimiser N dpdte}. By uniqueness of solutions to this minimisation problem, it thus corresponds to the 
density of equilibrium measure. As a consequence, there exists at most one solution $(a_N,b_N)$ to the system of equations \eqref{ecriture system equations pour endpoints}. 

Assume that the sequence $a_N$ and $b_N$ are bounded in $N$. Then, the leading asymptotic expansion of the two functionals in 
\eqref{ecriture system equations pour endpoints} yields 
\beq
\left\{  \ba{ccc} V^{\prime}(b_N) \, + \,   V^{\prime}(a_N) & = &  \e{O}\big( N^{-\a} \big)  \\
 V^{\prime}(b_N) \, - \,   V^{\prime}(a_N) & = &  u_1^{-1} \, + \, \e{O}\big( N^{-\a} \big) \ea \right. 
\qquad viz. \qquad \left( \ba{cc} 1 & 1 \\ 
				  1 & -1  \ea \right) \cdot \left( \ba{c} V^{\prime}(b_N) - V^{\prime}(b) \\ 
									  V^{\prime}(a_N) - V^{\prime}(a)  \ea \right) \; = \; \e{O}\big( N^{-\a} \big) \;. 
\label{ecriture eqn pour DA aN bN 1er ordre}
\enq
Note that the control on the remainder follows from the fact that $|a_N|$ and $|b_N|$ are bounded by an $N$-independent constant. 
Also, $(a,b)$ appearing above corresponds to the unique solution  to the system 
\beq
V^{\prime}(b) \, + \,   V^{\prime}(a) \, = \,  0 \quad \e{and}  \quad V^{\prime}(b) \, - \,   V^{\prime}(a) \, = \,  u_1^{-1} \;. 
\enq
We do stress that the existence and uniqueness of this solution is ensured by the strict convexity of $V$. 

The smoothness of the remainder in $(a_N, b_N)$ away from $0$, the control on its magnitude (guaranteed by the boundedness of $a_N$ and $b_N$)
as well as the strict convexity of $V$ and the invertibility of the matrix occurring in \eqref{ecriture eqn pour DA aN bN 1er ordre} ensure the 
existence of solutions $(a_N,b_N)$ by the implicit function theorem, this provided that $N$ is large enough. 
Hence, since a solution to \eqref{ecriture system equations pour endpoints} with $a_N$ and $b_N$ bounded in $N$ does exists, by uniqueness
of the solutions to \eqref{ecriture system equations pour endpoints}, it is the one that defines the endpoints of the support of the equilibrium measure. 
\qed

\begin{cor}
\label{Corollaire DA des bornes aN et bN}

Let the pair $(a,b)$ correspond to the unique solution  to the system 
\beq
V^{\prime}(b) \, + \,   V^{\prime}(a) \, = \,  0 \quad and  \quad V^{\prime}(b) \, - \,   V^{\prime}(a) \, = \,  u_1^{-1} \;. 
\label{definition des points a et b via contrainte}
\enq
Then the endpoints $(a_N,b_N)$ of the support of the equilibrium measure admit the asymptotic expansion
\beq
a_N \; = \;  \sul{\ell=0}{k-1} \f{ a_{N;\ell} }{ N^{\ell \a}  }  \; + \; \e{O}\big(N^{-k\a} \big) \qquad and \qquad
b_N \; = \;  \sul{\ell=0}{k-1} \f{ b_{N;\ell} }{ N^{\ell \a}  } \; + \; \e{O}\big(N^{-k\a} \big) \;, 
\label{ecriture DA aN et bN up to ordre k}
\enq
where $a_{N;0}=a$ and $b_{N;0}=b$. 

\end{cor}
Note that the existence and uniqueness of solutions to the system \eqref{definition des points a et b via contrainte} follows from the strict convexity of the potential $V$. 

\Proof   The invertibility of the matrix occurring in \eqref{ecriture eqn pour DA aN bN 1er ordre}  as well as the strict convexity of the potential $V$ ensure that $a_N$ and $b_N$ admit the
expansion \eqref{ecriture DA aN et bN up to ordre k} for $k=1$, \textit{viz}. up to  $\e{O}\big( N^{-\a} \big)$ corrections. Now suppose that this expansion holds up to 
$\e{O}\big( N^{-(k - 1)\a} \big)$. It follows from Lemma~\ref{Proposition DA integrale contrainte} and Proposition~\ref{Theorem DA tout ordre integrale 1D contre WN}
that the asymptotic expansion of $\mc{X}_N[V^{\prime} ]$ and $\mf{I}_{s}[1,V^{\prime}]$ up to $\e{O}\big( N^{-k\a} \big)$ can be recast as
\beq
\left( \ba{c}  \mc{X}_N[V^{\prime}] \cdot \daleth_0^{-1}  \\ 
		 \mf{I}_{\e{s}}[1,V^{\prime}]   \cdot u_1^{-1}  \ea \right)  \; = \; 
\left( \ba{c} V^{\prime}(b_N) + V^{\prime}(a_N) \; + \; \mc{B}_{1;k-1}[  V^{\prime}  ]  \; + \; \daleth_0^{-1}\cdot \De_{[k]} \mc{X}_{N}[V^{\prime}]  \\
 V^{\prime}(b_N) - V^{\prime}(a_N)\; + \; \mc{B}_{2;k-1}[  V^{\prime}  ] \, + \,  u_1^{-1}\cdot \De_{[k]} \mf{I}_{\e{s}}[1,V^{\prime}]  \ea \right) \;.  
\enq
In this expression, we have  $\big| \daleth_0^{-1}\cdot\De_{[k]} \mc{X}_{N}[V^{\prime}]  \big| \, + \,  \big|u_1^{-1}\cdot \De_{[k]} \mf{I}_{\e{s}}[1,V^{\prime}]  \big| \leq C N^{-k\a}$ since $a_N$ and $b_N$
are bounded uniformly in $N$, while 
\beq
\left( \ba{c} \mc{B}_{1;k-1}[ V^{\prime} ]   \\  \mc{B}_{2;k-1}[ V^{\prime} ]  \ea \right)  =  \sul{p=1}{k-1} \f{1}{  N^{p\a} } 
\left( \ba{c} 
 \i^p\cdot\daleth_p\daleth_0^{-1}  \cdot \Big( V^{(p+1)}(a_N) \, + \, (-1)^p  V^{(p+1)}(b_N) \Big)  \vspace{1mm} \\
\big( u_{p+1}+\daleth_{0,p-1} \big)u_1^{-1}\cdot V^{(p+1)}(b_N) 
			   -    \big( u_{p+1}+(-1)^{p}\daleth_{0,p-1} \big)u_1^{-1} \cdot  V^{(p+1)}(a_N)   \ea \right) \;. 
\enq
We remind that $\daleth_p$ was introduced in Definition~\ref{firstdalet}, $u_{p}$ in Definition~\ref{oinoin}, and $\daleth_{0,p}$ in Definition~\ref{secondal}.

Since both $\mc{B}_{1;k-1}[ V^{\prime} ]$ and $\mc{B}_{2;k-1}[ V^{\prime} ]$ have $N^{-\a}$ as a prefactor, by composition of asymptotic expansions, 
there exist functions $B_{p;\ell}\big(b_{N;1},\dots, b_{N;\ell-1} \,  \mid   a_{N;1},\dots, a_{N;\ell-1} \big) $,
indexed by $p \in \{1,2\}$ and  $\ell \in \intn{1}{k-1}$, independent of $k$, such that 
\beq
\left( \ba{c} \mc{B}_{1;k-1}[ V^{\prime} ]   \\  \mc{B}_{2;k-1}[ V^{\prime} ]  \ea \right)  \; = \; 
\sul{  \ell =1 }{k-1 } 
\f{ 1 }{ N^{ \ell \a} } \left( \ba{c} B_{1;\ell}\big(b_{N;1},\dots, b_{N;\ell-1} \, \mid  a_{N;1},\dots, a_{N;\ell-1} \big)  \\
			      B_{2;\ell}\big(b_{N;1},\dots, b_{N;\ell-1} \,  \mid  a_{N;1},\dots, a_{N;\ell-1} \big) \ea \right)   \; + \; \e{O}\big( N^{-\a k} \big)\;. 
\enq
As a consequence, we have the relation:
\beq
 \left( \ba{cc}  1 & 1  \\  
	       1 &  -1  \ea \right)  \left( \ba{c}  V^{\prime}(b_N) - V^{\prime}(b) \\ 
						 V^{\prime}(a_N) - V^{\prime}(a)  \ea \right)	\; = \; \sul{  \ell =1 }{k-1 } 
\f{ - 1 }{ N^{ \ell \a } } \left( \ba{c} B_{1;\ell}\big(b_{N;1},\dots, b_{N;\ell-1} \, \mid  a_{N;1},\dots, a_{N;\ell-1} \big)  \\
			      B_{2;\ell}\big(b_{N;1},\dots, b_{N;\ell-1} \,  \mid  a_{N;1},\dots, a_{N;\ell-1} \big) \ea \right)   \; + \; 
\e{O}\big( N^{-k\a}  \big) \;. 
\enq
This implies the existence of an asymptotic expansion of $a_N$ and $b_N$ up to a remainder of the order $\e{O}\big( N^{-k\a} \big)$. \qed

\section{Asymptotic evaluation of the double integral}
\label{Section asymptotic analysis of double integrals}

In this section we study the large-$N$ asymptotic expansion for the double integral in: 
\begin{defin}
\beq
\mf{I}_{\e{d}}[H,V] \; = \; \Int{a_N}{b_N} \mc{W}_N\circ\wt{\mc{X}}_N\Big[ 
\partial_{\xi}\big\{ S\big(N^{\a}(\xi-*)\big) \cdot \mc{G}_N\big[ \wt{\mc{X}}_N[H],V\big](\xi,*)     \big\}   \Big](\xi)\, \dd \xi\;,
\label{definition integrale double du DA correlateur a un point}
\enq
with 
\beq
 \mc{G}_N\big[H,V\big](\xi,\eta) \; = \; \f{ \mc{W}_N[H](\xi) }{ \mc{W}_N[V^{\prime}](\xi)  }
 \; - \; \f{ \mc{W}_N[H](\eta) }{ \mc{W}_N[V^{\prime}](\eta)  } \;. 
\label{definition integrande GN cal}
\enq
\end{defin}
We remind that * indicates the variable on which the operator $\mc{W}_N$ acts. The asymptotic analysis of the double integral $\mf{I}_{\e{d};\be}$ arising in the $\be\not=1$ large-$N$
asymptotics (\textit{cf}. \eqref{definition integrale double N dpdt beta non egal 1}) can be carried out within the setting of the method developed 
in this section. However, in order to keep the discussion minimal, we shall not present this calculation here. 

In order to carry out the large-$N$ asymptotic analysis of $\mf{I}_{\e{d}}[H,V]$, it is convenient to write down a decomposition for $ \mc{G}_N\big[H,V\big]$ 
ensuing from the decomposition of ${\cal W}_N$ that has been described  in Propositions~\ref{Proposition decomposition op WN en diverses sous parties} and \ref{Proposition Ecriture reguliere uniforme des divers const de WN}. 
We omit the proof since it consists of straightforward algebraic manipulations.
\begin{lemme}
The function $ \mc{G}_N\big[H,V\big](\xi,\eta)$ can be recast as
\bem
 \mc{G}_N[H,V](\xi,\eta) \; = \; \mc{G}_{\e{bk};k}[H,V](\xi,\eta) \; + \; \mc{G}_{R;k}^{(\e{as})}[H,V](x_R,y_R;\xi,\eta)  \\
 \; - \;  \mc{G}_{R;k}^{(\e{as})}\big[H^{\wedge},V^{\wedge}\big](x_L,y_L;a_N+b_N-\xi,a_N+b_N-\eta) \; + \; \De_{[k]} \mc{G}_N\big[H,V\big](\xi,\eta) \;. 
\label{ecriture decomposition GN en terme bk bd et reste}
\end{multline}
The functions arising in this decomposition read
\beq
\label{Gkbj}\mc{G}_{\e{bk};k}[H,V](\xi,\eta) \; = \;
 \f{ \mc{W}_{\e{bk};k}[H](\xi) }{ \mc{W}_{\e{bk};k}[V^{\prime}](\xi)  } \; -  \; (\xi \leftrightarrow \eta) \;, 
\enq
and
\beq
 \label{75}\mc{G}_{R;k}^{(\e{as})}[H,V](x,y;\xi,\eta)  \; = \; 
 \Bigg\{  \f{ \mc{W}_{R;k}^{(\e{as})}[H](x) }{ \mc{W}_{\e{bk};k}[V^{\prime}](\xi)  }  \; - \;
    \f{   \mc{W}_{R;k}^{(\e{as})}[V^{\prime}](x)  }{ \mc{W}_{\e{bk};k}[V^{\prime}](\xi)  } \cdot 
 \f{  \big( \mc{W}_{\e{bk};k}^{(\e{as})} + \mc{W}_{R;k}^{(\e{as})}\big) [H](x) }
 						{  \big( \mc{W}_{\e{bk};k}^{(\e{as})} + \mc{W}_{R;k}^{(\e{as})}\big)[V^{\prime}](x)  }   \Bigg\} 
\; - \; \Bigg( \ba{c} \xi \leftrightarrow \eta \\ x  \leftrightarrow y \ea \Bigg) \;. 
\enq
Finally, the remainder $\De_{[k]} \mc{G}_N$ takes the form 
\bem
\De_{[k]}\mc{G}_N[H,V](\xi,\eta)  \; = \; \f{1}{ \mc{W}_{\e{bk};k}[V^{\prime}](\xi)  }
 \Bigg\{   \De_{[k]}\mc{W}_N[H](\xi)   
 \; - \; \De_{[k]}\mc{W}_N [V^{\prime}](\xi)\cdot  \f{ \mc{W}_N[H](\xi) }{ \mc{W}_N[V^{\prime}](\xi)  }  \Bigg\} 
 \; - \; \Big(  \xi \leftrightarrow \eta \Big)\\
\; + \; \Delta_{[k]}\mc{G}_{N}^{(\e{as})}[H,V](x_R, y_R; \xi,\eta) \; - \;   \Delta_{[k]}\mc{G}_{N}^{(\e{as})}\big[H^{\wedge},V^{\wedge}](x_L, y_L; a_N+b_N-\xi,a_N+b_N-\eta)  \;. 
\end{multline}
The reminder $\De_{[k]}\mc{W}_N$ of order $k$ has been introduced in \eqref{definition reste asympt ordre k pour WN}, while 
\bem
\label{988a} \Delta_{[k]}\mc{G}_{N}^{(\e{as})}[H,V](x, y; \xi,\eta) \; = \; \f{1}{ \mc{W}_{\e{bk};k}[V^{\prime}](\xi)  } \Bigg\{ \Delta_{[k]}\mc{W}_{R}^{(\e{as})}[H](x)
					 \, - \,  \Delta_{[k]} \mc{W}_{R}^{(\e{as})}[V^{\prime}](x) \cdot  \f{ \mc{W}_N[H](\xi) }{ \mc{W}_N[V^{\prime}](\xi)  } \\
- \bigg[  \big(\Delta_{[k]}\mc{W}_N\big)_R[H](\xi)  
    \, - \,  \big(\Delta_{[k]}\mc{W}_N\big)_R[V^{\prime}](\xi) \cdot  \f{ \mc{W}_N[H](\xi) }{ \mc{W}_N[V^{\prime}](\xi)  }   \bigg]  
\cdot  \f{  \mc{W}_{R;k}^{(\e{as})}[V^{\prime}](x) }
			{   \big( \mc{W}_{\e{bk};k}^{(\e{as})} + \mc{W}_{R;k}^{(\e{as})}\big)[V^{\prime}](x)  } \Bigg\}
\; - \; \Bigg( \ba{c} \xi \leftrightarrow \eta \\ x  \leftrightarrow y \ea \Bigg) \;. 
\end{multline}
The local right boundary remainder arising above is defined as
\beq
\label{iugfgd}\big(\Delta_{[k]}\mc{W}_N\big)_{R} \; = \; \mc{W}_N \; - \; \mc{W}_{R;k}^{(\e{as})}\, -\, \mc{W}_{\e{bk};k}^{(\e{as})} \;.  
\enq
\end{lemme}
Note that the two terms $\mc{G}_{R;k}^{(\e{as})}$ present in \eqref{ecriture decomposition GN en terme bk bd et reste} 
correspond to the parts of $\mc{G}_N$ that localise at the right and left boundary. The way in which they appear
is reminiscent of the symmetry satisfied by $\mc{W}_{N}$:
\beq
\mc{W}_{N}[H](a_N + b_N-\xi)  \; = \; -  \mc{W}_{N}[H^{\wedge}](\xi)  \;. 
\enq
%
%
%

%%%%%%%%%%%%%%%%%%%%%%%%%%%%%%%%%%%%%%%%%%%%%%%%%%%%%%%%%%%%%%%%%%%%%%%%%%%%%%%%%%%%%%%%%%%%%%%%%%%%%%%%%%%%%%%%%%%%%%%%%%%%%%%%%%%%%%%%%%%%%%
%%%%%%%%%%%%%%%%%%%%%%%%%%%%%%%%%%%%%%%%%%%%%%%%%%%%%%%%%%%%%%%%%%%%%%%%%%%%%%%%%%%%%%%%%%%%%%%%%%%%%%%%%%%%%%%%%%%%%%%%%%%%%%%%%%%%%%%%%%%%%%

%%%%%%%%%%%%%%%%%%%%%%%%%%%%%%%%%%%%%%%%%%%%%%%%%%%%%%%%%%%%%%%%%%%%%%%%%%%%%%%%%%%%%%%%%%%%%%%%%%%%%%%%%%%%%%%%%%%%%%%%%%%%%%%%%%%%%%%%%%%%%%
%%%%%%%%%%%%%%%%%%%%%%%%%%%%%%%%%%%%%%%%%%%%%%%%%%%%%%%%%%%%%%%%%%%%%%%%%%%%%%%%%%%%%%%%%%%%%%%%%%%%%%%%%%%%%%%%%%%%%%%%%%%%%%%%%%%%%%%%%%%%%%

\begin{lemme}
\label{doubdub}The double integral $\mf{I}_{\e{d}}[H,V]$ can be recast as 
\bem
\mf{I}_{\e{d}}[H,V] \; = \; \mf{I}_{\e{d};\e{bk}}\Big[ \mc{G}_{\e{bk};k}[H,V] \Big] \; + \; 
\mf{I}_{\e{d};\e{bk}}\Big[ \mc{G}_{R;k}^{(\e{as})}[H,V] \; + \; \mc{G}_{R;k}^{(\e{as})}[H^{\wedge},V^{\wedge}] \Big]  \\ 
\; + \; \mf{I}_{\e{d};R}\Big[  \big(\mc{G}_{\e{bk};k}+\mc{G}_{R;k}^{(\e{as})}\big)[H,V] 
    					\; + \; \big(\mc{G}_{\e{bk};k} +\mc{G}_{R;k}^{(\e{as})}\big)[H^{\wedge},V^{\wedge}] \Big] 
    					\; + \; \De_{[k]}\mf{I}_{\e{d}}\Big[ \wt{\mc{X}}_N[H],V\Big] \; .
\end{multline}
The bulk part of the double integral is described by the functional 
\beq
\mf{I}_{\e{d};\e{bk}}[ F] \; = \; \f{-N^{2\a}}{4\pi \be} \Int{\intff{a_N}{b_N}^2}{} \!  J\big(N^{\a}(\xi -\eta) \big)
\cdot \big( \partial_{\xi} \, - \, \partial_{\eta}\big)
\big\{ S\big(N^{\a}(\xi -\eta) \big)  F(\xi,\eta)  \big\}\,\dd \xi \dd \eta\;. 
\enq
The local (right) part of the double integral is represented as 
\beq
\mf{I}_{\e{d};R}[ F] \; = \;- \f{N^{2\a} }{ 2 \pi \be} \Int{ \msc{C}^{(+)}_{\e{reg}} }{  } \f{\dd \la }{2 \i \pi} 
\Int{ \msc{C}^{(-)}_{\e{reg}}  }{} \f{\dd \mu }{2 \i \pi} \Int{a_N}{b_N}  
 \f{ \dd \xi \,\ex{ \i\la N^{a}(b_N - \xi)} }{ (\mu-\la)R_{\da}(\la)R_{\ua}(\mu) } 
 \Int{a_N}{b_N}\! \dd \eta\,\ex{-\i \mu N^{\a}(b_N - \eta)}\,\partial_{\xi}\big\{S\big(N^{\a}(\xi -\eta) \big)  F(\xi,\eta) \big\} \;. 
\label{definition integrale droite AA integrale double}
\enq
Finally, $\De_{[k]} \mf{I}_{\e{d}}$ represents the remainder which decomposes as 
\begin{eqnarray}
\De_{[k]}\mf{I}_{\e{d}}[H,V] & = & \sul{p=1}{4} \De_{[k]}\mf{I}_{\e{d};p}[H,V]  \\
\De_{[k]}\mf{I}_{\e{d};1}[H,V]  & = &  \Int{a_N}{b_N} \mc{W}_{\e{exp}}\Big[ 
\partial_{\xi}\big\{ S\big(N^{\a}(\xi-*)\big) \cdot \big( \mc{G}_N - \De_{[k]} \mc{G}_N \big) \big[H,V\big](\xi,*)     \big\}   \Big](\xi) \,\dd \xi  
\label{definition De k de mfracI double 1} \\
\De_{[k]}\mf{I}_{\e{d};2}[H,V] & = & \Int{a_N}{b_N}  \mc{W}_N\Big[ \partial_{\xi}\big\{ S\big(N^{\a}(\xi-*)\big) \cdot \De_{[k]} \mc{G}_N\big[H,V\big](\xi,*)    \big\}   \Big](\xi)      \,  \dd \xi
\label{definition De k de mfracI double 2} \\
\De_{[k]}\mf{I}_{\e{d};3}[H,V] & = & - \Int{a_N}{b_N}  \mc{W}_N[1](\xi) \cdot \mc{X}_N\Big[ \partial_{\xi}\big\{ S\big(N^{\a}(\xi-*)\big) 
\cdot \mc{G}_N\big[H,V\big](\xi,*)     \big\}   \Big](\xi)\,  \dd \xi \;. 
\label{definition De k de mfracI double 3} \\
\De_{[k]}\mf{I}_{\e{d};4}[H,V] & = & - \mf{I}_{\e{d};R}\Big[ \big(\mc{G}_{R;k}^{(\e{as})}[H,V]\big)^{\wedge} \; + \; \big(\mc{G}_{R;k}^{(\e{as})}[H^{\wedge},V^{\wedge}]\big)^{\wedge} \Big]
\label{definition De k de mfracI double 4}
\end{eqnarray}
where $\mc{W}_{\e{exp}}$ is as defined in \eqref{ecriture decomposition W infty sur ops type 0}.
\end{lemme}

\Proof  We first invoke the Definition~\ref{defXN} of the operator $\wt{\mc{X}}_N$ so as to recast  $\mf{I}_{\e{d}}[H,V]$
as an integral involving solely $\mc{W}_N$, and another one containing the action of $\mc{X}_N$. 
Then, in the first integral, we decompose the operator $\mc{W}_N$ arising in the "exterior" part of the double integral $\mf{I}_{\e{d}}[H,V]$ as 
$\mc{W}_N= (\mc{W}^{(0)}_{R}+\mc{W}^{(0)}_{\e{bk}}+\mc{W}_L^{(0)}+ \mc{W}_{\e{exp}})$, 
\textit{cf}. \eqref{ecriture decomposition W infty sur ops type 0}.
Then, it remains to observe that 
\beq
\Int{a_N}{b_N}   \mc{W}_L^{(0)}\Big[ 
\partial_{\xi}\big\{ S\big(N^{\a}(\xi-*)\big) \cdot \mc{G}_N\big[H,V\big](\xi,*)     \big\}   \Big](x_L)\,\dd \xi
\; = \; \Int{a_N}{b_N}   \mc{W}_R^{(0)}\Big[ 
\partial_{\xi}\big\{ S\big(N^{\a}(\xi-*)\big) \cdot \mc{G}_N\big[H^{\wedge},V^{\wedge}\big](\xi,*)     \big\}   \Big](x_R)\,\dd \xi
\enq
and that
\beq
- \mf{I}_{\e{d};\e{bk}}\Big[\big(\mc{G}_{R;k}^{(\e{as})}[H^{\wedge},V^{\wedge}]\big)^{\wedge} \Big] \; =  \; \mf{I}_{\e{d};\e{bk}}\Big[  \mc{G}_{R;k}^{(\e{as})}[H^{\wedge},V^{\wedge}] \Big]  \;. 
\enq
Putting all these results together, and using that the functions $\mc{G}_{\e{bk};k}[H,V]$ and $\mc{G}_{R;k}^{(\e{as})}[H,V] $
solely involve derivatives of $H$ which implies:
\beq
\mc{G}_{\e{bk};k}\big[ \wt{\mc{X}}_N[H],V\big] \; = \; \mc{G}_{\e{bk};k}[H,V]  \qquad \e{and} \qquad 
\mc{G}_{R;k}^{(\e{as})}\big[ \wt{\mc{X}}_N[H],V\big] \; = \;  \mc{G}_{R;k}^{(\e{as})}[H,V]\;,
\enq
we obtain the desired decomposition of the double integral.  \qed

\subsection{The asymptotic expansion related to $\mf{I}_{\e{d};\e{bk}}$ and $\mf{I}_{\e{d}; R }$}

Once again, we need to introduce new constants:
\begin{defin}
\label{gim}If $\ell \geq 0$ is an integer, we set:
\beq
\gimel_{2\ell} \;= \; \Int{\R }{} \f{J(u)\,u^{2\ell}}{4\pi \be\,(2\ell)!}\,\big[u S^{\prime}(u) + S(u)\big]\,\dd u 
\qquad and \qquad 
\gimel_{2\ell+1} \;= \; \Int{\R }{} \f{J(u)\,S(u)\,u^{2(\ell+1)} }{ 4\pi\be\,(2\ell+1)!}\,\dd u
\label{ecriture definition suite dn et fct S1}
\enq
where the function $J$ comes from Definition~\ref{defKKK} and $S$ is the kernel of ${\cal S}_N$ and appears lately in \eqref{82}.
\end{defin}
They are useful in the following:
\begin{lemme}
\label{kpkpk}Assume $F \in \mc{C}^{2k+2}( \intff{a_N}{b_N}^2)$ and antisymmetric \textit{viz}. $F(\xi,\eta)=-F(\eta,\xi)$. We have the asymptotic expansion:
\beq
\mf{I}_{\e{d};\e{bk}}[F] \; = \; - N^{\a}\cdot \gimel_0 \cdot \mc{T}_{\e{even}}[F](0)
\, - \, \sul{\ell = 1 }{k} \f{ 1 }{  N^{(2\ell-1)\a} } 
\bigg\{ \gimel_{2\ell} \cdot \mc{T}_{\e{even}}^{(2\ell)}[F](0)  \, + \,  \gimel_{2\ell-1}\cdot\mc{T}_{\e{odd}}^{(2\ell-1)}[F](0)  \bigg\}
\; + \; \e{O}\big( N^{-2k \a}\big) 
\label{ecriture DA integrale double bulk}
\enq
in terms of the integral transforms:
\beq
\label{eve}\mc{T}_{\e{even}}[F](s) \; =   \f{1}{s} \hspace{-2mm} \Int{2a_N-|s|}{2b_N-|s|} \hspace{-2mm}  F\big[(v + s)/2,(v-s)/2\big]\,\dd v 
\qquad \e{and} \qquad 
\mc{T}_{\e{odd}}[F](s) \; =  \hspace{-2mm} \Int{2a_N-|s|}{2b_N-|s|} \hspace{-2mm} \partial_{s} \big\{ s^{-1}\,F\big[(v+s)/2,  (v-s)/2\big] \big\}\,\dd v 
\;. 
\enq
\end{lemme}
The integral transforms $\mc{T}_{\e{even}}, \mc{T}_{\e{odd}}$ can be slightly simplified in the case of specific examples of the 
function $F$. In particular, if $F$ takes the form  $F(\xi,\eta)=g(\xi)-g(\eta)$ for some sufficiently regular function $g$, then we have:
\beq
\mc{T}_{\e{even}}[F](0) \; = \; \Int{2a_N}{2b_N} g^{\prime}(v/2)\,\dd v \; = \; 2\big[ g(b_N) - g(a_N) \big]\;. 
\enq

\Proof  We first implement the change of variables 
\beq
\left\{\begin{array}{lcl} u & = & N^{\a}(\xi-\eta) \\ v & = & \xi+\eta \end{array}\right.\,\qquad \textit{i.e.} \qquad \left\{\begin{array}{lcl} \xi & = & (v + N^{-\a}u)/2 \\ \eta & = & (v-N^{-\a}u)/2 \end{array}\right. 
\enq
in the integral representation for $\mf{I}_{\e{d};\e{bk}}[F]$.  This recasts the integral as 
\bem
\mf{I}_{\e{d};\e{bk}}\big[F\big] \; = \; -\f{N^{2\a} }{4\pi \be} \Int{ - \ov{x}_N }{ \ov{x}_N } \dd u\,J(u)
\Int{ 2a_N + |u|N^{-\a} }{ 2b_N -  |u|N^{-\a} } 
\partial_{u}\bigg\{ S(u)\cdot F\Big[\f{v+u N^{-\a}}{2}, \f{v-u N^{-\a}}{2} \Big] \bigg\}\,\dd v  \\ 
\; = \; -\f{N^{\a} }{4\pi \be} \Int{ - \ov{x}_N }{ \ov{x}_N }   \Bigg( J(u)\, \big[ u S^{\prime}(u)+S(u) \big]\, \mc{T}_{\e{even}}[F](uN^{-\a})
\, + \,  N^{-\a} J(u)\cdot u S(u)\cdot \mc{T}_{\e{odd}}[F](uN^{-\a}) \Bigg)\dd u 
\end{multline}
Both $J(u)\cdot u S(u)$ and $J(u)\big[ u S^{\prime}(u)+S(u) \big]$ decay exponentially fast at infinity. Hence,
the expansion \eqref{ecriture DA integrale double bulk} readily follows by using
the Taylor expansion with integral remainder for the functions $\mc{T}_{\e{even}/\e{odd}}[F](uN^{-\a})$ around $u=0$, and the parity properties of $\mc{T}_{\e{even}/\e{odd}}[F]$. \qed

\begin{lemme}
\label{Proposition Asympt Integrale bk vs fct localisees}

Let $F(x,y;\xi,\eta)$ be such that
\begin{itemize}
\item  $F(x,y;\xi,\eta)=-F(y,x;\eta,\xi)$ ; 
\item the map $ (x,y;\xi, \eta) \mapsto F(x,y;\xi,\eta)$ is $\mc{C}^{3}\big(\R^+\times \R^+ \times \intff{a_N}{b_N}^2 \big)$ ;
\item $F$ -- and any combination of partial derivatives of total order at most $3$ -- decays exponentially fast in $x,y$ uniformly in  $(\xi, \eta) \in \intff{a_N}{b_N}$ , \textit{viz}.
\beq
\max \bigg\{ \big| \partial_{x}^{p_1}\partial_{y}^{p_2}\partial_{\xi}^{p_3}\partial_{\eta}^{p_4} F(x,y;\xi,\eta) \big|   \; : \; \sul{a=1}{4}p_a \, \leq \, 3 \bigg\} 
\; \leq \; C\,\ex{-c \min(x,y)} \;. 
\label{ecriture hypothese sur rest dans DA pour integrable bulk}
\enq
\item the following asymptotic expansion holds uniformly in $(x,y) \in \intff{0}{ \eps N^{\a} }$, for some $\eps>0$
and with a differentiable remainder in the sense of \eqref{ecriture hypothese sur rest dans DA pour integrable bulk}.
\beq
F\big( x , y;b_N-N^{-\a} x, b_N - N^{-a}\,y\big) \; = \; 
\sul{\ell=1}{k} \f{ f_{\ell}(x,y)   }{ N^{\ell \a} } \; + \; \e{O}\Big( \f{ C_k\,\ex{-c \min(x,y)} }{ N^{(k+1) \a} } \Bigg) \;, 
\label{Hypothese sur DA fct F integrale bk}
\enq
where $f_{\ell} \in \mc{C}^3(\R^+\times\R^+)$ for $\ell \in \intn{1}{k}$ while
\beq
\max \bigg\{ \big| \partial_{x}^p\partial_{y}^{q} f_{\ell}(x,y) \big|   \; : \; p+q \, \leq \, 3 \;\;\;  and  \; \; \; \ell \in \intn{1}{k} \bigg\} 
\; \leq \; C_{k}\,\ex{-c \min(x,y)} \;. 
\enq
\end{itemize}
Then, denoting $F_{N}(\xi,\eta) \, = \,  F\big( N^{\a}(b_N -\xi),N^{\a}(b_N -\eta);\xi,\eta\big)$, we have an asymptotic expansion:
\beq
\mf{I}_{\e{d};\e{bk}}[F_N] %\; \equiv \;  \f{-N^{2\a}}{4\pi \be} \Int{a_N}{b_N} \!  J\big(N^{\a}(\xi -\eta) \big)
%
%\cdot \Big( \f{ \Dp{} }{ \Dp{}\xi } \, - \, \f{ \Dp{} }{ \Dp{}\eta }  \Big)
%
%\bigg\{ S\big(N^{\a}(\xi -\eta) \big)  F(x_R,y_R,\xi,\eta)  \bigg\} \cdot \dd \xi \dd \eta \\
%
%
\; = \; - \sul{\ell=1}{k}  \f{1}{N^{(\ell-1) \a}} \Int{ \R }{} \f{ \dd u\,J(u)}{4\pi \be } \Int{ |u| }{+\infty}  \hspace{-1mm} \dd v \; 
\partial_{u}\bigg\{ S(u)\cdot f_{\ell}\big[(v-u)/2,(v+u)/2\big]\bigg\}
\; + \; \e{O}\Bigg( \f{ 1 }{ N^{k\a}  } \Bigg) \;, 
\enq
\end{lemme}
Note that, necessarily, $f_{\ell}$ are antisymmetric functions of $(x,y)$.

\Proof  The change of variables 
\beq
\left\{\begin{array}{lcl} u & = & N^{\a}(\xi-\eta)  \\ v & = & N^{\a}\big(2b_N - \xi - \eta \big) \end{array}\right.\,\qquad \textit{i.e.} \qquad
\left\{\begin{array}{lcl} \xi & = & b_N-N^{-\a}(v-u)/2 \\ \eta & = & b_N-N^{-\a}(v+u)/2 \end{array}\right.
\enq
recasts the integral as
\beq
\mf{I}_{\e{d}; \e{bk}} [ F ]  \; = \; -  N^{\a}  \Int{ -\ov{x}_N }{ \ov{x}_N } \f{ \dd u\,J(u)}{4\pi \be } 
\Int{ |u| }{2\ov{x}_N-|u|}  \hspace{-1mm} \dd v \; 
\partial_{u}\bigg\{ S(u) \cdot
F\Big[ \f{v-u}{2},\f{v+u}{2} ; b_N- \f{v-u}{2N^{\a}} , b_N -\f{v+u}{2 N^{\a}}  \Big] \bigg\} \;. 
\enq
At this stage, we can limit all the domains of integration to $|u|, |v| \leq \eps N^{\a}$, this for the 
price of exponentially small corrections. Then, we insert the asymptotic expansion \eqref{Hypothese sur DA fct F integrale bk}
and extend the domains of integration up to $+\infty$ this, again, for the price of exponentially small
corrections, and we get the claim. \qed

\vspace{0.2cm}

Very similarly, but under slightly different assumptions on the function $F$, we have the large-$N$ asymptotic expansion of the right edge double integral.

\begin{lemme}
\label{Proposition Asympt Integrale R vs fct qcq}

Let $F(x,y;\xi,\eta)$ be such that
\begin{itemize}
\item  $F(x,y;\xi,\eta)=-F(y,x;\eta,\xi)$; 
\item the map $ (x,y;\xi, \eta) \mapsto F(x,y;\xi,\eta)$ is $\mc{C}^{3}\big(\R^+\times \R^+ \times \intff{a_N}{b_N}^2 \big)$;
\item $F$ decays exponentially fast in $x,y$ this uniformly in  $(\xi, \eta) \in \intff{a_N}{b_N}$ and for 
any combination of partial derivatives of total order at most $3$, \textit{viz}.:
\beq
\max \bigg\{ \big|\partial_{x}^{p_1} \partial_{y}^{p_2} \partial_{\xi}^{p_3} \partial_{\eta}^{p_4} F(x,y;\xi,\eta) \big|   \; : \; \sul{a=1}{4}p_a \, \leq \, 3 \bigg\} 
\; \leq \; C\,\ex{-c \min(x,y)} \;. 
\label{ecriture hypothese sur rest dans DA pour integrable bord}
\enq
\item the following asymptotic expansion holds uniformly in $(x,y) \in \intff{0}{ \eps N^{\a} }$, for some $\eps>0$
and with a differentiable remainder in the sense of \eqref{ecriture hypothese sur rest dans DA pour integrable bord}.
\beq
F\big( x , y;b_N-N^{-\a}x,b_N - N^{-\a}y\big) \; = \; 
\sul{\ell=1}{k} \f{ f_{\ell}(x,y)   }{ N^{\ell \a} } \; + \; \e{O}\bigg( \f{ C_k\,(x^k+y^k+1) }{ N^{(k+1) \a} }  \bigg) \;, 
\enq
where $f_{\ell} \in \mc{C}^3(\R^+\times\R^+)$ for $\ell \in \intn{1}{k}$ while 
\beq
\max \bigg\{ \big| \partial_{x}^{p} \partial_{y}^{q} f_{\ell}(x,y) \big|   \; : \; p+q \, \leq \, 3 \;\;\;  and  \; \; \; \ell \in \intn{1}{k}\bigg\} 
\; \leq \; C_{k}\,(x^k+y^k+1) \;. 
\enq
\end{itemize}
Then, we have the following asymptotic expansions
\beq
\mf{I}_{\e{d};R}[F_N] \; = \;    \sul{\ell =1 }{k}  \f{1}{N^{(\ell-1)\a}}
 \Int{ \msc{C}^{(+)}_{\e{reg}} }{  }\hspace{-2mm}\f{\dd \la }{2\i \pi} \Int{ \msc{C}^{(-)}_{\e{reg}}  }{} \hspace{-2mm} \f{\dd \mu }{2\i \pi} 
\f{  (2 \pi \be)^{-1}  }{ (\la-\mu)R_{\da}(\la)R_{\ua}(\mu) }   \Int{0}{+\infty} \!   \ex{ \i \la x - \i\mu y}  
\partial_{x} \big\{ S(x-y) \cdot f_{\ell}( x , y) \big\} \dd x \, \dd y 
 \; + \; \e{O}\Bigg( \f{ 1 }{ N^{\a k} } \Bigg)\;. 
\enq
The function $F_N$ occurring above is as defined in the previous Lemma.
\end{lemme}

\Proof  The change of variables $x = N^{\a} (b_N-\xi)$ and $y =N^{\a}(b_N-\eta)$ recasts the integral in the form
\beq
\mf{I}_{\e{d};R}\big[F\big] \; = \;  N^{\a}
 \!\!\!\Int{ \msc{C}^{(+)}_{\e{reg}} }{  }\hspace{-2mm}\f{\dd \la }{ 2  \i \pi } 
 \Int{ \msc{C}^{(-)}_{\e{reg}}  }{} \hspace{-2mm} \f{\dd \mu }{ 2 \i \pi } 
\f{  (2 \pi \be)^{-1}  }{ (\la-\mu) R_{\da}(\la) R_{\ua}(\mu) }   \Int{0}{ \ov{x}_N } \!   \ex{\i\la x-\i\mu y}  
\partial_{x}\big\{ S(x-y)\cdot F\big( x , y; b_N-N^{-\a}x, b_N - N^{-\a}y\big) \big\} \dd x \dd y \;.\nonumber 
\enq
We can then conclude exactly as in the proof of Lemma~\ref{Proposition Asympt Integrale bk vs fct localisees}. \qed

\subsection{Estimation of the remainder $\De_{[k]} \mf{I}_{\e{d}}[H,V]$.}

\begin{lemme}
\label{Proposition estimation reste integrale double}

Let $k \geq 1$ be an integer. Given $C_V>0$, assume $V$ strictly convex, smooth enough and $\norm{ V_{\mf{e}} }_{ W_3^{\infty}(\R) }<C_V$. 
There exists $C >0$ such that, for any $H \in \mathfrak{X}_{s}(\R)$ smooth enough, 
the remainder integral $\De_{[k]} \mf{I}_{\e{d}}[H,V]$ satisfies:
\beq
\left|\De_{[k]} \mf{I}_{{\rm d}}[H,V] \right|\; \leq \; C\, N^{-(k-5)\a}\cdot \mf{n}_{k+4}[V_{\mf{e}}] \cdot
\norm{ H_{\mf{e}} }_{ W_{ \max\{k,5\} + 4 }^{\infty}(\R) } \;. 
\label{estimation reste de l'integrale double}
\enq %A:ajoute valeurs absolutes

\end{lemme}

\Proof   It follows from Lemma \ref{Lemme structure locale au bord pour WR et Wbk} and \ref{Lemme comportement fonction a goth}, as well as $V^{\prime \prime}(b_N) \not=0$ by strict convexity, that:
\beq
x \mapsto  \f{ \big( \mc{W}_{\e{bk};k}^{(\e{as})} \, + \, \mc{W}_{R;k}^{(\e{as})} \big)[H](x) }
				{  \big( \mc{W}_{\e{bk};k}^{(\e{as})} \, + \, \mc{W}_{R;k}^{(\e{as})} \big)[V^{\prime}](x) }
\enq
is smooth at $x=0$. As a consequence, the function 
\bem
(\xi, \eta) \mapsto \mc{G}_{\e{bk};k}\big[H,V](\xi,\eta)  + \mc{G}_{R;k}^{(\e{as})}\big[H,V](x_R,y_R;\xi,\eta) \\
\; = \; \f{  \Delta_{[k]}\mc{W}_{\e{bk}}^{(\e{as})}\big[H](\xi) }{ \mc{W}_{\e{bk};k}\big[V^{\prime}](\xi) }
\; + \; \f{ \mc{W}_{\e{bk};k}^{(\e{as})}\big[V^{\prime}]\big( x_{R} \big) }{ \mc{W}_{\e{bk};k}\big[V^{\prime}](\xi) }
\cdot \f{ \big( \mc{W}_{\e{bk};k}^{(\e{as})} \, + \, \mc{W}_{R;k}^{(\e{as})} \big)[H]\big( x_{R} \big) }
				{  \big( \mc{W}_{\e{bk};k}^{(\e{as})} \, + \, \mc{W}_{R;k}^{(\e{as})} \big)[V^{\prime}]\big( x_{R} \big) }
\; - \; \Big( \xi \leftrightarrow \eta \Big) 				 				
\end{multline}
is smooth in $(\xi,\eta)$.
Furthermore, it follows from Theorem \ref{Proposition characterisation operateur UN} that $(\xi, \eta) \mapsto \mc{G}_N[H,V](\xi, \eta)$ 
is smooth on $\intff{a_N}{b_N}$ as well. Since  $\mc{G}_{R;k}^{(\e{as})}\big[H,V](x_R,y_R;\xi,\eta)$ is smooth in $\xi$ -- respectively $\eta$ --
as soon as the latter variable is away from $b_N$ or $a_N$, it follows that 
$(\xi, \eta) \mapsto \De_{[k]}\mc{G}_N[H,V](\xi, \eta)$ is smooth as well. 

The remainder $\Delta_{[k]}{\cal G}_{N}$ described in \eqref{ecriture bornage operateur integral SN loin en dehors support} involves 
the remainders $\Delta_{[k]}{\cal W}^{(\e{as})}_{R/{\rm bk}}$ studied in Lemma~\ref{Lemme structure locale au bord pour WR et Wbk}, and $(\Delta_{[k]} \mc{W}_{N})_{R}$ defined in \eqref{iugfgd} 
and for which Proposition~\ref{Proposition Ecriture reguliere uniforme des divers const de WN} and Lemma~\ref{Lemme structure locale au bord pour WR et Wbk} also provide estimates. 
Using the properties of the $\mathfrak{a}$'s obtained in Lemma~\ref{Lemme comportement fonction a goth} and involved in the asymptotic expansion of the $^{\e{(as)}}$ quantities, 
it shows the existence of constants $c_{\ell;k}^{(0)}, c_{\ell;k}^{(1/2)}$ and of functions $f_{m;k} \in W^{\infty}_{m}\big(\R^+\big)$ bounded uniformly in $N$ and satisfying 
$f_{m;k}(x) = \e{O}\big(x^{m+1/2}\big)$ such that 
\beq
 \De_{[k]}\mc{G}_N[H,V](\xi, \eta) \; = \; \f{1}{ N^{k\a} }\sul{\ell=0}{m} \big( c_{\ell;k}^{(0)}\,x_R^{\ell}  \, + \, 
    c_{\ell;k}^{(1/2)}\,x_R^{\ell-1/2} \big) \; + \; \f{f_{m;k}(x_R) }{ N^{k\a} }
\; - \; \Big( x_R \leftrightarrow y_R \Big)     \;, 
\label{ecriture developpement local reste DekGN en bN}
\enq
for $(x_R,y_R) \in \intff{0}{\eps}^2$. Since $ \De_{[k]}\mc{G}_N[H,V](\xi, \eta) $ is smooth, we necessarily have that $c_{\ell;k}^{(1/2)}=0$ for $\ell \in \intn{0}{m}$.  
The representation \eqref{ecriture developpement local reste DekGN en bN} thus ensures that 
\beq
\label{spii}\max_{0\leq \ell+p \leq n} \max_{ \substack{ (x_R,y_R)  \\ \in \intff{0}{\eps}^2} } \big| 
\partial_{\xi}^{\ell}\partial_{\eta}^{p} \De_{[k]}\mc{G}_N[H,V](\xi, \eta) \big| \; \leq \;
  \f{C_{n}}{ N^{(k-n)\a}  } \cdot \mf{n}_{n+k}[V] \cdot \norm{ H_{ \mf{e } } }_{W^{\infty}_{n+1+k}(\R) } \;. 
\enq
Here, the explicit control on the dependence of the bound on $V$ and $H$ issues from the control on the remainders entering in the expression for 
 $\De_{[k]}\mc{G}_N[H,V]$.
 
 \noindent Similar types of bounds can, of course, be obtained for  $(x_L,y_L) \in \intff{0}{\eps}^2$. 
Finally, as soon as a variable, be it $\xi$ or $\eta$, is uniformly (in $N$) away from an immediate neighbourhood of 
the endpoints $a_N$ and $b_N$, we can use more crude expressions for the remainders so as to bound 
derivatives of the remainder $\De_{[k]}\mc{G}_N[H,V]$. This does not spoil \eqref{spii} and we conclude:
\beq
\max_{0\leq \ell+p \leq n} \max_{ \substack{ (\xi,\eta)  \\ \in \intff{a_N}{b_N}^2} } \big| 
\partial_{\xi}^{\ell} \partial_{\eta}^p \De_{[k]}\mc{G}_N[H,V](\xi, \eta) \big| \; \leq \;
 \f{C_{n}}{ N^{(k-n)\a}  } \cdot \mf{n}_{n+k}[V] \cdot \norm{ H_{ \mf{e } } }_{W^{\infty}_{n+1+k}(\R) } \;. 
\label{ecriture bornes sur derivees de Dek GN}
\enq
Having at disposal such a control on the remainder $\De_{[k]}\mc{G}_N[H,V]$, we are in position 
to bound the double integral of interest. The latter decomposes into a sum of four terms 
\beq
\De_{[k]}\mf{I}_{\e{d}}[H,V] \, = \, \sul{p=1}{4} \De_{[k]}\mf{I}_{\e{d};p}[H,V] 
\enq
that have been defined in \eqref{definition De k de mfracI double 1}-\eqref{definition De k de mfracI double 4}. 

\subsubsection*{Bounding $\De_{[k]}\mf{I}_{\e{d};1}[H,V]$}

Let 
\beq
\tau(\xi,\eta) \, = \, \partial_{\xi} \Big\{ S\big(N^{\a}(\xi-\eta)\big) \cdot \mc{G}_N\big[H,V\big](\xi,\eta)   \Big\} 
\quad \e{and} \quad \De_{[k]}\tau(\xi,\eta) \, = \, \partial_{\xi} \Big\{ S\big(N^{\a}(\xi-\eta)\big) \cdot \De_{[k]}\mc{G}_N\big[H,V\big](\xi,\eta)   \Big\} \;. 
\label{definition fct tau et delta k tau}
\enq
Observe that given $(\xi,\eta) \mapsto f(\xi,\eta)$ sufficiently regular, we have the decomposition:
\beq
\Int{a_N}{b_N}\mc{W}_{\e{exp}}\big[ f(\xi,*) \big](\xi)\,\dd \xi \; = \; 
\Int{a_N}{b_N} \mc{W}_{\e{exp}}\big[ f(a_N,*) \big](\xi)\,\dd \xi  \; + \; 
\Int{a_N}{b_N}\! \dd \xi \Int{a_N}{\xi} \! \dd \eta \; \mc{W}_{\e{exp}}\big[ \Dp{\eta} f(\eta,*) \big] (\xi) \;. 
\enq
The latter ensures that 
\beq
\Big| \ \Int{a_N}{b_N} \mc{W}_{\e{exp}}\big[ f(\xi,*) \big](\xi)\,\dd \xi \Big| \; \leq  \; 
\Norm{ \mc{W}_{\e{exp}}\big[ f(a_N,*) \big] }_{ L^{1}(\intff{a_N}{b_N}) } \; + \;  
  (b_N - a_N) \sup_{\eta \in \intff{a_N}{b_N}} \Norm{ \mc{W}_{\e{exp}}\big[ \partial_{\eta}f(\eta,*)\big] }_{ L^{1}(\intff{a_N}{b_N}) }\;. 
\enq
The two terms can be estimated directly using the $L^1$ bound \eqref{ecriture bornage norme L1 de WN} obtained in 
Lemma \ref{Lemme bornage norme L1 de WN}. For the first one:
\beq
\Norm{ \mc{W}_{\e{exp}}\big[ f(a_N,*) \big] }_{ L^{1}(\intff{a_N}{b_N}) } 
\; \leq \;C_1 \ex{-C_2 N^{\a} }  \Norm{ f_{\mf{e}}(a_N,*) }_{W^{\infty}_1(\R)} \leq C_1\,\ex{-C_2 N^{\a}} \Norm{f}_{W^{\infty}_{1}(\R^2)}
\enq
for some $C_1,C_2 > 0$ independent of $N$ and $f$, and likewise for the second term. But the $W^{\infty}_{p}(\R^2)$ norm of $f_{\mf{e}}$ is also bounded by a 
constant times the $W^{\infty}_{p}(\intff{a_N}{b_N}^2)$ norm of $f$, and we can make the constant depends only on the compact support of the extension $f_{\mf{e}}$. Therefore:
\beq
\Norm{ \mc{W}_{\e{exp}}\big[ f(a_N,*) \big] }_{ L^{1}(\intff{a_N}{b_N}) } \leq C_1^{\prime}\,\ex{-C_2^{\prime}N^{\a}}  \norm{ f}_{W^{\infty}_1(\intff{a_N}{b_N}^2)} 
\enq
for some $C_1^{\prime},C_2^{\prime} > 0$.
%
%
%\beq
%
%\bigg| \Int{a_N}{b_N}\! \dd \xi \Int{a_N}{\xi} \! \dd \eta \; \mc{W}_{\e{exp}}\big[ \Dp{\eta} f(\eta,*) \big] (\xi) \bigg|
%
%\; \leq \; C_1^{\prime} \ex{-C_2 N^{\a} }   \norm{ f}_{W^{\infty}_2(\intff{a_N}{b_N}^2)}   \;. 
%
%\enq
%
%
%
Taking $f = \tau - \Delta_{[k]}\tau$ to match the definition \eqref{definition De k de mfracI double 1} of $\De_{[k]}\mf{I}_{\e{d};1}$, this implies:
\beq
\label{949}\big| \De_{[k]}\mf{I}_{\e{d};1}[H,V] \big| \; \leq  \;C_1^{\prime} \ex{-C_2 N^{\a} }\,
\Big\{  \norm{ \tau }_{W^{\infty}_2(\intff{a_N}{b_N}^2)} + \norm{ \De_{[k]}\tau }_{W^{\infty}_2(\intff{a_N}{b_N}^2)}  \Big\} \;. 
\enq
It solely remains to bound the $W^{\infty}_2(\intff{a_N}{b_N}^2)$ norm of $\tau$ and $\De_{[k]}\tau$. We remind that, for $\xi \in \intff{a_N}{b_N}$, 
we have from the definition \eqref{definition integrande GN cal} and the expression of ${\cal U}_{N}^{-1}$ given in \eqref{mimim}:
\beq
{\cal G}_{N}[H,V](\xi) = {\cal U}_{N}^{-1}[H](\xi) - {\cal U}_{N}^{-1}[H](\eta) \;. 
\enq
By invoking the mean value theorem and the estimate of Proposition~\ref{Theorem bornes sur norme inverse UN via estimation fines locales} for $W_{\ell}^{\infty}$ norm of ${\cal U}_{N}^{-1}[H]$, 
we obtain:
\begin{eqnarray}
\norm{ \tau }_{ W^{\infty}_{ \ell }(\intff{a_N}{b_N}^2) } & \leq & C\,N^{\a}\,
\Norm{(\xi,\eta) \mapsto \f{ \mc{G}_N\big[H,V\big](\xi,\eta) }{ \xi-\eta}    }_{ W^{\infty}_{ \ell +1 }(\intff{a_N}{b_N}^2) }
\; \leq \; C^{\prime}\,N^{\a}\, \Norm{  \mc{U}_N^{-1}[H]   }_{ W^{\infty}_{ \ell +2 }(\intff{a_N}{b_N}) } \\
& \leq & C^{\prime}_{\ell} \cdot (\ln N )^{2\ell+5} \cdot N^{ (\ell+4) \a } \cdot \mf{n}_{\ell+2}[V] \cdot \mc{N}_{N}^{(2\ell + 5)}\big[\mc{K}_{\kappa}[H]\big] \\
& \leq & C^{\prime\prime}_{\ell} \cdot (\ln N)^{2\ell + 5}\cdot N^{(\ell + 4)\a} \cdot \mf{n}_{\ell + 2}[V]\cdot \norm{H_{\mf{e}}}_{W_{2\ell + 5}^{\infty}(\R)}
\label{ecriture borne sur la fct tau}
\end{eqnarray}
where the last step comes from domination of the weighted norm by the $W^{\infty}$ norm of the same order -- and the exponential regularisation can easily be traded for a 
compactly supported extension up to increasing the constant prefactor. Similarly, in virtue of the bounds \eqref{ecriture bornes sur derivees de Dek GN}, we get: 
\beq
\norm{ \De_{[k]} \tau }_{ W^{\infty}_{ \ell }(\intff{a_N}{b_N}^2) } 
\; \leq \; C^{\prime} \cdot N^{(\ell+3-k)\a} \cdot \mf{n}_{k	+\ell+2}[V] \cdot \norm{ H_{\mf{e}} }_{ W_{k+\ell+3}^{\infty}(\R) } \; .
\label{ecriture borne de Dek de tau}
\enq
Putting these two estimates back in \eqref{949} with $\ell = 2$, we see that:
\beq
\big| \De_{[k]}\mf{I}_{\e{d};1}[H,V] \big| \; \leq  \;C_1^{\prime}\cdot N^{6\a}\cdot \ex{-C_2 N^{\a} }  \mf{n}_{k+4}[V] \cdot \norm{ H_{\mf{e}} }_{ W_{ \max\{k,5\}+4}^{\infty}(\R) }
  \;. 
\enq
which is exponentially small when $N \rightarrow \infty$.

\subsubsection*{Bounding $\De_{[k]}\mf{I}_{\e{d};2}[H,V]$}

$\De_{[k]}\mf{I}_{\e{d};2}[H,V]$ has been defined in \eqref{definition De k de mfracI double 2} and can be bounded by repeating the previous handlings. 
Indeed, using \eqref{ecriture bornage norme L1 de WN} on the $L^1$ norm  of $\mc{W}_N$ and then following the previous steps, one finds:
\beq
 \Big| \De_{[k]}\mf{I}_{\e{d};2}[H,V] \Big| \; \leq \;  \norm{ \De_{[k]}\tau }_{ W^{\infty}_2(\intff{a_N}{b_N}^2) }
 \enq
 with $\De_{[k]}\tau$ defined in \eqref{definition fct tau et delta k tau} and bounded in $W_{\ell}^{\infty}$ norms in \eqref{ecriture borne de Dek de tau}. Hence, we find:
\beq
\big| \De_{[k]}\mf{I}_{\e{d};2}[H,V] \big| \; \leq  \; C_1^{\prime}\cdot N^{(5-k)\a}\cdot \mf{n}_{k+4}[V] \cdot \norm{ H_{\mf{e}} }_{ W_{k+5}^{\infty}(\R) }
 \;. 
\enq

\subsubsection*{Bounding $\De_{[k]}\mf{I}_{\e{d};3}[H,V]$}

This quantity is defined in \eqref{definition De k de mfracI double 3}, and it follows from the explicit expression for $\mc{W}_N[1](\xi)$ given in \eqref{forumule explicite pour WN de 1ter}  that 
\beq
\big| \De_{[k]}\mf{I}_{\e{d};3}[H,V]  \big| \; \leq \; C\,N^{\a} \norm{ \tau }_{W^{\infty}_0(\intff{a_N}{b_N}^2)}\cdot \big| \chi_{12;+}(0) \big| \cdot 
|b_N-a_N| \cdot  \sup_{\xi \in \intff{a_N}{b_N} }
\Bigg| \Int{ \R +\i \eps^{\prime} }{} \f{ \chi_{11}(\la) }{ \la }\ex{-\i N^{\a} \la (\xi-a_N) }\cdot \f{ \dd \la }{ 2\i \pi } \Bigg|
\enq
where $\tau$ is as defined in \eqref{definition fct tau et delta k tau}. The decomposition \eqref{ecriture decompositon asymptotique chi} for $\chi$ and its properties show the existence of $C,C^{\prime} > 0$ such that:
\beq
\forall \la \in \R + \i\epsilon^{\prime},\qquad |\chi_{11}(\la)| \; \leq \;  C\,|\la|^{-1/2},\quad \qquad \e{and} \qquad 
| \chi_{12}(\la) | \; \leq \; C^{\prime}\,\ex{-N^{\a} \varkappa_{\epsilon^{\prime}} } \;. 
\enq
Hence, by invoking the bounds \eqref{ecriture borne sur la fct tau} satisfied by $\tau$, we get: 
\beq
\Big| \De_{[k]}\mf{I}_{\e{d};3}[H,V]  \Big| \; \leq \; C^{\prime\prime} \cdot N^{5\a}\cdot ( \ln N)^5 \cdot \ex{-N^{\a} [ \varkappa_{\eps^{\prime}} -\eps^{\prime}(b_N - a_N)]}
\cdot \mf{n}_{2}[V] \cdot \norm{ H_{\mf{e}} }_{ W_5^{\infty}(\R) } \;. 
\enq
Since $\varkappa_{\eps^{\prime}} > 0$ is bounded away from $0$ when $\eps^{\prime} \rightarrow 0$ according to its definition \eqref{definition constante c epsilon de Pi}, we also have $ \varkappa_{\eps} -\eps^{\prime} x_N>0$ uniformly in $N$ for some choice of $\eps^{\prime}$ small enough but independent of $N$.

\subsubsection*{Bounding $\De_{[k]}\mf{I}_{\e{d};4}[H,V]$}

This quantity is defined in \eqref{definition De k de mfracI double 4}, and it involves integration of:
\beq
\tau_{L}(\xi,\eta) \; = \;  \partial_{\xi} \Big\{ S\big(N^{\a}(\xi-\eta)\big) 
\cdot \mc{G}_{R;k}^{(\e{as})}\big[H,V\big](x_L,y_L;b_N+a_N-\xi,b_N+a_N-\eta)   \Big\}
\enq
where $\mc{G}_{R;k}^{(\e{as})}$ was defined in \eqref{75}. It only involves the operators $\mc{W}^{(\e{as})}_{\e{bk};k}$ and $\mc{W}^{(\e{as})}_{R;k}$, whose expression is given in Lemma~\ref{Lemme structure locale au bord pour WR et Wbk}. Let us fix $\epsilon > 0$. Straightforward manipulations show that, for $(\xi,\eta) \in \intff{a_N+ \eps }{b_N}^2$, we have: 
\beq
\big| \tau_{L}(\xi,\eta) \big| \; \leq \; C N^{3 \a}\ex{-C^{\prime} \min(x_L,y_L)}\cdot  \mf{n}_{k+1}[V] \cdot \norm{ H_{\mf{e}} }_{ W_{k}^{\infty}(\R) }
\; \leq\; \wt{C}\cdot N^{3 \a}\cdot \ex{-\eps C^{\prime}N^{\a} } \cdot  \mf{n}_{k+1}[V] \cdot \norm{ H_{\mf{e}} }_{ W_{k}^{\infty}(\R) } 
\enq
which is thus exponentially small in $N$. Similar steps show that, for 
\beq
(\xi,\eta) \in \Big\{ \intff{a_N+ \eps }{b_N}\times \intff{a_N}{a_N+ \eps } \Big\} \cup 
 \Big\{ \intff{a_N}{a_N+ \eps }\times \intff{a_N+ \eps }{b_N} \Big\} \cup 
  \Big\{ \intff{a_N}{a_N+ \eps }\times \intff{a_N}{a_N+ \eps } \Big\}  \;,
\enq
we have:
\beq
\big| \tau_{L}(\xi,\eta) \big| \; \leq \; C\,N^{3 \a} \mf{n}_{k+1}[V] \, \norm{H}_{ W^{\infty}_k(\R) } \;. 
\enq
Here, the exponential decay in $N$ will come after integration of $\tau_{L}$ as it appears in \eqref{definition De k de mfracI double 4}. Indeed, given ${\rm Im}\,\la > 0$ and ${\rm Im}\,\mu<0$ we have:
\bem
\bigg| \Int{a_N}{b_N}    \ex{\i \la x_R} \ex{-\i\mu y_R} \tau_{L}(\xi,\eta)\, \dd \xi \dd \eta  \bigg| \leq 
C\,N^{3 \a}\ex{- \eps C^{\prime} N^{\a} } \mf{n}_{k+1}[V] \, \norm{H}_{ W^{\infty}_k(\R) } 
\Int{a_N+\eps }{b_N}    \ex{-|{\rm Im}\,\la|N^{\a}(b_N-\xi) -|{\rm Im}\,\mu| N^{\a}(b_N-\eta)}\,\dd \eta \, \dd \xi     \\
+ C N^{3 \a} \mf{n}_{k+1}[V] \, \norm{H}_{ W^{\infty}_k(\R) } \Bigg\{ \Int{a_N+\eps }{b_N}\hspace{-2mm} \dd \xi \Int{a_N}{a_N+\eps } \hspace{-2mm} \dd \eta \;  + \;
\Int{a_N}{a_N+\eps } \hspace{-2mm} \dd \xi \Int{a_N+\eps }{b_N} \hspace{-2mm} \dd \eta  \; + \; \Int{a_N}{a_N+\eps } \Int{a_N}{a_N + \eps} \dd \xi\dd \eta  \Bigg\}
 \ex{-|{\rm Im}\,\la|N^{\a}(b_N-\xi) -|{\rm Im}\,\mu| N^{\a}(b_N-\eta)}  \\
 \; \leq \;  \mf{n}_{k+1}[V] \, \norm{H}_{ W^{\infty}_k(\R) }\cdot \f{\wt{C}\,N^{3 \a} \ex{-\wt{C}^{\prime} N^{\a} }   }{ |\la \cdot \mu | }  \;. 
\end{multline}
Note that, above, we have used that for $\la \in \mathscr{C}_{{\rm reg}}^{(+)}$ and $\mu \in \mathscr{C}_{{\rm reg}}^{(-)}$, we can bound:
\beq
|{\rm Im}\,\la|^{-1} \leq c_1 |\la|^{-1}\,,\qquad |{\rm Im}\,\mu|^{-1} \leq c_1 |\mu|^{-1}
\enq
for some constant $c_1 > 1$.
Hence, all in all, we have:
\begin{eqnarray}
\Big| \mf{I}_{\e{d};R}\Big[\big(\mc{G}_{R;k}^{(\e{as})}[H,V]\big)^{\wedge}\Big] \Big| & \leq & 
C^{\prime \prime}  N^{3 \a} \ex{-\wt{C}^{\prime} N^{\a} }  \cdot \Int{ \msc{C}^{(+)}_{\e{reg}}  }{} |\dd \la | 
\Int{ \msc{C}^{(-)}_{\e{reg}}  }{} |\dd \mu | \cdot  \f{ \mf{n}_{k+1}[V] \, \norm{H}_{ W^{\infty}_k(\R) }  }{ |\mu-\la| \cdot |\la R_{\da}(\la) R_{\ua}( \mu)\mu|} \nonumber \\
& \leq & C^{\prime\prime\prime}\,N^{-3\a}\,\ex{-\wt{C}^{\prime}N^{\a}} \cdot \mf{n}_{k+1}[V] \, \norm{H}_{ W^{\infty}_k(\R) } \;. 
\end{eqnarray}
Then, putting together all of the results for each $\Delta_{[k]}\mf{I}_{{\rm d};p}$ for $p \in \intn{1}{4}$ entails the global bound \eqref{estimation reste de l'integrale double}. \qed

\subsection{Leading asymptotics of the double integral}

We need to introduce two new quantities before writing down the asymptotic expansion of the double integral $\mathfrak{I}_{{\rm d}}$.
\begin{defin}
\label{ralph} We define the function:
\beq
\mathfrak{c}(x) = \frac{\mathfrak{b}_1(x) - \mathfrak{b}_0(x)\mathfrak{a}_1(x)}{u_1}
\enq
and the constant:
\bem
\aleph_0 \; = \; - 
\Int{ \R }{} \frac{\dd u \,J(u)}{4\pi\be} \Int{ |u| }{+\infty}  \hspace{-1mm} \dd v \; 
\partial_{u}\bigg\{ S(u)\cdot\Big( \mathfrak{c}\big[ \f{v-u}{2} \big] \, - \,  \mathfrak{c}\big[\f{v+u}{2} \big]  \Big) \bigg\} \\
\, + \,  \f{1}{2\pi \be}  \Int{ \msc{C}^{(+)}_{\e{reg}} }{  }\hspace{-2mm}\f{\dd \la }{2 \i \pi} \Int{ \msc{C}^{(-)}_{\e{reg}}  }{} \hspace{-2mm} \f{\dd \mu }{2 \i \pi} 
\f{  1 }{ (\la-\mu)R_{\da}(\la)R_{\ua}(\mu) }   \Int{0}{+\infty} \!   \ex{ \i \la x - \i \mu y}  
 \partial_{x} \bigg\{ S(x-y)
 \big[\mathfrak{c}(x)-\mathfrak{c}(y) \big] 
  \; - \; x+y  \bigg\} \dd x \, \dd y 
\end{multline}
\end{defin} %A change sign y

\begin{prop}
\label{Theorem DA ordre dominant integrale double} We have the large-$N$ behaviour:
\beq
\mf{I}_{\e{d}}[H,V] \; = \; -2\gimel_0\cdot N^{\a}\cdot 
\bigg\{  \f{ H^{\prime}(b_N)  }{ V^{\prime\prime}(b_N) } \, - \,  \f{ H^{\prime}(a_N)  }{ V^{\prime\prime}(a_N) }  \bigg\}
\, + \, \aleph_0 \cdot \bigg\{  \Big(\f{ H^{\prime}  }{ V^{\prime\prime} } \Big)^{\prime}(b_N) \, + \, 
 \Big(\f{ H^{\prime}  }{ V^{\prime\prime} } \Big)^{\prime}(a_N)  \bigg\} \; + \; \Delta\mf{I}_{\e{d}}[H,V] 
\enq
and the remainder is bounded as:
\beq
\Delta \mf{I}_{\e{d}}[H,V] \; \leq \; 
\f{C}{ N^{\a} } \cdot \mf{n}_{10}[V_{\mf{e}}] \cdot \norm{ H_{\mf{e}} }_{ W_{11}^{\infty}(\R) } \;. 
\enq
\end{prop}

\Proof  We first need to introduce two universal sequences of polynomials $\mathpzc{P}_{\ell}\big(\{x_p\}_1^{\ell} \big)$ 
and $\mathpzc{Q}_{\ell}\big(\{y_p\}_1^{\ell};\{a_p\}_1^{\ell}\big)$. Given formal power series
\beq
f(z)=1 \, + \, \sul{\ell \geq 1}{} f_{\ell}\,z^{\ell} \qquad  \e{and} \qquad 
g(z)=1 \, + \, \sul{\ell \geq 1}{} g_{\ell}\,z^{\ell}
\enq
they are defined to be the coefficients arising in the formal power series
\beq
\label{PlQl}\f{ 1 }{ f(z) } \; = \; 1+\sul{\ell \geq 1}{} \mathpzc{P}_{\ell}\big(\{f_p\}_1^{\ell}\big)\,z^{\ell} \quad \e{and} \quad
\f{ g(z) }{ f(z) } \; = \; 1+\sul{\ell \geq 1}{} \mathpzc{Q}_{\ell}\big(\{g_p\}_1^{\ell};\{f_p\}_1^{\ell}\big)\,z^{\ell} \;. 
\enq
Note that 
\beq
\mathpzc{Q}_{\ell}\big( \{ g_p \}_1^{\ell} ; \{ f_p \}_1^{\ell} \big) \; = \; \sul{ \substack{ r+s = \ell \\ r,s\geq 0} }{}
g_r \cdot \mathpzc{P}_{s}\big( \{ f_p \}_1^{s} \big)  \;, 
\enq
where we agree upon the convention $\mathpzc{P}_{0}=1$ and $g_0=1$. This notation is convenient to write down the large-$N$ expansion of $\mc{G}_{{\rm bk};k}$ -- defined in \eqref{Gkbj} -- ensuing from the large $N$-expansion of $\mc{W}_{{\rm bk};k}$ provided by Lemma~\ref{Lemme structure locale au bord pour WR et Wbk}. We find, uniformly in $(\xi,\eta) \in \intff{a_N}{b_N}^2$:
\beq
\mc{G}_{\e{bk};k}[H,V](\xi,\eta)  \; = \; \sul{\ell = 0}{ k-1 } 
\f{ \mf{G}_{\e{bk};\ell}[H,V](\xi,\eta) }{ N^{\a \ell} } \; + \; \e{O}\big( N^{-k\a}\big)
\enq
where
\beq
 \mf{G}_{\e{bk};\ell}[H,V](\xi,\eta)  \; = \; \mf{g}_{\e{bk};\ell}[H,V](\xi)\; - \; \mf{g}_{\e{bk};\ell}[H,V](\eta)
\enq
with
\beq
\label{gggg}\mf{g}_{\e{bk};\ell}[H,V](\xi) \; = \; \f{ H^{\prime}(\xi) }{ V^{\prime\prime}(\xi) }
 \cdot \mathpzc{Q}_{\ell}\Bigg(  \bigg\{ \, \f{H^{(\ell+1)}(\xi)}{H^{\prime}(\xi)}\,\frac{u_{\ell+1}}{u_1} \bigg\}_{\ell};
 \bigg\{  \, \f{V^{(\ell+2)}(\xi)}{V^{\prime\prime}(\xi) }\,\frac{u_{\ell+1}}{u_1} \bigg\}_{\ell} \Bigg) \;. 
\enq
Also, in the case of a localisation of the variables around $b_N$, we have:
\bem
\label{8j1}\mc{G}_{\e{bk};k}[H,V]( b_N-N^{-\a}x, b_N  -N^{-a}y)  \; = \; 
\f{ H^{\prime}(b_N) }{ V^{\prime\prime}(b_N) } \sul{\ell=1}{k} N^{-\ell\a}\cdot \mathpzc{Q}_{\ell}\Bigg(  \bigg\{ \frac{H^{(p + 1)}(b_N)}{H^{\prime}(b_N)}\,\frac{\mathfrak{u}_{p}(x)}{u_1}\bigg\}\,;\,  \bigg\{ \frac{V^{(p + 2)}(b_N)}{V^{\prime\prime}(b_N)}\,\frac{\mathfrak{u}_{p}(x)}{u_1}\bigg\} \Bigg)  \\
\, - \, (x \leftrightarrow y)  + \; \e{O}\Bigg( \f{x^k+y^k+1}{ N^{(k+1) \a} }  \Bigg) \;. 
\end{multline}
Finally, we also have the expansion,
\beq
\label{8j2}\mc{G}_{R;k}^{(\e{as})}[H,V](x,y ;  b_N-N^{-\a}x, b_N  -N^{-\a}y)  \; = \; 
\sul{\ell=1}{k} \f{ \mf{G}_{R;\ell}[H,V](x,y) }{ N^{\a \ell} } \; + \; \e{O}\Bigg( \f{ \ex{-c\min(x,y)} }{ N^{(k+1)\a} } \Bigg)
\enq
where $\mf{G}_{R;\ell}[H,V](x,y)  \; = \; \mf{g}_{R;\ell}[H,V](x)- \mf{g}_{R;\ell}[H,V](y)$ and 
\bem
\label{gggg2}\mf{g}_{R;\ell }[H,V](x) \; = \; \f{1}{ u_1 V^{\prime\prime}(b_N) }\sul{ \substack{ m+s=\ell \\  m,s \geq 0} }{}
\mathpzc{P}_{m}\Bigg(  \bigg\{ \frac{V^{(q + 2)}(b_N)}{V^{\prime\prime}(b_N)}\,\frac{\mathfrak{u}_{q}(x)}{u_1}\bigg\}_{q}\Bigg)  \cdot  \frac{ H^{(s + 1)}(b_N)}{H^{\prime}(b_N)}\cdot\mathfrak{b}_{s}(x) \\
\; - \; \f{ H^{\prime}(b_N)  }{ u_1 \big[ V^{\prime\prime}(b_N) \big]^2 } \sul{ \substack{ m+s+p=\ell \\  m,s ,p \geq 0} }{}
\mathpzc{P}_{m}\Bigg(  \bigg\{\frac{V^{(q + 2)}(b_N)}{V^{\prime\prime}(b_N)}\,\frac{\mathfrak{u}_{q}(x)}{u_1}\bigg\}_{q}\Bigg) \\
\times
\mathpzc{Q}_{p}\Bigg(  \bigg\{ \frac{H^{(q + 1)}(b_N)}{H^{\prime}(b_N)}\,\mathfrak{a}_{q}(x)\bigg\}_{q}\,;\,\bigg\{\frac{V^{(q + 2)}(b_N)}{V^{\prime\prime}(b_N)}\,\mathfrak{a}_{q}(x)\bigg\}_{q}\Bigg)\cdot\frac{ V^{(s + 2)}(b_N)}{V^{\prime\prime}(b_N)}\cdot\mathfrak{b}_{s}(x)\;.
\end{multline}

We can now come back to the double integral $\mathfrak{I}_{{\rm d}}$. It has been decomposed in Lemma~\ref{doubdub}. If we want a remainder $\Delta_{[k]}\mathfrak{I}_{{\rm d}}$ decaying with $N$, we should take $k = 6$ in Lemma~\ref{Proposition estimation reste integrale double}. Then, up to $\e{O}(N^{-\a})$, we are thus left with operators $\mathfrak{I}_{{\rm d};{\rm bk}}$ and $\mathfrak{I}_{{\rm d};R}$, and Lemmas~\ref{kpkpk} and \ref{Proposition Asympt Integrale R vs fct qcq} describe for us their asymptotic expansion knowing the asymptotic expansion of the functions to which they are applied. Here, they are applied to the various functions involving $\mc{G}_{{\rm bk};k}$ and $\mc{G}_{{\rm R};k}^{(\e{as})}$ whose expansion has been described in \eqref{8j1} and \eqref{8j2}. As these expression shows, in order to get $\mathfrak{I}_{{\rm d}}$ up to $\e{O}(N^{-\a})$, one just need the expressions of $\mf{g}_{\e{bk};0 }[H,V](\xi)$ from \eqref{gggg} and  $\mf{g}_{R;1 }[H,V](x)$ from \eqref{gggg2}. These only involve the universal polynomials $\mathpzc{P}_{1}$ and $\mathpzc{Q}_{1}$, whose expression follows from their definitions in \eqref{PlQl}:
\beq
\mathpzc{P}_{1}\big(\{f_1\}\big) \, = -f_1\,\qquad \mathpzc{Q}_{1}\big(\{g_1\};\{f_1\}\big) \, = \, g_1-f_1 \;. 
\enq
Therefore, we get
\beq
\mf{g}_{\e{bk};1 }[H,V](\xi) \; = \; \f{ H^{\prime}(\xi) }{ V^{\prime\prime}(\xi) }
 \qquad \e{and} \qquad 
\mf{g}_{R;1 }[H,V](x) \; =  \; \f{ \mf{b}_1(x)\, - \, \mf{a}_1(x)\mf{b}_0(x) }{ u_1 }\cdot\Bigg( \f{ H^{\prime} }{ V^{\prime\prime} } \Bigg)^{\prime}(b_N) \;. 
\enq
and we recognize in the prefactor of the second equation the function $\mathfrak{c}(x)$ of Definition~\ref{ralph}. Finally, we remind that we take the remainder at order $k = 6$. The claim then follows upon recognising the constant $\aleph_0$ from Definition~\ref{gim} in the computation of the leading term by Lemma~\ref{Proposition Asympt Integrale R vs fct qcq}. \qed

%%%%%%%%%%%%%%%%%%%%%%%%%%%%%%%%%%%%%%%%%%%%%%%%%%%%%%%%%%%%%%%%%%%%%%%%%%%%%%%%%%%%%%%%%%%%%%%%%%%%%%%%%%%%%%%%%%%%%%%%%%%%%%%%%%%%%%%%%%%%%%%%%%%%%%%%%%%%%%%%%%%%%%%%%%%
%%%%%%%%%%%%%%%%%%%%%%%%%%%%%%%%%%%%%%%%%%%%%%%%%%%%%%%%%%%%%%%%%%%%%%%%%%%%%%%%%%%%%%%%%%%%%%%%%%%%%%%%%%%%%%%%%%%%%%%%%%%%%%%%%%%%%%%%%%%%%%%%%%%%%%%%%%%%%%%%%%%%%%%%%%%

%%%%%%%%%%%%%%%%%%%%%%%%%%%%%%%%%%%%%%%%%%%%%%%%%%%%%%%%%%%%%%%%%%%%%%%%%%%%%%%%%%%%%%%%%%%%%%%%%%%%%%%%%%%%%%%%%%%%%%%%%%%%%%%%%%%%%%%%
%%%%%%%%%%%%%%%%%%%%%%%%%%%%%%%%%%%%%%%%%%%%%%%%%%%%%%%%%%%%%%%%%%%%%%%%%%%%%%%%%%%%%%%%%%%%%%%%%%%%%%%%%%%%%%%%%%%%%%%%%%%%%%%%%%%%%%%%

%%%%%%%%%%%%%%%%%%%%%%%%%%%%%%%%%%%%%%%%%%%%%%%%%%%%%%%%%%%%%%%%%%%%%%%%%%%%%%%%%%%%%%%%%%%%%%%%%%%%%%%%%%%%%%%%%%%%%%%%%%%%%%%%%%%%%%%%
%%%%%%%%%%%%%%%%%%%%%%%%%%%%%%%%%%%%%%%%%%%%%%%%%%%%%%%%%%%%%%%%%%%%%%%%%%%%%%%%%%%%%%%%%%%%%%%%%%%%%%%%%%%%%%%%%%%%%%%%%%%%%%%%%%%%%%%%

%%%%%%%%%%%%%%%%%%%%%%%%%%%%%%%%%%%%%%%%%%%%%%%%%%%%%%%%%%%%%%%%%%%%%%%%%%%%%%%%%%%%%%%%%%%%%%%%%%%%%%%%%%%%%%%%%%%%%%%%%%%%%%%%%%%%%%%%
%%%%%%%%%%%%%%%%%%%%%%%%%%%%%%%%%%%%%%%%%%%%%%%%%%%%%%%%%%%%%%%%%%%%%%%%%%%%%%%%%%%%%%%%%%%%%%%%%%%%%%%%%%%%%%%%%%%%%%%%%%%%%%%%%%%%%%%%

%%%%%%%%%%%%%%%%%%%%%%%%%%%%%%%%%%%%%%%%%%%%%%%%%%%%%%%%%%%%%%%%%%%%%%%%%%%%%%%%%%%%%%%%%%%%%%%%%%%%%%%%%%%%%%%%%%%%%%%%%%%%%%%%%%%%%%%%
%%%%%%%%%%%%%%%%%%%%%%%%%%%%%%%%%%%%%%%%%%%%%%%%%%%%%%%%%%%%%%%%%%%%%%%%%%%%%%%%%%%%%%%%%%%%%%%%%%%%%%%%%%%%%%%%%%%%%%%%%%%%%%%%%%%%%%%%

\appendix
 
%%%%%%%%%%%%%%%%%%%%%%%%%%%%%%%%%%%%%%%%%%%%%%%%%%%%%%%%%%%%%%%%%%%%%%%%%%%%%%%%%%%%%%%%%%%%%%%%%%%%%%%%%%%%%%%%%%%%%%%%%%%%%%%%%%%%%%%% 
%%%%%%%%%%%%%%%%%%%%%%%%%%%%%%%%%%%%%%%%%%%%%%%%%%%%%%%%%%%%%%%%%%%%%%%%%%%%%%%%%%%%%%%%%%%%%%%%%%%%%%%%%%%%%%%%%%%%%%%%%%%%%%%%%%%%%%%%

%%%%%%%%%%%%%%%%%%%%%%%%%%%%%%%%%%%%%%%%%%%%%%%%%%%%%%%%%%%%%%%%%%%%%%%%%%%%%%%%%%%%%%%%%%%%%%%%%%%%%%%%%%%%%%%%%%%%%%%%%%%%%%%%%%%%%%%%
%%%%%%%%%%%%%%%%%%%%%%%%%%%%%%%%%%%%%%%%%%%%%%%%%%%%%%%%%%%%%%%%%%%%%%%%%%%%%%%%%%%%%%%%%%%%%%%%%%%%%%%%%%%%%%%%%%%%%%%%%%%%%%%%%%%%%%%%

%%%%%%%%%%%%%%%%%%%%%%%%%%%%%%%%%%%%%%%%%%%%%%%%%%%%%%%%%%%%%%%%%%%%%%%%%%%%%%%%%%%%%%%%%%%%%%%%%%%%%%%%%%%%%%%%%%%%%%%%%%%%%%%%%%%%%%%%
%%%%%%%%%%%%%%%%%%%%%%%%%%%%%%%%%%%%%%%%%%%%%%%%%%%%%%%%%%%%%%%%%%%%%%%%%%%%%%%%%%%%%%%%%%%%%%%%%%%%%%%%%%%%%%%%%%%%%%%%%%%%%%%%%%%%%%%%

%%%%%%%%%%%%%%%%%%%%%%%%%%%%%%%%%%%%%%%%%%%%%%%%%%%%%%%%%%%%%%%%%%%%%%%%%%%%%%%%%%%%%%%%%%%%%%%%%%%%%%%%%%%%%%%%%%%%%%%%%%%%%%%%%%%%%%%%
%%%%%%%%%%%%%%%%%%%%%%%%%%%%%%%%%%%%%%%%%%%%%%%%%%%%%%%%%%%%%%%%%%%%%%%%%%%%%%%%%%%%%%%%%%%%%%%%%%%%%%%%%%%%%%%%%%%%%%%%%%%%%%%%%%%%%%%%

%%%%%%%%%%%%%%%%%%%%%%%%%%%%%%%%%%%%%%%%%%%%%%%%%%%%%%%%%%%%%%%%%%%%%%%%%%%%%%%%%%%%%%%%%%%%%%%%%%%%%%%%%%%%%%%%%%%%%%%%%%%%%%%%%%%%%%%%
%%%%%%%%%%%%%%%%%%%%%%%%%%%%%%%%%%%%%%%%%%%%%%%%%%%%%%%%%%%%%%%%%%%%%%%%%%%%%%%%%%%%%%%%%%%%%%%%%%%%%%%%%%%%%%%%%%%%%%%%%%%%%%%%%%%%%%%%

\chapter{Several theorems and properties of use to the analysis}
\label{Appendix Reminder on several useful theorems}

\begin{theorem}[Hunt, Muckenhoupt, Wheeden \cite{HuntMuckenhouptWheedenBoundednessHilbertTransformOnWeightedSpaces}]
\label{Theorem conte transfo Cauchy sur Fourier espaces Hs} The Hilbert transform, defined as an operator $$\mc{H} \, : \, L^{2}(\R, w(x) \dd x ) \; \tend \; L^{2}(\R, w(x) \dd x ) $$ is bounded if and
only if there exists a constant $C > 0$ such that, for any interval $I \subseteq \mathbb{R}$:
\beq
\bigg\{ \f{1}{|I|} \Int{ I }{} w(x)\dd x \bigg\} \cdot \bigg\{  \f{1}{|I|} \Int{ I }{} \frac{\dd x}{w(x)}  \bigg\} \;  < \; C 
\label{Condition for continuity of the Hilbert transform}
\enq
In particular, the operators "upper/lower boundary values" $\mc{C}_{\pm} \; :  \; \mc{F}\big[ H_{s}(\R) \big] \; \tend \; \mc{F}\big[ H_{s}(\R) \big]$
are bounded if and only if $|s|<1/2$.

\end{theorem}

A less refined version of this theorem takes the form :

\begin{prop}
 \label{Proposition continuite operateurs cauchy shiftes}

 For any $\ga>0$, the shifted Cauchy operators  $\mc{C}_{\ga} : f \mapsto \mc{C}_{\ga}[f]$ with $\mc{C}_{\ga}[f](\la) = \mc{C}[f](\la+\i\ga)$ are continuous on 
 $\mc{F}\big[ H_{s}(\R) \big]$  with $|s|<1/2$. 
 
\end{prop}

\begin{theorem}[Calderon \cite{CalderonContinuityCauchyTransformLipschitzCurves}]
\label{Theorem conte transfo Cauchy sur courbes Lipschitz} 
Let $\Sg$ be a non-self intersecting Lipschitz curve in $\Cx$ and $C_{\Sg}$ the Cauchy transform on $L^{2}(\Sg,\dd s)$:
\beq
\forall f \in L^{2}(\Sg,\dd s) \qquad \mc{C}_{\Sg}[f](z) \; = \; \Int{ \Sg }{} \f{ f(s)  }{s-z} \cdot \f{ \dd s }{2\i \pi }  \in \mc{O}\big( \Cx \setminus \Sg \big) \;. 
\enq
For any $f \in L^{2}(\Sg,\dd s)$, $\mc{C}_{\Sg}[f]$ admits $ L^{2}(\Sg,\dd s)$ $\pm$ boundary values $\mc{C}_{\Sg;{\pm}}[f]$. 
The operators $\mc{C}_{\Sg;{\pm}}[f]$ are continuous operators on $ L^{2}(\Sg,\dd s)$ which, furthermore, satisfy $\mc{C}_{\Sg;+}-\mc{C}_{\Sg;-}=\e{id}$. 
\end{theorem}

\begin{theorem}[Paley, Wiener \cite{PaleyWienerTFinC}]
\label{PaleyWie}%
Let $u \in L^2(\R^{\pm})$. Then $\mc{F}[u]$ is the $L^2(\R)$
boundary value on $\R$ of a function $\wh{u}$ that is holomorphic on $\mathbb{H}^{\pm}$, and there exists a constant $C > 0$ such that:%
\beq
\forall \mu > 0,\qquad \Int{\R}{} \big|\wh{u}(\la \pm \i \mu) \big|^2 \cdot \dd \la \; < \; C
\label{ecriture borne norme Hardy}
\enq
Reciprocally, every holomorphic function on $\wh{u}$ on $\mathbb{H}^{\pm}$ that satisfies the bounds \eqref{ecriture borne norme Hardy} and admits $L^{2}(\R)$ $\pm$ boundary values $\wh{u}_{\pm}$ on $\R$, is the Fourier transform of a function $u \in L^2(\R^{\pm})$, \textit{viz}. $\wh{u}(z)=\mc{F}[u](z)$, $z\in \mathbb{H}^{\pm}$. 

\end{theorem}

%%%%%%%%%%%%%%%%%%%%%%%%%%%%%%%%%%%%%%%%%%%%%%%%%%%%%%%%%%%%%%%%%%%%%%%%%%%%%%%%%%%%%%%%%%%%%%%%%%%%%%%%%%%%%%%%%%%%%%%%%%%%%%%%%%%%%%%% 
%%%%%%%%%%%%%%%%%%%%%%%%%%%%%%%%%%%%%%%%%%%%%%%%%%%%%%%%%%%%%%%%%%%%%%%%%%%%%%%%%%%%%%%%%%%%%%%%%%%%%%%%%%%%%%%%%%%%%%%%%%%%%%%%%%%%%%%%

%%%%%%%%%%%%%%%%%%%%%%%%%%%%%%%%%%%%%%%%%%%%%%%%%%%%%%%%%%%%%%%%%%%%%%%%%%%%%%%%%%%%%%%%%%%%%%%%%%%%%%%%%%%%%%%%%%%%%%%%%%%%%%%%%%%%%%%%
%%%%%%%%%%%%%%%%%%%%%%%%%%%%%%%%%%%%%%%%%%%%%%%%%%%%%%%%%%%%%%%%%%%%%%%%%%%%%%%%%%%%%%%%%%%%%%%%%%%%%%%%%%%%%%%%%%%%%%%%%%%%%%%%%%%%%%%%

%%%%%%%%%%%%%%%%%%%%%%%%%%%%%%%%%%%%%%%%%%%%%%%%%%%%%%%%%%%%%%%%%%%%%%%%%%%%%%%%%%%%%%%%%%%%%%%%%%%%%%%%%%%%%%%%%%%%%%%%%%%%%%%%%%%%%%%%
%%%%%%%%%%%%%%%%%%%%%%%%%%%%%%%%%%%%%%%%%%%%%%%%%%%%%%%%%%%%%%%%%%%%%%%%%%%%%%%%%%%%%%%%%%%%%%%%%%%%%%%%%%%%%%%%%%%%%%%%%%%%%%%%%%%%%%%%

%%%%%%%%%%%%%%%%%%%%%%%%%%%%%%%%%%%%%%%%%%%%%%%%%%%%%%%%%%%%%%%%%%%%%%%%%%%%%%%%%%%%%%%%%%%%%%%%%%%%%%%%%%%%%%%%%%%%%%%%%%%%%%%%%%%%%%%%
%%%%%%%%%%%%%%%%%%%%%%%%%%%%%%%%%%%%%%%%%%%%%%%%%%%%%%%%%%%%%%%%%%%%%%%%%%%%%%%%%%%%%%%%%%%%%%%%%%%%%%%%%%%%%%%%%%%%%%%%%%%%%%%%%%%%%%%%

%%%%%%%%%%%%%%%%%%%%%%%%%%%%%%%%%%%%%%%%%%%%%%%%%%%%%%%%%%%%%%%%%%%%%%%%%%%%%%%%%%%%%%%%%%%%%%%%%%%%%%%%%%%%%%%%%%%%%%%%%%%%%%%%%%%%%%%%
%%%%%%%%%%%%%%%%%%%%%%%%%%%%%%%%%%%%%%%%%%%%%%%%%%%%%%%%%%%%%%%%%%%%%%%%%%%%%%%%%%%%%%%%%%%%%%%%%%%%%%%%%%%%%%%%%%%%%%%%%%%%%%%%%%%%%%%%

\chapter{Proof of Theorem~\ref{Proposition Asympt non rescaled part fct}  }
\label{Appendix Section preuve asympt dom part fct unrescaled}

We denote by $\mf{p}_N$ the rescaled probability density on $ \R^N$  associated with $\mf{z}_N$, namely 
\beq
\mf{p}_N ( \bs{\la} ) \;  =  \;  \f{ N^{\a_q N} }{ \mf{z}_N[W] } 
\pl{a<b}{N} \Big\{ \sinh\big[\pi\om_1 N^{\a_q} (\la_a-\la_b)\big] \sinh\big[\pi\om_2 N^{\a_q} (\la_a- \la_b)\big]  \Big\}^{\be}
\cdot \pl{a=1}{N}\ex{- W( N^{\a_q} \la_a) } \qquad \e{with} \quad \a_q \, = \, \f{1}{q-1} \;. 
\nonumber
\enq
To obtain the above probability density, we have rescaled in the variables in \eqref{ecriture MI qSoV comme exemple} as $y_a=N^{\a_q} \la_a$ with the value of $\a_q$ guided by the 
heuristic arguments that followed the statement of Theorem~\ref{Proposition Asympt non rescaled part fct}. 
We shall denote by $\mc{P}_N$ the probability measure on $\mc{M}^{1}(\R)$ induced by $\mf{p}_N$, \textit{viz}. the measurable sets in $\mc{M}^{1}(\R)$
are generated by the Borel $\sigma$-algebra for the weak topology, and for any open subset in $\mc{M}^{1}(\R)$, we have: 
\beq
\mc{P}_N \big[ O \big] \; = \; \Int{\{L^{(\bs{\la})}_N \in O\}}{}  \mf{p}_N(\bs{\la})\,\dd^N \bs{\la} \;.  
\enq
The strategy of the proof consists in proving that $\mc{P}_N$ is exponentially tight and then establishing a weak 
large deviation principle, namely upper and lower bounding $\mc{P}_N$ on balls of shrinking radius this for balls relatively to 
the bounded Lipschitz topology, see \textit{e.g.} \cite{AndersonGuionnetZeitouniIntroRandomMatrices}.

\subsection*{Exponential tightness}

\begin{lemme}
\label{tighti}The sequence of measures ${\cal P}_{N}$ is exponentially tight, \textit{i.e.}:
\beq
\limsup_{L \rightarrow +\infty} \limsup_{N \rightarrow \infty} N^{-(2 + \a_q)}\,\ln {\cal P}_{N}\big[K_{L}^{c}\big] = -\infty\;.
\enq
where $K_L = \big\{ \mu \in \mc{M}^1(\R )  \; : \;   \Int{\R}{} \abs{x}^q \dd \mu(x)  \leq L  \big\}$.
\end{lemme}

\Proof  By the monotone convergence theorem, 

\beq
%]
\Int{\R}{} \abs{x}^q\,\dd \mu(x)  = \sup_{ M \in \mathbb{N} }  \Int{\R}{} \min(\abs{x}^q,M)\,\dd \mu(x).
\enq
The left-hand side is lower semi-continuous as a supremum of 
a continuous family of functionals on $\mc{M}_1\pa{\R}$. Thus, $K_L$ is closed as a level set of a lower semi-continuous function. 
For any $\mu \in K_L$, we have by Chebyshev inequality:
\beq
\mu\big[\intff{-M}{M}^{c}\big] \; \leq  \; \f{1}{M^q} \hspace{-3mm} \Int{ \intff{-M}{M}^{c} }{}  \hspace{-4mm} \abs{x}^q \dd \mu (x) \leq \f{L}{M^q} \;. 
\enq
As a consequence, 
\beq
K_L \subseteq \bigcap_{ M\in \mathbb{N} }  \bigg\{ \mu \in \mc{M}^1(\R) \; : \; 
\mu\big[\intff{-M}{M}^{c}\big] \leq \f{L}{M^q}  \bigg\} \;. 
\enq
The right-hand side is uniformly tight, by construction and is closed as an intersection of level sets of lower semi-continuous
functions on $\mc{M}^1(\R)$. Thence by Prokhorov theorem, it is compact. As $K_L$ is closed, it must be as well compact.

We now estimate $\mc{P}_N\big[  K_L^{c} \big]$. We start by a rough estimate for the partition function. 
It follows by Jensen inequality applied to the probability measure of $\R^N$
\beq
\pl{a=1}{N} \f{ \ex{-W(\la_a)}  \dd\la_a }{ \Int{\R}{} \ex{-W(\la)}\dd \la }  \;, 
\enq
that
\bem
 \ln \big[  \mf{z}_N[W] \big] \geq  N \ln \Big[ \Int{}{} \ex{-W(\la)}\dd \la \Big] 
\; + \; \Int{\R^N}{} \sul{a<b}{} \be \ln \big\{ \sinh\big[ \pi\om_1 (\la_a-\la_b)\big] \sinh\big[\pi\om_2  (\la_a-\la_b)\big]  \big\}
\pl{a=1}{N} \f{ \ex{-W(\la_a)}  \dd\la_a }{ \Int{}{} \ex{-W(\la)}\dd \la } \\
%
%
%\leq - N \ln \Big[ \Int{}{} \ex{-V(\la)}\dd \la \Big] 
%
%\; - \; \sul{a<b}{} c  \Int{\R^N}{} \ln \abs{\la_a-\la_b)} 
%
%\pl{a=1}{N} \f{ \ex{-V(\la_a)}  \dd\la_a }{ \Int{}{} \ex{-V(\la)}\dd \la }  \\
%
%
\geq   N \ln \Big[ \Int{}{} \ex{-W(\la)}\dd \la \Big]  +
\f{\beta N(N-1)}{2} \Int{\R^2}{}  \ln \big\{ \sinh\big[\pi\om_1 (\la_1-\la_2)\big] \sinh\big[\pi\om_2  (\la_1-\la_2)\big]  \big\}  
\cdot \f{ \ex{-W(\la_1)-W(\la_2) } \dd\la_1\dd\la_2}{ \Bigg(  \Int{}{} \ex{-W(\la)}\dd \la   \Bigg)^2  }
\end{multline}
As a consequence, $ \mf{z}_N[W] \geq \ex{-N^2 \kappa}$ for some $\kappa \in \R$. It now remains to estimate the integral arising from the integration over $K_L^{\e{c}}$.  Using that $\abs{\s{\la}} \leq \ex{ \abs{\la} }$  we get:
\bem
\pl{a<b}{N} \Big\{ \sinh\big[\pi\om_1 N^{\a_q}(\la_a-\la_b)\big] \sinh\big[\pi\om_2 N^{\a_q} (\la_a-\la_b)\big]  \Big\}^{\be} \;  \leq \;
\pl{a<b}{N} \exp\Big\{ \pi \be (\om_1+\om_2) N^{\a_q} |\la_a-\la_b |  \Big\} \\
 \leq  \pl{a<b}{N} \exp\Big\{ \pi \be (\om_1+\om_2) N^{\a_q} \big( |\la_a| +|\la_b | \big)  \Big\}  \; \leq \; 
\pl{a=1}{N}  \exp\Big\{ \pi \be (\om_1+\om_2) N^{\a_q+1}  |\la_a|  \Big\}  \;. 
\end{multline}
Hence,
\beq
\mc{P}_N \big[ K_L^{\e{c}} \big] \leq \ex{\kappa N^2} N^{\a_q N} \Int{  \big\{ L_N^{(\bs{\la})} \in K_L^{\e{c}}\big\} }{} 
\pl{a=1}{N} \exp\Big\{ \pi \be (\om_1+\om_2) N^{\a_q+1} \abs{\la_a} - W(N^{\a_q}\la_a) \Big\}\cdot \dd^N \bs{\la}
\enq
Since $ \abs{\xi}^{1-q}  \limit{\abs{\xi}}{+\infty} 0$ there exists a constant $C \in \R$ such that 
\beq
\label{fqeu}\forall \xi \in \mathbb{R},\qquad \pi \be (\om_1+\om_2)|\xi| \,  \leq \, \f{c_q \abs{\xi}^q}{2} + C  \, .
\enq
Likewise it follows from \eqref{conditon cptmt asympt V}
that given any $\eps >0$ there exists $\tau_{\eps} \in \R^+$ such that 
\beq
\forall \xi \in \R,\qquad -c_q (1+\eps) \abs{\xi}^q  - \tau_{\eps} \, \leq \,  -W(\xi)  \, \leq \,  -c_q (1 - \eps) \abs{\xi}^q  + \tau_{\eps}\;.%
\label{ecriture bornage inf et sup du potentiel}
\enq
In the following, $\eps$ will be taken small. Taking into account that $q\a_q = \a_q +1$, \eqref{fqeu} and the upper bound of \eqref{ecriture bornage inf et sup du potentiel} lead to:
\begin{eqnarray}
\mc{P}_N\big[ K_L^{\e{c}} \big] & \leq & \ex{\kappa N^2}N^{\a_q N} \Int{ \big\{ L_N^{(\bs{\la})} \in K_L^{\e{c}}\big\} }{} 
\pl{a=1}{N} \exp\Bigg\{ N^{\a_q+1} C  \; +  \; \f{c_q}{2} N^{\a_q+1}  \abs{\la_a}^q + \tau_{\eps} \; - \; 
 c_q(1-\eps) N^{q\a_q}  \abs{\la_a}^q \Bigg\} \cdot \dd^N \bs{\la}  \nonumber \\
& \leq & N^{\a_q N}  \ex{\kappa N^2 + C N^{2+\a_q} + \tau_{\eps}N  } 
\hspace{-5mm}
\Int{ \big\{ L_N^{(\bs{\la})} \in K_L^{\e{c}}\big\} }{} 
\Bigg(\pl{a=1}{N} \ex{- \eps c_q N^{\a_q+1}  \abs{\la_a}^q }\Bigg) 
\exp \Bigg\{-\f{c_{q}(1-4\eps)}{2} N^{2+\a_q} \Int{\R}{} \abs{x}^q  \dd L_N^{(\bs{\la})}(x) \Bigg\}\cdot\dd^N \bs{\la} \nonumber \\
 & \leq & N^{\a_q N}  \ex{C^{\prime} N^{2+\a_q} + \tau_{\eps}N  -[c_q(1-4\eps)/2]LN^{2+\a_q}}
\Bigg( \Int{ \R  }{}    \ex{- \eps c_q   \abs{\la_a}^q } \dd \la \Bigg)^N 
\end{eqnarray}
for some constant $C^{\prime} > C$ and $N$ large enough. As a consequence, 
\beq
\limsup_{N \tend +\infty} N^{-(2+\a_q)}\,\ln \mc{P}_{N}\big[ K_L^{\e{c}} \big] \leq  C - L c_q(1-4\eps)/2 \; ,
\nonumber
\enq
and this upper bound goes to $-\infty$ when $L \rightarrow +\infty$. \qed

\subsection*{Lower bound}

In the following we focus on the renormalised measure on $\mc{M}_1(\R)$ defined as $\ov{ \mc{P} }_N \, =\,  \mf{z}_N[W] \cdot \mc{P}_N$.  
We will now derive a lower bound for the $\ov{\cal P}_{N}$ volume of small Vasershtein balls, in terms of the energy functional ${\cal E}_{({\rm ply})}$ 
of \eqref{ecriture fct taux pour LDP de zN ac pot ply}, namely:
\beq
\label{Regsaf}\mc{E}_{(\e{ply})}[ \mu ] \; = \; \Int{}{} E(\xi,\eta)\,\dd\mu(\xi)\dd\mu(\eta),\qquad E(\xi,\eta) = \frac{c_q}{2}\big(\abs{\xi}^q + \abs{\eta}^{q}\big) - \f{ \be \pi (\om_1+\om_2) }{ 2 }\,\abs{\xi - \eta}
\enq

\begin{lemme}
\label{lowerbou} Let $B_{\delta}(\mu)$ be the ball in $\mc{M}^{1}(\R)$ centred at $\mu$ and of radius $\de$ with respect to $D_V$. Then, for any $\mu \in {\cal M}^{1}(\R)$, it holds
\beq
\label{desire}\liminf_{\delta \rightarrow 0} \liminf_{N \rightarrow \infty} N^{-(2 + \a_q)}\,\ln \ov{\mc{P}}_{N}\big[B_{\delta}(\mu)\big] \; \geq \;  -{\cal E}_{({\rm ply})}[\mu]
\enq
\end{lemme}

\Proof Let $\mu \in \mc{M}^1(\R)$ and $\de>0$. If $\int |x|^q\,\dd \mu(x) \, = \,  + \infty$, then $\mc{E}_{(\e{ply})}[\mu]=+\infty$ and there is nothing to prove. 
Thus we may assume from the very beginning that $\int |x|^q\,\dd \mu(x) < + \infty$. If $M > 0$ is large enough, we have $\mu(\intff{-M}{M}) \neq 0$, and we can introduce:
\beq
\mu_{M} \; = \;  \frac{\mathbf{1}_{\intff{-M}{M}}\cdot \mu}{\mu\big(\intff{-M}{M}\big)}
\enq
which is now a compactly supported measure. We will obtain the lower bound for $\ov{\mc{P}}_{N}\big[B_{\delta}(\mu)\big]$ by restricting to configurations close enough to the classical positions of $\mu_M$, 
and only at the end, see how the estimate behaves when $M \rightarrow \infty$. For any given integer $N$, we define:
\beq
\forall a \in \llbracket 1,N \rrbracket,\qquad  x_{a}^{N,M} =  \inf \Bigg\{x \in \R\; :\quad \Int{-\infty}{x} \dd\mu_M \geq \frac{a}{N + 1}\Bigg\} .%
\enq
When $N \rightarrow \infty$, $L_N^{(\bs{x}^{N,M})}$ approximates $\mu_{M}$ for the Vasershtein distance, so there exists $N_{\delta}$ such that, for any $N \geq N_{\delta}$, we have the inclusion:
\beq
\Om_{\de} := \bigg\{  \bs{\la} \in \R^N \; : \;\;\forall a \in \llbracket 1,N \rrbracket,\;\abs{\la_a - x_a^{N,M}} < \de/2 \bigg\}
\;\subseteq \; \bigg\{  \bs{\la} \in \R^N \; : \;  D_{V}\big(\mu_M, L_N^{(\bs{\la})} \big) < \de  \bigg\} \;. 
\enq
Subsequently:
\beq
\ov{ \mc{P} }_N \big[ B_{\de}(\mu_M)  \big] \; \geq  \;  N^{N\a_q}  \Int{\Om_{\de}}{}  
\pl{a<b}{N} \bigg\{ \sinh\big[\pi\om_1 N^{\a_q} (\la_a-\la_b) \big] \sinh\big[\pi\om_2 N^{\a_q} \abs{\la_a-\la_b}\big]  \bigg\}^{\be}
\pl{a=1}{N} \ex{-W(N^{\a_q} \la_a)}\cdot \dd^{N} \bs{\la}\;. 
\enq
It follows from the lower bound 
\beq
\abs{\s{x}} \geq  \f{ \ex{\abs{x}} }{ 2 }  \f{ \abs{x}}{1+\abs{x}}  
\enq
from the lower bound for $W$ in \eqref{ecriture bornage inf et sup du potentiel}, and $q\a_q = \a_q + 1$, that:
\beq
%
%\ov{ \mc{P} }_N \Big[ B_{2\de}( \wt{\mu} )  \Big]%
\ov{ \mc{P} }_N \big[ B_{\de}(\mu)  \big] \geq  \f{e^{N(\a_q\ln N + \tau_{\epsilon})}}{ 2^{\beta N(N-1) }}  \Int{\Om_{\de}}{} 
\exp\Bigg\{ \pi \be (\om_1+\om_2)  N^{\a_q} \sul{a<b}{N} \abs{\la_a-\la_b}  -  N^{\a_q + 1}c_q (1 + \epsilon)\sul{a=1}{N} \abs{\la_a}^{q}  \Bigg\}
\pl{a<b}{N} \Big\{ g_N (\la_a-\la_b)  \Big\}^{\be}
\cdot \dd^{N}\bs{\la} \;.
\enq
where we have set 
\beq
g_N(\la) =  \f{ \pi \om_1 N^{\a_q} \abs{\la}  }{ 1+ \pi \om_1 N^{\a_q} \abs{\la}  } \cdot 
\f{ \pi \om_2 N^{\a_q} \abs{\la}  }{ 1+ \pi \om_2 N^{\a_q} \abs{\la}  }
\enq
Now, we would like to replace $\la_{a}$ by $x_a^{N,M}$. Since the configurations $\bs{\la} \in \Om_{\de}$ satisfy $|x_a^{N,M} - \la_a| < \delta/2$, we have:
\beq
\sum_{a < b} \abs{\la_a - \la_b} \; \geq \; -  N(N - 1) \f{\delta}{2} \;  + \;  \sum_{a < b} (x_{b}^{N,M} - x_{a}^{N,M})
\enq
Since $q > 1$, we also deduce from the mean value theorem:
\beq
\abs{\la_a}^{q} \leq \abs{x_a^{N,M}}^{q} + \frac{q\delta}{2}\big(|x_a^{N,M}| + \delta/2\big)^{q - 1}
\enq
and thus
\beq
\label{B23} -(1+\eps) |\la_{a}|^q  \; = \; - (1 + \epsilon) \,\abs{x_a^{N,M}}^{q} \, + \, 
\f{\delta}{c_q} \,h_{\eps, \delta}(x_a^{N,M}) \,\qquad h_{\epsilon,\delta}(x) \, = \,  \f{qc_{q}}{2}(1 + \epsilon)\cdot \big(\abs{x} + \delta/2\big)^{q - 1}
\enq
These inequalities yield the lower bound:
\beq
\ov{\mc{P}}_{N}\big[B_{\delta}(\mu)\big]  \geq \exp\Bigg\{C\,N^2 - N^{2 + \a_q}\Bigg(\delta\Big\{C^{\prime} 
+ \Int{}{} h_{\epsilon,\delta}(\xi)\dd L_{N}^{(\bs{x}^{N,M})}(\xi)\Big\} 
+ {\cal E}_{({\rm ply})}[L_N^{(\bs{x}^{N,M})}] + \epsilon c_q \Int{}{} \abs{\xi}^{q}\,\dd L_N^{(\bs{x}_{N,M})}(\xi)\Bigg)\Bigg\}\cdot G_{N,\delta}
\enq
for some irrelevant, $N$ and $\delta$ independent, constants $C,C^{\prime} > 0$. Furthermore, the factor $G_{N,\delta}$ reads 
\beq
G_{N,\delta} \; = \;  \Int{\Om_{\de}^{{\rm ord}}}{} \prod_{a > b}^{N} \big\{g_{N}(\la_b - \la_a)\big\}^{\beta}\cdot \dd^{N} \bs{\la} 
\enq
in which $\Om_{\de}^{{\rm ord}}=\Om_{\de}\cap \{\bs{\la}\in \R^N \; : \; \la_1<\dots < \la_N \}$.

To find a lower bound for $G_{N,\delta}$, we can restrict further to configurations such that $u_a = \la_a - x_a^{N,M}$ increases with $a \in \llbracket 1,N \rrbracket$, 
and satisfies $|u_1| < \delta/(2N)$ and $|u_{a + 1} - u_{a}| \leq \delta/2N$ for any $a \in \llbracket 1, N - 1 \rrbracket$. Using that $\xi \mapsto g_N(\xi)$ is increasing on $\mathbb{R}_+$, we have:
\beq
G_{N,\delta} \geq \Int{[-\delta/2,\delta/2]^{N}}{} \prod_{a = 1}^{N - 1} \big\{g_{N}(u_{a + 1} - u_{a})\big\}^{\beta(N - a)}\cdot \dd^N u \;\geq \; 
\Int{[0,\delta/2N]^{N}}{} \prod_{a = 2}^N \big\{g_N(v_a)\big\}^{\beta(N - a + 1)} \cdot \dd^N \bs{ v}
\enq
Now, using an arithmetic-geometric upper bound for the denominator in $g_N(v)$, we can write:
\beq
\forall v \in [0,\delta/2N],\qquad g_N(v) \geq \frac{N^{\a_q + 1}\pi\sqrt{\omega_1\omega_2}}{2\delta}\,\cdot |v|^2 \geq \wt{C} N^{\a_q + 1}\cdot |v|^2
\enq
for some irrelevant $C^{\prime} > 0$ independent of $\delta$ provided that $\delta < 1$. So, we arrive to:
\beq
G_{N,\delta} \geq \frac{\delta}{2N}\cdot \big(\wt{C}\,N^{\a_q - 1}\big)^{\beta N(N - 1)/2} \cdot \prod_{a = 2}^{N - 1} \frac{(\delta/2N)^{2\beta(N - a + 1)}}{2\beta(N - a + 1) + 1} \geq e^{\wt{C}^{\prime}\,N^2 \ln N}
\enq
for some $\wt{C}^{\prime} > 0$ independent of $\delta$. Hence, ultimately
\beq
\label{iniugrw}\ov{\mc{P}}_{N}\big[B_{\delta}(\mu)\big] \geq e^{\wt{C}^{\prime\prime} N^2\ln N} 
\exp\Bigg\{-N^{2 + \a_q}\Bigg[\delta\bigg(C^{\prime} + \Int{}{} h_{\epsilon,\delta}(\xi)\,\dd L_N^{(\bs{x}^{N,M})}(\xi)\bigg) 
+ {\cal E}_{({\rm ply})}\big[L_N^{(\bs{x}^{N,M})}\big] + \epsilon c_q \Int{}{} \abs{\xi}^{q}\,\dd L_N^{(\bs{x}^{N,M})}(\xi)\Bigg]\Bigg\}
\enq
To establish the desired result \eqref{desire}, we only need to focus on the last exponential. If $\phi$ is a ${\cal C}^1$ function of $p$ real variables, we denote:
\beq
\label{truncs}\phi^{[M]}(\xi) = \min\Big[\phi(\xi)\,;\,\norm{\phi}_{L^{\infty}(\intff{-M}{M}^{p})}\Big]
\enq
which has the advantage of being bounded and Lipschitz. Since $\mu_M$ is supported on $\intff{-M}{M}$, so must be the classical positions $x_{a}^{N,M}$, 
and we can apply the truncation to all the functions against which $L_N^{(\bs{x}^{N,M})}$ is integrated. In particular, we make appear the truncated functional:
\beq
{\cal E}_{({\rm ply})}^{[M]}[\mu] \;=  \; \Int{}{} E^{[M]}(\xi,\eta)\,\dd\mu(\xi)\dd\mu(\eta) \; . 
\enq
The advantage is that now, all functions to be integrated are Lipschitz bounded. Since, $D_{V}\big(\mu_{M},L_N^{(\bs{x}^{N,M})}\big) \rightarrow 0$ when $N \rightarrow \infty$, we get:
\beq
\liminf_{N \rightarrow \infty} \f{ \ln \ov{\mc{P}}_{N}\big[B_{\de}(\mu)\big] }{ N^{2 + \a_q} }  \; \geq \; 
-\delta\,\bigg(C + \Int{}{} h_{\epsilon,\delta}(\xi)\,\dd\mu_{M}(\xi)\bigg) - {\cal E}_{({\rm ply})}^{[M]}[\mu_{M}] - \epsilon c_q \Int{}{} \big\{\max(|\xi|,M)\big\}^{q}\,\dd\mu_M(\xi) \;. 
\enq
The right-hand is an affine function of $\epsilon$, and at this stage, we can send $\epsilon$ to $0$:
\beq
\liminf_{N \rightarrow \infty} N^{-(2 + \a_q)}\,\ln \ov{\mc{P}}_{N}\big[B_{\de}(\mu_M)\big] \geq - \delta\,\bigg(C + \Int{}{} h_{0,\delta}(\xi)\,\dd\mu_{M}(\xi)\bigg) - {\cal E}_{({\rm ply})}[\mu_M] \;. 
\enq
Now, for any fixed $\delta$, there exists $M_{\delta}$ such that, for any $M \geq M_{\delta}$, $D_{V}(\mu,\mu_M) \leq \delta$, and consequently:
\beq
\liminf_{N \rightarrow \infty} N^{-(2 + \a_q)}\,\ln \ov{\mc{P}}_{N}\big[B_{2\delta}(\mu)\big] \geq -\delta\,\bigg(C + \Int{}{} h_{0,\delta}(\xi)\,\dd\mu_{M}(\xi)\bigg) - {\cal E}_{({\rm ply})}[\mu_M] \;. 
\enq
We could replace ${\cal E}_{({\rm ply})}^{[M]}$ by ${\cal E}_{({\rm ply})}$ here because $\mu_M$ is supported on $\intff{-M}{M}$. Now, we can consider sending $M \rightarrow \infty$. 
Since we have the bound:
\beq
\forall \xi,\eta \in \R,\qquad E(\xi,\eta) \leq C^{\prime}\,\big(1 + \abs{\xi}^{q} + \abs{\eta}^{q}\big),\qquad h_{0,\delta} \leq C^{\prime}\,\big(1 + \abs{\xi}^{q}\big)
\enq
and we assumed that $\int \abs{\xi}^{q}\,\dd\mu(\xi) < +\infty$, we get by dominated convergence:
\beq
\liminf_{N \rightarrow \infty} N^{-(2 + \a_q)}\,\ln \ov{\mc{P}}_{N}\big[B_{2\delta}(\mu)\big] \geq -\delta\,\bigg(C + \Int{}{} h_{0,\delta}(\xi)\,\dd\mu(\xi)\bigg) - {\cal E}_{({\rm ply})}[\mu] \;. 
\enq
Last but not least, sending $\delta \rightarrow 0$, the first term disappears and we find:
\beq
\liminf_{\delta \rightarrow 0} \liminf_{N \rightarrow \infty} N^{-(2+\a_q)}\,\ln \ov{\mc{P}}_{N}\big[B_{2\delta}(\mu)\big] \geq - {\cal E}_{({\rm ply})}[\mu] \;. 
\enq
\qed

\subsection*{Upper bound}

In this paragraph, we complete our estimate by an upper bound on the probability of small Vasershtein balls:

\begin{lemme}
\label{upperbou2}\beq
\limsup_{\delta \rightarrow 0} \limsup_{N \rightarrow \infty} N^{-(2 + \a_q)}\,\ln \ov{\mc{P}}_{N}\big[B_{\delta}(\mu)\big] \leq - {\cal E}_{({\rm ply})}[\mu]
\enq
\end{lemme}

\Proof Let $\mu \in \mc{M}^1(\R)$. In order to establish an upper bound, we use that $\abs{\s{x}} \leq \ex{ \abs{x} } $ and the upper bound in \eqref{ecriture bornage inf et sup du potentiel} for the potential $W$. This makes appear again the function ${\cal E}_{({\rm ply})}$ of \eqref{Regsaf}:
\beq
\ov{\mc{P}}_N  \big[ B_{\de}(\mu)  \big] \leq e^{N(\a_q \ln N + \tau_{\epsilon})} \Int{L_N^{(\bs{\la})} \in B_{\delta}(\mu)}{} 
\exp\Bigg\{-N^{2 + \a_q}\,\Bigg(- 2c_{q}\epsilon \Int{}{} \abs{\xi}^{q}\,\dd L_N^{(\bs{\la})} + {\cal E}_{({\rm ply})}[L_N^{(\bs{\la})}]\Bigg)\Bigg\}\prod_{a = 1}^N e^{-N^{\a_q + 1}\,c_q\epsilon \abs{\la_{a}}^{q}}\cdot\dd^{N}\bs{\la}
\enq
where we have put aside one exponential decaying with rate $\epsilon$ to ensure later convergence of the integral. If $M > 0$, let us define the truncated functional:
\beq
{\cal E}^{\{M,\epsilon\}}_{({\rm ply})}[\mu] = \Int{}{} E^{\{M,\epsilon\}}(\xi,\eta)\,\dd\mu(\xi)\dd\mu(\eta),\qquad E^{\{M,\epsilon\}} = \min\big[M\,;\,E(\xi,\eta)  - c_{q}\epsilon\,\big(\abs{\xi}^{q} + \abs{\eta}^{q}\big)\big] \;. 
\enq
Since $E^{\{M,\epsilon\}}$ is a Lipschitz function bounded by $M$, with Lipschitz constant bounded by $O(M^{1 - 1/q})$, 
we deduce the following bounds when the event $L_N^{(\bs{\la})} \in B_{\delta}(\mu)$ is realised:
\beq
\big|{\cal E}^{\{M,\epsilon\}}_{({\rm ply})}[L_{N}^{(\bs{\la})} ] - {\cal E}^{\{M,\epsilon\}}_{({\rm ply})}[\mu]\big| \leq C\,\delta\,M\;,
\enq
for some constant $C > 0$ independent of $N$, $\delta$ and $\epsilon$. Therefore:
\beq
\ov{\mc{P}}_{N}\big[B_{\de}(\mu)\big] \leq \exp\Bigg\{C^{\prime} N\ln N + N^{\a_q + 2}\,\Bigg(CM\cdot\delta 
- {\cal E}_{({\rm ply})}^{\{M,\epsilon\}}[\mu]\Bigg)\Bigg\}\cdot \Bigg(\Int{\R}{} \ex{-c_q\epsilon\,|\la|^{q} }\,\dd\la\Bigg)^{N} 
\enq
It follows that:
\beq
\limsup_{N \rightarrow \infty} N^{-(2 + \a_q)}\,\ln \wt{\mc{P}}_{N}\big[B_{\delta}(\mu)\big] \leq CM\cdot \delta - {\cal E}^{\{M,\epsilon\}}_{({\rm ply})}[\mu]
\enq
We observe that $-E^{\{M,\epsilon\}}$ is an increasing function of $\epsilon$. We can now let $\epsilon \rightarrow 0$ by applying the monotone convergence theorem:
\beq
\limsup_{N \rightarrow \infty} N^{-(2+\a_q)}\,\ln \wt{\mc{P}}_{N}\big[B_{\de}(\mu)\big] \leq C\, M\cdot \delta - {\cal E}^{\{M,0\}}_{({\rm ply})}[\mu] \;. 
\enq
Then, sending $\delta \rightarrow 0$ erases the first term, and finally letting $M \rightarrow \infty$ using again monotone convergence:
\beq
\limsup_{\delta \rightarrow 0} \limsup_{N \rightarrow \infty} N^{-(2 + \a_q)}\,\ln \wt{\mc{P}}_{N}\big[B_{\delta}(\mu)\big] \leq -{\cal E}_{({\rm ply})}[\mu]\;,
\enq
Notice that monotone convergence proves this last inequality even in the case where ${\cal E}_{({\rm ply})}[\mu] = +\infty$. \qed

\subsection{Partition function and equilibrium measure}
\label{PLs}
By applying the reasoning described in \cite{DemboZeitouniLargeDeviationTechniques}, 
to the lower bounds (Lemma~\ref{lowerbou}) and upper bounds (Lemma~\ref{upperbou2}), along with the 
property of exponential tightness (Lemma~\ref{tighti}), we deduce that ${\cal E}_{({\rm ply})}$ is a good rate function for large deviations, \textit{i.e.}
\begin{eqnarray}
\mathrm{for}\,\,\mathrm{any}\,\,\mathrm{open}\,\,\mathrm{set}\,\,\Omega \subseteq \mc{M}^1(\R), & & \qquad 
\liminf_{N\tend +\infty}  N^{-(2+\a_q)}\,\ln\ov{\mc{P}}_N [ \Omega ] \;  \geq \;  - \inf_{\mu \in \Omega} \mc{E}_{(\e{ply})}[\mu] \;, \nonumber \\
\mathrm{for}\,\,\mathrm{any}\,\,\mathrm{closed}\,\,\mathrm{set}\,\,F \subseteq \mc{M}^1(\R) & & \qquad \limsup_{N\tend +\infty}  N^{-(2+\a_q)}\,\ln \ov{\mc{P}}_N [ F ] \; \leq  \; - \inf_{\mu \in F}  \mc{E}_{(\e{ply})}[\mu]  \;. 
\end{eqnarray}
These two estimates, taken for $\Omega=F=\mc{M}^1(\R)$, lead to 
\beq
\lim_{N \rightarrow \infty} N^{-(2 + \a_q)}\,\ln \mf{z}_N = - \inf_{{\cal \mu} \in \mc{M}^1(\R)} {\cal E}_{({\rm ply})}[\mu] \;. 
\enq
The proof of the statements relative to the existence, uniqueness and characterisation of the minimiser of ${\cal E}_{({\rm ply})}$ is identical to those for the usual logarithmic energy 
\cite{SaffTotikLogarithmicPotential} -- and even simpler since there is no log singularity here. The minimiser is denoted $\mu^{({\rm ply})}_{{\rm eq}}$ 
and it is characterised by the existence of a constant  $C_{{\rm eq}}^{({\rm ply})}$ such that:
\begin{eqnarray}
c_{q}\abs{\xi}^{q} - \pi\beta(\omega_1 + \omega_2)\Int{}{} |\xi - \eta|\,\dd\mu_{{\rm eq}}^{({\rm ply})}(\eta) & = & C_{{\rm eq}}^{({\rm ply})}
	    \quad \mathrm{for}\,\,\xi\,,\quad \mu_{{\rm eq}}^{({\rm ply})}\,\,\mathrm{everywhere} \\
c_{q}\abs{\xi}^{q} - \pi\beta(\omega_1 + \omega_2) \Int{}{} |\xi - \eta|\,\dd\mu_{{\rm eq}}^{({\rm ply})}(\eta)
	    & \geq & C_{{\rm eq}}^{({\rm ply})}\quad \mathrm{for}\,\,\mathrm{any}\,\,\xi \in \R 
\end{eqnarray}
The construction of the solution of this regular integral equation is left as an exercise to the reader. We only give the final result in the announcement
of Theorem~\ref{Proposition LDP order dominant fct part rescalee ZN}. Actually, the fact that \eqref{2344} is a solution can be checked directly by integration by parts,
and we can conclude by uniqueness.

\qed

\chapter{Properties of the $N$-dependent equilibrium measure}

\label{Appendix minimisation de la mesure equilibre}

We give here elements for the proof of Theorem~\ref{Proposition caracterisation rudimentaire mesure equilibre}, which establishes the main properties of the minimiser of:
\beq
\mc{E}_N [ \mu  ] \; =\f{1}{2} \; \Int{}{} \left( V(\xi)\,+\,V(\eta)\; - \;  \f{ \be  }{  N^{\a} } \ln\bigg\{ \pl{p=1}{2} \sinh\big[ \pi N^{\a} \om_p(\xi-\eta) \big] \bigg\}\right)\,\dd \mu(\xi)\dd \mu(\eta) \;.
\enq
among probability measure $\mu$ on $\R$, with $N$ considered as a fixed parameter. %We can write:
%\beq
%\mc{E}_{N}[\mu] = \frac{\be}{2N^{\a}}\Bigg(\mc{E}_{{\rm C}}^{W_{N;1}}[\phi_{N;1}^*\,\mu] + \mc{E}_{{\rm C}}^{W_{N;2}}[\phi_{N;2}^*\,\mu] -2\ln 2\Bigg)\;,
%\enq
%where we have introduced the change of variables and the new potentials
%\beq@@
%p \in \{1,2\},\qquad \phi_{N;p}(\xi) = e^{2\pi\omega_{p}N^{\a} \xi},\qquad W_{N;p}(\xi) = \frac{2N^{\a}}{\beta}\,V(\xi) - \ln|\phi_{N;p}(\xi)|\;.
%\enq
%In this formula, the Coulomb energy functional:
%\beq
%\mc{E}_{{\rm C}}^{W}[\mu] = \Int{}{} W(\xi)\,\dd\mu(\xi) - \f{\be}{N^{\a}} \Int{}{} \ln|\xi - \eta\|
%\enq
%is considered for at least continuous, and confining potentials $W$, and is notoriously strictly convex over $\mc{M}^{1}(\R)$.%Pas comprise ca !!
As for any probability measures $\mu,\nu$ and $\alpha\in [0,1]$,
$$\mc{E}_{N}[\alpha \mu+(1-\alpha)\nu]-\alpha \mc{E}_{N}[ \mu]-(1-\alpha)\mc{E}_{N}[\nu]=-\alpha(1-\alpha)\mf{D}^2 \big[ \mu-\nu, \mu-\nu \big]\,,$$ 
where $\mf{D}$ is given in Definition \ref{dddef}, 
$\mc{E}_{N}$  is  strictly convex, and the standard arguments of potential theory \cite{LandkofFoundModPotTheory,SaffTotikLogarithmicPotential} show that it admits a unique minimiser, 
denoted $\mu_{{\rm eq}}^{(N)}$. 
More precisely, one can prove that $\mu_{{\rm eq}}^{(N)}$ has a
continuous density $\rho_{{\rm eq}}^{(N)}$ (as soon as $V$ is ${\cal
C}^2$) and is supported on a compact of $\R$ (since the potential here
is confining for any given value of $N$) a priori depending on $N$, see
\textit{e.g.} \cite[Lemma 2.4]{KozBorotGuionnetLargeNBehMulIntMeanFieldTh}. What
we really need to justify in our case is that:
\begin{itemize}
\item[$(0)$] the support of $\mu_{{\rm eq}}^{(N)}$ is contained in a compact independent of $N$ ;
\item[$(i)$] $\mu_{{\rm eq}}^{(N)}$ is supported on a segment ;
\item[$(ii)$] $\rho_{{\rm eq}}^{(N)}$ does not vanish on the interior of this segment and vanishes like a square root at the edges.
\end{itemize}

As a preliminary, we recall that the characterisation of the equilibrium
measure is obtained by writing that
$\mc{E}_N[\mu_{\e{eq}}^{(N)}+\epsilon \nu] \, \ge \,
\mc{E}_N[\mu_{\e{eq}}^{(N)}]$ for all $\epsilon>0$, all
measures $\nu$ with zero mass and such that $\mu_{\e{eq}}^{(N)}+\epsilon
\nu$ is non-negative. The resulting condition can be formulated in terms
of the effective potential introduced in \eqref{definition fonction controle deviation vp sur bord}:
\beq
V_{N;\e{eff}}(\xi) \; =\; U_{N;\e{eff}}(\xi) - \inf_{\R}
U_{N;\e{eff}},\qquad U_{N}(\xi) = V(\xi) \; - \; \Fint{}{}
s_N(\xi-\eta)\,\dd \mu_{\e{eq}}^{(N)}(\eta)
\enq
with the two-point interaction kernel:
\beq
s_N(\xi) \; = \; \f{ \be }{ 2 N^{ \a } } \ln \Big[ \sinh\big(\pi \om_1
N^{\a}\xi \big) \,\sinh\big(\pi \om_2 N^{\a}\xi \big) \Big] \;. 
\enq
The equilibrium measure is characterised by the condition:
\beq
V_{N;\e{eff}}(\xi) \geq
0\,,\qquad\,\,\mathrm{with}\,\,\mathrm{equality}\,\,\mu_{{\rm
eq}}^{(N)}\,\,\mathrm{almost}\,\,\mathrm{everywhere}
\enq

\Proof of $(0)$. Let $m_N > 0$ such that the support of $\mu_{{\rm eq}}^{(N)}$ is contained in $[-m_N,m_N]$. For $|\xi| > 2m_N$, we have an
easy lower bound:
\beq
\Big|\int s_N(\xi - \eta)\,\dd\mu_{{\rm eq}}^{(N)}(\eta)\Big| \geq
\frac{\beta}{2N^{\a}} \ln \Big[ \sinh\big(\pi \om_1 N^{\a}m_N\big)
\,\sinh\big(\pi \om_2 N^{\a}m_N \big) \Big] \geq \frac{\beta
\pi(\omega_1 + \omega_2)}{2}\,m_N + \e{O}(1)
\enq
where the remainder is bounded uniformly when $N \rightarrow \infty$ and
$m_N \rightarrow \infty$. By the growth assumption on the potential, there exists constant $C,C' > \epsilon > 0$ such that
\beq
V(\xi) \geq C |\xi|^{1 + \epsilon} + C'
\enq
Therefore, we can choose $m:=2m_N$ large enough and independent of $N$
such that $V_{N;{\rm eff}}(\xi) > 0$ for any $|\xi|>m$. This guarantees that the support of $\mu_{{\rm eq}}^{(N)}$ is included in the compact $\intff{-m}{m}$ for any $N$.

\Proof of $(i)$. 

We observe that $-s_N$ is strictly convex:
\beq
s_N''(\xi) \;=\; -\,\frac{\beta N^{a}}{2} \sum_{p = 1}^2 \frac{(\pi\omega_{p})^2}{\big(\sinh \pi\omega_{p}\xi\big)^2} \;<\; 0\;.
\enq
Since $V$ is assumed strictly convex and $\mu_{{\rm eq}}^{(N)}$ is a positive measure, it implies that $V_{N;{\rm eff}}$ is strictly convex. Therefore, the locus where it reaches 
its minimum must be a segment. So, there exists $a_N < b_N$ such that $\intff{a_N}{b_N}$ is the support of $\mu_{{\rm eq}}^{(N)}$. This strict convexity also ensures that 
\beq
V_{N;\e{eff}}^{\prime}(\xi) \, > \, 0 \qquad \e{for} \; \e{any} \qquad \xi >b_N, \quad V_{N;\e{eff}}^{\prime}(\xi) \, < \, 0 \qquad \e{for} \; \e{any} \qquad \xi <a_N
\;. 
%A:correct
\label{ecriture positivite stricte de V eff en dehors support mu eq}
\enq

\Proof of $(ii)$. 

This piece of information is enough so as to build the representation:
\beq
\label{ouech}\rho_{{\rm eq}}^{(N)}(\xi) = {\cal W}_{N}[V^{\prime}]\cdot\mathbf{1}_{\intff{a_N}{b_N}}(\xi)
\enq
for the equilibrium measure. Indeed, we constructed $\mc{W}_N[H]$ in Section~\ref{Sous section construction inverse a l'operateur SN} so that it provides the unique solution to:
\beq
\forall \xi \in \intoo{a_N}{b_N},\qquad \Fint{a_N}{b_N} s_N[N^{a}(\xi - \eta)]\,\dd\mu_{{\rm eq}}^{(N)}(\eta) \; = \;  V^{\prime}(\xi) 
\enq
which extends continuously on $\intff{a_N}{b_N}$, and this was only possible when $\mc{X}_{N}[V^{\prime}] = 0$ in terms of the linear form introduced in Definition~\ref{defXN}. 
Since the equilibrium measure exists, this imposes the constraint:
\beq
\label{zc1}\mc{X}_N[V^{\prime}]= 0 \;. 
\enq
Besides, since the total mass of \eqref{ouech} must be $1$, we must also have:
\beq
\label{zc2}\Int{a_N}{b_N} \mc{W}_{N}[V^{\prime}] = 1 \;. 
\enq
At this stage, we can use Corollary~\ref{Corollaire DA des bornes aN et bN}, which shows that \eqref{zc1}-\eqref{zc2} determine uniquely the large-$N$ asymptotic expansion of $a_N$ and $b_N$, 
in particular there exists $a < b$ such that $(a_N,b_N) \tend (a,b)$ with rate of convergence $N^{-\a}$. Besides, the leading behaviour at $N \rightarrow \infty$ of $\mc{W}_{N}$ is described by 
Proposition~\ref{Proposition decomposition op WN en diverses sous parties} and \ref{Proposition Ecriture reguliere uniforme des divers const de WN}. 
It follows from the reasonings outlined in the proof of Proposition \ref{Theorem bornes sur norme inverse UN via estimation fines locales}
that
\beq
\label{863A}\rho_{{\rm eq}}^{(N)}(\xi) \; =  \; \mc{W}_{N}[V^{\prime}](\xi) =  \left\{ \ba{cc}  \frac{V^{\prime\prime}(\xi)}{2\pi\beta(\omega_1 + \omega_2)} + \e{O}(N^{-\a}) &
												  \xi \in \Big[ a_N+ \f{(\ln N)^2}{N^{\a}} ; b_N -  \f{(\ln N)^2}{ N^{\a} }  \Big] \\
V^{\prime\prime}(b_N) \, \mf{a}_0\big( N^{\a} (b_N-\xi) \big) \; + \; \e{O}\bigg( \f{(\ln N)^3}{ N^{\a} } \sqrt{N^{\a} (b_N-\xi)} \bigg)   & 
									\xi \in \intff{ b_N -  (\ln N)^2 \cdot N^{-\a} }{b_N}    \\

V^{\prime\prime}(a_N) \, \mf{a}_0\big( N^{\a} (\xi-a_N) \big) \; + \; \e{O}\bigg( \f{(\ln N)^3}{ N^{\a} } \sqrt{N^{\a} (\xi-a_N)} \bigg)   & 
									\xi \in \intff{a_N}{ a_N+ (\ln N)^2 \cdot N^{-\a} } 
\ea \right. 
\enq

Therefore, for $N$ large enough, $\rho_{{\rm eq}}^{(N)}(\xi) > 0$ on $\intff{a_N}{b_N}$.
The vanishing like a square root at the edges then follows from he properties of the $\mathfrak{a}$'s 
established in Lemma~\ref{Lemme comportement fonction a goth}. In fact, one even has

\beq
\label{miumum}\lim_{\xi \rightarrow b_N^{-}} \frac{ \rho_{{\rm eq}}^{(N)}(\xi)  }{\sqrt{b_N - \xi}} \; = \; 
N^{\a/2}\Bigg(V^{\prime\prime}(b_N)\cdot \lim_{x \rightarrow 0} x^{-1/2}\mathfrak{a}_0(x)  \; +  \; \e{O}(N^{-\a})\bigg\} 
\; = \;  \frac{N^{\a/2} V^{\prime\prime}(b)}{\pi\beta\sqrt{\pi(\omega_1 + \omega_2)}} + \e{O}(N^{-\a/2}) \;. 
\enq
 \qed

\chapter{The Gaussian potential}
\label{Appendix Section etude de l'integrale Gaussien}

In this appendix we focus on the case of a Gaussian potential and
establish two results. On the one hand, we establish in Lemma~
\ref{Proposition pot Gaussien avec meme support mesure eq} that, for $N$
large enough, there exists a unique sequence of Gaussian potential
$V_{G;N} \, = \, g_N \la^2+t_N \la$ such
that their associated equilibrium measure has support
$\sg_{\e{eq}}^{(N)} = \intff{a_N}{b_N}$.
On the other hand we show, in Proposition \ref{Proposition calcul
explicite fct part Gaussienne},
that the partition function associated with any Gaussian potential can
be explicitly evaluated, and thus is amenable to a direct asymptotic
analysis when $N \rightarrow \infty$. Note that when $\tf{\om_1}{\om_2}$ is rational,
such Gaussian partition functions have been evaluated in \cite{DolivetTierzGaussianPartitionFunction} by using biorthogonal 
analogues of the Stieltjes-Wigert orthogonal polynomials.

\begin{lemme}
\label{Proposition pot Gaussien avec meme support mesure eq}

There exists a unique sequence of Gaussian potentials
\beq
V_{G;N}  \, = \, g_N \la^2+ t_N \la
\enq
such that their associated equilibrium measure has support
$\sg_{\e{eq}}^{(N)} = \intff{a_N}{b_N}$. The coefficients $g_N,t_N$ take the form
\beq
g_N \;  = \;  \pi \be (\om_1+\om_2) \Bigg\{ b_N-a_N \; + \; N^{-\a}\sul{p=1}{2} \f{1}{\pi \om_p} \ln\Big( \f{ \om_1 \om_2 }{ \om_p (\om_1+\om_2) }  \Big)  \Bigg\}^{-1}
 \; + \; \e{O}\big(N^{-\infty})
\enq
and
\beq
t_N \; = \; -(a_N+b_N)g_N \; + \; \e{O}\big(N^{-\infty}\big) \;. 
\enq

\end{lemme}

\Proof Let $V_G(\la)= g \la^2 + t \la$ be any Gaussian potential. Since
it strictly convex, all previous results apply.
Suppose that $V_G$ gives rise to an equilibrium measure supported on
$\sg_{\e{eq}}^{(N)} = \intff{a_N}{b_N}$. This means that the potential $V_G$
has to satisfy the system of two  equations that are linear in
$V_G^{\prime}$:
\beq
\Int{a_N}{b_N} \mc{W}_N[V_G^{\prime}](\xi)\,\dd \xi \; = \; 1
\qquad \e{and} \qquad
\Int{\R +\i \eps^{\prime} }{} \f{\dd \mu }{2 \i \pi }\,\chi_{11}(\mu)
\Int{a_N}{b_N} V^{\prime}_G(\eta) \ex{ i N^{\a} \mu (\eta-b_N)}\,
\dd \eta  \; = \; 0 \;.
\enq
It follows from the multi-linearity in $(g,t)$ of $V_G$ and from
the evaluation of single integrals carried out
in Lemma~\ref{Proposition DA integrale contrainte} and Proposition~\ref{Theorem DA tout ordre integrale 1D contre WN} that there exist two
linear forms $L_1, L_2$ of $(g,t)$ whose norm is a 
$\e{O}(N^{-\infty})$ and such that
\beq
1 \; = \; \f{ g }{ \pi \be (\om_1+\om_2) }
\Bigg\{ (b_N-a_N) + \f{1}{ N^{\a}} \cdot \sul{p=1}{2} \f{1}{\om_p \pi} \ln\Big( \f{ \om_1 \om_2 }{ \om_p (\om_1+\om_2) }  \Big)  \Bigg\} 
\; + \; L_1(g,t)
\enq
where we have used that
\beq
\Int{0}{+\infty} \mf{b}_0(x)\,\dd x \; = \; \f{1}{2\pi \be
(\om_1+\om_2) } \cdot
\sul{p=1}{2} \f{1}{\om_p \pi} \ln\Big( \f{ \om_1 \om_2 }{ \om_p
(\om_1+\om_2) }  \Big)
\enq
a formula that can be established with the help of \eqref{definition
fcts a goth} and \eqref{expression integrale pour a goth 0}.
One also obtains that
\beq
0 \; = \; \f{2}{N^{\a} \sqrt{\om_1+\om_2}} \Big(  g (b_N+a_N) + t  \Big)  \; + \; L_2(g,t) \;.
\enq
In virtue of the unique solvability of perturbations of linear solvable
systems, the existence and uniqueness of the
potential $V_{G;N}$ follows. \qed

\begin{prop}
\label{Proposition calcul explicite fct part Gaussienne}
The partition function $Z_N[V_G]$ at $\be=1$ associated with the
Gaussian potential $V_{G}(\la) =  g  \la^2 +  t \la$
can be explicitly computed as
\beq
\label{ecriture expression explicite fct part Gaussienne beta=1} %A below changed V_0 in V_G
Z_N[V_G]_{\mid \be=1} \; = \;  \f{ N! }{ 2^{N(N-1)} } \,\Bigg(\frac{\pi}{g N^{1 + \alpha}}\Bigg)^{N/2}\,
\exp\bigg\{\frac{N^{2 + \alpha}t^2}{4g} + \frac{\pi^2(\omega_1 + \omega_2)^2}{ 12 g }N^{\alpha}(N^2 - 1)\bigg\}\,
\pl{j = 1}{N } \big(  1 -  \ex{- \f{2 j N^{\a }}{g N } \pi^2 \omega_1\omega_2 }\big)^{N - j} \;. 
\enq
\end{prop}

\Proof We can get rid of the linear term in the potential by a
translation of the integration variables.
Then
\beq
{Z_N[V_G]}_{\mid \be=1} \; = \; \exp\bigg\{ \frac{N^{2+\a}t^2}{ 4 g } \bigg\} \cdot { Z_N[ \wt{V}_G ] }_{\mid \be=1} \quad
\e{where} \qquad \wt{V}_G(\la) =  g \la^2  \;.
\enq
Further, the products over hyperbolic sinh's can be recast as two
Vandermonde determinants
\beq
\pl{a<b}{N} \Big\{ \sinh\big[\pi\om_1 N^{\a}(\la_a-\la_b)\big]
\sinh\big[\pi\om_2 N^{\a} (\la_a-\la_b)\big]  \Big\} \; = \;
\pl{a=1}{N} \bigg\{ \f{ \ex{-\pi(\om_1+\om_2)N^{\a} (N-1)\la_a} }{
2^{N-1} }\bigg\}  \, \cdot  \,
\pl{p=1}{2} \det_N\Big[  \ex{2\pi \om_p N^{\a} \la_j (k-1) }  \Big]   \;.
%\cdot  \det_N\Big[  \ex{2\pi \om_2 \la_a (k-1)}  \Big]  
%
\enq
%
%-
%

Inserting this formula into the multiple integral representation for
$Z_N[V_G]_{\mid \be=1}$ and using the symmetry of the integrand, one can
replace one of the determinants by $N!$
times the product of its diagonal elements. Then, the integrals separate
and one gets:
\beq
{Z_N[\wt{V}_G]}_{\mid \be=1}  \; = \; \f{ N! }{ 2^{N(N-1)}  }
\cdot\det_N \bigg[   \Int{ \R }{}    \ex{ -\pi(\om_1+\om_2)N^{\a}(N-1)\la }  
\ex{ 2\pi N^{\a} [ \om_1 (k-1) + \om_2(j-1) ] \la }
 \cdot \ex{-g N^{1+\a}\la^2}       \dd  \la \bigg] \;.
\enq
The integral defining the $(k,j)^{{\rm th}}$ entry of the determinant
is Gaussian and can thus be computed. This yields, upon factorising the
trivial terms arising in the determinant,
\beq
{Z_N[\wt{V}_G]}_{\mid \be=1}  \; = \; \Big(  \f{\pi}{ g N^{1+\a}} \Big)^{\f{N}{2}}  
\f{  N! }{  2^{N(N-1)} }  \cdot
\pl{ j=1 }{ N }   \ex{  \f{  \pi^2 }{4g} N^{\a-1} (\om_1^2 +
\om_2^2) (2j-1-N)^2   } \cdot D_N \; ,
\enq
 where
\beq
D_N \; = \; \det_N \bigg[   \exp\Big\{     \f{\pi^2}{2g} N^{\a-1}
\om_1\om_2   (2k-N-1)(2j-1-N) \Big\}    \bigg]  \;.
\enq
The last determinant can be reduced to a Vandermonde. Indeed, we have:
\bem
D_N \; = \;   \exp\Big\{    \f{\pi^2}{2g}\om_1\om_2   (N-1)^2
N^{\a}  \Big\}   \;  \cdot \;
\pl{j=1}{N}  \Big\{   \ex{  - 2 \f{ \pi^2}{g} \om_1\om_2 N^{\a-1}(N-1)(j-1)   }    \Big\}   \; \cdot \;
  \det_N \bigg[   \exp\Big\{     2 \f{\pi^2}{g} N^{\a-1} \om_1\om_2   (k-1)(j-1) \Big\}    \bigg]   \\
\; = \;   \exp\bigg\{    - \f{\pi^2}{2g}\om_1\om_2   (N-1)^2 N^{\a}  \bigg\}   \;  \cdot \;
\pl{k>j}{N} \bigg(   \ex{ 2  \f{ \pi^2}{g} N^{\a-1}\om_1\om_2 (k-1) }
                 \; - \; \ex{  2  \f{ \pi^2}{g} N^{\a-1} \om_1\om_2 (j-1) }  \bigg) \;.
\end{multline}
In order to present the last product into a convergent form, we factor
out the largest exponential of each term.
The product of these contributions is computable as
\beq
\pl{k>l}{N} \bigg(   \ex{ 2  \f{ \pi^2}{g} N^{\a-1} \om_1\om_2 (k-1) } \bigg) \; = \;
\pl{k=1}{N-1} \ex{ 2\f{ \pi^2}{g} N^{\a-1}\om_1\om_2 k^2 }  \; = \;
 \exp\Big\{  \f{ \pi^2}{3 g } \om_1\om_2  N^{\a} (N-1)(2N-1) \Big\} \;,
\enq
where we took advantage of
\beq
\sul{p=1}{N}  p^2  \; = \; \f{  N(N+1)(2N+1) }{ 6 } \;.
\enq
Putting all of the terms together leads to the claim. \qed

The large-$N$ asymptotic behaviour of the partition function at $\be=1$ and associated to a Gaussian potential 
can be extracted from \eqref{ecriture expression explicite fct part Gaussienne beta=1}.

\begin{prop}
Assume $0 < \alpha < 1$. We have the asymptotic expansion:
\bem
\ln Z_{N}[V_G]|_{\beta = 1} \;  =  \; N^{2 + \a}  \cdot  \Big[  \f{ t^2 }{4 g } \, + \,  \frac{ \pi^2(\omega_1 + \omega_2)^2 }{12 g }  \Big]  
\; - \;  N^2 \cdot \ln 2 \; - \;  N^{2 - \a} \cdot  \frac{g }{12 \om_1 \om_2 }   \\
\; + \;    N^{2 - 2\a} \cdot  \frac{ g^2 \,\zeta(3) }{ \big( 2\pi^2 \om_1 \om_2 \big)^2 } 
\; +\;  (1 -\a)\,N \ln N  \;  +  \; N \cdot \ln \Big( \f{2/ \ex{} }{\sqrt{\om_1\om_2} }\Big)  \\
\; - \;  N^{\a} \cdot \f{ \pi^2 (\om_1 + \om_2)^2 }{ 12 g }   %A changed factor in \om1 and \om_2
 \; + \; \ln N \cdot   \f{\a + 5}{12}
\;  + \;  \f{1}{12}\ln\Bigg(\frac{128\pi^{8}\omega_1\omega_2}{g} \Bigg) \, + \,  \zeta^{\prime}(-1)  \; + \; \e{o}(1)\;. 
\label{ecriture asymptotique fct part gaussienne a beta egal 1}
\end{multline}

\end{prop}

\Proof The sole problematic terms demanding some further analysis is the last product in  \eqref{ecriture expression explicite fct part Gaussienne beta=1}. 
The latter can be recast as :
\beq
\label{produh}
\prod_{\ell = 1}^{N } (1 - \ex{-\tau_N \ell} )^{N - \ell} \; = \; \Bigg[ \frac{ M_0\big( \ex{-N\tau_N} ; \ex{-\tau_N} \big) }{ M_0(1;\ex{-\tau_N}) } \Bigg]^{N} \cdot 
\frac{ M_1(1;\ex{-\tau_N}) }{ M_1\big(\ex{-N\tau_N} ; \ex{-\tau_N} \big) }
\qquad \e{where} \quad \tau_N \;=\;   \ \f{ 2  N^{\a } }{g N } \pi^2 \omega_1\omega_2 
\enq
and $M_{r}(a,q)$ corresponds to the infinite products $M_r(a;q) = \prod_{\ell = 1}^{\infty}
(1 - aq^{\ell})^{-\ell^{r}}$. 

We will exploit the fact that asymptotics of $M_r(a;\ex{-\tau})$ when $\tau \rightarrow 0^+$ up to $o(1)$ can be read-off from the singularities of the
Mellin transform of its logarithm 
\beq
\mathfrak{M}_{r}(a;s) = \int_{0}^{\infty} \ln M_r(a;\ex{-t})\,t^{s - 1}\dd t \qquad \e{where} \qquad  
\ln M_r(a;q) \; \equiv \; -\sul{\ell=1}{+\infty}\ell^r  \ln\big( 1 - aq^{\ell}  \big) \;. 
\enq
\noindent The above Mellin transform  is well-defined for $\mathrm{Re}(s) > r + 1$ and can be easily computed. For any $|a| \leq 1$, we have:
\beq
\label{pilyh}
\mathfrak{M}_{r}(a;s) \; = \; \sum_{\ell = 1}^{\infty}\sum_{m = 1}^{\infty}
\frac{\ell^{r}\,a^{m}}{m}\,\int_{0}^{\infty} t^{s - 1}\, \ex{-t\ell m}\,\dd t = \Gamma(s)\,\zeta(s - r)\,\mathrm{Li}_{s + 1}(a) \;. 
\enq
Above, $\zeta$ refers to the Riemann zeta function whereas ${\rm Li}_{s}(z)$ is the polylogarithm which is defined 
by its series expansion in a variable $z$ inside the unit disk:
\beq
\label{polygs}\mathrm{Li}_{s}(z) = \sum_{k \geq 1} \frac{z^k}{k^{s}}
\enq
Note that, when $\mathrm{Re}\,s > 1$, the series also converges uniformly up to the boundary of
the unit disk. We remind that the first two polylogarithms can be expressed in terms of elementary functions:
\beq
\mathrm{Li}_{0}(z) \; =\;  \f{z}{1 - z} \qquad  \e{and} \qquad \mathrm{Li}_{1}(z) \;  = \; -\ln(1 - z)\;.
\enq
In both cases $|a|<1$ or $a=1$, $\mathfrak{M}_{r}(a;s)$ admits a meromorphic extension from $\mathrm{Re}\,s > 1$ to  $\Cx$. 
When $|a|<1$ this is readily seen at the level of the series expansion of the polylogarithm whereas
when $a=1$, this follows from $\mathrm{Li}_{s + 1}(1) = \zeta(s + 1)$. 
Furthermore, this meromorphic continuation is such that  $\mathfrak{M}_{r}(a; x + \i y) = \e{O}(e^{-c|y|})$, $c>0$, when $y \rightarrow \pm\infty$. This estimate  is uniform for $a$ in compact subsets of the open unit disk and for $x$ belonging to compact subsets of $\mathbb{R}$.
The same type of bounds also holds for $a=1$, namely  $\mathfrak{M}_{r}(1; x + \i y) = \e{O}(e^{-c|y|})$, $c>0$, when $y \rightarrow \pm\infty$ 
 for $x$ belonging to a compact subset of $\mathbb{R}$.
This is a consequence of three facts: 
\begin{itemize}
\item[$\bullet$] $\Ga(x+ \i y)$ decays exponentially fast when $|y|\tend +\infty$ and $x$ is bounded, as follows from the Stirling
formula ;
\item[$\bullet$] $|\zeta(x+\i y)| \leq C |x+\i y|^{c}$ for some $c > 0$ valid provided that $x$ is bounded \cite{TitchmarshHeatBrownTheoryZetaFunction} ;
\item[$\bullet$]  $\mathrm{Li}_{x+\i y }(a)$ is uniformly bounded for $x$ in compact subsets of $\R$ and $a$ in compact subsets of the open unit disk, as is readily inferred from the series representation \eqref{polygs}.
\end{itemize}
Thanks to the inversion formula for the Mellin transform 
\beq
\ln M_r(a,\ex{-\tau})  \; = \; \Int{ c-\i \infty }{ c+\i \infty }  \tau^{-s}\,\mathfrak{M}_{r}(a;s)\,\frac{\dd s}{2{\rm i}\pi}
\qquad \e{with} \quad c> r+1 \;, 
\enq
we can compute the $\tau \rightarrow 0$ asymptotic expansion of $\ln M_r(a,\ex{-\tau})$ -- this principle 
is the basis of the transfer theorems of \cite{FlajoletGourdonDumasMellinTransformsAsymptotics}. To do so, we deform the contour of integration to the region $\mathrm{Re}\,s < 0$.
The residues at the poles of $\mathfrak{M}_{r}(a;s)$ are picked up in the process.  There are two cases to distinguish since the polylogarithm factor in \eqref{pilyh} is entire if $|a| < 1$,
while for $a=1$ one has  $\mathrm{Li}_{s + 1}(1) = \zeta(s + 1)$ what generates an additional pole at $s = 0$. We remind that:
\beq
\Gamma(s) \; \mathop{=}_{s \rightarrow 0} \;  \frac{1}{s} \, - \,  \gamma_{E}  \, + \,  \e{O}(s)\, , \qquad \zeta(s)  \; =  \; \frac{1}{s - 1}  \, + \,  \gamma_{E} \, + \, \e{O}(s)
\enq
where $\gamma_{E}$ is the Euler constant. For $a < 1$, $\mathfrak{M}_{r}(a;s)$ has simple poles at $s = 1 + r$ and $s = 0$:
\beq
\mathfrak{M}_{r}(a;s) \; = \,  \frac{\mathrm{Li}_{2 + r}(a)\,r!}{s - (1 + r)}  \, + \,  \e{O}(1) \;, 
\qquad \mathfrak{M}_{r}(a;s) \;  = \; \frac{-\zeta(-r)\ln(1 - a)}{s}  \, + \,  \e{O}(1) \; .
\enq
Notice that here $r \in \{0,1\}$ and the Riemann zeta function has the special values $\zeta(0) = -1/2$ and $\zeta(-1) = -1/12$. Therefore, 
\beq
\label{Masy2} \ln M_{r}(a;e^{-\tau}) \; =  \; \frac{r!\,\mathrm{Li}_{2 +r}(a)}{\tau^{1 + r}} \, - \, \zeta(-r)\ln(1 - a)  \, + \,  \e{o} (1)\, ,\qquad \tau \rightarrow 0^+
\enq
and the remainder is uniform for $a$ uniformly away from the boundary of the unit disk. 
For $a = 1$,  $\mathfrak{M}_{r}(a;s)$ has the same simple pole at $s = 1 + r$ with residue $r!\,\zeta(2 + r)$, but now a double pole at $s = 0$:
\beq
\mathfrak{M}_{r}(1;s) \; = \;  \frac{\zeta(-r)}{s^2} \, + \,  \frac{\zeta^{\prime}(-r)}{s}  \, + \, \e{O}(1) \, , 
\qquad \mathfrak{M}_{r}(1;s)  \; = \;  \frac{r!\,\zeta(2 + r)}{s - (1 + r)} \, + \,  \e{O}(1)
\enq
and we remind the special value $\zeta^{\prime}(0) = -\ln(2\pi)/2$. In this case, we thus have:
\beq
\label{Masy1}M_{r}(1;e^{-\tau}) \;  = \;  \frac{r!\,\zeta(2 + r)}{\tau^2}  \, - \,  \zeta(-r)\ln \tau  \, + \,  \zeta^{\prime}(-r) \, + \,  \e{o}(1),\qquad \tau \rightarrow 0^+ \;. 
\enq
Collecting all the terms from \eqref{Masy2}-\eqref{Masy1}, we obtain the asymptotics of the product \eqref{produh} that are uniform in $a$
belonging to compact subsets of the unit disk:
\bem
\label{sfad}\ln\Bigg[\prod_{\ell = 1}^{N - 1} (1 - e^{-\tau_N})^{N -\ell}\Bigg] \, = \, 
\frac{\zeta(3) - \mathrm{Li}_{2}\big( \ex{-N\tau_N} \big) }{ \tau_N^2 }  \; + \;  \frac{ N }{  \tau_N } \Big(\mathrm{Li}_{1}\big( \ex{-N\tau_N} \big) - \f{\pi^2}{6} \Big) \\
\, + \,  \Bigg( \frac{N}{2} - \frac{1}{12} \Bigg) \ln\Bigg( \frac{1 -  \ex{-N\tau_N} }{\tau_N} \Bigg) \, + \,  \frac{ N\ln(2\pi) }{ 2 }  \, + \,  \zeta^{\prime}(-1)  \, + \, \e{o}(1) \;. 
\end{multline}
Here, we have used the special value $\zeta(2) = \pi^2/6$. It remains to insert in \eqref{sfad} the value of the 
parameter of interest $\tau_N = N^{\alpha -1}\,2\pi^2\omega_1\omega_2/g$, and return to the original formula. The
announced result for the Gaussian partition function \eqref{ecriture asymptotique fct part gaussienne a beta egal 1} follows, upon using the Stirling approximation $N!\sim
\sqrt{2\pi}N^{N + 1/2}e^{-N}$ for the factorial prefactor.

We remark that for $\a \geq 1$, $\tau_N \geq 0$ is not anymore going to $0$ when $N \rightarrow \infty$, therefore the asymptotic regime will be different.

\qed

\chapter{Summary of symbols}
\label{Appendix Section liste des formule}

\subsection*{Empirical and equilibrium measures}

\begin{tabular}{lll}
$\mathcal{E}_{(\e{ply})}[\mu]$ & \eqref{Regsaf} & energy functional for the baby integral of \S~\ref{baby} \\
$E(\xi,\eta)$ & \eqref{Regsaf} & its kernel function \\
$\mu_{\e{eq}}^{(\e{ply})}$ & \S~\ref{PLs} & minimiser of $\mathcal{E}_{(\e{ply})}$ \\
$\mathcal{E}_{N}[\mu]$ & \eqref{definition fnelle a minimiser N dpdte} &  $N$-dependent energy functional \\
$\mathcal{E}_{\infty}[\mu]$ & \eqref{definition rate fct pour fct part ZN rescalee} & same one at $N = \infty$ \\
$\mathfrak{D}[\mu,\nu]$ & Def.~\ref{dddef} & pseudo-distance between probability measures induced by $\mathcal{E}_N$ \\ 
$\mu_{{\rm eq}}^{(N)}$ & \eqref{definition de la cste Ceq par eqn int eq meas}-\eqref{ecriture condition negativite dehors support mu eq} & $N$-dependent equilibrium measure (maximiser of $\mathcal{E}_N$)\\
$\rho_{{\rm eq}}^{(N)}$ & Thm.~\ref{Proposition caracterisation rudimentaire mesure equilibre} & density of $\mu_{{\rm eq}}^{(N)}$ \\
$[a_N,b_N]$ & Thm.~\ref{Proposition caracterisation rudimentaire mesure equilibre} & support of $\mu_{{\rm eq}}^{(N)}$ \\
$x_{a}^N$ & Def.~\ref{clapo} & classical positions for $\mu_{{\rm eq}}^{(N)}$ \\ 
$V_{N;{\rm eff}}$ & \eqref{definition fonction controle deviation vp sur bord} & effective potential \\
$L_N^{(\bs{\lambda})}$ & \eqref{definition mesure empirique} & empirical measure \\
$\tilde{\bs{\lambda}}$ & Def.~\ref{definition suite lambda's tildes} & deformation of $\bs{\lambda}$ enforcing a minimal spacing \\
$L_{N;u}^{(\bs{\lambda})}$ & Def.~\ref{definition suite lambda's tildes} & convolution of $L_N^{(bs{\lambda})}$ with uniform law of small support \\
$\mathcal{L}_{N}^{(\bs{\lambda})}$ & Def.~\ref{Lcurl} & centred empirical measure with respect to $\mu_{{\rm eq}}^{(N)}$ \\
$\mathbb{M}_{N;\kappa}^{(n)}$ & Def.~\ref{Definition regularisation exponentielle} & probability measure including exponential regularization of $n$ variables
\end{tabular}

\subsection*{Partition functions}

\begin{tabular}{lll}
$Z_{N}[V]$ & \eqref{ecriture premiere intro dans texte de fct part rescale} & partition function of the sinh model with potential $V$ \\
$V_{G;N}$ & Lemme~\ref{Proposition pot Gaussien avec meme support mesure eq} & Gaussian potential leading to support $[a_N;b_N]$
\end{tabular}

\subsection*{Pairwise interactions}

\begin{tabular}{lll}
$s_N(\xi)$ & \eqref{definition fonction sN noyau integral eqn mes eq} & pairwise interaction kernel \\
$S(\xi)$ & \eqref{ecriture eqn int sing de depart} & derivative of $\beta\ln\big|{\rm sinh}(\pi\omega_1\xi){\rm sinh}(\pi\omega_2\xi)\big|$ \textit{viz}.  $\frac{1}{2}\partial_{\xi} s_N(N^{-\alpha}\xi)$ \\
$S_{{{\rm reg}}}(\xi)$ & \eqref{definition noyau integral VN} & $S$ minus its pole at $0$ \\ 
$\mathcal{S}_{N}$ & \eqref{ecriture eqn int sing de depart} & integral operator with kernel $S(N^{\alpha}(\xi_1 - \xi_2))$ \\
$\mathcal{S}_{N;\gamma}$ & \eqref{definition operateur regularise S N gamma} & same one with extended support \\
$\msc{S}_{N;\gamma}$ & \eqref{ecriture equation sing en variables reduites} & same one in rescaled and centered variables
\end{tabular}

\subsection*{Operators}

\begin{tabular}{lll}
$\mathcal{K}_{\kappa}$ & Def.~\ref{Definition regularisation exponentielle} & multiplication by a decreasing exponential \\
$\Xi^{(p)}$ & Def.~\ref{Xipdef} & operator inserting a copy of $\xi_1$ at position $p$ \\
$\mathcal{U}_{N}$ & \eqref{definition noyau integral operateur S driven by mu eq} & master operator \\
$\mathcal{D}_{N}$ & \eqref{noncommet} & hyperbolic analog of the non-commutative derivative \\
$\mathcal{V}_{N}$ & Prop \ref{Proposition characterisation operateur UN} & building block of  $\mathcal{U}_{N}^{-1}$ \\
$\mathcal{W}_N$ & \eqref{definition operateur WN} & inverse of $\mathcal{S}_{N}$  \\
$\mc{X}_{N}$ & Def.~\ref{defXN} & linear form related to $\mathscr{I}_{11}$ \\
$\wt{\mc{X}}_{N}$ & Def.~\ref{defXN} & projection to the hyperplane $\mathfrak{X}_{s}(\intff{a_N}{b_N}) = \mathrm{Ker}\,\mc{X}_N$ \\
$\wt{\mathcal{U}}_{N}^{-1}$, $\wt{\mathcal{W}}_{N}$ & \eqref{tildeopXN} & operators composed to the right with $\wt{\mc{X}}_N$ \\
$\msc{W}_{N}$ & \eqref{definition de wN espace transforme} & operator ${\cal W}_{N}$ in rescaled and centered variables \\
$\wt{\msc{W}}_{\vartheta; z_0}$ & \eqref{ecriture TF de solution fondamentale} & a pseudo-inverse of $\msc{S}_{N;\gamma}$. \\
$\mathscr{I}_{11},\mathscr{I}_{12}$ & Prop.~\ref{Theorem caracterisation sous class reguliere solutions}& functionals appearing in the inversion of $\mathscr{S}_{N;\gamma}$ \\
$\mathscr{J}_{1a}(\lambda)$ & \eqref{definition integral J a1 caligraphe} & related functionals \\
%
%
%$\mathscr{J}$ & 7.1 & a functional proportional to $\mathcal{I}_{11}$ \\
%
%
$\mpzc{w}_{k;a}^{(1/2)}$, $\mpzc{w}_{k;a}^{(1)}$ & \eqref{definition fonctions wak} & functionals appearing in the large $\lambda$ expansion of the latter  \\
$H^{\wedge}$ & Def.~\ref{definition tilde fonction} & reflection of the function $H$ (exchanging left and right boundary) \\
%
%
%$\widehat{a}_{\ell},\widehat{b}_{\ell}$ & 6.107-6.99 & linear combinations of derivatives at boundaries \\
%
%
%$\widehat{u}_{\ell}$ & 6.99 & another linear combination of derivatives at boundaries divided by the first derivative \\
%
%
$\mathcal{G}_N$ & \eqref{definition integrande GN cal} & 2-variable operator related to $\mathcal{W}_{N}$ \\
$\mathcal{T}_{\e{even}},\mathcal{T}_{\e{odd}}$ & \eqref{eve} & some even/odd averaging operator
\end{tabular}

\subsection*{Decomposition of operators for asymptotic analysis}

\begin{tabular}{lll}
$\mathcal{W}^{(\infty)},\delta\mathcal{W}$ & \eqref{ecriture decomposition operateur WN} & leading and subleading terms in $\mathcal{W}_{N}$ when $N \rightarrow \infty$ \\
$\mathcal{W}_{R},\mathcal{W}_{L}$ & \eqref{ecriture decomposition WN en drte gche bk exp rudimentaire} & contribution of the right/left boundary to $\mathcal{W}_{N}$ \\
$\mathcal{W}_{R;k}$ & Prop.~\ref{Proposition Ecriture reguliere uniforme des divers const de WN} & terms contributing to the latter up to $O(N^{-k\alpha})$ \ldots \\
$\Delta_{[k]}\mathcal{W}_{R}$ & Prop.~\ref{Proposition Ecriture reguliere uniforme des divers const de WN} & \ldots and the remainder \\
$\mathcal{W}_{R;k}^{({\rm as})},\Delta_{[k]}\mathcal{W}_{R;k}^{(\e{as})}$ & Lemma~\ref{Lemme structure locale au bord pour WR et Wbk} & putting aside exponentially small terms in $\mc{W}_{R;k}$\\
$\mathcal{W}_{{\rm bk}}$ & \eqref{ecriture decomposition WN en drte gche bk exp rudimentaire} & contribution of the bulk to $\mathcal{W}_{N}$ \\
$\mathcal{W}_{{{\rm bk};k}}$ & Prop.~\ref{Proposition Ecriture reguliere uniforme des divers const de WN} & the terms contributing to the latter up to $O(N^{-k\alpha})$ \ldots \\
$\Delta_{[k]}\mathcal{W}_{{\rm bk}}$ & Prop.~\ref{Proposition Ecriture reguliere uniforme des divers const de WN} & \ldots and the remainder \\

$\mathcal{W}_{{\rm bk};k}^{({\rm as})},\Delta_{[k]}\mathcal{W}_{{\rm bk};k}^{(\e{as})}$ &  Lemma~\ref{Lemme structure locale au bord pour WR et Wbk} &  putting aside exponentially small terms in the bulk operator \\
$\mathcal{W}_{{\rm exp}}$ & \eqref{ecriture decomposition WN en drte gche bk exp rudimentaire} & exponentially small contribution \\
$(\Delta_{[k]}\mathcal{W}_{N})_{R}$ & \eqref{iugfgd} & local right boundary remainder
\end{tabular}

\vspace{0.1cm}

\noindent Similar notations are used throughout Section~\ref{Section asymptotic analysis of double integrals} for the decompositions of $\mathcal{G}$ and the various $\mathfrak{I}$.

\subsection*{Riemann-Hilbert problems}

\begin{tabular}{lll}
$R(\lambda)$ & \eqref{defrtr} & reflection coefficient \\
$\kappa_N$ & \eqref{ecriture forme precise TF gN} & coefficient of $1/\lambda$ term \\
$R_{\ua/\da}(\lambda)$ & \eqref{ecriture explicite R plus}-\eqref{ecriture explicite R moins} & Wiener-Hopf factors of $R(\lambda)$ \\

$\upsilon(\la)$ & \eqref{definition fonction alpha} & related, piecewise holomorphic function \\
$\Phi$ & Lemma~\ref{Proposition corresp operateur et WH factorisation} & two-dimensional vector in correspondence with solutions of $\msc{S}_{N;\gamma}[f] = g$. \\
$\chi(\lambda)$ & Prop.~\ref{Theorem ecriture forme asymptotique matrice chi} & $2 \times 2$ matrix solution of the homogeneous Riemann--Hilbert problem with jump $G_{\chi}$ \\
$\chi^{(\e{as})}_{\ua/\da}(\lambda)$ & \eqref{777} & leading part of $\chi(\lambda)$ when $N \rightarrow \infty$ \\
$\chi_{\ua/\da}^{(\e{exp})}(\lambda)$ & \eqref{778} & exponentially small part of $\chi(\lambda)$ \\
$\chi_{k}$ & \eqref{cninfr} & matrix coefficients in the large $\la$ expansion of $\chi(\la)$ \\

$G_{\chi}$ & \eqref{definition matrice de saut G chi} & jump matrix of the Riemann--Hilbert problem of $\Phi$ and $\chi$ \\
$\Psi(\lambda)$ & \eqref{ecriture matrice Psi finale et cste cR} and Fig.~\ref{contour pour le RHP de Phi} & $2 \times 2$ matrix related to $\chi(\la)$ \\
$\Pi(\lambda)$ & \eqref{ecriture rep int matrice Pi} and Fig.~\ref{Figure definition sectionnelle de la matrice chi} & related $2 \times 2$ matrix  \\
$\Delta\Pi(\la)$ & \eqref{778} & difference between $\Pi(\lambda)$ minus identity \\
$G_{\Psi}$ & \eqref{ecriture saut Psi hors de R}-\eqref{ecriture saut Psi sur de R} &  jump matrix of the auxiliary Riemann--Hilbert problem \\
$\varkappa_{\epsilon}$ & \eqref{definition constante c epsilon de Pi} & rate of exponential decay of $G_{\Psi} - I_2$ \\
$\mathcal{R}_{\ua/\da}(\lambda)$ & \eqref{h} & some factors of the jump matrix \\
$\mathcal{R}_{\ua/\da}^{(\infty)}$ & \eqref{h2} & their non-oscillatory parts \\
$M_{\ua/\da}(\lambda)$ & \eqref{h3} & some factors of the jump matrix \\
$P_{R}(\lambda) ,P_{L;\ua/\da}(\lambda)$ & \eqref{PRla} & some factors in the auxiliary Riemann--Hilbert problem \\
$\theta_R$ & \eqref{ecriture matrice Psi finale et cste cR} & a constant involved in the auxiliary Riemann--Hilbert problem \\
$\Upsilon(\lambda)$ & \eqref{55}-\eqref{A55} & polynomial remainder in the inhomogeneous Riemann--Hilbert problem \\
$\bs{H}(\lambda)$ & \eqref{definition matrice de saut G chi}& two-dimensional vector on the right-hand side of the inhomogeneous Riemann--Hilbert problem \\
$\wh{\bs{H}}(\lambda)$ & \eqref{ecriture solution RHP vectoriel sur Hs general s negatif} & related quantity \\
\end{tabular}

\subsection*{Auxiliary functions, contours, and constants}

\begin{tabular}{lll}
$J_{k}(\lambda)$ & \eqref{Jk} & model integral appearing in the asymptotics of $\mathcal{J}_{1a}(\lambda)$ \\
$x_{R},x_{L}$ & Def.~\ref{xrxl} & reduced variables centered at the right and left boundary \\
$\Gamma_{\ua/\da}$ & Figure~\ref{Figure definition des courbes C pm reg} & contours in the upper/lower half-plane \\
$\mathscr{C}_{{\rm reg}}^{(\pm)}$ & Def.~\ref{defKKK} and Figure~\ref{Figure definition des courbes C pm reg} & contours between $\Gamma_{\ua/\da}$ and $\R$ \\
$J(x)$ & Def.~\ref{defKKK} & related to the Fourier transform of $1/R(\lambda)$ \\
$\varrho_0(x)$ & \eqref{definition fonction varrho0}-\eqref{explitrho} & proportional to a primitive of $J(x)$ \\
$\varrho_{\ell}(x)$ & Def.~\ref{oinoin} & related to higher primitives \\
$\varpi_{\ell}(x)$ & Def.~\ref{oinoin} & integrals of $x^{\ell}J(x)$ from $x$ to $\infty$ \\
$u_{\ell}$ & Def.~\ref{oinoin} & coefficients in the Taylor expansion of $1/R(\lambda)$ at $\lambda = 0$ \\
$\mathfrak{u}_{\ell}(x)$ & Def.~\ref{giu}, \eqref{u1eq} & related to the $\ell^{{\rm th}}$ order truncation of the Taylor series of $1/R$ \\
$\mathfrak{a}_{\ell}(x)$,$\mathfrak{b}_{\ell}(x)$ & Def.~\ref{giu}  & combinations of the above, involved in asymptotics of $\mc{W}_N$ \\
$\daleth_p$ & Def.~\ref{firstdalet} & negative moments of $1/R_{\da}$ \\
$\daleth_{s,\ell}$ & Def.~\ref{secondal} &  $s^{{\rm th}}$ order moment of $\mathfrak{b}_{\ell}$ \\
$\gimel_{\ell}$ & Def.~\ref{gim} & $\ell^{{\rm th}}$ order moments related to $J$ and $S$ \\
$\mathpzc{P}_{\ell}, \mathpzc{Q}_{\ell}$ & \eqref{PlQl} & some universal multivariable polynomial \\
$\mathfrak{g}_{R;\ell},\mathfrak{g}_{{\rm bk};\ell}$ & \eqref{gggg}-\eqref{gggg2} & a specialisation of the latter involving the functions above 
\end{tabular}

\subsection*{Answer for the partition function}
\begin{tabular}{lll}
$\mathfrak{I}_{{\rm s}}[H,G]$ & Def.~\ref{simpleJ} & bilinear pairing induced by $\mathcal{W}_{N}$ \\
$\mathfrak{I}_{{\rm s};\beta}^{(1)}[H,G]$ & \eqref{ididi} & related expression appearing only for $\beta \neq 1$ \\
$\mathfrak{I}_{{\rm s};\beta}^{(1)}[H,G]$ & \eqref{ididi2} & related expression appearing only for $\beta \neq 1$ \\
$\mathfrak{I}_{{\rm d}}[H,G]$ & \eqref{definition integrale double du DA correlateur a un point} & related expression \\
$\mathfrak{I}_{{\rm d};\beta}[H,G]$ & \eqref{definition integrale double N dpdt beta non egal 1} & related expression appearing only for $\beta \neq 1$ \\
$\leo[V,V_0]$ & \eqref{leoDef} & a functional appearing in the interpolation \\
$\mathfrak{c}(x)$ & Def.~\ref{ralph} & a function involving the $\mathfrak{a}$'s and $\mathfrak{b}$'s appearing in expansion of $\mathfrak{I}_{{\rm d}}$ \\
$\aleph_{0}$ & Def.~\ref{ralph} & a constant involving integrals of $J$, $S$ and $R_{\ua/\da}$, appears in expansion of $\mathfrak{I}_{{\rm d}}$ \\

\end{tabular}

\subsection*{Norms}
\begin{tabular}{lll}
$\mathcal{N}_{N}^{(\ell)}[\phi]$ & Def.~\ref{weiei} & weighted norms involving $W_{k}^{\infty}$ norms for $k \in \intn{0}{\ell}$ \\
$\mathfrak{n}_{\ell}[V]$ & Def.~\ref{weie2} & some estimates for the magnitude of potential
\end{tabular}

\subsection*{Miscellaneous}
\begin{tabular}{lll}
$q(z)$ & \eqref{sqtu} & squareroot  \\
$q_{R}(z)$ & \eqref{Rgith} & squareroot at the right boundary
\end{tabular}

\bibliographystyle{amsplain}
%\bibliography{/home/kkozlows/Documentos/integrabilite/Bibliography/bibliotemple.bib}

\providecommand{\bysame}{\leavevmode\hbox to3em{\hrulefill}\thinspace}
\providecommand{\MR}{\relax\ifhmode\unskip\space\fi MR }
\providecommand{\MRhref}[2]{%
  \href{http://www.ams.org/mathscinet-getitem?mr=#1}{#2}
}
\providecommand{\href}[2]{#2}

\end{document}